\pdfoutput=1
\documentclass[11pt,openright,oneside,letterpaper,onecolumn]{report}  

\newcommand{\thesistitle}{Jet Quenching in Relativistic Heavy Ion Collisions at the LHC}
\newcommand{\thesisauthor}{Aaron Angerami}
\newcommand{\thesisyear}{2012}

\usepackage{amsmath}
\usepackage{amssymb}
\usepackage{mathtools}
\usepackage{fancyhdr}
\usepackage{afterpage}
\usepackage{atlasphysics} 

\usepackage{HIJet}
\usepackage{slashed}
\usepackage{mathrsfs}
\usepackage{graphicx} 
\usepackage{multirow}
\usepackage{braket}
\usepackage{subfig}
\usepackage[numbers,sort&compress]{natbib}
\usepackage[pdftitle={\thesistitle},pdfauthor={\thesisauthor},pdfpagemode={UseOutlines},letterpaper,bookmarks,bookmarksopen=true,pdfstartview={FitH},bookmarksnumbered=true]{hyperref}

\newcommand{\singlespace}{\renewcommand{\baselinestretch}{1.15} \small \normalsize}

\newcommand{\doublespace}{\renewcommand{\baselinestretch}{1.5} \small \normalsize}
\newcommand{\normalspace}{\doublespace}
\newcommand{\thesistitlepage}{
    \normalspace
    \thispagestyle{empty}
    \begin{center}
        \textbf{\LARGE \thesistitle} \\[1cm]
        \textbf{\LARGE \thesisauthor} \\[8cm]
        Submitted in partial fulfillment of the \\
        requirements for the degree \\
        of Doctor of Philosophy \\
        in the Graduate School of Arts and Sciences \\[4cm]
        \textbf{\Large COLUMBIA UNIVERSITY} \\[5mm]
        \thesisyear
    \end{center}
    \clearpage
}

\newcommand{\thesiscopyrightpage}{
    \thispagestyle{empty}
    \strut \vfill
    \begin{center}
      \copyright \thesisyear \\
      \thesisauthor \\
      All Rights Reserved
    \end{center}
    \cleardoublepage
}

\begin{document}

\pagestyle{empty}

\thesistitlepage
\thesiscopyrightpage

    \thispagestyle{empty}
    \begin{center}
    \textbf{\LARGE ABSTRACT} \\[1cm]
     \textbf{\LARGE \thesistitle} \\[1cm]
     \textbf{\LARGE \thesisauthor} \\[1cm]
    \end{center}
    Jet production in relativistic heavy ion collisions is studied using
\PbPb\ collisions at a center of mass energy of 2.76~\TeV\ per
nucleon. The measurements reported here utilize data collected with
the ATLAS detector at the LHC from the 2010 Pb ion run corresponding
to a total integrated luminosity of $7~\mu\mathrm{b}^{-1}$. The
results are obtained using fully reconstructed jets using the
anti-\kt\ algorithm with a per-event background subtraction
procedure. A centrality-dependent modification of the dijet asymmetry distribution is
observed, which indicates a higher rate of asymmetric dijet pairs
in central collisions relative to peripheral and \pp\ collisions. Simultaneously the dijet
angular correlations show almost no centrality dependence. These results provide the first direct observation of
jet quenching. Measurements of the single inclusive jet spectrum, measured
with jet radius parameters $R=0.2, 0.3, 0.4$ and 0.5, are also presented. The spectra are unfolded to
correct for the finite energy resolution introduced by both detector
effects and underlying event fluctuations. Single jet production,
through the central-to-peripheral ratio \Rcp, is found to be
suppressed in central collisions by approximately a factor of two,
nearly independent of the jet \pt. The \Rcp\ is
found to have a small but significant increase with increasing
$R$, which may relate directly to aspects of radiative energy loss.
    \cleardoublepage


\pagenumbering{roman}
\pagestyle{plain}

\setlength{\footskip}{0.5in}

\setcounter{tocdepth}{3}
\renewcommand{\contentsname}{Table of Contents}
\tableofcontents
\cleardoublepage


\listoffigures
\cleardoublepage

\listoftables 
\cleardoublepage

~\\[1in] 
\textbf{\Huge Acknowledgments}\\

\noindent 
I would like to thank CERN and my collaborators in ATLAS for
successful operation of the LHC and data taking with our detector. 
I am indebted my thesis committee -- Gustaaf Brooijmans, Jianwei 
Qiu, Jamie Nagle, Miklos Gyulassy and Brian Cole -- for their
participation in my most significant educational experience. I would
also like to thank the various faculty of the Physics Department 
at Columbia who have taught me and who have been extremely
accessible to curious students. I would like to thank my high school
physics teacher, Mr. Arnold, for introducing me to physics. I firmly believe
that you learn the most from your peers, and I would like to thank the
many physics students from whom I have had a chance to learn
something. Special thanks are due for my classmates and friends -- 
Anshuman Sahoo, Rob Spinella, David Tam and Fabio Dominguez.

The story of the heavy ion program in ATLAS is very much a story of an
underdog. I am grateful to Professors Zajc and Hughes for their
unwavering support of this effort, especially when things looked bleak. I would like to thank
Martin Spousta and Martin Rybar, my teammates working on
jets through this whole experience. I owe a special thanks to Peter Steinberg for providing me with guidance
during my time as a graduate student and for his commitment to the
ATLAS heavy ion effort; any success from this effort would not have been
possible without his extraordinary determination. I owe a great debt of
gratitude to my thesis adviser Brian Cole, who took a chance on
mentoring a precocious student many years ago. He has continually
inspired me with his extraordinary commitment to science. From him, not only have
I learned a great deal of physics, but more importantly I have learned
what it means to be a physicist.

I have the benefit of having a large family that has provided me with
unconditional support and encouragement for as long as I can
remember. My grandparents have always shown so
much pride in my accomplishments; seeing this pride has brought me great
joy. I would like to thank my dog Tito for being unconditional and
uncomplicated. My brother Matteo has supported me by showing a fierce sense of
loyalty and pride for his big brother. My parents have always made
sure I had what I needed to pursue my dreams, from 
indulging my obsessions and interests when I was young, to building
my sense of self-belief by radiating pride in my achievements. Lastly, I want to thank my girlfriend Shaina Rubin for supporting me
throughout this challenging time in my life. It is not easy living with a
physicist, and she has been here with me every day through the ups and
downs, embracing my idiosyncrasies and month-long trips to
CERN. Because of her presence in my life, I have been able to achieve far
more than I thought possible.

\cleardoublepage

\thispagestyle{plain}
\strut \vfill
\centerline{\LARGE 
For Vincent Angerami,\newline my grandfather and my biggest fan.
}
\vfill \strut
\cleardoublepage


\pagestyle{headings}
\pagenumbering{arabic}

%
%
\setlength{\textheight}{8.5in}
\setlength{\footskip}{0in}

\fancypagestyle{plain} {%
\fancyhf{}
\fancyhead[LE,RO]{\thepage}
\fancyhead[RE,LO]{\itshape \leftmark}
\renewcommand{\headrulewidth}{0pt}
}
\pagestyle{plain}

\chapter{Introduction}
\label{section:intro}
The subject of this thesis is a series of experimental measurements of jets in relativistic heavy ion collisions which are related to the phenomenon of jet quenching. The high energy densities created in these collisions allow for the formation of a hot, evanescent medium of deconfined quarks and gluons termed the quark-gluon plasma (QGP). The QGP is a fundamental physical system described by quantum chromodynamics (QCD), in which degrees of freedom of the QCD Lagrangian are in principle manifest. The transition from ordinary nuclear matter to the QGP phase may provide critical insight into the dynamics of the early Universe. Experimental searches of such a medium in nucleus-nucleus collisions have been made at the AGS, SPS, RHIC and most recently in \PbPb\ collisions at the LHC. Measurements of collective phenomena such as elliptic flow have led to the interpretation that the produced medium rapidly equilibrates after which it is well described by near-ideal hydrodynamics. Highly collimated clusters of particles produced from hard partonic scatterings known as jets are produced in these collisions. Jets, which are not in thermal equilibrium with the rest of the plasma, have long been thought to be sensitive to detailed medium properties as they provide an external probe of the system.  The process by which a quark or gluon loses energy and suffers a modification of its parton shower in a medium of high color charge density is referred to as jet quenching.

The measurement of the suppression of high transverse momentum (\pt) hadrons and the structure of dihadron correlations at RHIC provided indirect experimental evidence for jet quenching and was one of the early successes of the RHIC program. These measurements indicated that the formulation of QCD factorization used in the calculation of hard scattering rates in hadronic collisions could not be applied to nuclear collisions, implying that dynamics at new scales are an essential feature of the medium. However, the utility of these results was limited as single particle observables do not provide the full details of the jet-medium interaction. To relate partonic energy loss to single particle measurements a fragmentation function must be applied. This procedure loses sensitivity to the angular pattern of medium-induced radiation and forces the assumption of vacuum fragmentation. Furthermore, the limited rapidity coverage and low rate at RHIC for jets clearly above the underlying event fluctuations prevented more detailed conclusions from being drawn about the quenching mechanism. 

\begin{figure}[tbh]
\centering
\includegraphics[width=0.99\textwidth]{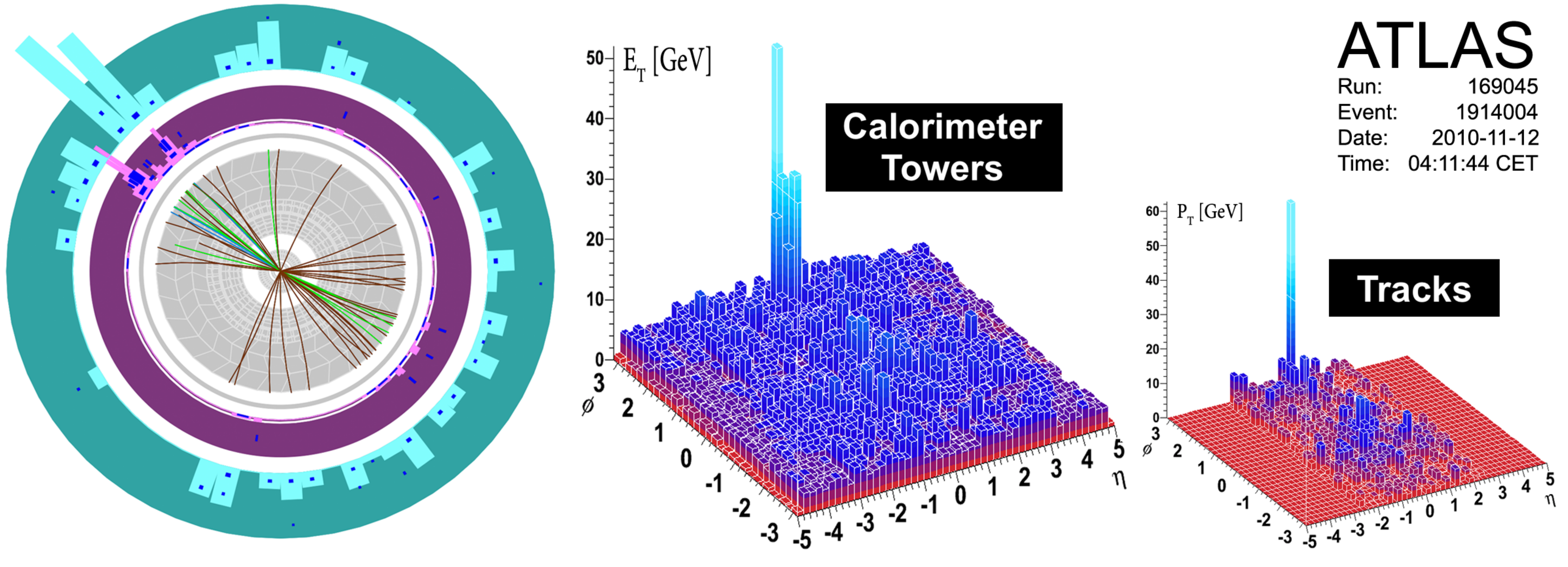}
\caption{A highly asymmetric dijet event recorded by ATLAS during the early portion of the 2010 \PbPb\ run. A view along the beam axis is shown on the left, with high-\pt\ charged particle tracks indicated by lines and energy deposits in the calorimeter by colored bars. The center figure shows the calorimeter \et\ distribution in $\eta-\phi$, indicating a highly energetic jet with no balancing jet opposite in azimuth. A similar distribution is shown (right) for the inner detector charged particle \pt, which is consistent with the calorimeter signal.}
\label{fig:intro:event_display}
\end{figure}

More recently, fully reconstructed jets, which carry the full information of the partonic energy loss, have been measured at the LHC. Dijet events were observed to become increasingly asymmetric in more central collisions, while retaining an angular correlation consistent with back-to-back jets. This result~\cite{Aad:2010bu} has generated a great deal of excitement, as it strongly suggests the interpretation that back-to-back jets traverse different path lengths in the medium resulting in events with large dijet asymmetry where one jet leaves the medium mostly unscathed while the other experiences significant quenching. An event display of a highly asymmetric event is shown in Fig.~\ref{fig:intro:event_display}. This observation is the first direct evidence for jet quenching and is one of two results presented in this thesis. 

The asymmetry is sensitive to the energy loss of one parton relative to one another, and additional insight into the effects of quenching can be provided by observables that are sensitive to the quenching on a per jet basis. Energy loss of single partons could result in jets emerging from the medium with significantly less energy, resulting in a different final \pt\ spectrum for events in which the medium is produced. Due to the steeply falling nature of the unquenched \pt\ spectrum, a systematic downward shift in jet energies due to quenching causes
a reduction in the total number of jets at a given \pt. The second result presented in this thesis is a measurement inclusive production of single jets. Ratios of jet spectra in different centrality intervals relative to a peripheral reference were used to construct an observable, \Rcp, quantifying the extent to which jet production is suppressed by the presence of the QGP through quenching. The \Rcp\ has been measured as a function of jet \pt\ and size as well as collision centrality, which should provide the ability to distinguish between different models of the quenching mechanism.

This thesis is organized as follows. Chapter~\ref{section:background} provides background on both experimental and theoretical developments leading to the conclusion that QCD is the correct theory to describe the strong nuclear force. These ideas are developed to motivate the use of heavy ion collisions as an experimental tool, and various formulations of the jet quenching problem are presented. Chapter~\ref{section:detector} describes the experimental apparatuses, the LHC and the ATLAS detector, used to provide the measurements presented in this thesis. Chapter~\ref{section:jet_rec} discusses the experimental techniques used to separate underlying event contributions from the jet signal and perform jet energy scale calibrations. Analysis of reconstructed jets is discussed in Chapter~\ref{section:analysis}, including unfolding of the jet spectrum to account for detector effects and estimates of systematic uncertainties. The final results of this analysis, the jet \Rcp\ and \dijet\ asymmetry, are presented in Chapter~\ref{section:results} with conclusions following in Chapter~\ref{section:conclusion}.

\clearpage
\chapter{Background}
\label{section:background}
Nuclear physics deals with the study of the fundamental
properties of nuclear matter and the character of the strong
nuclear force. By the 1960's the concept of nuclear matter had been extended
from the nucleons making up atomic nuclei to a vast number of strongly
interacting particles known as hadrons. The flavor symmetries among
various types of hadrons suggested that these particles are composite
particles with their quantum numbers carried by common, sub-hadronic
objects, and that these objects possessed a new quantum number known as
color~\cite{Han:1965pf}. A group theoretic approach to hadron taxonomy led to the
development of the Quark Model~\cite{GellMann:1962xb,GellMann:1964nj,Kokkedee:936332,Zweig:352337,Zweig:570209}
which predicted the $\Omega^{-}$ baryon observed at BNL in
1964~\cite{Barnes:1964pd}.

Developments in current algebra led to a proposed scaling behavior of the structure functions in deep inelastic
scattering. Bjorken Scaling~\cite{Bjorken:1968dy} implies that at high
momentum transfer, $Q^2$, the hadronic structure functions are independent of
$Q^2$ and functions of a single scaling variable, $x$. Scaling behavior consistent with this relation was observed in a series of
measurements in deep inelastic scattering (DIS) at SLAC~\cite{Bloom:1969kc,Friedman:1972sy}, leading to the
interpretation of point-like constituents within the
nucleon known as the Parton
Model~\cite{Bjorken:1969ja,Feynman:1969ej,Feynman:1969wa}.

Advances in
renormalization group
techniques~\cite{GellMann:1954fq,Wilson:1969zs,Callan:1970yg,Symanzik:1970rt}
led to the conclusion that only a theory possessing asymptotic freedom, a
coupling strength that becomes arbitrarily weak at high energies, could
be consistent Bjorken Scaling~\cite{Callan:1973pu,Christ:1972ms}. The symmetry considerations of the Quark
Model suggested that non-Abelian gauge theories~\cite{Yang:1954ek} would be potential
candidates for such a theory. These theories were found to possess the
required renormalization criteria with the $\beta$-function in the
calculation of the running of the coupling being negative for an SU(3)
gauge symmetry~\cite{Politzer:1973fx,Gross:1973id}. Furthermore, explicit
calculations of the anomalous dimensions of the DIS structure
functions showed that these non-Abelian gauge theories possessed scaling
properties consistent with experimental observations ~\cite{Gross:1973ju,Gross:1974cs,Politzer:1974fr,Georgi:1951sr}.

This description of the strong interaction as an SU(3) (color)
gauge symmetry between spin-$\dfrac{1}{2}$ quarks and mediated by
massless spin-1 gluons is known as Quantum Chromodynamics
(QCD). At short distance scales, asymptotic freedom ensures that the coupling becomes
weak and perturbative techniques can be used to calculate
observables. The validity of QCD as the correct theory of the strong
interaction has been rigorously demonstrated by a variety of
experimental results that are well described by perturbative
calculations. A summary of this evidence is presented in Section~\ref{section:bkgr:QCD_support}.

Color charges are never observed as isolated particles but rather as
constituents of color neutral hadrons, an aspect of QCD known as
confinement. Furthermore, the QCD Lagrangian possess chiral symmetry
which is dynamically broken through the interactions that bind the
quarks into hadrons and is responsible for the vast majority of the
mass of ordinary matter\footnote{This symmetry is only approximate as it
  is broken explicitly by the quark masses. Thus the pions
  are pseudo-Goldstone bosons.}. However, at soft momentum scales the
coupling becomes increasingly large and at hadronic energies the
theory is non-perturbative. This leads to significant challenges in
theoretical calculations and interpreting experimental results, and
much of the work done since the inception of QCD has been to navigate
these challenges to develop a comprehensive picture of the strong interaction.

\section{Experimental Evidence for QCD}
\label{section:bkgr:QCD_support}
\begin{figure}[htb]
\centering
\includegraphics[width=0.5\textwidth]{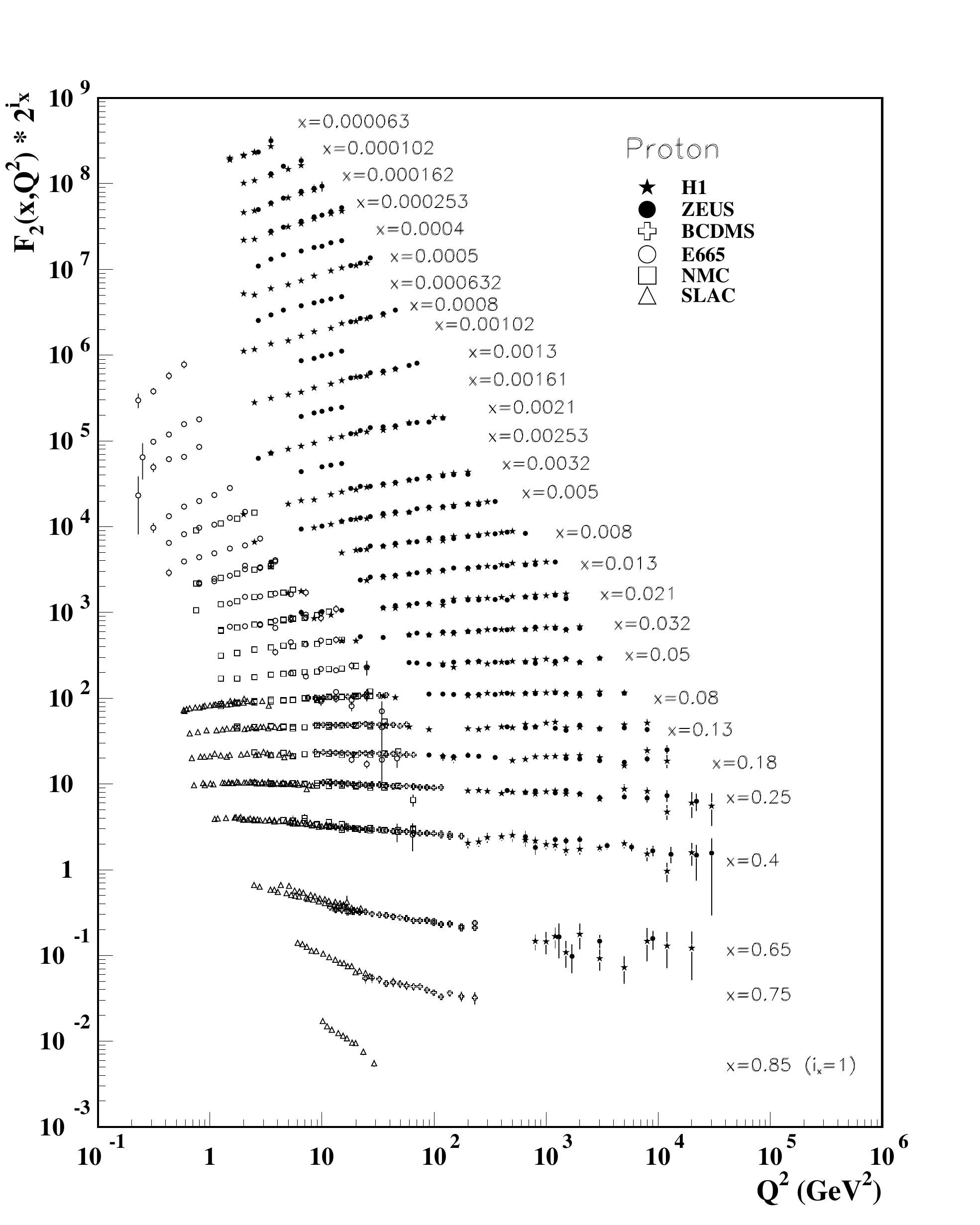}
\caption{The proton structure function $F_2$ vs $Q^2$ at fixed $x$
  measured by a variety of experiments (see
  \protect\cite{Nakamura:2010zzi} and references therein). $F_2$ is
  approximately independent of $Q^2$. The observed logarithmic violations of this
  scaling are well-described by the QCD phenomenon of parton evolution.}
\label{fig:bkgr:f2}
\end{figure}

Initial experimental evidence for QCD as a theory of the
strong interaction was provided in DIS through the observation of Bjorken
Scaling. Measurements of the proton structure function $F_2$ are shown
in Fig.~\ref{fig:bkgr:f2}, and DIS structure functions in the context of
the Parton Model and collinear factorization are further
discussed in Section~\ref{section:bkgr:pQCD}. Subsequent tests of the theory have been performed in
\ee\ collisions and hadronic collisions, particularly through the Drell-Yan process, the inclusive production of lepton
pairs in hadronic scatterings, $A+B\rightarrow
\ll+X$~\cite{Drell:1970wh}. These results have provided overwhelming
experimental support for QCD~\cite{Brock:1994er,Nakamura:2010zzi}, and
a brief survey is discussed below.

\begin{figure}[htb]
\centering
\includegraphics[width=0.5\textwidth]{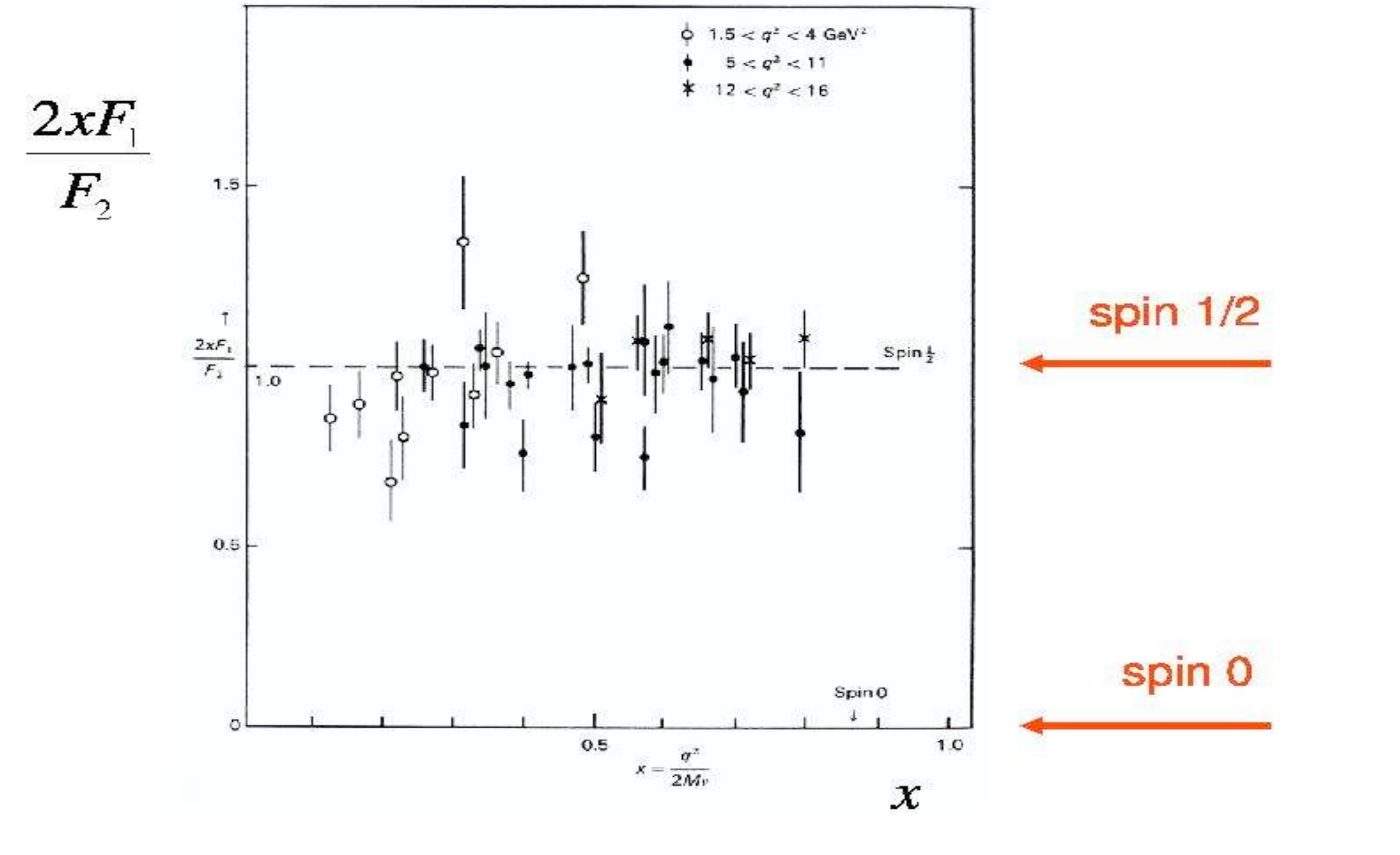}
\caption{The ratio $2xF_1/F_2$ as a function of
  $x$~\protect\cite{Bodek:1979rx}. The ratio is approximately
  independent of $x$, consistent with the Callan-Gross relation, which is the result of scattering
  off of spin-$\dfrac{1}{2}$ constituents within the proton.}
\label{fig:bkgr:callan_gross}
\end{figure}

\begin{figure}[htb]
\centering
\includegraphics[width=0.7\textwidth]{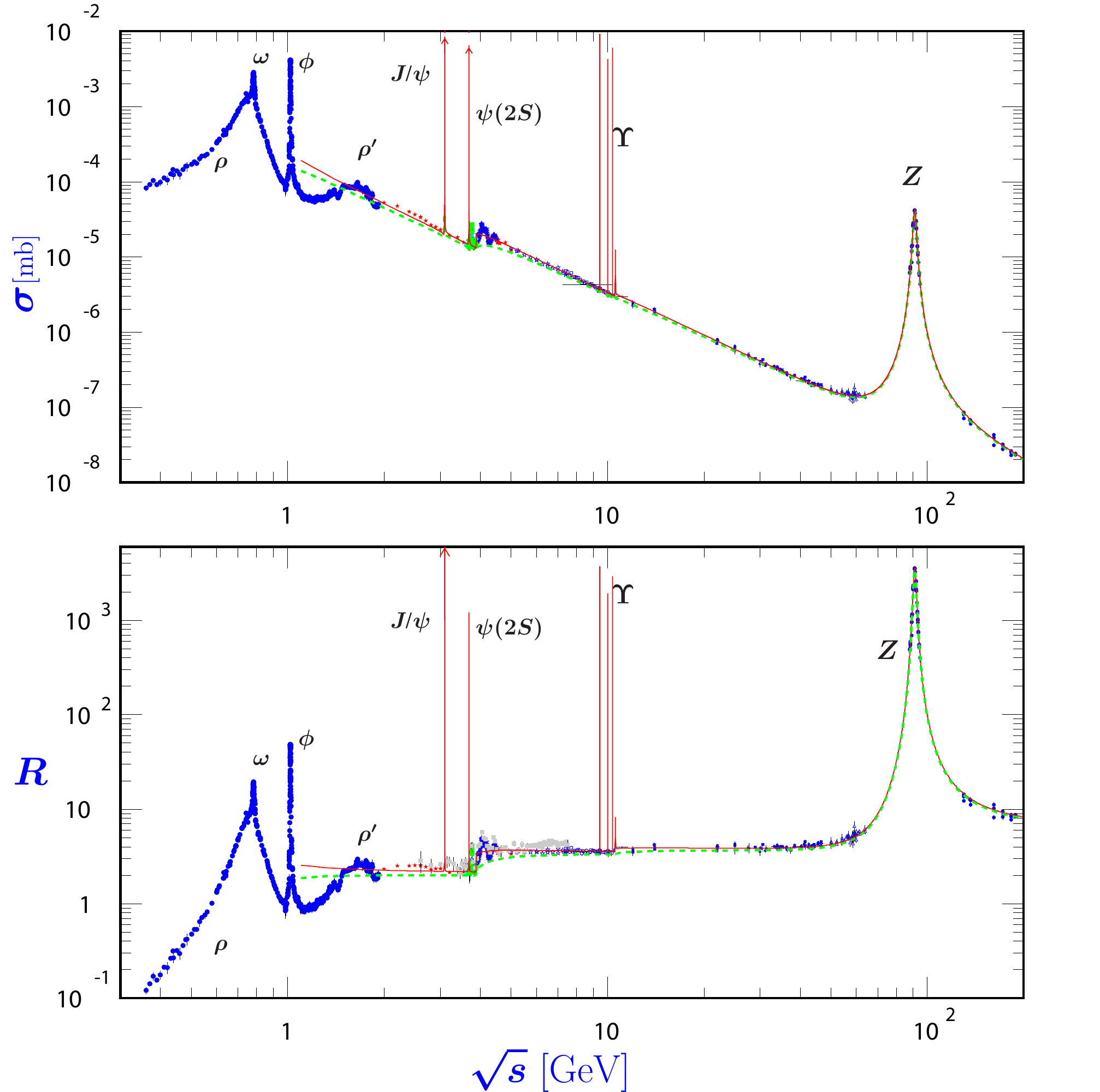}
\caption{World data on inclusive $\ee\rightarrow hadrons$ cross
section (top) and ratio to muon pair production cross
sections (bottom) \protect\cite{Nakamura:2010zzi,Ezhela:2003pp}.}
\label{fig:bkgr:cross_section_ratio}
\end{figure}

The Callan-Gross~\cite{Callan:1969uq} relation between the DIS structure
functions, $2xF_1=F_2$, arises due to scattering off
spin-$\dfrac{1}{2}$, point-like objects; the measured behavior of
the structure function, shown in Fig.~\ref{fig:bkgr:callan_gross}, is
approximately consistent with this relation. 
In \ee\ collisions the ratio of cross sections for the production of
inclusive hadrons to muon pairs is given to leading
order in QCD by
\begin{equation}
R(s)=\dfrac{\sigma(\ee\rightarrow hadrons,s)}{\sigma(\ee\rightarrow
\mumu,s)}=N_c\displaystyle\sum_{f}^{N_q}Q_f^{2},
\end{equation}
where $s$ is the center of mass energy, $N_c$ is the number of colors
($N_c=3$ for QCD),
and $Q_f$ is the charge of quark $f$. The sum runs over the quark
flavors up to the heaviest quark capable of being produced at that
$s$. This ratio, aside from resonance peaks and away from the \Zzero\ threshold, is measured to be
approximately constant with discrete jumps at the quark mass
thresholds as shown in Fig.~\ref{fig:bkgr:cross_section_ratio}.

\begin{figure}[htb]
\centering
\includegraphics[width=0.4\textwidth]{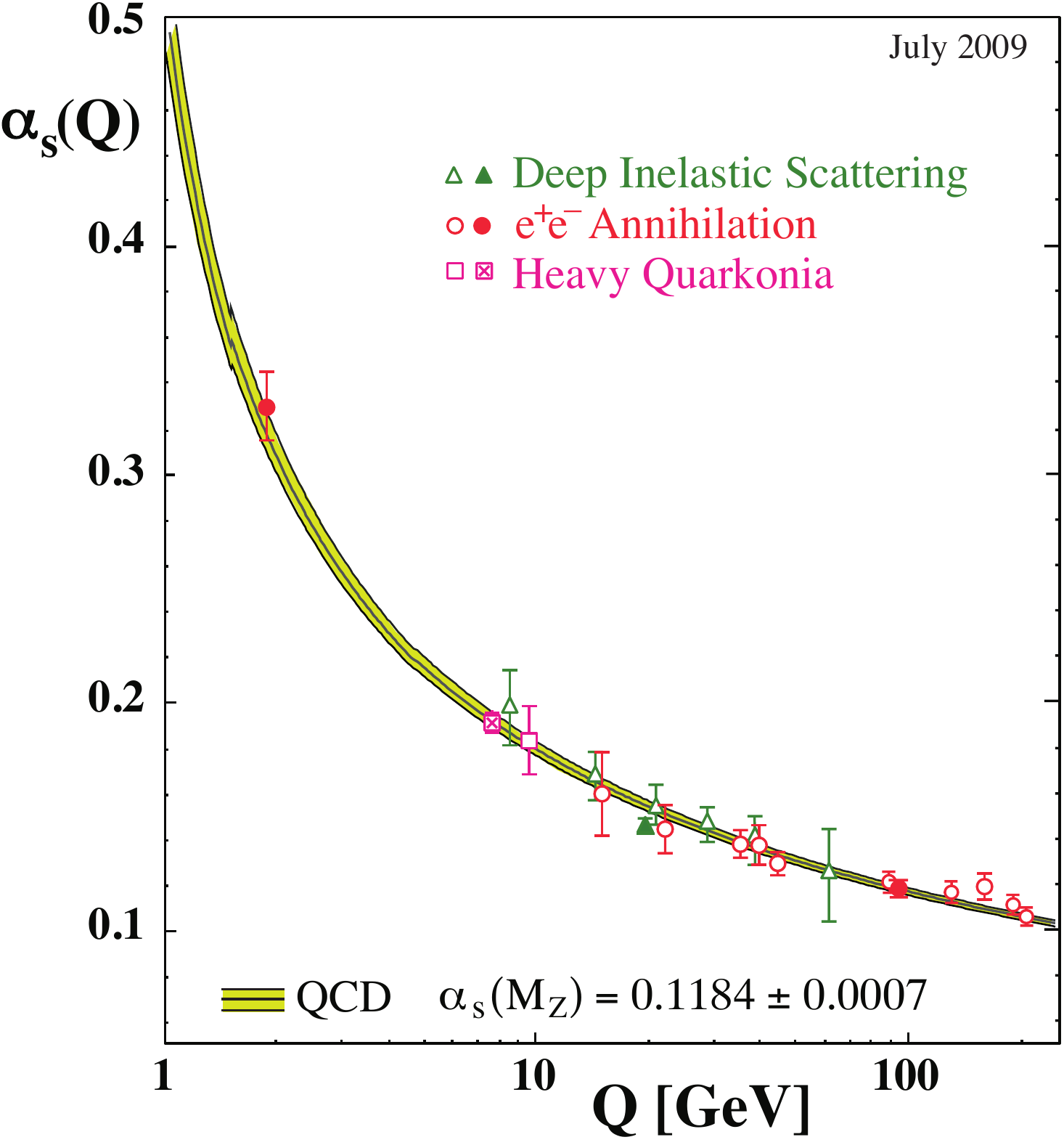}
\includegraphics[width=0.4\textwidth]{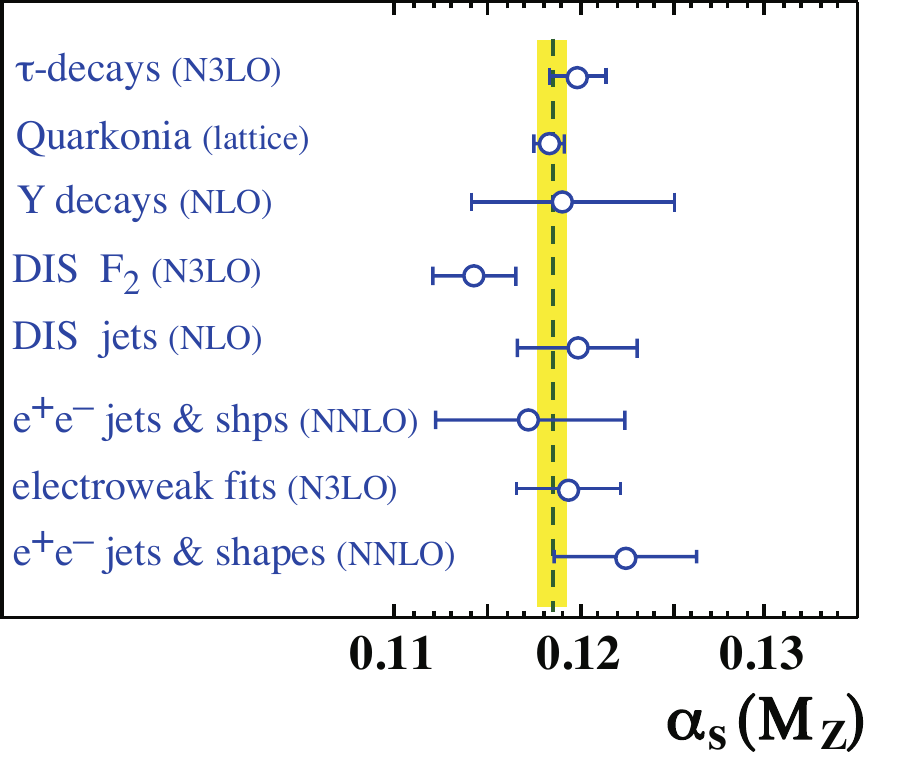}
\caption{The strong coupling constant, \alphas, as a function of $Q^2$
  (left) and fixed at $Q^2=\mZ^2$ (right). Figure adapted from
  Ref.~\protect\cite{Bethke:2009jm}. From these results, \alphas\ was
  determined to have a world average of $\alphas(\mZ)=0.1184\pm0.0007$.}
\label{fig:bkgr:alpha_s_summary}
\end{figure}

Highly collimated, energetic sprays of particles balanced in momentum known as jets were
observed in \ee\ collisions at SLAC~\cite{Hanson:1975fe},
consistent with the hypothesis that hadrons are produced in these
collisions through the production of \qqbar\ pairs of opposite
momentum. The observation of three jet events at
PETRA through the process $\ee\rightarrow \qqbar g$
provided experimental evidence for the gluon as well as measurements
of the strong coupling constant $\alphas$~\cite{Wu:1984ik}. Further precision measurements of the strong coupling constant have shown
remarkable consistency~\cite{Bethke:2009jm} and a summary is shown in
Fig.~\ref{fig:bkgr:alpha_s_summary}.

\section{Fundamentals of QCD}
\label{section:bkgr:basic_QCD}
QCD is the quantum field theory that describes the strong nuclear
force, and is expressed
in terms of quarks and gluons which contain color charge described by
an SU(3) gauge symmetry. The spin-$\dfrac{1}{2}$ quarks fields,
$\psi_{a}$, transform under the fundamental, three-dimensional,
representation of the gauge group,
\begin{equation}
\psi^{\prime}_a(x) =e^{i\alpha^{C}(x)t^C_{ab}}\psi_b(x)=U_{ab}(x)\psi_b(x)\,,
\end{equation}
with the subscript $a$ describing the
quark color.
The $t^C_{ab}$ are the generators of the gauge group, represented by eight 3$\times$3 matrices ($C=1,\cdots 8$, $a,b=1,2,3$), and obey the Lie Algebra \footnote{In this section the generator/adjoint indices
are denoted by capital, Roman script, as a superscript with the summation over repeated
indices is implied. The fundamental, matrix multiplication indices
are denoted by lower case, Roman script appearing as subscripts. The
Lorentz indices are indicated by Greek characters and the standard
Einstein summation convention applies.}
\begin{equation}
[t^A,t^B]=f^{ABC}t^C\,,
\end{equation}
with $f^{ABC}$ the structure constants of SU(3).
The local gauge symmetry is imposed by replacing the
derivatives appearing in the usual Dirac Lagrangian
with the
gauge-covariant derivative, $D_{\mu,\,ab}$,
\begin{equation}
D_{\mu,\,ab}=\partial_{\mu}\delta_{ab}+i g A_{\mu}^Ct^C_{ab}\,.
\end{equation}
The color indices $a$
and $b$ on $D$ are explicit here to indicate that the covariant derivative is Lie
Algebra-valued (i.e. matrix-valued).
The covariant derivative introduces a
non-Abelian coupling, of strength $g$, between the
quarks and massless, spin-1 gauge fields $A_{\mu}^{C}$ representing
gluons. These transform under the adjoint, eight-dimensional, representation
of the gauge group.
Including the pure Yang-Mills action for the gauge fields, the full
QCD Lagrangian is
given by
\begin{equation}
\mathcal{L}=
\displaystyle\sum_q
\bar{\psi}_q(i\gamma^{\mu}D_{\mu}-m_q)\bar{\psi}_q
-\dfrac{1}{4}F_{\mu\nu}^AF^{A\,\mu\nu},
\label{eqn:bkgr:QCD_lagrangian}
\end{equation}
with the index $q$ denoting the quark flavor with mass $m_q$.
The field strength tensor $F_{\mu\nu}^A$ is expressed in terms of the
gauge field via
\begin{equation}
F_{\mu\nu}^A=\partial_{\mu}A_{\nu}^A-\partial_{\nu}A_{\mu}^A-gf^{ABC}A_{\mu}^BA_{\nu}^C\,.
\end{equation}

\begin{figure}[htb]
\centering
\includegraphics[width=0.9\textwidth]{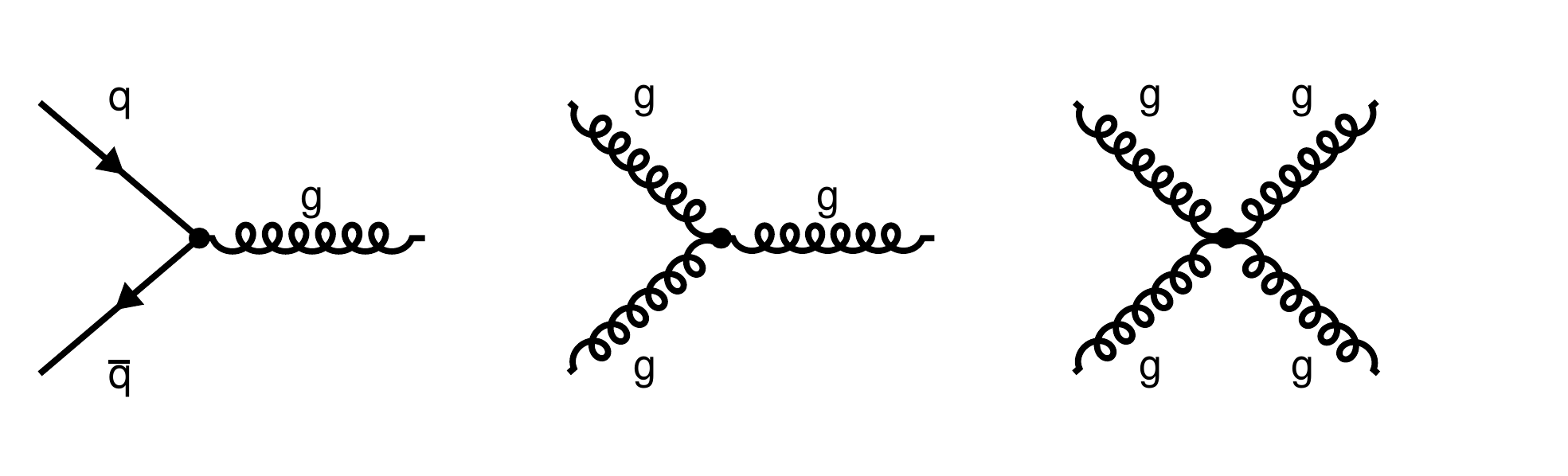}
\caption{Possible couplings allowed by the QCD Lagrangian. The
quark-gluon (left) and three-gluon (center) vertices are proportional
to $g$. The four-gluon vertex is proportional to $g^2$.}
\label{fig:bkgr:QCD_vertices}
\end{figure}

The QCD vertices are shown in
Fig.~\ref{fig:bkgr:QCD_vertices}. In addition to the analog of the QED
gauge coupling shown on the left, the non-Abelian structure of the
theory allows for coupling between the gauge fields themselves.
Color factors, associated with different processes, arise from the
group structure in the Feynman rules,
\begin{equation}
t^A_{cd}t^B_{cd}=T_F\delta^{AB},\,\,\,\,\,
f^{ACD}f^{BCD}=C_A\delta^{AB},\,\,\,\,\,
t^A_{bc}t^A_{cd}=C_F\delta_{bd}.
\end{equation}
In QCD, these have values 
\begin{equation}
T_F=\dfrac{1}{2},\,\,\,\,\,
C_A=3,\,\,\,\,\,
C_F=4/3,\,\,\,\,\,
\label{eqn:bkgr:color_factors}
\end{equation}
and are typically associated
with the splittings $g\rightarrow q\bar{q}$, $g\rightarrow gg$ and $q\rightarrow
qg$ respectively.

The quantization of this theory is typically performed using the path
integral formalism. In this technique, the gauge-fixing condition
imposes a constraint on the functional integral which can be removed
by the introduction of Faddeev-Popov ghost
fields~\cite{Faddeev:1967fc}. These unphysical fields appear as an
artificial mechanism for preserving gauge invariance. Their
contributions, and the contribution of the gluon propagator, will
depend on the choice of gauge, with this dependence dropping out in the
computation of any gauge invariant quantity.
\subsection{Ultraviolet Behavior}
\label{section:bkgr:uv_qcd}
The ultraviolet (UV) behavior of QCD was the main reason it was originally
proposed as the theory of the strong interaction. The relationship among the UV
divergences is encoded in the Callan-Symanzik equation. In particular,
the derivative of the coupling constant with respect to the
renormalization scale, $\mu$, is defined by the $\beta$-function,
\begin{equation}
\dfrac{\partial g}{\partial \ln\mu}=\beta(\mu)\,.
\end{equation}
The one-loop $\beta$-function for SU($N$) non-Abelian gauge theories was first
computed by Wilczek, Gross and Politzer in
1973~\cite{Politzer:1973fx,Gross:1973id}. For $N=3$ it is given by
\begin{equation}
\beta(\alphas)=-g\left(\dfrac{\alphas}{4\pi}\beta_1+\left(\dfrac{\alphas}{4\pi}\right)^2\beta_2+\cdots\right),\,
\end{equation}
with $\alphas=g^2/4\pi$. The one and two-loop coefficients are
\begin{equation}
\beta_1=11-\dfrac{2}{3}N_{f}\,,\,\,\,\,\,\,\,
\beta_2=102-\dfrac{38}{3}N_{f}\,.
\end{equation}
To one-loop, the solution is
\begin{eqnarray}
&\alphas(\mu)=\dfrac{\alphas(\mu_0^2)}{1+\dfrac{\beta_1}{4\pi}\alphas(\mu_0^2)\ln(\mu^2/\mu_0^2)}=\dfrac{4\pi}{\beta_1\ln(\mu^2/\Lambda_{\mathrm{QCD}}^2)}\\ \nonumber
&\Lambda_{\mathrm{QCD}}=\mu_0 e^{-2\pi/(\beta_1\alphas(\mu_0^2))}\,,
\end{eqnarray}
where $\Lambda_{\mathrm{QCD}}$ is used to set the scale of the strong
coupling. To higher order in $\ln\mu^2/\Lambda$ this is
\begin{equation}
\alphas(\mu)=\dfrac{4\pi}{\beta_1\ln(\mu^2/\Lambda^2)}-
\dfrac{\beta_2\ln[\ln(\mu^2/\Lambda^2)]}{\beta_1^3\ln^2(\mu^2/\Lambda^2)}+
\mathcal{O}\left((
\dfrac{1}{\beta_1^3\ln^2(\mu^2/\Lambda^2)}
\right)\,,
\end{equation}
where the definition of $\Lambda$ becomes dependent on the
renormalization scheme. The result above is for the
$\overline{\mathrm{MS}}$ scheme~\cite{Brock:1994er}. As the renormalization scale is increased
the coupling decreases, as indicated by the leading $-$ sign in the
$\beta$-function. This property is asymptotic freedom. For general
SU($N$), $\beta_1=\dfrac{1}{3}(11N_{c}-2N_{f})$. The
leading, positive term is due to the gluons, which reduce the coupling
at large $\mu$. There is competition, which weakens the asymptotic freedom, from the second term which is due
to fermion loops and is proportional to the number of flavors.
\subsection{Infrared Behavior}
\label{section:bkgr:ir_qcd}
Although QCD is well behaved at large momenta, the infrared
behavior must be handled with care in perturbative calculations. Since gluons (and to a good approximation the
light quarks) are massless, any sensitivity to the long range behavior
of QCD appears in perturbation theory as an infrared
divergence. Fortunately, the configurations that introduce this
sensitivity have been systematically analyzed and a formal procedure
exists to define quantities which are infrared
safe~\cite{Sterman:1977wj,Dokshitzer:1978hw,Sterman:1978bi}. These divergences are
associated with the contributions to the momentum integration where
the massless particle has zero momentum (soft) or at zero angle
(collinear). Observables that are safe from these divergences must be
insensitive with respect to the emission of an additional soft or
collinear gluon. Formally, this means that for an inclusive quantity
$\mathcal{I}$, defined by the functions $\mathcal{S}_n$,
\begin{equation}
\mathcal{I}=\dfrac{1}{2!}\int d\Omega_2 \dfrac{d\sigma_2}{d\Omega_2}
\mathcal{S}_2(p_1^{\mu},p_2^{\mu})+
\dfrac{1}{3!}\int d\Omega_3 \dfrac{d\sigma_3}{d\Omega_3} \mathcal{S}_3(p_1^{\mu},p_2^{\mu},p_3^{\mu})+\cdots,
\end{equation}
the quantity is said to be infrared and collinear (IRC) safe if
\begin{equation}
\mathcal{S}_{n+1}(p_1^{\mu},p_2^{\mu},\cdots,(1-\lambda)p_n^{\mu},\lambda
p_{n+1}^{\mu})=\mathcal{S}_{n}(p_1^{\mu},p_2^{\mu},\cdots,p_n^{\mu}).
\end{equation}
This idea is expanded in the discussion of applications of
perturbative QCD in the following section.
\subsection{Non-Perturbative Dynamics}
\label{section:bkgr:np_qcd}
\begin{figure}[htbp]
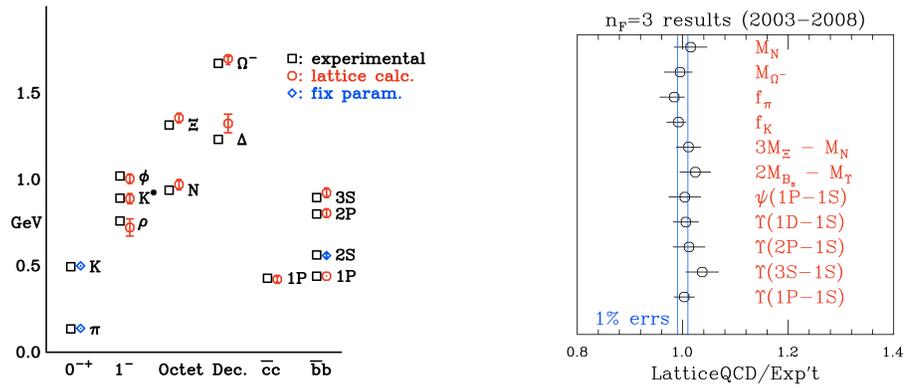

\centering
\includegraphics[width=0.4\textwidth]{background/lattice_hadron_spectra.pdf}
\includegraphics[width=0.4\textwidth]{background/lattice_hadron_spectra_data_theory.pdf}
\caption{Lattice calculation of the hadron spectrum with comparison to
  measured values (left) and ratios (right)\protect~\cite{Bazavov:2009bb}.}
\label{fig:bkgr:lattice_hadrons}
\end{figure}

Inspired by Wilson's picture of renormalization, a discrete lattice
formulation of QCD was developed to supplement perturbation theory
where the latter is inapplicable~\cite{Wilson:1974sk}. This procedure
uses the path integral formalism in Euclidean space-time, where the
field configurations are explicitly integrated over. The integration is
accomplished by recasting the quantity of interest in terms of gauge links between
adjacent lattice sites. The finite
lattice spacing, $a$, serves
as an ultraviolet cutoff, and a valid lattice formulation must respect all
of the features of QCD in the limit $a\rightarrow0$. Technical
challenges arise as spurious fermion states can be produced in this
limit as lattice artifacts~\cite{Nielsen:1980rz,Nielsen:1981hk}. This
can be mitigated, although not completely removed, while maintaining
chiral symmetry, by introducing staggered
fermions~\cite{Kogut:1974ag}. Furthermore, the finite spacing
introduces integration errors in the evaluation of the action. These can be minimized by using
improved definitions of the action~\cite{Symanzik:1983dc}. Advances in
computing power and improved actions have led to highly accurate
calculations, such as decay constants, form factors and the spectrum
of hadrons, which is shown in Fig.~\ref{fig:bkgr:lattice_hadrons}. Additionally, the
Euclidean path integral formulation is amenable to calculating
thermodynamic quantities in QCD, a topic discussed in Section~\ref{section:bkgr:lattice_thermo}.

\begin{figure}[htbp]
\centering
\includegraphics[width=0.45\textwidth]{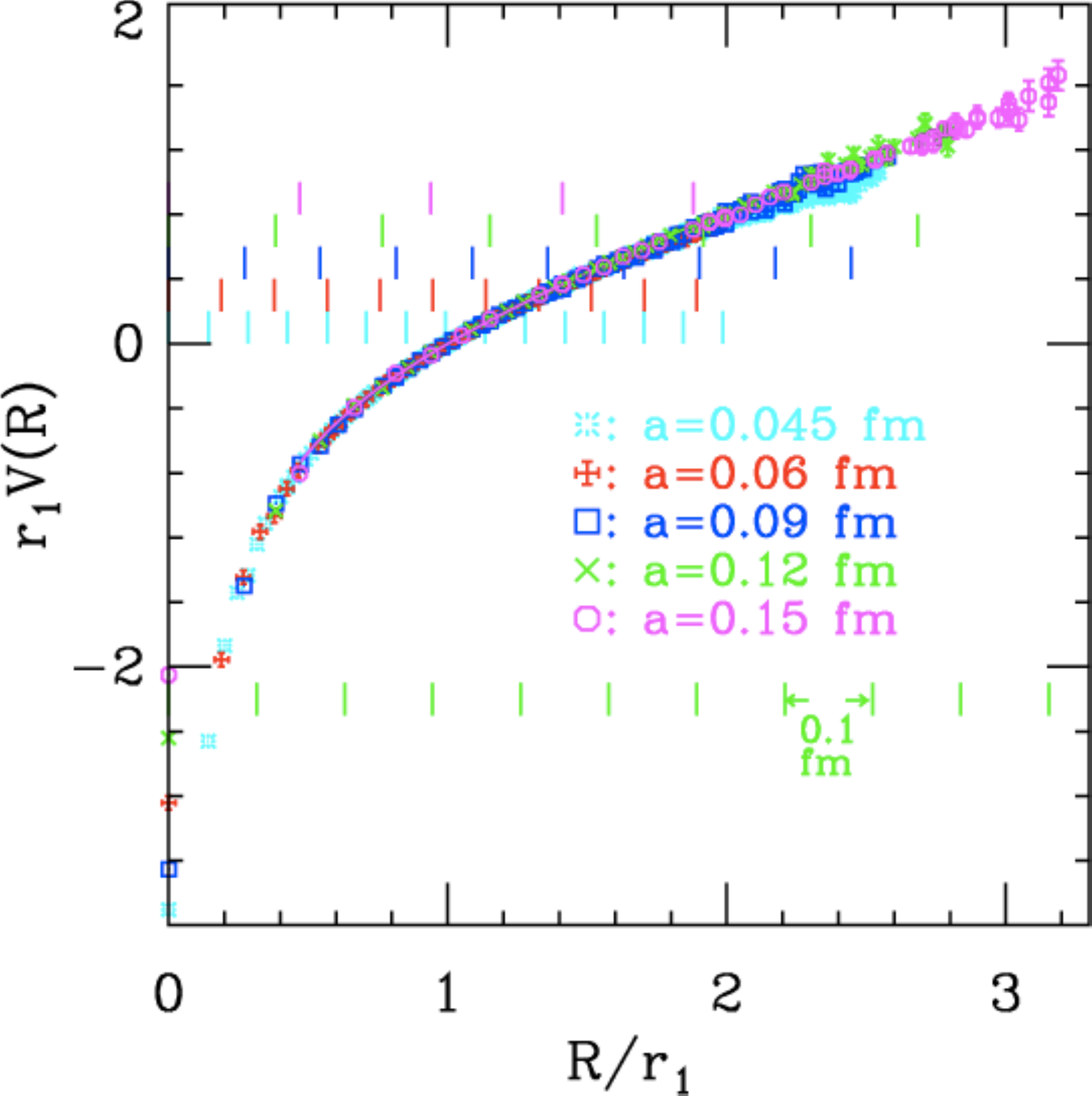}
\caption{Static quark potential as a function of $r$ calculated with different values
  lattice spacings \protect~\cite{Bazavov:2009bb}. The distance is
  expressed in units of the lattice size, with the colored
  gradations indicating the scale in physical units.}
\label{fig:bkgr:lattice_quark_potential}
\end{figure}

\begin{figure}[htb]
\centering
\includegraphics[width=0.8\textwidth]{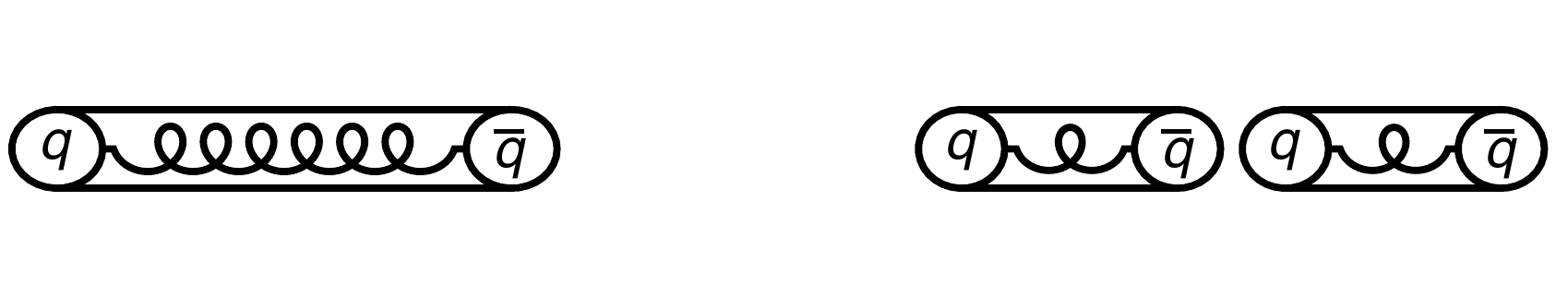}
\caption{Flux tube representing static \qqbar\ potential (left)
  fragmenting into two color-neutral objects (right).}
\label{fig:bkgr:flux_tube}
\end{figure}

A qualitative picture of confinement is to view the static \qqbar\ potential
as Coulombic at short distances, but growing at long range,
\begin{equation}
V_{\mathrm{QCD}}(r)=-\dfrac{4}{3}\dfrac{\alpha}{r}+kr.
\label{eqn:bkgr:qcd_potential}
\end{equation}
A recent lattice calculation of the static quark potential is shown in
Fig.~\ref{fig:bkgr:lattice_quark_potential}, which indicates behavior
approximately consistent with this analytic form.
As the separation between the \qqbar\ pair increases the force remains
constant, and an increasing amount of energy is stored in the
stretched gluon field called a flux tube. At some separation it
becomes energetically favorable to break the flux tube and produce a
new \qqbar\ pair from the
vacuum. This process, known as string fragmentation, is shown schematically in
Fig.~\ref{fig:bkgr:flux_tube}. Such a mechanism provides a heuristic
construct for modeling confinement; no matter how much energy is
applied only color-neutral objects can be created from the system. The
flux tube picture has led to developments in hadron phenomenology by
considering hadrons as relativistic
strings~\cite{Nambu:1970aaz,Goto:1971ce}, and the constant, $k$, in Eq.~\ref{eqn:bkgr:qcd_potential} can
be interpreted as a string tension. Monte Carlo (MC) event generators for
simulating the production of hadrons have had a long history of
success using string models~\cite{Andersson:1983ia,Sjostrand:1993yb}.
\section{Applications of Perturbative QCD}
\label{section:bkgr:pQCD}
\subsection{The Parton Model and Factorization}
\label{section:bkgr:parton_model}
QCD has had great success in providing reliable perturbative
calculations of experimental observables. The most basic application of
QCD uses the parton model, which is a tree-level, impulse approximation
to the full perturbation theory. The scattering process is
formulated in terms of point-like
constituents integrated over
probability distributions for a given parton to have momentum fraction
$x$, 
\begin{equation}
\sigma_{AB}(p_A,p_B) \sim \displaystyle\sum_{i,j} \int
dx_idx_j\hat{\sigma}(x_ip_A,x_jp_B)\phi_i^A(x_i)\phi_j^B(x_j)\,,
\label{eqn:bkgr:parton_factorization}
\end{equation}
Here $A$ and $B$ denote the colliding hadrons and $i$ and $j$ denoting
partons of a particular type. The functions $\phi_i^B(x_i)$ give the
probability density of finding a parton of type $i$ in hadron $A$ with momentum
$p_i=x_i p_A$. While
asymptotic freedom ensures that at sufficiently hard scales the partonic
level matrix elements are calculable in perturbation theory, almost any
real world observable will involve hadronic initial and/or
final-states, where the theory becomes non-perturbative. The
applicability of Eq.~\ref{eqn:bkgr:parton_factorization} in certain
kinematic regimes suggests that calculations can be performed by
separating the short distance behavior of QCD, encoded in the partonic matrix
element $\hat{\sigma}$, from the long range behavior represented by
the probability distributions. The formal apparatus through which this
is accomplished is known as factorization. Factorization theorems take a
similar form to Eq.~\ref{eqn:bkgr:parton_factorization} where each
term in the integrand is also dependent on a scheme-dependent
factorization scale, $\mu_f$. This parameter sets the cutoff for which aspects
of the dynamics are being included in the description of the long and
short range components. The form of the factorization ensures that
there is no quantum mechanical interference
between the long and short range behavior. Furthermore, the
probability distributions, which are known as parton distribution
functions (PDFs) in the general factorized formulation, are independent of the specific
scattering process. Although they are not
calculable by perturbation theory, once these functions are determined
experimentally they can be applied to a calculation regardless of the
details of the partonic level scattering, thus making them universal
features of the hadrons. Factorization theorems have been
proven for a variety of processes, $A+B\rightarrow C+D$ among particles
$A, B, C$ and $D$, with the schematic form of the theorem being
\begin{equation}
d\sigma(A+B\rightarrow
C+D)=d\hat{\sigma}\otimes\Phi_{A}\otimes\Phi_{B}\otimes\Delta_{C}\otimes\Delta_{D}+\mathrm{p.s.c}.
\label{eqn:bkgr:general_factorization}
\end{equation}
Here $\Phi$ and $\Delta$ denote the parton distribution and
fragmentation functions (see Section~\ref{section:bkgr:frag_function}) and $\otimes$ represents a convolution over
parton momentum fraction, trace over color indices and sum over parton
species. This form is traditionally referred to as a twist expansion,
with the first term known as the leading twist. The remainder terms,
the power suppressed corrections, are suppressed by powers of a hard scale
and are referred to as higher-twist effects. Most factorization
theorems have been proven using a collinear factorization scheme,
where the transverse momentum of the parton has been
integrated over. These theorems have been proven in the cases of DIS
and \ee, but the only case involving hadron-hadron scattering to have
been rigorously proven is the Drell-Yan
process~\cite{Collins:1985ue,Collins:1988ig,Collins:1989gx}. 

Observables such as spin asymmetries in hadronic 
collisions provide access to the spin structure of the
proton. However, the extension of the usual collinear formalism in
these cases has been
problematic~\cite{Brodsky:2002cx,Brodsky:2002ue,Collins:2002kn,Boer:2003cm}
as the observables contain sensitivity to the transverse momentum dependence (TMD) of the
PDFs. Without integrating over the transverse dependence of the
PDFs, gauge links couple soft
and collinear portions of diagrams to the hard sub-graphs, breaking the
factorization. Attempts to construct TMD factorization schemes, such as
\kt-factorization~\cite{Collins:1981uk,Collins:1984kg}, have proven
difficult and it has been shown explicitly
that \kt-factorization is violated in high-\pt\ hadron production in
hadronic collisions~\cite{Collins:2007nk}. Building a framework to
perform these types of calculations is still a subject under active development~\cite{Rogers:2010dm}.

\subsection{Deep Inelastic Scattering}
\label{section:bkgr:DIS}
The parton model was first developed in DIS, and this collision system
will be used here to illustrate the utility of the factorized approach. In
these collisions an electron interacts with a hadronic target of mass $M$ via the exchange
of a photon with large virtuality $q^2=-Q^2$. In addition to $Q^2$,
the system is typically described in terms of the variables $x$ and
$y$
\begin{equation}
x=\dfrac{Q^2}{2q\cdot P},\,\,\,\,\,y=\dfrac{E-E^{\prime}}{E},
\label{eqn:bkgr:dis_kine_def}
\end{equation}
where $P$ is the four-momentum of the target and $E$ and $E^{\prime}$
are the initial and final energies of the electron as measured in the
rest frame of the target. In the parton model, $x$ corresponds to the
fraction of target's momentum carried by the struck quark. The
differential cross section is usually expressed in terms of structure
functions $F$,
\begin{equation}
\dfrac{d\sigma}{dxdy}=\dfrac{4\pi\alpha^2_{\mathrm{EM}}}{xyQ^2}\left\{y^2xF_1(x,Q^2)+\left(1-y-\dfrac{2Mxy}{2E}\right)F_2(x,Q^2)\right\}.
\label{eqn:bkgr:dis_factorization}
\end{equation} 
The observation of Bjorken scaling~\cite{Bloom:1969kc,Friedman:1972sy}, that for large $Q^2$, $F_1$ and
$F_2$ are independent of $Q^2$, was the inspiration for the parton
model. For spin-$\dfrac{1}{2}$ (charged) partons, the structure functions obey the
Callan-Gross relation $2xF_1=F_2$. The experimental confirmation
of this result provided support for the interpretation of
spin-$\dfrac{1}{2}$ quarks.
In collinear factorized form the structure functions have the form,
\begin{subequations}
\begin{equation}
F_1(x,Q^2)=\displaystyle \sum_a \int_x^1
\dfrac{d\xi}{\xi}f_{a/A}(\xi,\mu^2,\mu_f)C_{1\,a}\left(\dfrac{x}{\xi},\dfrac{Q^2}{\mu^2},\dfrac{\mu^2_f}{\mu^2},\alpha(\mu)\right)+\mathcal{O}(Q^{-2})\,,
\label{eqn:bkgr:f1_dis_factorization}
\end{equation}
\begin{equation}
F_2(x,Q^2)=\displaystyle \sum_a \int_x^1 d\xi
f_{a/A}(\xi,\mu^2,\mu_f)C_{2\,a}\left(\dfrac{x}{\xi},\dfrac{Q^2}{\mu^2},\dfrac{\mu^2_f}{\mu^2},\alpha(\mu)\right)
+\mathcal{O}(Q^{-2})\,.
\label{eqn:bkgr:f1_dis_factorization}
\end{equation}
\label{eqn:bkgr:f1f2_dis_factorization}
\end{subequations}
Here the functions $f_{a/A}$ are the PDFs for
parton $a$ in hadron $A$. The coefficient functions, $C_a$, are
calculable in perturbation theory and are IRC safe. They are dominated by contributions of
order $Q$; propagators off shell by $\mu^2_f$ will contribute to $C_a$
while contributions below this scale are grouped into
$f_{a/A}$. The result depends on both the factorization and
renormalization scales, which need not be set equal. 

Shortly after the computation of the $\beta$-function,
similar renormalization group techniques were applied to the DIS
structure functions, by considering moments of the PDFs (specifically non-singlet PDF, $f_{q}-f_{\qbar}$),
\begin{equation}
f^{(n)}(\mu^2)=\int_0^1z^{n-1}f(z,\mu^2)dz.
\end{equation}
These are related to the anomalous dimensions,
\begin{equation}
\gamma_n=\dfrac{d}{d\ln\mu}f^{(n)}(\mu^2)\,,
\end{equation}
which appear to give dimensions to dimensionless, but scale-dependent quantities through
renormalization (e.g. $\ln f^{(n)}\sim \mu^{-\gamma_n}$), and can be computed in perturbation
theory~\cite{Christ:1972ms}. The $Q$-dependence of the PDF moments can be used to determine
the $Q$-dependence of the moments of the structure functions,
\begin{equation}
\int x^{n-1}F_1(x,Q^2)dx=C_1^{(n)}(\alpha(Q^2))f^{(n)}(Q^2_0)\times\mathrm{exp}
\left[-\dfrac{1}{2}\int_0^{\ln Q^2/Q^2_0}dt \gamma_n(\alpha(Q^2_0e^t))
\right]\,.
\end{equation}
This relation predicts power-law scaling, $F_1(x,Q^2)\sim
(Q/Q_0)^{-\alpha_0\gamma_n/\pi}$, for theories that do not possess
asymptotic freedom and posed a problem reconciling potential theories with the
observed scaling prior to the demonstration of asymptotic freedom in
QCD. To one-loop in QCD the scaling is ~\cite{Gross:1973ju,Gross:1974cs,Politzer:1974fr,Georgi:1951sr}
\begin{equation}
  F_1(x,Q^2)\propto\left[\dfrac{\ln Q^2/\Lambda^2}{\ln Q^2_0/\Lambda^2}\right]^{-2\gamma_{n}/4|\beta_1|}.
\end{equation}
This weak $Q$-dependence correctly described the early measurements of
scaling in DIS, further building support for QCD as the correct theory
of the strong interaction. 

\subsection{DGLAP and Parton Evolution}
\label{section:bkgr:DGLAP}
\begin{figure}[htb]
\centering
\includegraphics[width=0.7\textwidth]{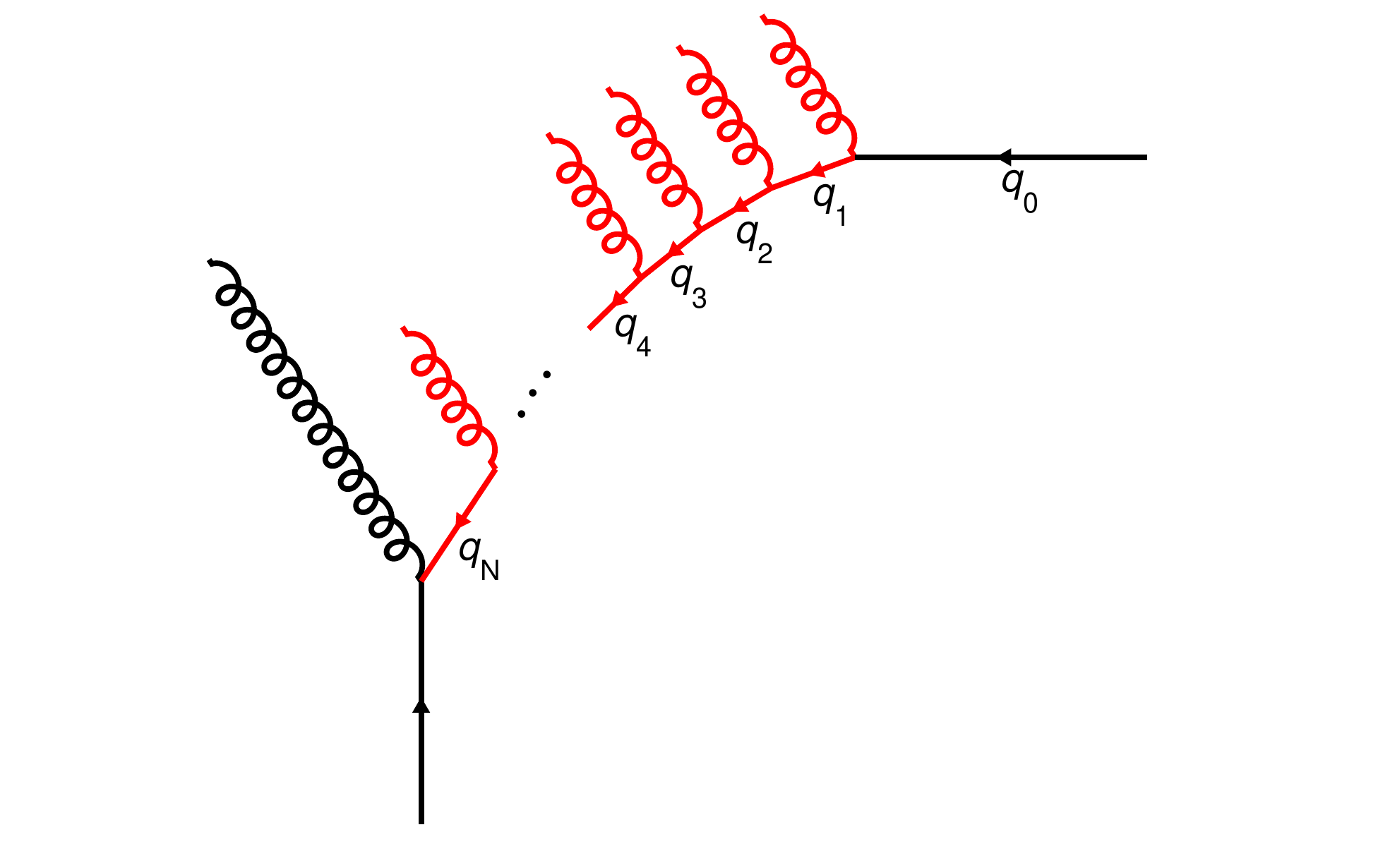}
\caption{Schematic representation of multiple collinear
  splittings. The portion of the diagram shown in red is interpreted
  as part of the structure of the struck hadron instead of a correction
  to the hard matrix element.}
\label{fig:bkgr:evolution_structure}
\end{figure}

\begin{figure}[htb]
\centering
\includegraphics[width=0.95\textwidth]{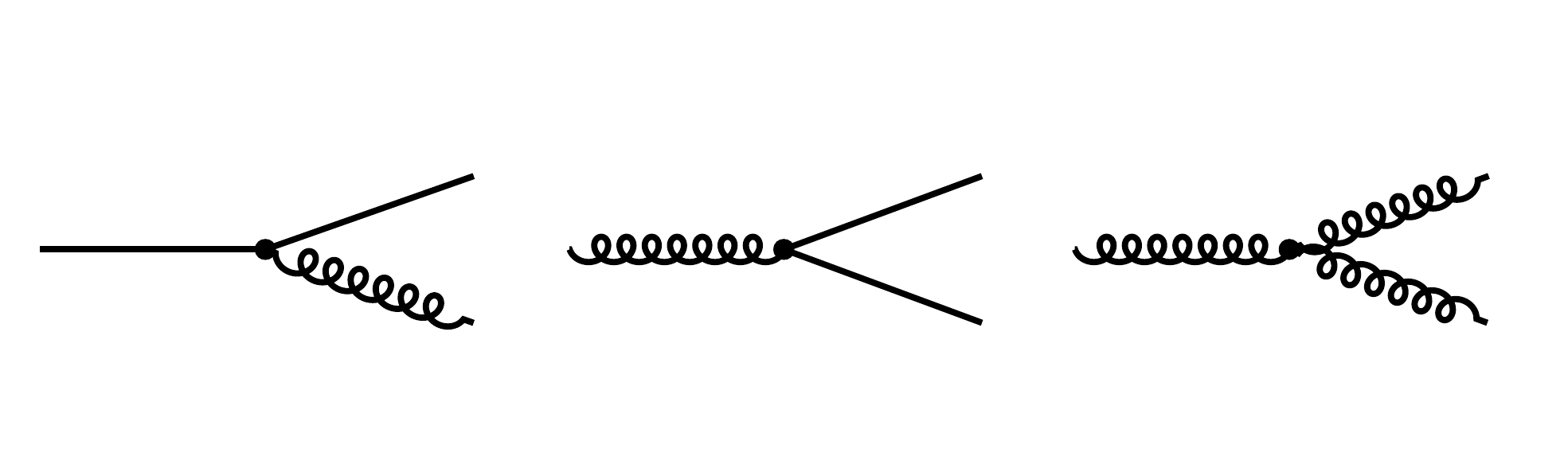}
\caption{Collinear QCD processes used to compute splitting functions.}
\label{fig:bkgr:evolution_diagram}
\end{figure}

When calculating $C_a$, logarithms can arise due to the collinear 
emission at some scale $\Lambda$, which gives a factor of approximately
\begin{equation}
\alpha(Q^2)\ln\dfrac{Q^2}{\Lambda^2}.
\end{equation}
These logarithms can become large such that the above product becomes
of order unity. The result is that $C_a$ will have large contributions from all
orders and is shown schematically in
Fig~\ref{fig:bkgr:evolution_structure}. This multiple emission is
enhanced by $\alpha(Q^2)^N\ln^N\dfrac{Q^2}{\Lambda^2}$, but only for
the case where the virtualities are strongly ordered, $q_n^2 <
q_{n-1}^2 < \cdots < q_0^2$, with other orderings appearing with fewer
powers of $\ln\dfrac{Q^2}{\Lambda^2}$. 
The large logarithms are a symptom of interactions far away
from the scale at which the coupling was fixed. Fortunately, these
collinear contributions can be resummed by renormalization group
methods. As most of the collinear emissions are well separated in
scale from the probe, $q_0$, these emissions can be
reinterpreted as modifying the hadron structure as opposed to
corrections to $C_a$. This results in a $Q$-dependence of the PDF that
evolves the probe from the hard scale to lower momentum scales
(indicated by the red sub-diagram in Fig~\ref{fig:bkgr:evolution_structure}). 
For the
change $Q\rightarrow Q+\Delta Q$ the differential probability of
an emission with energy fraction $z$ and transverse momentum
$Q < p_{\perp} < Q+\Delta Q$ is given by
\begin{equation}
\dfrac{\alpha}{2\pi}\dfrac{dp^2_{\perp}}{p^2_{\perp}} P_{a\leftarrow
  b}(z)\simeq\dfrac{\alpha}{\pi}\dfrac{\Delta Q}{Q} P_{a\leftarrow
  b}(z),
\label{eqn:bkgr:splitting_probability}
\end{equation}
where $P_{a\leftarrow b}(z)$ is the splitting function for parton of
$b$ splitting into type $a$, and can be
computed from the diagrams shown in Fig~\ref{fig:bkgr:evolution_diagram}.
Changes in the distribution of parton $a$ at momentum fraction $x$ can come from splittings
of other partons at $x^{\prime}=x/z$, and can be written as
\begin{eqnarray}
\Delta f_a(x,Q)&=&\displaystyle\sum_b 
\int_0^1 dx^{\prime} \int_0^1 dz\dfrac{\alpha}{\pi}\dfrac{\Delta
  Q}{Q} P_{a\leftarrow b}(z)
f_b(x^{\prime},Q)\delta(x-zx^{\prime})\\ 
&=&\Delta \ln Q\displaystyle\sum_b  \dfrac{\alpha}{\pi}\int_x^1 \dfrac{dz}{z}
f_b(\dfrac{x}{z},Q^2)  P_{a\leftarrow b}(z). \nonumber
\end{eqnarray}
These lead to the DGLAP evolution equations ~\cite{Altarelli:1977zs,Gribov:1972ri,Dokshitzer:1977sg} which are of
the form
\begin{equation}
\dfrac{\partial}{\partial\ln Q} f_a(x,Q)=\displaystyle\sum_b
\dfrac{\alpha}{\pi}\int_x^1 \dfrac{dz}{z}f_b(\dfrac{x}{z},Q^2)
P_{a\leftarrow b}(z).
\label{eqn:bkgr:dglap}
\end{equation}
See Ref.~\cite{Nakamura:2010zzi} for a complete tabulation of the
splitting functions and full evolution equations. This result indicates that if $f_a$ is measured as a function of $x$ at a
given value of $Q_0$, the PDF at any other scale $Q$
can be determined by evolving the PDF according to
Eq.~\ref{eqn:bkgr:dglap}. The $Q^2$ dependence of the structure
function $F_2$ at fixed $x$ is shown in
Fig.~\ref{fig:bkgr:f2}. This dependence agrees well with the dependence
predicted by the DGLAP equations over a range of $Q^2$ and $x$ values.
\subsection{Fragmentation Functions}
\label{section:bkgr:frag_function}
\begin{figure}[htb]
\centering
\includegraphics[width=0.49\textwidth]{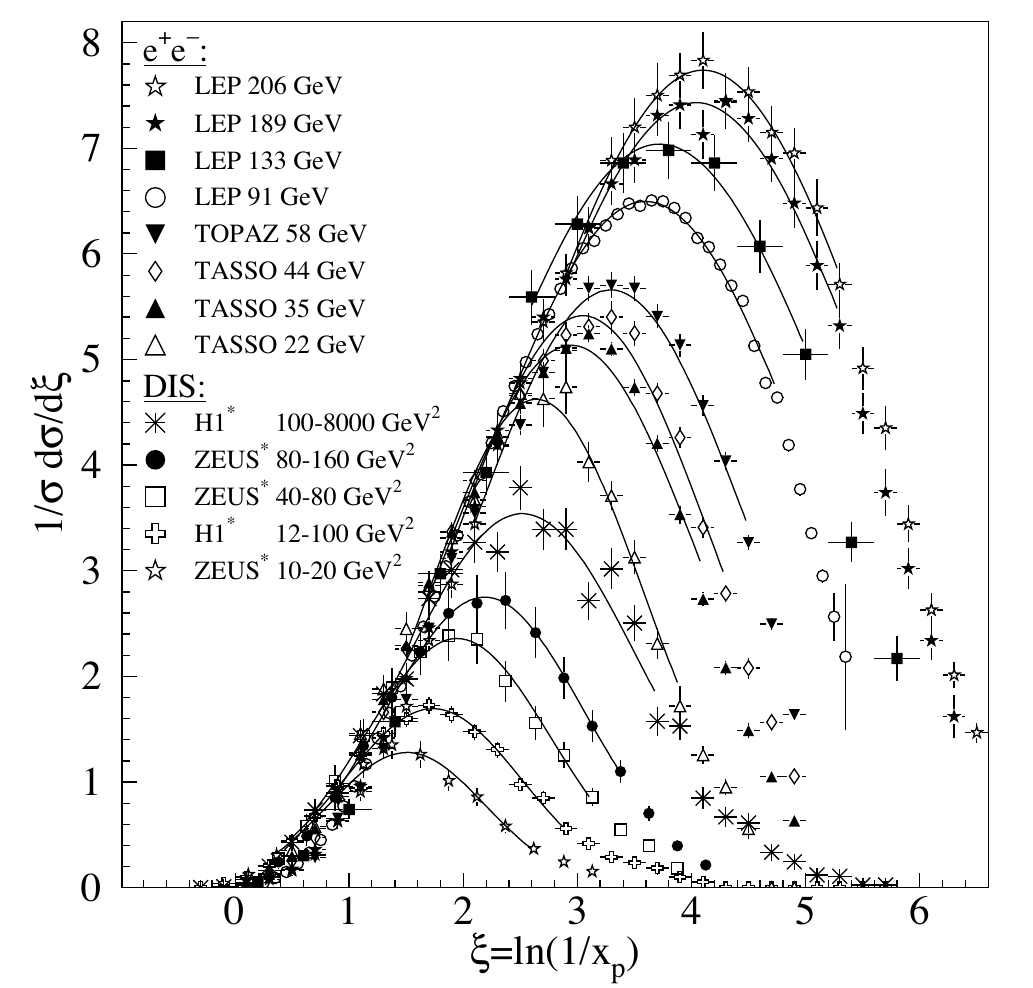}
\includegraphics[width=0.49\textwidth]{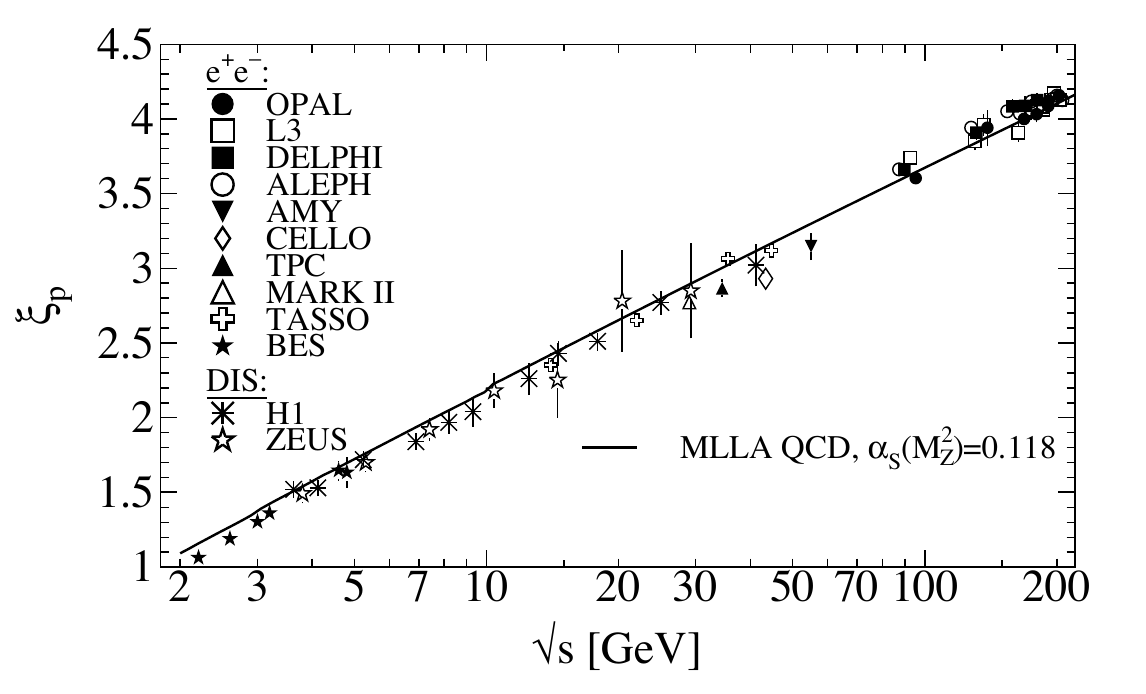}
\caption{The hump-backed plateau structure of fragmentation
  functions through the measurement of single inclusive hadron cross
  sections (right). The cross sections are fit with a Gaussian and the
  extracted peak position compares well with the MLLA QCD prediction. Figure adapted from~\protect\cite{Nakamura:2010zzi}.}
\label{fig:bkgr:hump_back_plateau}
\end{figure}

The final-state analog of the parton distribution function is known as
a fragmentation function. The fragmentation function $D_i^h(z,\mu^2)$
encapsulates the probability that a parton of type $i$ will fragment
into a hadron of type $h$ with momentum fraction $z$ of the original
parton. In \ee\ collisions, where there are no initial
parton distributions, cross sections for the production of a particular species of
hadron can be written as,
\begin{equation}
\dfrac{d\sigma^h_{\ee}}{dx}=\displaystyle\sum_i\int_x^1
\dfrac{dz}{z}C_{i}(z,\alphas(\mu^2),\dfrac{s}{\mu^2}) D_i^h(z,\mu^2) +\mathcal{O}\left(\dfrac{1}{\sqrt{s}}\right).
\end{equation}
For situations with arbitrary numbers initial/final-state hadrons, the
collinear factorized relation takes the form of
Eq.~\ref{eqn:bkgr:general_factorization}. The \ee\ measurements
provide the cleanest access to the fragmentation functions since there
is no integration over PDFs. The fragmentation functions
obey an evolution equation in $\mu^2$ identical to
Eq.~\ref{eqn:bkgr:dglap}, with the inverse splitting functions
(i.e. $P_{a\leftarrow b}(z)$ is replaced with $P_{a\rightarrow
  b}(z)$). In the parton model these obey the momentum sum rule,
\begin{equation}
\displaystyle \sum_h \int_0^1 dz\,z\,D_i^h(z,\mu^2)  = 1\,.
\end{equation}
The higher-order corrections to the splitting functions can see
logarithmic enhancements at low $x$, an effect which causes the
leading order approximation of the evolution to break down much
sooner in \ee\ than DIS. This has led to development of calculations
using a modified leading log approximation (MLLA), which include
corrections for next-to-leading effects~\cite{Mueller:1982cq,Dokshitzer:1991ej,Dokshitzer:238162,Khoze:1996dn,Fong:1990nt}. In general, the behavior of
these enhancements is to cause the ``hump-backed plateau'' behavior,
\begin{equation}
xD(x,s)\propto \exp\left\{
-\dfrac{1}{2\sigma^2}(\xi-\xi_p)^2
\right\},
\end{equation}
with $\xi=\ln 1/x$ and the width and peak approximately given by~\cite{Dokshitzer:1982xr},
\begin{equation}
\xi_p\simeq \dfrac{1}{4} \ln s/\Lambda^2,\,\,\,\,\,\,
\sigma \propto \left(\ln s/\Lambda^2\right)^{3/4}.
\end{equation}
This behavior is shown in Fig.~\ref{fig:bkgr:hump_back_plateau}, with
the fragmentation functions fit with a Gaussian for a variety of
$\sqrt{s}$ and $Q^2$ values shown on the right and
a comparison of the extracted peak position to the MLLA value shown on
the left.

\begin{figure}[htb]
\centering
\includegraphics[width=0.9\textwidth]{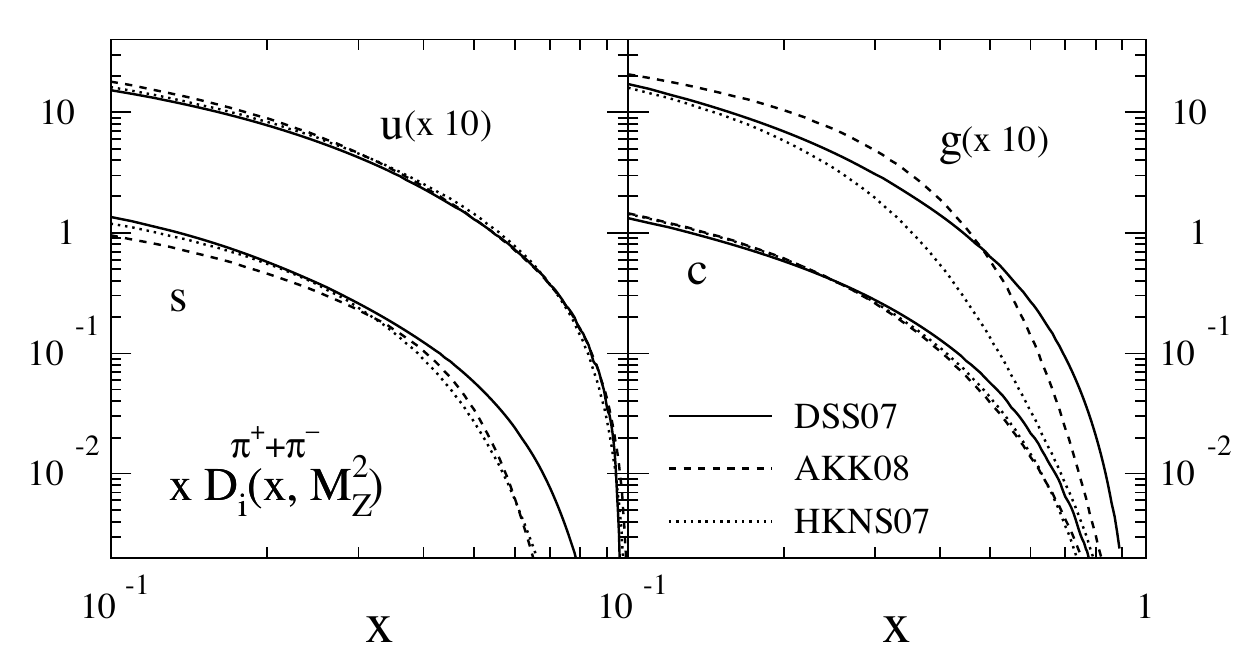}
\caption{Contributions to the $\piplus+\piminus$ NLO fragmentation
  functions from up, strange (left), gluon and charm (right) from
  three different global analyses~\protect\cite{Florian:2007hc,Florian:2007aj,Hirai:2007cx}, with $Q^2=\mZ^2$. Figure adapted from~\protect\cite{Nakamura:2010zzi}.}
\label{fig:bkgr:ff_global}
\end{figure}

To span a large kinematic range and provide separation between quark
and anti-quark contributions, the \ee\ data has been supplemented by
DIS and hadronic collider data in recent global NLO
extractions~\cite{Florian:2007hc,Florian:2007aj,Hirai:2007cx} (see~\ref{section:bkgr:fixed_order} for discussion of fixed
order calculations). A comparison of results of some such analyses are shown in
Fig~\ref{fig:bkgr:ff_global}, where the various partonic contributions to the
charged pion fragmentation function are shown.
\subsection{Fixed Order Calculations}
\label{section:bkgr:fixed_order}
Perturbative calculations are typically performed at fixed order and
it is useful to consider IRC safety in this
context~\cite{Nakamura:2010zzi}. For an $n$-particle process, the cross section for an
observable, $\mathcal{O}_n$, can be constructed at leading order (LO) as
 \begin{equation}
 \sigma_{\mathcal{O},\,\mathrm{LO}}=\alphas^{n-2}(\mu_{\mathrm{R}})\int
 d\Omega_n |M^2_{n, 0}|(p_1,\cdots,p_n)\mathcal{O}_n(p_1,\cdots,p_n),
\end{equation}
 where $M^2_{n, 0}$ is the $n$-particle matrix element at tree-level,
 $d\Omega_n$ is the measure over the $n$-particle phase space and
 $\mu_{\mathrm{R}}$ is the renormalization scale. Non-perturbative
 corrections have been omitted here. The tree-level amplitudes are finite but the
 integral will diverge in the soft and collinear regions of the
 momentum integration. Therefore this calculation is only safe when
$\mathcal{O}_n$ also vanishes in this limit (e.g. jet cross sections). At next-to-leading-order (NLO) the $n+1$-particle matrix element must be
 considered along with the interference between the tree-level and
 1-loop $n$-particle processes,
\begin{eqnarray}
 \sigma_{\mathcal{O},\,\mathrm{NLO}}&=&\sigma_{\mathcal{O},\,\mathrm{LO}}
 \\ \nonumber
&+&\alphas^{n-1}(\mu_{\mathrm{R}})\int
 d\Omega_{n+1} |M^2_{n+1,
   0}|(p_1,\cdots,p_{n+1})\mathcal{O}_{n+1}(p_1,\cdots,p_{n+1}) \\ \nonumber
&+&
\alphas^{n-1}(\mu_{\mathrm{R}})\int
 d\Omega_{n} 2\mathrm{Re}\left[M_{n,0}M_{n,1}^{*}\right]
(p_1,\cdots,p_n)\mathcal{O}_n(p_1,\cdots,p_n).
\end{eqnarray}
Here $M_{n,1}$ is the 1-loop $n$-particle amplitude, which diverges in the soft
and collinear limits. This divergence will cancel with the aforementioned tree-level
divergence in the integration if the observable is IRC safe,
\begin{eqnarray}
&&\lim_{p_i\rightarrow0} 
\mathcal{O}_{n+1}(p_1,\cdots,p_i,\cdots,p_{n+1})=\mathcal{O}_n(p_1,\cdots,p_{n})
\\ \nonumber
&& \lim_{p_j \parallel \,p_i} 
\mathcal{O}_{n+1}(p_1,\cdots,p_i,p_j,\cdots,p_{n+1})=\mathcal{O}_n(p_1,\cdots,p_i,\cdots,p_{n})\,.
\end{eqnarray}
A similar procedure applies in extending to NNLO calculations, which
involve additional contributions the tree-level $n+2$ contribution, interference between
tree-level and 1-loop $n+1$-particle diagrams, interference between
the $n$-particle tree-level and 2-loop diagrams and the squared
contribution of the 1-loop $n$-particle diagram. These calculations
become increasingly difficult to compute, especially in an
automated fashion, due to the intricate higher-loop diagrams.


Often these calculations are supplied in the form of Monte Carlo event
generators. These tools typically compute a hard scattering matrix
element and then evolve the partons by means of a parton
shower. This technique constructs a probability for a parton not
split from the splitting functions defined in
Eq.~\ref{eqn:bkgr:splitting_probability}. The probability that a
parton of type $a$ will not split between scales $Q^2$ and $Q_0^2$, $\Delta_a(Q^2,Q_0^2)$, known as a Sudakov
form factor is given by,
\begin{equation}
\Delta_a(Q^2,Q_0^2)=\mathrm{exp}\left\{-\int_{Q_0^2}^{Q^2} \dfrac{dk_t^2}{k_t^2}\int_0^1
  dz \dfrac{\alphas}{2\pi}\displaystyle\sum_b P_{a\leftarrow  b}(z)\right\}.
\end{equation}
MC sampling determines the scale of the first splitting, and this
process is repeated down to some hadronization scale $Q\sim
1$~\GeV. At this point a hadronization model~\cite{Andersson:1983ia,Sjostrand:1993yb,Webber:1983if,Marchesini:1991ch} converts the evolved
partons into final-state hadrons which can be used to simulate
response of a detector to an input physics signal. Leading order event
generators (e.g. PYTHIA) typically give the correct description of the
soft and collinear emission, but fail to describe wide angle emissions
due to the lack of multi-parton matrix elements. This can be improved
by generators which use this higher order matrix elements and
merge the parton showers in such a way as to avoid double counting the
real and virtual
contributions~\cite{Catani:2001cc,Alwall:2007fs,Gleisberg:2003xi,Mangano:2002ea}.

In general LO calculations for applications at hadron colliders are
accurate to approximately a factor of two. A $K$-factor, which
represents the ratio of NLO to LO contributions, is typically used to
tune the cross sections.

\subsection{Jets}
\label{section:bkgr:jets}
Highly collimated sprays of particles known
as jets are a ubiquitous feature of high energy particle collisions. These objects are the
experimental signature of partonic hard scattering processes and have been used to test the
theoretical application of the theory since their initial observation
at SLAC~\cite{Hanson:1975fe}. However, the multi-particle final states
associated with jet production are sufficiently complicated that any
experimental measurement must have a precise definition of an
observable~\cite{Sterman:1977wj}. Furthermore, these definitions must be relatable to
quantities that are well-defined theoretically, particularly with
regard to the IRC safety issues discussed previously.

\begin{figure}[htb]
\centering
\includegraphics[width=0.7\textwidth]{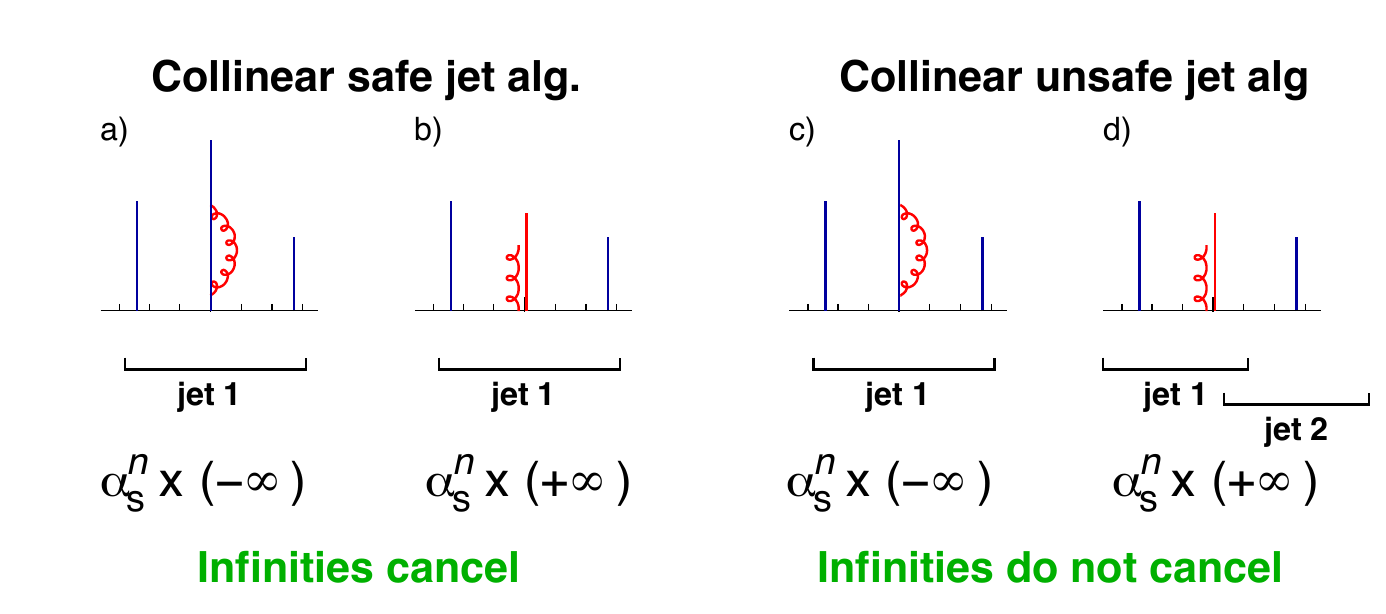}
\caption{Jet reconstruction using two algorithms on a system with and
  without additional collinear gluon radiation. The IRC safe algorithm
reconstructs the two scenarios as the same jet (a,b), allowing for
cancellation of divergences. The unsafe algorithm gives different
results in the two cases (c,d). Figure adapted from Ref\protect~\cite{Cacciari:2008gp}.}
\label{fig:bkgr:irc_divergence}
\end{figure}

Most early jet algorithms were based on associating particles or
energy deposits that were nearby in angle, typically referred to as cone
algorithms. The Snowmass workshop in 1990~\cite{Huth:1990mi} proposed a
standard set of criteria for jet algorithms~\cite{Ellis:1990ek},
however implementations of these criteria were not immediately
obvious. Many cone algorithms used at the time suffered from varying degrees of IRC
unsafety, which limited their relatability to
theory~\cite{Salam:2009jx}. The effects of unsafety are illustrated in
Fig~\ref{fig:bkgr:irc_divergence}, where two algorithms, one safe
(left) and one unsafe (right) are compared. In the IRC safe
algorithm, the emission of a collinear gluon does not affect the
clustering (b), allowing the divergence in this graph to cancel the
1-loop (a) divergence. However, this emission causes different
clustering in the unsafe algorithm (d), resulting in a failure in
cancellation of the divergences. In general, issues of split or
overlapping jets are not handled in a natural way by cone algorithms,
and some additional criteria, often introducing undesired artifacts,
are required to handle such cases.

Sequential clustering algorithms, which
perform pairwise clustering of particles~\cite{Bartel:1986ua,Bethke:1988zc}, showed
particular promise as they were IRC safe. This led to the proposal for
the \kt\ algorithm in \ee\ collisions
~\cite{Catani:1991hj},
which was formulated to be boost-invariant and IRC safe and was later
adapted for hadronic
collisions~\cite{Catani:1993hr}. The inclusive
formulation of this algorithm~\cite{Ellis:1993tq}  uses the measure,
\begin{equation}
d_{ij}=\min(p_{\mathrm{T}\,i}^2, p_{\mathrm{T}\,j}^2)\dfrac{\Delta
  R_{ij}^2}{R^2},
\,\,\,\,\,\,\,
\Delta R_{ij}^2=(y_i-y_j)^2+(\phi_i-\phi_j)^2,
\end{equation}
where $y$ and $\phi$ are the particles' rapidity and azimuthal angle
respectively, to cluster nearby particles. The algorithm works using the following
steps:
\begin{enumerate}
\item Calculate $d_{ij}$ for all pairs and also $d_{iB}=p_{\mathrm{T}\,i}^2$.
\item Find the minimum of the $d_{ij}$ and $d_{iB}$.
\item If the minimum is a $d_{ij}$, combine $i$ and $j$ and start
  over.
\item If the minimum is a  $d_{iB}$, call $i$ a final-state jet and
  remove it from subsequent clustering and start over.
\item Finish clustering when no particles remain.
\end{enumerate}
The parameter $R$ controls the size of the jets in analogy with the
radius parameter in cone jets. In this inclusive form, all particles
are clustered into final-state jets; thus some final discriminating
criteria, often a minimum cut on \et\ or \pt, must be applied. This
algorithm has the feature that for soft or collinear pairs the distance
measure is inversely proportional to the differential probability of
collinear emission,
\begin{equation}
\dfrac{dP_{k\rightarrow ij}}{dE_id\theta_{ij}}\sim \dfrac{\alphas}{\min(E_j,E_j)\theta_{ij}},
\end{equation}
which is easily adaptable to theoretical
calculations~\cite{Salam:2009jx}. This algorithm was not originally
favored by experiment due to its slow processing time and
geometrically irregular jets, which made experimental corrections more
difficult. However, a faster implementation by the \verb=FastJet= package($\mathcal{O}(N\ln N)$
vs. $\mathcal{O}(N^3)$ ) has significantly improved the former
issue~\cite{Cacciari:2005hq}.

\begin{figure}[htb]
\centering
\includegraphics[width=0.7\textwidth]{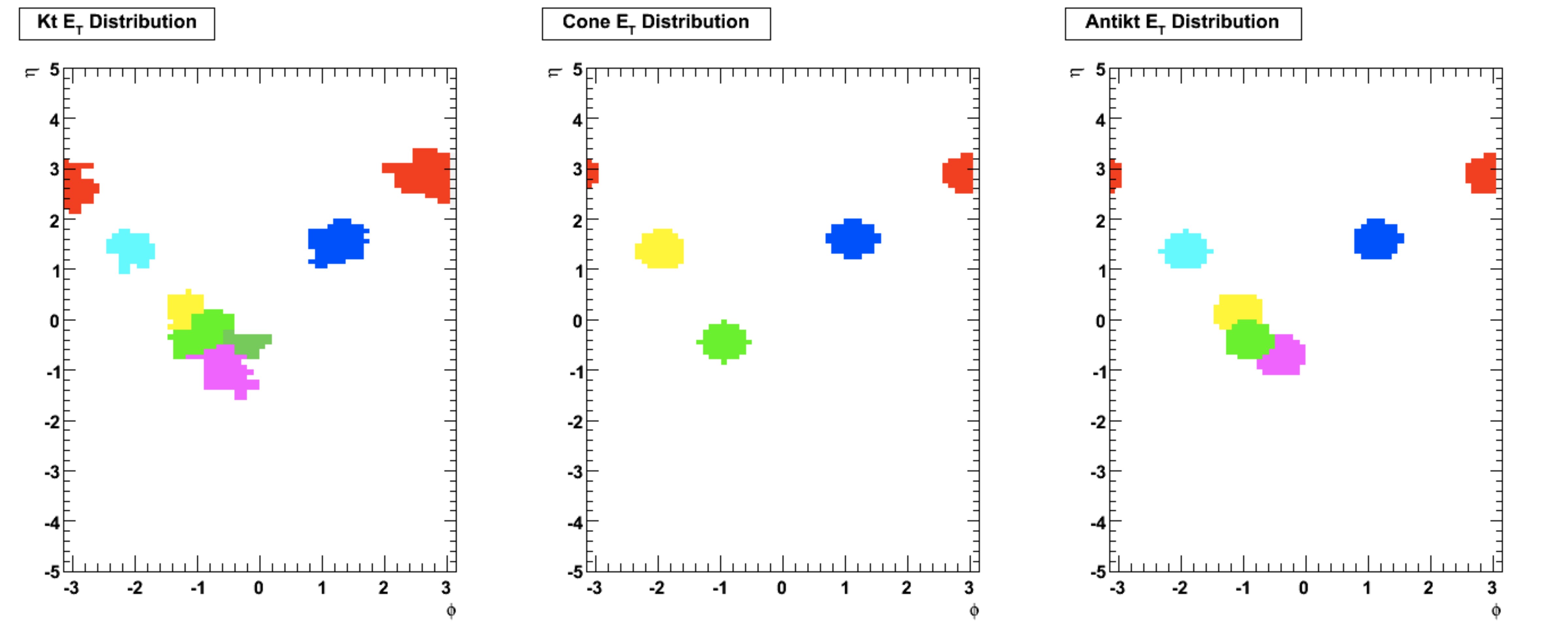}
\caption{Jet clustering on an event in $\eta-\phi$ using three
  different jet algorithms: \kt\ (left), ATLAS Cone (center), and
  anti-\kt\ (right) all with \RFour. The cone algorithm does a poor
  job of resolving the split jet. The \kt\ and anti-\kt\ resolve this
  as two separate jets, but the \kt\ clustering is strongly influenced
  by other soft particles in the event.}
\label{fig:bkgr:jet_alg_clustering}
\end{figure}

The distance measure of the \kt\ algorithm can be generalized in the
following way:
\begin{equation}
d_{ij}=\min(p_{\mathrm{T}\,i}^{2p}p_{\mathrm{T}\,j}^{2p})\dfrac{\Delta R_{ij}^2}{R^2}.
\end{equation}
The choice $p=1$ corresponds to the usual \kt\ algorithm, while $p=0$
corresponds to the Cambridge/Aachen
algorithm~\cite{Dokshitzer:1997in,Wobisch:1998wt}, which uses only
geometric considerations when clustering objects. The generalization
was used to define the anti-\kt\ algorithm
($p=-1$), which clusters hard particles first
and produces jets similar with regular, cone-like
geometry~\cite{Cacciari:2008gp}. Examples of the clustering behavior
of the \kt, ATLAS cone~\cite{ATL-PHYS-PUB-2009-012} and anti-\kt\
algorithms are shown
Fig.~\ref{fig:bkgr:jet_alg_clustering}. The \kt\
allows jets clustered from soft background particles to compete with
the real jet signal, a feature which is particularly problematic in
heavy ion collisions. This is contrasted by the behavior of the
anti-\kt\ which preferentially clusters particles with the harder of
two jets.

\begin{figure}[htb]
\centering
\includegraphics[width=0.7\textwidth]{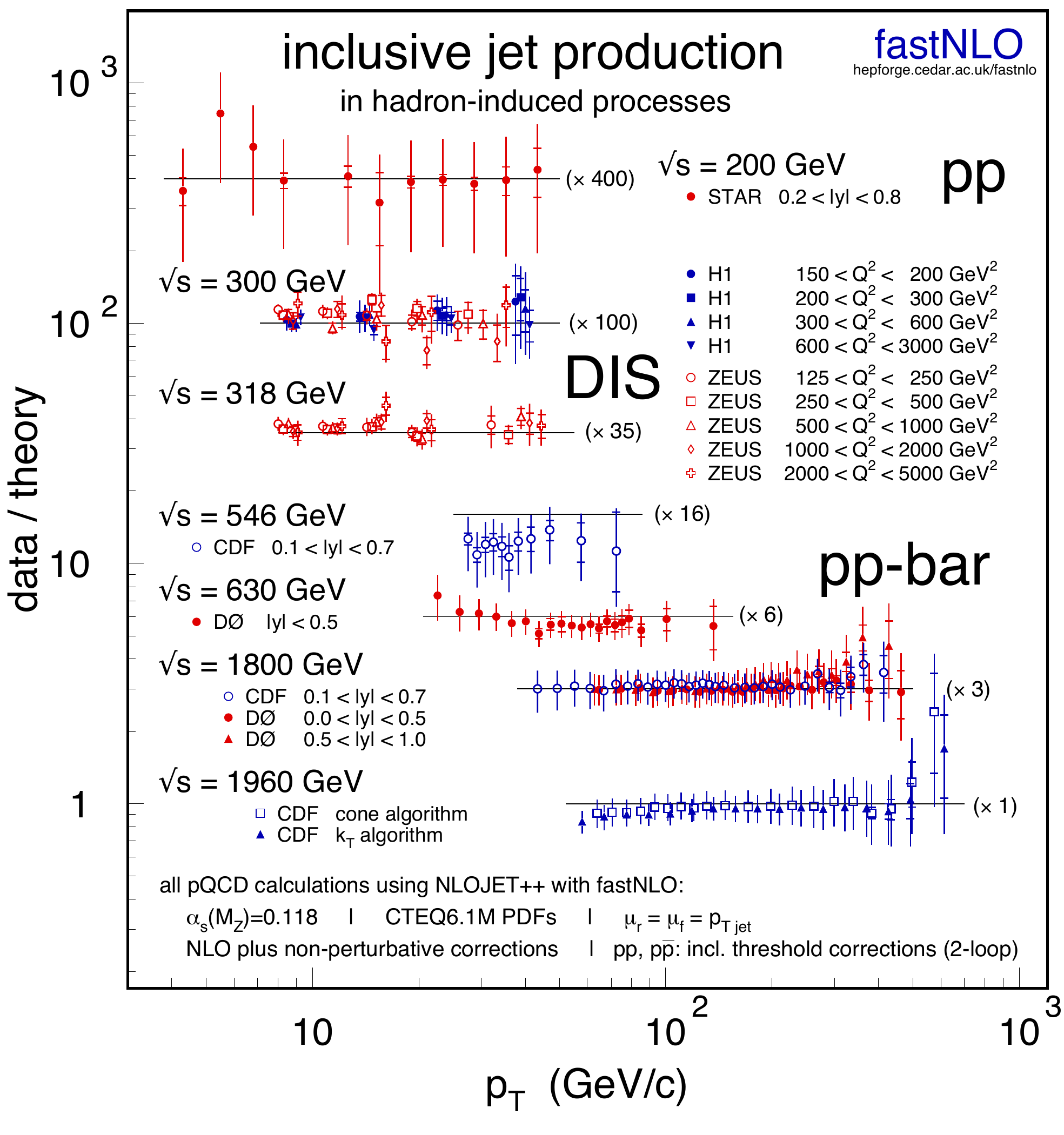}
\caption{Ratio of data to theory for the single inclusive jet cross
  section at a variety of energies~\protect\cite{Kluge:2006xs,Nakamura:2010zzi}.}
\label{fig:bkgr:jet_cross_section}
\end{figure}

Jets have become an essential part of high energy experiments, both as
tools for testing QCD, but also as input into reconstructing more
complicated physics objects. Figure~\ref{fig:bkgr:jet_cross_section}
shows the data/theory ratio for the inclusive jet cross
section as a function of jet \pt\ for a wide range of experiments. The
theoretical calculation is provided by \verb=fastNLO= with \verb=NLOJET++=~\cite{Kluge:2006xs}.
\section{Phase Structure of Nuclear Matter}
\label{section:bkgr:thermal_QCD}
Prior to the advent of QCD, attempts were made to apply statistical
methods to the large particle multiplicities in high energy
collisions~\cite{Fermi:1950jd,Landau:1953gs}. This led to the analysis of the high temperature limit of
nuclear matter in the context of hadronic resonances. The growth in the number of these resonance
states with increasing energy led Hagedorn to propose the
Statistical Bootstrap Model, where resonances are thought of as being
composed of excited lower mass resonance constituents. In the high
temperature limit the density of resonant states and the thermodynamic
energy density of states approach each other and the partition
function diverges. The temperature
cannot be increased beyond some limiting value, the Hagedorn
temperature, as adding energy will only excite more resonances, not
add to the kinetic energy of the system~\cite{Hagedorn:1965st}.

Asymptotic freedom indicates that this will not occur, but rather 
that at sufficiently high temperatures the coupling will become weak
leading to nuclear matter with dynamics described by perturbative
QCD~\cite{Collins:1974ky}. This suggests that there is a phase
transition where the degrees of freedom evolve from hadrons to
quarks and gluons. This state of matter the Quark Gluon Plasma (QGP),
represents the form of nuclear matter in the early stages of the
Universe~\cite{Shuryak:1980tp}. 

In the weak coupling limit the Stefan-Boltzmann relation between
energy density, $\varepsilon$, and pressure, $p$, applies:
\begin{equation}
\varepsilon-3p=0.
\end{equation}
Each is proportional to $T^4$ with the constant of
proportionality indicating the degeneracy of particles obeying
Bose-Einstein, $g_{\mathrm{BE}}$, and Fermi-Dirac, $g_{\mathrm{FD}}$, statistics,
\begin{equation}
\varepsilon=\left(g_{\mathrm{BE}}+\dfrac{7}{8}g_{\mathrm{FD}}\right)\dfrac{\pi^2}{30}T^4.
\end{equation}
At the lowest temperatures the system is hadron gas
containing only the lowest state, the pions ($g_{\mathrm{BE}}=3_{\mathrm{isospin}}$,
$g_{\mathrm{FD}}=0$). At high temperature the system is a gas of deconfined
quarks and gluons ($g_{\mathrm{BE}}=8_{\mathrm{color}}\times2_{\mathrm{spin}}$,
$g_{\mathrm{FD}}=3_{\mathrm{color}}\times2_{\mathrm{spin}}\times2_{\qqbar}\times
N_f$). The total energy density is
\begin{equation}
\varepsilon=\dfrac{\pi^2}{30}T^4\left\{\begin{array}{cc}3 & T\sim0 \\16+\dfrac{21}{2}N_f & T\rightarrow\infty\end{array}\right. \,.
\end{equation}
For two flavor QCD the difference in phases corresponds to a factor of
9 increase in $\varepsilon/T^4$, indicating a dramatic change in
energy density.

The temperature range over which this
transition occurs, specifically the possible divergences of thermodynamic
variables, can characterize the transition. In the limit of zero quark
mass, the QCD Lagrangian possesses chiral symmetry, which is broken by
non-perturbative effects resulting in three massless pions. The high
temperature limit exhibits both deconfinement and restoration of
chiral symmetry, and it is unclear to what extent these two phenomena
are interrelated and whether QCD may exhibit a distinct phase transition associated
with each of these phenomena~\cite{Shuryak:1981fz}.

\subsection{Thermal Field Theory}
\label{section:bkgr:tQCD}
The extension to finite-temperature field theory is accomplished by using the
relationship between the path integral formulation in quantum field
theory and the partition function of statistical
mechanics~\cite{lebellac,Blaizot:2001nr}. Explicitly, the path integral for a single quantum
mechanical particle with Hamiltonian $H$ is given by
\begin{equation}
\bra {q'(t)} e^{-i\hat{H}t}\ket{q(0)}=\int_{q(0)}^{q'(t)}\mathcal{D}[q(t)]e^{i\int_0^t dt' L},
\end{equation}
which denotes the amplitude for a particle at $q$ at $t=0$ to go to
$q'$ at $t$, and $L=\dfrac{1}{2}m\dot{q}(t)^2-V(q(t))$. This functional
integral can be analytically continued to imaginary time via
$t\rightarrow -i\tau$. This generally provides a more rigorous
definition for the path integral and its convergence. In the imaginary
time formulation the action in the exponent is replaced by the
Euclidean action
\begin{equation}
S_{E}=\int_0^{\tau} d\tau^{\prime}L_{E}(\tau^{\prime}),
\end{equation}
where $L_E=\dfrac{1}{2}m\dot{q}(t)^2+V(q(t))$. This can be compared
directly to the partition function of the canonical ensemble, with
temperature $T$ and $\beta=1/kT$,
$Z(\beta)=\mathrm{Tr}[e^{-\beta\hat{H}}]$. In the non-diagonal basis
$\ket{q}$ this can be written in the same form as the usual path integral,
\begin{equation}
Z(\beta)= \int dq \bra{q} e^{-\beta\hat{H}} \ket{q}=\int_{q(0)=q(\tau)}\mathcal{D}[q(\tau)]e^{-\int_0^{\beta} dt' L_E}
\end{equation}
Thermal expectation values of time-ordered products of fields
$\langle\cdots\rangle_{\beta}$ can be computed in the same fashion as
zero-temperature field theory, but using the imaginary time/Euclidean
action. For example, a generating functional, $Z(\beta , \mathcal{J})$, can be constructed by
introducing a source $\mathcal{J}$, and adding the term
$\mathcal{J}\hat{q}$ to the Lagrangian. A two-point correlation
function can be evaluated with functional derivatives with respect to $\mathcal{J}$,
\begin{equation}
\Delta(\tau)=\langle T\, \hat{q}(-i\tau)\hat{q}(0) \rangle_{\beta}= \dfrac{1}{Z(\beta,0)}\dfrac{\delta^2 Z(\beta , \mathcal{J} )}{\delta \mathcal{J} (\tau) \delta \mathcal{J} (0)}\Big |_{\mathcal{J}=0}\,\,.
\label{eq:bkgr:func_deriv_eqn}
\end{equation}
The cyclic nature of the trace imposes the constraint
$\Delta(\tau-\beta)=\pm\Delta(\tau)$, with the sign depending on whether the field
obeys commutation or anti-commutation relations. Thus the
periodic (anti-periodic) boundary conditions on the correlation
functions enforce the Bose-Einstein (Fermi-Dirac) statistics and $\Delta$
can be described as a Fourier series
\begin{equation}
\displaystyle \Delta(\tau)=\beta \sum_{n=0}^{\infty}e^{-i\omega_n \tau}\Delta(i\omega_n).
\end{equation}
Here $\omega_n=\dfrac{2\pi n}{\beta}$ for bosons and
$\omega_n=\dfrac{\pi (2n+1)}{\beta}$ for fermions and are known as the
Matsubara frequencies. The full real-time gauge theory propagators at tree level are
\begin{subequations}
\begin{equation}
D_F(k_0)=\dfrac{i}{k^2_0-\omega^2+i\epsilon}+2\pi\delta(k^2_0-\omega^2)\dfrac{1}{e^{\beta k_0}-1},
\end{equation}
\begin{equation}
S(p)=\dfrac{i}{\slashed{p}-m}+2\pi\delta(p^2-m^2) (\slashed{p}+m) \dfrac{1}{e^{\beta p_0}+1},
\end{equation}
\end{subequations}
for the bosons and fermions respectively. In both expressions, the first term describes the usual vacuum propagation while the second
represents the disconnected part of the propagation which counts the
modes absorbed by the medium.

\subsection{Lattice Thermodynamics}
\label{section:bkgr:lattice_thermo}
The lattice formulation of QCD discussed previously can be naturally applied to
calculate thermodynamic variables in QCD, since it is formulated as a Euclidean space-time
evaluated path integral. The four-dimensional Euclidean space-time is
represented by $N_{\sigma}^3\times N_{\tau}$ lattice sites with
spacing $a$. The system is described by volume $V$ and temperature,
$T$,
\begin{equation}
V=(N_{\sigma}a)^3,\,\,\,\,\, T=\dfrac{1}{N_{\tau} a},
\end{equation}
and the thermodynamic limit is taken with $N_{\tau}\rightarrow\infty$,
$a\rightarrow 0$ but holding $T$ fixed. Statistical errors can enter
since the integration is performed over a finite number of gauge field
configurations, thus making calculations with large $N_{\tau}$
difficult. Recent developments in improved actions with reduced
discretization errors allow for accurate calculations without
resorting to large $N_{\tau}$, and have facilitated advances in
lattice thermodynamic calculations.

Typically the action is separated
into a contribution from the gauge fields and a fermionic
contribution. The latter is expressed as a functional
determinant in a way that removes or minimizes fermion doubling
problems. The action and determinant are written in terms of link
variables, which involve the evaluation of gauge field configurations
at adjacent lattice sites. The numerical integration techniques
require a probabilistic interpretation, and problems occur when the
determinant becomes negative. For zero chemical potential, $\mu$, this is not
a problem, however the determinant becomes complex-valued for
$\mu\neq0$. This makes calculations at finite chemical potential more
difficult, although recent advances have been made on this subject.

\begin{figure}
\centering
\includegraphics[width=0.75\textwidth]{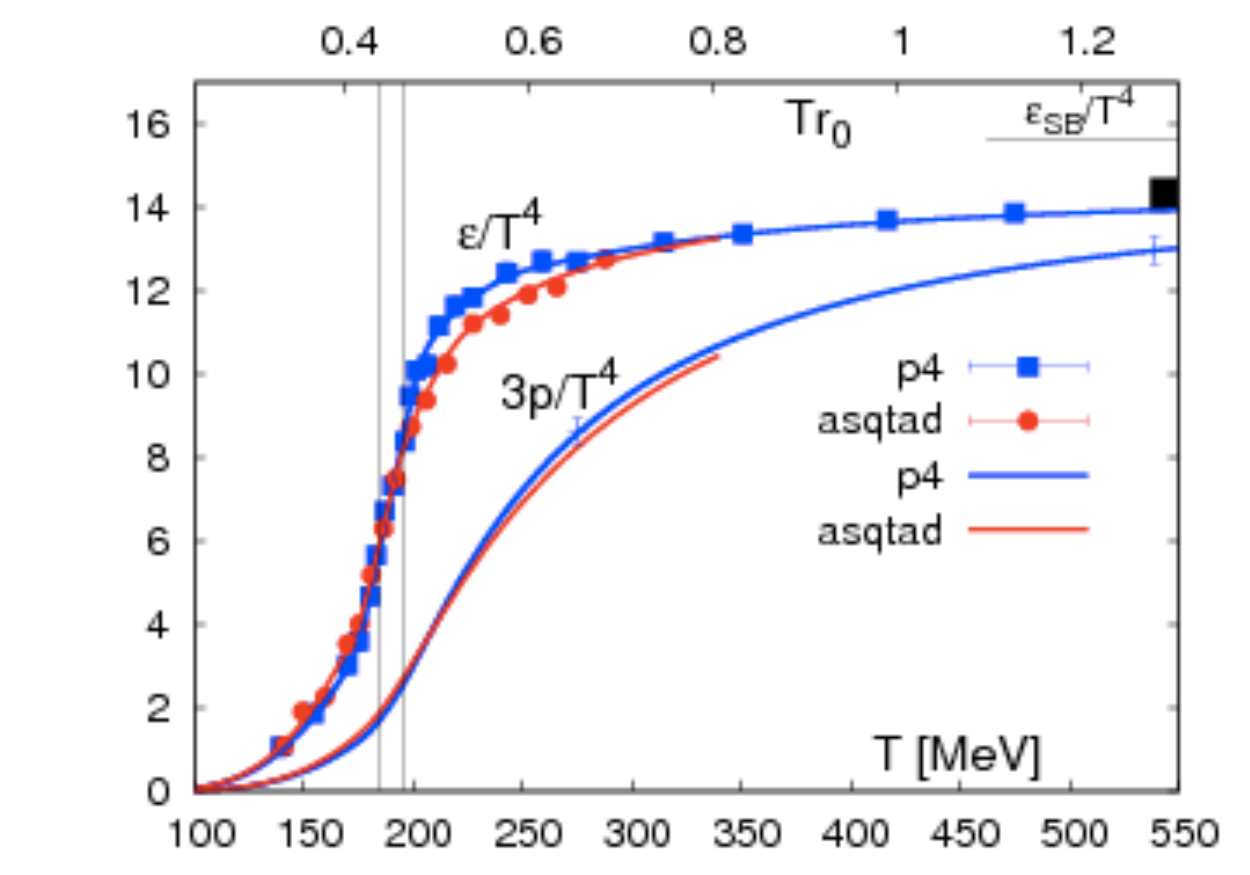}
\caption{Pressure and energy density divided by $T^4$ as a function of temperature in a
  lattice calculation using physical quark masses~\protect\cite{Bazavov:2009zn}.}
\label{fig:bkgr:lattice_eos}
\end{figure}

\begin{figure}
\centering
\includegraphics[width=0.45\textwidth]{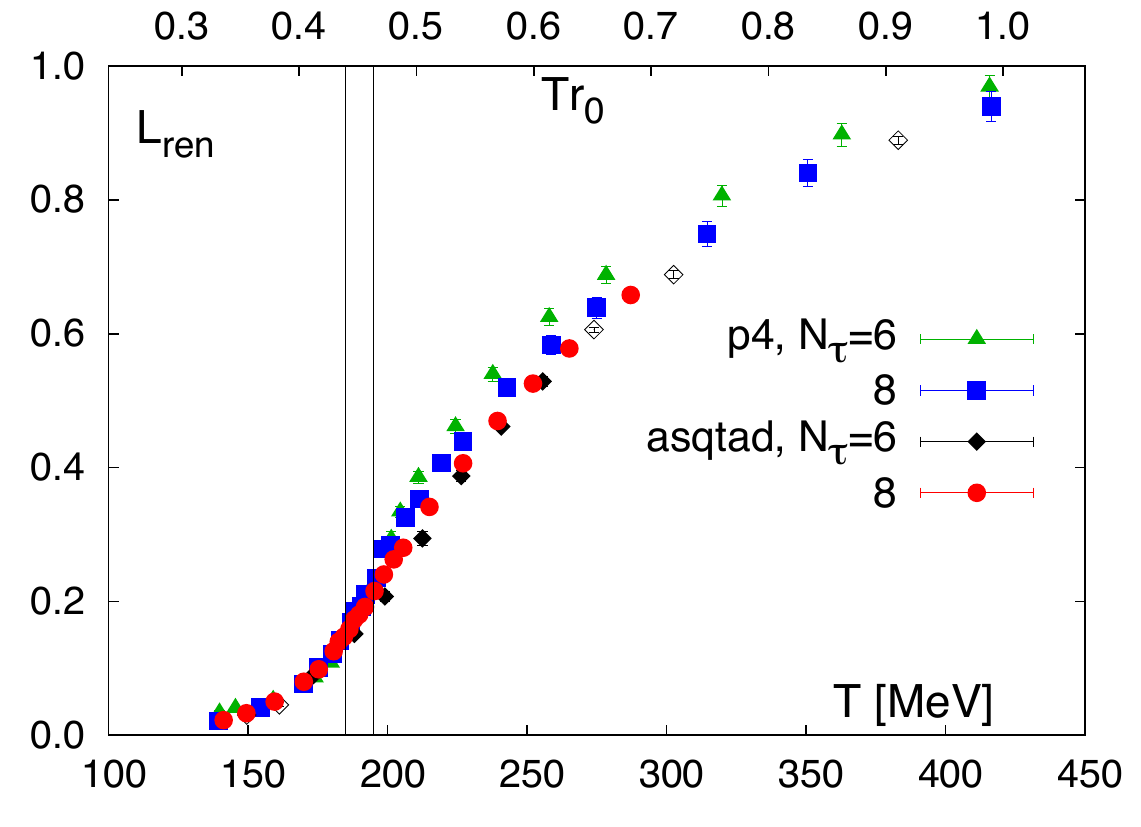}
\includegraphics[width=0.45\textwidth]{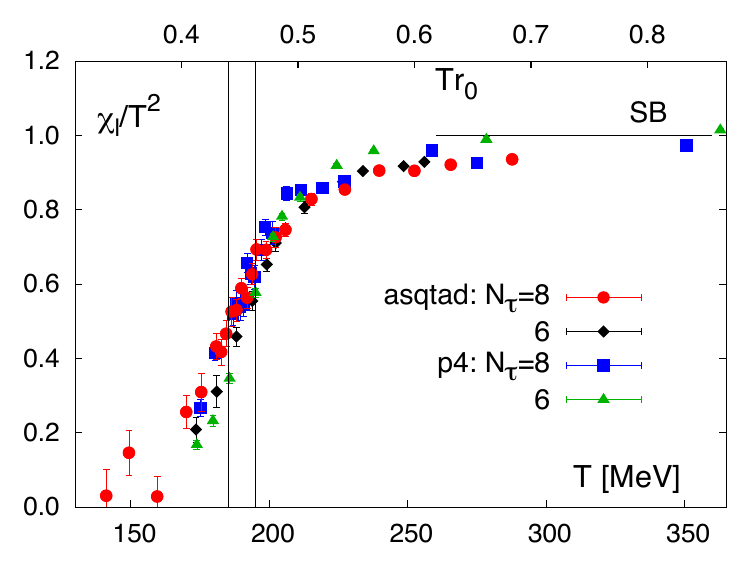}
\caption{Lattice calculations of the $\langle L \rangle$ (left) and
  $\chi/T^2$ (right) for light quarks as functions of
  temperature. These are order parameters for the deconfinement and
  chiral symmetry restoration transitions respectively. The vertical
  lines indicate the temperature range over which the rapid cross over
  is inferred to occur from the equation of state. Figure adapted
  from Ref.~\protect\cite{Bazavov:2009zn}.}
\label{fig:bkgr:lattice_PT}
\end{figure}

The QCD equation of state can be determined on the lattice by
evaluating the trace anomaly (the
trace of the stress-energy tensor $\Theta^{\mu\nu}(T)$) in the grand
canonical ensemble,
\begin{equation}
\dfrac{\Theta^{\mu\mu}}{T^4}=\dfrac{\varepsilon-3p}{T^4}=T\dfrac{\partial}{\partial
  T} \left(\dfrac{p}{T^4}\right)\,.
\end{equation}
Since only expectation values can be calculated, the variables must be
re-expressed in this form. The pressure can be evaluated directly from
integrating the trace anomaly. The entropy density, $s$, and speed
of sound $c_{\mathrm{s}}$, can likewise be calculated from the
relations
\begin{equation}
\dfrac{s}{T^3}=\dfrac{\varepsilon+p}{T^4},\,\,\,\, c_{\mathrm{s}}^2=\dfrac{dp}{d\varepsilon}\,.
\end{equation}
The pressure and energy density divided by $T^4$ from a recent
calculation~\cite{Bazavov:2009zn} are shown in
Fig.~\ref{fig:bkgr:lattice_eos}.
The lattice results indicate a rapid, but continuous transition in
$\varepsilon$ at $T_{\mathrm{C}}\sim 170$~\MeV, which is not
indicative of a first or second order phase transition, but rather a smooth crossover
between phases. Additionally, order parameters describing the
transition to deconfinement and the restoration of chiral symmetry
have been formulated on the lattice. The deconfinement order parameter
is the expectation value of the Polyakov loop, $\langle L
\rangle\sim\exp(-F_{\mathrm{q}}(T)/T)$ where $F_{\mathrm{q}}$ is the
static free quark energy. The chiral transition is described by the
chiral susceptibility $\chi=-\dfrac{\partial}{\partial m} \langle \qqbar
\rangle \Big |_{m=0}$. These quantities involve subtleties relating to
their renormalization and corrections due to finite quark masses. A
calculation of these two quantities~\cite{Bazavov:2009zn} is
shown in Fig.~\ref{fig:bkgr:lattice_PT}. The results indicate that both of
these transitions occur smoothly over the same temperature region
where the equation of state shows a rapid change in energy density.

\subsection{The Hard Thermal Loop Approximation}
\label{section:bkgr:htl}
When using perturbation theory at finite temperature, including the effects of loops becomes problematic as the finite scale
introduced by $T$ breaks the usual perturbative expansion. If the
external momenta are hard, all loops will contribute to the same order
as the tree level diagram. However if the external momenta are soft,
$P\sim gT$, only the contribution from Hard Thermal Loops (HTLs) will
compete with the tree level term. These are cases where the ``hard'' loop
momenta are much greater than the ``soft'' external momenta and the dominant
contribution to the loop comes from momenta of order $T$. In gauge
theories, the HTLs
can be identified by power counting and an effective theory of
resummed HTLs can be derived without affecting the vacuum
renormalizability properties of the theory~\cite{Braaten:1989mz,Braaten:1991gm}. These loops lead to the generation of
thermal masses $m$ and $m_f$ for the gauge bosons and fermions respectively. The static component of the gauge boson propagator
takes the form
\begin{equation}
D_{00}=\dfrac{-i}{\vec{q}^2+2m^2},
\end{equation}
where the thermal mass gives rise to Debye screening. In the static
limit, $\omega/q\rightarrow 0$, the magnetic behavior is similar to
the above expression, but with 
\begin{equation}
m^2\rightarrow\dfrac{\pi m^2}{2}\dfrac{\omega}{q}.
\end{equation}
This can be interpreted as a frequency-dependent dynamical screening
with cutoff $m\sqrt{\dfrac{\pi\omega}{2q}}$, which is in some cases
effective in protecting infrared divergences~\cite{weldon}. The full
thermal masses including the effects from a finite chemical potential,
$\mu$, are given by
\begin{subequations}
\begin{equation}
m^2=\dfrac{1}{6}g^2T^2C_A+\dfrac{1}{12}g^2C_F(T^2+\dfrac{3}{\pi^2}\mu^2),
\end{equation}
\begin{equation}
m_f^2=\dfrac{1}{8}g^2(T^2+\dfrac{1}{\pi^2}\mu^2)C_F,
\end{equation}
\end{subequations}
with the color factors given previously in
Eq.~\ref{eqn:bkgr:color_factors}.

\begin{figure}
\centering
\includegraphics[width=0.75\textwidth]{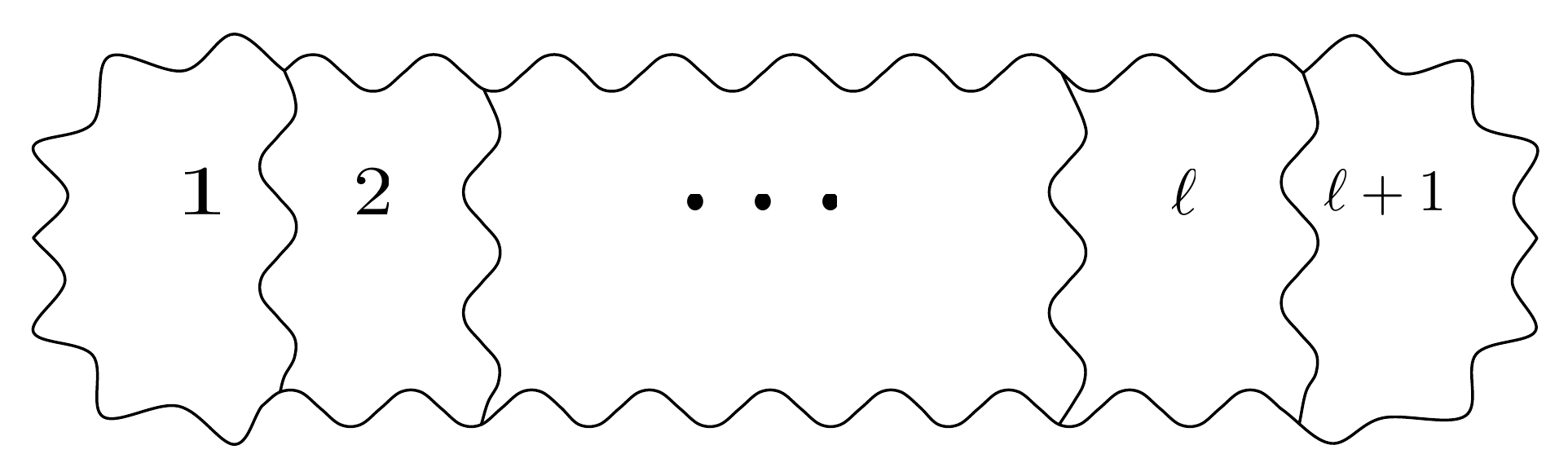}
\caption{Diagram containing $\ell+1$ gluon loops}
\label{fig:bkgr:gluon_ring}
\end{figure}

One issue that is still not fully understood is the screening of the
chromomagnetic fields. Using power counting arguments it can be shown
that a diagram containing $\ell+1$ gluon loops, shown in
Fig.~\ref{fig:bkgr:gluon_ring}, with $\ell>3$, has infrared
divergences of the form~\cite{linde}
\begin{equation}
g^6T^4\left(\dfrac{g^2T}{m}\right)^{\ell-3}.
\end{equation}
For the longitudinal components this is regulated by the electric
mass and gives behavior $\sim g^{\ell+3}T^4$, which differs from the
$g^{2\ell}$ behavior expected from perturbation theory. Based on
arguments from theories that are infrared-equivalent to QCD, the
magnetic mass is of order $\sim g^2T$~\cite{gross:1981py}. This leads to infrared
divergences in the transverse part of the $\ell+1$ gluon loop diagram
to go as $\sim g^{6}T^4$ and signals a total breakdown of perturbation
theory. This remains an outstanding problem, with thermal field theory
not on rigorous theoretical ground except at asymptotically high temperatures.

\subsection{Heavy Ion Collisions}
\label{section:bkgr:HIC}
\begin{figure}[hbtp]
\centering
\includegraphics[width=0.5\textwidth]{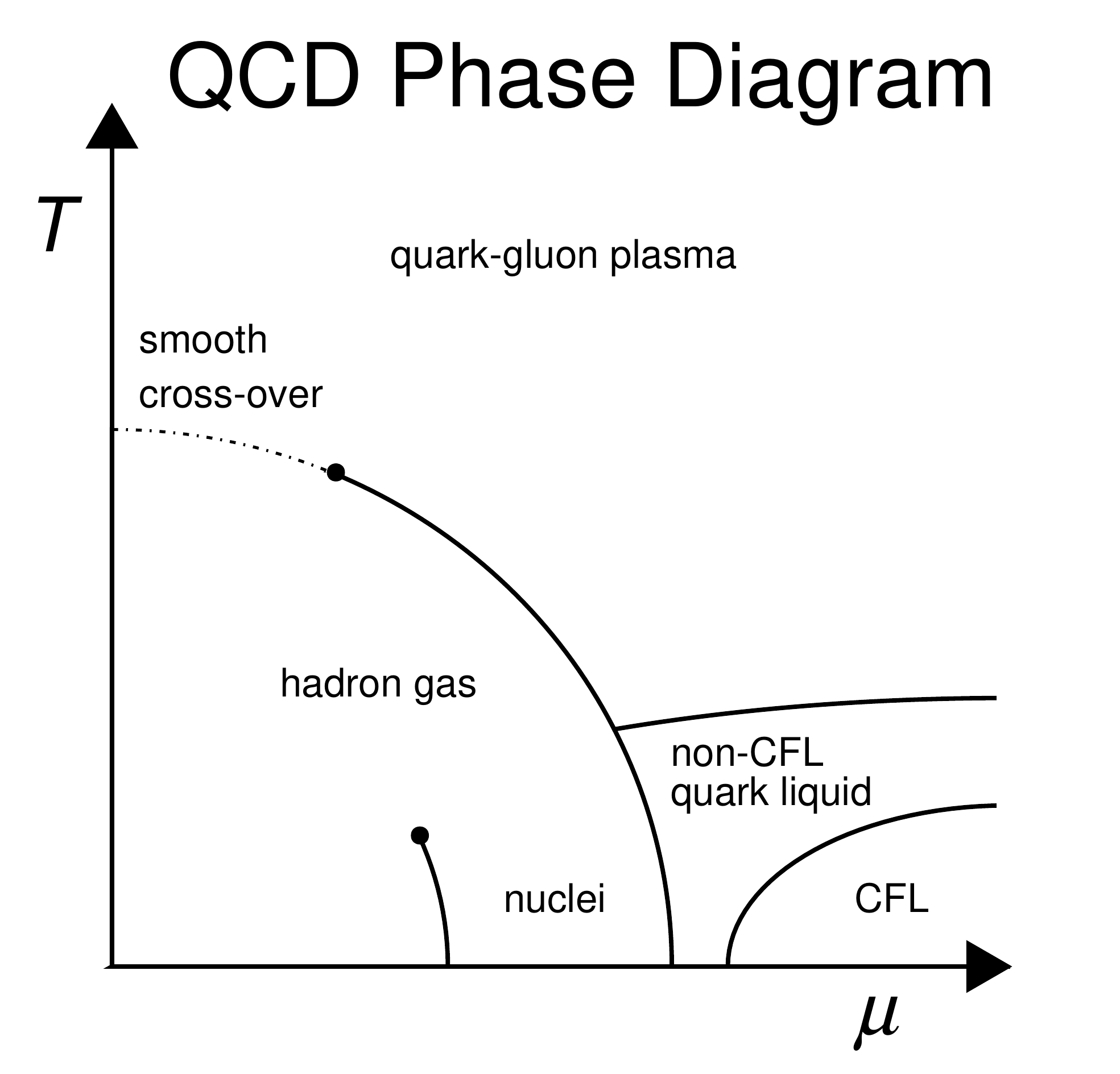}
\caption{The QCD phase diagram. The transition is believed to possess
  a critical point where the first-order phase transition between
  hadron gas to QGP changes to a smooth cross over.}
\label{fig:bkgr:qcd_phase_diagram}
\end{figure}

The phase structure predicted by the lattice indicates that at low
baryon chemical potential and high temperature, hadronic matter
undergoes a rapid, but continuous increase in the number of degrees
of freedom. This transition is illustrated in a possible QCD phase diagram, shown in
Fig.~\ref{fig:bkgr:qcd_phase_diagram}.
 In addition to this transition, abnormal forms
of nuclear matter can exist at high densities~\cite{Lee:1974ma,Lee:1974kn}. Astrophysical objects such as neutron
stars are stabilized against gravitational collapse by the
degeneracy pressure created by an extremely high density of
neutrons. At the highest densities novel states such as quark liquids and
color superconductors (color flavor locked) are thought to
exist~\cite{Rajagopal:2000wf}. 

It is believed that the QGP transition is experimentally accessible in
collisions among heavy ions at high energies. In these collisions the
nuclei appear as highly Lorentz contracted
``pancakes''~\cite{Bjorken:1982qr}. Particle production occurs as the two
pancakes overlap leaving a ``central'' region of high energy density as they
recede and carry away the net baryon density (leading baryon effect). If the system is formed
after some time $t_{f}$, the energy density in
this region can be estimated following the Bjorken procedure by considering the total energy
per unit rapidity in nucleus-nucleus collisions, $\dfrac{d\et}{dy}$. The
energy density is then
\begin{equation}
\varepsilon=\dfrac{d\et}{dy}\Delta y \dfrac{1}{V}.
\end{equation}
For a small slice along the beam direction $\Delta
y=\dfrac{\Delta z}{t_{f}}$, and $V=\pi
R^2_{\mathrm{A}}\Delta z$. $R_{\mathrm{A}}$ is the nuclear radius and
behaves as $R_{\mathrm{A}}\simeq 1.2A^{1/3}$~fm. The energy density
can be expressed as
\begin{equation}
\varepsilon=\dfrac{1}{\pi R_{\mathrm{A}}^2 t_{f}}\dfrac{d\et}{dy}\,.
\end{equation}
For $\mathrm{Au}^{197}$ collisions at RHIC, $\dfrac{d\et}{dy}$ was measured
by PHENIX to be 688~\GeV\ at
$\sqrtsnn=130$~\GeV~\cite{Adcox:2001ry}. If the formation time is on the
order of $0.1-1$~fm, this procedure gives an energy density estimate of
$4.6-46\,\GeV/\mathrm{fm}^3$. The Stefan-Boltzmann equation of
state can be used to infer a temperature,
\begin{equation}
T\simeq\left(\dfrac{\varepsilon (\hbar c)^3}{\alpha}\right)^{1/4},
\end{equation}
where
$\alpha=\dfrac{30}{\pi^2}(16+\dfrac{21}{2}N_f)$ is a constant of
proportionality determined by the effective number of
degrees of freedom. Since the lattice
results indicate that this limit is not reached a more appropriate
estimate would be to use the lattice value of $\alpha=\varepsilon/T^4\simeq
13$, just above the transition region. This estimate gives a
temperature of $128\lesssim T \lesssim 229$~\MeV. The transverse
energy density in \PbPb\ collisions at the LHC at
$\sqrtsnn=2.76$~\TeV\ is larger than at RHIC by roughly a factor of $5$,
indicating a temperature range of $191\lesssim T \lesssim
342$~\MeV. As the temperature region of the phase transition is
spanned by these estimates, it is likely that such a transition is
probed in relativistic heavy ion collisions.

If the mean free path in the interacting system is small compared to the
system size a hydrodynamic description can be applied to the
created nuclear matter~\cite{Landau:1953gs}. The evolution of the
system in the absence of viscosity or heat conduction is described by ideal hydrodynamics,
\begin{equation}
\partial_{\mu}T^{\mu\nu}=0,\,\,\,\,\,\, T^{\mu\nu}=(\varepsilon+p)u^{\mu}u^{\nu}-pg^{\mu\nu},
\end{equation}
where $T^{\mu\nu}$ is the relativistic stress-energy tensor,
$u^{\mu}$ is the local four-velocity of a fluid element and the system
of equations is closed by a thermodynamic equation of state. The Bjorken picture
considers the longitudinal expansion of this medium which leads to a
slow decrease in the temperature, $T\sim
\tau^{-1/3}$, where $\tau$ is the local fluid proper time, and
constant entropy inside the co-moving fluid volumes.

Local anisotropies in the transverse density of produced particles can
be produced through initial-state fluctuations or more commonly
through the elliptical geometry of the collision zone when there is
incomplete nuclear overlap. As the particle production is isotropic,
the only way this can be converted to a momentum-space anisotropy, is if these
density fluctuations form pressure gradients, which result in radial
flow. Such an effect should be observable in the angular distribution
of hadrons and may allow for a determination of the fluid's shear
viscosity, $\eta$.

Calculational techniques in string theory have established a
correspondence between quantum field theories and their dual gravity
theories. This AdS/CFT correspondence facilitates a translation between
quantities in a gravity calculation in the weakly coupled region to
quantities in the strongly coupled regime of the appropriate dual~\cite{Maldacena:1997re,Aharony:1999ti}. In
particular the thermodynamics of black hole can be mapped to 
thermodynamics in QCD-like $\mathcal{N}=4$ supersymmetric Yang-Mills
theory. Some time ago, arguments based on the uncertainty principle suggested that
there is a quantum lower limit to the viscosity~\cite{Danielewicz:1984ww}. This has been
demonstrated explicitly in the AdS/CFT picture where a limiting value
of the viscosity to entropy density ratio was found to be $\eta/s \leq 1/4\pi$~\cite{Policastro:2001yc,Kovtun:2003wp}.

This application of hydrodynamics illustrates how the transport properties of the system can provide
further insight into the dynamics of the medium, in particular through
transport coefficients such as the shear viscosity, which are
sensitive to the microscopic behavior. As a transport phenomenon, hydrodynamics coincides with the long-wavelength
behavior of the system and the relevant transport and other phenomena such as radiation and
diffusion can provide access to other aspects of the medium.

\section{Hard Processes in Nuclear Collisions}
\label{section:bkgr:hard_AA_collisions}
\subsection{The Glauber Model}
\label{section:bkgr:glauber}
In collider experiments, the measured rates of quantities
are proportional to the luminosity, or instantaneous flux of colliding
particles. At high energies, the nuclear size is easily resolved, and
a beam constituent not only sees a flux of nuclei but an instantaneous flux of nucleons/partons determined
by the geometric overlap between the colliding nuclei. For two nuclei, $\mathrm{A}$ and
$\mathrm{B}$, with centers separated by impact parameter $\mathbf{b}$, this flux
is represented by the factor $\TAB(\mathbf{b})$. Many nuclear effects are
expected to vary with $\mathbf{b}$, and \TAB\ is often used to normalize
quantities to remove any variation due to trivial geometric
effects. Experimentally, events are grouped into different
centrality classes, which are determined to have a similar amount of
geometric overlap based on a set of global event criteria. The average
\TAB\ is used to normalize quantities observed in those events to
assess the centrality dependence of an observable. This procedure
makes use of a framework known as the Glauber Model~\cite{Glauber:1959wd,Glauber:1970jm,Czyz:1969jg} and
is employed by nearly all heavy ion experiments~\cite{Miller:2007ri}.

\begin{figure}[hptb]
\centering
\includegraphics[width=0.75\textwidth]{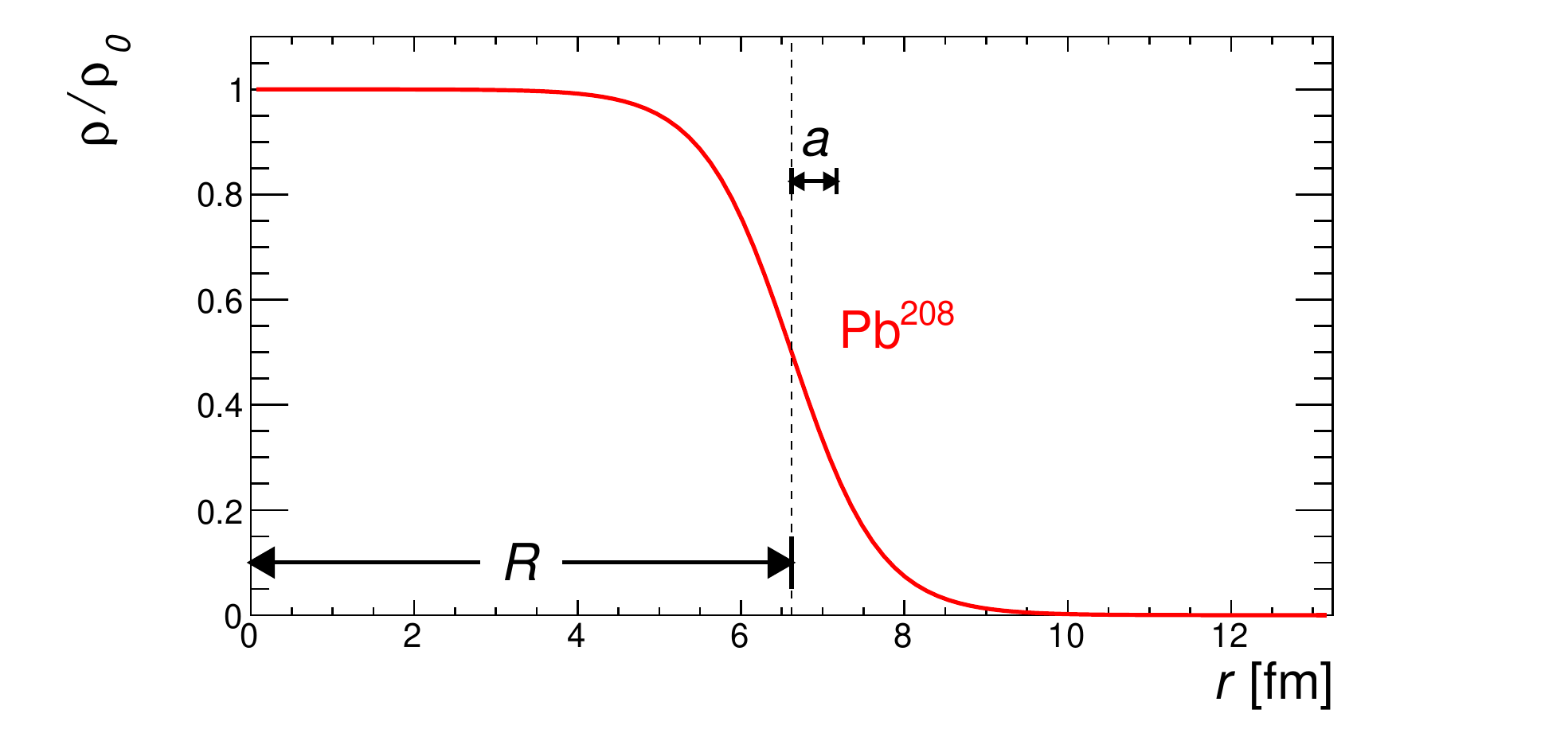}
\caption{Woods-Saxon distribution for $\mathrm{Pb}^{208}$.}
\label{fig:bkgr:pb_woods_saxon}
\end{figure}

The Glauber Model treats the incoming nuclei as smooth distributions
of nucleons, each traveling on independent linear trajectories. It is
formulated in the optical limit, in which the overall phase shift is
evaluated by summing over all per-nucleon pair phase shifts. It relies
on a parametrization of the nuclear density, $\rho$, typically a two-parameter Woods-Saxon distribution,
\begin{equation}
\rho(r)=\rho_0\dfrac{1}{1+\mathrm{exp}\left(\dfrac{r-a}{R}\right)}\,,
\label{eqn:bkgr:woods_saxon}
\end{equation}
where $R$ and $a$ are experimentally determined parameters,
describing the radius and skin depth of the nucleus. The
constant $\rho_0$ is an overall normalization factor ensuring the
distribution is normalized to the number of nucleons. For $\mathrm{Pb}^{208}$ these
parameters are $R=6.62\pm0.06$~fm, $a=0.546\pm0.01$~fm~\cite{DeJager:1987qc}
and the distribution is shown in Fig~\ref{fig:bkgr:pb_woods_saxon}.

\begin{figure}[hptb]
\centering
\includegraphics[width=0.75\textwidth]{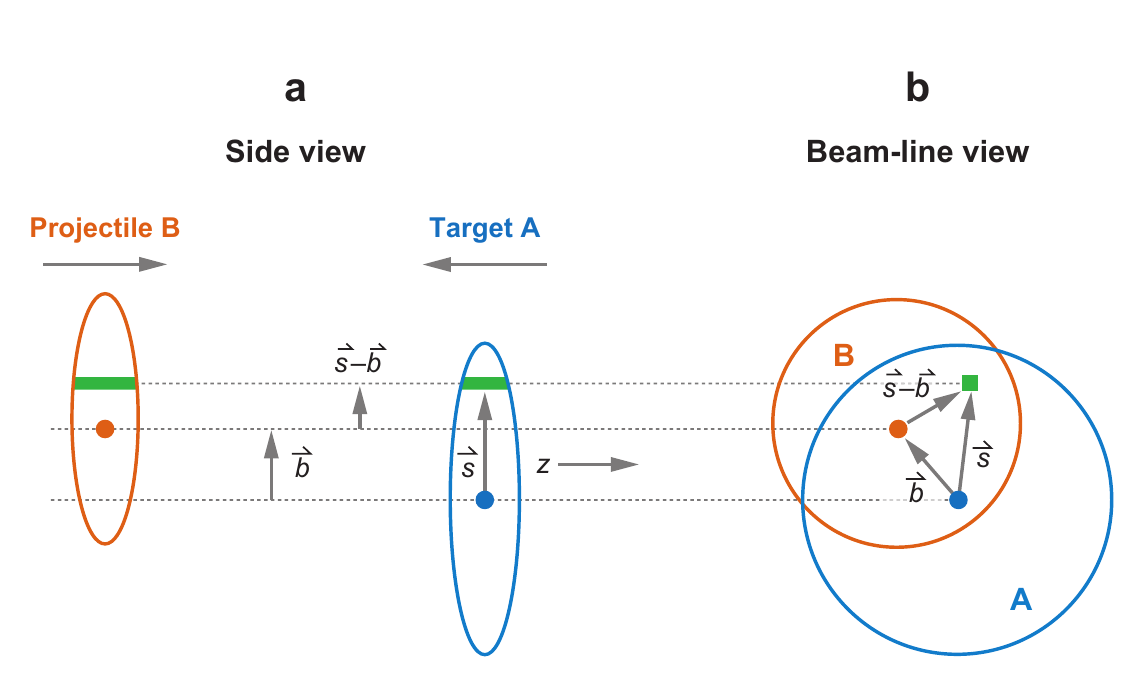}
\caption{Views of the collision system transverse (left) and parallel
(right) to the beam axis~\protect\cite{Miller:2007ri}.}
\label{fig:bkgr:glauber_geometry}
\end{figure}

In the Glauber formalism the transverse density,
\begin{equation}
\TA(\mathbf{b^{\prime}})=\int dz_{\mathrm{A}} \rho_{\mathrm{A}}(\mathbf{b^{\prime}},z_{\mathrm{A}})
\end{equation}
is the expected number of nucleons at position
$\mathbf{b^{\prime}}$. Then \TAB\ is defined as $AB$ times the probability to
simultaneously find nucleons
in nuclei $\mathrm{A}$ and $\mathrm{B}$ at the same position $\mathbf{b}$ as shown in
Fig.~\ref{fig:bkgr:glauber_geometry},
\begin{equation}
\TAB(\mathbf{b})=\int d^2 b^{\prime}\TA(\mathbf{b^{\prime}}) \TB(\mathbf{b^{\prime}}-\mathbf{b})\,.
\end{equation}
Each nucleon in $\mathrm{A}$ can interact with a nucleon in $\mathrm{B}$ with
probability $p_{\mathrm{AB}}=\dfrac{\TAB\sigmaNNinelastic}{AB}$, so the total probability of $n$
collisions is binomial and is given by
\begin{equation}
P(n,\mathbf{b})=\binom{n}{k} p_{\mathrm{AB}}^n \left(1-p_{\mathrm{AB}}\right)^{AB-n}.
\end{equation}
This allows for the definition of the expected number of collisions as
$\Ncoll=\TAB\sigmaNNinelastic$, and the expected number of participants \Npart, the
total number of nucleons that participated in any scatterings
(sometimes referred to as wounded nucleons) as
\begin{eqnarray}
\Npart&=&
\int
d^2b^{\prime}\TA(\mathbf{b^{\prime}})\left[1-\left(1-\dfrac{\TB(\mathbf{b^{\prime}}-\mathbf{b})}{B}\right)^B\right]\\ \nonumber
&+&\int d^2b^{\prime}\TB(\mathbf{b^{\prime}})\left[1-\left(1-\dfrac{\TA(\mathbf{b^{\prime}}-\mathbf{b})}{A}\right)^A\right]\,.
\end{eqnarray}

\begin{figure}[hptb]
\centering
\includegraphics[width=0.75\textwidth]{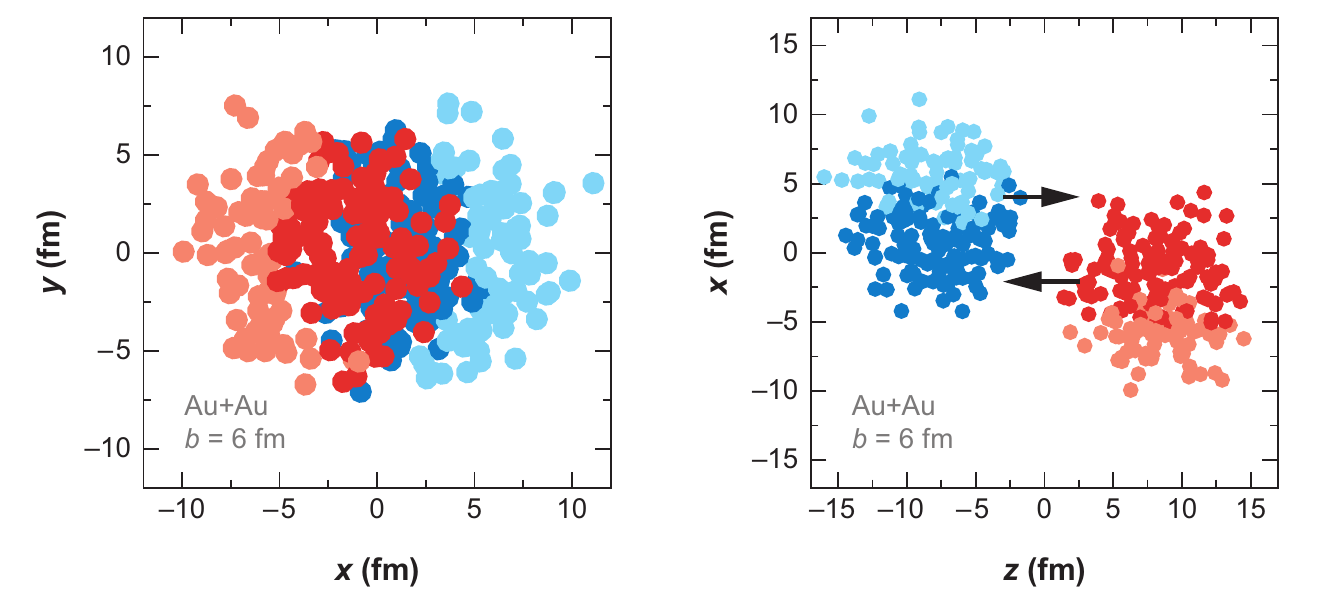}
\caption{Distributions of nucleons generated with the MC Glauber
procedure in the $x-y$ (left) and $z-x$ (right) planes for
$\mathrm{Au}^{197}$. The two nuclei are shown in different colors,
with the participants shown in a darker color~\protect\cite{Miller:2007ri}.}
\label{fig:bkgr:glauber_mc}
\end{figure}

\begin{figure}[hptb]
\centering
\includegraphics[width=0.75\textwidth]{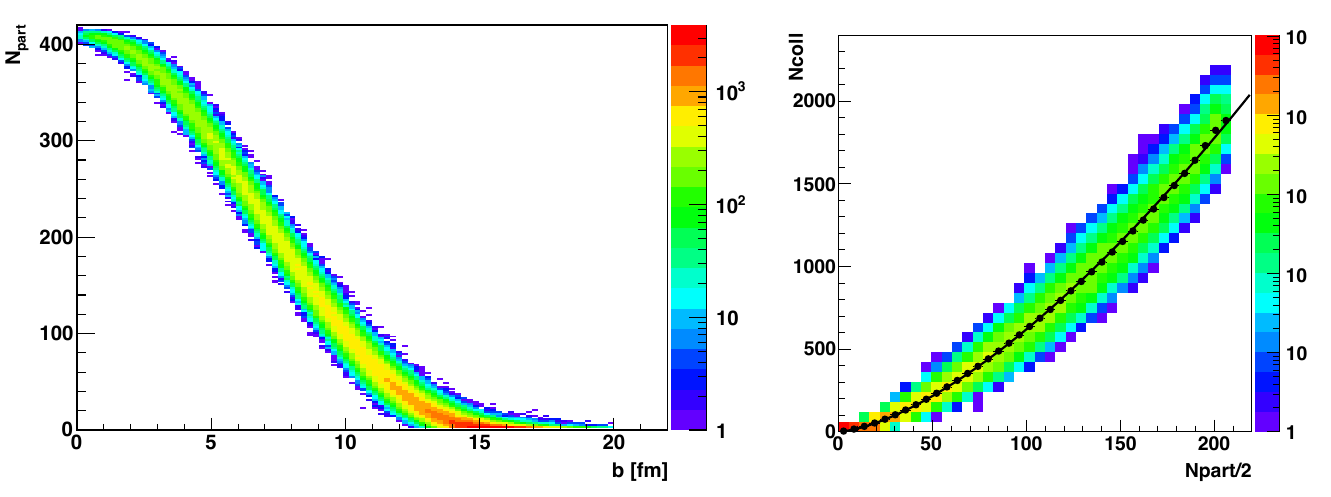}
\caption{Distribution of \Npart\ values as a function of $b$ (left)
and the correlation between \Npart\ and \Ncoll\ (right) for \PbPb\ collisions.}
\label{fig:bkgr:glauber_b_npart_ncoll}
\end{figure}

An alternative to performing the analytic integrals is to use Monte
Carlo techniques. This has the advantage of including terms neglected
in the optical approximation that incorporate local per-event density
fluctuations~\cite{Bialas:1976ed,Miller:2007ri}. This method is
performed by sampling the full Woods-Saxon distribution in
Eq.~\ref{eqn:bkgr:woods_saxon} $\mathrm{A}$ times to populate positions for
nucleus $\mathrm{A}$. To prevent overlap, a position is regenerated if it is
found to be within some minimum distance of a previously generated
nucleon. Once the positions have been generated for both nuclei, a
random impact parameter vector is generated defining an offset between
the nuclear centers. The transverse position for all nucleons
in nucleus $\mathrm{A}$ is compared to each of the analogous nucleons in $\mathrm{B}$. If the distance between the pair is $\Delta r <
\sqrt{\sigmaNNinelastic/\pi}$, the nucleons are considered to have
participated. \Ncoll\ is defined as the number of times this condition
is satisfied, with \Npart\ defining the number of nucleons for which
this condition was satisfied at least once. An example of an event
generated with this technique is shown for an \AuAu\ collision in
Fig~\ref{fig:bkgr:glauber_mc}.
The distribution of \Npart\ values for different impact parameters is
shown in Fig.~\ref{fig:bkgr:glauber_b_npart_ncoll}, as well as the
relationship between \Ncoll\ and \Npart.

\begin{figure}[hptb]
\centering
\includegraphics[width=0.45\textwidth]{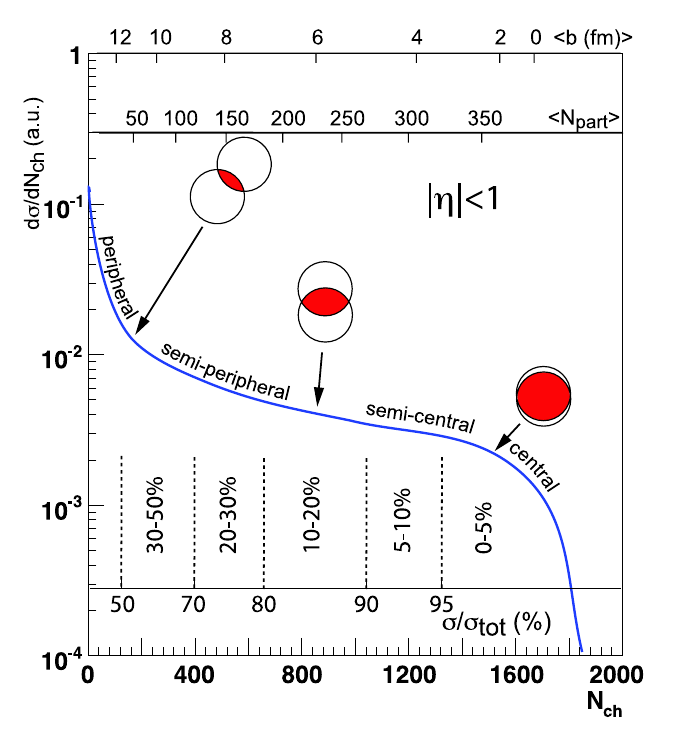}
\caption{Schematic diagram from Ref.~\protect\cite{Miller:2007ri} of a
  distribution of a centrality variable, in this case the number of
  charged particles in the interval $|\eta|<1$. The events contributing
to the upper 5\% of the integral of the distribution are the 0-5\%
centrality bin, with near complete nuclear overlap. The parameters
from a Glauber simulation, $\langle b\rangle$, \ANpart, are shown as different horizontal scales.}
\label{fig:bkgr:centrality_schematic}
\end{figure}

Experimentally, the per-event impact parameter is not measurable so a
procedure must be performed to relate a distribution of some
measurable quantity
to the Glauber parameters. Variables like the total charged particle
multiplicity or transverse energy typically have distributions similar
to the \Npart\ and \Ncoll\ distributions. This feature, combined with
the fact that these are global variables and therefore less sensitive
to detailed features and fluctuations make them excellent choices for
centrality variables. The centrality determination procedure considers
the minimum bias distribution of such a variable, $\zeta$, and divides the range
of observed values into sub-ranges where the integral of the
distribution over that range is some percentage of the total; an example of this division is shown in
Fig~\ref{fig:bkgr:centrality_schematic}. These
sub-ranges are called centrality intervals, and centrality-dependent
observables are usually calculated by averaging over all events in the
same interval.

The $i^{\mathrm{th}}$ centrality bin, $\zeta_i < \zeta < \zeta_{i+1}$,
is typically defined in terms of percentages, $a-b\%$, such that:
\begin{equation}
\int_{\zeta_{\mathrm{min}}}^{\zeta_{i}} d\zeta \dfrac{1}{\Nevt}\dfrac{d\Nevt}{d\zeta}
=a\%,\,\,\,\,\,
\int_{\zeta_{i+1}}^{\zeta_{\mathrm{max}}} d\zeta \dfrac{1}{\Nevt}\dfrac{d\Nevt}{d\zeta} =b\%\,.
\end{equation}
The Glauber model parameters must then be related to the class of events in a
particular centrality bin. This can be accomplished by constructing a
new variable, $\xi$, from the Glauber variables \Ncoll\ and \Npart\ such
that the $\xi$ distribution is similar to the experimentally observed
$\zeta$ distribution. The $\zeta$ distributions tend to be similar enough to
the \Ncoll\ and \Npart\ distributions that a simple linear combination
of the two, a two-component model, is a suitable choice for $\xi$,
\begin{equation}
\xi=\xi_0\left(x\dfrac{\Npart}{2}+(1-x)\Ncoll\right) .
\label{eqn:background:two_comp_model}
\end{equation}
The parameters $\xi_0$ and $x$ can be
determined from fitting the measured $d\Nevt/d\zeta$ distribution with
$d\Nevt/d\xi$ from an MC Glauber sample. The centrality bins
for $\xi$ are defined in terms of integral fractions just as for
$\zeta$. In the $i^{\mathrm{th}}$ centrality bin, the \Ncoll\ and
\Npart\ are averaged over all events for which  $\xi_i < \xi <
\xi_{i+1}$. These \ANcoll\ and \ANpart\ values are then associated
with the $i^{\mathrm{th}}$ centrality bin in the data.

\subsection{Nuclear Modification}
\label{section:bkgr:nuclear_modification}
The first evidence for modification in nuclear collisions at high
energy was observed \pA\ collisions, where it was found that the production of particles at large transverse momentum ( $2\lesssim \pt\lesssim
6$~\GeV) was enhanced in these collisions relative to \pp~\cite{Cronin:1974zm,Antreasyan:1978cw}. This effect, Cronin
enhancement, was originally interpreted as additional transverse
momentum, $\kt^2\propto L$, imparted by additional, independent, elastic interactions
with multiple nucleons. Here $L$ is the average path length, which
scales with atomic number as $L\propto A^{1/3}$.

The nuclear parton
distribution functions (NPDFs), show both a suppression at lower
values of $x$ and an enhancement with increasing $x$ with respect to
the nucleon PDFs, termed shadowing and
anti-shadowing respectively. This nuclear modification is quantified
through the ratio,
\begin{equation}
R^{\mathrm{A}}_{i}=\dfrac{f^{\mathrm{A}}_i}{f_i},
\label{eqn:bkgr:NPDF_modification}
\end{equation}
where $f$ are the PDFs for the nucleus $\mathrm{A}$ and the nucleon, and $i$ is
the parton species: valence quark, sea quark or gluon.
When viewed in the rest
frame of the target nucleus, the shadowing/anti-shadowing is
the result of multiple scattering processes that
destructively/constructively interfere at the amplitude level~\cite{Qiu:2004da}. In
the collinear factorized approach coherent multiple scattering terms
are suppressed by powers of $1/Q^2$, however for a large nucleus these
contributions receive an enhancement of $A^{1/3}$, leading to
sensitivity to higher twist effects~\cite{Qiu:2001hj}.

When viewed in the infinite
momentum frame, the shadowing effects arise from recombination of
low-$x$ gluons in the nuclear wave
function~\cite{Mueller:1985wy}. This phenomenon is known as saturation
and sets in as unitarity requirements force the nominal evolution
equations to be modified at small $x$. The gluon distribution,
$xG(x,Q^2)$, represents the number of gluons per unit rapidity in a region of
transverse size $1/Q^2$ and grows as $\ln 1/x$ for small $x$ at fixed
$Q^2$ without this modification. However, as $Q^2R^2\lesssim xG(x,Q^2)$, where $R$ is the radius
of a nucleon, the gluons will begin to overlap and recombination of
these gluons will limit the growth of $xG(x,Q^2)$ with decreasing $x$~\cite{Gribov:1981ac,Gribov:1984tu}.

\begin{figure}[tbhp]
\centering
\includegraphics[width=0.75\textwidth]{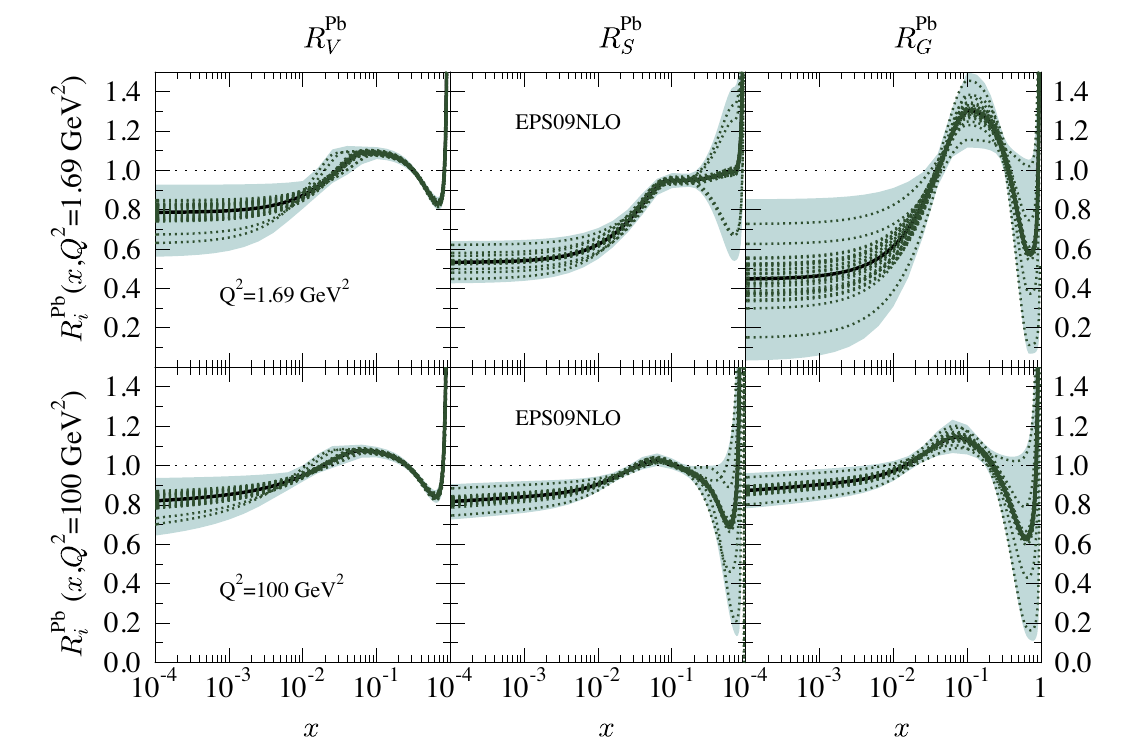}
\caption{Nuclear modification ratios $R^{\mathrm{Pb}}_{\mathrm{v}}$,
$R^{\mathrm{Pb}}_{\mathrm{s}}$ and $R^{\mathrm{Pb}}_{\mathrm{g}}$ for
the valance, sea and gluon PDFs for two values of
$Q^2$~\protect\cite{Eskola:2009uj}. For Pb the shadowing region is 
$x\lesssim 0.05$, the anti-shadowing region is $0.05\lesssim
x\lesssim 0.5$ and the EMC region is $x\gtrsim 0.5$.}
\label{fig:bkgr:EPS09}
\end{figure}

At larger values of $x$, away from the shadowing/anti-shadowing
region ($x\gtrsim 0.5$), the NPDFs exhibit additional suppression which is known as the
EMC effect~\cite{Geesaman:1995yd}. The origin of this effect may be related to non-nucleon degrees of freedom in the
nucleus~\cite{Frankfurt:1988nt} and has led to increased interest in
short-range correlations between nucleons. A sharp enhancement of the
NPDF at the largest $x$ values is thought to be described by the Fermi
motion of the nucleons. Global analyses at NLO have been performed
using data from DIS, Drell-Yan and \dAu\ collisions at RHIC to extract
the $x$-dependence of $R^{\mathrm{A}}_{\mathrm{v}}$,
$R^{\mathrm{A}}_{\mathrm{s}}$ and $R^{\mathrm{A}}_{\mathrm{g}}$ at
multiple $Q^2$ values~\cite{Eskola:2009uj}. These distributions, shown
for Pb in Fig.~\ref{fig:bkgr:EPS09}, are a crucial input to any
interpretation of any high-\pt\ phenomena observed in nucleus-nucleus collisions.
\subsection{HIJING}
The phenomenology underlying the
implementation of \pp\ event generators such as PYTHIA, requires extensions to be
appropriate for nucleus-nucleus collisions. An important step in
achieving this was the development of the HIJING MC event
generator~\cite{Wang:1991hta}. To correctly model the multiplicity fluctuations the soft production
mechanism combines elements of different models. It introduces
multiple, sea \qqbar\ strings as in the dual parton model~\cite{Capella:1992yb,Ranft:1987xn,Ranft:1986ut}, but
also allows for induced gluon bremsstrahlung by introducing string
kinks as in the Lund FRITIOF procedure~\cite{Andersson:1986gw,NilssonAlmqvist:1986rx}. Additional transverse
momentum kicks are applied that are dependent on the particle's \pt\
proportional to $(\pt^2+a^2)^{-1}(\pt^2+p_0^2)^{-1}$.

The production
of ``minijets'' with $\pt \geq p_0 \sim2$~\GeV\ is expected to play an
important role in the total energy and particle
production~\cite{Kajantie:1987pd,Eskola:1988yh}. These processes take
place at an intermediate momentum scale, lower than that associated
with typical jet production, but still describable by perturbative
QCD. Production of multiple jets in HIJING is implemented through a
probabilistic model of independent jet (pair) production. The average
number of minijets at impact parameter $b$ is given by
$\sigma_{\mathrm{jet}}T_{\mathrm{N}}(b)$, where
$\sigma_{\mathrm{jet}}$ is the
inclusive cross section for jets in nucleon-nucleon collisions, integrated above some threshold
$p_0$. The probability for multiple independent minijet production is given by~\cite{Wang:1990qp},
\begin{eqnarray}
&g_j(b)=\dfrac{(\sigma_{\mathrm{jet}}T_{\mathrm{N}}(b))^j}{j!}e^{-\sigma_{\mathrm{jet}}T_{\mathrm{N}}(b)},\,\,
j\geq 1\,,\\
&g_0(b)=\left[1-e^{-\sigma_{\mathrm{soft}}T_{\mathrm{N}}(b)}\right]
e^{-\sigma_{\mathrm{jet}}T_{\mathrm{N}}(b)}\,,
\end{eqnarray}
where $\sigma_{\mathrm{soft}}$ is the non-perturbative inclusive 
cross section for soft processes and $T_{\mathrm{N}}(b)$ is the
partonic overlap function between two nucleons separated by impact
parameter $b$. An eikonal picture is used to relate the inelastic,
elastic and total cross sections and to define the multi-jet probabilities
for minijets in nucleon-nucleon collisions,
\begin{eqnarray}
&G_0=\dfrac{\pi}{\sigma_{\mathrm{in}} }\int_0^{\infty} d^2b
\left[1-e^{-2\chi_{\mathrm{S}}(b,s)}\right]e^{-2\chi_{\mathrm{H}}(b,s)}
\\
&G_0=\dfrac{\pi}{\sigma_{\mathrm{in}} }\int_0^{\infty} d^2b
\dfrac{\left[2\chi_{\mathrm{H}}(b,s)\right]^j}{j!}
e^{-2\chi_{\mathrm{H}}(b,s)} \,.
\end{eqnarray}
Here the eikonal factors are related by
$\chi_{\mathrm{S}}+\chi_{\mathrm{H}}=\chi_{0}(1+\sigma_{\mathrm{jet}}/\sigma_{\mathrm{soft}})$,
and ensure geometric scaling. The independence of the multi-jet
production is an appropriate ansatz if $\sigma_{\mathrm{jet}} \lesssim
2 A^{-2/3} (p_0R_{\mathrm{N}})^2\sigma_{\mathrm{inel}}$. The overall
jet cross section is reduced from the nucleon-nucleon case to include
the effects of nuclear shadowing, including the impact parameter
dependence which has not been measured by experiment. The modification factor defined in
Eq.~\ref{eqn:bkgr:NPDF_modification} can be generalized to include
impact parameter dependence,
\begin{equation}
R_{\mathrm{A}}(x,Q^2,e)=R_{\mathrm{A}}^0 (x,Q^2)+\alpha(r)R_{\mathrm{A}}^{\mathrm{S}}(x,Q^2),
\end{equation}
where
$\alpha(r)\propto(A^{1/3}-1)\sqrt{1-r^2/R_{\mathrm{A}}^2}$,
models the impact parameter dependence~\cite{Wang:1991hta}. Then the
effective jet cross section is a function of the transverse positions
of each of the colliding nucleons in the binary system,
\begin{equation}
\sigma_{\mathrm{jet}}^{\mathrm{eff}}(r_{\mathrm{A}},r_{\mathrm{B}})
=\sigma_{\mathrm{jet}}^0
+\alpha_{\mathrm{A}}(r_{\mathrm{A}})\sigma_{\mathrm{jet}}^{\mathrm{A}}
+\alpha_{\mathrm{B}}(r_{\mathrm{B}})\sigma_{\mathrm{jet}}^{\mathrm{B}}
+\alpha_{\mathrm{A}}(r_{\mathrm{A}}) \alpha_{\mathrm{B}}(r_{\mathrm{B}})\sigma_{\mathrm{jet}}^{\mathrm{AB}}.
\end{equation}

The event generation proceeds by using an MC Glauber setup to
determine the set of colliding nucleon pairs. For each of these binary
collisions, the probability of scattering and number of jets is
determined, along with whether the collision is elastic or
inelastic. Hard scattering partons are treated separately, and their
energies are subtracted from the nucleons with the remaining energy
used in processing soft string excitations. The resulting scattered
gluons are ordered in rapidity and color-connected to the valence
quark/di-quark of the nucleon. The correlated semi-hard particle
production mechanism is the key feature of HIJING and is why it has
remained a useful tool long after its inception.

\subsection{Jet Quenching}
\label{section:bkgr:quenching}
A significant form of nuclear modification occurs in heavy ion
collisions where the byproducts of hard scatterings can interact with the
QGP through the phenomenon of jet quenching. The character of this
interaction provides key insight into the dynamics of the medium. The
most important is the identification of the relevant scales for the jet-medium interaction and whether it can be described
by perturbative QCD. If so, it provides a key testing ground for
thermal, perturbative QCD and the HTL formulation as well as an
example of the transport phenomena of radiation and diffusion in a
fundamental physical system. 

The potential for jets as a tool to study the plasma was first
recognized by Bjorken~\cite{Bjorken:1982tu}, who suggested that events
with back-to-back jets would be sensitive to differential energy
loss if the two partons had different in-medium path lengths. In
extreme cases a highly energetic jet may emerge while its partner
deposits all of its energy in the medium, which would represent a
striking experimental signature.

The energy loss mechanism originally proposed by Bjorken was through
elastic collisions with the medium constituents. This was originally given in
Ref.~\cite{Bjorken:1982tu} as, 
\begin{equation}
\dfrac{dE}{dx}=C_{\mathrm{R}}\pi
\alphas^2T^2\left(1+\dfrac{N_f}{6}\right)\ln \dfrac{4ET}{m_{\mathrm{D}}^2}\,,
\end{equation}
with more intricate forms of the term inside the logarithm due to
improvements in the collision integral given in
Refs.~\cite{Thoma:1990fm,Braaten:1991we,Thoma:1991ea}. In a QCD
plasma, the Debye screening mass, $m_{\mathrm{D}}$, is given by,
\begin{equation}
m_{\mathrm{D}}=(1+\dfrac{1}{6}N_{f})g^2T^2\,.
\label{eqn:bkgr:debye_def}
\end{equation}

Some of the first predictions of jet quenching signatures involve a
modification of the dijet acoplanarity distribution in heavy ion
collisions~\cite{Appel:1985dq,Blaizot:1986ma}. These calculations
were formulated in terms of the probability distribution for dijet
pairs to have momentum imbalance $K_{\eta}$,
\begin{equation}
\dfrac{dP}{dK_{\eta}}\equiv\dfrac{1}{\sigma_0(\pt)}\dfrac{1}{\pt}\dfrac{d\sigma}{d\phi}\,.
\label{eqn:bkgr:acoplanarity}
\end{equation}
The acoplanarity was expected to show a
temperature-dependent modification due to elastic collisions with the medium
constituents. It was shown that similar effects could be
produced through collisional energy loss in a hadronic resonance
gas~\cite{Rammerstorfer:1990js}, and thus an observed modification
would not prove the presence of a QGP phase.

Interactions with the medium can also induce radiative energy loss through
the emission of bremsstrahlung gluons. In QED, energy loss for high
energy electrons is typically in the Bethe-Heitler regime
$\dfrac{dE}{dx}=-\dfrac{E}{L}$, where $L$ is a characteristic
length. Classically, the total energy loss per scattering is the
integral of the bremsstrahlung spectrum
\begin{equation}
\Delta E=\int d\omega d\kt^2 \omega\dfrac{dI}{d\omega d\kt^2}\,
\end{equation}
with multiple scatterings adding incoherently to give a total energy
loss
\begin{equation}
\Delta E^{\mathrm{tot}}=N\int d\omega d\kt^2 \omega \dfrac{dI}{d\omega d\kt^2},
\end{equation}
where $N=L/\lambda$ is the medium opacity and $\lambda$ is the mean
free path between scattering centers.

Most models use a formalism that treats the medium as a series of
static scattering centers with the parton and radiated gluons
with energies, $E$
and $\omega$ respectively, traveling along eikonal trajectories. This
is combined with the kinematic limits $q_{\perp} \muchless \omega \muchless E$,
where  $q_{\perp}$ is the momentum transfer with the medium. This
kinematic regime is
referred to as the soft eikonal limit~\cite{Armesto:2011ht}. The
radiation spectrum is typically derived by using the single gluon emission
kernel as a Poisson probability for multi-gluon emission. 

At high energies, significant
interference occurs between the parton and quanta emitted at small angles. This
results in a finite formation time for the radiation and suppresses this
contribution relative to incoherent radiation, known as the LPM
effect~\cite{Landau:1953um,Migdal:1956tc} in QED. The QCD analog of
this phenomenon has been
proposed as an important feature of the quenching
mechanism~\cite{Gyulassy:1990ye,Gyulassy:1993hr} and was shown by
BDMPS~\cite{Baier:1996kr} and independently by
Zakharov~\cite{Zakharov:1996fv} to give an energy loss that grows
quadratically with path length in the medium $\Delta E\propto
L^2$. 

As the interference
suppresses the coherent radiation, the emission spectrum will be
dominated by those quanta which have decohered. These are gluons which
have acquired a phase, $\varphi$, of order unity~\cite{Baier:2002tc,Salgado:2003gb},
\begin{equation}
\varphi=\Big \langle \dfrac{q_{\perp}^2 \Delta z}{2\omega} \Big \rangle \sim
\dfrac{\hat{q}L^2}{2\omega}\,,
\end{equation}
and thus appear with a characteristic energy,
\begin{equation}
\omega_{\mathrm{c}}\equiv \dfrac{1}{2}\hat{q}L^2.
\end{equation}
Here the transport coefficient $\hat{q}$ has been introduced, which in
this picture represents the mean squared transverse momentum imparted to the parton
per unit length, $\hat{q}=\langle
q_{\mathrm{\perp}}^2\rangle/L$.

In this picture the energy loss is determined by soft multiple
scattering. In the coherent limit the parton undergoes Brownian motion
with a Gaussian elastic cross section $\propto
\dfrac{1}{\hat{q}L}e^{-q_{\perp}^2/\hat{q}L}$, and the scattering
centers behave as a single source of radiation. The bremsstrahlung
spectrum is given by
\begin{equation}
\omega\dfrac{dI}{d\omega}\simeq
\dfrac{2\alphas C_{\mathrm{R}}}{\pi}
\left\{\begin{array}{cc}
\sqrt{ \dfrac{\omega_{\mathrm{c}}}{\omega} }&  \omega < \omega_{\mathrm{c}} \\
\dfrac{1}{12}\left(\dfrac{ \omega_{\mathrm{c}} }{\omega}\right)^2&  \omega > \omega_{\mathrm{c}}
\end{array}\right. \,,
\end{equation}
which results in energy loss $\Delta E\sim
\alphas\omega_{\mathrm{c}}=\dfrac{1}{2}\alphas\hat{q}L^2$. The $L^2$ dependence is
qualitatively different than both the Bethe-Heitler energy loss and
the LPM effect in QED ($\omega \dfrac{dI}{d\omega}\sim
\sqrt{\omega}$). 

Other models frame the energy loss
as an expansion in opacity. These models are not restricted to small
momentum transfers, including the power-law tail in the scattering
cross section, but the coherence effects of
BDMPS-Z formulation must be enforced order-by-order in opacity. The GLV~\cite{Gyulassy:1999zd,Gyulassy:2000er}
model is an example of this approach, which models the scattering
centers as screened Yukawa potentials with screening length,
$\mu$. The gluon spectrum is constructed by integrating first over the
longitudinal direction to enforce the LPM interference at a given
order. The momentum transfer $q_{\perp}$ is then averaged over giving
the double-differential gluon distribution, $\omega \dfrac{dI}{d\omega
  d\kt^2}$. This can be analyzed to give information about the
transverse pattern of radiation before integrating over $\kt$ and
$\omega$ to give the full energy loss.

The ASW formalism~\cite{Wiedemann:2000ez,Salgado:2003gb} is a path integral
formulation that can be
applied to both the multiple soft scattering (MS) or single hard (SH)
dominated scenarios. In the SH approximation the gluon distribution
agrees exactly with the GLV formula to first order in opacity, however the assumptions of the two
models and ranges of integration differ, giving different results for
the total energy loss. In the limit where these differences can be
neglected (ignoring the kinematic constraints) the radiation spectrum
is given by~\cite{Salgado:2003gb},
\begin{equation}
\omega\dfrac{dI}{d\omega}\simeq
\dfrac{2\alphas C_{\mathrm{R}}}{\pi} \dfrac{L}{\lambda}
\left\{\begin{array}{cc}
\dfrac{\pi}{4}\dfrac{ \overline{\omega}_{\mathrm{c}} }{\omega}&
\omega < \overline{\omega}_{\mathrm{c}} \\
\ln \dfrac{\overline{\omega}_{\mathrm{c}}}{\omega} &  \omega > \overline{\omega}_{\mathrm{c}} 
\end{array}\right. \,,
\end{equation}
where $\overline{\omega}_{\mathrm{c}}=\dfrac{1}{2}\mu^2L$ is a
characteristic frequency, different than the BDMPS-Z case. In the SH
scenario, the radiation is dominated by $\omega >
\overline{\omega}_{\mathrm{c}}$, also different from BDMPS-Z. The total
energy loss,
\begin{equation}
\Delta E\simeq \dfrac{2\alphas C_{\mathrm{R}}}{\pi} \dfrac{L}{\lambda} \overline{\omega}_{\mathrm{c}}\ln\dfrac{E}{\overline{\omega}_{\mathrm{c}}},
\end{equation}
is enhanced by $\ln\dfrac{E}{\overline{\omega}_{\mathrm{c}}}$ relative to
the region $\omega < \overline{\omega}_{\mathrm{c}}$. Despite the
differences from BDMPS-Z, this limit also gives $\Delta E\simeq L^2$.

The ASW approach allows for the calculation of quenching weights,
which give the probability for gluon splitting in the medium. With the
assumption of an ordering principle in the virtualities of the fast
parton, these weights can be combined with the vacuum DGLAP splitting
functions. These modified splitting functions can be used to define a
medium-evolved fragmentation function~\cite{Armesto:2007dt} and can be
used to calculate the effect of quenching on jet shape observables~\cite{Polosa:2006hb}.

Some of the difficulties occurring in these analytic calculations can
be alleviated by considering an MC approach. These typically
supplement the generation of hard scattering processes used in
generators by an additional step to include the quenching
effects, and allow for the application of medium effects to the entire
jet, not just the leading parton (democratic
treatment)~\cite{Zapp:2012nw}. Generators like
Q-PYTHIA~\cite{Armesto:2009fj}, PQM~\cite{Dainese:2004te,Loizides:2006cs} and 
JEWEL~\cite{Zapp:2008gi}, replace the vacuum parton showers
implemented in PYTHIA with medium modified parton showers by altering
the Sudakov form factors to include the quenching weights. The PYQUEN model imposes radiative
energy loss of the BDMPS-Z type as well as collisional energy loss
before applying vacuum fragmentation~\cite{Lokhtin:2005px}. 

The approach that most directly incorporates the thermal of the system
was developed by Arnold, Moore and Yaffe (AMY)~\cite{Arnold:2000dr,Arnold:2001ba,Arnold:2001ms,Arnold:2002ja,Jeon:2003gi,Turbide:2005fk}. This relaxes some of the assumptions used by
other models, and includes thermal partons
which are dynamical scattering centers. Furthermore, the requirement that the emitted
gluons be softer than the initial parton energy ($\omega \muchless E$) is
not required. The energy loss occurs through elastic scatterings with
differential collision rates of the form
\begin{equation}
\dfrac{d\overline{\Gamma}}{d^2q_{\perp}}=\dfrac{1}{2\pi^2}\dfrac{g^2Tm_{\mathrm{D}}^2}{\mathbf{q}_{\perp}^2(\mathbf{q}_{\perp}^2 +m_{\mathrm{D}}^2)}\,.
\end{equation}
Multiple soft interactions are resummed using the HTL
techniques discussed in Section~\ref{section:bkgr:htl}, leading to a differential emission rate
for each parton species $\dfrac{d\Gamma}{dk}$. These are used to
construct rate equations describing the evolution of the momentum
distributions. This setup treats collisional and
radiative energy loss in a consistent picture, but loses the distinction
between radiated and thermal gluons. Additionally, finite length
effects which result in finite formation times for the radiated gluons,
are not included in the formalism.

Although these formulations differ on a number of features, they do have some key elements
in common. The most important of these is the operating assumption of
factorization between the production cross section, energy loss
process and fragmentation. While there is no rigorous proof of
factorization in heavy ion collisions, experimental evidence supports
the interpretation that quenching is not an initial-state effect and
that final-state interactions are partonic in
nature~\cite{Armesto:2011ht}. In all of these models, the quenching
mechanism is not applied to the radiated quanta, thus the calculation is limited to the energy loss
of the leading parton, except in the MC implementations. Furthermore,
emitted quanta themselves are radiation sources. Coherence effects
between these sources establish the
angular ordering in the vacuum parton shower and allow for the
cancellation of infrared and collinear divergences. Such effects may
play an important role in medium-induced parton showers as
well~\cite{MehtarTani:2010ma}.

It should also be noted that all models
use repeated application of an inclusive single gluon emission kernel,
extending to multiple emissions either inferring a Poisson
distribution or using rate equations. However, this is not the same as
applying the exclusive multi-gluon emission. Implementing this type of emission forces
the parton kinematics to be updated or dynamical changes in the
medium. Including these local effects as well as the global
effects from the LPM-type interference poses a serious challenge~\cite{Armesto:2011ht}

\begin{figure}[htb]
\centering
\includegraphics[width=0.49\textwidth]{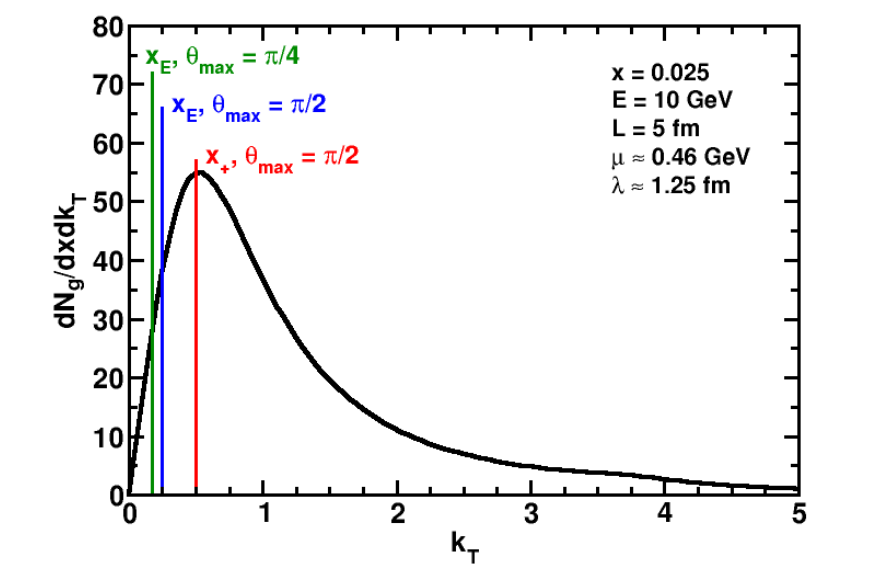}
\includegraphics[width=0.49\textwidth]{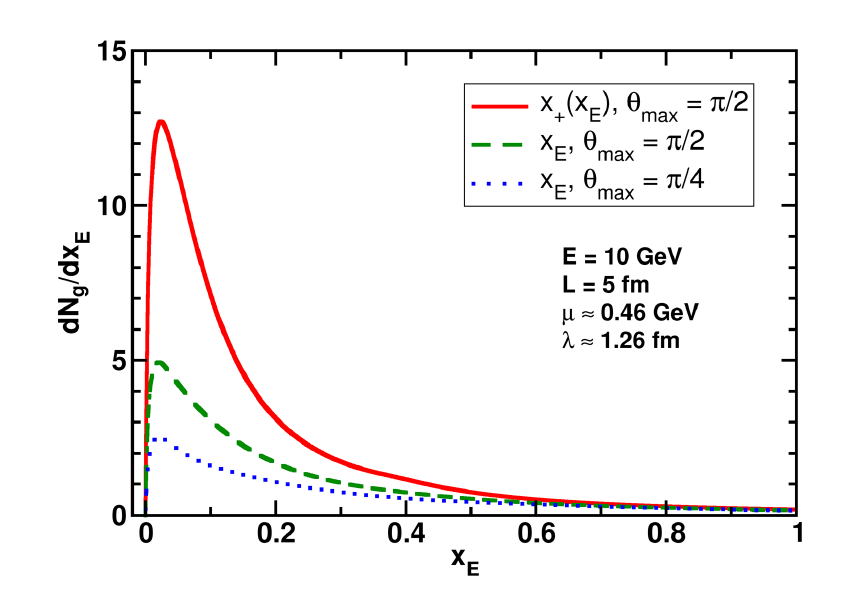}
\caption{The double differential gluon distribution (left),
  $\dfrac{dN}{dxdk_{\mathrm{t}}}$, calculated in the
  GLV formalism at fixed $x$. Here, $x=\omega/E$, is the fraction of
  the quenched parton's energy carried away by the radiated gluon. The
  effect of different kinematic cutoffs in the \kt\ integration is
  illustrated by the vertical lines. The effects of these cutoffs on
  the integration is shown on the right. Figures adapted from Ref.~\protect\cite{Armesto:2011ht}.}
\label{fig:bkgr:brick_kt}
\end{figure}

\begin{figure}[htb]
\centering
\includegraphics[width=0.9\textwidth]{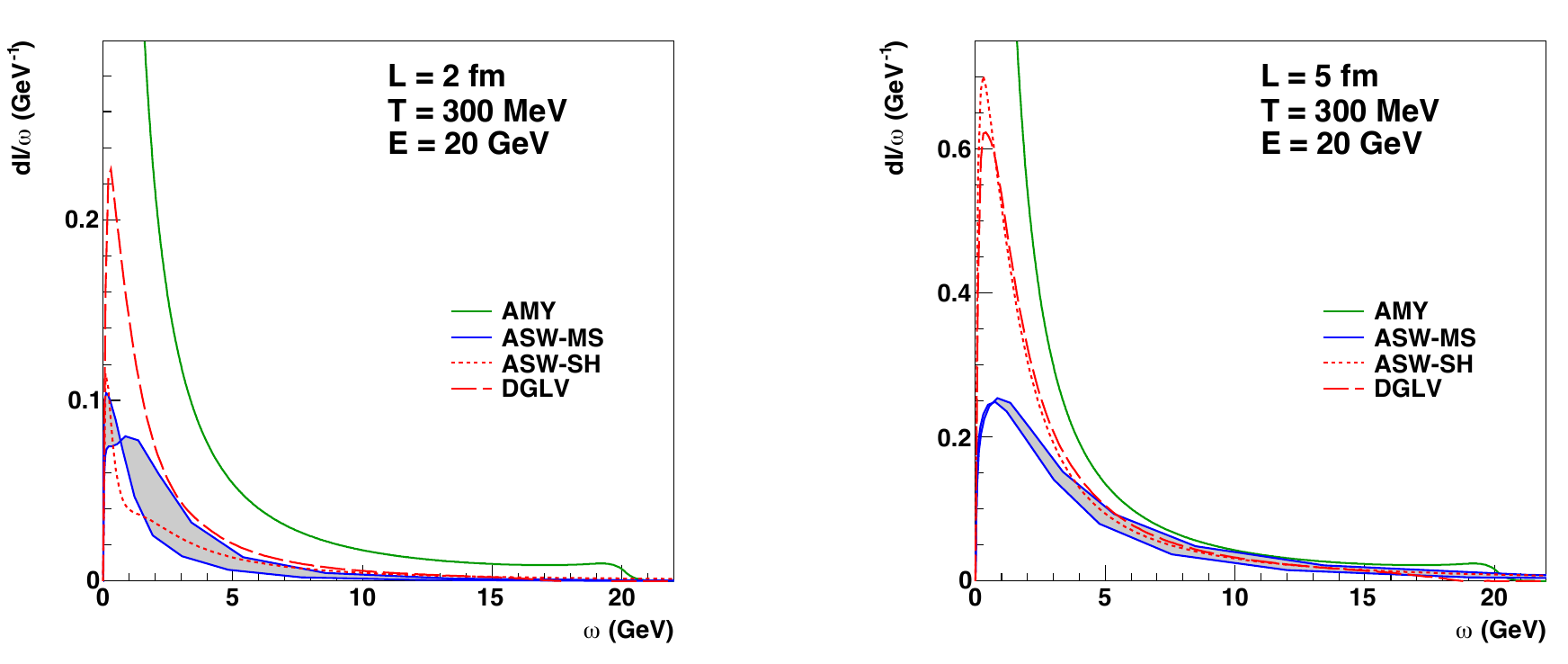}
\caption{Single gluon emission spectrum $\dfrac{dI}{d\omega}$ as a
  function of gluon energy $\omega$ for the AMY, GLV, ASW-MS and
  ASW-SH formalisms. The calculations are for a 20~\GeV\ parton
  passing through QGP bricks at  $T=300$~\MeV\ with 
  size $L=2$~fm and  $L=5$~fm, shown on the left and right
  respectively. Figure adapted from Ref.~\protect\cite{Armesto:2011ht}.}
\label{fig:bkgr:brick_rad_spectrum}
\end{figure}

\begin{figure}[htb]
\centering
\includegraphics[width=0.9\textwidth]{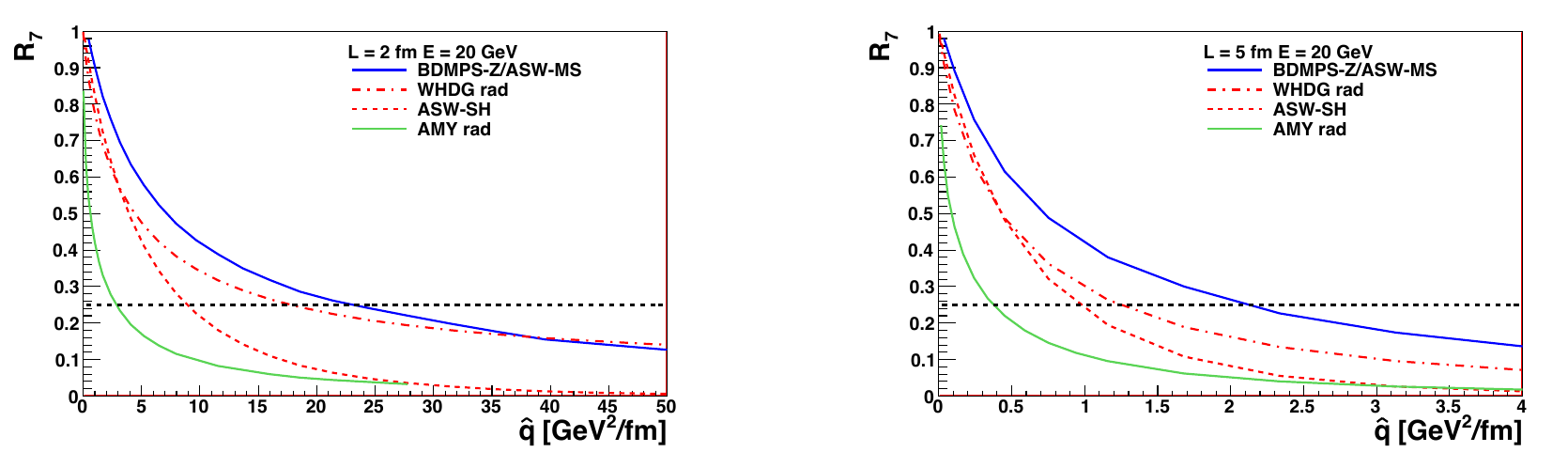}
\caption{Comparison between different models of the single hadron suppression factor, $R_7$, of a $\pt^{-7}$
  spectrum as a function of $\hat{q}$. 
  respectively. Figure adapted from Ref.~\protect\cite{Armesto:2011ht}.}
\label{fig:bkgr:brick_compare}
\end{figure}

There are also issues with these approaches relating to the treatment of large angle
radiation. The BDPMS-Z formalism receives its dominant contribution
from radiation that is outside the valid kinematical range of the
approximation~\cite{Wiedemann:2000tf}. In ASW and GLV, the transverse
distribution of emitted gluons is calculable. For energy loss, this
distribution is integrated over \kt\ and the behavior at
large angles is linked to how the kinematic constraints on the parton
are enforced and the assumption of collinearity. It has been shown that the choice of maximum opening
angle $\theta_{\mathrm{max}}$ can change the results appreciably,
leading to at the least a large systematic
error~\cite{Armesto:2011ht,Horowitz:2009eb}. The effects of different
transverse cutoffs are shown in Fig.~\ref{fig:bkgr:brick_kt}.
The different physical pictures have made it difficult to consistently
fix parameters for a direct comparison of models. However a recent
comparison was performed by applying the GLV, AMY and ASW (both MS and
SH) formalisms to a ``brick'' of QGP matter of fixed
length~\cite{Armesto:2011ht}. A comparison of medium-induced gluon
emission spectra and nuclear suppression
factors as a function of $\hat{q}$ for each of the different models
are shown in Figs.~\ref{fig:bkgr:brick_rad_spectrum} and
\ref{fig:bkgr:brick_compare} respectively.

Depending on the formalism, the quenching mechanism is sensitive to
the system size as well as some combination of the intensive parameters $\hat{q}$, $\mu$ and
$\lambda$, which provide information about the microscopic medium
dynamics. The soft multiple scattering approximation is only sensitive to
$\hat{q}$, which is defined as the mean-squared momentum transfer per
unit length,
\begin{equation}
\hat{q}=\rho\int d^2q_{\mathrm{T}}
q_{\mathrm{T}}^2\dfrac{d\sigma}{d^2q_{\mathrm{T}} } \equiv
\int_0^{q_{\mathrm{max}}}d^2q_{\perp} \mathbf{q}_{\perp}^2\dfrac{d\Gamma_{\mathrm{el}}}{d^2q_{\perp} },
\end{equation}
where $\dfrac{d\Gamma_{\mathrm{el}}}{d^2q_{\perp}}$ is the differential
elastic scattering rate for a hard parton in a thermal medium. This
behaves as $\dfrac{1}{\mathbf{q}^4_{\perp}}$ at high temperatures, but is
screened at low temperatures by $m^2_{\mathrm{D}}$, given by Eq.~\ref{eqn:bkgr:debye_def}. The
form,
\begin{equation}
\dfrac{d\Gamma_{\mathrm{el}}}{d^2q_{\perp}}\simeq\dfrac{C_{\mathrm{R}}}{(2\pi)^2}
\dfrac{g^4 \mathcal{N}}
{\mathbf{q}^2_{\perp}(\mathbf{q}^2_{\perp}+m^2_{\mathrm{D}})},
\label{eqn:bkgr:scattering_rate_elastic}
\end{equation}
interpolates smoothly between these limits~\cite{Armesto:2011ht}, and the number density
$\mathcal{N}$ has been introduced~\cite{Arnold:2008vd},
\begin{equation}
\mathcal{N}=\dfrac{\zeta(3)}{\zeta(2)}(1+\dfrac{1}{4}N_{f})T^3.
\end{equation}
The leading coefficient is the ratio of Riemann zeta function values,
$\dfrac{\zeta(3)}{\zeta(2)}\approx 0.731$. The value of $\hat{q}$ is
given by,
\begin{equation}
\hat{q}(T)=\dfrac{C_{\mathrm{R}} g^4 \mathcal{N}(T)}{4\pi}\ln\left(1+\dfrac{q^2_{\mathrm{max}}(T)}{m^2_{\mathrm{D}}(T)}\right),
\end{equation}
where $q^2_{\mathrm{max}}$ is the largest transverse momentum transfer allowed in
the elastic scatterings and is a function of $T$ and in principle the
parton energy as well. This introduces a logarithmic dependence of
$\hat{q}$ on $E$, which is an effect of approximating the collision
integral in the evaluation of the elastic scattering rate. This value
is sometimes taken as $q^2_{\mathrm{max}}=g^2ET$ although
$q^2_{\mathrm{max}}=g^2\sqrt{ET^3}$ may be more
appropriate~\cite{CaronHuot:2008ni}. The value,
\begin{equation}
\hat{q}\approx\dfrac{m_{\mathrm{D}}^2}{\lambda}=\dfrac{C_{\mathrm{R}}
  g^4 \mathcal{N}(T)}{4\pi}\propto T^3,
\label{eqn:bkgr:qhat_est}
\end{equation}
and $\lambda$ calculated below, is a commonly used, energy
independent expression and is equivalent to omitting the logarithmic
variation of $\hat{q}$. These estimates can differ by up to 40\% for
large parton energies.

The opacity can be calculated as
\begin{equation}
n=\dfrac{L}{\lambda}=L\int d^2q_{\perp} \dfrac{d\Gamma_{\mathrm{el}}}{d^2q_{\perp}},
\end{equation}
which in turn, determines $\lambda$. However the form of the
scattering rate here is slightly different than the one given in
Eq.~\ref{eqn:bkgr:scattering_rate_elastic}, and is chosen to be
consistent with the model used in the opacity expansion
framework~\cite{Armesto:2011ht}. This gives
\begin{equation}
\lambda=\dfrac{4\pi m^2_{\mathrm{D}}}{C_{\mathrm{R}} g^4 \mathcal{N} },
\end{equation}
which is consistent with Eq.~\ref{eqn:bkgr:qhat_est}.

\begin{figure}[htb]
\centering
\includegraphics[width=0.49\textwidth]{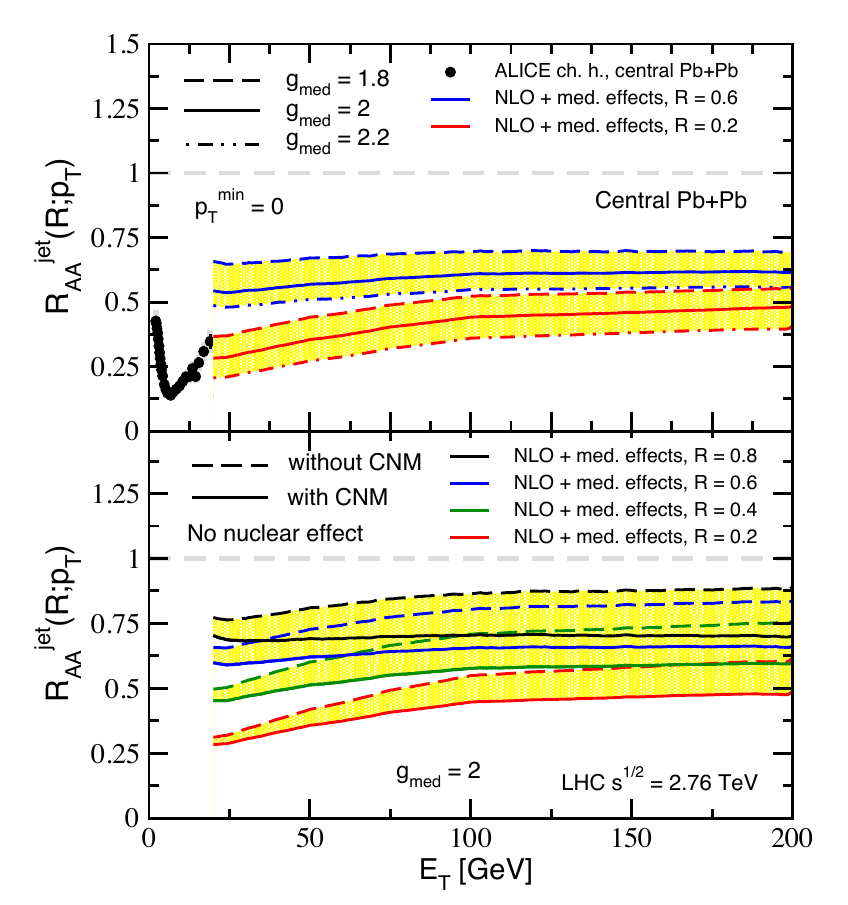}
\caption{Calculation of jet \RAA\ for various jet radii as a function of
  jet \et\ (bottom). The top panel shows a comparison to single
  particle \RAA. Predictions with and without NPDF effects are shown
  in the solid and dashed lines. Figure adapted from Ref.~\protect\cite{He:2011pd}}
\label{fig:bkgr:vitev_raa}
\end{figure}

\begin{figure}[htb]
\centering
\includegraphics[width=0.9\textwidth]{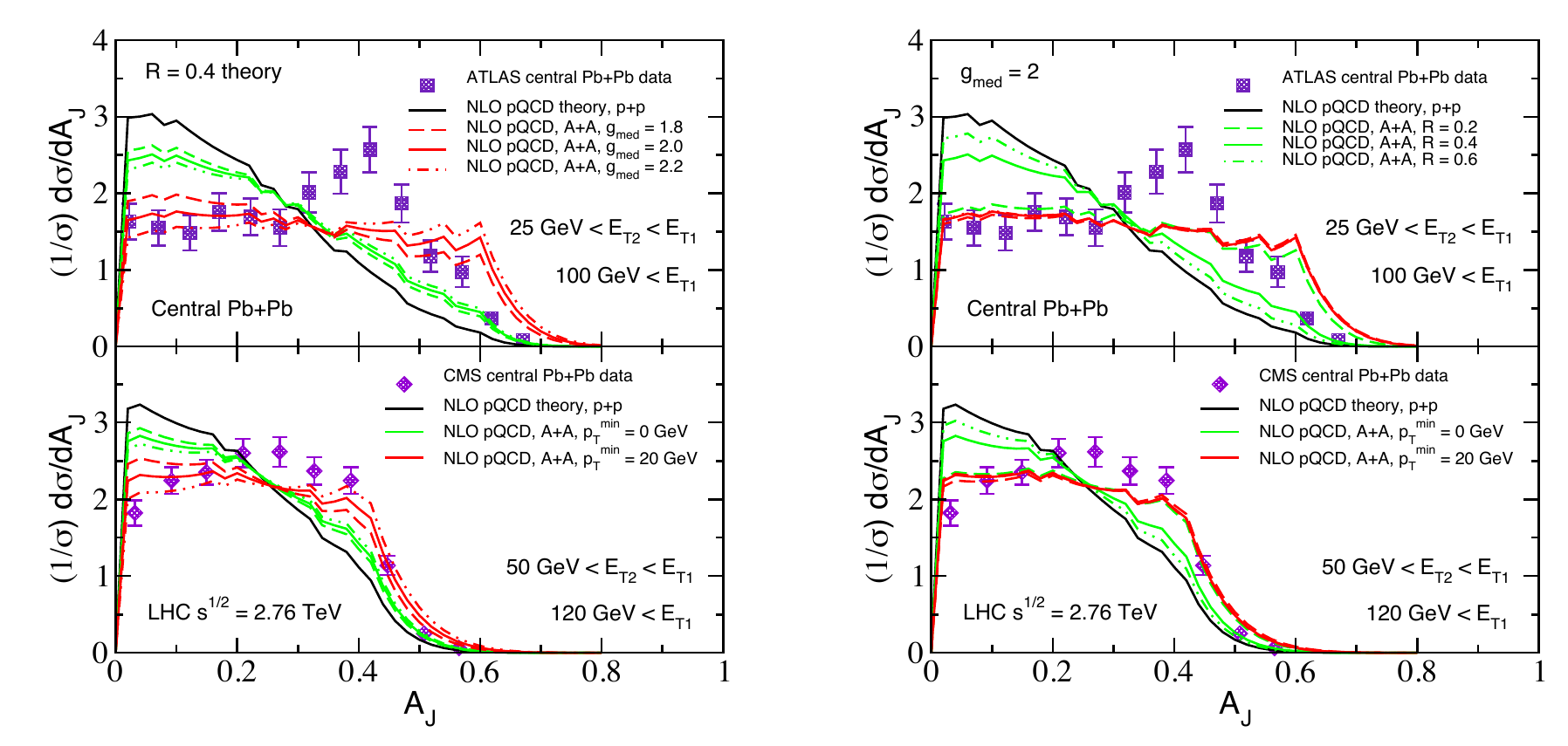}
\caption{Calculation of jet dijet asymmetry for various model
  parameters using kinematic cuts chosen to match measurements from
  ATLAS (left) and CMS (right). Figure adapted from Ref.~\protect\cite{He:2011pd}}
\label{fig:bkgr:vitev_asym}
\end{figure}

Motivated by new experimental capabilities at the LHC, extensions of
energy loss calculations to full jets have recently been
made~\cite{Vitev:2008rz,Vitev:2009rd,He:2011pd}. These function by
expressing the differential jet cross section as,
\begin{equation}
\dfrac{d\sigma^{\mathrm{AA}}_{\mathrm{jet}}}{d\et dy}=\ANcoll
\displaystyle \sum_q \int dx P_q(x,\et)
\dfrac{d\sigma_q}{d\et^{\prime} dy}|\mathcal{J}_q(x)|,
\label{eqn:bkgr:quenching_factorization}
\end{equation}
where $P_q(x)$ is the probability a jet will lose a fraction $x$ of
its energy. The Jacobian $|\mathcal{J}(x)|$ relates the initial and
final parton \et\ by $\et^{\prime}=|\mathcal{J}_q(x)|\et$, and is given
by
\begin{equation}
|\mathcal{J}_q(x)|=\left[1-x(1-f_q(R,p_{\mathrm{T}}^{\mathrm{min}})\right]^{-1}\,.
\end{equation}
The factor $f_q(R,p_{\mathrm{T}}^{\mathrm{min}})$ is the fraction of the
emitted radiation above $p_{\mathrm{T}}^{\mathrm{min}}$ and inside the
jet cone defined by $R$,
\begin{equation}
f_q(R,p_{\mathrm{T}}^{\mathrm{min}})=\dfrac{
\int_0^{R} dr \int_{p_{\mathrm{T}}^{\mathrm{min}}}^{\et} d\omega
\dfrac{dI_q}{d\omega dr}
}{
\int_0^{R^{\infty}} dr \int_{0}^{\et} d\omega
\dfrac{dI_q}{d\omega dr}
}\,.
\end{equation}
While this form still only considers the energy loss of the leading
parton, it does take into account the extent to which radiated energy
is not lost, but only redistributed within the jet. Furthermore, it facilitates the usage
of NLO and NPDF effects in the pQCD calculation of
the unmodified jet cross section,
$\dfrac{d\sigma_q}{d\et^{\prime} dy}$. The factorization of
medium-induced radiation from the production cross section, which is
used in Eq.~\ref{eqn:bkgr:quenching_factorization} has been proven in
the context of an effective theory. This uses Soft Collinear Effective
Theory (SCET) techniques to decouple the hard and soft components of
the gluon fields and has been shown to reproduce the GLV expression
when expanded to first order in opacity~\cite{Ovanesyan:2011xy}. This
factorization is also applicable to multi-jet systems, allowing
calculations of differential energy loss in dijet
systems. Calculations of the inclusive jet suppression factor and
dijet asymmetry in this framework are shown in
Figs.~\ref{fig:bkgr:vitev_raa} and ~\ref{fig:bkgr:vitev_asym}, respectively.

The AdS/CFT correspondence principle was discussed in
Section~\ref{section:bkgr:HIC}, in the context of estimating the shear
viscosity. Similar techniques have also been applied to study the
energy loss of fast partons. While many of these results have focused
on the specific case of heavy quarks~\cite{Gubser:2006bz,Herzog:2006gh,CasalderreySolana:2006rq,CasalderreySolana:2007qw}, some work has been done
to provide an estimate of
$\hat{q}$~\cite{Liu:2006ug,Liu:2006he,Armesto:2006zv}. These results
agree qualitatively with the range of $\hat{q}$ estimates from experiment.

\section{Relativistic Heavy Ion Collisions: RHIC to LHC}
\label{section:bkgr:HI}
\begin{figure}[htb]
\centering
\includegraphics[width=0.49\textwidth]{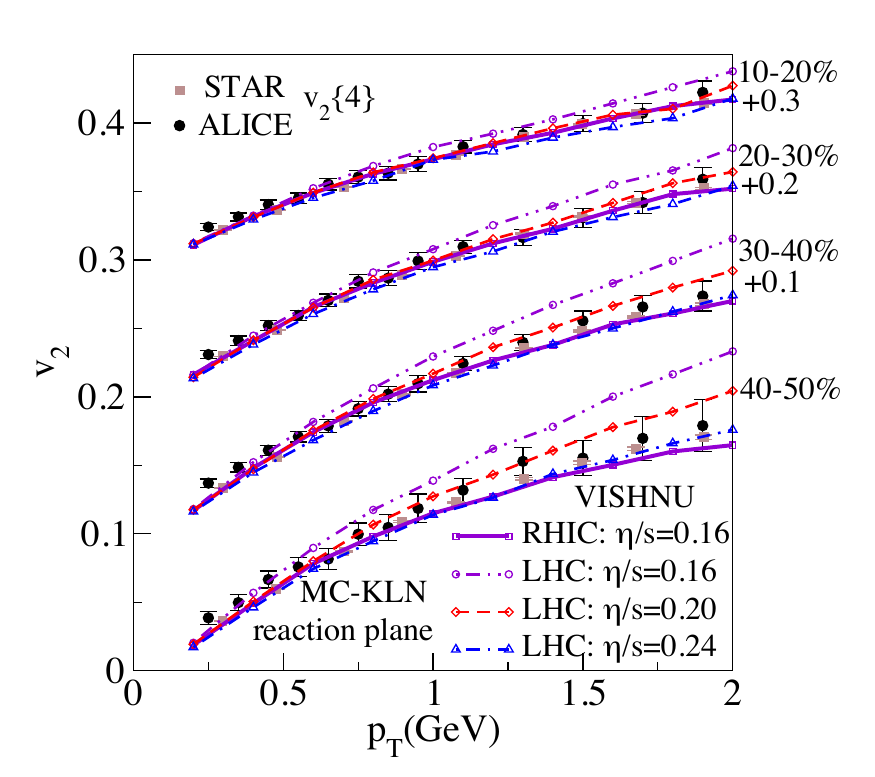}
\includegraphics[width=0.49\textwidth]{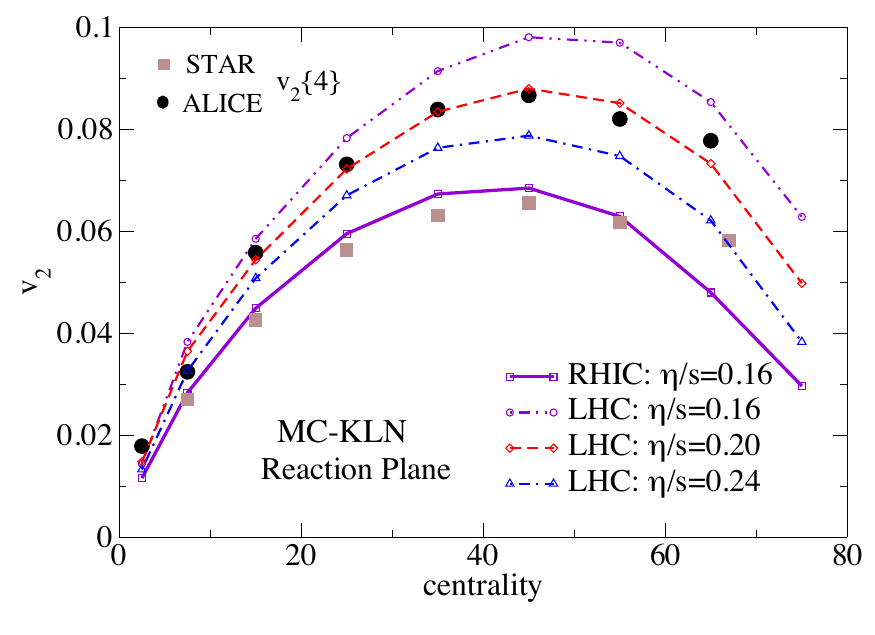}
 \caption{Values of $v_2$ as a function of \pt\ (left) and centrality
 (right) measured by the STAR (\AuAu,
 $\sqrtsnn=200$~\GeV)~\protect\cite{Abelev:2008ae} and ALICE (\PbPb,
 $\sqrtsnn=2.76$~\TeV)~\protect\cite{Aamodt:2010pa} collaborations are
 compared with values calculated from the VISHNU viscous hydro model
 with hadron cascade~\protect\cite{Song:2011qa}. The elliptic flow is measured using a four-particle cumulant technique.}
\label{fig:bkgr:v2_compare}
\end{figure}

With the commencement of the RHIC program, many of the speculated
features of heavy ion collisions became experimentally
accessible. The collective behavior of the system can be studied by
measuring the angular distribution of particles, $\dfrac{dN}{d\phi}$. As discussed previously, in the absence of
hydrodynamical expansion, this distribution is expected to be
isotropic. The anisotropy can be quantified by considering the Fourier
decomposition of the distribution,
\begin{equation}
\dfrac{dN}{d\phi}=f\left\{1+\displaystyle \sum_{n=1}^{\infty}
2 v_{n}\cos[n(\phi-\Psi_n)]
\right\}\,,
\end{equation}
where $\Psi_n$ are the event plane angles, and $v_n$ describe the
magnitude of the modulation. In all but the most central collisions,
the direction of the impact parameter vector between the nuclear centers defines
a reaction plane. The overlap region is ellipsoidal and symmetric
about this reaction plane. The second Fourier coefficient, $v_2$, is
an observable with sensitivity to how the initial-state anisotropy is
converted to the final-state particle distribution, a phenomenon known as
elliptic flow. Measurements from the
RHIC program show that the elliptic flow, in particular the magnitude
of the azimuthal modulation
of particles as a function of their \pt, is
well-described by ideal, relativistic hydrodynamics. The
near-vanishing viscosity indicates that the system is strongly
coupled. This came as a surprise when compared to the prediction of asymptotically
free quarks and gluons and which has led to monikers such as the strongly-coupled QGP (sQGP) and ``the perfect
liquid'' applied to the medium created at RHIC. Since early RHIC
results, significant advances have been made in the formulation and
numerical implementation of viscous hydrodynamics. RHIC and LHC $v_2$
values are compared with a recent viscous hydrodynamical calculation
in Fig.~\ref{fig:bkgr:v2_compare}.

\begin{figure}[htb]
\centering
\includegraphics[width=0.75\textwidth]{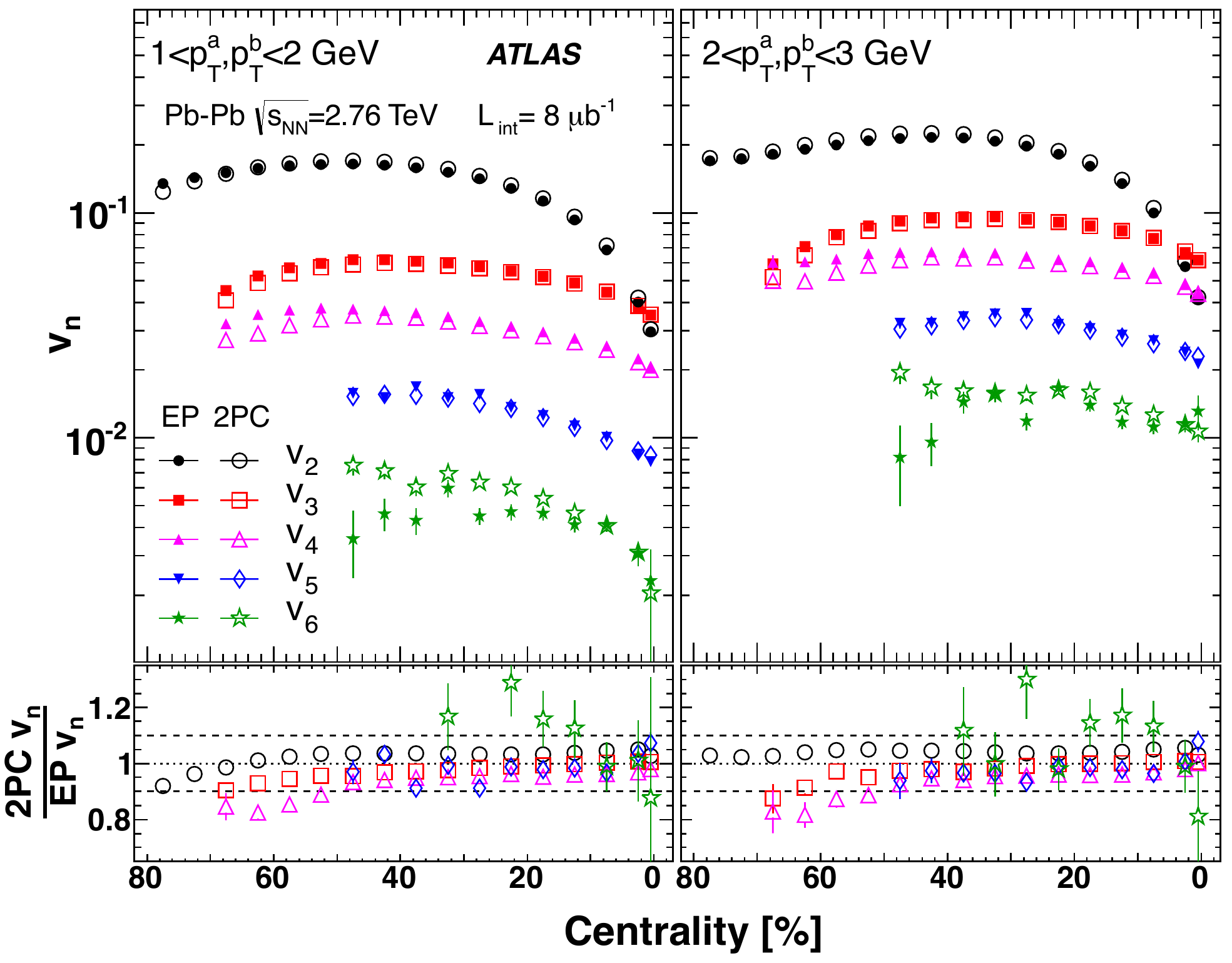}
\caption{Values of $v_1-v_6$ as a function of centrality measured by
ATLAS~\protect\cite{Aad:2012bu}. Two techniques, two-particle correlations (open markers) and the event plane
method (solid markers), were used and were found to be in good
agreement. The ratio of the results from the two methods is shown in
the bottom panel.}
\label{fig:bkgr:atlas_vn}
\end{figure}

More recently, it was recognized that the event-by-event fluctuations
in collision geometry could drive higher flow harmonics. These provide
additional information about the initial-state geometry and may be
able to provide additional constraints on the hydrodynamical
formulation or the value of the shear viscosity. The first six
harmonics as measured by ATLAS ~\cite{Aad:2012bu} are shown in
Fig.~\ref{fig:bkgr:atlas_vn}.

\subsection{Hard Processes}
Indirect experimental evidence for jet quenching was first established
by two important measurements. The single hadron spectrum was found
to be heavily suppressed at high-\pt\ in central
collisions~\cite{Adcox:2001jp,Adler:2003qi,Adler:2002xw,Adams:2003kv}. Furthermore,
the away-side correlation in the dihadron angular distribution was
found to be heavily modified in central \AuAu\ relative to
\pp~\cite{Adler:2002tq,Adler2005ee,Adams:2005ph}. These two
measurements provide slightly different handles on the phenomenon as
the former is an inclusive measurement of the total effect of the
suppression on the \pt\ spectrum. The latter provides a differential
measurement of the quenching as it is sensitive to the quenching of
one jet relative to another.

The interpretation of these observations as evidence for quenching was
further supported by results from the \dAu\ run which established a
critical baseline. The lack of suppression at
high-\pt~\cite{Adler:2003ii,Adams:2003im} combined with the unmodified
dihadron correlation indicated that modification of the NPDFs
was not responsible for the observations in \AuAu. Subsequent
measurements of the direct photon \RAA\ were consistent with no
suppression and provided further systematic control on the quenching
effects~\cite{Adler:2005ig}.

\begin{figure}[htb]
\centering
\includegraphics[width=0.7\textwidth]{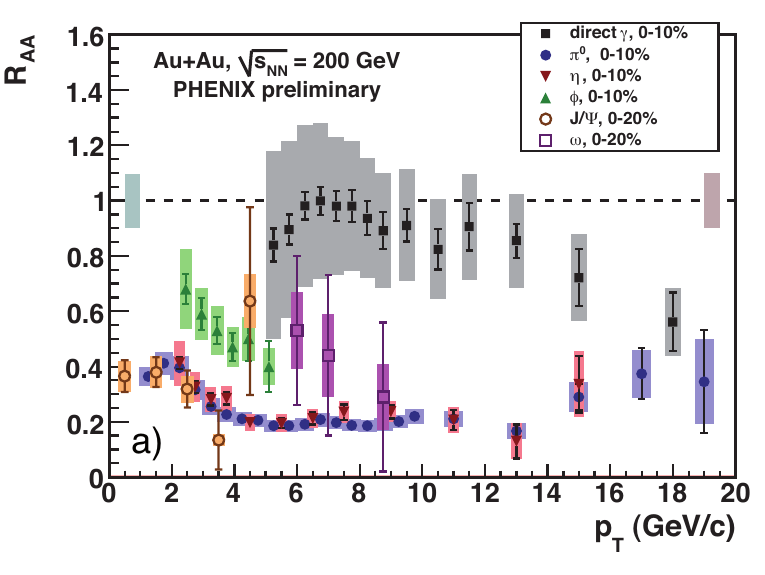}
\caption{PHENIX \RAA\ measurements for \pizero, $\eta$, $\phi$,
  $\omega$, \jpsi\ and direct
\gam~\protect\cite{Reygers:2008pq}.}
\label{fig:bkgr:raa_summary}
\end{figure}

The degree of suppression is typically quantified by the nuclear
modification factor, \RAA,
\begin{equation}
\RAA = \dfrac{
E\dfrac{d^3n_{\mathrm{jet}}^{\mathrm{cent}}}{dp^3}
}
{
\TAA E\dfrac{d^3\sigma_{\mathrm{jet}}^{\mathrm{p+p}}}{dp^3}
}\,.
\label{eqn:bkgr:raa_def}
\end{equation}
A summary of the \RAA\ measurements from PHENIX is shown in
Fig.~\ref{fig:bkgr:raa_summary}. These results established that the production of high-\pt\ particles is
strongly influenced by the produced medium in heavy ion
collisions.
However, the single particle measurements (with the
exception of the photon), are limited in their utility. The medium
effects involve a high momentum parton, not a final-state hadron. Thus
single particle observables can only be connected to the process of
interest through a fragmentation function. This necessity forces the
interpretation of the results in the context of strict factorization
between the medium effects and fragmentation. While this factorization
may ultimately prove to be an appropriate assumption, it is desirable
and more objective to work in a more general paradigm for the
jet-medium interaction, namely one that does not enforce strict
separation between medium effects and the jet
fragmentation. Furthermore, the single hadron observables are only
linked to the parton level quantities in an average sense; there
is no guarantee that the highest energy hadron in an event came from
the leading jet, or the jet that suffered the least energy loss. This
limitation is removed if per-jet fragmentation distribution is
eliminated from the observable.

It is less restrictive, although more experimentally challenging, to
construct observables from fully reconstructed jets. These quantities
provide direct sensitivity to quenching effects. Furthermore, the
possibility of using jets as input objects into physics analyses opens
many new possibilities, which will be discussed in detail in
Chapter~\ref{section:conclusion}. Measurements of full jets at RHIC
have been attempted, but are limited by a number of factors and as of
this time no such measurement has been published. The main
complication is due to the production rates of jets that are easily
detectable above the medium background. Measurements are further
constrained by the limited acceptance of the PHENIX and STAR
detectors. Although highly collimated, particles from jets can still
be distributed over a substantial angular range. Typical sizes for the
cone radii and $R$ parameters in the sequential clustering algorithms
used in \pp\ experiments are on the order of $0.5$. Thus larger
acceptance detectors are preferred for jet measurements.

 Many of these issues are not present at the LHC, in particular the ATLAS experiment,
discussed at length in Chapter~\ref{section:detector}. At these high
energies, the first \PbPb\ run was at $\sigmaNN=2.76$~\TeV, the rate
for producing jets well above the background from the underlying event
is much higher; the differential single jet inclusive cross section
$d\sigma/d\ET d\eta\sim 1$~nb for 100~\GeV\ jets in \pp\ collisions
at these energies. Furthermore, the high quality calorimetry covering
10 units in $\eta$ enables precise measurements of jets and their properties.


\clearpage

\chapter{Experimental Setup}
\label{section:detector}
\section{The Large Hadron Collider}
\label{section:det:LHC}
The Large Hadron Collider (LHC) is a particle accelerator at CERN
outside of Geneva, Switzerland. Although it was primarily designed to
collide protons, the machine is also capable of colliding heavy
ions. This program began in November 2010 with the first lead ion
collisions. 

The LHC machine~\cite{Evans:2008zzb} was constructed using the tunnel
originally used for the LEP experiment. It consists of two parallel beam lines
circulating particles in opposite directions, intersecting at
designated interaction points (IPs). The ring is 26.7 km in
circumference and contains
eight arcs and straight sections. The particle orbits are primarily
controlled by the 1232 dipole magnets, while strong transverse
focusing of the beam constituents is maintained by alternating field
gradients supplied by 392 quadrupole magnets. Almost all of these
devices use superconducting NbTi cabling operating in a cryogenic
system maintained by superfluid He II at 1.9 K.

\begin{figure}[h]
\centering
\includegraphics[width=0.9\textwidth]{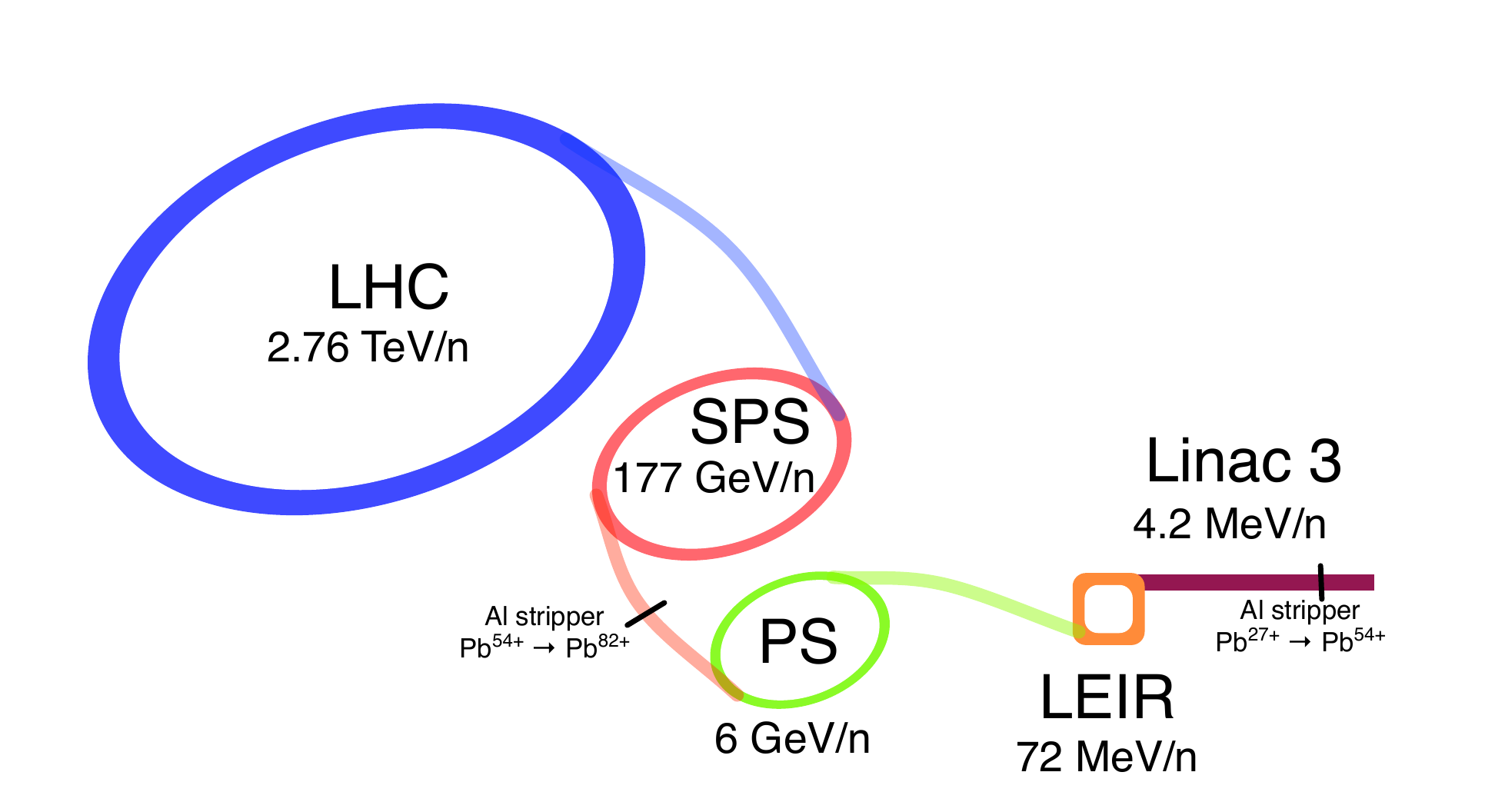}
\caption{LHC ion injection chain.}
\label{fig:det:injection_diagram}
\end{figure}

$\mathrm{Pb}^{208}$ ions are extracted from a source and processed by a sequence of injection chain elements before being injected
into the LHC. This sequence is shown in Fig.~\ref{fig:det:injection_diagram}
consists of: Linac3, the Low Energy Ion Ring (LEIR), the Proton
Synchrotron (PS) and the Super Proton
Synchrotron (SPS). In this process the ions are stripped of
electrons, squeezed into longitudinal bunches via the application of a
radio frequency electric field (RF) and accelerated to a beam energy of 177~\GeV\ per nucleon.

\subsection{Performance and Luminosity}
\label{section:det:LHC:luminosity}
The most important factor in determining a collider's performance is
the luminosity. The instantaneous luminosity of a particle beam, $ \mathscr{L}$, is the
flux of scattering particles per unit area per unit time. For a
process with cross section $\sigma$ the interaction rate is given by
\begin{equation}
\frac{dN}{dt}=\sigma \mathscr{L}.
\label{eqn:det:rate_cross_seciton_lumi}
\end{equation}

Therefore to enhance the rate of rare processes, it is a design goal
of an accelerator to maximize this quantity. The colliding beams consist of
bunches of ions with densities $n_{1}$ and $n_{2}$. Along the beam
direction, $s$, the bunches can be organized into various patterns
according to an injection scheme in which each colliding bunch pair
collides regularly with frequency $f$. In the transverse direction,
the beams have profiles characterized by $\sigma_{x}$ and
$\sigma_{y}$. For $N_{b}$ bunches the instantaneous luminosity can be computed via
\begin{equation}
\mathscr{L}=N_{b}f\frac{n_{1}n_{2}}{4\pi\sigma_{x}\sigma_{y}}.
\label{eqn:det:lumi_def_1}
\end{equation}

The beam's transverse focusing is controlled by applying alternating
 gradient fields forcing the beam to converge. The transverse motion of single particles within the beam 
is then a sinusoid with an envelope modulated by the beta-function, $\beta(s)$
\begin{equation}
x(s)=A\sqrt{\beta(s)}\cos(\psi(s)+\delta)
\label{eqn:det:ags}
\end{equation}
where $A$ and $\delta$ are integrals of motion, describing the area
and angle of an ellipse in the phase space ($x$, $x'=dx/ds$) inhabited by beam particles~\cite{Nakamura:2010zzi}. To maximize
luminosity in the neighborhood of the interaction zones, the beams are squeezed in the transverse direction by
focusing magnets~\cite{Jackson:1998hs}. The squeezing is parabolic in
the longitudinal displacement and is controlled by the parameter $\beta^{*}$
\begin{equation}
\sigma(s)=\sigma(0)\left(1+\frac{s^2}{\beta^{*2}}\right).
\label{eqn:det:beta_star_def}
\end{equation}

In addition to $\beta^{*}$, the other parameter affecting luminosity
is the emittance, $\epsilon$. In the absence of beam losses, the phase
space ellipse has fixed area $\pi A^2$. For a beam with a Gaussian
transverse profile, the emittance is the area containing one standard
deviation $\sigma_x$:
\begin{equation}
\epsilon_{x}=\pi\frac{\sigma_{x}^{2}}{\beta_{x}}.
\label{eqn:det:emittance_def}
\end{equation}
In terms of these parameters, the luminosity can be expressed as
\begin{equation}
\mathscr{L}=N_{b}f\frac{n_{1}n_{2}}{4\sqrt{\epsilon_{x}\beta^{*}_{x}\epsilon_{y}\beta^{*}_{y}}}\,.
\label{eqn:det:lumi_beta_star}
\end{equation}

\begin{figure}[h]
\centering
\includegraphics[width=0.5\textwidth]{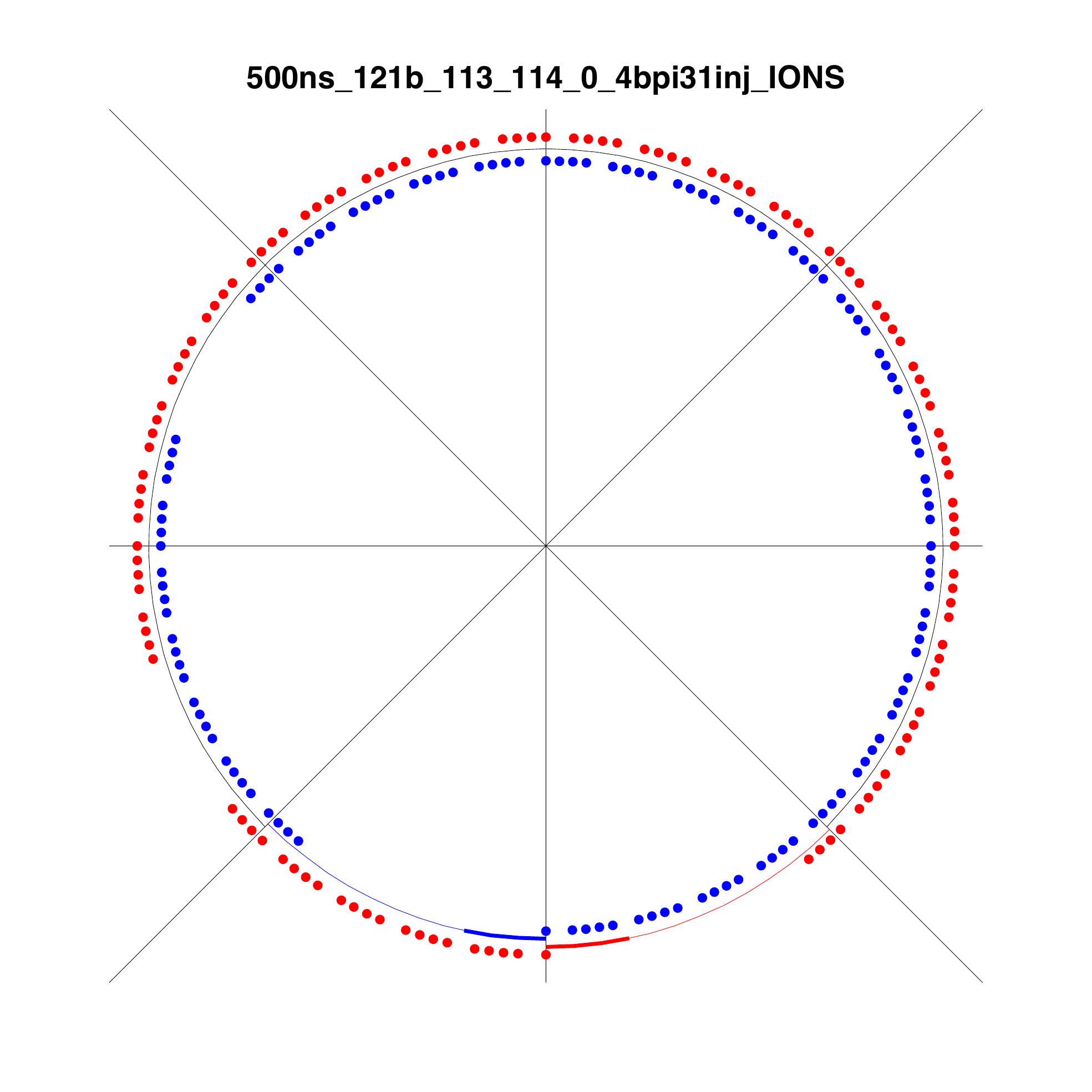}
\caption{Injection scheme for the 500 ns bunch spacing fills in 2010
  ion run. The blue and red points correspond to the filled BCIDs in beams 1
  (clockwise) and 2 (anti-clockwise) respectively. Of the 121 filled
  bunches 113 are configured to collide. This scheme was used to
  achieve the maximum instantaneous luminosity for Pb ions in 2010: $2.88\times10^{25}
\instlumi$.}
\label{fig:det:injection_2010}
\end{figure}

\begin{figure}
\centering
\includegraphics[width=0.5\textwidth]{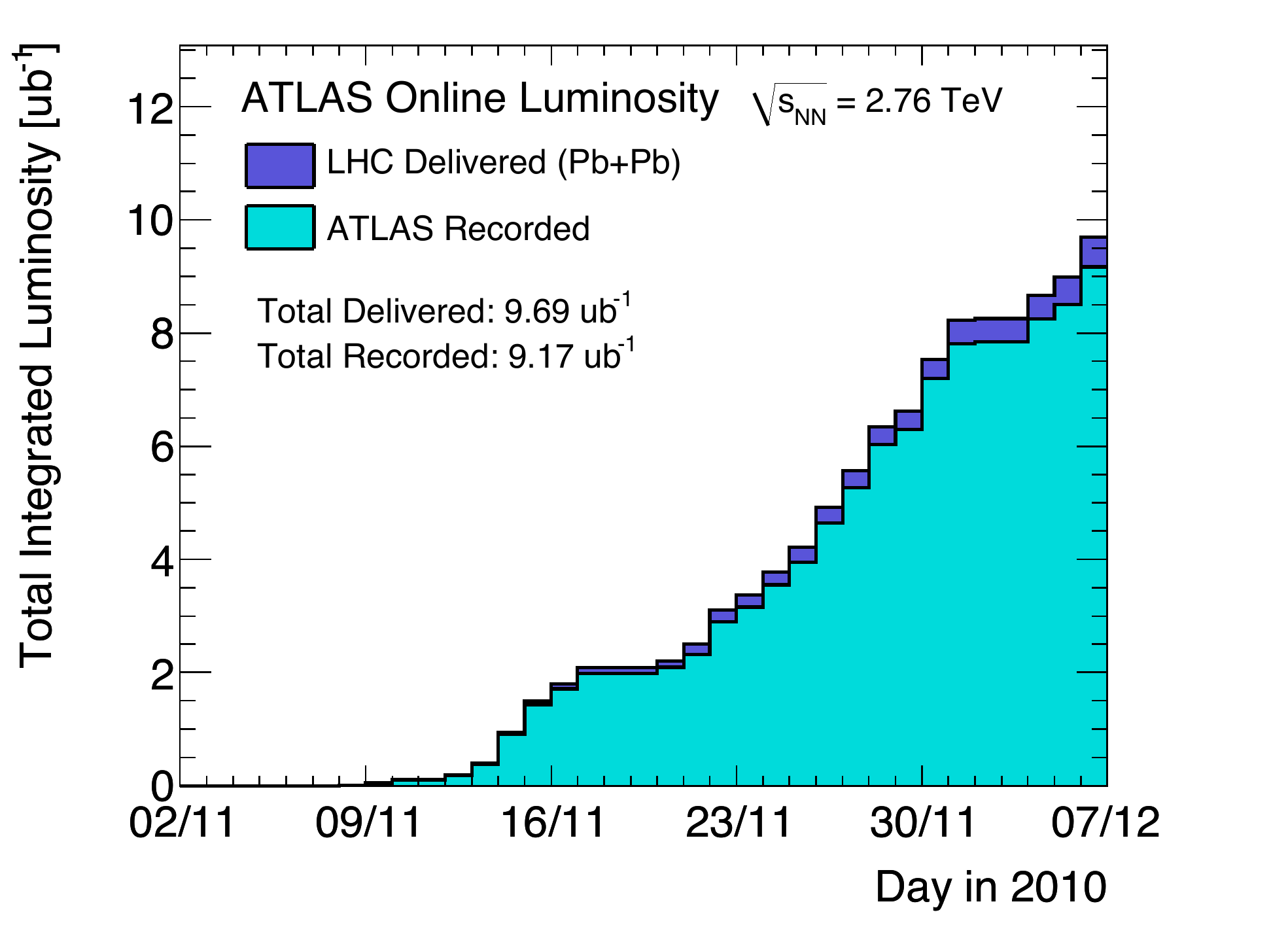}
\caption{Total integrated luminosity delivered by the LHC (dark blue)
  and recorded by the ATLAS detector (light blue) as a function of day
  during the 2010 Pb ion run. The total delivered luminosity was 9.69~\invmub.}
\label{fig:det:lumi_per_day}.
\end{figure}

The longitudinal beam structure is governed by the bunch injection
scheme~\cite{Bailey:691782}. The SPS is capable of injecting proton bunches at 25~ns spacing or 40
MHz into the LHC, corresponding to a total 2,808 bunch crossing slots
(BCIDs). As of yet,  this design limitation has not been reached. The end of the
2011 proton run injected 1380 bunches, 1331 of which were brought into
collision for a luminosity of $3.65\times10^{33}~\instlumi$. The 2010 Pb ion run saw a maximum luminosity of $2.88\times10^{25}
\instlumi$  corresponding 121 bunches per beam with 113
colliding bunches and a 500 ns bunch spacing shown in Fig.~\ref{fig:det:injection_2010}.
For the entire 2010 Pb ion run, the LHC delivered a total integrated luminosity of
9.69~\invmub. This quantity, as well as the total integrated
luminosity recorded by the ATLAS detector as a function of time is
shown in Fig.~\ref{fig:det:lumi_per_day}. Although not analyzed here,
the 2011 Pb ion run recorded a total integrated luminosity of approximately 140~\invmub.

\begin{figure}[h]
\centering
\includegraphics[width=0.5\textwidth]{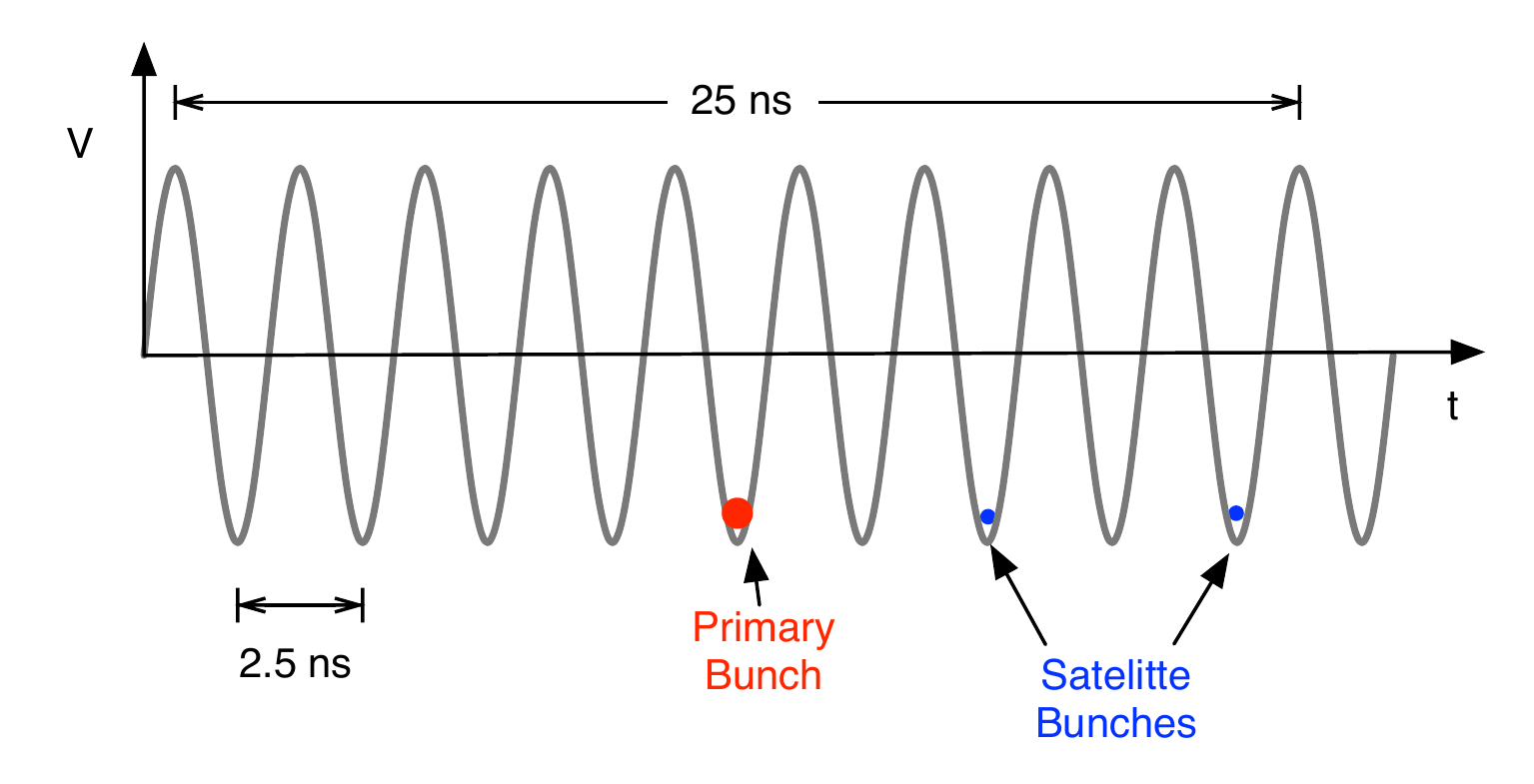}
\caption{A schematic view of the RF bucket structure corresponding to
  a single bunch fill. The red point indicates the filled bucket in
  time with the LHC clock. The blue points indicate satellite bunches
  which can cause out of time collisions.}
\label{fig:det:rf_buckets}
\end{figure}

In addition to the discrete bunch structure, RF cavities are used to focus the bunches in the
longitudinal direction. The 400 MHz oscillating field causes
particles to clump together around the ring in each of the 35640
RF buckets. Superimposing the RF structure with the injection
bunch spacing causes 10 RF buckets to be associated with each
BCID. Ideally all particles in the injected bunch are forced into a
single RF bucket, however a few stray particles become trapped in
adjacent buckets as indicated in Fig.~\ref{fig:det:rf_buckets}. Collisions among these particles, called satellite bunches, will appear out of
time with the LHC clock in discrete 2.5 ns intervals.

\section{ATLAS Overview}
\label{section:det:overview_ATLAS}
\begin{figure}[htb]
\centering
\includegraphics[width=1\textwidth]{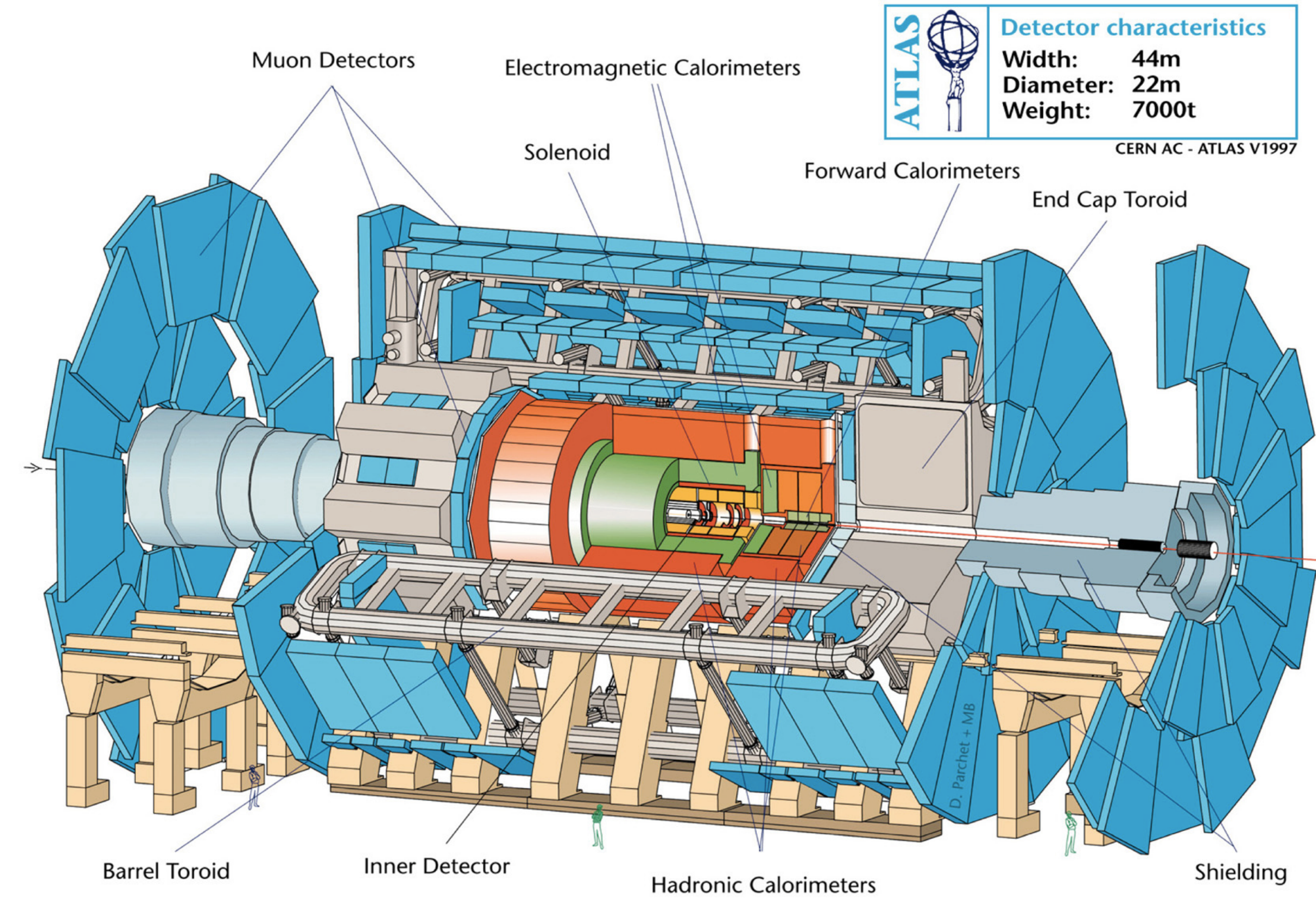}
\caption{A diagram of the ATLAS detector showing the major detector
systems.}
\label{fig:det:atlas_main}
\end{figure}

\begin{figure}[htb]
\centering
\includegraphics[width=0.55\textwidth]{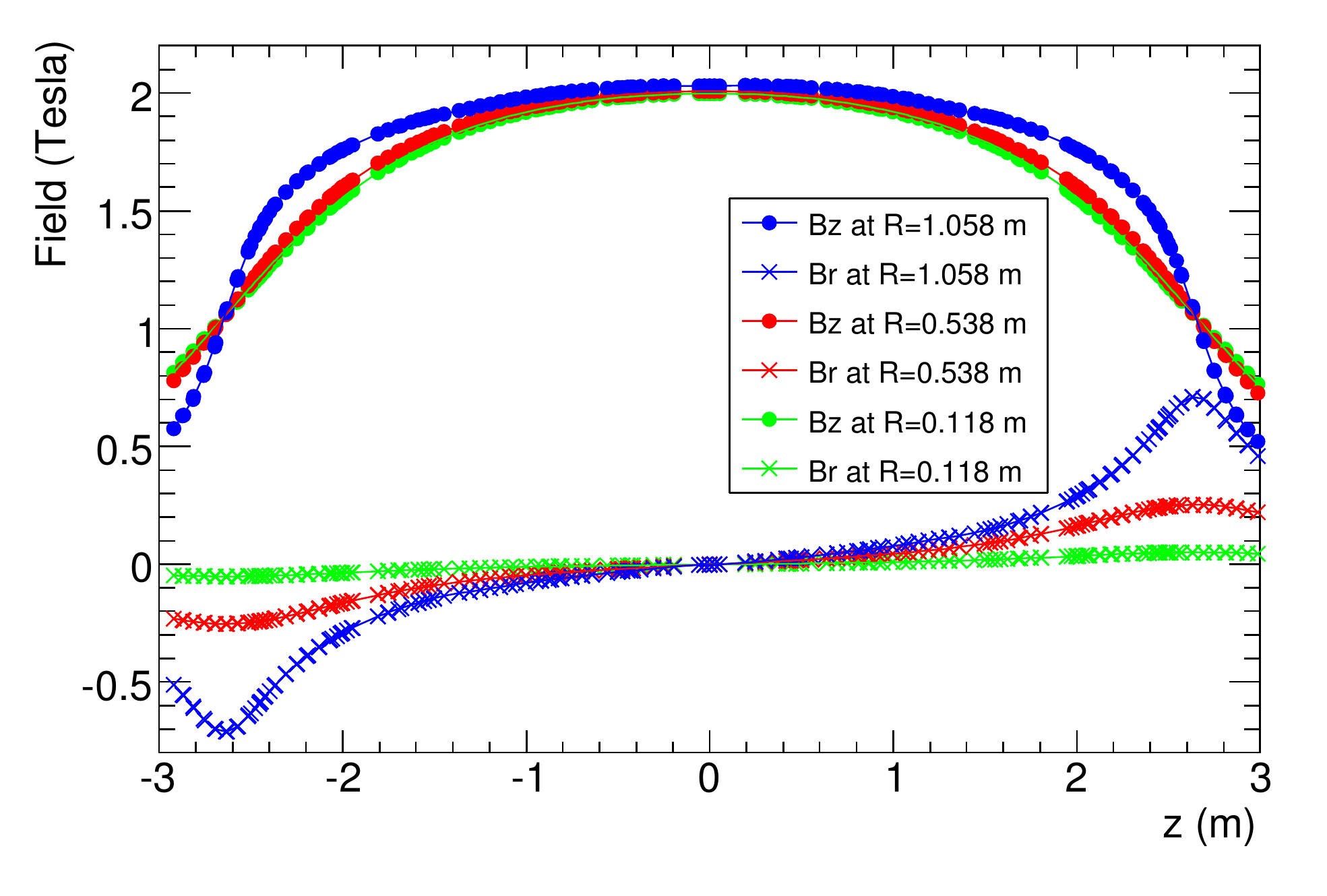}
\caption{The longitudinal and radial components of the solenoidal magnetic field as a function of $z$
at various radii.}
\label{fig:det:mag_field}
\end{figure}

The ATLAS experiment is a multi-purpose particle detector
situated at interaction point 1 (IP1) of the ATLAS ring~\cite{Aad:2008zzm}. It is
forward-backward symmetric covering the full 2$\pi$ in
azimuth. Charged particle tracking is provided by the inner detector, covering $\eta<2.5$,
immersed in a 2 T solenoidal magnetic field which is shown in Fig.~\ref{fig:det:mag_field}. Energy measurements are
provided by a combination of electromagnetic and hadronic calorimeters covering
$|\eta| < 4.9$ enclosing the inner detector. A dedicated muon
spectrometer is positioned beyond the calorimeter utilizing a toroidal
field maintained by a barrel and two end-cap toroidal magnets. This
system allows muon measurements over the range $|\eta| < 2.7$
utilizing a variety of subsystems.
Forward detectors such as the Zero
Degree Calorimeter ($|\eta| > 8.3$) and Minimum Bias Trigger
Scintillators ($2.09 < |\eta| < 3.84$) provide minimum bias event
triggering and event selection capabilities. A graphical
representation of ATLAS highlighting the various subsystems is shown
in Fig.~\ref{fig:det:atlas_main}.

This analysis makes extensive use of the calorimetry, which is
described in detail in Section~\ref{section:det:calorimeter}. The
remaining systems are described in less detail in
Sections~\ref{section:det:MBTS},~\ref{section:det:ZDC}~and~\ref{section:det:ID},
except the muon spectrometer which is not used at all in this analysis
and is mentioned for completeness.

\section{Trigger}
\label{section:det:tigger}
A trigger is a combination of hardware and software elements designed
to select which collision candidates are recorded by the data
acquisition (DAQ) system. When few colliding bunches are circulating
in the machine, a simple requirement that the given pair of bunches is intended to collide in the filling scheme is the most basic criterion for triggering. However this is insufficient as many of the recorded events may contain no real collisions or detector signals from beam and cosmic backgrounds. Additionally, at higher luminosity the DAQ typically cannot record events at this rate and a trigger selection must be employed to ensure that only events with desired physics signals are recorded. In general, implementing a sophisticated trigger strategy to solve this problem is a significant experimental effort, requiring multiple triggers sensitive to different physics signals as well as using prescales to reduce the rate of less interesting triggers relative to the rarer ones. However, the luminosity of the 2010 ion run was low enough that a set of minimum bias trigger items could be selected without need for prescale.

The ATLAS trigger system is composed of trigger items on three different levels. The Level 1 (L1) trigger is entirely hardware-based. In addition to the normal data readout path, detectors integrated into L1 have parallel paths for data readout, often involving coarser and faster signal sums. The specialized trigger signals are sent to the Central Trigger Processor (CTP), which combines the information into a set of L1 bits. The other two levels Level 2 (L2) and Event Filter (EF) are both software based and are collectively referred to as the High Level Trigger (HLT). L1 regions of interest (ROIs) are used to seed algorithms run as part of L2. Finally, L2 items feed full scan EF algorithms, which use higher level detector signals from the offline readout path and are not constrained to the L1 ROIs. A trigger chain is defined as an EF item that is seeded by various L1 and L2 items. Multiple chains can be seeded by the same L1 and L2 items and prescales can be applied at any of the three levels. Trigger chains are grouped together by type to form data streams; an event is recorded and reconstructed if it has been selected by one or more of its chains after all prescales have been applied.

During the 2010 ion run, a single minimum bias stream,
\verb=physics_bulk=, was used for physics analyses. While some chains
using HLT items were used for efficiency and background studies, this
stream was  primarily composed of raw L1 items passed through the HLT
without prescale.

\section{Calorimetry}
\label{section:det:calorimeter}
\begin{figure}[htb]
\centering
\includegraphics[width=0.49\textwidth]{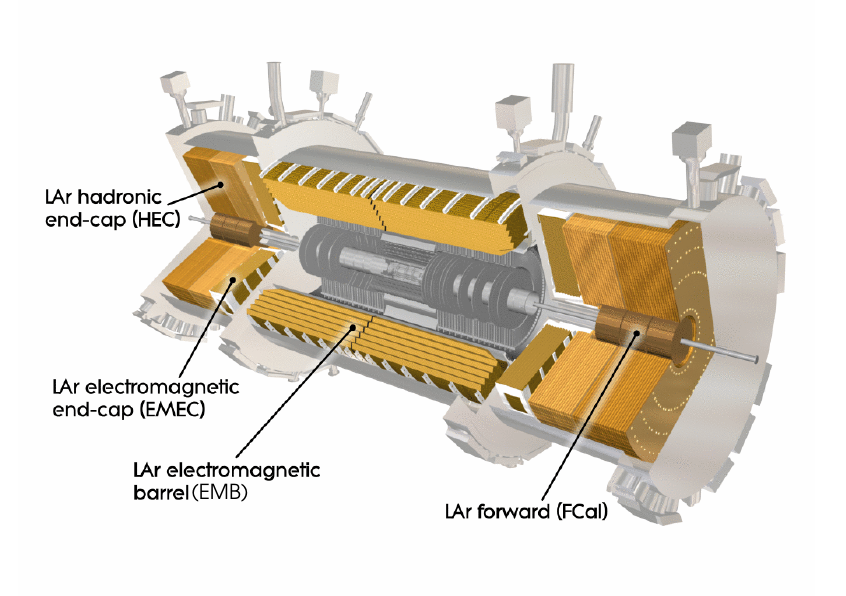}
\includegraphics[width=0.49\textwidth]{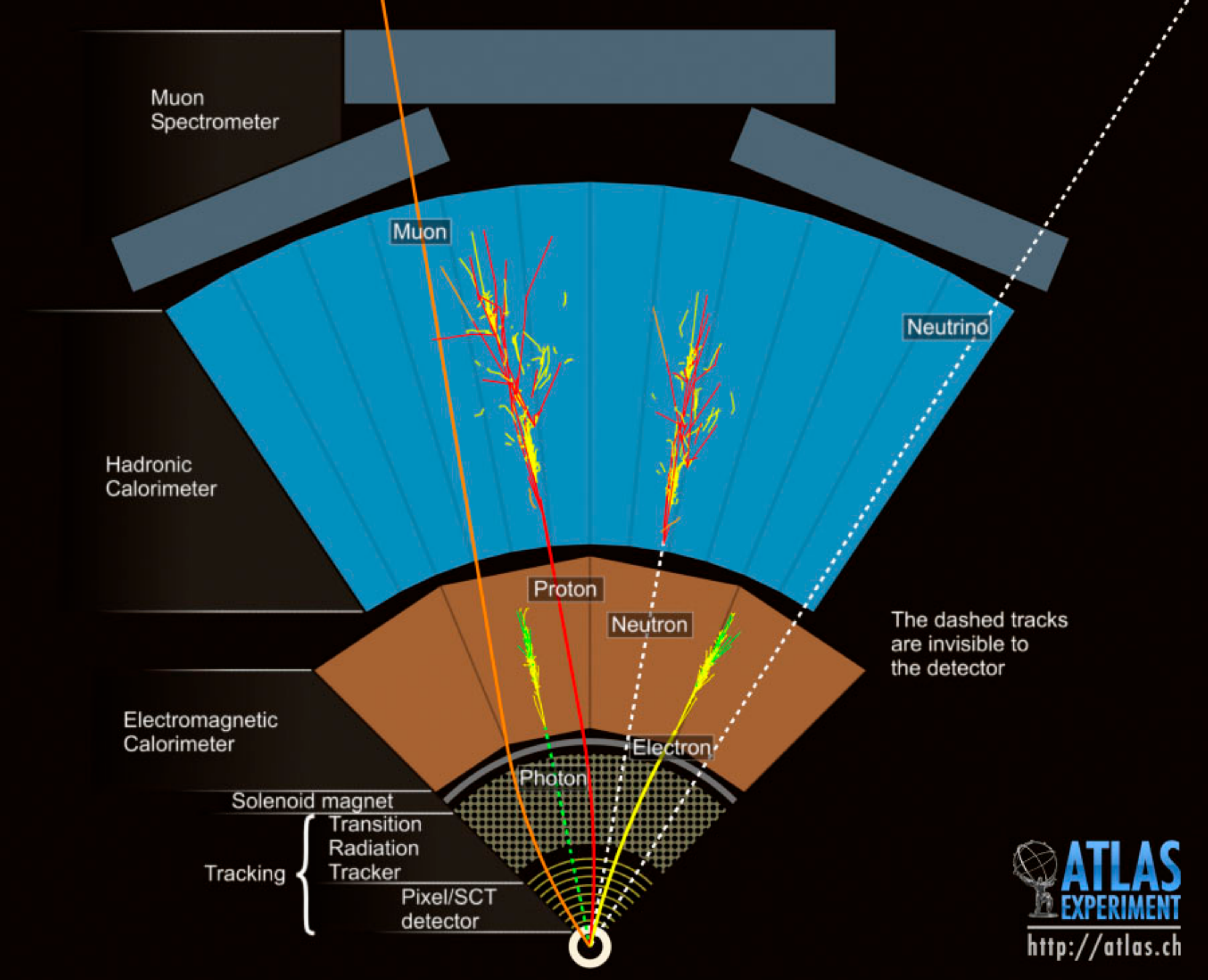}
\caption{Drawing of ATLAS calorimeter system is shown on the left. The
right figure shows how different particle species are detected by the
calorimeter system.}
\label{fig:det:shower}
\end{figure}

The ATLAS calorimeter system is composed of electromagnetic and
hadronic calorimetry using sampling calorimeters based on two distinct
technologies: liquid Argon (LAr) and scintillating tiles. The \eta\ coverage and segmentation of each subsystem is
summarized in table~\ref{tbl:det:calo_summary}. Energies of particles
well above the ionization regime are measured by a sampling
technique. Electromagnetic and hadronic showers are initiated when the
particle strikes an absorber. The details of these showering mechanism
are discussed in Sections~\ref{section:det:calo:EM_showers} and \ref{section:det:calo:hadronic_showers}
respectively. As the showers develop the energy of the
incident particle is increasingly spread among an ensemble of particles of lower
energy. An active material is placed behind the absorber which
collects some fraction of energy of the lower energy particles, either
through ionization (LAr) or scintillation (tile). Alternating layers of absorber
and active material are placed in succession and the
shower-sampling is repeated. A pictorial representation of the
different calorimeters and the sampling
of the different shower types is shown in Fig.~\ref{fig:det:shower}.
\begin{table}
\footnotesize
\centering
\begin{tabular}{|c|c|c|c|c|c|c|} \hline
Type & Sub-detector & Absorber & Layer & \abseta-range
&$\Delta\eta\times\Delta\phi$& Channels \\ \hline
 \multirow{13}{*}{LAr} & Barrel Presampler& None & 1&$\abseta<1.52 $& $0.025\times0.1$ &7808 \\ \cline{2-7}
&\multirow{3}{*}{EM Barrel}
& \multirow{3}{*}{Steel }& 1&\multirow{2}{*}{$\abseta < 1.475$}& $0.003\times0.1$ &57216 \\\cline{4-4}\cline{6-7}
&& &2&& $0.025\times0.025$ &28672\\ \cline{4-7}
&& &3& $\abseta < 1.35$& $0.050\times0.025$ &13824 \\  \cline{2-7}
&End-cap Presampler&None& 1&$1.5<\abseta<1.8$& $0.025\times0.1$ &1536 \\  \cline{2-7}
&\multirow{3}{*}{EM End-cap}
&\multirow{3}{*}{Steel }& 1&\multirow{2}{*}{$1.375 < \abseta < 3.2$}&-&28544\\ \cline{4-4}\cline{6-7}
&&&2& &$0.025\times0.025,\,0.1\times0.1$&23424\\ \cline{4-7}
&&& 3&$1.5   < \abseta < 2.5$& $0.050\times0.025$ &10240\\ \cline{2-7}
&\multirow{2}{*}{Hadronic End-cap}
&\multirow{2}{*}{Copper }& 1&\multirow{2}{*}{$1.5 < \abseta < 3.2$}& \multirow{2}{*}{$0.1\times0.1,\,0.2\times0.2$}  &3008\\ \cline{4-4}\cline{7-7}
&&& 2&& &2624\\ \cline{2-7}
&\multirow{3}{*}{FCal}
& Copper & 1&\multirow{3}{*}{$3.1 <\abseta <4.9$}&\multirow{3}{*}{$\sim0.2\times0.2$}
&1008\\ \cline{3-4}\cline{7-7}
&&\multirow{2}{*}{Tungsten }& 2&& &500\\ \cline{4-4}\cline{7-7}
&&& 3& & &254\\ \hline
\multirow{6}{*}{Tile}&\multirow{3}{*}{Tile Barrel}
& \multirow{3}{*}{Steel }& 1&\multirow{3}{*}{$ \abseta < 1.0$}& \multirow{2}{*}{$0.1\times0.1$} &\multirow{3}{*}{5760}\\ \cline{4-4}
&&& 2&& &\\ \cline{4-4}\cline{6-6}
&&&  3&&$0.2\times0.1$ &\\ \cline{2-7}
&\multirow{3}{*}{Tile Extended}
& \multirow{3}{*}{Steel }& 1&\multirow{3}{*}{$0.8 < \abseta < 1.7$}&\multirow{2}{*}{$0.1\times0.1$} &\multirow{3}{*}{4092}\\ \cline{4-4}
&&& 2&& &\\ \cline{4-4}\cline{6-6}
&&&  3&&$0.2\times0.1$ &\\ \hline
\end{tabular}
\caption{Description of coverage and segmentation of each calorimeter
  sampling layer. The $\Delta\eta\times\Delta\phi$, correspond to the segmentation applying to most of that layer,
  although not necessarily constant over the full layer. The
  segmentation of the EM
  end-cap, with \eta-dependent segmentation in $\Delta\eta$ but
  constant $\Delta\phi=0.1$, and the FCal which has non-projective
  geometry onto $\eta-\phi$ coordinates, are not given. Additional
  tiles (not shown) are interspersed irregularly in the gaps between the cryostats
and support structures.}
\label{tbl:det:calo_summary}
\end{table}

\begin{figure}
\centering
\includegraphics[width=0.4 \textwidth]{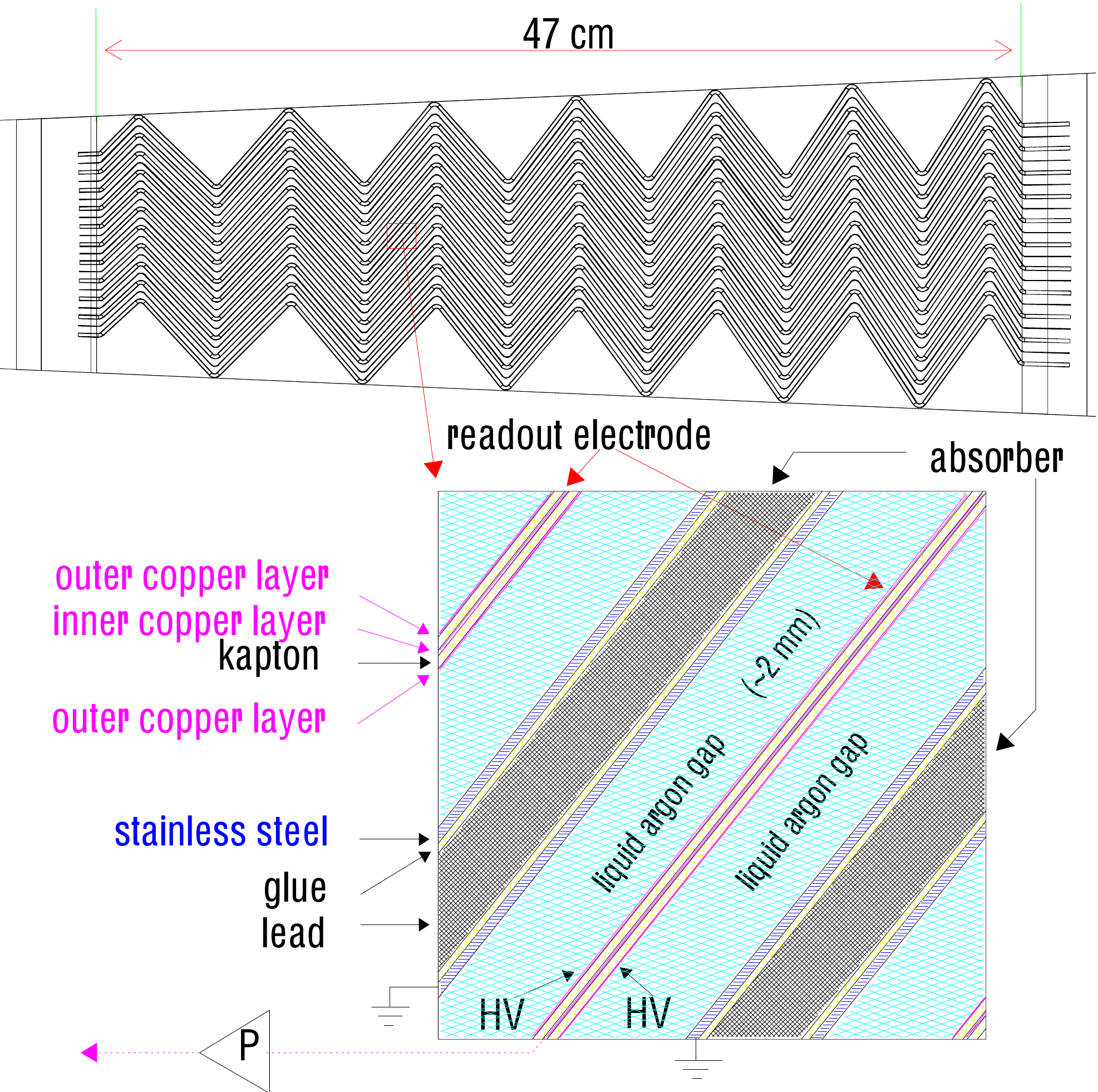}
\includegraphics[width=0.4 \textwidth]{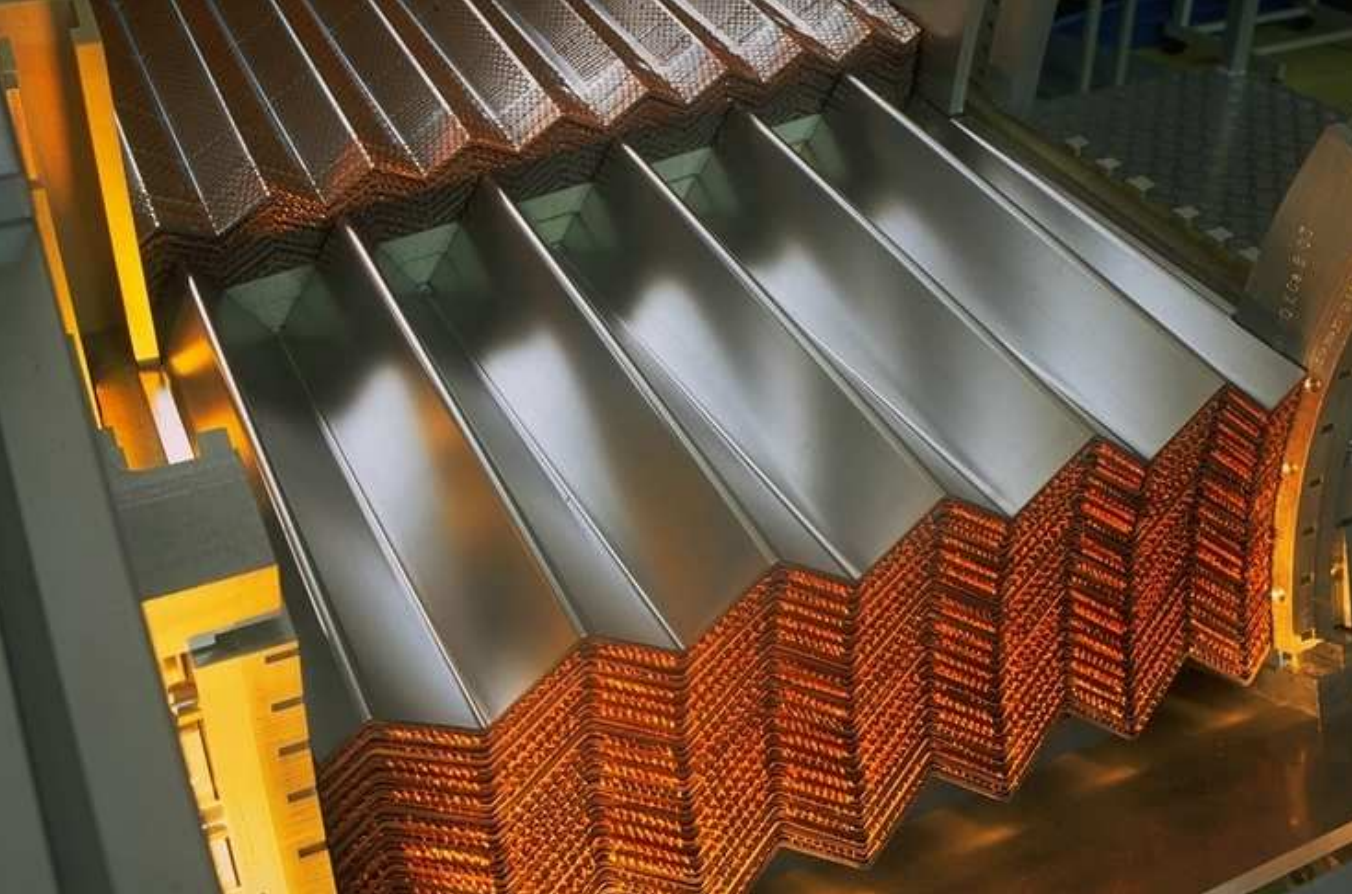}
\caption{Accordion structure of the EM barrel (left) shown as a cross sectional
  slice transverse to the beam direction with particles incident
from the left. The region between two sheets is zoomed in to show the
position of the electrodes and the liquid Argon gap. A photograph of
the accordion structure of the electromagnetic end-cap is shown on the
right.}
\label{fig:detector:calo:accordion_geometry}
\end{figure}

The EM barrel and end-cap sub-detectors constitute the first sampling
layers over the central portion of the detector ($\abseta < 3.2$) and
possess fine segmentation for high-precision measurements. Both of
these sub-detectors use sheets of steel-reinforced lead as absorbers, between 1.53 and 1.7~mm thick, folded into an
accordion shape as shown in
Fig.~\ref{fig:detector:calo:accordion_geometry}.
 The sheets are
stacked and interleaved with readout electrodes positioned in the
middle of the gaps on
honeycomb spacers. The barrel consists
of two separate half-barrels ($z<0$ and $z>0$) each containing
1024 sheets with the accordion pattern extending radially, stacked to
form cylinders covering full $2\pi$ in azimuth. The cylinders each
have a length 3.2~m with inner and outer diameters of 2.8~m and 4~m
respectively. The end-caps consist of two co-axial wheels, of inner
and outer radii 330~mm and 2098~mm respectively, composed stacked of absorber
sheets and interspaced electrodes with the accordion pattern running parallel to the beam
direction. 

\begin{figure}[htb]
\centering
\includegraphics[width=0.9\textwidth]{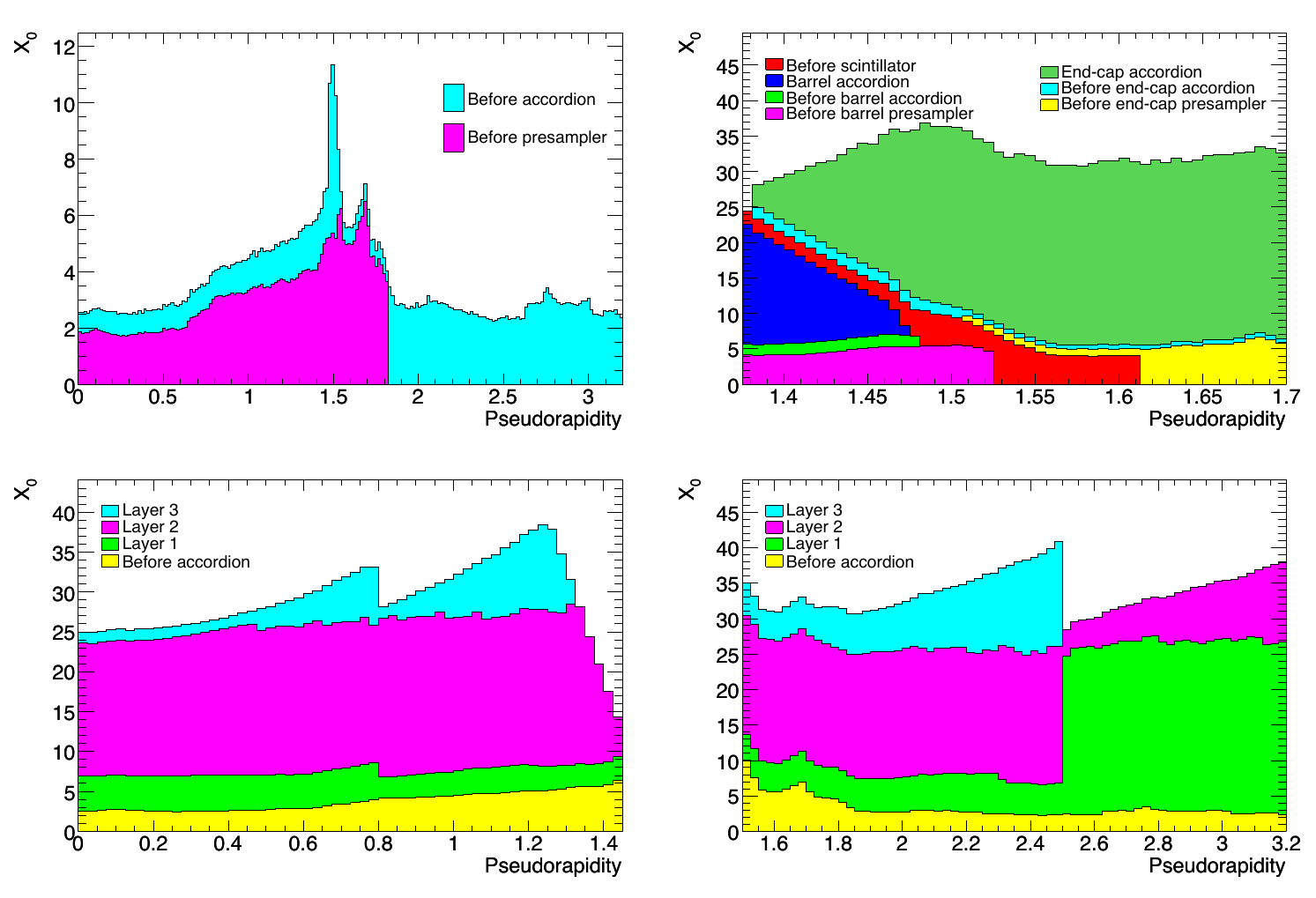}
\caption{Material budget as a function of $\eta$ in front of various
  layers of the calorimeter in terms of the radiation length
  $X_0$. The total amount of material before the calorimeter and
  presampler is shown on the upper right. A complete breakdown of the
  number of radiation lengths before the presampler, before accordion
  sampling and after accordion sampling is shown in the upper
  right. The number of radiation lengths broken down by calorimeter
  sampling layer is shown for the barrel and end-cap in the bottom
  left and right plots respectively.}
\label{fig:det:material_budget}
\end{figure}

The absorber structures are sealed inside cryostats filled with liquid
Argon at 88.5~K. The absorbers initiate electromagnetic showers, with the lower
energy particles in these showers producing ionization electrons in
the LAr. In addition to the sampling layers, separate ``presampler'' layers of
LAr are placed in front of the first barrel and end-cap layers. These
modules allow for the collection of energy from electromagnetic
showers that start early due to the material in front of the
calorimeter. The amount of material, as indicated by the number of
radiation lengths as a function of $\eta$ is shown in
Fig.~\ref{fig:det:material_budget}. The sharp peaking in the upper
left figure is caused by the edge of the cryostat, which projects to a
narrow range in $\eta$.

The electrodes are held at high voltage and collect the
drifting ionization electrons as a signal. This signal is processed by
cold electronics mounted inside the cryostat. These signals are
passed out of the cryostat through dedicated signal feed-throughs to a
front-end crate, which provides signal amplification, pulse shaping
and analog-to-digital conversion (ADC), and output with optical links to
the main DAQ system. The front-end electronics possess an
independent readout path for the L1 trigger where analog sums of
adjacent cells are taken before ADC. The trigger readout possesses a
dedicated readout path. This has the benefit that a correction for ``dead cells'' in the offline readout due
to a failure of the optical link is possible using the coarser trigger
information (dead OTX correction).

\begin{figure}[htb]
\centering
\includegraphics[width=0.45\textwidth]{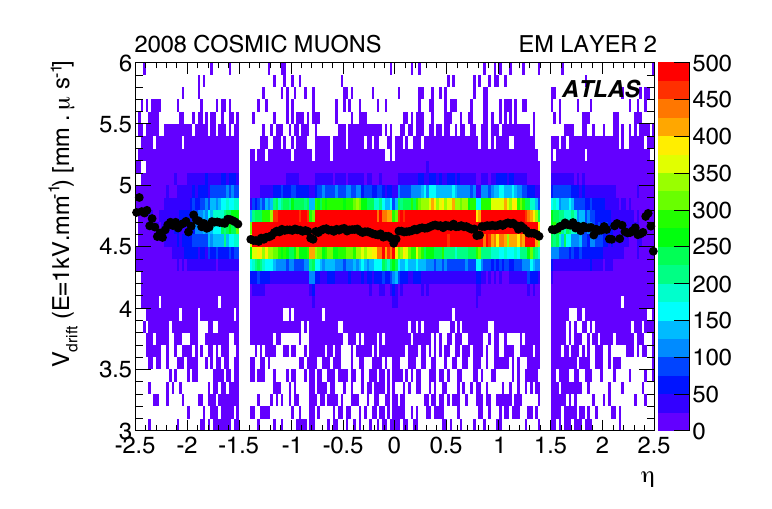}
\includegraphics[width=0.45\textwidth]{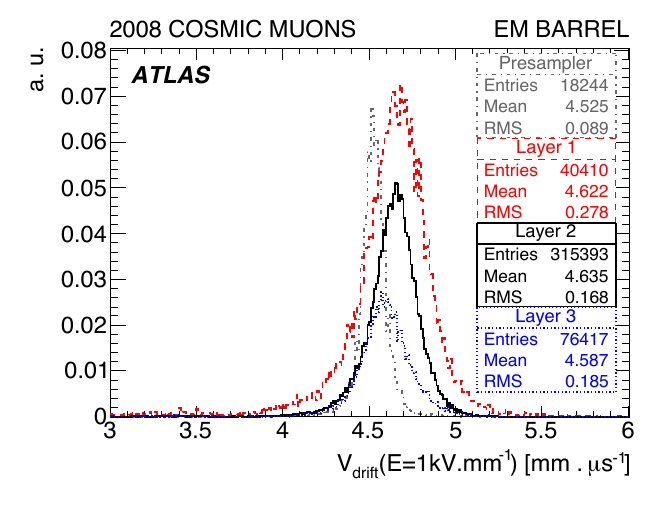}
\caption{Distribution of electron drift velocities in the second
  electromagnetic barrel and end-cap layer (left) as functions of $\eta$. Black points
  denote the mean drift velocity. The $\eta$-averaged drift velocities
  for each of the electromagnetic barrel layers is shown on the left.}
\label{fig:det:lar_electron_drift}
\end{figure}

The high luminosity requires short
drift times ($\sim450$~ns). In the barrel the geometry allows for gaps
of fixed size between the absorber and electrodes, 2.1~mm, which
corresponds to an operating voltage of 2000~V. The end-cap geometry
does not allow this, thus the applied voltage is $\eta$-dependent,
varying from 1000-2500~V. The signal current is proportional to the electron drift velocity,
which is both a function of the applied voltage and the temperature of
the cryogenic liquid, thus the temperature and purity are monitored closely.
Measurements of the drift velocity performed using cosmic muons
prior to the start of LHC operations are shown in Fig~\ref{fig:det:lar_electron_drift}; good uniformity was found in
the measured drift velocities at the operating voltages and temperature~\cite{Aad:2010zh}.

\begin{figure}[htb]
\centering
\includegraphics[width=0.35\textwidth]{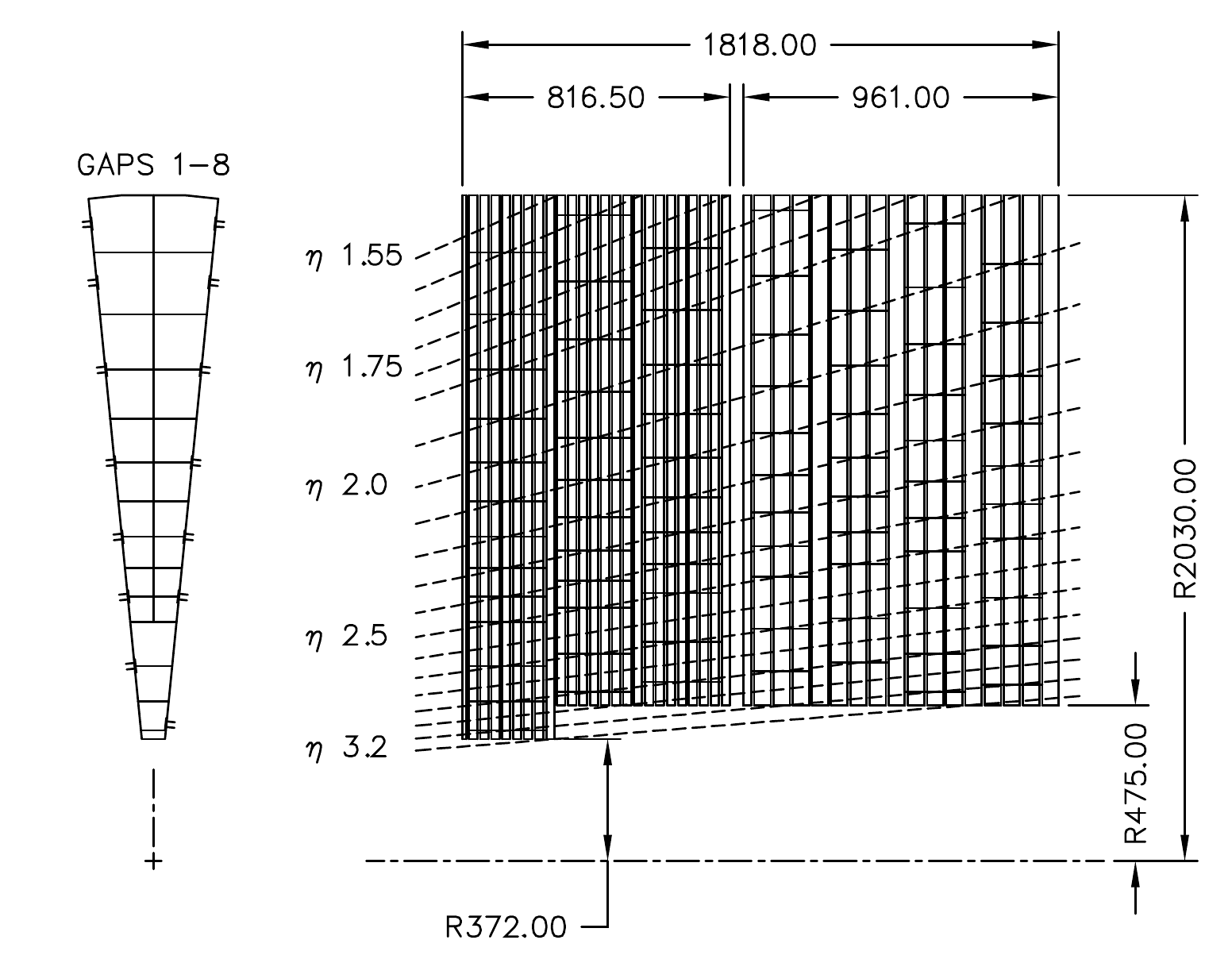}
\caption{Schematic diagram of the HEC showing views in the $R-\phi$ (left)
  and $R-z$ (right). Dimensions are in mm.}
\label{fig:det:HEC}
\end{figure}

The hadronic end-cap calorimeter (HEC) also uses liquid Argon as an active
medium, however it uses copper plates to increase the
number of interaction lengths and thus functions as a hadronic
calorimeter. The end-caps are $\pm z$ symmetric, and each side is
composed of a front and rear wheel module. The copper plates are oriented perpendicular to the beam axis
and there are 24 plates each  25~mm thick in the front wheel modules,
and 16 plates each 50~mm thick in the rear
modules. The wheels each have a front plate of half the nominal plate
thickness, 12.5 and 25 mm respectively. The LAr gaps are fixed at 8.5~mm, and are divided by three
electrodes into four drift zones of 1.8~mm. The readout cells are
defined by pads which are etched on the surfaces of the
electrodes. A schematic drawing of the HEC including dimensions is
shown in Fig.~\ref{fig:det:HEC}.

\begin{figure}[htb]
\centering
\includegraphics[width=0.45\textwidth]{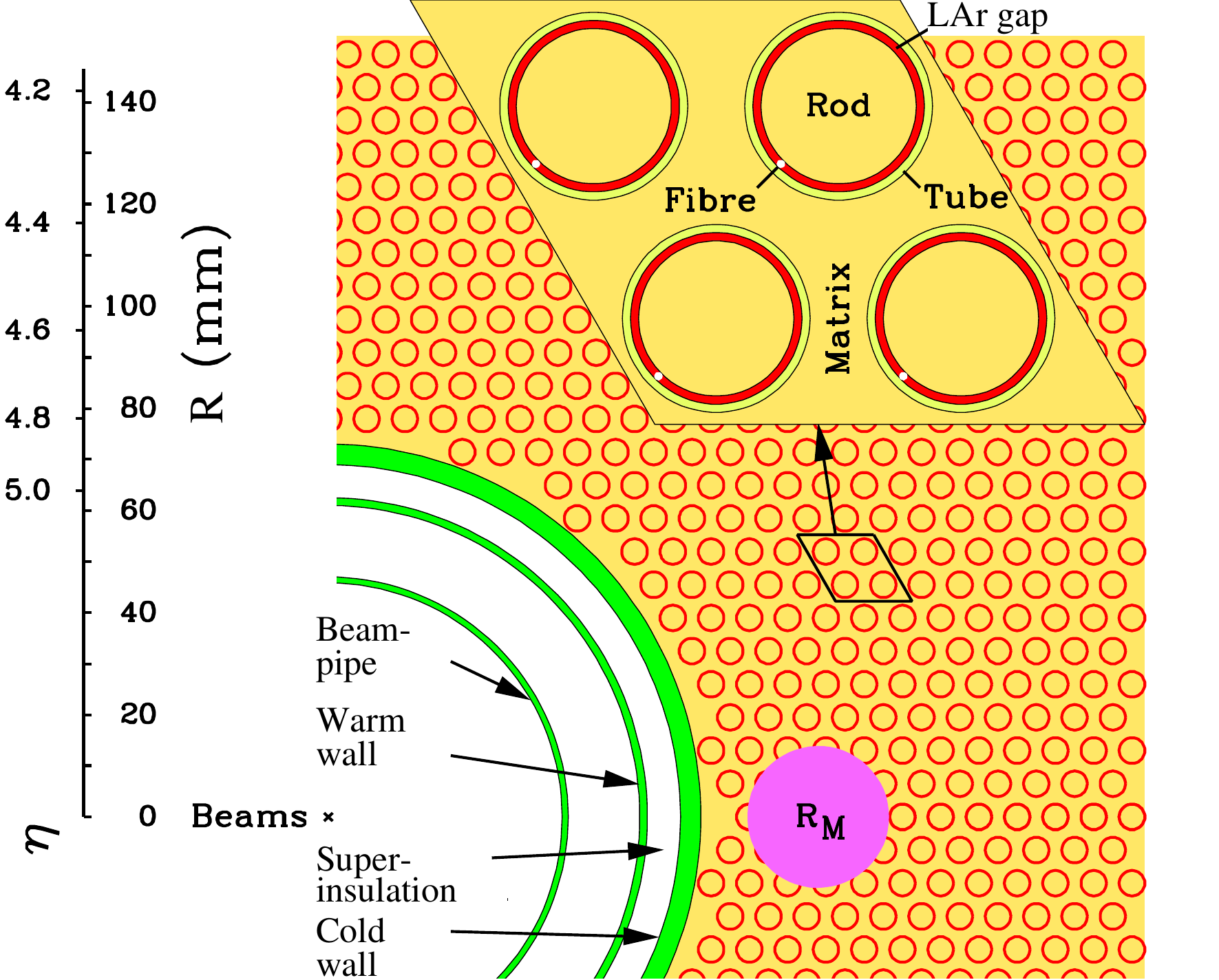}
\includegraphics[width=0.45\textwidth]{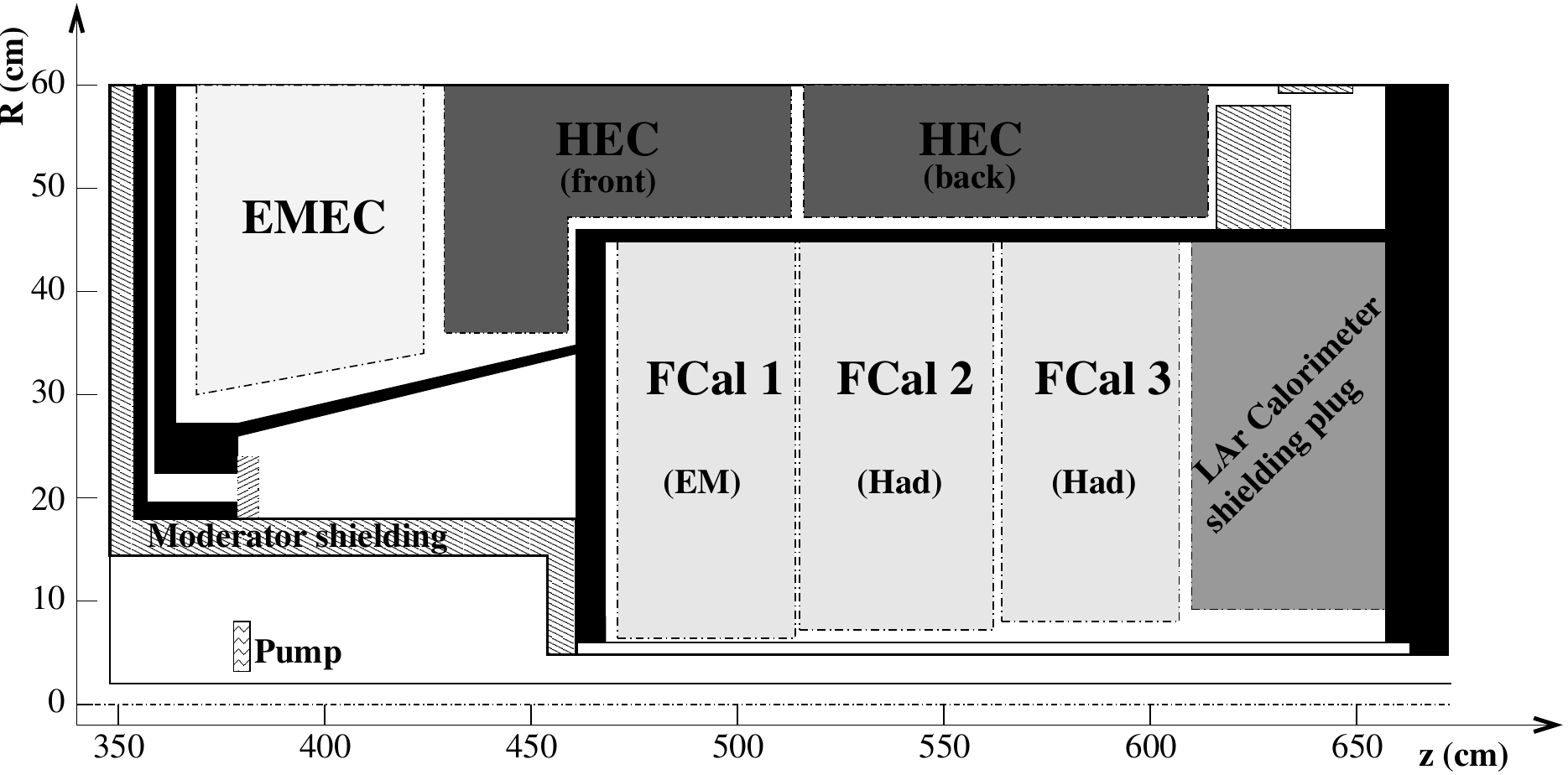}
\caption{The structure of the first FCal module is shown on the
  left. The red circles indicate the LAr gap. The Moli\`{e}re radius is
  indicated by a pink circle, showing the scale of a single shower
  relative to the sampling matrix. The positioning of the FCal modules
  within the end-cap cryostat is shown on the right. 
}
\label{fig:det:fcal} 
\end{figure}

The final element of LAr-based technology are the FCal modules. These
modules cover $3.1 < | \eta| < 4.9$ and are divided into three
longitudinal layers. All three layers use absorber plates
perpendicular to the beam axis. These plates have tubes bored into
them in the beam direction. Each tube is filled with a rod containing
a coaxial cathode and anode electrodes. The space between electrodes is filled with LAr forming a small, 0.249~mm
gap. The first layer is an electromagnetic calorimeter and uses copper
for the absorber. The second and third layers have similar geometry
but are constructed from 2.35~cm copper front and back plates which
support tungsten rods. The region in between rods is filled by filleted
tungsten slugs to maximize the number of interaction lengths. The
layout of the absorber matrix for the first FCal layer is shown in
the left panel of Fig.~\ref{fig:det:fcal}. The FCal, HEC and EM end-cap modules are
all situated inside the same cryostat with the positioning of the FCal
modules shown on the right of Fig.~\ref{fig:det:fcal}.

\begin{figure}[htb]
\centering
\includegraphics[width=0.4\textwidth]{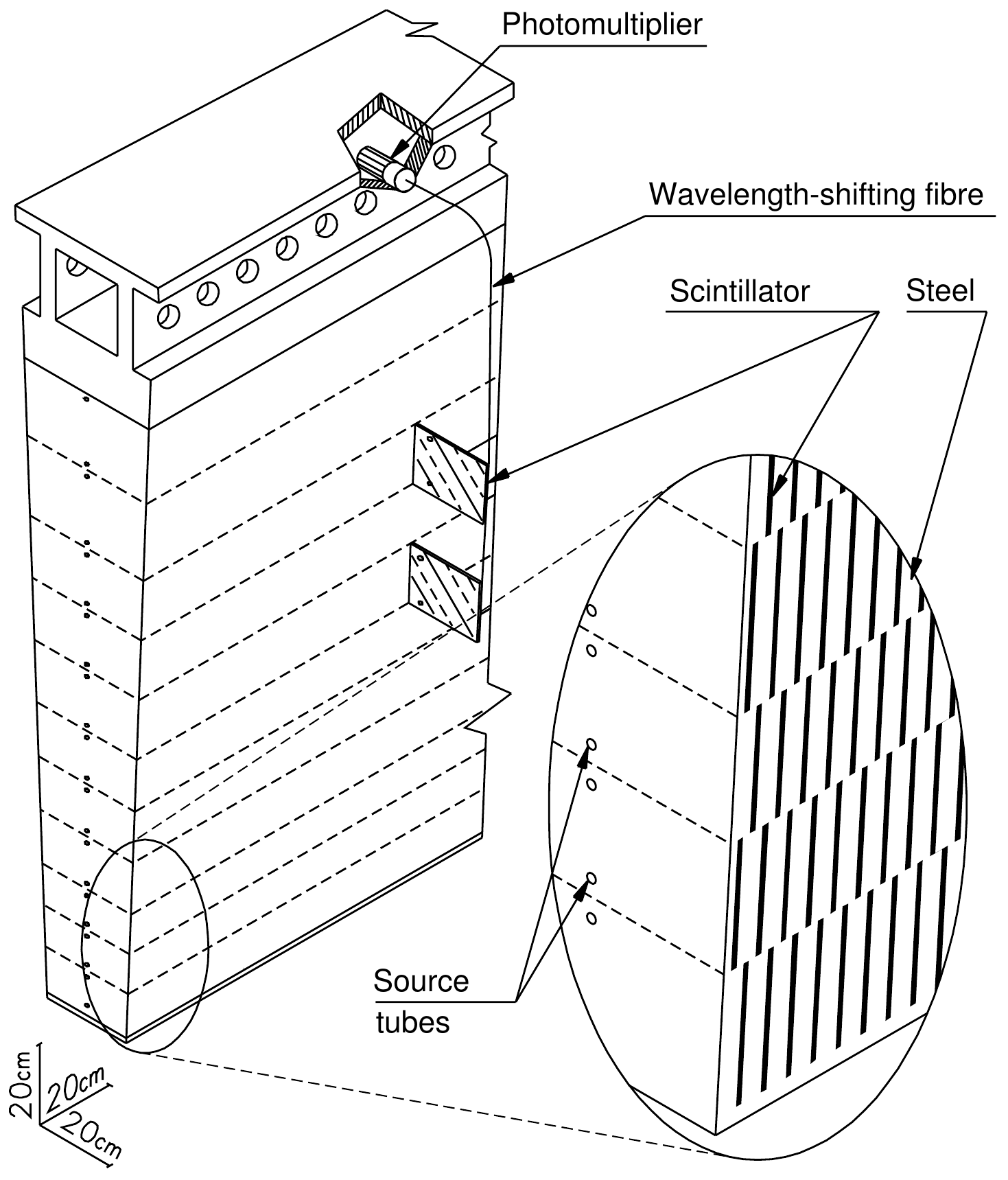}
\caption{Diagram of a tile drawer.}
\label{fig:det:tile_drawer}
\end{figure}

\begin{figure}[htb]
\centering
\includegraphics[width=0.7\textwidth]{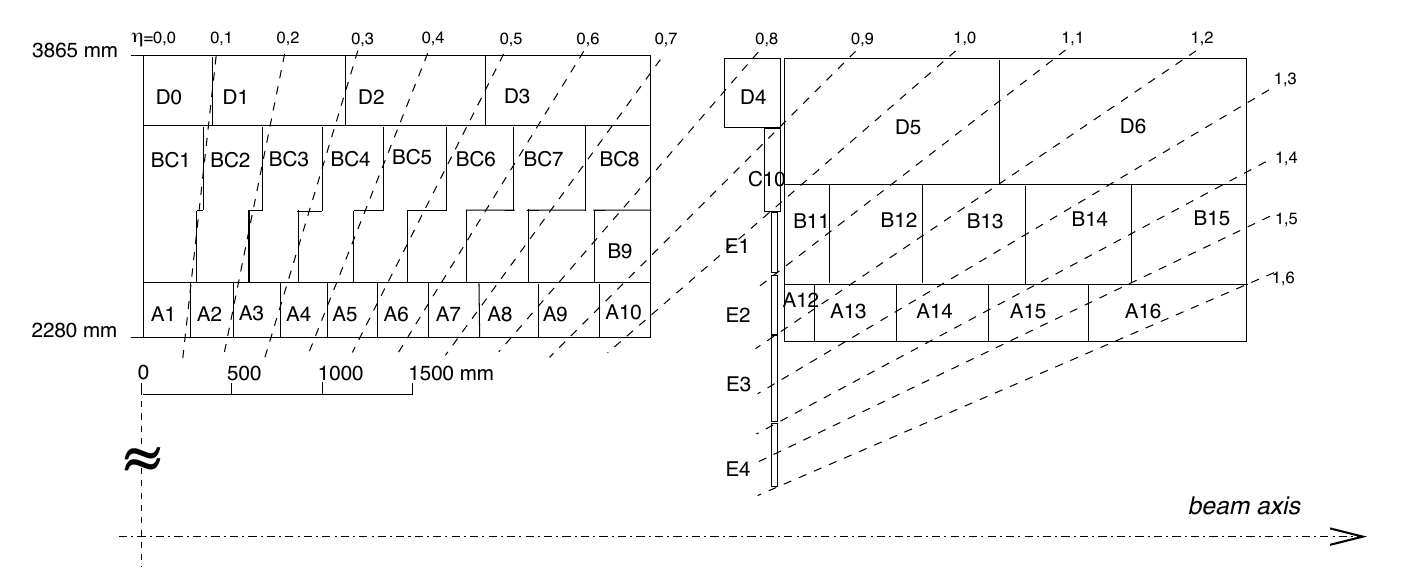}
\caption{The segmentation of the central (left) and extended (right)
  tile barrel calorimeters. The segmentation is chosen to be
  approximately projective, and lines of constant $\eta$ are shown. 
}
\label{fig:det:tile}
\end{figure}

\begin{figure}[htb]
\centering
\includegraphics[width=0.7\textwidth]{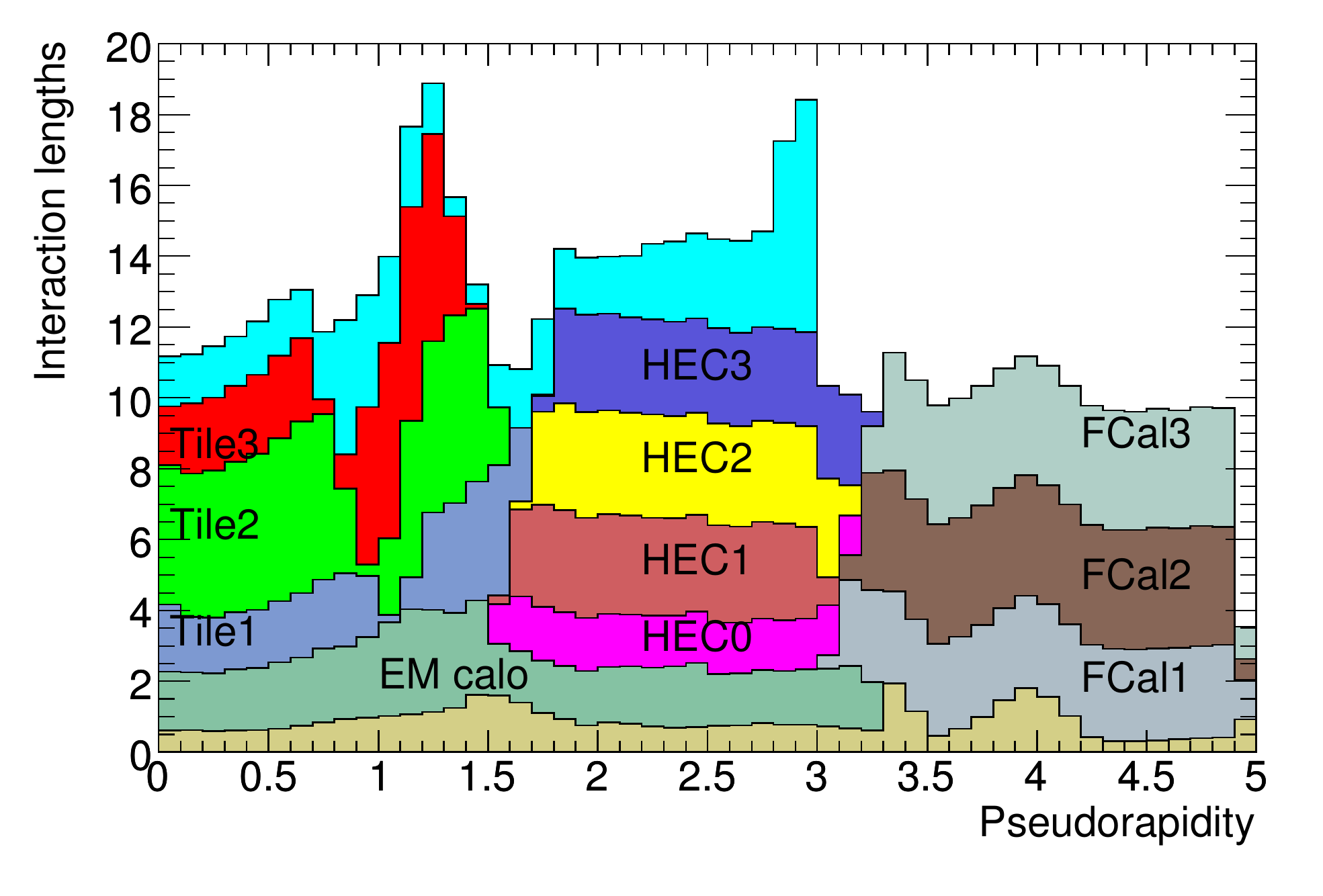}
\caption{Number of nuclear interaction lengths as a function of $\eta$
 with the contribution from each sub-detector and layer shown as a
 different color. The contributions without labels (blue and light
 brown) are due to material. An increase is seen near the cryostat boundaries at
 $|\eta|\sim 1$ and $|\eta|\sim 2.8$.}
\label{fig:det:interaction_lengths}
\end{figure}

The remainder of the calorimetry in ATLAS is provided by the tile
system, which uses alternating steel plates and polystyrene scintillating
tiles. This system is composed of central ($|\eta| < 1$) and
extended ($0.8 < | \eta |  < 1.7$) barrel sections. Each consists of
64 self-supporting, wedge-shaped modules, partitioned in azimuth. A diagram of one such tile drawer is shown in
Fig.~\ref{fig:det:tile_drawer}. Scintillation light is collected in
wavelength-shifting fibers and fed to photomultiplier tubes mounted at
the edge of the modules. The fibers are grouped to provide three
radial sampling depths of thicknesses 1.5, 4.1 and 1.8 interaction
lengths (at $\eta=0$). In addition to the nominal central and
extended barrel regions, a series of tile scintillators are placed in
the geometrically-irregular gap regions between the cryostat. The
segmentation of the tile modules and positioning of the gap
scintillators is shown in Fig.~\ref{fig:det:tile}.
A summary of the amount of hadronic sampling as a function of
calorimeter layer an $\eta$, as indicated by the number of nuclear
interaction lengths (see Section~\ref{section:det:calo:hadronic_showers})
is shown in Fig.~\ref{fig:det:interaction_lengths}.

\subsection{Electromagnetic Showers}
\label{section:det:calo:EM_showers}
In the low to moderate energy regime, energy loss is
described by the Bethe formula which describes the interaction
of charged particles and material through the excitation and ionization
of atoms. Typically the lost energy is extractable by collection of
ionization electrons or scintillation light. Above some energy, called
the critical energy $E_{\mathrm{c}}$, the energy loss is dominated by bremsstrahlung radiation, or pair
production in the case of photons. The length scale associated with
the energy loss rate is known as the radiation length, $X_0$, and
describes the mean distance over which a particle will lose $1/e$ of
its initial energy through radiation. This quantity
depends on the detailed structure of the material including the atomic charge
screening effects on the bremsstrahlung cross section as well as the
density of scattering centers. It has approximate dependence on the
nuclear charge and atomic number of the material as $X_0\sim
AZ^{-2}$. The photon pair-production cross section is controlled by
physics at a similar scale, and at high energies $X_0$ is
approximately $7/9$ of the mean free path for a photon to travel
before pair production.

When a high energy electron or photon enters material, the induced
radiation or splitting results in the initial energy being distributed
among two lower energy particles. These two particles lose energy
through the same mechanism, and this process continues developing an
ensemble of particles among which the initial energy is shared. This
process is known as an electromagnetic shower or cascade. Once
particles in this ensemble become lower than the critical energy, they
no longer contribute to the shower's development, which eventually
dies off. The longitudinal development of the shower can be described
by a simple model. Since each step in the shower development is
binary, after a number of interaction lengths $t=x/X_0$ the shower
will contain
approximately $2^t$ particles, and the average energy is $E=E_0/2^t$,
where $E_0$ is the incident electron energy. The system should freeze
out when the mean energy is of order $E_{\mathrm{c}}$, which
corresponds to a shower depth of
\begin{equation}
x=X_0\frac{\ln(E_0/E_{\mathrm{c}})}{\ln 2}\,.
\end{equation}
More sophisticated models describe the average evolution by a gamma
distribution, which allows for more parameters in describing the
medium dependence of the shower. In the transverse direction, the
shower width is well described by the Moli\`{e}re radius,
$R_{\mathrm{M}}=X_0E_{\mathrm{s}}/E_{\mathrm{c}}$, where
$E_{\mathrm{s}}\approx 21$~\MeV.

\subsection{Hadronic Showers}
\label{section:det:calo:hadronic_showers}
Electrons and photons, as well as particles that decay into them such
as the \pizero, are measured exclusively through electromagnetic
showers. Occasionally \pipm\ will undergo quasi-elastic charge exchange, resulting
in an electromagnetically detectable \pizero. The bremsstrahlung rate
for heavier particles is suppressed by the large particle masses. Energy
measurements of other particles typically occurs through hadronic
showers, which are considerably more complicated than their
electromagnetic counterparts. The exceptions are muons and neutrinos,
neither of which interact hadronically. The former are typically
detected as minimum ionizing particles and often require a dedicated
muon tracking system to measure. The neutrino
energy can be inferred from so-called missing \ET\ variables in
hermetic detectors.

Inelastic collisions of hadrons incident on atomic nuclei
result in a wide range of byproducts resulting in large fluctuations
in the measurable energy in a hadronic shower. As the shower develops
some fraction of energy becomes electromagnetically visible in the
form of produced \pizero's, which start their own electromagnetic
showers. While some of the collisional debris may be detectable
through ionization, many of the struck nuclei become fragmented resulting in
spallation neutrons, which are difficult to detect. Although some neutrons
thermalize producing photons which can be measured through late
neutron capture, much of the shower energy in spallation
neutrons and nuclear recoil cannot be collected. A material's ability
to initiate hadronic showers and appear opaque to hadronically
interacting matter is given by the nuclear interaction length
$\lambda_{\mathrm{I}}$, which describes the mean free path for
undergoing inelastic hadronic interactions in a medium.

Hadronic showers typically develop a substantial electromagnetic
component. Thus hadronic calorimeters have separate responses to
electromagnetic and hadronic processes. If these responses are
significantly different, an effect known as non-compensation, fluctuations in the electromagnetic/hadronic
composition in a shower will significantly increase the energy
resolution and cause a response that is not linear in the incident
particle energy. Both of these features are undesirable and represent
key constraints when designing a hadronic calorimeter. Typically these
features are minimized by decreasing the electromagnetic sensitivity (lower
$Z$) while increasing the hadronic (higher $A$).

\section{Minimum Bias Trigger Scintillators}
\label{section:det:MBTS}
The ATLAS Minimum Bias Trigger Scintillators (MBTS) detector is used
to both select collision candidates online and reject background
events in offline event selection. This detector is composed of 32
modules, each a 2\cm\ thick polystyrene scintillator embedded with
wavelength shifting fibers for readout. The modules are located $\pm
3.6$~m from the detector center with a total coverage of $2.09 <|\eta|< 3.84$ and $2\pi$ in azimuth. Each side (A and C) contains 16 modules in two groups in $\eta$:  $2.09 < |\eta| < 2.82$, $2.82<|\eta|<3.84$. Each group contains 8 wedges which together span the full azimuth.

A trigger readout is implemented by applying a leading-edge
discriminator to the signal pulse sending a hit to the CTP for each of
the 32 modules that is over a threshold. Additional L1 items are built
from these bits including the coincidence triggers,
\verb=MBTS_N_N=, which are fired if $N$ or more modules fired on each of the A and C sides.

Aside from the L1 trigger, the MBTS is primarily used offline for
timing. From the pulse sampling, a time measurement relative to the LHC clock can
be determined for each side. The MBTS $\Delta
t_{\mathrm{MBTS}}=t_{A}-t_{C}$ can be used to reject out-of-time
signals corresponding to non-collision background or collisions
between satellite bunches.

\section{Zero Degree Calorimeters}
\label{section:det:ZDC}

Positioned $\pm 140$~m from the interaction points are the
ATLAS Zero Degree Calorimeters (ZDC). The detector measures neutral particles with $|\eta| >
8.3$, in particular spectator neutrons in heavy ion collisions;
all charged nuclear fragments are swept away by magnetic fields before
reaching the ZDC. Each side contains four rectangular modules,
the first of which is an electromagnetic module with finer
readouts. Each module contains 11 plates perpendicular to the
beam direction made from a combination of tungsten and stainless
steel absorbers. Between the plates are $1.55$~mm quartz
strips. Hadronic showers of incident particles are initiated by the absorber. \v{C}erenkov light from the showering
particles is produced in the quartz and fed through air light-guides
into photomultiplier tubes situated on top of the detector. Some of
the modules contain specialized
quartz rods to provide position measurements, although this aspect of
the detector capability is not used in minimum bias event
selection for heavy ion collisions.

For the 2010 ion run, the ZDC was used primarily as a minimum bias
trigger. Each side is capable of producing a L1 trigger signal by passing the
analog channel sums through a discriminator to the CTP. The
coincidence trigger, \verb=ZDC_A_C=,
is a logical \verb=AND= of the one-sided triggers,
\verb=ZDC_A= and \verb=ZDC_C=. The requirement
of a single neutron on both sides is effective in rejecting against
photo-nuclear collisions, which are typically asymmetric and are a
large background to the most peripheral collisions.

\section{Inner Detector}
\label{section:det:ID}
\begin{figure}[htb]
\centering
\includegraphics[width=0.9\textwidth]{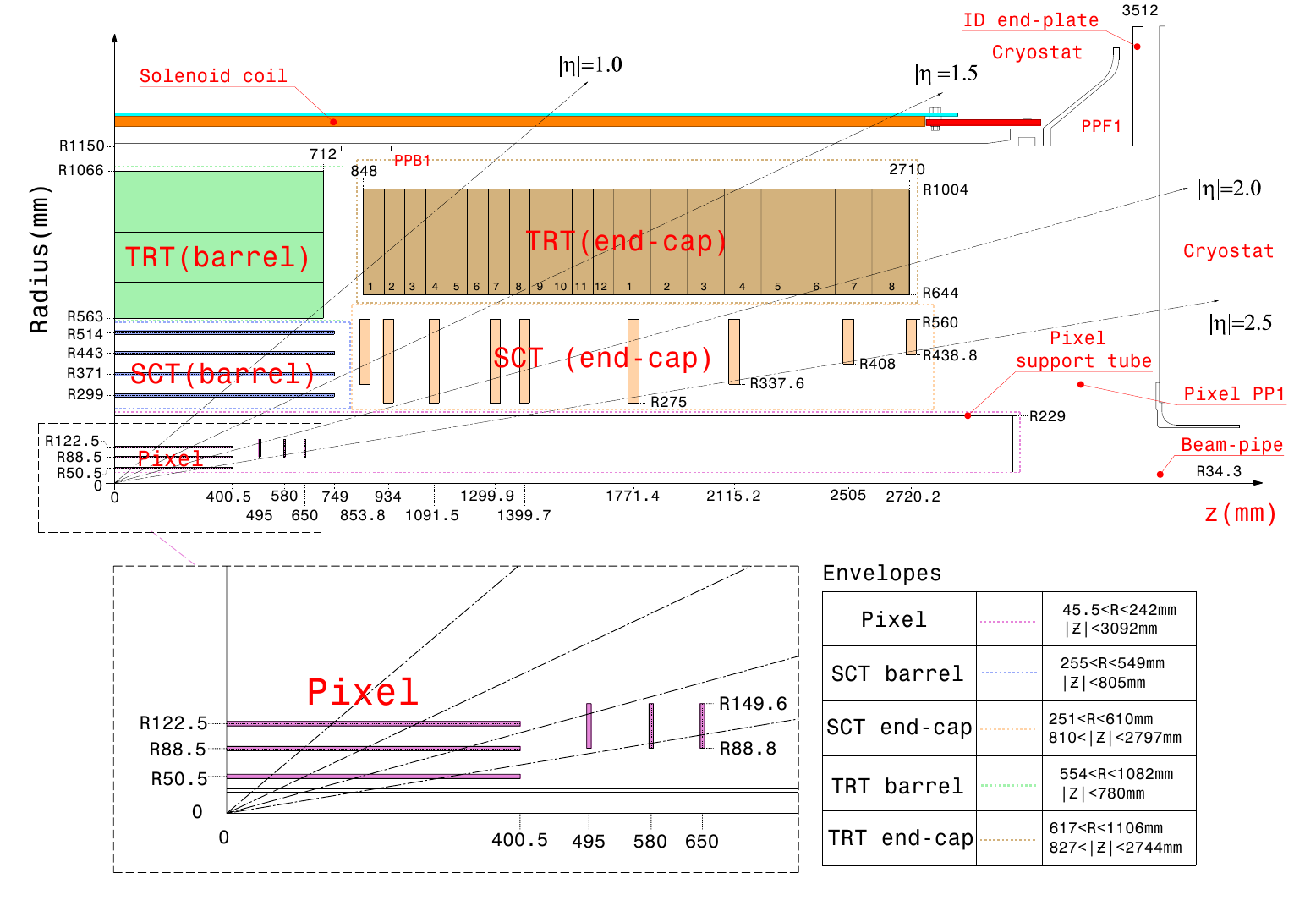}
\caption{Schematic view of the inner detector showing the positions of
the various modules in $R$ and $z$.}
\label{fig:det:id_slice}
\end{figure}

\begin{figure}[htb]
\centering
\includegraphics[width=0.49\textwidth]{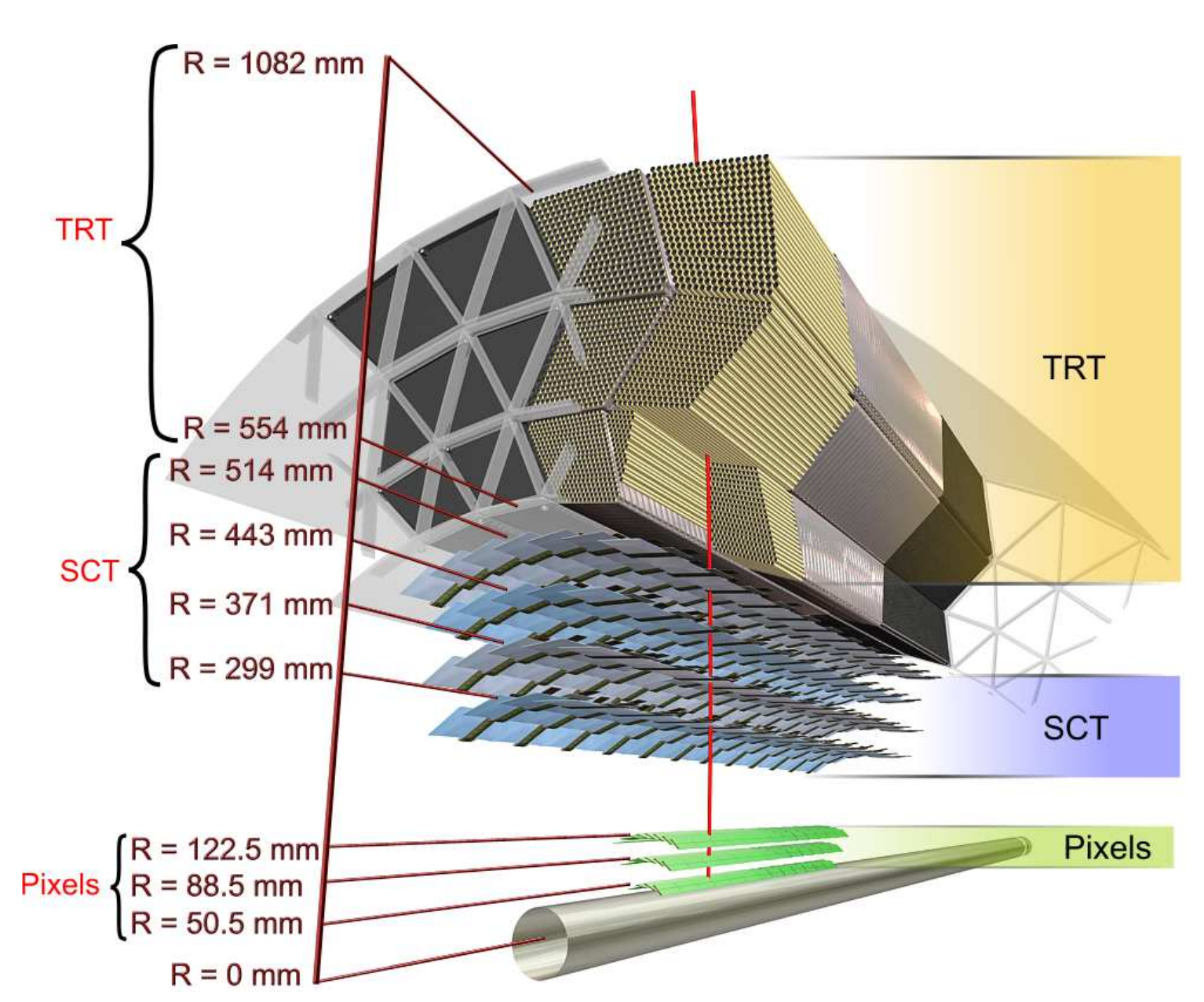}
\includegraphics[width=0.49\textwidth]{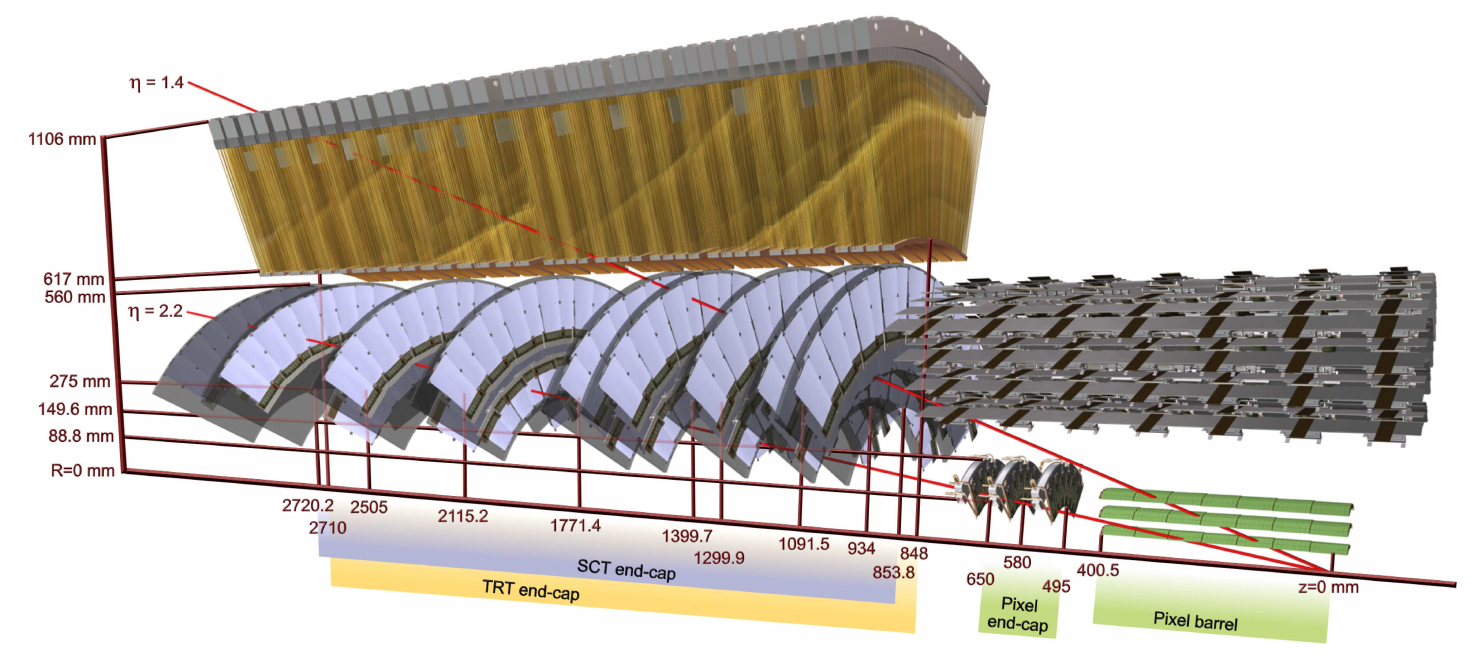}
\caption{Perspective view of the inner detector showing the barrel
  (left) and end-cap (right) regions.}
\label{fig:det:id_perspective}
\end{figure}

The ATLAS inner detector occupies a
cylindrical volume around the detector center spanning $\pm3512$~mm in
the $z$ direction and 1150~mm in radius. Space-point measurements are
provided by three separate sub-detectors over the radial extent of the system, each divided into barrel and
end-cap modules. A slice of the detector in the $R-z$ plane showing
the various components is shown in
Fig.~\ref{fig:det:id_slice}. Additional figures showing perspective
views of the inner detector system in the barrel and end-cap region
are shown in Fig.~\ref{fig:det:id_perspective}.

The silicon pixel detector is composed of three
cylindrical layers beginning at a radial distance of 50.5~mm and three
end-cap disks per side. It contains 1744 identical pixel sensors which
are 250~\mumeter\ thick oxygenated n-type wafers with readout
pixels on the $\mathrm{n}^{+}$-implanted side of the detector.
Each sensor contains 47232 pixels for a total of approximately
80~million readout channels (the total number of channels is reduced
by the ganging of adjacent pixels on front-end chips). The pixels have
a nominal size of
 50~\mumeter\ in $R-\phi$ and 400~\mumeter\ in $z$ providing measurements with intrinsic inaccuracies of
 10~\mumeter\ in $R-\phi$ for all modules and 115~\mumeter\ in $z$ and $R$ for
 the barrel and disk modules respectively.

The silicon microstrip detector (SCT) consists of four cylindrical
layers and nine disks per side each composed of small angle stereo
strips positioned beyond the pixel detector. The barrel covers the
radial region $299<R < 514$~mm, and $|z| < 805$~mm. The end-cap
 modules are positioned $251 < R < 610$~mm and $810 < | z| < 2797$~mm. The strips are composed of two 6.4~cm daisy-chained sensors,
 $285\pm15$~\mumeter\ thick,
with a strip pitch of 80~\mumeter. In the barrel these strips are oriented parallel to the beam
direction, measuring $R-\phi$, while in the end-caps they are
trapezoidal wedges and are oriented radially.

The Transition Radiation Tracker (TRT) is the outermost component of the
tracking system. The barrel spans radial region $554 < R < 1066$~mm
and $|z| < 780$~mm. The end-cap wheels span $615 < R < 1106$~mm,
positioned between $ 827 < |z| < 2744$~mm. It is composed of
polymide drift tubes, 4~mm in diameter and 71.2~cm in length. The tube wall is composed of
two 35~\mumeter\ multi-layer films. The cathode is a 0.2~\mumeter\ Al
film on the inner surface. The anode wires are 31~\mumeter\
gold-plated tungsten, positioned at the nominal center of the tube. To
ensure stable operation the wires are required to have an offset with
respect to the tube center of less than 300 $\mu\mathrm{m}$. The tubes are filled with a gas mixture of approximately 70\%
Xe, 27\% $\mathrm{CO}_2$ and 3\% $\mathrm{O}_2$, held slightly
over-pressure. Photons from transition radiation of high energy
electrons can be distinguished
from normal ionization tracking signals on a per tube basis using
separate high and low thresholds on the readout electronics. The wires
are divided into three sections. They are read out at each end, with
an inefficient center section of approximately 2~cm, which contains a mid-wire
support structure. The TRT barrel is composed of 73
layers of tubes oriented along the beam direction, while the end-caps
consist of 160 tube planes. Charged tracks will cross at least 36
tubes except in the barrel/end-cap transition region ($0.8 < |\eta| <
1.0$) where this minimum drops to 22.

\clearpage
\chapter{Jet Reconstruction}
\label{section:jet_rec}
\section{Subtraction}
\label{section:jet_rec:subtraction}
For a calorimetric measurement, the input to the jet reconstruction 
procedure is the measured \et\ distribution, which contains both the
jet signal and contributions from the underlying event,
\begin{equation}
\dfrac{d\et^{\mathrm{total}}}{d\eta d\phi}=\dfrac{d\et^{\mathrm{UE}}}{d\eta d\phi}+\dfrac{d\et^{\mathrm{jet}}}{d\eta
d\phi}\,.
\end{equation}
The magnitude of the underlying event can vary over several orders of
magnitude depending on the collision geometry, thus an accurate procedure requires that a
background estimation be performed on an event-by-event
basis. Correlations in the underlying event,
specifically $v_2$, must be incorporated into the subtraction
procedure as well. The background energy density is estimated by
\begin{equation}
\dfrac{d\et^{\mathrm{UE}}}{d\eta d\phi}\simeq B(\eta,\phi)=\rho(\eta)\left[1+2v_{2}\cos(2(\phi-\Psi_{2}))\right]\,,
\end{equation}
where $\Psi_2$ is the event plane angle and
$\rho(\eta)=\Big\langle\dfrac{d\et}{d\eta d\phi}\Big\rangle$  is the
average \et\ density taken over the full $2\pi$ in azimuth in strips
of constant \eta\ with width 0.1. To account for variation of detector
response in the different
longitudinal sampling layers, the background determination and
subtraction is performed per layer.

The possibility exists for the jets to bias the determination of
$\rho$. If a jet's energy is included in the background calculation,
the background is overestimated, resulting in an over-subtraction being
applied to the jet. This effect, called the self-energy bias,
will bias the jet energy after subtraction by approximately 10\% and will distort
the $v_2$ as well. To address this, the jet reconstruction used in
this analysis employed an iterative procedure where jets were reconstructed and
identified as jet seeds. $B$ was
constructed with the seeds excluded from the determination of $\rho$. This reduced the bias in the background
and the subtracted jets. These jets were then be used as seeds to
iterate the procedure, by constructing new estimates of $\rho$ and $v_2$.

\subsection{Seeds}
\label{reco:seeds}
In clustering algorithms, such as 
anti-$k_{t}$, the entire $\eta-\phi$ distribution is
tessellated by reconstructed jets. While some of these jets likely
correspond to a true jet signal, the clustering of many of the
background jets is due to the underlying event. If a set of
criteria can be chosen to partition the full set of jets output by the
reconstruction into one that characterizes the background
distribution and one that does not, the latter can be used as jet
seeds. A simple \et\ threshold cannot be used to define the seeds at
this stage since the subtraction has not yet been performed. For the
first step in the iterative procedure, a discriminant was constructed by considering
the distribution of tower \et's inside a jet and taking the ratio of
the maximum and mean of this distribution:
\begin{equation}
D=\dfrac{\mathrm{max}(\et^{\mathrm{tower}})}{\langle \et^{\mathrm{tower}}\rangle}\,.
\label{eqn:discriminant_def}
\end{equation}
A cut of $D>4$ identifies jets with a dense core spanning a few
calorimeter towers, which are typical of the true signal. This
requirement fails
when calorimeter noise allows the denominator of the
discriminant to become arbitrarily small resulting in a
large discriminant even when the distribution maximum is also
small. To prevent this from occurring, a threshold on $D$ was be applied in conjunction with a minimum requirement on the
tower distribution maximum or mean. This cut was taken to be a minimum value
of $\mathrm{max}(\et^{\mathrm{tower}})>$ 3~\GeV.

In subsequent stages of the iteration procedure the jets
that were the output of the previous iteration step were reused as
seeds. Since these jets had already been subtracted, an \et\ cut of
25~\GeV\ could be applied to define the seeds. In addition to these jets,
track jets with $\pt > 10\GeV$ were also added to the list of seeds.

\subsection{Background Determination}
\label{reco:background_determination}
\begin{figure}[htb]
\centering
\includegraphics[width =0.49\textwidth] {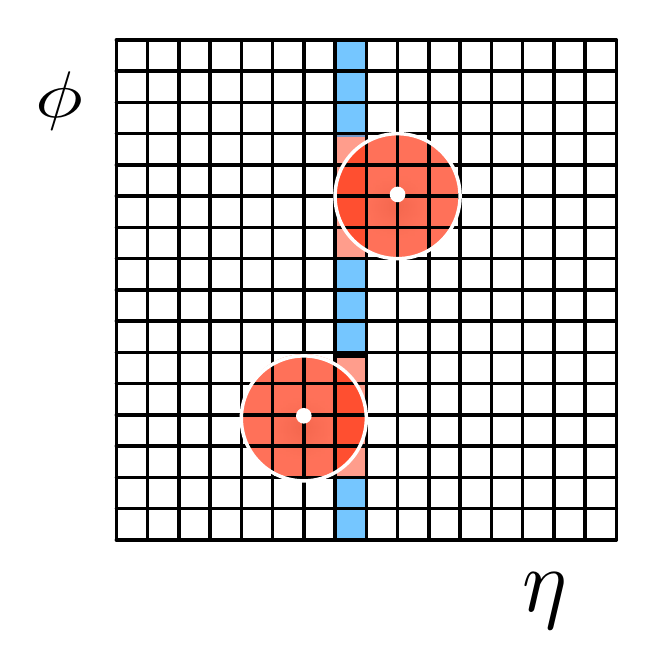}
\includegraphics[width =0.49\textwidth] {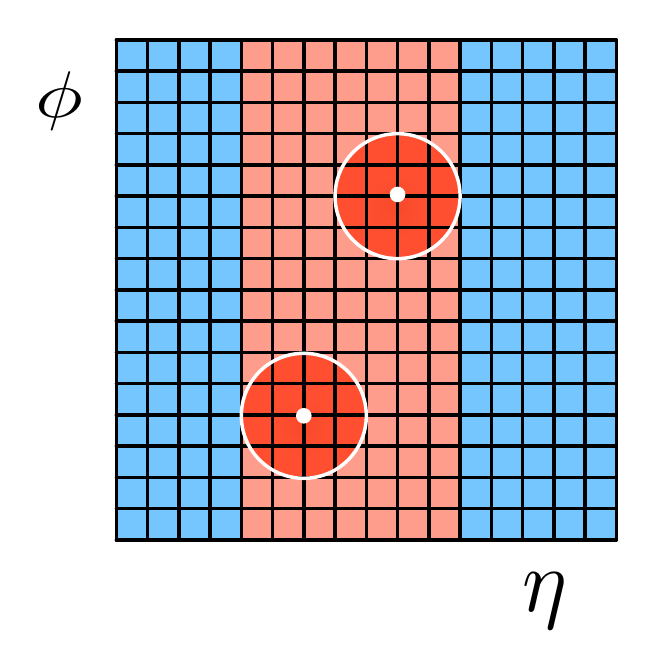}
\caption{An example of how seeds are excluded from the $\rho$ (left) and $v_2$
  (right) calculations. The positions of the seeds are indicated by
  white dots. The red regions are excluded from the averages and the
  blue regions are included.
}
\label{fig:reco:exclusion}
\end{figure}

The average density was computed by considering all calorimeter cells and
calculating an average transverse energy density per cell. 
For a sampling layer $i$  the average is defined as
\begin{equation}
\rho_{i}(\eta)=\dfrac{1}{N_{i}}\displaystyle \sum_{j}^{N_{i}} \dfrac{{\et}_j}{\Delta\eta_j \Delta\phi_j}\dfrac{1}{1+2v_{2\,i}\cos(2(\phi_j-\Psi_2))}\,.
\label{eqn:rho_with_flow}
\end{equation}
The average was only taken over cells that are in the same pseudorapidity
bin as \eta\ and are not associated with a
seed. In the discriminant-based approach, the seed-associated cells were
defined as the constituents of the jets that pass the discriminant cut. In all other
cases they were defined as being within $\Delta R < 0.4$ of a seed
position as shown in Fig.~\ref{fig:reco:exclusion}. This exclusion
radius was used regardless of the
$R$ value used in the anti-\kt\ algorithm. The $\phi$-dependent demodulation factors were required since
the average was not performed over the full $\phi$ range and thus elliptic flow
effects do not cancel. 

\subsection{Elliptic Flow Correction}
\label{reco:v2_correction}
The event plane angle was determined using the
first layer of the FCal using the same procedure as
in Refs.~\cite{flowpaper}. This angle was then used to determine the
$v_2$ per calorimeter sampling layer $i$,
\begin{equation}
v_{2\,i}=\langle\cos 2(\phi-\Psi_2)\rangle_{i}\,,
\label{eqn:eta_averaged_v2}
\end{equation}
where the brackets denote an \et\ weighted average over all cells in
layer $i$.

As was the case for $\rho$, it is possible for the jet signal to bias
the $v_2$. This effect is worsened by the fact dijet signals have
the same $\pi$ symmetry as the flow modulation. 
\begin{figure}[t]
\centering
\includegraphics[width=5 in]{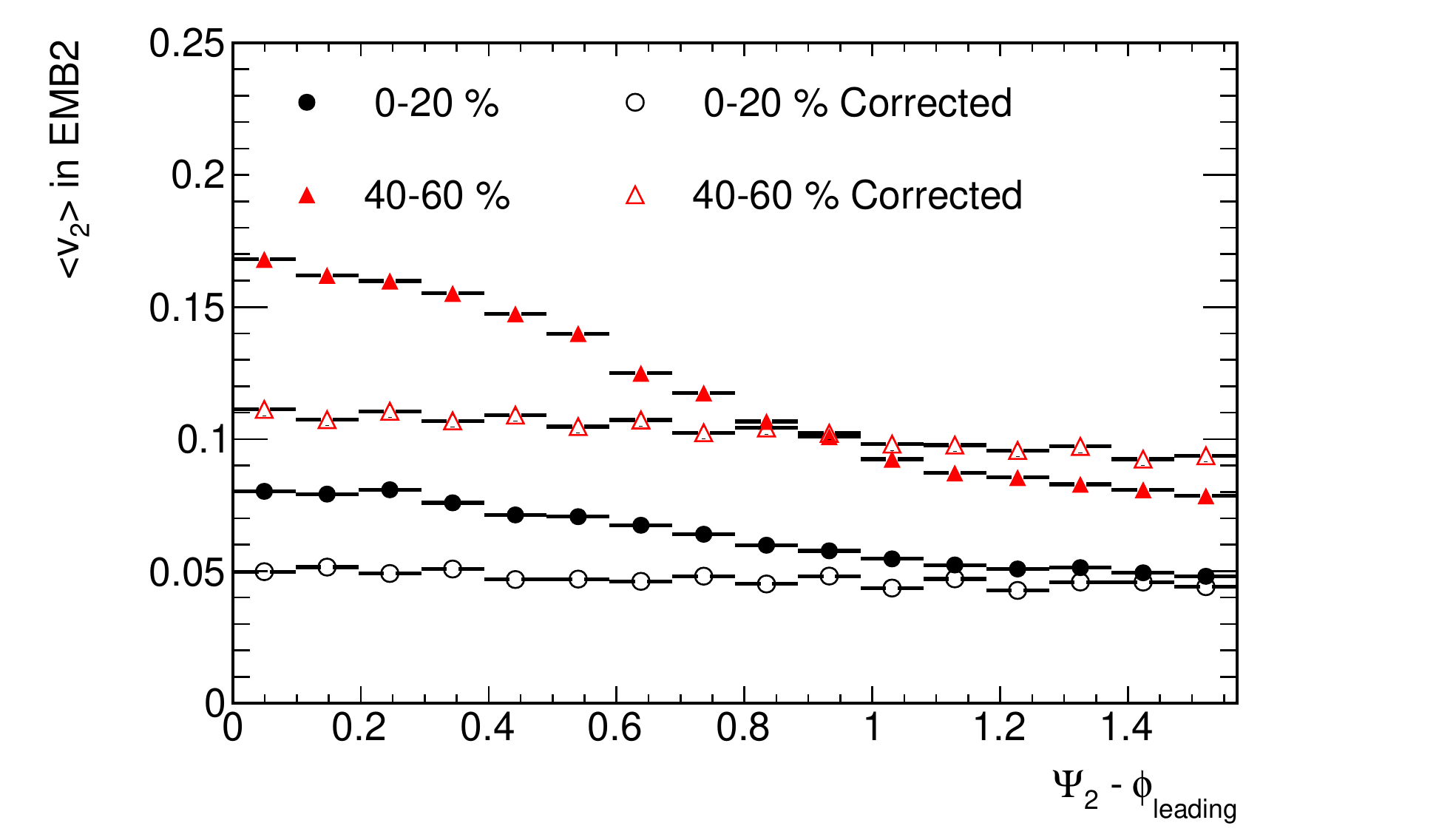}
\caption{
 $v_2$ computed from the second layer of the
  electromagnetic barrel as a function of the leading jet's angle with
  respect to the event plane $\Psi_2$. Two centrality bins (0-20\% and
  40-60\%) are shown before (solid circle/triangle) and after (hollow circle/triangle)
  the iteration step. Without removing the seeds from the calculation,
  the $v_2$ is enhanced in events where the leading jet is aligned with the
  event plane and reduced when the leading jet is out of plane. The
  correlation is removed by the iteration step.
 }
\label{fig:reco:v2_bias}
\end{figure}
Jet seeds can be used to exclude
regions containing potential jets from the averaging and eliminating
the bias. However,
excluding small regions around a seed as described in Section~\ref{reco:background_determination}
cannot be done. In a strip of constant \eta, excluding some of
the cells from the full azimuthal interval will cause an
anti-bias; the $v_2$ for events where jets are aligned with the event
plane will be reduced. To prevent this overcompensation, the entire
\eta\ interval was considered to be biased by the jet. Such \eta\
intervals were flagged using the jet seeds and excluded from the
average in Eq.~\ref{eqn:eta_averaged_v2}. This exclusion is shown
schematically in
Fig.~\ref{fig:reco:exclusion}. The effect of jets on $v_2$ as measured in the second sampling layer
of the barrel electromagnetic calorimeter in MC simulations is shown
in Fig.~\ref{fig:reco:v2_bias}. Without excluding jets from
the $v_2$ determination (solid markers), the calculated $v_2$ is larger in events where
the leading jet is aligned with the event plane. This bias is removed
once the jet seeds are excluded from the calculation as indicated by
the open markers in the figure.

\subsection{Subtraction}
\label{reco:subtraction}
Once the background and flow parameters were determined, a
subtraction was applied at the cell level. For a cell at position $\eta,\phi$ in layer $i$, the corrected \et\ was determined from the raw \et\ via
\begin{equation}
\et=\et^{\mathrm{raw}}-\rho_{i}(\eta)\Delta\eta\Delta\phi[1+2v_{2\,i}\cos(2(\phi-\Psi_2))]\,.
\label{eqn:reco:subtraction}
\end{equation}
Each cell was considered as a massless four-vector using the \et\ after subtraction. The jet's kinematics were set as the four-vector sum of the constituent
cells.

This subtraction was applied to all cells except dead cells that receive no
correction from neighboring cells or the associated trigger tower. The
latter correction was only applied if the trigger tower was above a
2~\GeV\ noise threshold. Thus for the purposes of subtraction, dead
cells eligible for this correction were checked event-by-event. 

\subsection{Calibration}
\label{reco:calibration}
Jet collections were constructed using two different calibration
schemes. The EM+JES scheme uses the cell energies, after subtraction, at the EM scale and applies
a final, multiplicative calibration factor to the jet energy to
calibrate to the full hadronic scale. In the GCW scheme, a calibration is done at
the cell level, which designed to minimize variation in hadronic
response to jets. However, a multiplicative calibration factor is still
required to account for the mean hadronic response. In this scheme the $\rho$
determination and subtraction use calibrated cell energies\footnote{The $v_2$ was determined using cells at the
  electromagnetic scale. For jets using the GCW scheme, the $\et^{\mathrm{raw}}$ in
  Eq.~\ref{eqn:reco:subtraction} was first demodulated by the
  flow weight $1+2v_{2\,i}\cos(2(\phi-\Psi_2))$, calibrated and then
  re-weighted by the same factor.}. The GCW also accounts for energy lost due
to material in the cryostat by calculating a correction using the energy
deposition in the third electromagnetic and first hadronic sampling
layers after subtraction.

After subtraction the multiplicative jet energy calibration factor
was applied to correct for the overall jet energy. The factors
depend on both the jet's energy and \eta, and different constants are
used for the GCW and EM+JES schemes. The
numerical inversion constants were developed specifically for heavy
ion jets and are discussed in Section~\ref{sec:corrections:num_inv}.

\subsection{Track Jets}
\label{reco:track_jets}
Jets reconstructed with charged particles provide an invaluable
cross-check on the calorimetric jets. The anti-\kt\ algorithm
with \RFour\ was run using charged particles as input. Instead of performing a
subtraction, particles were required to have a minimum \pt\ of 4~\GeV\
to be included in a track jet in addition to the following track
selection requirements, where impact parameter variables are defined
with respect to the primary reconstructed vertex:
\begin{itemize}
\item $d_0,\,\, z_0\sin\theta < 1.0\,\mathrm{mm}$,
\item $\dfrac{d_0}{\sigma_{d_0}},\,\,\dfrac{z_0\sin\theta}{\sigma_{z_0\sin\theta}} < 3.0$,
\item $N_{\mathrm{Pixel}} \geq 2$,
\item $N_{\mathrm{SCT}} \geq 8$.
\end{itemize}
\subsection{Jet Collections}
\label{reco:collections}
With the exception of the track jets, all the results presented here use jet collections obtained with the
iterative method using a single iteration step applied to both the
background and flow. Seeds were constructed by running anti-\kt\
with \RTwo. The discriminant procedure was used to determine the
background, and the initial $v_2$ values were used with this background
to produce subtracted, flow corrected jets. The list of seeds for the iteration step was constructed from
jets with $\et>25\GeV$ and all track jets with $\et>10\GeV$. This list
of seeds was used to define new background and $v_2$
values. These were used to construct iterated jet collections
with four different values of $R$: $R=0.2,\,0.3,\,0.4$ and $0.5$. For each of these values of $R$,
two versions of the collections exist corresponding to the two
calibration schemes: GCW and EM+JES. The
25~\GeV\ cut defining the seeds was chosen based on the observation that the fake rate for
\RTwo\ jets above this \pt\ is not too high to cause biases in the
subtraction when a fake jet is used as a seed.

\section{Monte Carlo Sample}
\label{section:jet_rec:mc_sample}
The studies presented here use a Monte Carlo sample of minimum bias
HIJING~\PbPb\ events with embedded PYTHIA~\pp\ events at
$\sqrtsnn=2.76\TeV$. Events from both generators were produced and
the detector response to each event was independently simulated using a
full GEANT~\cite{Agostinelli:2002hh,Allison:2006ve} description of the ATLAS detector~\cite{Aad:2010ah}. The signals were combined during the
digitization stage and then reconstructed as a combined event. For
each HIJING sample a set of PYTHIA samples (referred to as J samples) was produced, each with a
fixed range set on the $\hat{p}_{\mathrm{T}}^{\mathrm{min}}$ and $\hat{p}_{\mathrm{T}}^{\mathrm{max}}$ in the PYTHIA hard
scattering. A single HIJING event was overlaid on an event from each
of the different J samples, with no reuse of HIJING events within a J
sample, i.e. $N_{\mathrm{evt}}^{\mathrm{HIJING}}=N_{\mathrm{evt}}^{\mathrm{J}}$. The different J samples
were then combined using a cross-section weighting obtained from PYTHIA
to build a combined sample with good counting statistics over a wide
range of jet~\pt. The definitions of these samples and the associated
cross sections are shown in Table~\ref{tbl:mc:pythia_J_samples}.
\begin{table}[h]
\centering
\begin{tabular}{| c | r | r | r@{}l|} \hline
J &	\multicolumn{1}{|c|}{$\hat{p}_{\mathrm{T}}^{\mathrm{min}}$ [\GeV] }&  \multicolumn{1}{|c|}{$\hat{p}_{\mathrm{T}}^{\mathrm{max}}$  [\GeV]} & \multicolumn{2}{|c|}{$\sigma$ [$\mathrm{nb}$]} \\ \hline
   1 & 17  &  35  &  157000& \\ \hline
   2 &  35  &  70  &  7090& \\ \hline
   3 &  70 &  140 &  258&\\ \hline
   4 &   140 &  280 &  5&.85\\ \hline
   5 &  280  &  560 &  0&.0612\\ \hline
\end{tabular}
\caption{Definitions of PYTHIA \pp\ samples used in embedding. For each, J
  value samples were produced with the same number of events to ensure
high statistical sampling for a jet~\pt\ out to 500~\GeV.}
\label{tbl:mc:pythia_J_samples}
\end{table}

As HIJING does not contain a mechanism for
simulating elliptic flow, an afterburner was applied to modulate
the azimuthal angles of the HIJING particles using a \pt, \eta\ and
centrality-dependent parametrization of existing $v_2$
data~\cite{Masera:2009zz}. Samples used in this study use a parametrization derived
from 2010~\PbPb\ data at the LHC. 

The goal of the embedding procedure is to study the combined effect of
the underlying event and detector response on the jet
signal. Truth jets were constructed from the generator-level particle content of the
PYTHIA event only. These jets were then compared to the output of
offline jet reconstruction to extract metrics of jet performance and
information for correcting measurements for these effects.

\begin{figure}[tb]
\centering
\includegraphics[height =0.5\textheight] {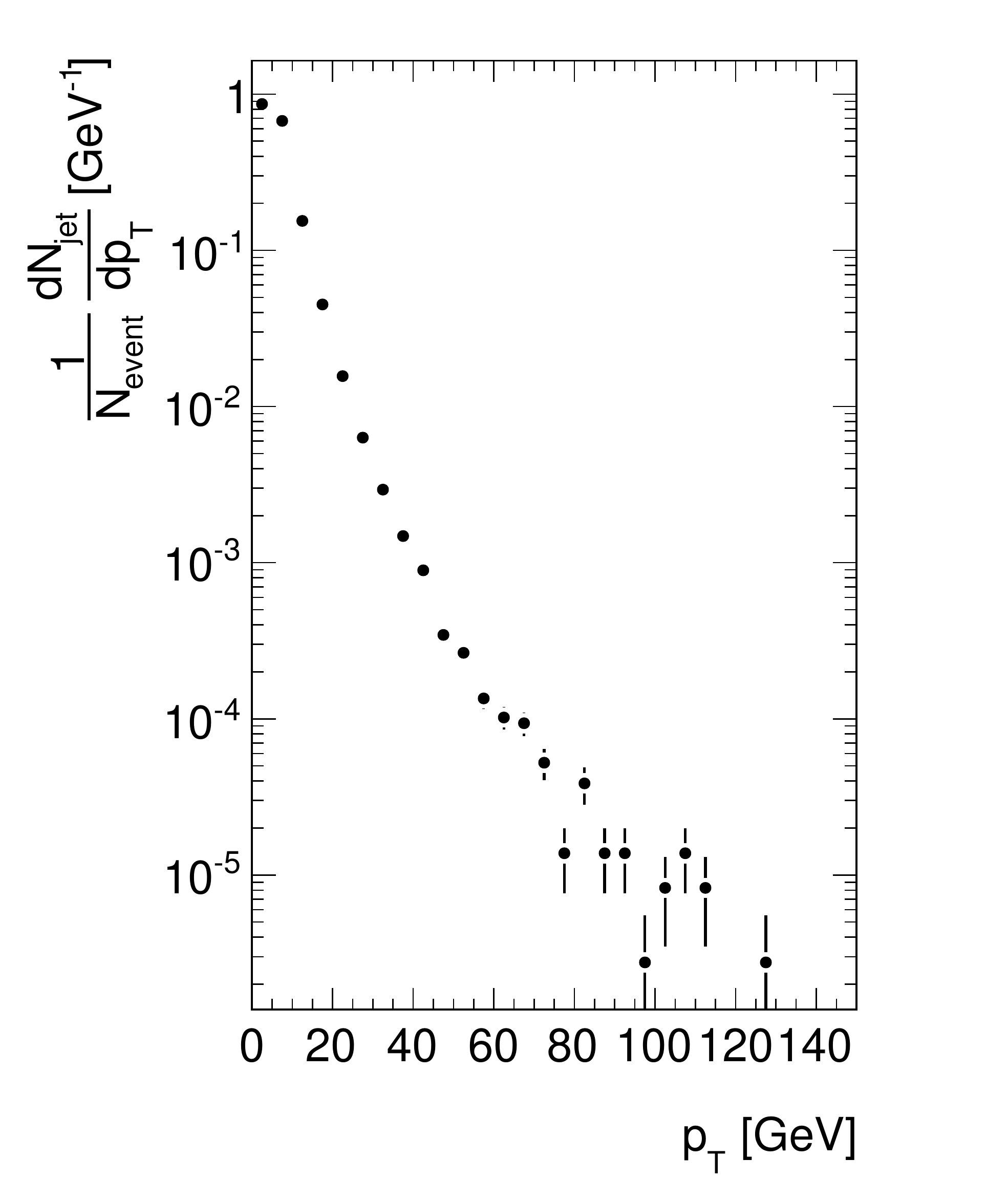}
\caption{A per event jet yield as a function of~\et\ for jets produced
with HIJING. Jets were reconstructed from a 75,000 event minimum bias
sample running the anti-\kt\ algorithm on particles with $\pt>4\GeV$.}
\label{fig:mc:hijing_spectrum}
\end{figure}

The HIJING events themselves contain jet production at a rate
consistent with binary scaling. A per-event jet yield for these jets
is shown in Fig.~\ref{fig:mc:hijing_spectrum}. A subtlety
arises in the embedding procedure when the particles from a hard
scattering in the HIJING background overlap with those from the PYTHIA
jets. The reconstructed jet signal will contain contributions from
both PYTHIA and HIJING, however the matching truth information will
only contain the jet energy from PYTHIA.
While this problem affects
truth jets of all energies, it is most easily seen in
samples of J1 events where the PYTHIA jets have relatively modest
energies. In a high-statistics sample, some of the HIJING events
contain jets in excess of 100~\GeV, resulting in many events where these
jets overlap in~$\eta$-$\phi$ with a PYTHIA jet of much lower~\pt\ of
order 10~\GeV. Given
the large weight of the J1 sample, without correcting for the HIJING
contamination the most probable way to get a 100~\GeV\ jet in the
full study is through this mechanism. The associations between 
10~\GeV\ truth jets and 100~\GeV\ reconstructed jets significantly
distort studies of jet performance and the response matrix.

Running jet reconstruction on the HIJING
particles leads to many of the challenges of background
separation/subtraction that are faced in the full offline
reconstruction. Uncertainties in identifying the truth jet energy
would reduce the effectiveness of any performance studies. It is
possible for multiple hard scatterings in the same heavy ion event to produce jets that overlap in
the calorimeter. This is likely to be a small effect and would
be best addressed by extending the embedding process to multiple
overlapping PYTHIA jets, where the identification of truth
energy is clear. The procedure used in this study to remove HIJING contamination is to
run the anti-\kt\ algorithm with~\RFour\ on HIJING particles with $\pt >
4\GeV$. The energies of these jets were not used in the performance study, but instead any
PYTHIA jet found within $\Delta R < 0.8$ of a HIJING jet with $\pt >
10 \GeV$ is removed from the sample. The fraction of jets removed
during this procedure in each \pt\ and centrality bin is recorded and
the final distributions are re-weighted by the inverse of this fraction.

\begin{figure}[t]
\centering
\includegraphics[width =5 in] {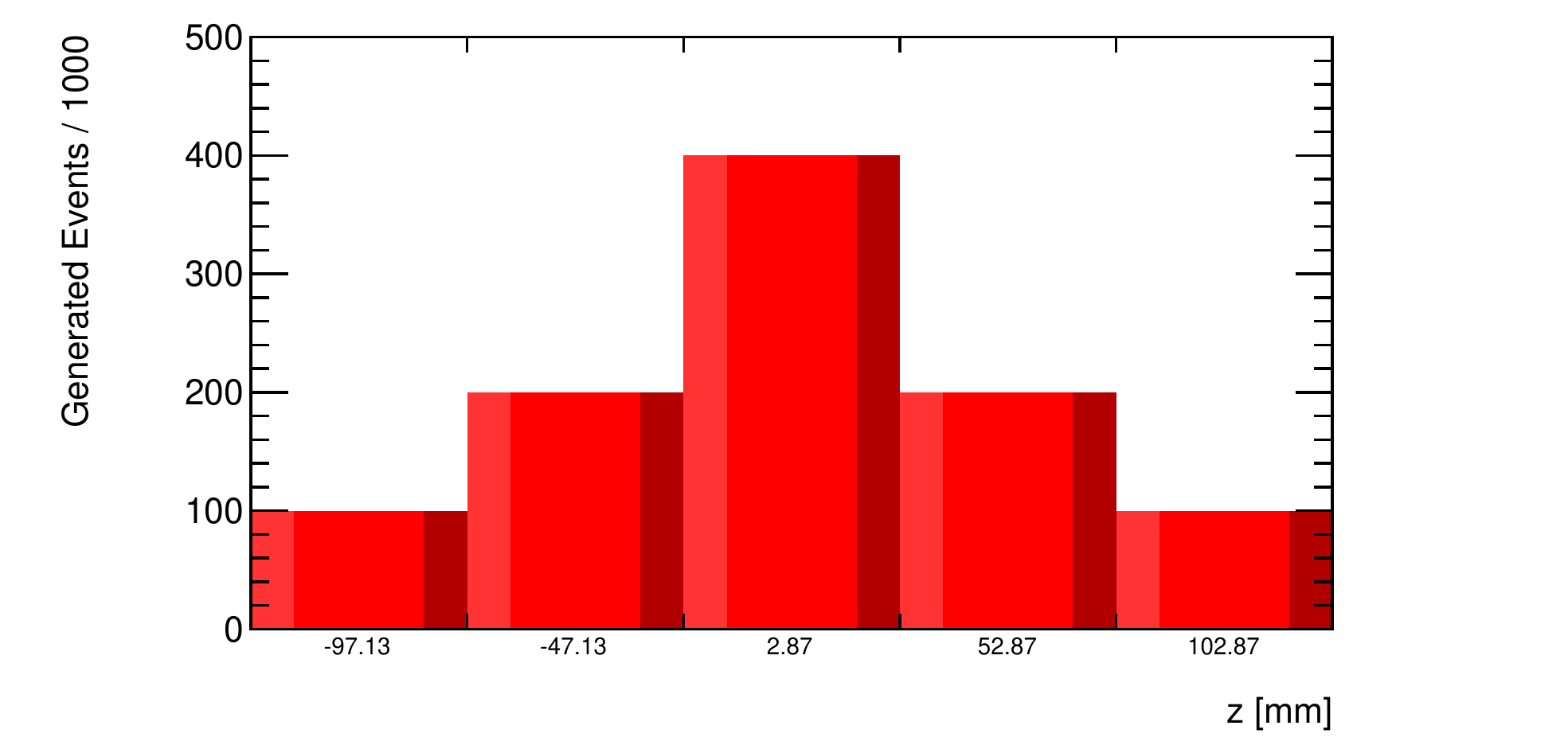}
\caption{Distribution of $z$ vertex position in combined MC
  sample corresponding to one million events for each J sample. Transverse position is fixed at $x=0.1352\,\mathrm{mm}$,
  $y=1.1621\,\mathrm{mm}$. 
}
\label{fig:mc:vertex_samples}
\end{figure}

The overlaid events must have the same vertex position at the generator
level. This constraint requires that multiple samples be produced at
different, fixed $z$ vertex positions, rather than the usual case of a
single production sampling a continuous distribution. The positions of
the samples, and the relative number of events produced per position
is shown in Fig.~\ref{fig:mc:vertex_samples}. The vertex $x$
and $y$ positions were fixed for all samples to match the mean beam-spot position from
the 2010~\PbPb\ run. A total of one million HIJING events were
generated along with one million PYTHIA events in each of the five different
J samples.

\section{Corrections}
\label{section:jet_reco:corrections}
\subsection{Self and Mutual Energy Biases}
\label{sec:corrections:seb}
As mentioned in Section~\ref{section:jet_rec:subtraction}, a self-energy bias
can occur when there is an incomplete match between the jets used as
seeds and the final jets after iteration. To address this possibility,
a variable was computed during reconstruction assuming the jet failed
to be excluded from the background and thus biased its own energy
subtraction. At the analysis stage, any jet not associated with a seed
has its energy corrected. The iteration step reduces this problem
considerably as all high energy jets ($\et>25\GeV$) are excluded from
the background and are unaffected. However for jets around this threshold the self-energy
bias correction must be included.

The above procedure assumes the bias is binary: either a jet is biased or it is not. However, if the region around a seed does not contain
the entire geometric area spanned by the jet, a differential bias can
occur determined by the region of non-overlap. Additionally, jets that
are not excluded from the background may cause over-subtraction in
another jet in the same~\eta\ interval (mutual energy bias). The
underlying event can produce calorimetric signals that appear as jets,
but are in fact uncorrelated with hard particle production in the
event. The contribution and systematic removal of such fake jets is
discussed extensively in Section~\ref{corrections:fake}. The set of jets
capable of biasing the background (or being biased themselves) must be
consistent with the definition of fake jets. The energy bias correction was applied by considering the
constituent towers of jets that pass fake rejection and calculating a bias
for each strip in~\eta\ by summing the energies of those towers
unassociated with a seed ($\Delta R > 0.4$). Each jet was then corrected at
the constituent level for biases introduced by constituents of other
jets that were not excluded by the seeds.

\subsection{Numerical Inversion}
\label{sec:corrections:num_inv}

Numerical inversion is a method to derive the \eta\ and \pt-dependent jet calibration. The method described below follows the procedure used in the jet energy scale determination used in \pp\ jet measurements~\cite{Aad:2011he}. The main motivation to derive calibration constants for heavy ion jet reconstruction is that 
no such constants exist at this time for $R=0.2, 0.3$, and $0.5$ in
\pp. Additionally, some aspects of the reconstruction differ
slightly between \pp\ and \PbPb, most notably the issue of noise
suppression. Thus the application of numerical inversion constants derived using
heavy ion jets will be more appropriate than the existing \RFour\ and
\RSix\ \pp\ constants.

The calibration constants are derived using 60-80\% peripheral events from PYTHIA+HIJING Monte Carlo 
samples described in Section~\ref{section:jet_rec:mc_sample}.
The basic quantity used in the derivation of the calibration 
constants is the response, 
\begin{equation}
\mathcal{R}=\dfrac{\ETemscale}{\ETtrue}\,,
\end{equation}
where \ETemscale\ is  reconstructed
\et\ at the EM scale and \ETtrue\ is the \et\ of the truth jet.
The response was evaluated for all calorimeter jets which match a truth jet and track jet 
simultaneously. The matching condition required a separation of $\Delta R \leq 0.3$ between the calorimeter jet axis and truth (track) jet 
axis. The calorimeter jet was required to be isolated, that is no 
other calorimeter jet within $\Delta R = 2.5 R$ must be present ($R$ is the parameter of the \antikt\ 
algorithm).

The response was evaluated in 12 bins in pseudorapidity of the reconstructed
calorimeter jet, $\eta_{\mathrm{det}}$, on the interval $|\eta_{\mathrm{det}}| < 2.8$
and 23 bins in the \ETtrue. For \ETtrue\ in bin $j$, the quantities $\langle 
\mathcal{R}\rangle_{j}$ and $\langle\ETemscale\rangle_{j}$ were determined
using a Gaussian fit and a statistical mean respectively. The set of points
$\left(\langle\ETemscale\rangle_j,\langle\mathcal{R}\rangle_j\right)$
was fit with the function:
\begin{equation}
 F_{\mathrm{calib}}(\ETemscale) = \sum_{i = 0}^{N_{\mathrm{max}}} a_i
\left(\ln(\ETemscale)\right)^i\,,
\label{corrections:num_inv:eqn_fit}
\end{equation}
where $N_{\mathrm{max}}$ is between 1 and 4. The actual value of $N_{\mathrm{max}}$ is optimized to provide 
the best $\chi^2/\mathrm{NDF}$. A separate set of points and fit
parameters were determined for each  $\eta_{\mathrm{det}}$. The
calibrated \et,  $E^{\mathrm{EM+JES}}_{\mathrm{T\, calo}}$,  was then
defined by
\begin{equation}
 E^{\mathrm{EM+JES}}_{\mathrm{T\, calo}} =
 \dfrac{\ETemscale}{F_{\mathrm{calib}}(\ETemscale)}\,.
\end{equation}

\begin{figure}[t]
\centering
   \includegraphics[scale=0.75]{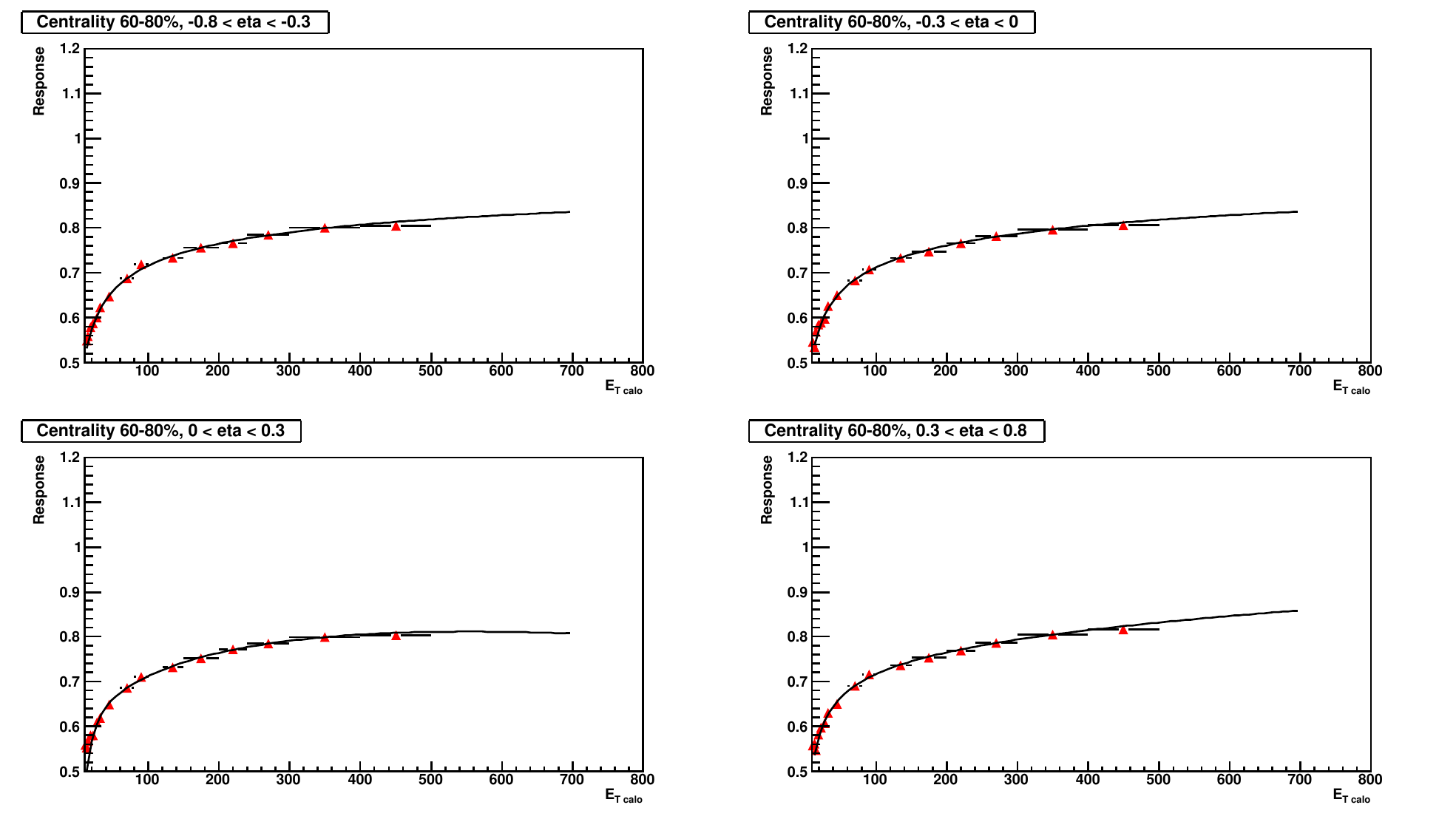}
\caption{
The $E_{\mathrm{T, calo}}^{\mathrm{EM}}$ dependence of the response 
with the logarithmic fit.}
\label{corrections:num_inv:fig_illu}
\end{figure}
\begin{figure}[h]
 \centering
  \includegraphics[scale=0.45]{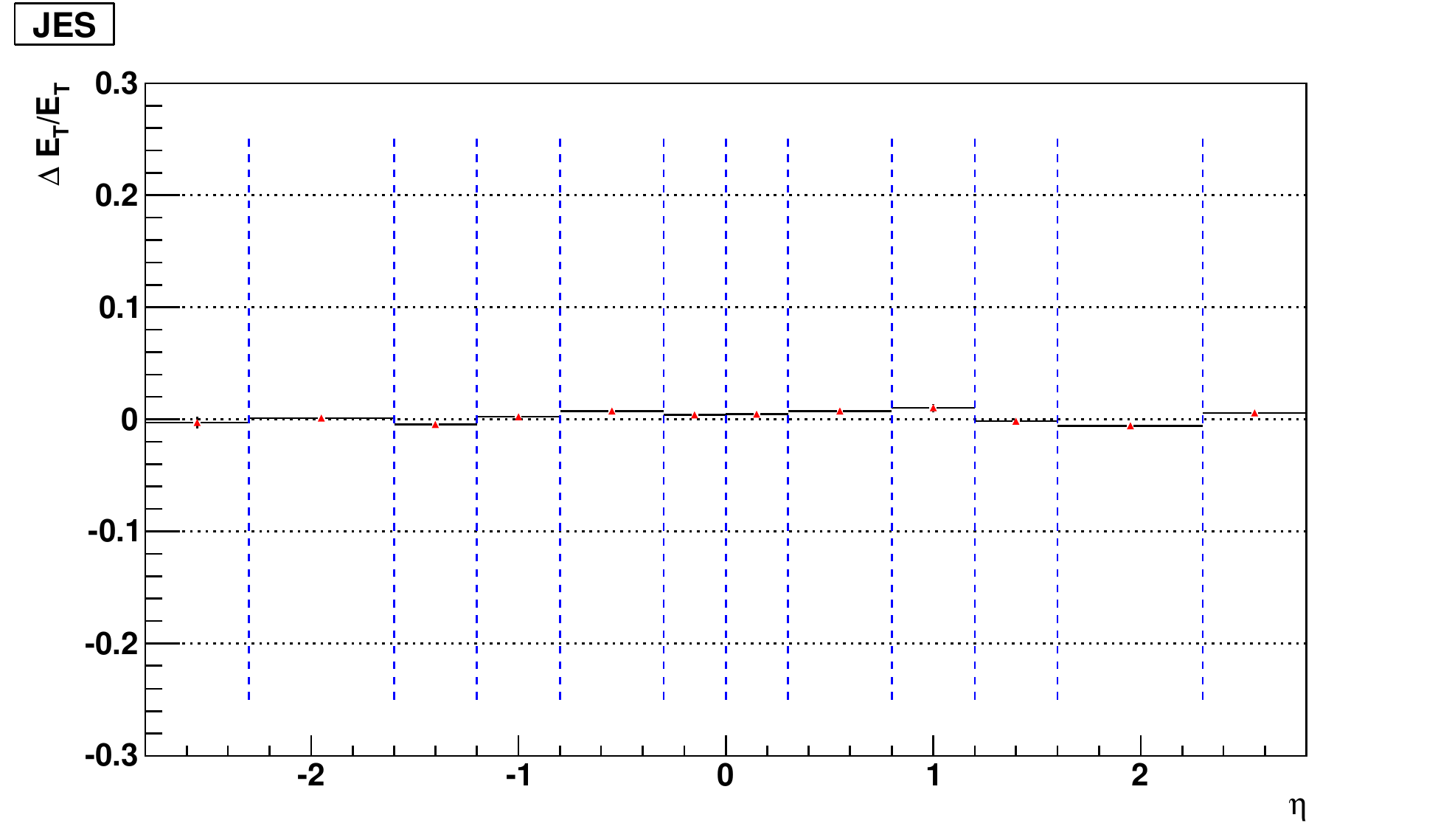}
\caption{$\eta$ dependence of the jet energy scale after applying the derived calibration constants.}
\label{corrections:num_inv:fig_eta}
\end{figure}

An example of the response as a function of \ETemscale\ is shown in 
Fig.~\ref{corrections:num_inv:fig_illu}, where the response is well
described by the functional form of
Eq.~\ref{corrections:num_inv:eqn_fit}. A similar level of agreement is
found for all $\eta_{\mathrm{det}}$
bins. Figure~\ref{corrections:num_inv:fig_eta} shows the $\eta$
dependence of the jet energy scale after applying the derived
calibration constants. A detailed closure test is discussed in
Section~\ref{performance:JES}. 

\subsection{Fake Rejection}
\label{corrections:fake}
Fluctuations of the underlying event that are unassociated with a hard scattering can create localized high~\et\ regions
in the calorimeter that may be reconstructed as jets. In the most
central collisions these fake jets contribute significantly to the
jet spectrum up to 80~\GeV. As most of this signal is caused by
accumulated energy from soft particles, requiring a physics signature
consistent with hard particle production in the neighborhood of a jet
can reject against this background. To minimize the contamination of the sample of 
reconstructed jets by fake jets different methods for
fake jet identification and rejection were studied. The general requirement of a coincidence of the reconstructed calorimeter jet with the track 
jet or an electromagnetic cluster was found to be efficient at
removing fakes without a severe impact on the overall efficiency. For
the duration of this thesis, the fake rejection, unless otherwise noted, is
defined as the requirement that a jet have a track jet or EM cluster with $\pt > 7\GeV$ within $\Delta R
\leq 0.2$ of the jet's axis. The effectiveness of this rejection
procedure is discussed in Section~\ref{sec:corrections:residual}.

\begin{figure}[hbt]
\centering
\includegraphics[scale=0.8]{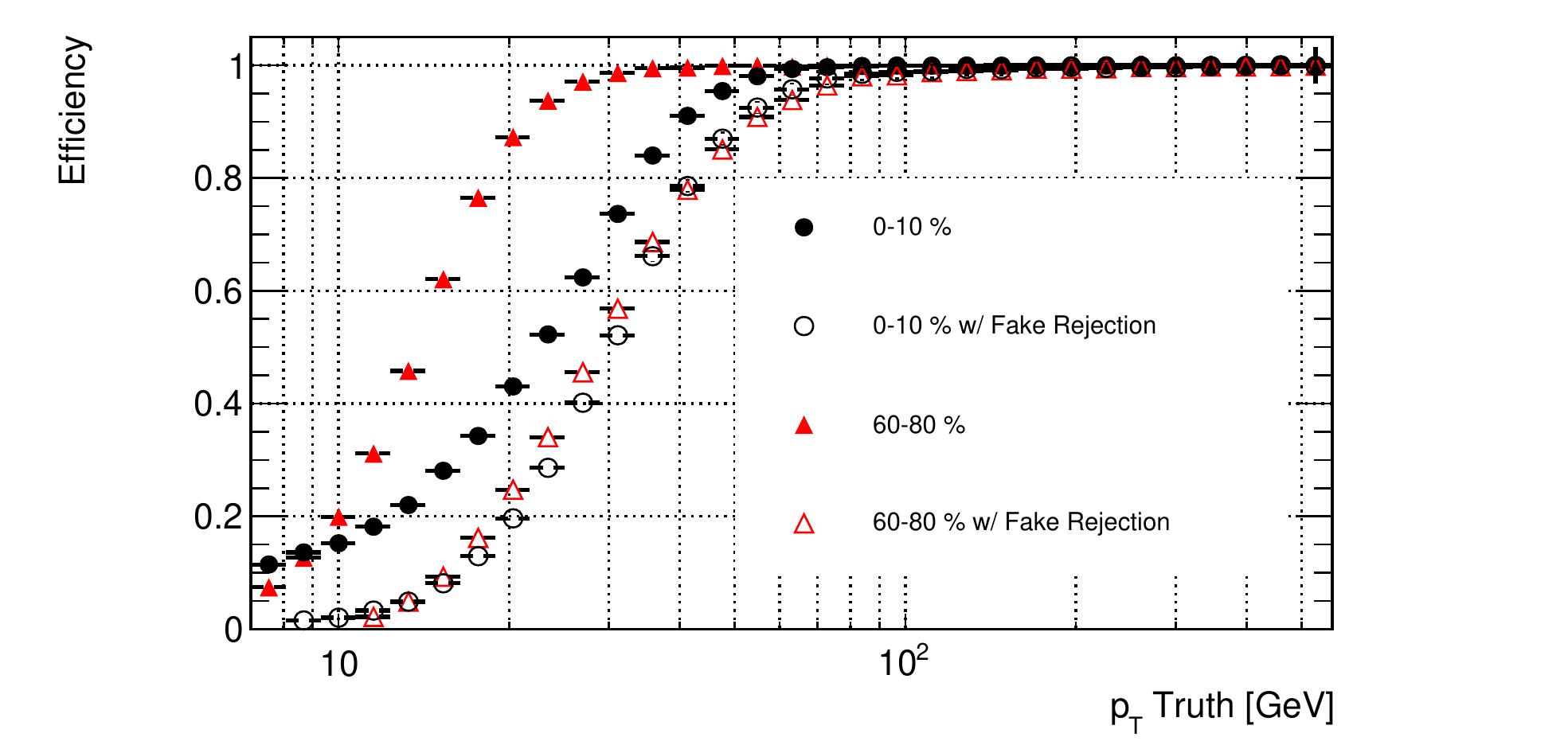}
\caption{The jet reconstruction efficiency in central (black) and
  peripheral (red) collisions both before (solid) and after (open)
  fake rejection for~\RFour\ jets.}
\label{fig:corrections:fake:efficiency}
\end{figure}

Fig.~\ref{fig:corrections:fake:efficiency} compares the jet reconstruction efficiency in central (0-10\%) and peripheral (60-80\%) 
collisions in two cases: before applying the fake rejection (solid symbols) and after applying the fake rejection (open 
symbols). The overall jet reconstruction efficiency is reduced by the
fake rejection requirement but still remains near 50\% for 30~\GeV\ jets. 

Corrections derived from MC for efficiency loss due to fake rejection (moving from
solid to open markers in Fig.~\ref{fig:corrections:fake:efficiency})
will be sensitive to how accurately the MC describes this effect. The
accuracy of the MC result can be checked in the
data by comparing the ratio of the spectrum after fake
rejection to the total spectrum, which shows the survival fraction of jets at a given~\pt\
that pass the rejection criteria. This comparison must be made
with caution and only in regions where a significant fake rate in the
data is not expected because the embedded MC sample is not expected to
accurately describe the fake rate. Since every event in the embedded
MC sample contains a PYTHIA hard scattering event the relative rates
of true jets (from the PYTHIA signal) to fake jets (from soft
particles in the HIJING background) will not be correct. A HIJING-only
sample is required for an evaluation of the absolute fake rate, a
procedure which is discussed in Section~\ref{sec:corrections:residual}.

\begin{figure}[hbt]
\centering
\includegraphics[scale=0.65]{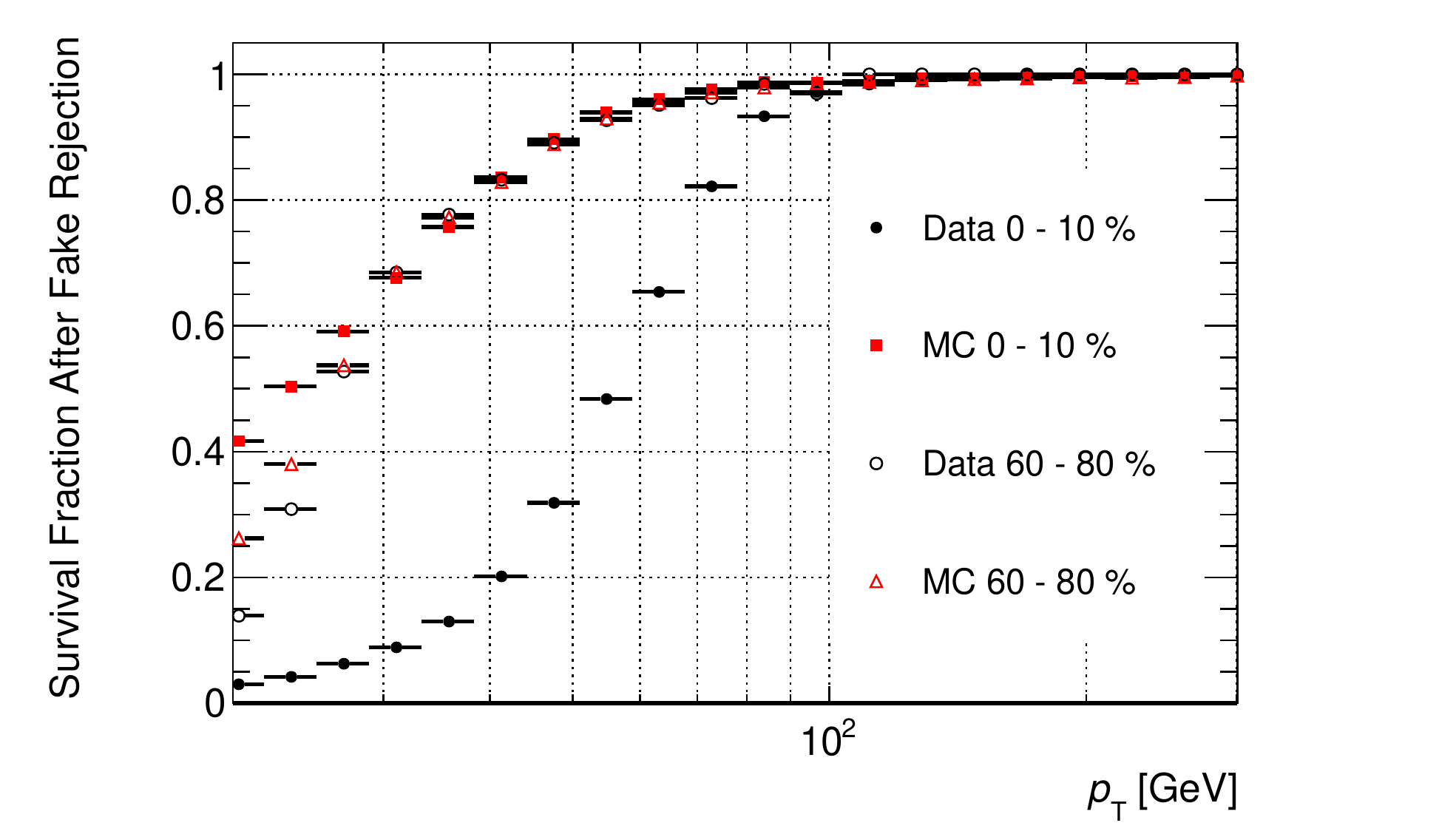}
\caption{The effect of the fake rejection data and MC in the 0-10\%
  and 60-80\% centrality bins as defined by survival fraction, the
  ratio of the spectra after and before fake rejection.}
\label{fig:corrections:fake:data_ratio}
\end{figure}

A comparison of the survival fraction\footnote{It should be emphasized
that this distribution is not trivially related to the efficiency shown in
Fig.~\ref{fig:corrections:fake:efficiency}. One important difference
is that the efficiency is shown as a function of \pt\ of the truth jets. The
survival fraction shown in Fig.~\ref{fig:corrections:fake:data_ratio}
is evaluated as a function of reconstructed \pt, since the data is
being used, which has the effects of upward feeding from finite
resolution.} in data and MC is shown in
Fig.~\ref{fig:corrections:fake:data_ratio}.
The fake rate is expected to be negligible in peripheral collisions and
good agreement is seen between data and MC in the survival fraction in the 60-80\%
centrality bin (red markers). This indicates that efficiency loss of true jets due to the fake
rejection requirement is well described by the MC, and that a MC-based
correction for this effect is appropriate. For $\pt>100$~\GeV, fakes are also expected to be a negligible
contribution to the spectrum in the 0-10\% bin. In this region, where
the efficiency approaches its asymptotic value, the agreement between
data and MC is better than 0.5\%.

\begin{figure}[h]
\centering
\includegraphics[width=0.7\textwidth]{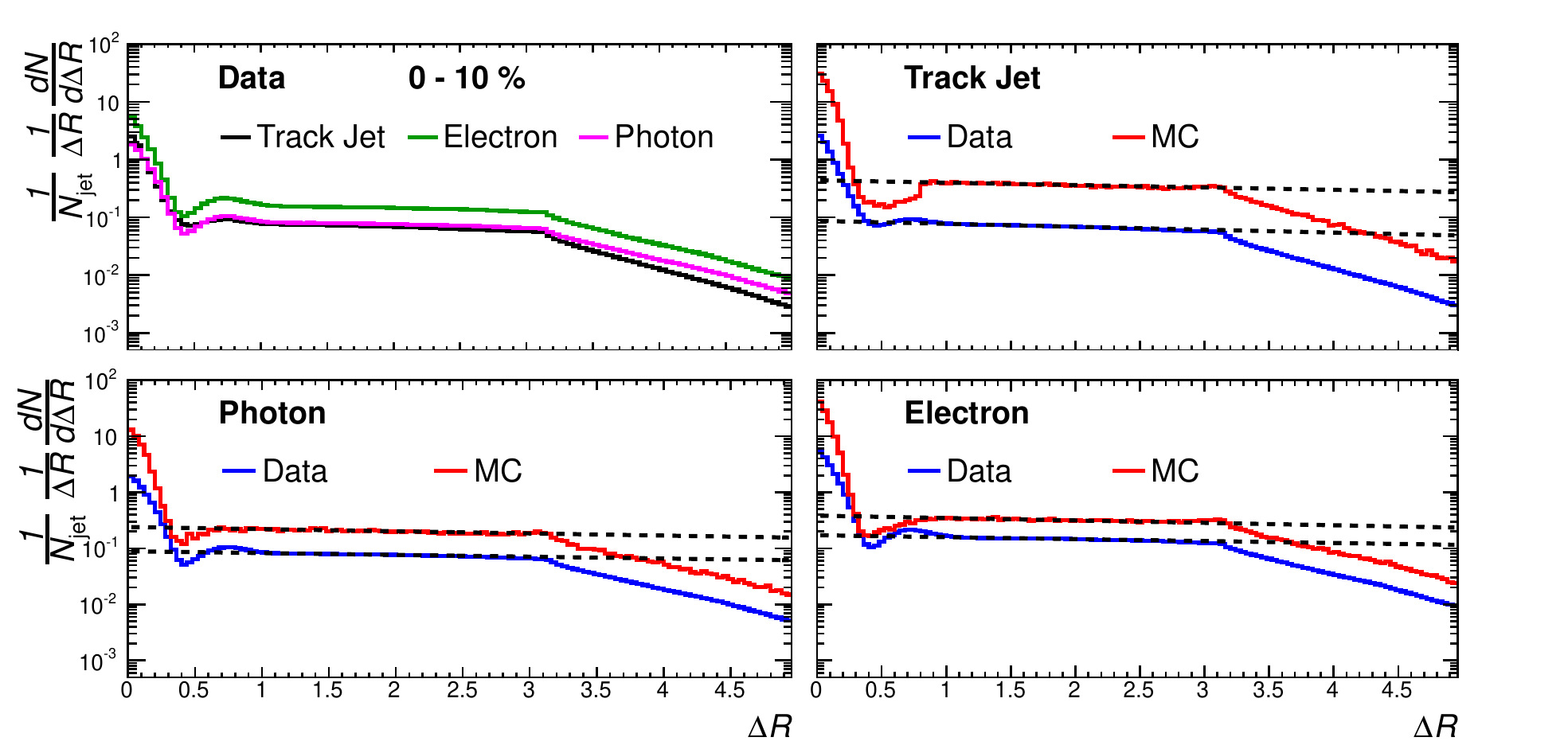}
\includegraphics[width=0.7\textwidth]{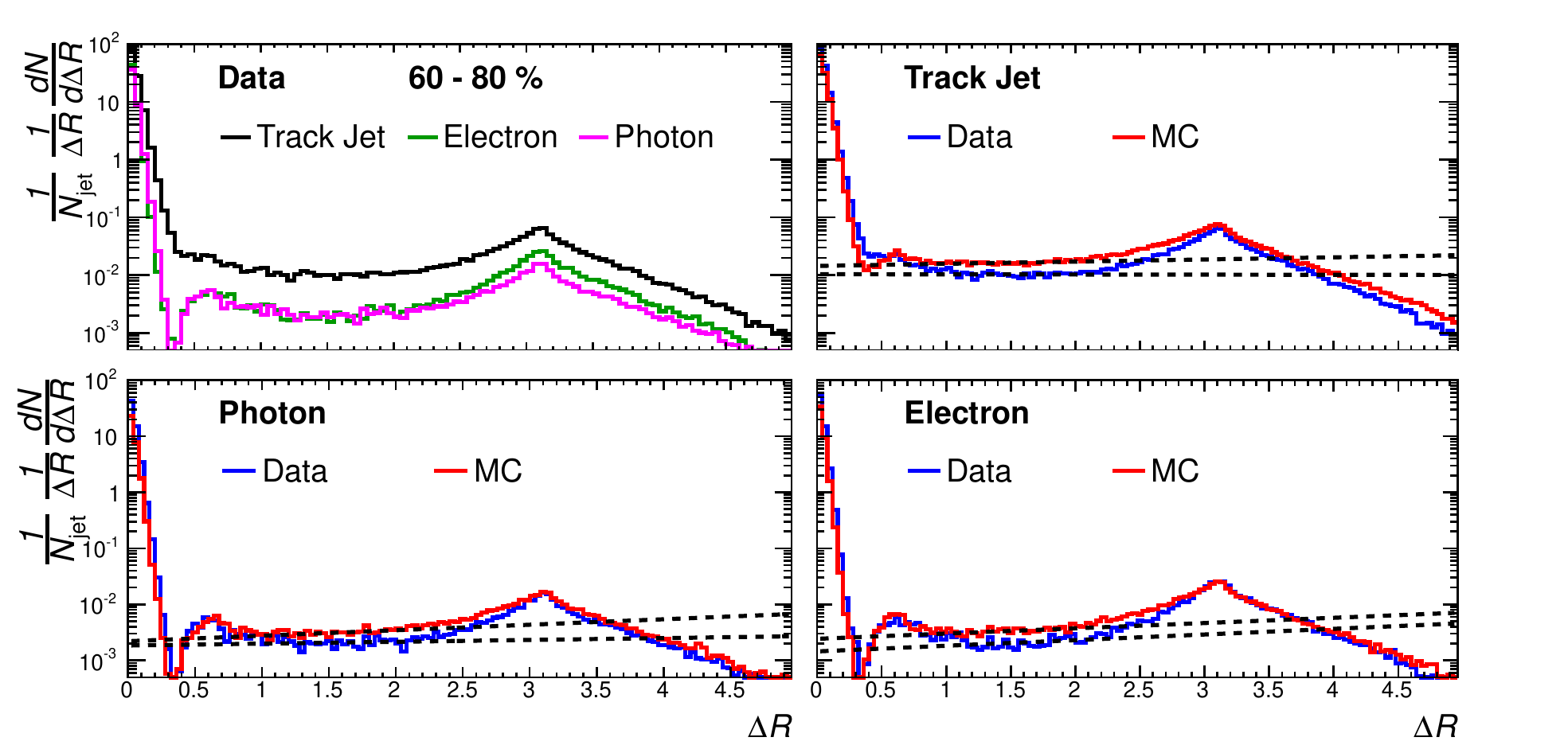}
\caption{Data to MC comparison of $\Delta R$ distributions between track jets, electrons and
  photons and \RFour\ jets in central (top) and peripheral (bottom)
  collisions. Distributions have been normalized per jet and have
  radial Jacobian removed to facilitate extraction of combinatoric
  backgrounds (dashed lines).}
\label{fig:corrections:dR}
\end{figure}

\begin{figure}[hbt]
\centering
\includegraphics[width=0.9\textwidth]{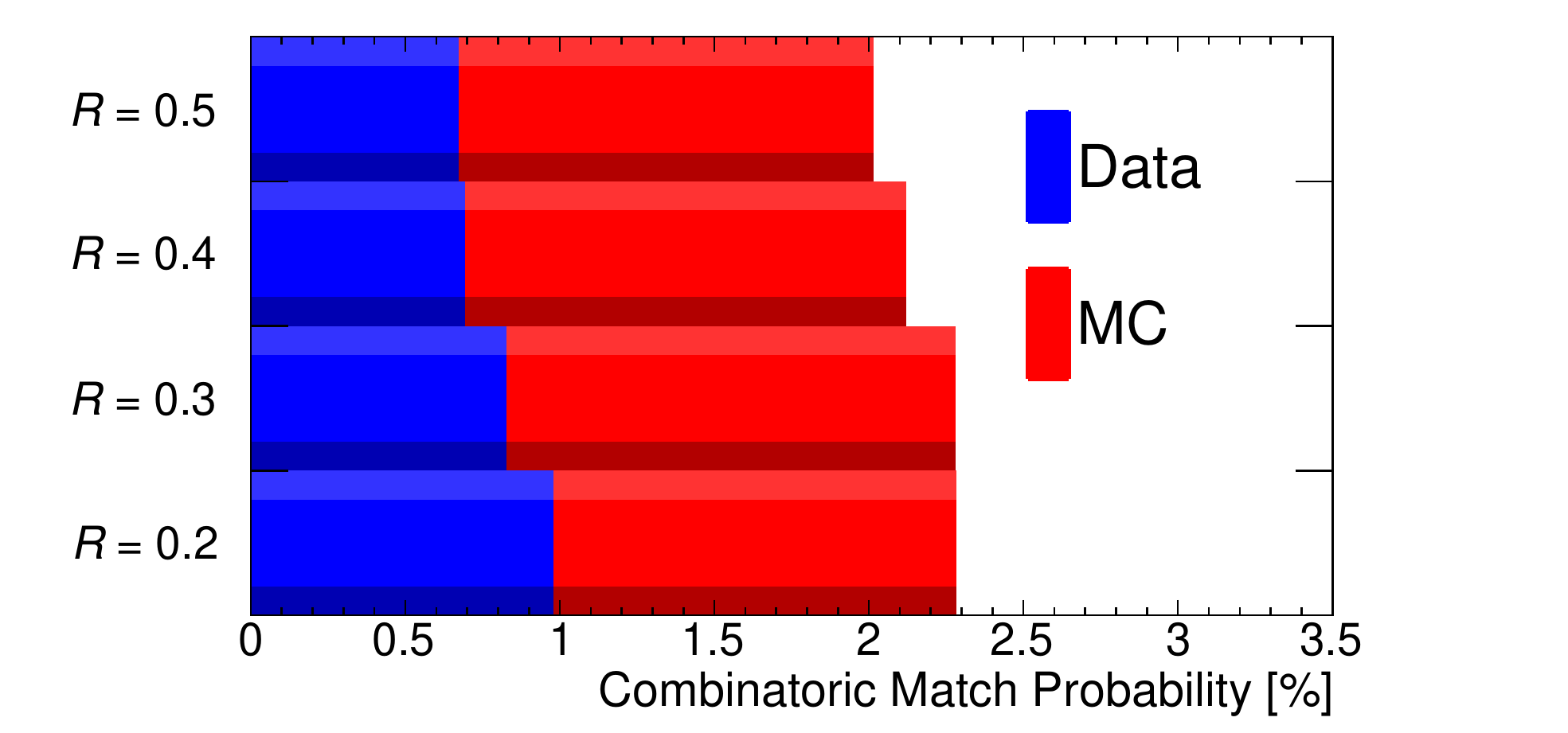}
\caption{Comparison between combinatoric matching probabilities
  between data and MC for different $R$ values. Rates are generally
  independent of $R$ and are a factor of 2-3 larger in MC, consistent
  with the absence of quenching.}
\label{fig:corrections:fake:combinatoric_rate}
\end{figure}

High-\pt\ tracks and clusters are expected to be associated with
particles within jets,
however it is possible for reconstructed jets to pass fake jet
rejection due to combinatoric overlap between the jet and the track
jet/cluster. This combinatoric rate is the mechanism by which fake
jets pass rejection, and a data to MC comparison of this rate can be
used to construct a more accurate estimate of the fake rate in data
from the residual fake rate obtained from the MC study. This rate was
evaluated by considering the $\Delta R$ distribution of all track jets,
electrons and photons in an event with respect to a jet's axis. To
simplify geometric effects from splitting of nearby jets, only
isolated jets were considered. These jets were required to be the
highest \pt\ jet in a $\Delta R <1$ cone about their axes. In the MC
sample jets were additionally forced to match a truth jet ($\Delta R < 0.2$) and
be isolated from any HIJING jets ($\Delta R > 0.8$ for $\ET^{\mathrm{HIJING}} >
10$~\GeV). Comparisons between the different objects (track jets,
electrons and photons) and between data and MC are shown for \RFour\
jets in central and peripheral collisions in
Fig.~\ref{fig:corrections:dR}. 
The distributions show strong correlation for matches at
small $\Delta R$ with a more diffuse peak at $\Delta R\sim\pi$ associated
with the dijet structure of the events. The remainder of the
distribution is expected to be determined purely
by geometry, thus the trivial geometric enhancement has been removed by
dividing by $R$. With this effect removed, the matching rate is
expected to be constant away from the jet and dijet peaks, which is
estimated by the background extrapolations shown in the figures. These
distributions have been normalized per jet, such that the background
distribution integrated over the matching region can be interpreted as
a per jet probability of combinatoric match, $P$, via
\begin{equation}
P=\int_0^{0.2} \left(\dfrac{1}{\Njet}\dfrac{1}{\Delta
    R}\dfrac{dN}{d\Delta R}\right) \Delta R^2 d\Delta R\simeq 0.02\times \left(\dfrac{1}{\Njet}\dfrac{1}{\Delta
    R}\dfrac{dN}{d\Delta R}\right)\,,
\label{eqn:corrections:fake:combinatoric_rate}
\end{equation}
where the quantity in parentheses is estimated from fitting the
distribution in the region $1.5 < \Delta R < 2$ and extrapolating
underneath the correlation peak at $\Delta R < 0$.
Estimates of these probabilities in central collisions are compared for data and MC in
Fig.~\ref{fig:corrections:fake:combinatoric_rate}.
The rates are nominally
independent of $R$, which is expected as the jet definition
does not explicitly enter into the matching. The MC are higher than
the data by factor of 2-3, which is qualitatively consistent with observed
single particle suppression factors, caused by a phenomenon
(quenching) present in the data but not the MC. 

\subsection{Residual Fake Rate}
\label{sec:corrections:residual}
The residual per-event rate of fake jets (underlying event
fluctuations) after fake rejection was determined from an analysis of
a HIJING-only MC sample. Each reconstructed jet passing the fake
rejection criteria but not within $\Delta R <0.2$ of a HIJING jet with
\ETtrue $> 10$~\GeV\ was considered a fake jet. The
residual fake rate is defined as the ratio of these unmatched jets to
the total number of reconstructed jets passing fake rejection and is a
function of \ETreco. 

Unlike the truth jet reconstruction in PYTHIA-embedded events, the
HIJING truth jet reconstruction was run with a $\pt> 4$~\GeV\
cut on the input particles to the reconstruction algorithm.  Because
of this, a hard-scattered parton may fragment in a way that is not
reconstructed by the HIJING truth jet reconstruction as a jet with
\ETtrue $> 10$~\GeV. A jet reconstructed from these fragments will
fail to match a truth jet causing the fake rate calculation to come out anomalously
high. Therefore, it is necessary to calculate the truth jet
reconstruction efficiency of a truth jet if the reconstruction had been
performed with a $\pt> 4$~\GeV\ cut on the constituents. After
correcting the fraction of jets that match to a HIJING truth jet by
the truth jet reconstruction efficiency, the difference between the
efficiency corrected rate and the full jet rate is the residual fake
rate.

Only reconstructed jets within $\left|\eta\right| < 2.0$ are
considered for the residual fake rate analysis. Reconstructed jets
that pass fake jet rejection are corrected for the self-energy bias as
described in Section~\ref{sec:corrections:seb}. The analysis is
repeated for three different fake jet rejection schemes:

\begin{itemize}
\item $\pt> 7$~\GeV\ track jet or $E_\mathrm{T} > 7$~\GeV\ cluster within $\Delta{R} < 0.2$, 
\item $\pt> 7$~\GeV\ track jet or $E_\mathrm{T} > 9$~\GeV\ cluster within $\Delta{R} < 0.2$,
\item $\pt> 7$~\GeV\ track jet within $\Delta{R} < 0.2$.
\end{itemize}

To calculate the truth jet reconstruction efficiency resulting from
the $\pt> 4$~\GeV\ constituent cut, a subset of the J1 and J2 PYTHIA-embedded
Monte Carlo samples were used ($400,000$ events $z=2.87$~mm
only). PYTHIA truth jets were matched to reconstructed jets with
$\Delta R <0.2$. To determine if this truth jet would be reconstructed by the
definition used in the HIJING only sample, the summed energy of PYTHIA truth particles with $\pt>
  4$~\GeV\ within a $\Delta R < 0.4$ radius of the truth jet axis was
  required to be greater than 10~\GeV. Because $>
4$~\GeV\ truth jets in the underlying HIJING event can combinatorially
overlap with a PYTHIA truth jet, the truth jet reconstruction
efficiency is actually higher for a PYTHIA truth jet at the same
\ETtrue\ in central than in peripheral collisions. To include this
effect, the \ET\ of HIJING jets within $\Delta R < 0.4$ of the truth
jet's associated reconstructed jet's axis was also added the PYTHIA truth particle sum.

As previously noted, a significantly
higher-\ET\ hard scattering in HIJING may combinatorially
overlap with an embedded PYTHIA truth jet. To prevent such scenarios from contaminating the efficiency
calculation, the same HIJING contamination removal condition is applied.  Even with this
constraint, there are cases when a HIJING jet overlaps a PYTHIA
jet. However, this effect of such overlaps is not obvious due to  truth jet
reconstruction inefficiency. Therefore, jets with $\ETreco-\ETtrue>
30$~\GeV\ (about $\sim2$-$2.5$ times the RMS energy from 
underlying event fluctuations in central events for \RFour\ jets) were
also excluded. In central events and peripheral events, $\sim75\%$
and $\sim99.5\%$ of jets, respectively, are properly isolated from
contamination from HIJING hard scatterings.

\begin{figure}[hbt]
\centering
\includegraphics[width=0.48\linewidth]{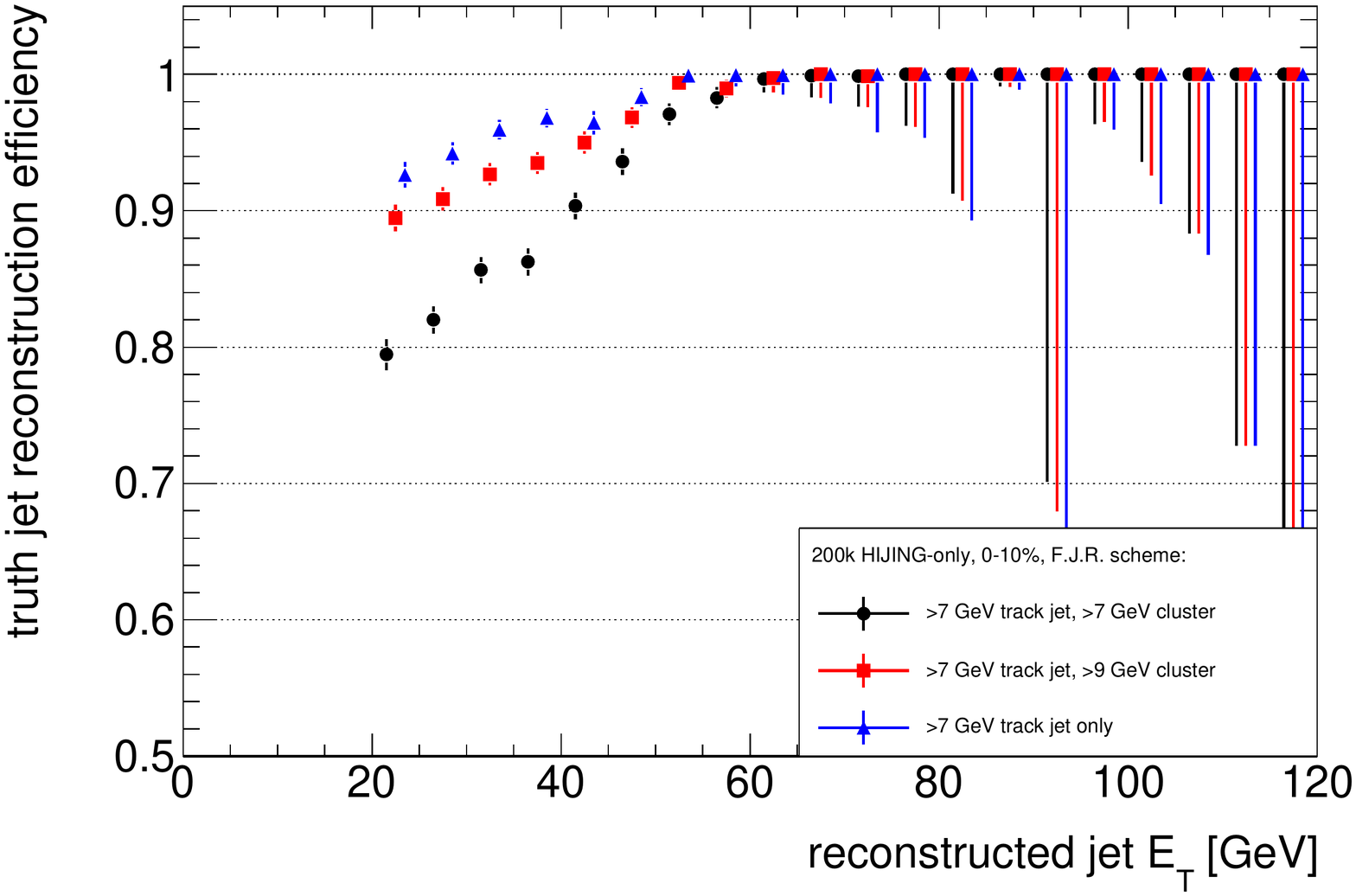}
\includegraphics[width=0.48\linewidth]{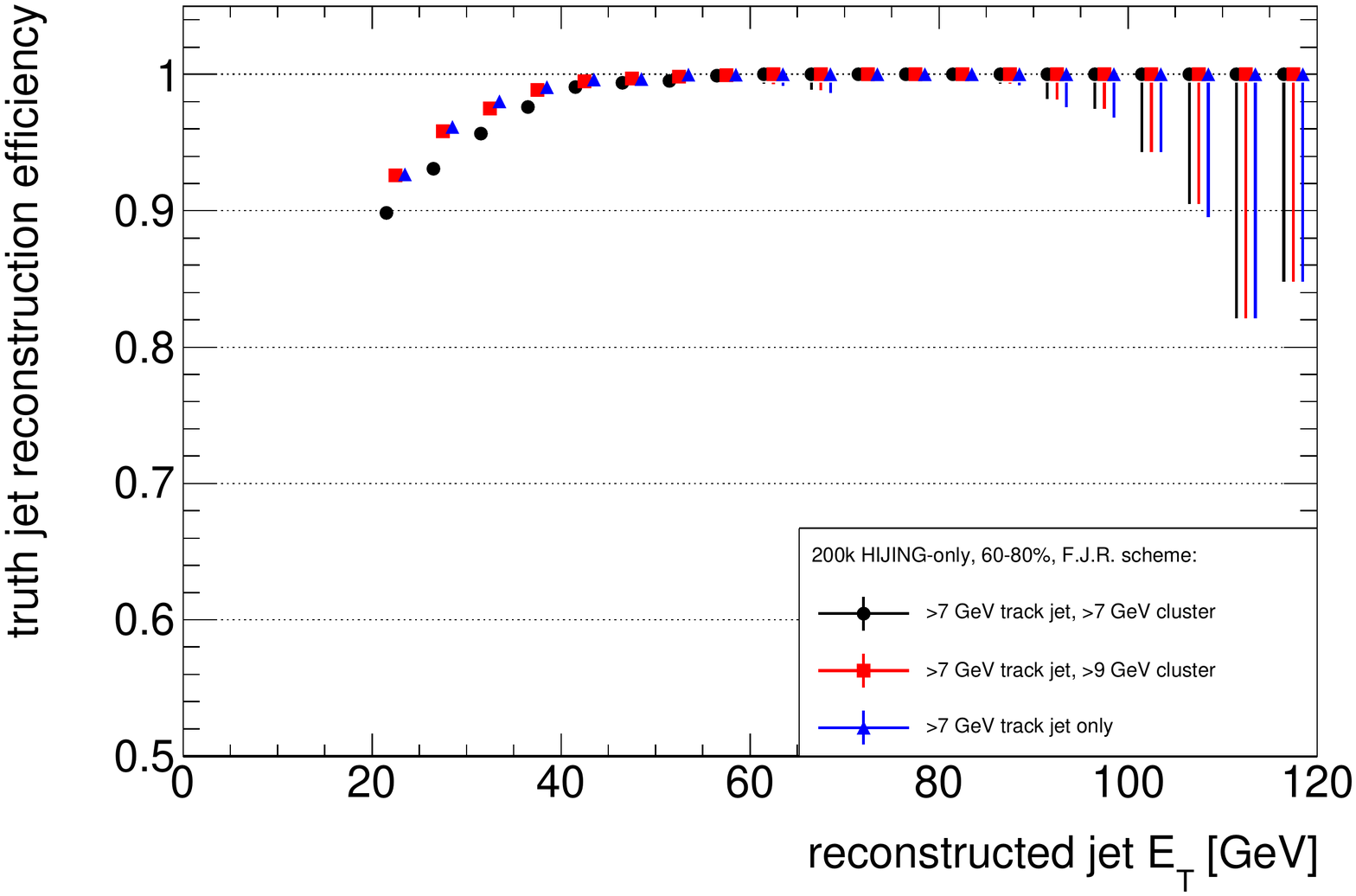}

\caption{Truth jet reconstruction efficiency as a function of \ETreco\ in 
0-10\% events (left) and 60-80\% events (right). The efficiency using
three different fake jet rejection schemes is shown.}

\label{fig:corrections:residual:truth_reco_eff}
\end{figure}

To determine the efficiency in each \ETreco\
bin, the J samples were combined in a weighted average,
\begin{equation}
\varepsilon(\ETreco) = \dfrac{
\displaystyle\sum_J
\varepsilon_J(\ETreco) \dfrac{1}{N_\mathrm{evt}^{J}}
\sigma^J N^{(J)}_\mathrm{jet}(\ETreco)
}
{\displaystyle\sum_J  \dfrac{1}{N_\mathrm{evt}^{J}}
\sigma^J N^{(J)}_\mathrm{jet}(E_\mathrm{T}^\mathrm{reco})}, 
\end{equation}
where $N_{\mathrm{evt}}^J$ is the number of generator events in that
centrality selection, $\sigma^J$ is the generator cross-section for
the J sample given in Table~\ref{tbl:mc:pythia_J_samples} and $N_{\mathrm{jet}}^{J}(\ETreco)$ is the number of jets
at the given reconstructed
\ETreco. Figure~\ref{fig:corrections:residual:truth_reco_eff} shows the truth jet
reconstruction efficiency in the most central and most peripheral
events for each of the three fake jet rejection schemes.

\begin{figure}[hbt]
\centering
\includegraphics[width=0.48\linewidth]{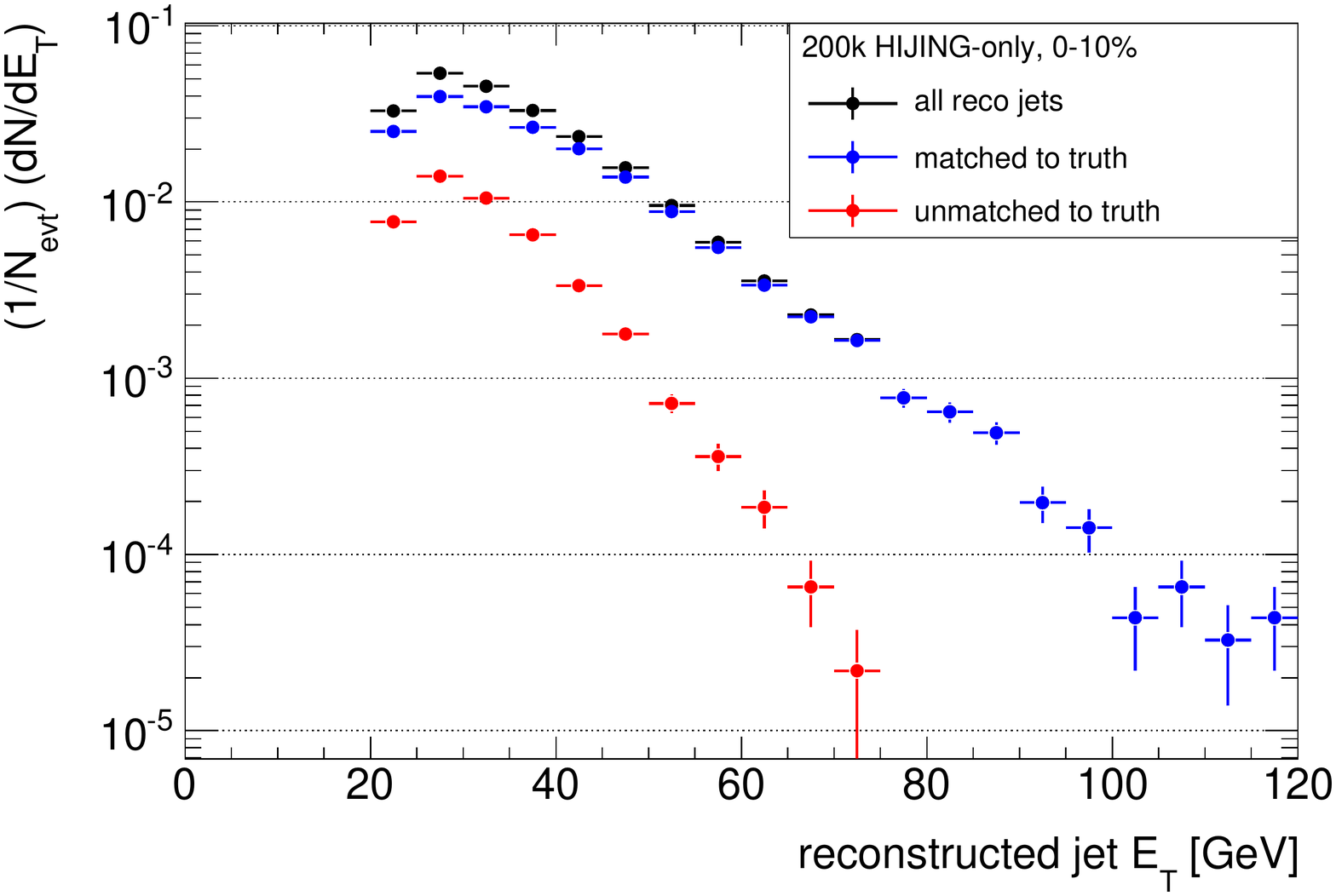}
\includegraphics[width=0.48\linewidth]{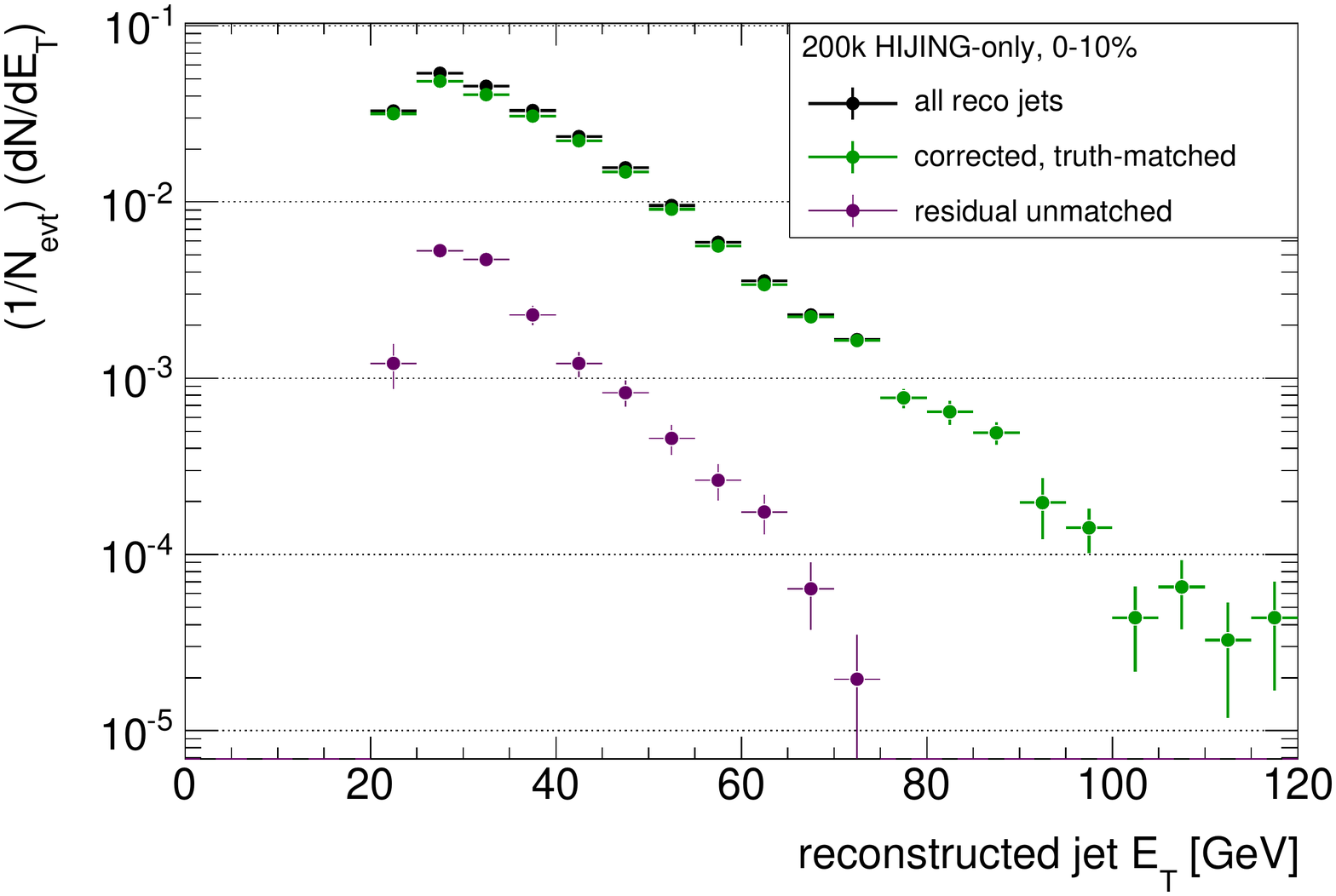}
\caption{Per-event yields of reconstructed in 0-10\% central events
  using the HIJING-only sample. The total yield (black) is the same in
  both figures. The unmatched contribution is what
  remains after subtracting the matched from the total. The left
  plot shows the matched (blue) and unmatched (red) as extracted from
  the MC. The right figure shows the matched (green) after correction
  for truth jet reconstruction inefficiency and the residual fake yield (purple).}
\label{fig:corrections:residual:example_spectra}
\end{figure}

The truth jet reconstruction efficiency was then used to
determine the residual fake rate in central events from the
HIJING-only data set.
Reconstructed jets that pass fake rejection were tested to see if they
match to a HIJING hard scattering. This is accomplished by considering
the summed \ET\ of all HIJING truth jets with $E_\mathrm{T} > 4$~\GeV\ within $\Delta{R}
< 0.4$ of the reconstructed jet. Reconstructed jets where this sum
exceeds 10~\GeV\ were considered a match, with the unmatched yield
defined as the remainder after the matched is subtracted from the total. The total per-event yield in
the 0-10\% centrality bin along with the matched and unmatched is shown on the left side of
Fig.~\ref{fig:corrections:residual:example_spectra}. The matched
distribution was then corrected for the truth jet inefficiency and the
difference between the total yield and this corrected, matched yield
provides an estimate of the absolute fake rate is shown on the right
side of Fig.~\ref{fig:corrections:residual:example_spectra}.

\begin{figure}[hbt]
\centering
\includegraphics[width=0.6\linewidth]{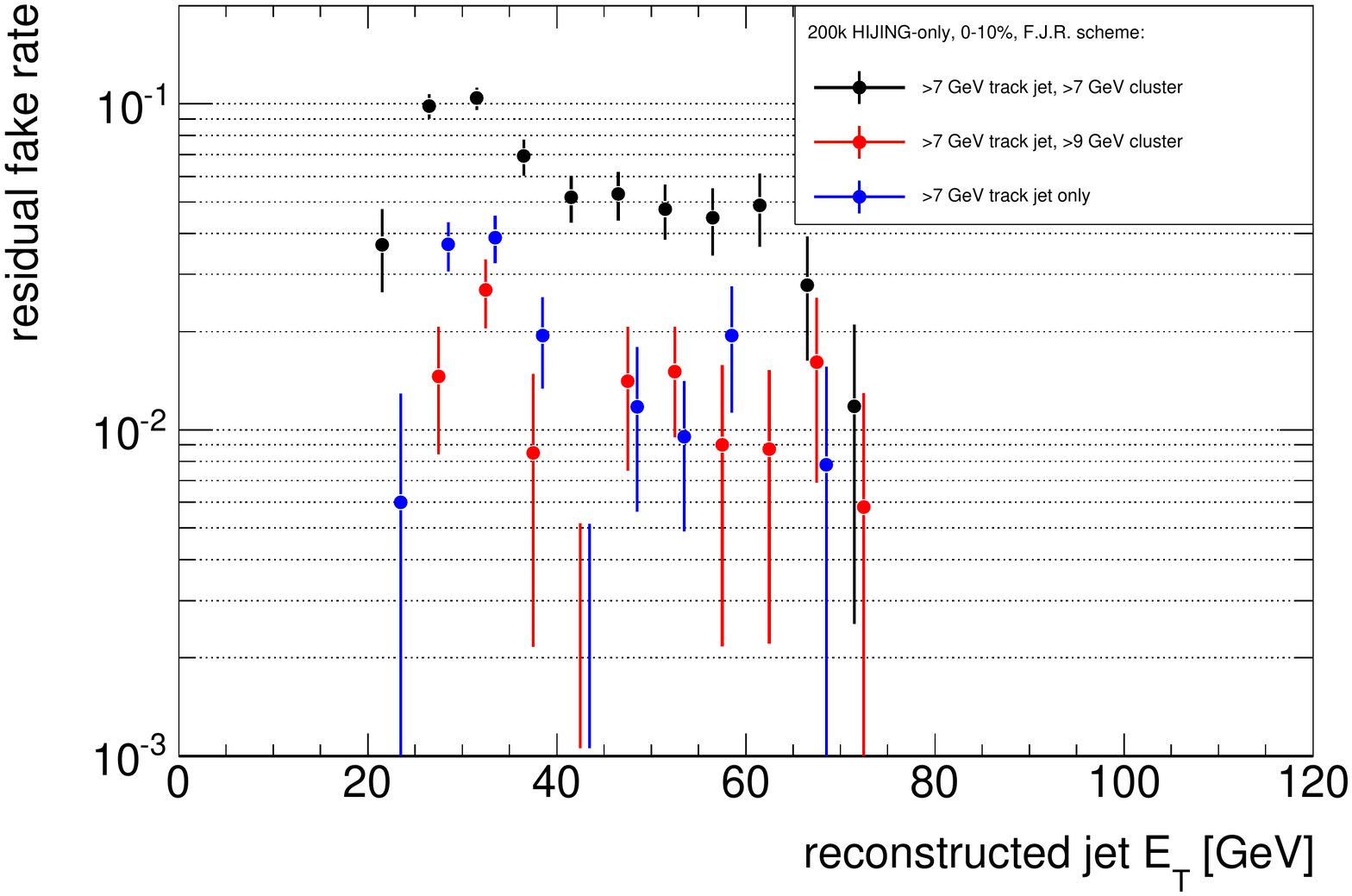}
\caption{Residual fake rate in 0-10\% events (as a fraction of the total 
reconstructed jet spectrum) shown for three different fake jet
rejection schemes.}

\label{fig:corrections:residual:contamination}
\end{figure}

For reconstructed jet at a given \pt, the probability that this jet is
unmatched defines the residual fake rate. This quantity was calculated by
normalizing corrected, unmatched yield by the total yield, and is shown
in Fig.~\ref{fig:corrections:residual:contamination} for the three
fake jet rejection schemes in 0-10\% centrality bin. For
$\ETreco\gtrsim40$~\GeV, this rate is $4$-$5$\% for the fake
jet rejection scheme used in this analysis: matching to track jet or
cluster with $\pt> 7$~\GeV. This rate drops to $1$-$2$\%
for the schemes with stricter requirements, although these were not
used in the jet spectrum analysis due to their effect on the
efficiency. If a fake jet passes the rejection it is due to a
combinatoric overlap between an underlying event fluctuation
reconstructed as a jet and a single track or cluster also from the
underlying event. The rate for this behavior was estimated in the
previous section, and the MC rate was found to exceed that of the data
by a factor of 2-3. The residual fake rate extracted from the MC
should be reduced by this factor when describing data. Thus the
absolute fake rate is estimated to be 2-3\% for $\et > 40$~\GeV\ in
the most central 10\% of events.

\section{Performance}
\label{section:jet_rec:performance}
To evaluate jet performance using the Monte Carlo samples, ``truth''
jets were matched to reconstructed jets by requiring an angular separation
between truth and reconstructed jet, $\Delta R \equiv \sqrt{\Delta
  \eta^2 + \Delta \phi^2} < 0.2$. If multiple reconstructed jets were
matched to a truth jet, the reconstructed jet with the smallest
$\Delta R$ was selected. For the results presented in this section,
performance plots are presented for all jets with $|\eta| <
2.8$ using the EM+JES calibration scheme.

\subsection{Jet Energy Scale}
\label{performance:JES}	
For each matched truth jet, the \ET\ difference,
\begin{equation}
\Delta \ET \equiv \ETreco - \ETtrue \,,
\label{eq:deltaetdiff}
\end{equation}
was calculated. The distribution of $\Delta \et$ was then evaluated as functions of ~\ETtrue, 
$\eta^{\mathrm{truth}}$ and centrality. To evaluate the jet energy scale (JES) as a function of~\ETtrue\ 
the $\Delta \et$ distributions were each fit by a Gaussian. At low
\ETtrue\ ( $\ETtrue \lesssim\ 50$~\GeV\ for \RFour\ jets) the $\Delta \et$
distributions are affected  
by a truncation of reconstructed jet \et\ at 10~\GeV. For finite bin size in \ETtrue, this results in a 
distortion of the shape of the $\Delta \et$ distribution at low \ETtrue. Therefore the $\Delta \et$ distribution was fit in the region which is
unaffected by this truncation. Above 50~\GeV\ the effect of the
truncation is minimal as indicated in the left plot of
Fig.~\ref{fig:JESFitExample}. 

\begin{figure}[htb] 
\centering
\includegraphics[width=0.49\textwidth]{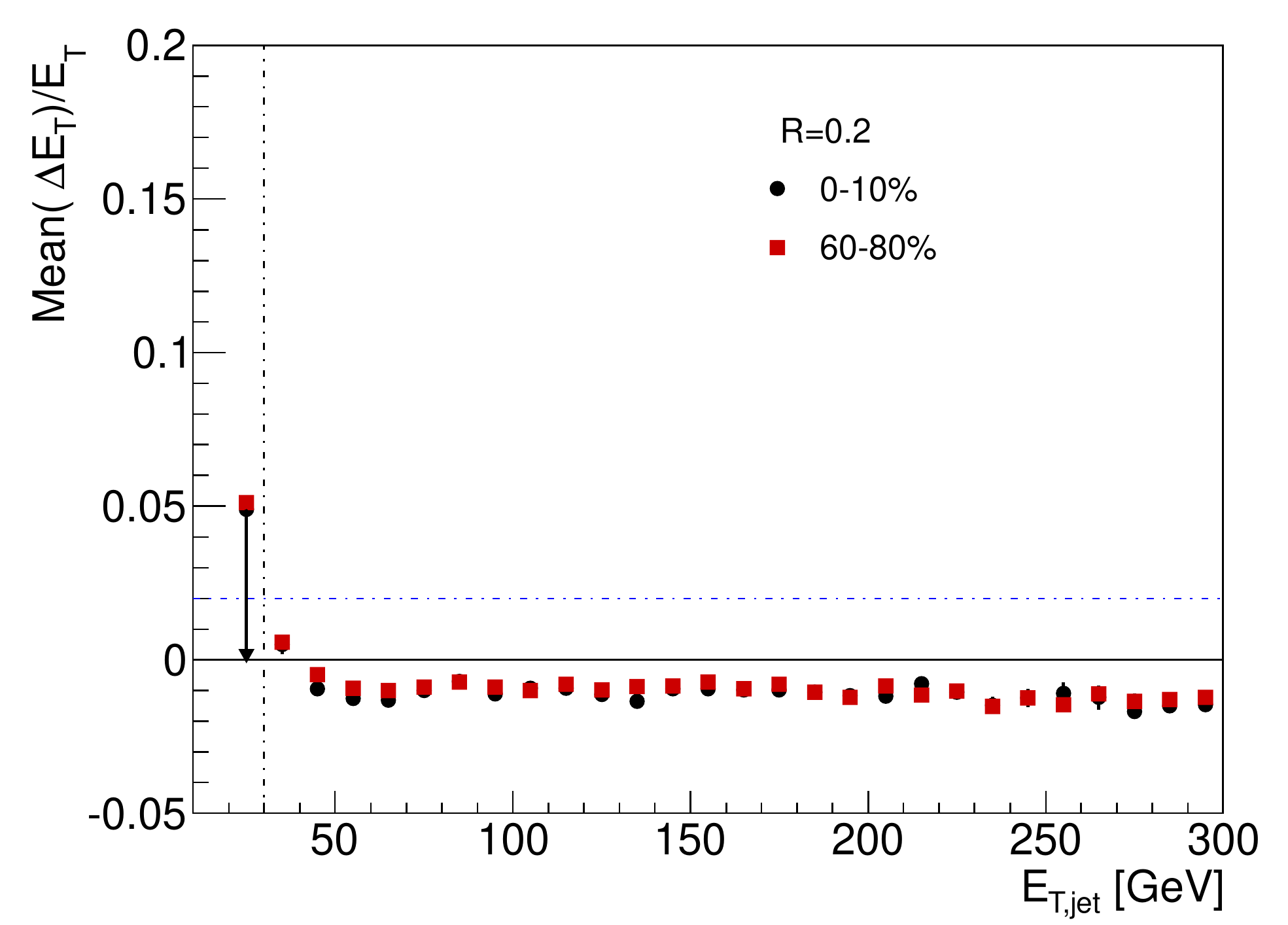}
\includegraphics[width=0.49\textwidth]{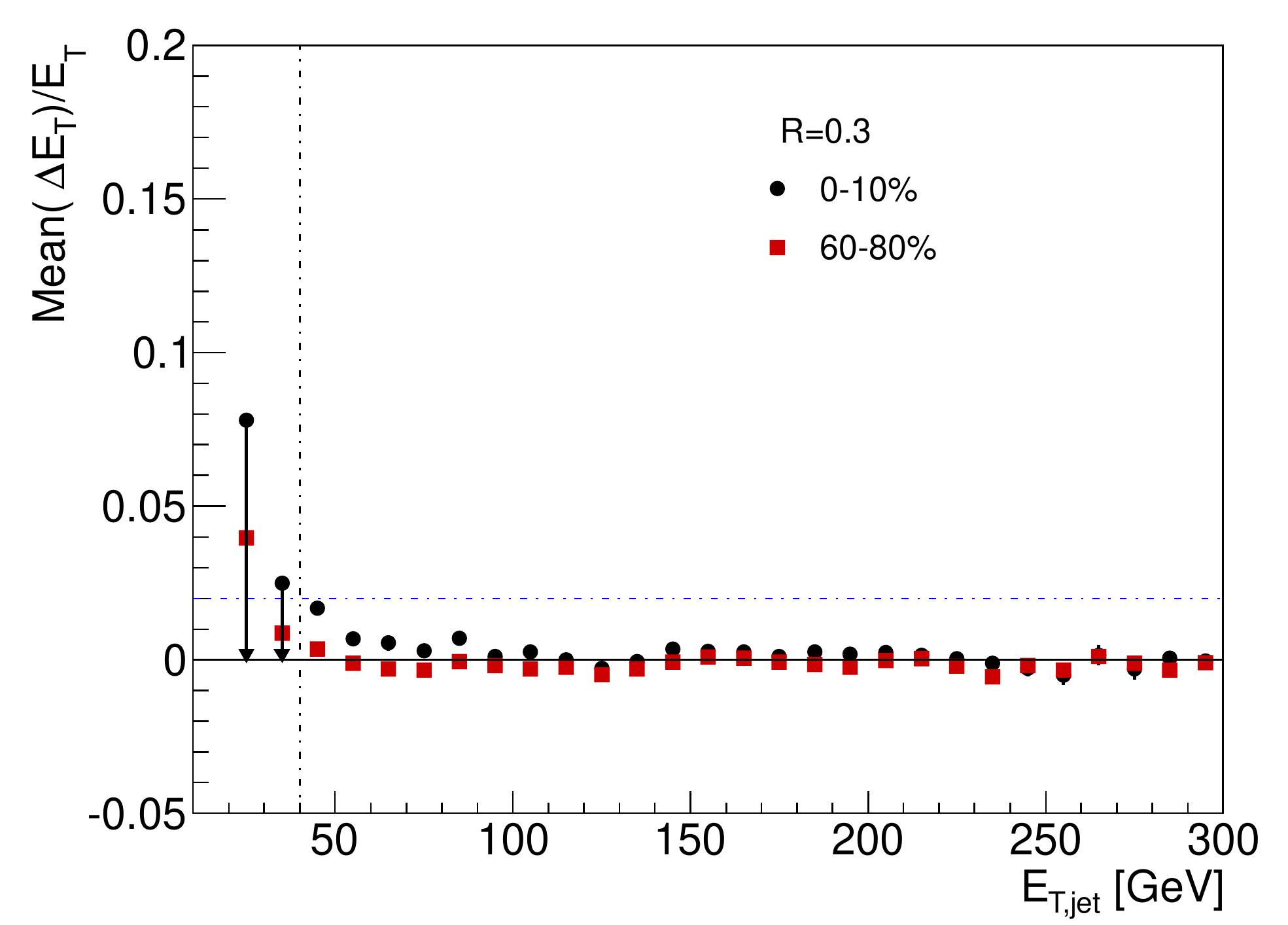}
\includegraphics[width=0.49\textwidth]{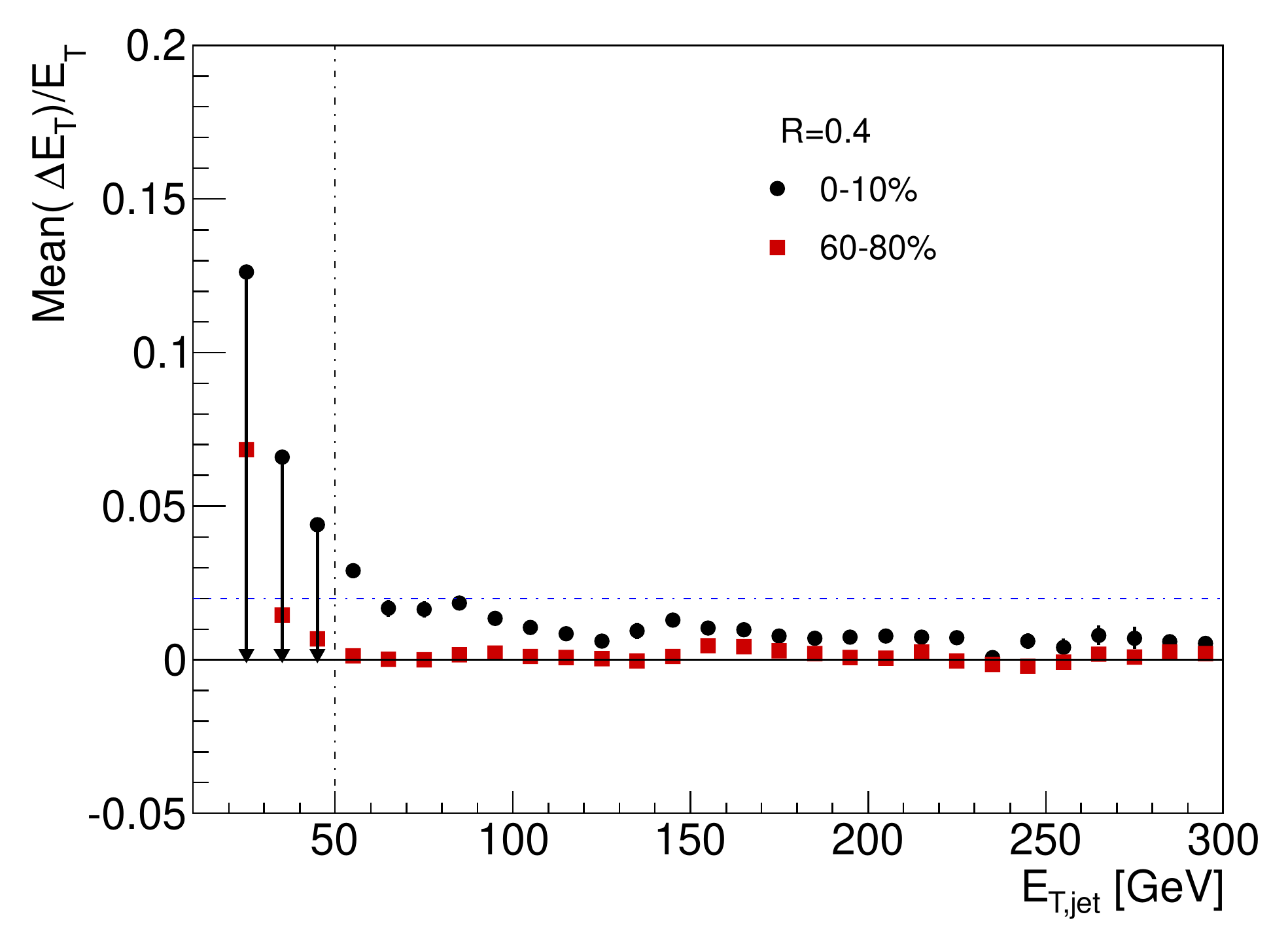}
\includegraphics[width=0.49\textwidth]{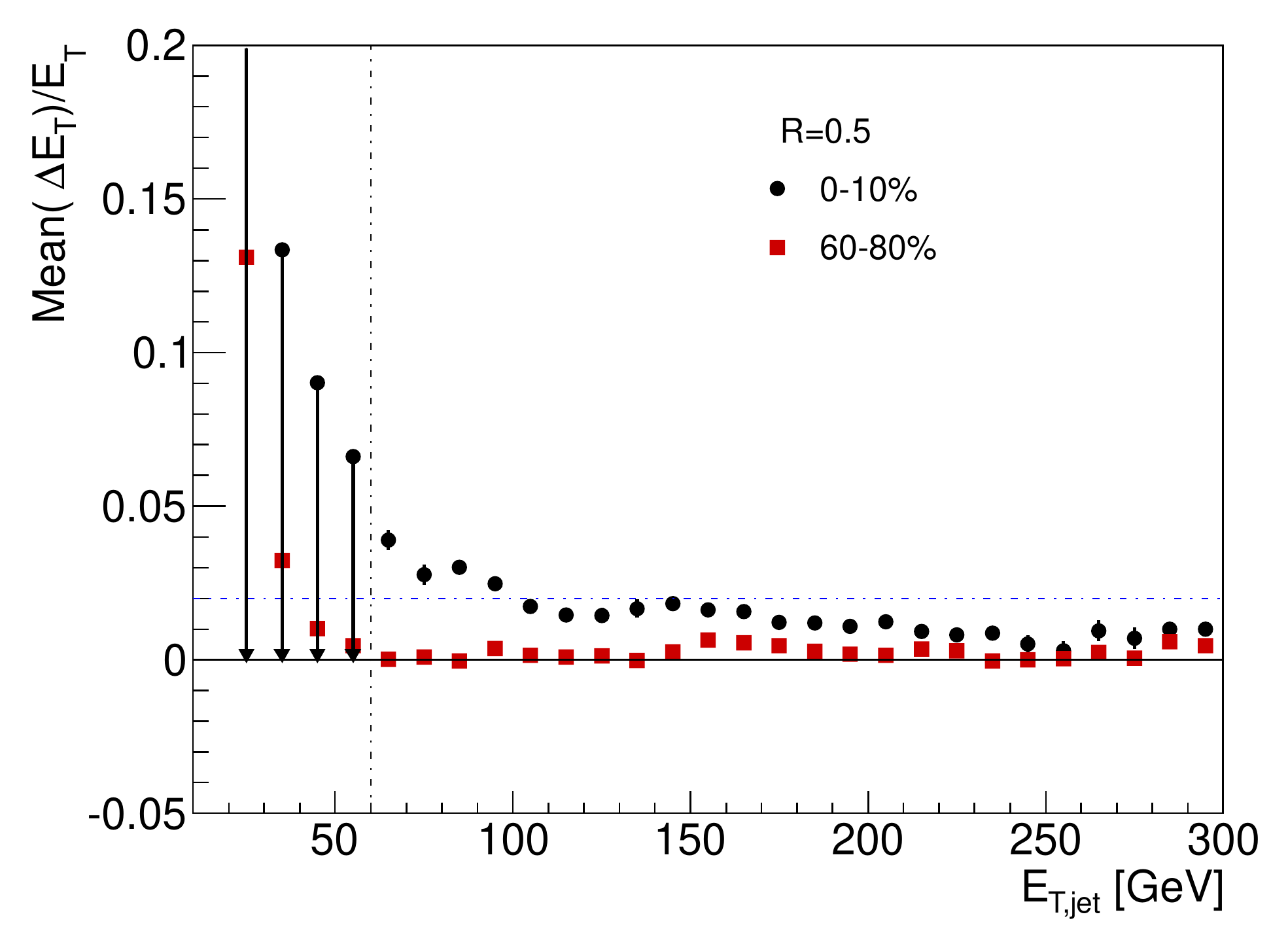}
\caption{Jet energy scale vs \ETtrue\ for different jet definitions compared between central and
peripheral collisions.}
\label{fig:JESEt}
\end{figure}

The evaluation of the JES as a function of \ETtrue\ is shown on Fig.~\ref{fig:JESEt}. Upper plots and lower 
two plots of Fig.~\ref{fig:JESEt} compare the JES for different jet
definitions in the 0-10\% and 60-80\% centrality bins. Lower right
plot then shows the JES in seven centrality bins for \RFour\ jets. At
low \ETtrue\ the accuracy of the evaluation of the JES is limited by precision of fitting of the truncated 
Gaussian distribution. In these cases, to the left of vertical lines are provided in the
figures, an upper limit on the JES (as indicated by the arrows on the
error bars) is estimated. The difference between central and peripheral collisions is at most 2\% for $\RTwo 
- 0.5$ collections. The one percent non-closure in \RTwo\ jets is due
to the fact that the derivation of numerical inversion constants
requires isolated jets, but the JES evaluation includes non-isolated
jets as well.

\begin{figure}[htb] 
\centering
\includegraphics[width=0.4\textwidth]{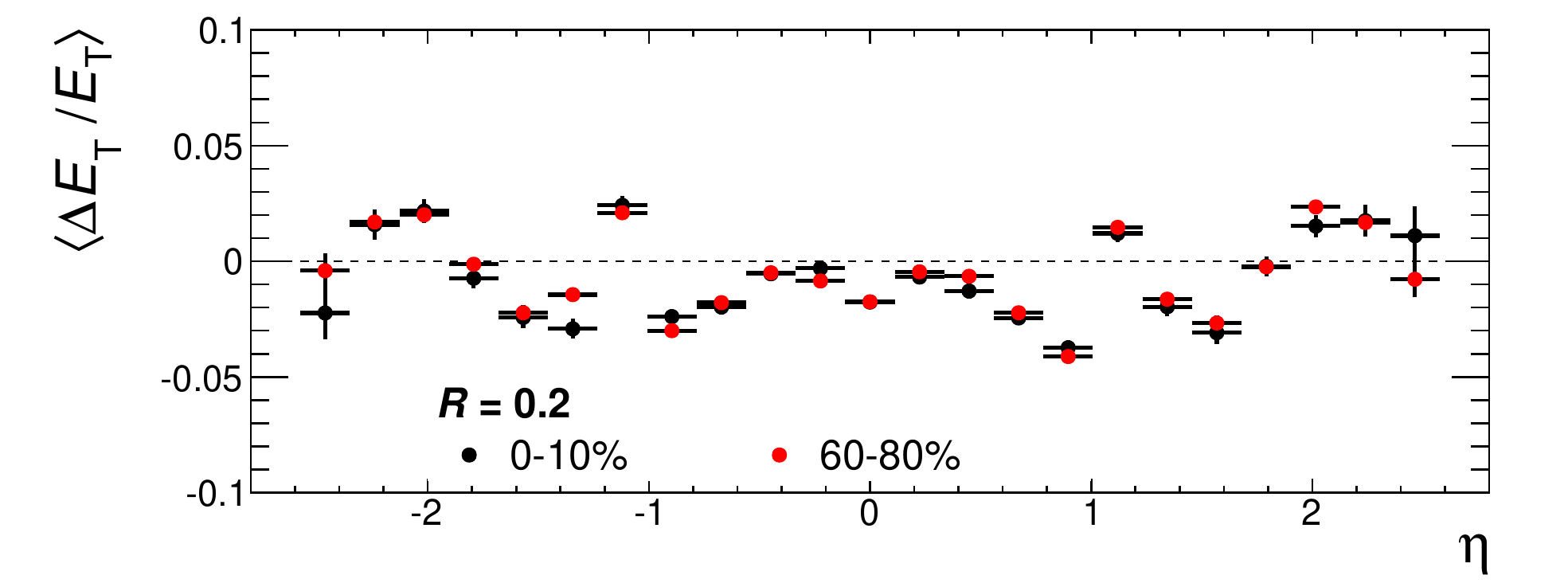}
\includegraphics[width=0.4\textwidth]{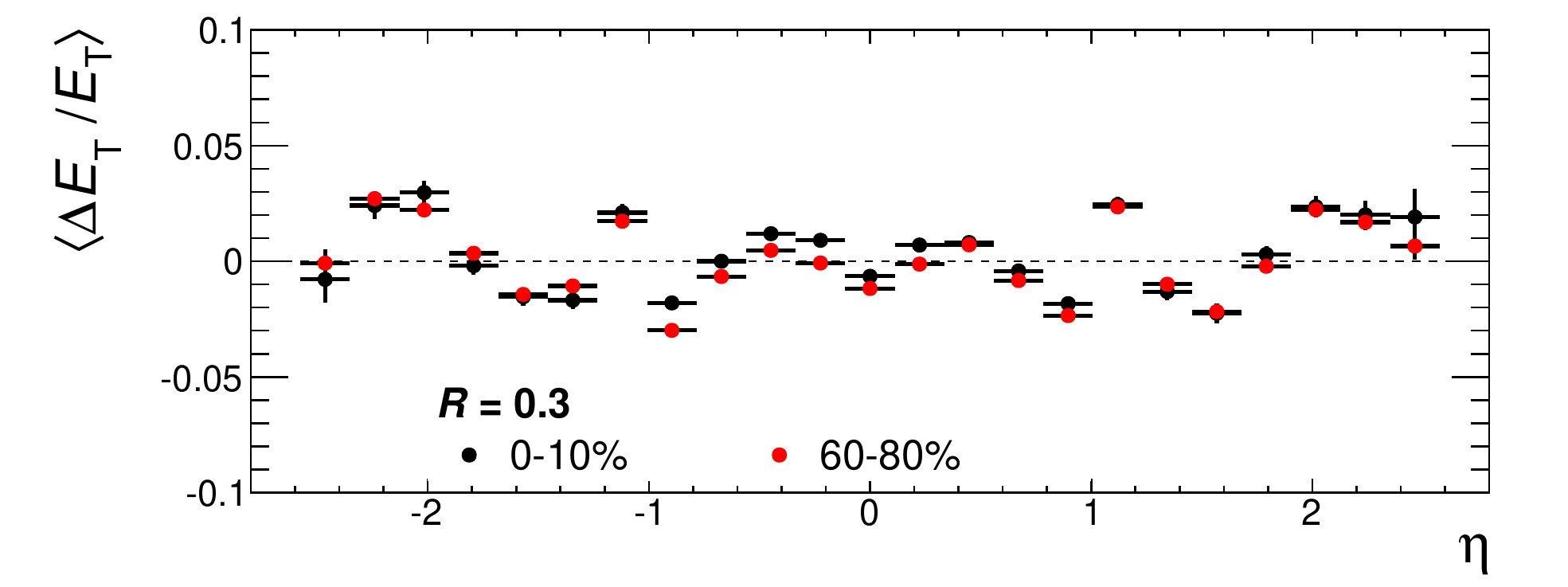}
\includegraphics[width=0.4\textwidth]{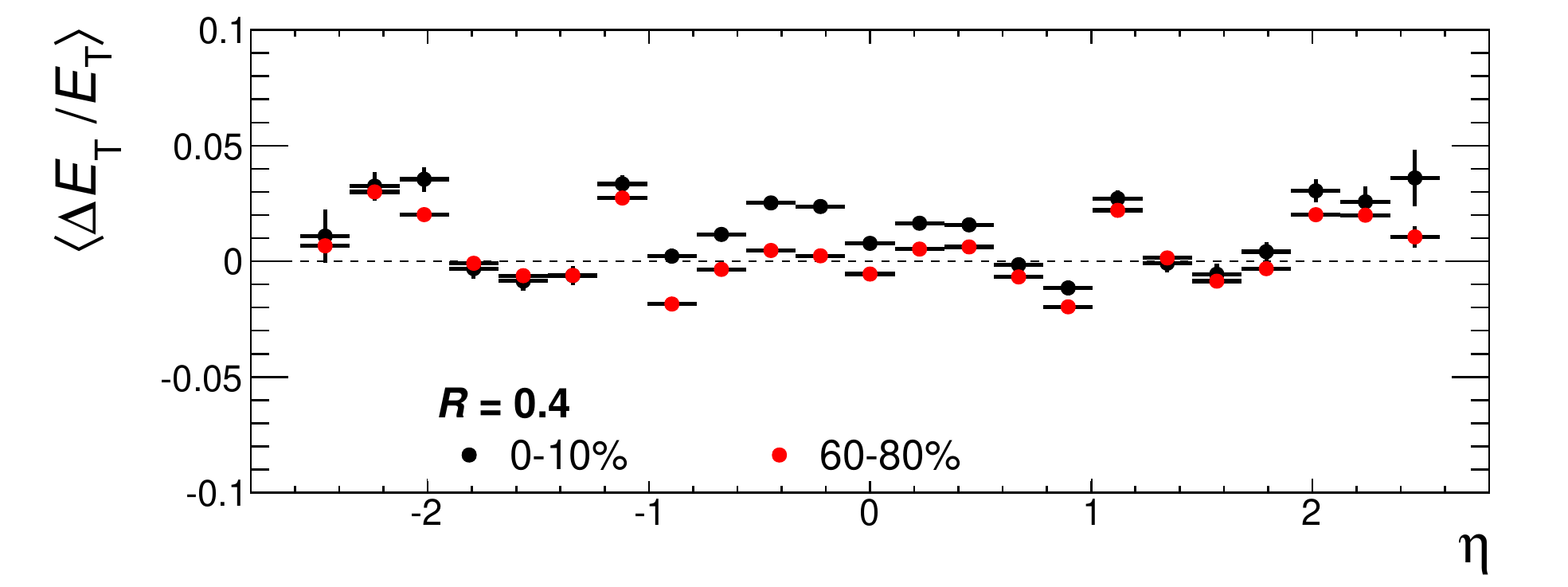}
\includegraphics[width=0.4\textwidth]{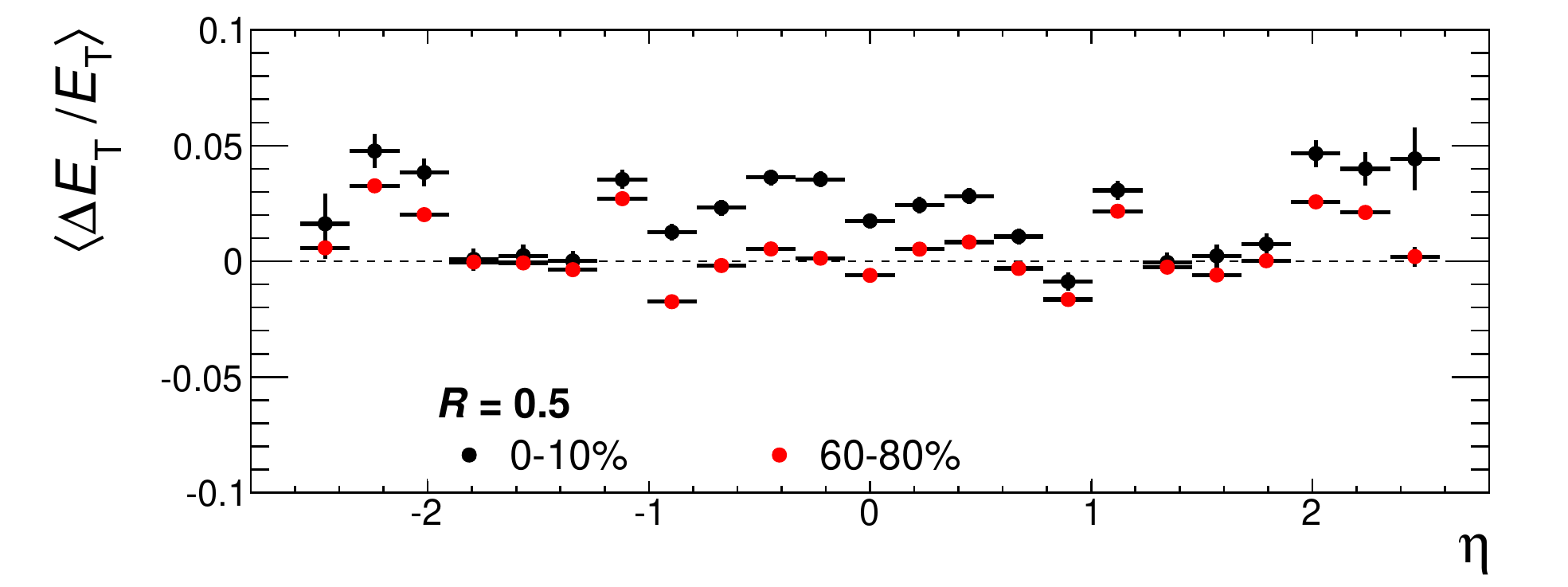}
\caption{Jet energy scale vs $\eta^{\mathrm{truth}}$ for different jet definitions compared between central and
peripheral collisions.}
\label{fig:JESEta}
\end{figure}

Comparison of JES as a function of $\eta^{\mathrm{truth}}$ is shown in Fig.~\ref{fig:JESEta}. The JES as a function of 
$\eta^{\mathrm{truth}}$ was evaluated by calculating the mean of the $\Delta \et / \et$ distribution of all 
reconstructed jets corresponding to truth jets with $\et > 90$~\GeV.

\begin{figure}
\centerline{
\includegraphics[width=0.7\textwidth]{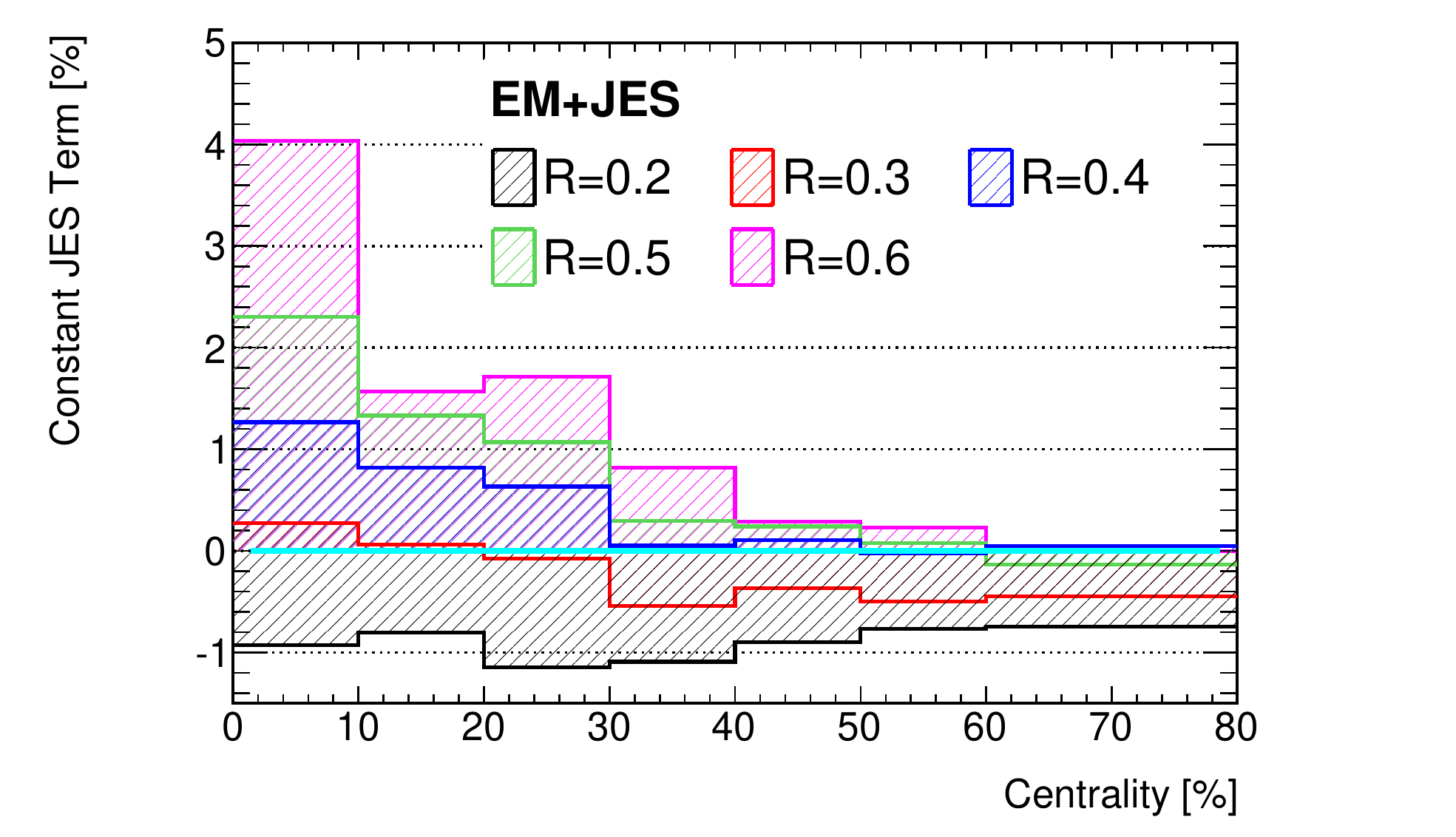}
}
\caption{Non-closure in JES calibration obtained from fits.}
\label{fig:JES_summary}
\end{figure}

The behavior of the JES as a function of~\ETtrue\ can be summarized
by fitting the distribution with a linear function:
\begin{equation}
\Delta \et / \et (\ETtrue)=a\ETtrue+b\,.
\end{equation}
The $a$ term is typically small, and the constant $b$ term represents the
non-closure in the JES calibration. The values of this constant term
for all centralities and $R$ values are shown in
Fig.~\ref{fig:JES_summary}.

\clearpage
\subsection{Jet Energy Resolution}

\begin{figure}[htb] 
\centering
\includegraphics[width=0.4\textwidth]{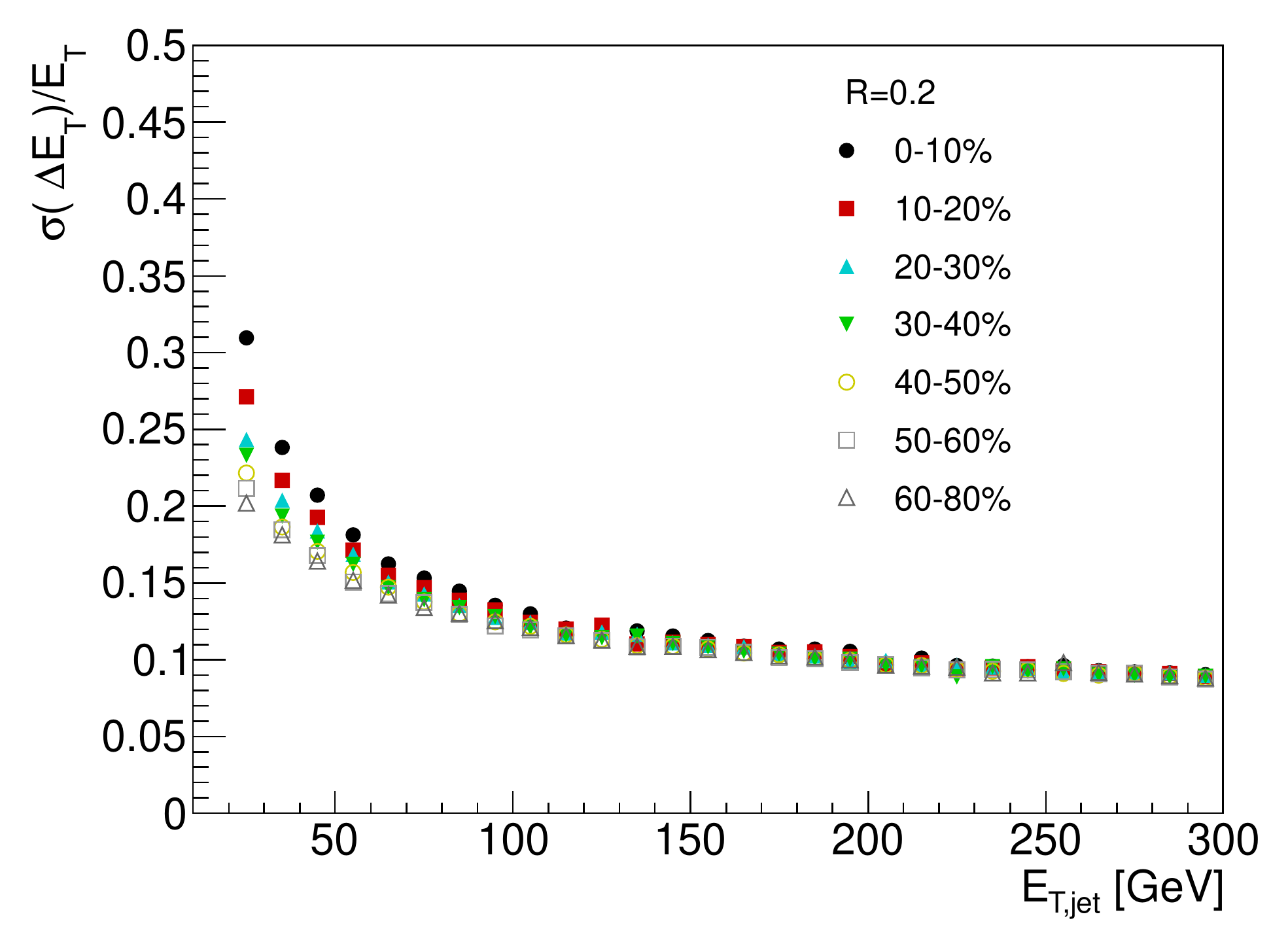}
\includegraphics[width=0.4\textwidth]{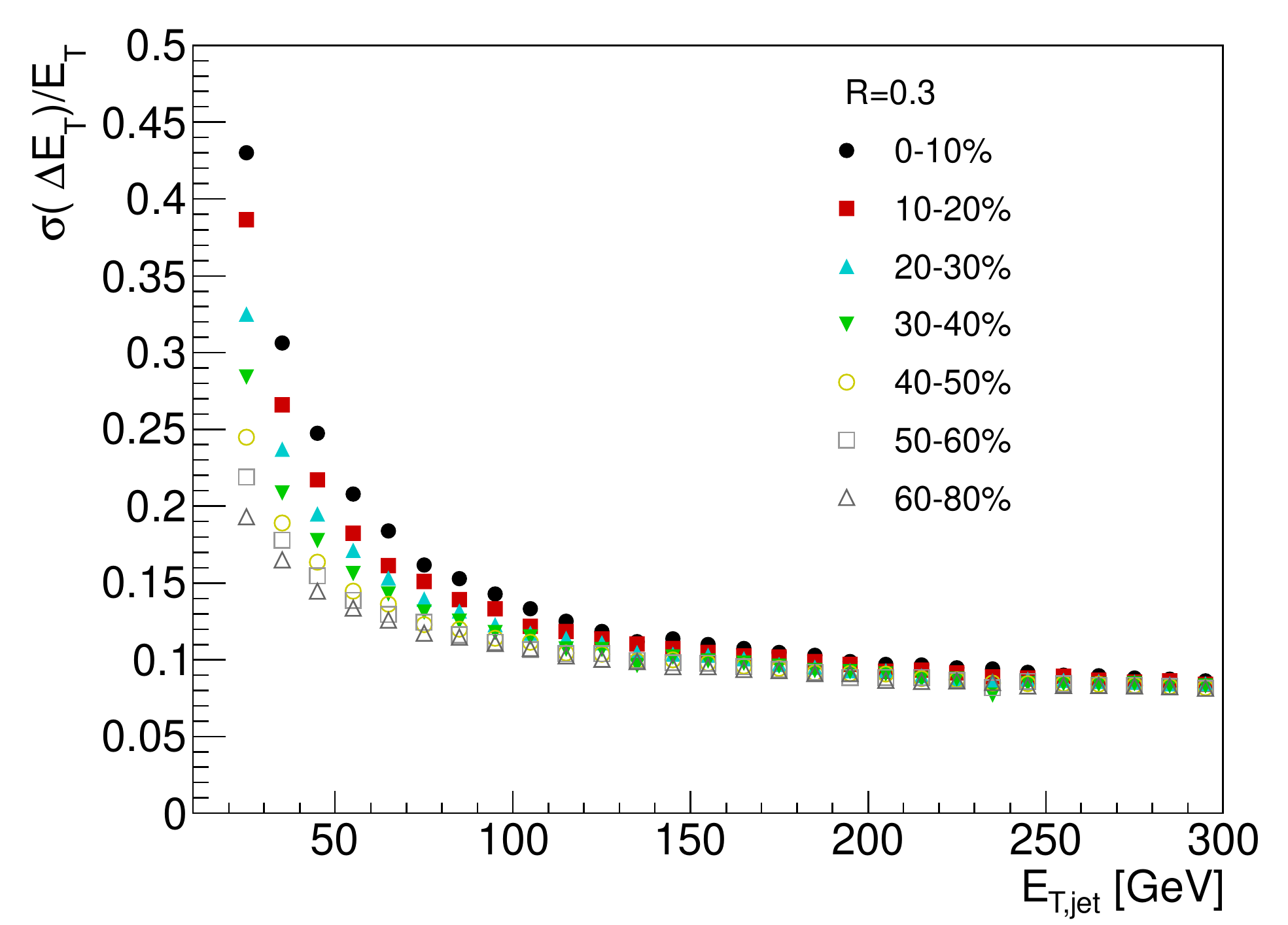}
\includegraphics[width=0.4\textwidth]{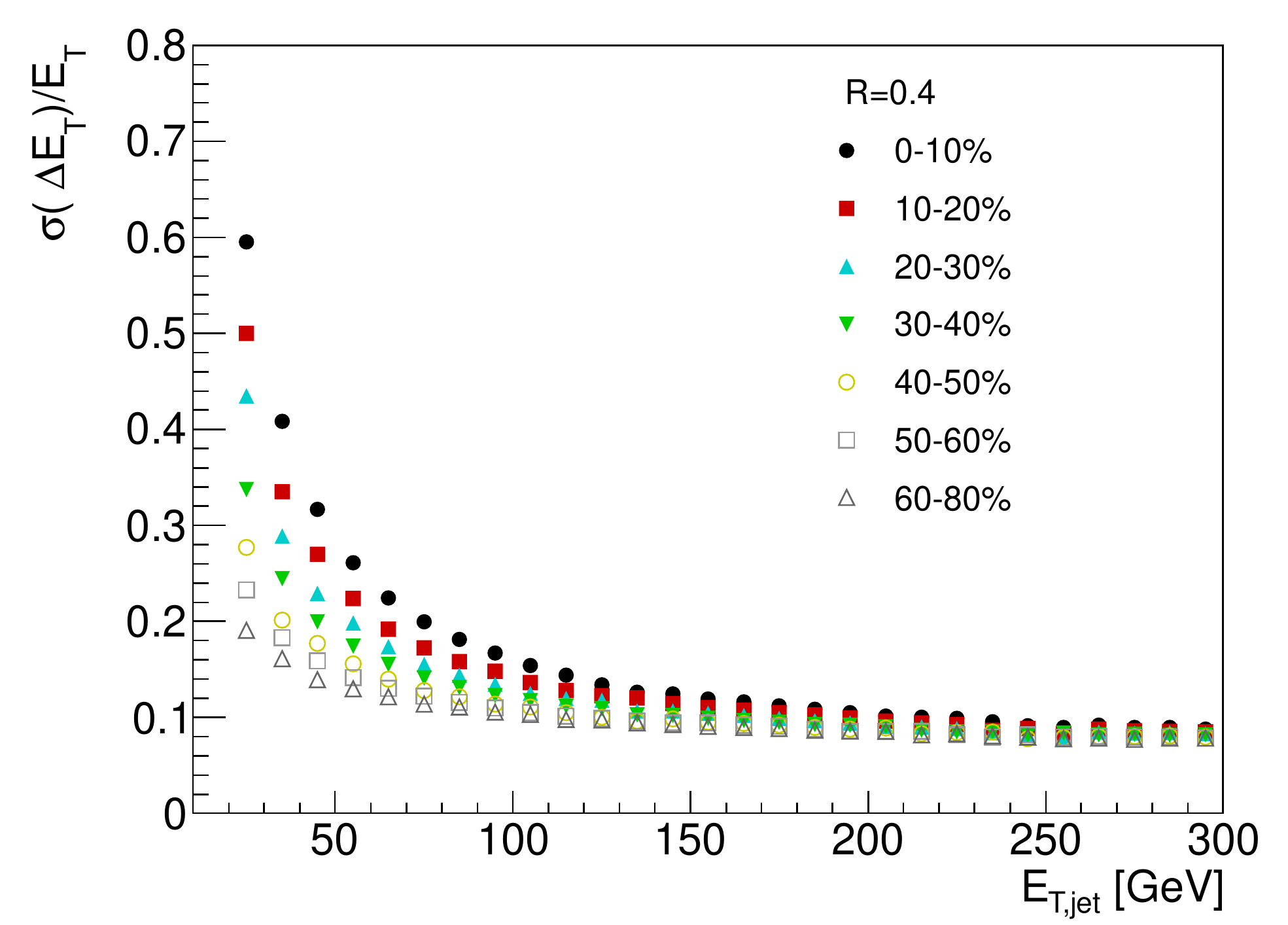}
\includegraphics[width=0.4\textwidth]{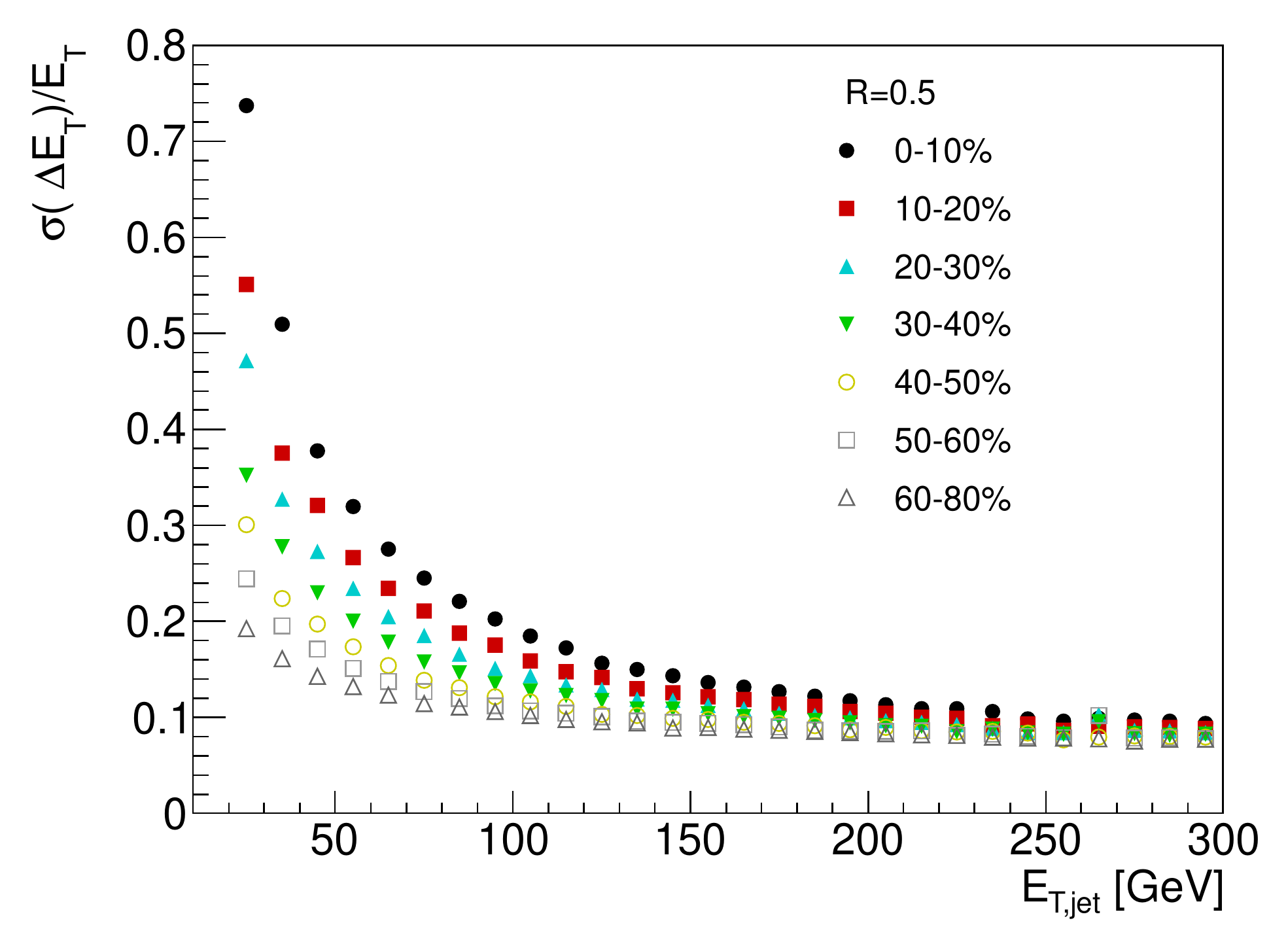}
\caption{Jet energy resolution vs \ETtrue\ for different jet definitions compared between central and
peripheral collisions.}
\label{fig:JEREt}
\end{figure}

The jet energy resolution (JER) can be defined using the 
width of the Gaussian fit discussed in the previous section. 
Figure~\ref{fig:JEREt} shows the evaluation of JER in different bins of centrality and for different jet 
definitions. The JER increases both with increasing centrality and
increasing jet radius. Validation studies of the JER using data are presented in Sec.~\ref{sec:validation:JER}.

\begin{figure}[htb] 
\centering
\includegraphics[width=0.5\textwidth]{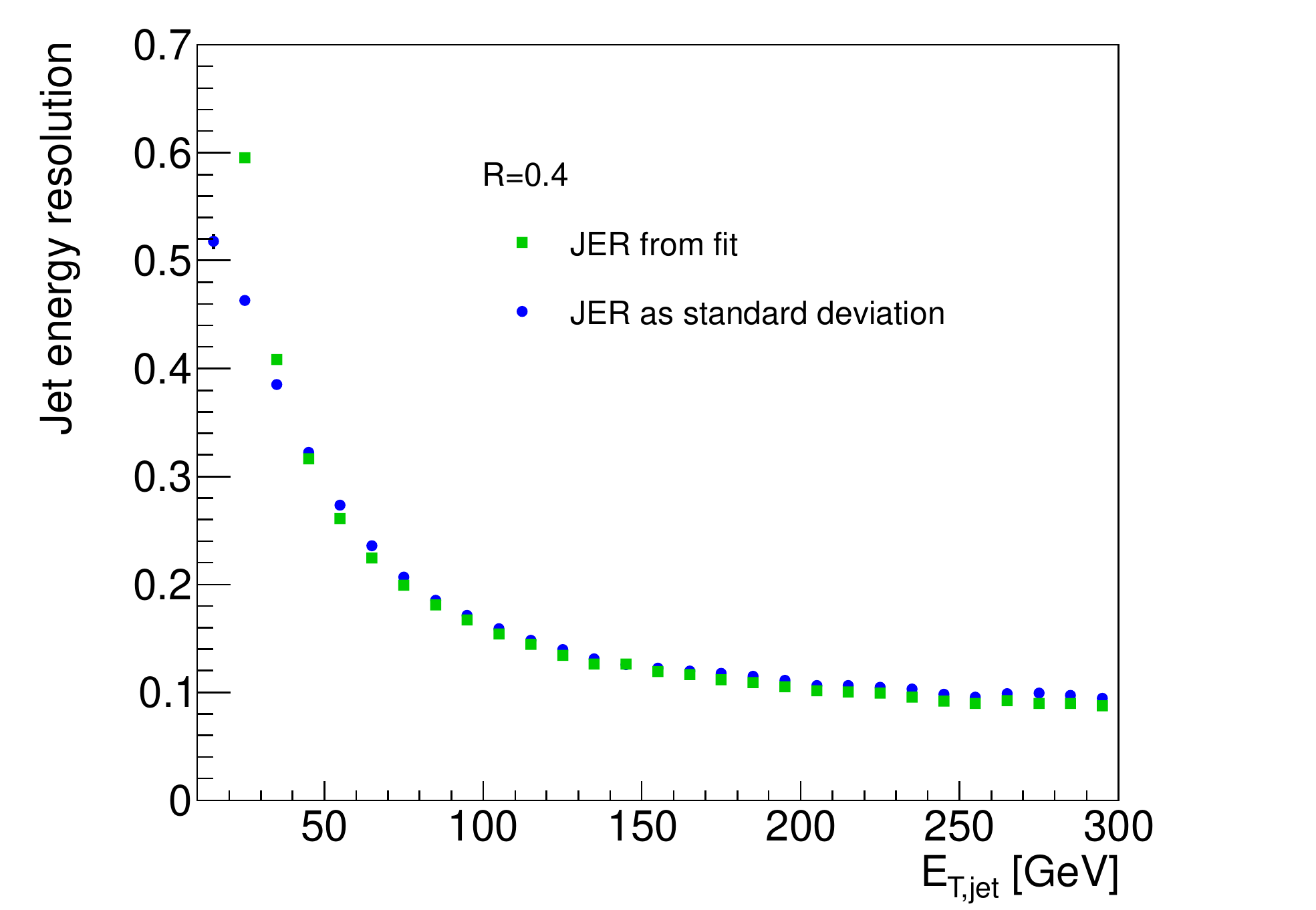}
\caption{Comparison of JER evaluated using the Gaussian fit and as a standard deviation of the $\Delta \et 
/ \et$ distribution.}
\label{fig:JERComparison}
\end{figure}

An alternative definition of the JER uses the standard 
deviation of the $\Delta\ET/\ET$ distribution in bins of \ETtrue\ ,i.e.
\begin{equation}
\sigma(\Delta\ET/\ET) \equiv \sqrt{ \left\langle\left(\dfrac{\Delta
    \ET}{\ETtrue}\right)^{2}\right\rangle - 
\left\langle\dfrac{\Delta \ET}{\ETtrue}\right\rangle^{2}}\,.
\end{equation}
Figure~\ref{fig:JERComparison} shows the evaluation of the difference between the two
JER definitions for \RFour\ jets and 0-10\% central collisions. The
Gaussian-fit JER  is slightly smaller for $\ETtrue \lesssim 50$~\GeV, which can be attributed to the
asymmetric $\Delta \ET$ distribution resulting from the minimum \et\
threshold on reconstructed jets.

\subsection{Jet Reconstruction Efficiency}
 
\begin{figure}[htb] 
\centering
\includegraphics[width=5 in]{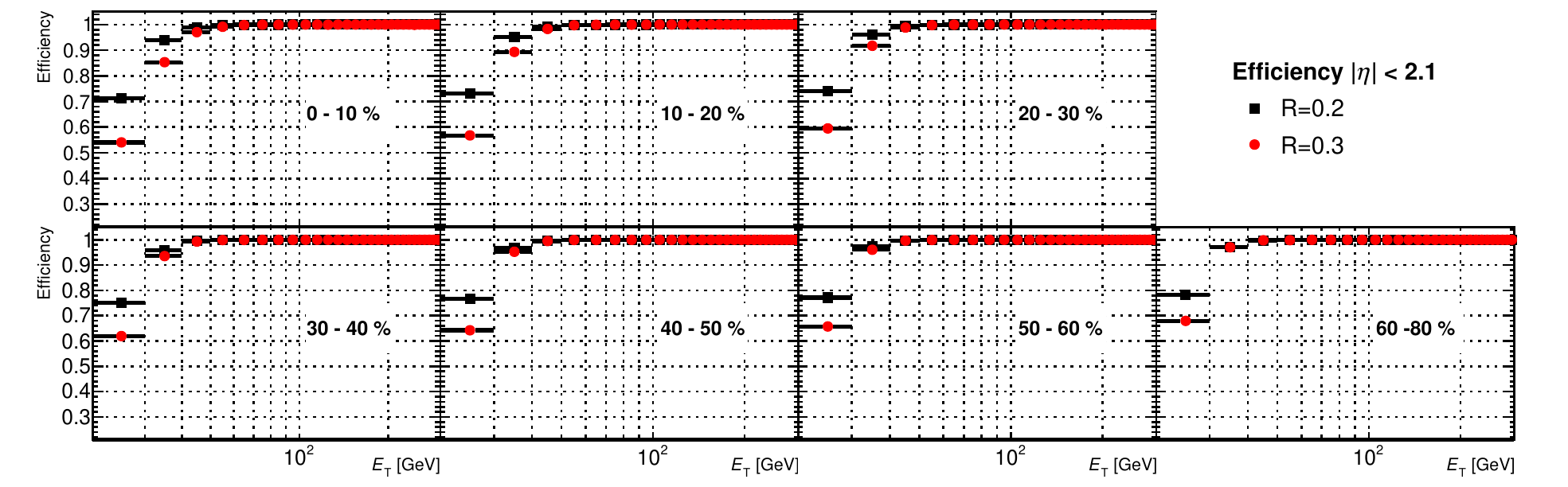}
\includegraphics[width=5 in]{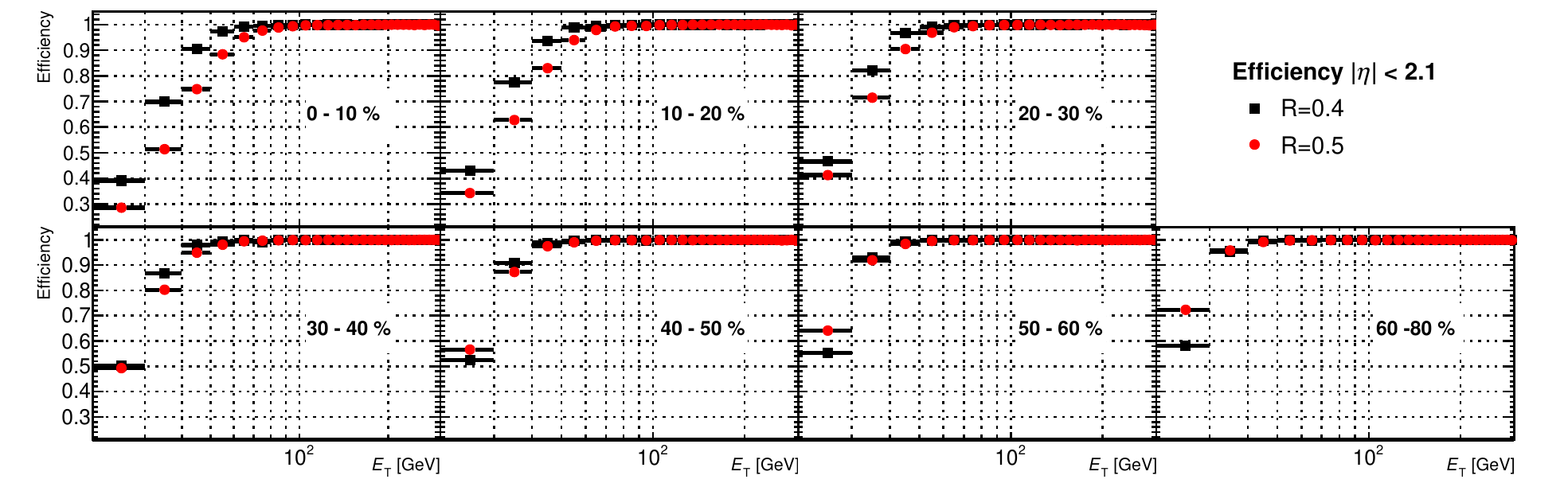}
\caption{Jet reconstruction efficiency vs $\et^{\mathrm{truth}}$ in
  the EM+JES scheme for jets with $|\eta^{\mathrm{truth}}| < 2.1$ for
  all $R$ values and centralities.}
\label{fig:perf:jre_et}
\end{figure}

The jet reconstruction efficiency is defined as:
\begin{equation}
\varepsilon(\ETtrue)  = \dfrac{\Delta N^{\mathrm{match}}}{\Delta N^{\mathrm{truth}}}\,,
\label{eqn:efficiency_definition}
\end{equation}
where $\Delta N^{\mathrm{truth}}$ and $\Delta N^{\mathrm{match}}$ are
the total number of truth jets and number of truth jets with a
matching reconstructed jet in a given \ETtrue\ bin. The
reconstruction efficiencies are shown in
Fig~\ref{fig:perf:jre_et}. The $R=0.4$ and $R=0.5$ jet
reconstruction efficiency reaches 95\% by 60~\GeV\ in the most
central (0-10\%) \PbPb\ collisions and 30~\GeV\ in the most peripheral
(60-80\%) collisions. The last 5\% increase in the efficiency takes
place near 70~\GeV\ for all centrality bins. For the $R=0.2$ and
$R=0.3$ jets, the efficiency reaches 95\% below 40~\GeV\ for all
collision centralities and is approximately $1$ above 50~\GeV.

\begin{figure}[htb] 
\centering
\includegraphics[width=5 in]{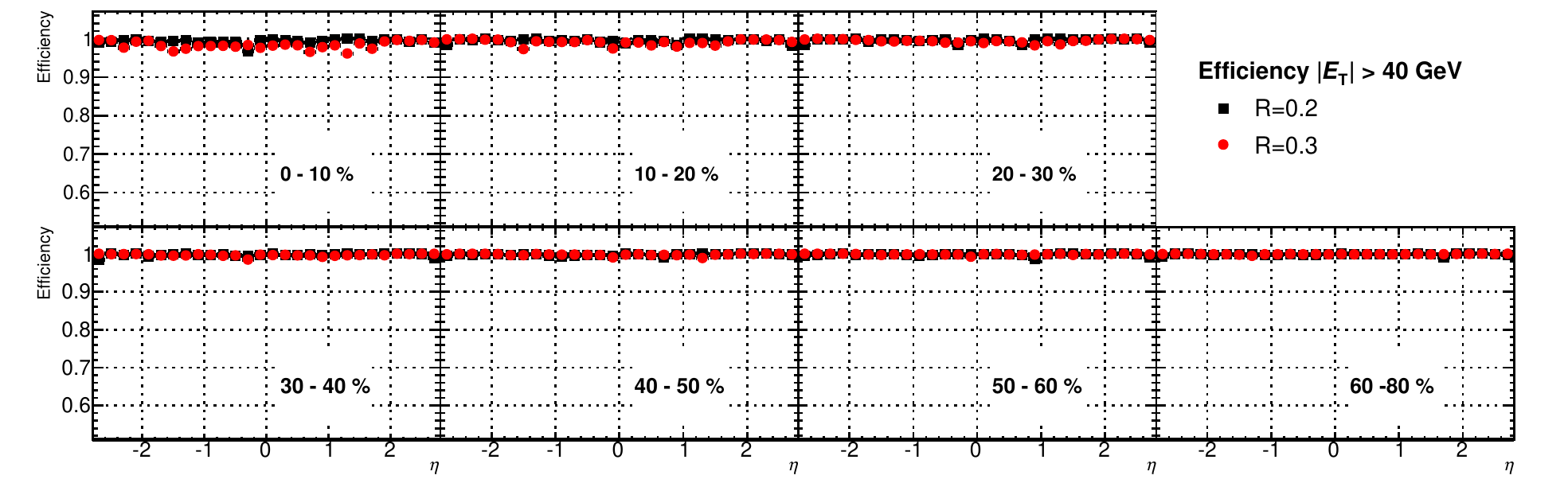}
\includegraphics[width=5 in]{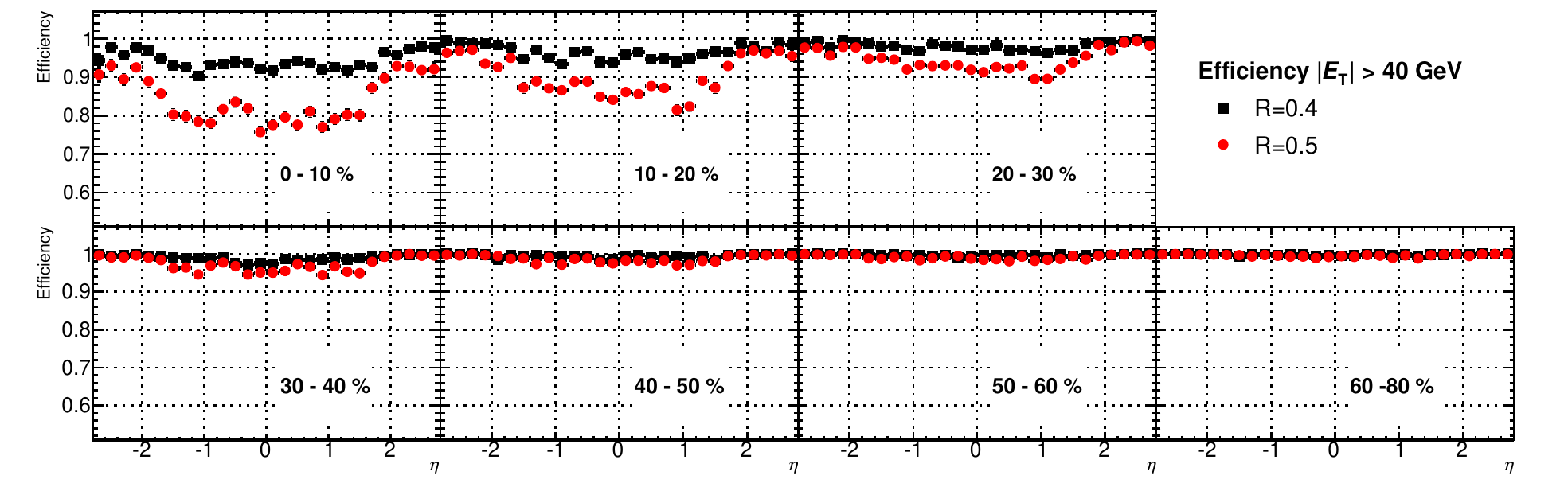}
\caption{Jet reconstruction efficiency vs $\eta^{\mathrm{truth}}$ in
  the EM+JES scheme for jets with $\et^{\mathrm{truth}} > 40\GeV$ for
  all $R$ values and centralities.}
\label{fig:perf:jre_eta}
\end{figure}

Figure~\ref{fig:perf:jre_eta} shows
the jet reconstruction efficiency a function of
$\eta^{\mathrm{truth}}$ for jets with $\et > 40\GeV$. The smaller
radii have saturated the efficiency turn on at this energy and show
no~\eta\ variation. The $R=0.4$ and $R=0.5$ jets are not full
efficient at this energy in the most central collisions and show a slight
reduction in the efficiency between the barrel ($|\eta| < 1.5$) and
end-cap ($1.5 < |\eta| < 3.2$) regions.

{
\clearpage
\chapter{Data Analysis}
\label{section:analysis}
\section{Data Set}
\label{section:analysis:data_set}
The data presented in the following section were recorded by ATLAS during the
2010 \PbPb\ run. During this run a total integrated luminosity of
approximately $\mathcal{L}=7 \mu \mathrm{b}^{-1}$ was recorded and determined to
be of sufficient quality for physics analysis. An additional sample of
$\mathcal{L}\sim 1\mu\mathrm{b}^{-1}$ was recorded with the solenoidal
magnetic field turned off; this data set will not be considered
here.

The events analyzed were required to satisfy a set of minimum bias
event selection criteria. These were chosen to require a physics
signature consistent with an inelastic \PbPb\ collision. Additional
conditions were imposed to reject photo-nuclear and non-collision
background. The requirements were as follows:
\begin{enumerate}
\item The event was recorded because it satisfied a minimum bias
trigger. Events were required to have fired one of the following
triggers after prescale and veto (events dropped due to dead time):
         \begin{itemize}
         \item MBTS coincidence: \verb=L1_MBTS_N_N=, where N $=$ 1, 2, 3 or 4.
         \item ZDC coincidence: \verb=L1_ZDC_A_C= or \verb=L1_ZDC_AND=.
        \end{itemize}
\item Good MBTS timing. The A and C side times were required to satisfy
$|\Delta t_{\mathrm{MBTS}}| < 3$~ns.
\item ZDC coincidence. Regardless of how the trigger selection was
made the event must have one of the ZDC coincidence triggers before prescale, a logical \verb=OR= between \verb=L1_ZDC_A_C= and
\verb=L1_ZCD_AND=. This was found to be essential for rejecting
against photo-nuclear events.
\item A good reconstructed vertex.
\end{enumerate}
Applying these event selection criteria, combined with selecting
luminosity blocks consistent with stable running and detector operation
resulted in a sample of 50 million events. These event selection
criteria were used in all heavy ion analyses within ATLAS using the
2010 data. 
\section{Centrality Determination}
\label{section:analysis:centrality}
The centrality determination used the total transverse energy as measured by the
ATLAS FCal system, \ETfcal. The pseudorapidity coverage of this system, $ 3.2 < | \eta | <
4.9$, is well separated from the central region of the detector,
meaning that specific, centrality-dependent physics processes will not
bias the centrality determination. The \ETfcal\ was found to be
strongly correlated with the total energy deposited in the rest of the
calorimeter, making it an excellent indicator of global event activity. The correlation between total energy in the
electromagnetic barrel, $|\eta| < 2.8$, and \ETfcal\ is shown in
Fig.~\ref{fig:analysis:et_correlation}.
\begin{figure}[tbh]
\centering
\includegraphics[width=0.7\textwidth]{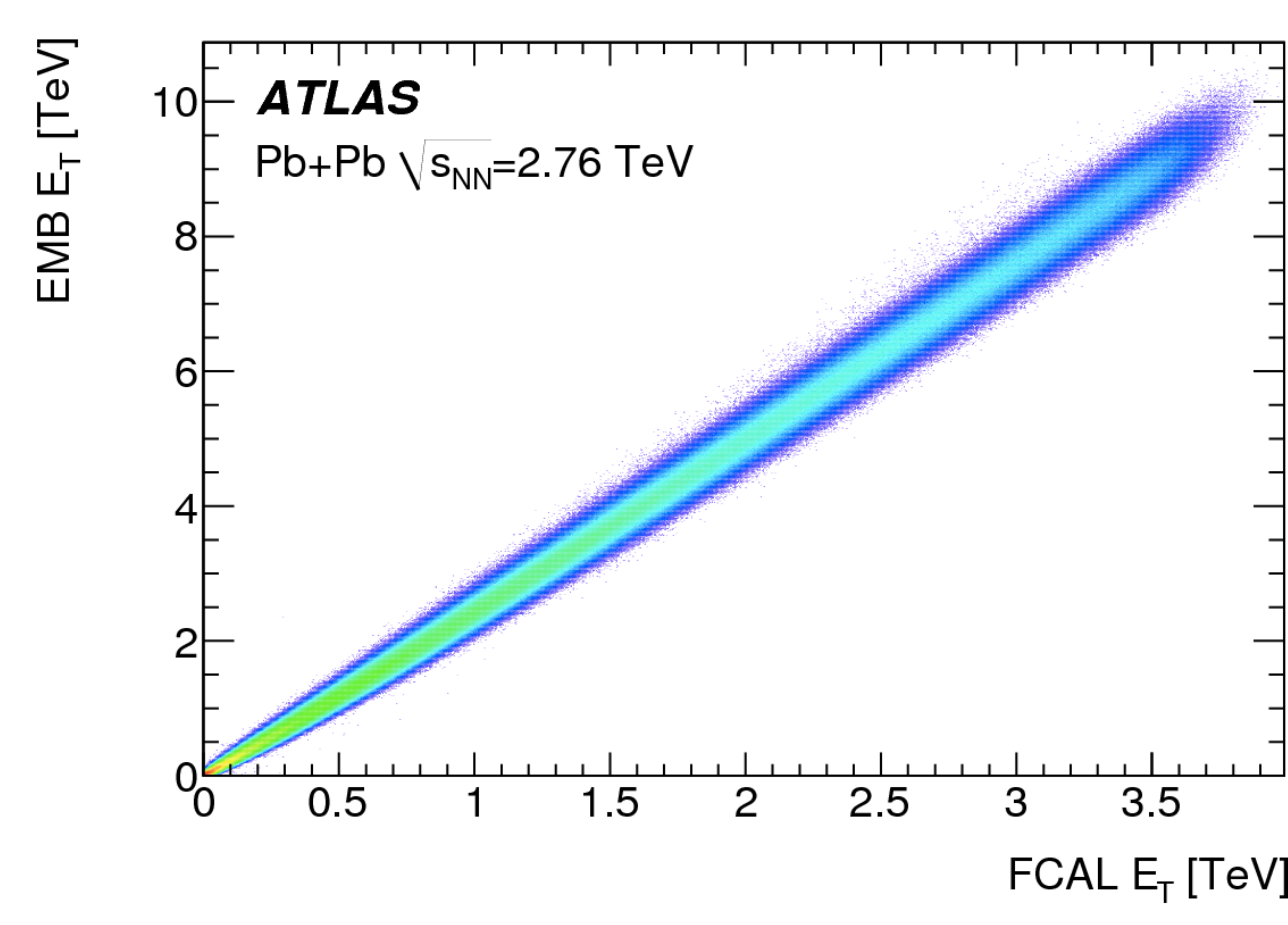}
\caption{Minimum bias EM barrel \ET\ vs \ETfcal\ correlation.}
\label{fig:analysis:et_correlation}
\end{figure}

The minimum bias \ETfcal distribution has a distribution typical of
other centrality variables such as the total number of charged
particles, as well as Glauber quantities \Ncoll\ and \Npart. This
distribution is shown in Fig.~\ref{fig:analysis:et_fcal}.
\begin{figure}[tbh]
\centering
\includegraphics[width=0.7\textwidth]{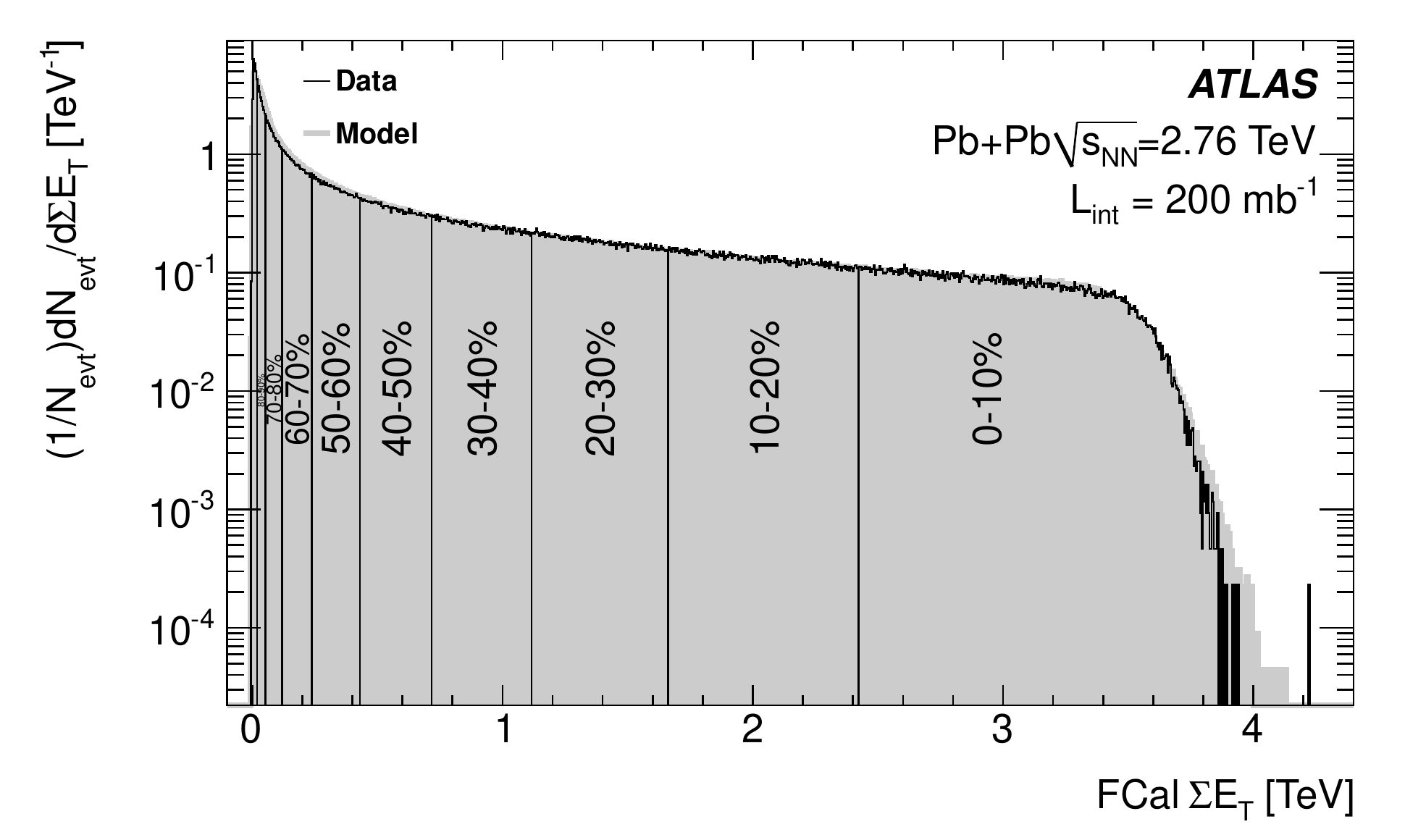}
\caption{Minimum bias \ETfcal\ distribution, with the labeled centrality bins
  corresponding to 10\% of the total integral of the distribution.}
\label{fig:analysis:et_fcal}
\end{figure} A two-component model as given by
Eq.~\ref{eqn:background:two_comp_model} was used to fit this
distribution. An additional check was performed by taking an n-fold
convolution of the \pp\
\ETfcal\ distribution based on a Glauber MC, which agreed well. A
complete analysis of the 2010 centrality determination is given in
Ref.~\cite{centralitynote}. The minimum bias event selection cuts were
determined to introduce a 2\% inefficiency, almost entirely from the MBTS
timing cut. The distribution of $\Delta t_{\mathrm{MBTS}}$ as a
function of \ETfcal\ is shown in
Fig.~\ref{fig:analysis:fcal_et_vs_mbts}. 
\begin{figure}[tbh]
\centering
\includegraphics[width=0.7\textwidth]{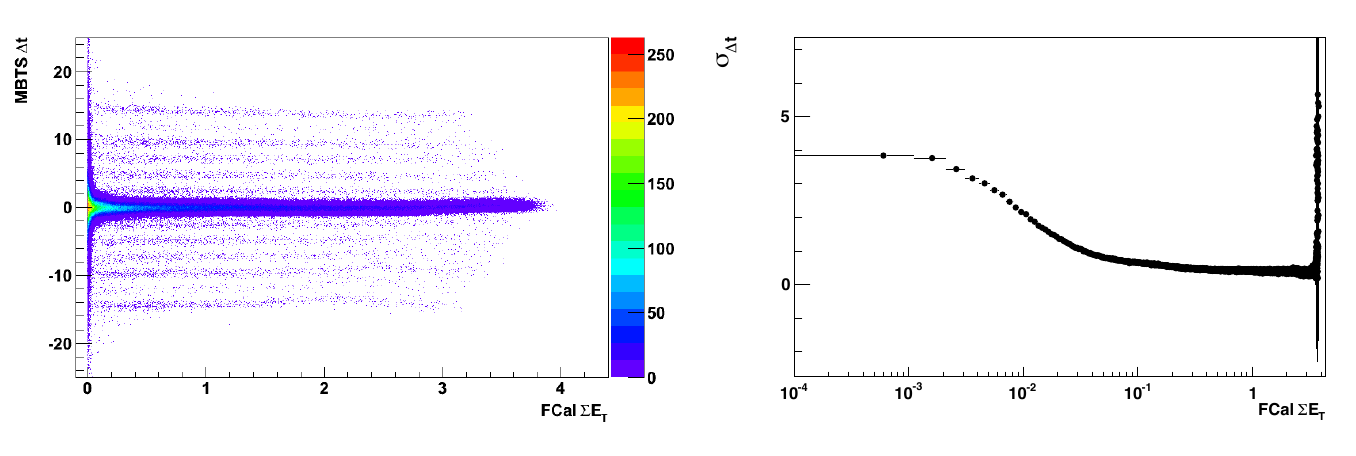}
\caption{$\Delta t_{\mathrm{MBTS}}$ distributions in bins of
\ETfcal\ is shown on the left. The structure of the satellite peaks is clearly visible, as is
the substantial broadening of the timing width at low \ETfcal. The
Gaussian sigma of these distributions as a function of \ETfcal\ is
shown on the right.}
\label{fig:analysis:fcal_et_vs_mbts}
\end{figure}

Under the assumption of 98\% efficiency, the \ETfcal\ distribution was divided into 10\% intervals, except for
the range 60-100\%, which was divided into two intervals of 20\% each,
in the fashion discussed in
Section~\ref{section:bkgr:glauber}. In each bin the
corresponding events in the Glauber MC were averaged to obtain \Ncoll\
and \Npart. These values are shown in
Table~\ref{tbl:centrality_bins}. This table also shows the ratios of
\Ncoll\ factors, \Rcollcent\, which are used directly in the \Rcp\ determination. The
uncertainties were evaluated by considering the effects of varying
ingredients into the Glauber calculation such as the nucleon-nucleon cross
section, the two Woods-Saxon parameters as well as a 2\% variation of
the inefficiency. The errors on \Ncoll\ in different centrality bins
are correlated so the uncertainties on \Rcollcent\ were constructed appropriately.
\begin{table}
\centering
\begin{tabular}{| c | c ||
r@{.}l|
r@{.}l||
r@{.}l|
r@{.}l|
r@{.}l|
r@{.}l|
r@{.}l|
r@{.}l|} \hline
\multicolumn{2}{|c||}{Centrality [\%]}&
 \multicolumn{4}{|c||}{\ETfcal[\TeV]}&
 \multicolumn{4}{c|}{\Ncoll}&
 \multicolumn{4}{c|}{\Rcollcent}&
 \multicolumn{4}{c|}{\Npart}\\ \hline
low&
high&
\multicolumn{2}{|c|}{low}&
\multicolumn{2}{|c||}{high}&
\multicolumn{2}{|c|}{$\langle \Ncoll\rangle$} &
\multicolumn{2}{|c|}{$\delta \Ncoll$} &
\multicolumn{2}{|c|}{$\Rcollcent$} &
\multicolumn{2}{|c|}{$\delta \Rcollcent [\%]$} &
\multicolumn{2}{|c|}{$\langle \Npart\rangle$} &
\multicolumn{2}{|c|}{$\delta \Npart$} 
\\ \hline \hline
0& 10 &   2&423 &   \multicolumn{2}{|l||}{$\infty$} &   1500&63 &
114&8 &56&7 & 11&4 &356&2 & 2&5    \\ \hline 
10& 20 &   1&661 &   2&423 &   923&29 &   68&0 & 34&9 & 10&5  &261&4 & 3&6 \\ \hline 
20& 30 &   1&116 &   1&661 &   559&02 &   40&5 & 21&1 & 9&4   &186&7 & 3&8 \\ \hline 
30& 40 &   0&716 &   1&116 &   322&26 &   23&9 & 12&2 & 7&9   &129&3 & 3&8 \\ \hline 
40& 50 &   0&430 &   0&716 &   173&11 &   14&1 & 6&5 &  6&1   &85&6 & 3&6  \\ \hline 
50& 60 &   0&239 &   0&430 &   85&07 &   8&4   &  3&2 &  3&8  &53&0 & 3&1    \\ \hline
60& 80 &   0&053 &   0&239 &   26&47 &   3&5 &                
\multicolumn{2}{|c|}{$-$} &  \multicolumn{2}{|c|}{$-$} &22&6 & 2&1  \\ \hline 
\end{tabular}
\caption{Centrality bins, \Ncoll, \Rcollcent\ values and  their fractional error evaluated directly from the Glauber Monte Carlo~\cite{centralitynote}.}  
\label{tbl:centrality_bins}
\end{table}\section{Validation}
\label{section:analysis:validation}
\subsection{Jet Energy Scale}
\label{sec:validation:JES}
A number of studies were performed using data to check the accuracy of
the jet energy scale calibration procedure which was derived from
MC.
\subsubsection{EM+JES vs GCW Comparison}
\label{sec:validation:JES:EMJES_GCW}
The jet reconstruction was performed with both the EM+JES and GCW
calibration schemes. A comparison of spectra using each of these
schemes is shown for~\RFour\ jets in
Fig.~\ref{fig:validation:calibration_comparison}. In general good
agreement is found between both schemes.
\begin{figure}[htb]
\centering
\includegraphics[width =0.49\textwidth] {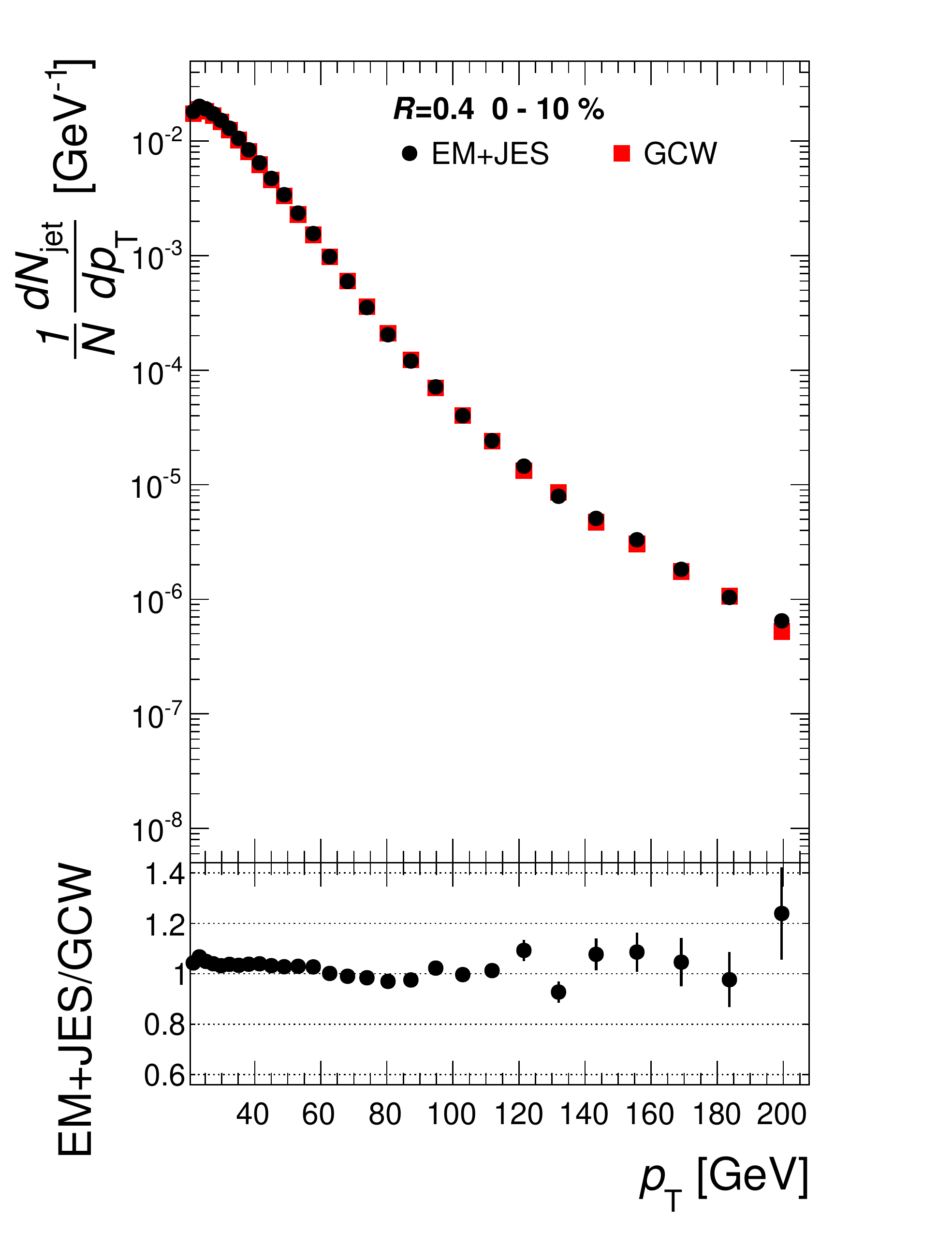}
\includegraphics[width =0.49\textwidth] {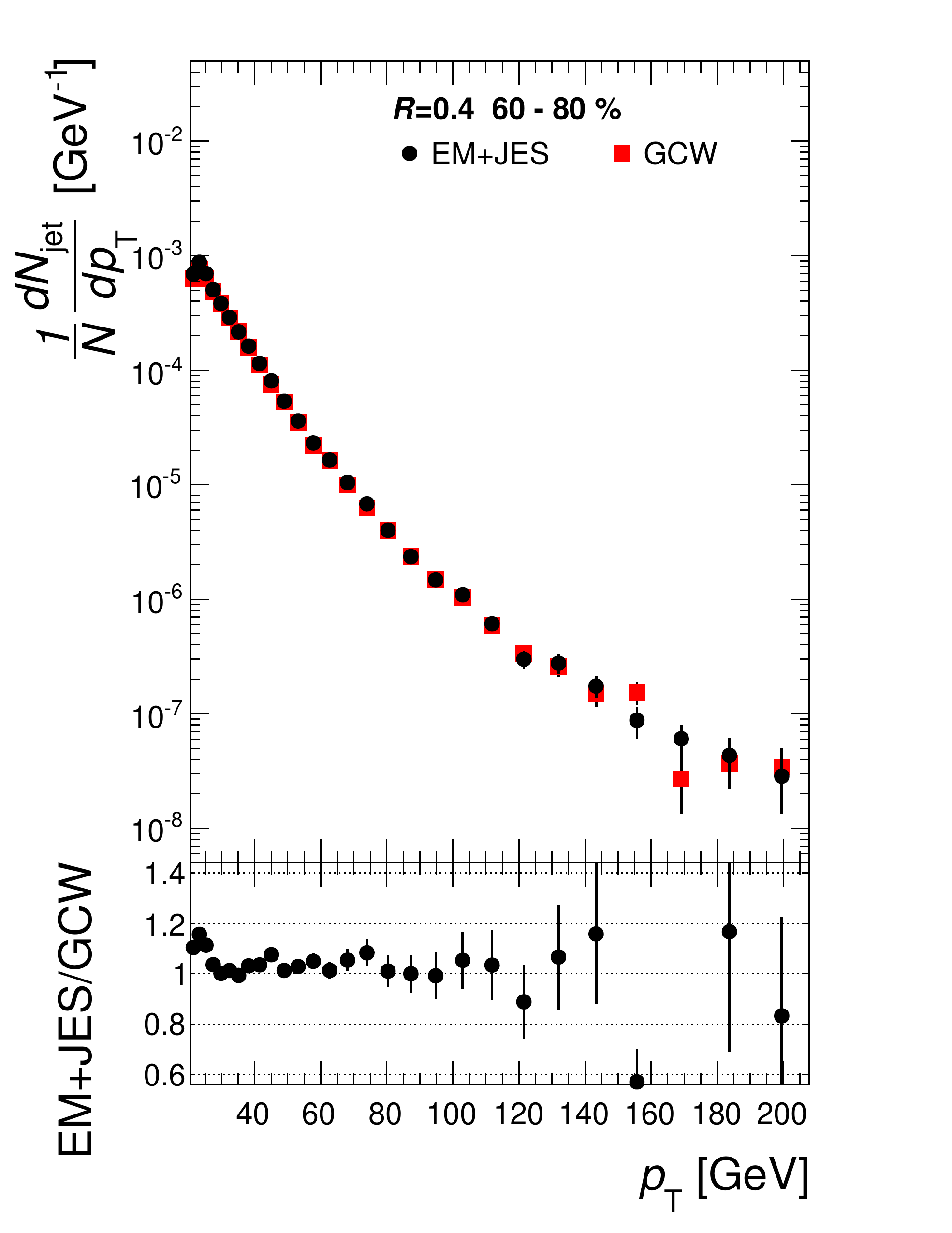}
\caption{\pt\ spectra for EM+JES (black) and GCW (red) calibration
  schemes for~\RFour\ jets and the ratio in central (left) and
  peripheral (right) collisions. Error bars in the ratio are
  constructed by considering which spectrum has the largest relative
  error and assuming the ratio has the same relative error.}
\label{fig:validation:calibration_comparison}
\end{figure}

\subsubsection{Track Jet Matching}
\label{sec:validation:JES:track_jet}

One of the methods to validate the performance of the jet finding, in particular the jet energy scale, is to match the calorimeter jets 
to track jets and compare their momenta. Such a comparison can be done
both in the MC and in the data. Calorimeter jets were matched to track
jets with $\pt> 7$~\GeV, with a matching requirement of $\Delta R < 0.2$ between the calorimeter and track jet axes. Fig.~\ref{fig_InsituJES_scatter} shows the calorimeter jet 
\pt\ as a function of a corresponding track jet \pt\, both for the
data and MC for the 0-10\% and 60-80\% centralities. For calorimeter
jets with  $\pt\ > 50$~\GeV,the mean calorimeter jet \pt, $\langle p_{\mathrm{T}}^{\mathrm{calo}}
\rangle$,  was computed as a 
function of the track jet \pt and is overlaid on
top of the two-dimensional distributions.

\begin{figure}[htb]
\centering
\includegraphics[width =0.49\textwidth] {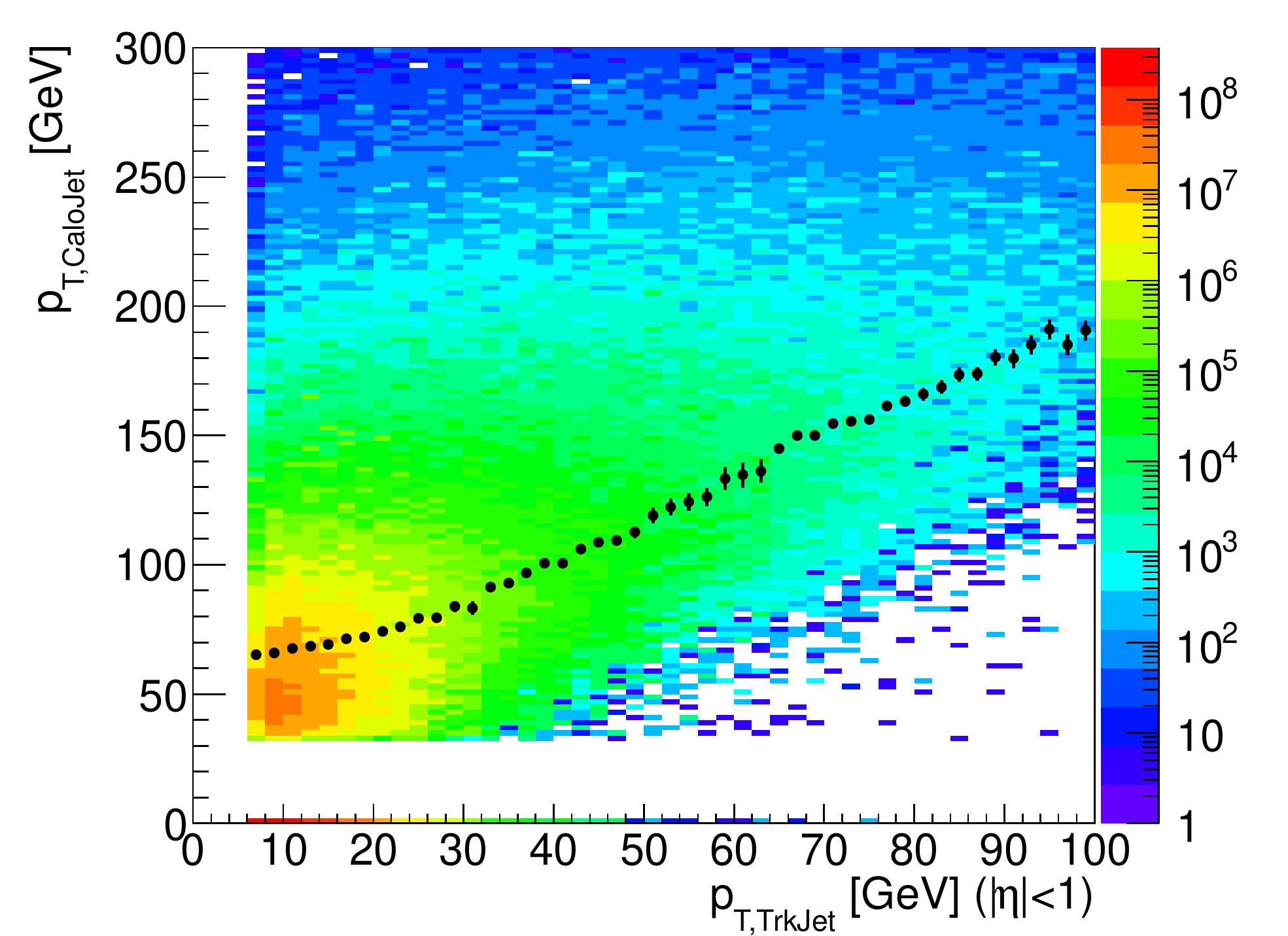}
\includegraphics[width =0.49\textwidth] {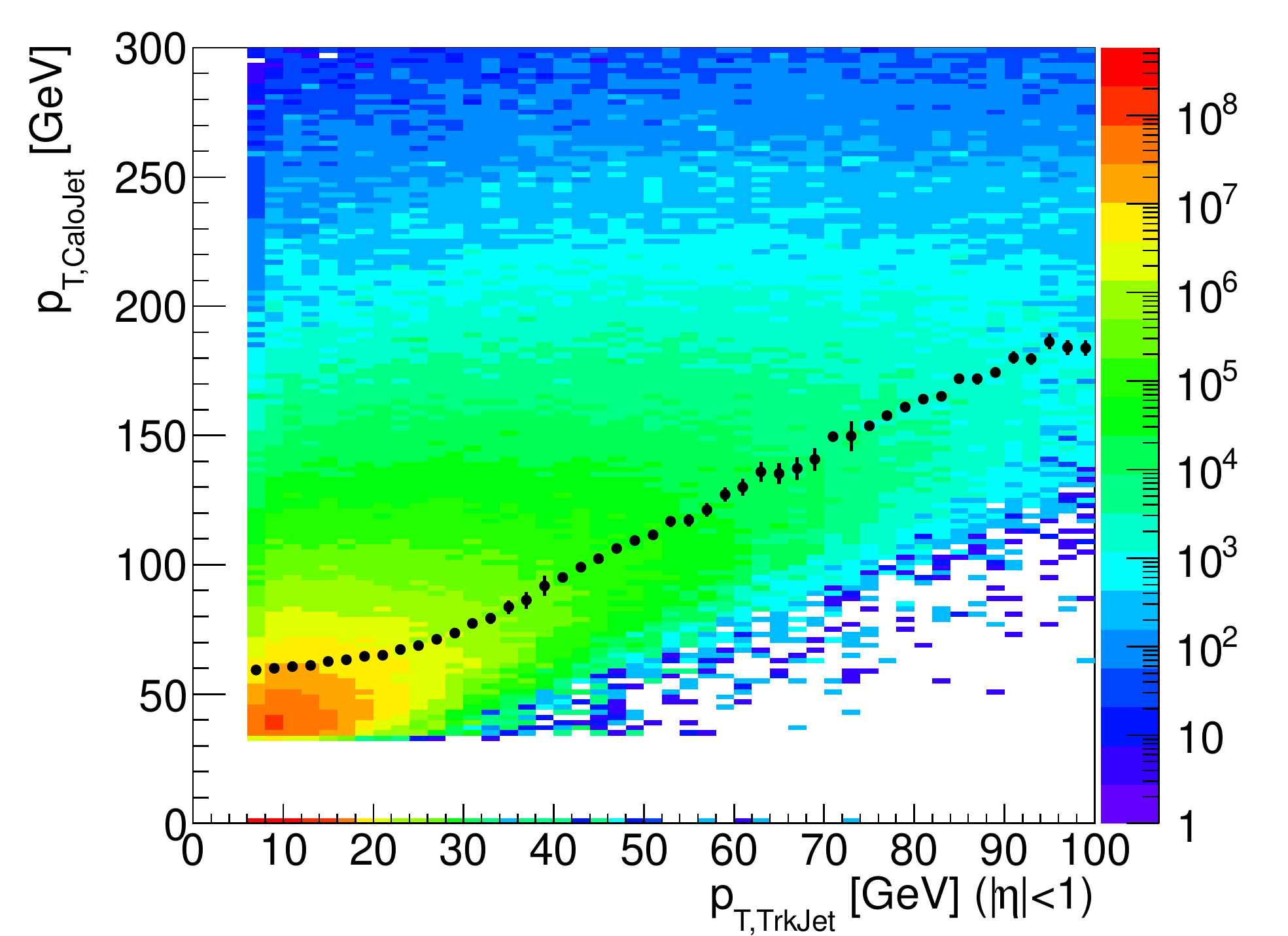}
\centering
\includegraphics[width =0.49\textwidth] {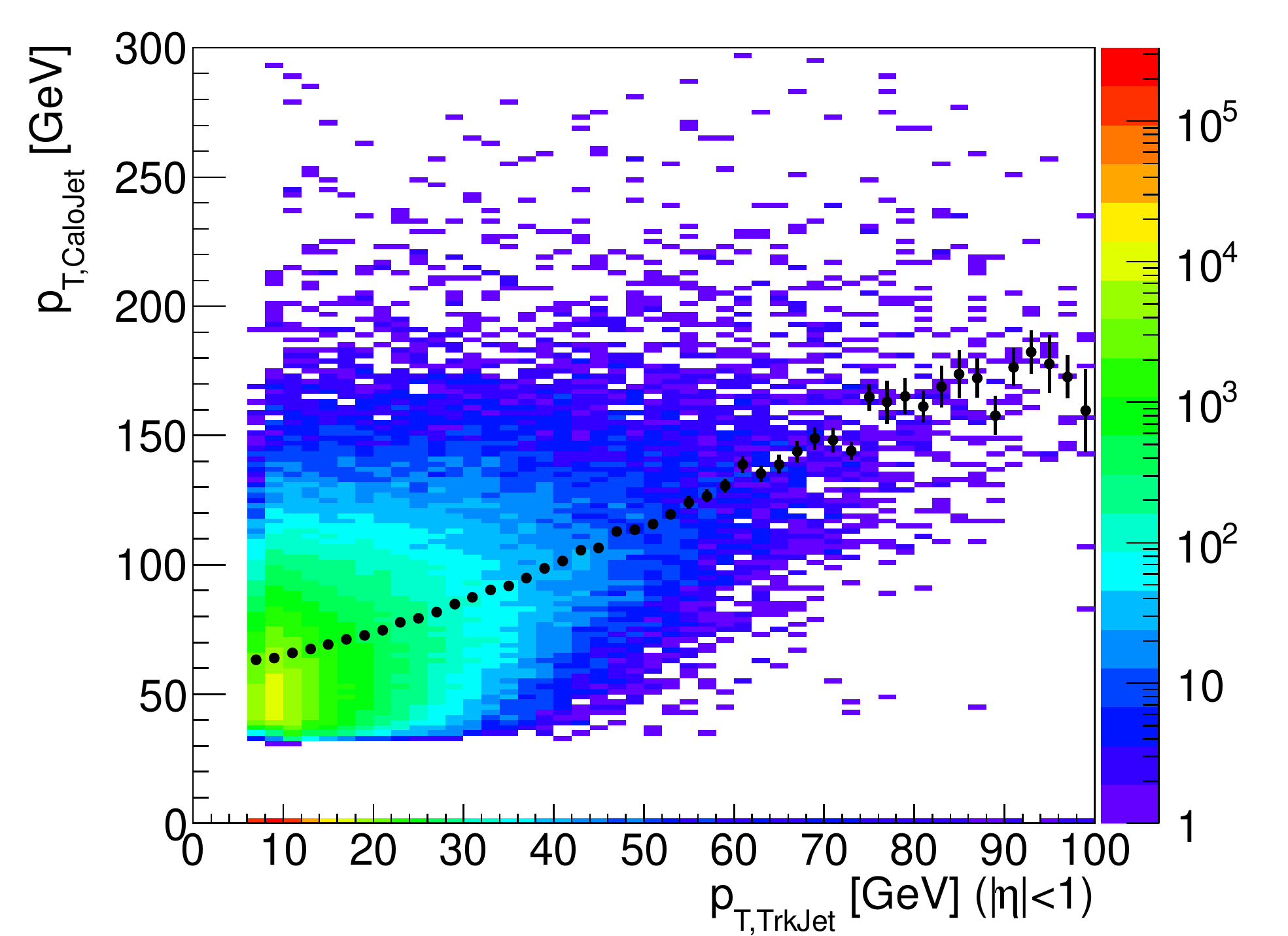}
\includegraphics[width =0.49\textwidth] {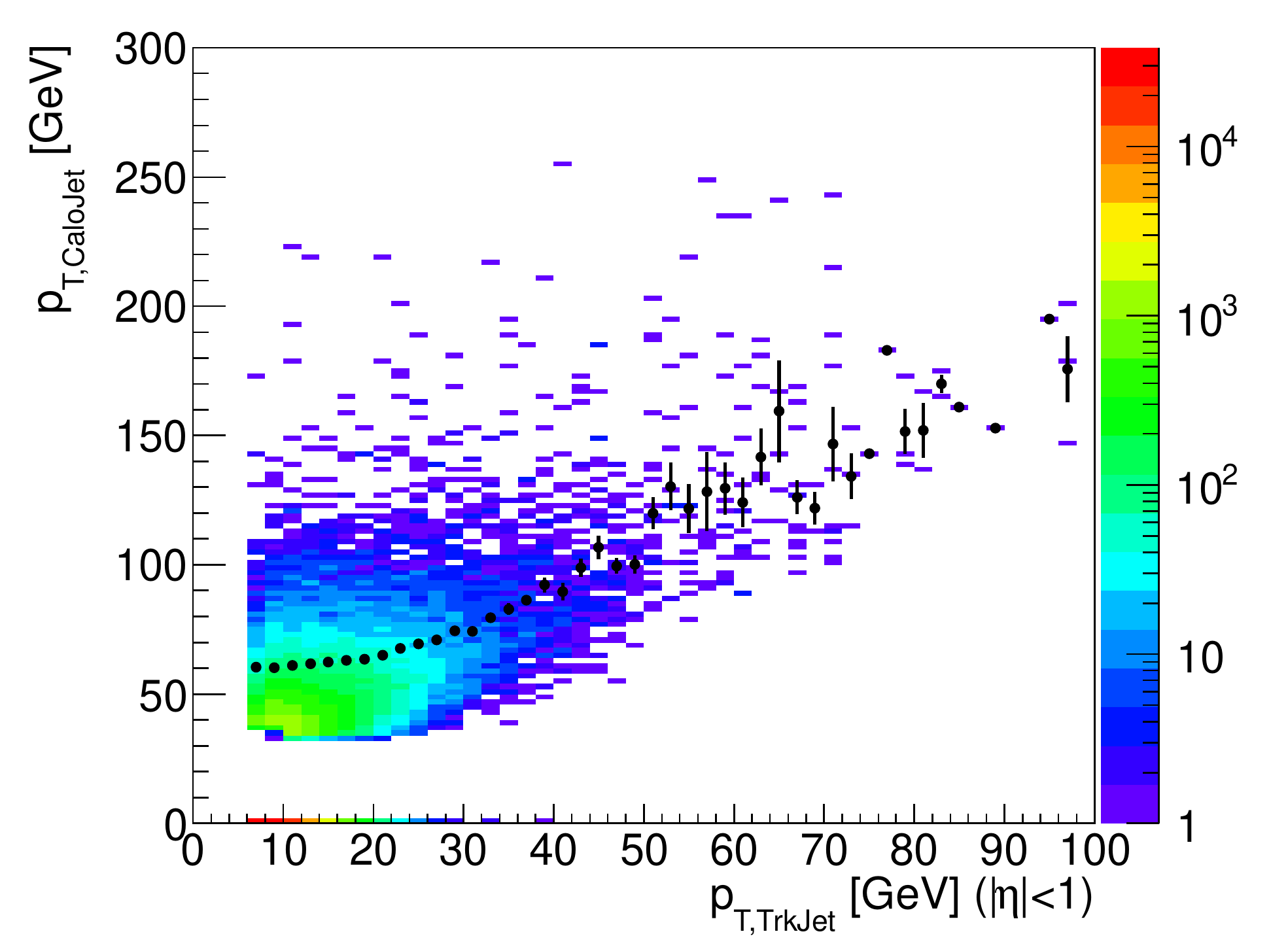}
\caption{
  Two-dimensional distribution of \pt\ of reconstructed calorimeter jets versus \pt\ of corresponding track jets for 0-10\% central (left) and 60-80\% 
peripheral (right) events for anti-\kt~\RFour\ jets. Upper plots show MC, lower plots show data. Markers indicate the average 
reconstructed calorimeter jet \pt\ as a function of the track jet \pt\ for calorimeter jets with $\pt\ > 50$~\GeV.
}
\label{fig_InsituJES_scatter}
\end{figure}

The centrality dependence of this energy scale was assessed directly
by constructing ratios of $\langle p_{\mathrm{T}}^{\mathrm{calo}} \rangle$ distributions in central and
peripheral collisions. Figure~\ref{fig:validation:insitu_CP} compares this ratio
as computed in data and MC. The ratios are roughly constant for track
jets with $\pt > 50$~\GeV, away from the minimum calorimeter jet \pt\
threshold, and each distribution is fit to a constant in this
region. The values and errors of these fit parameters are shown in Fig.~\ref{fig:validation:insitu_fit}
for different centrality bins. The differences in the fit constants
are evaluated directly in Fig.~\ref{fig:validation:insitu_err}. The comparison indicates that the
centrality dependence of the calorimetric response differs by no more
than 2\% between data and MC.
\begin{figure}[htb]
\centering
\includegraphics[width=0.45\textwidth]{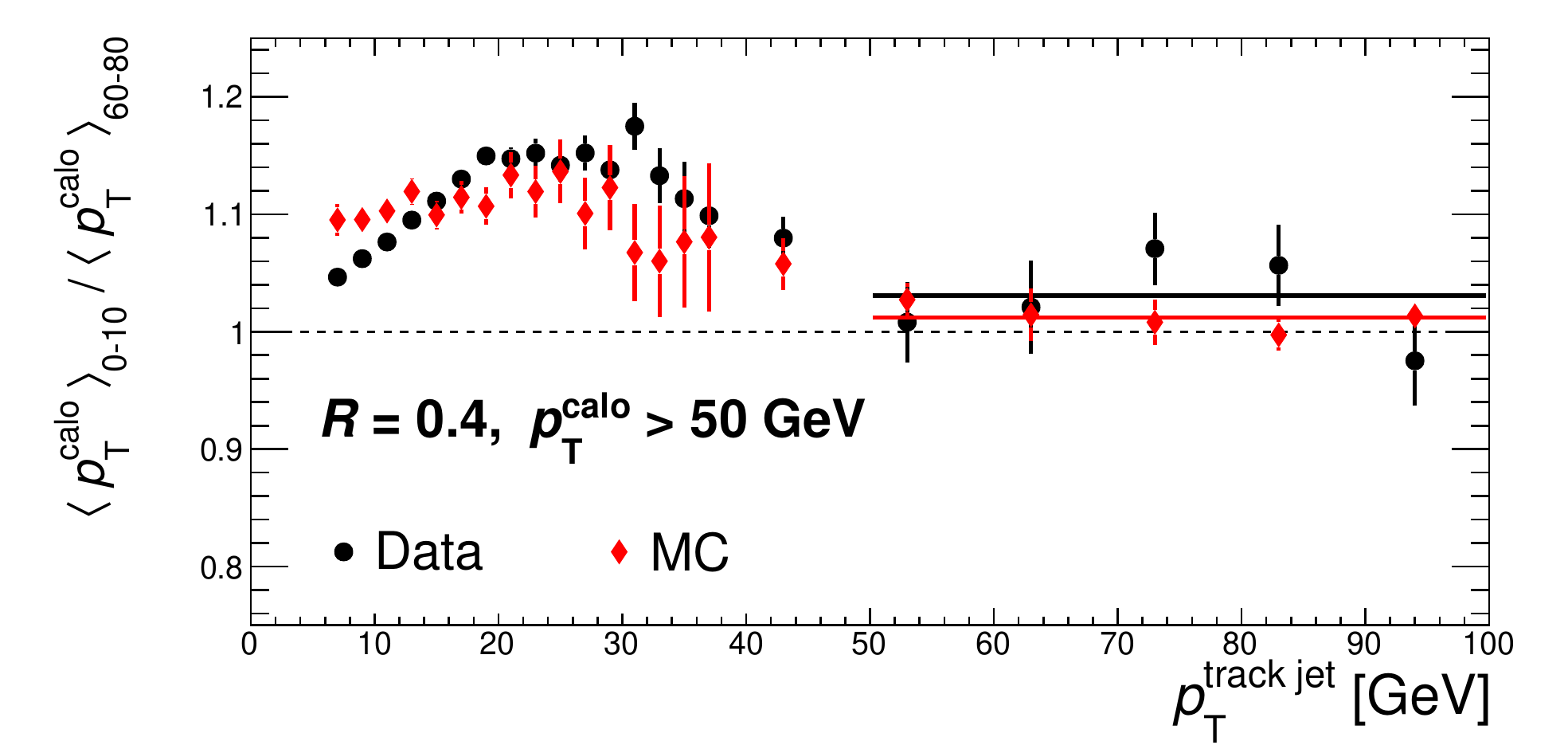}
\includegraphics[width=0.45\textwidth]{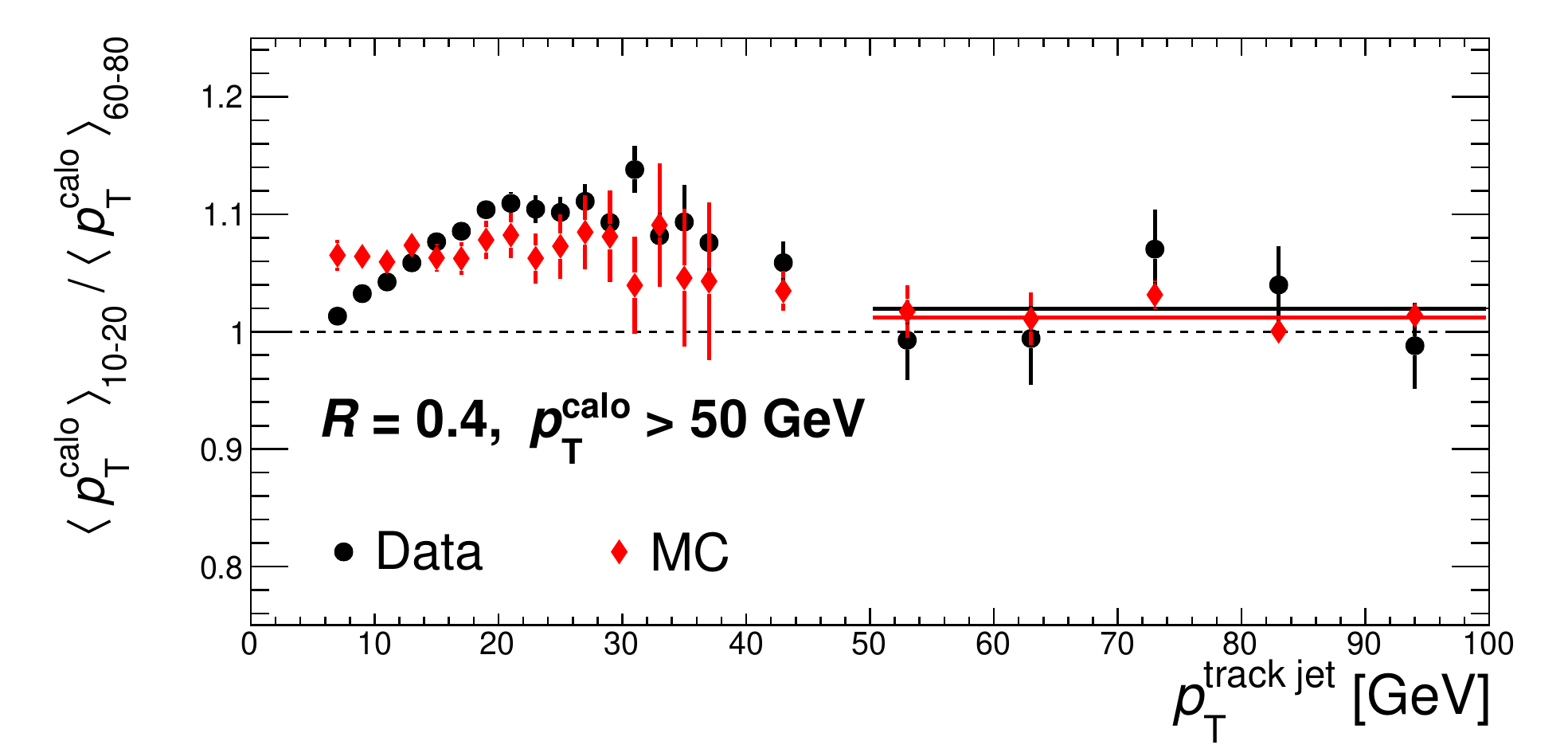}
\includegraphics[width=0.45\textwidth]{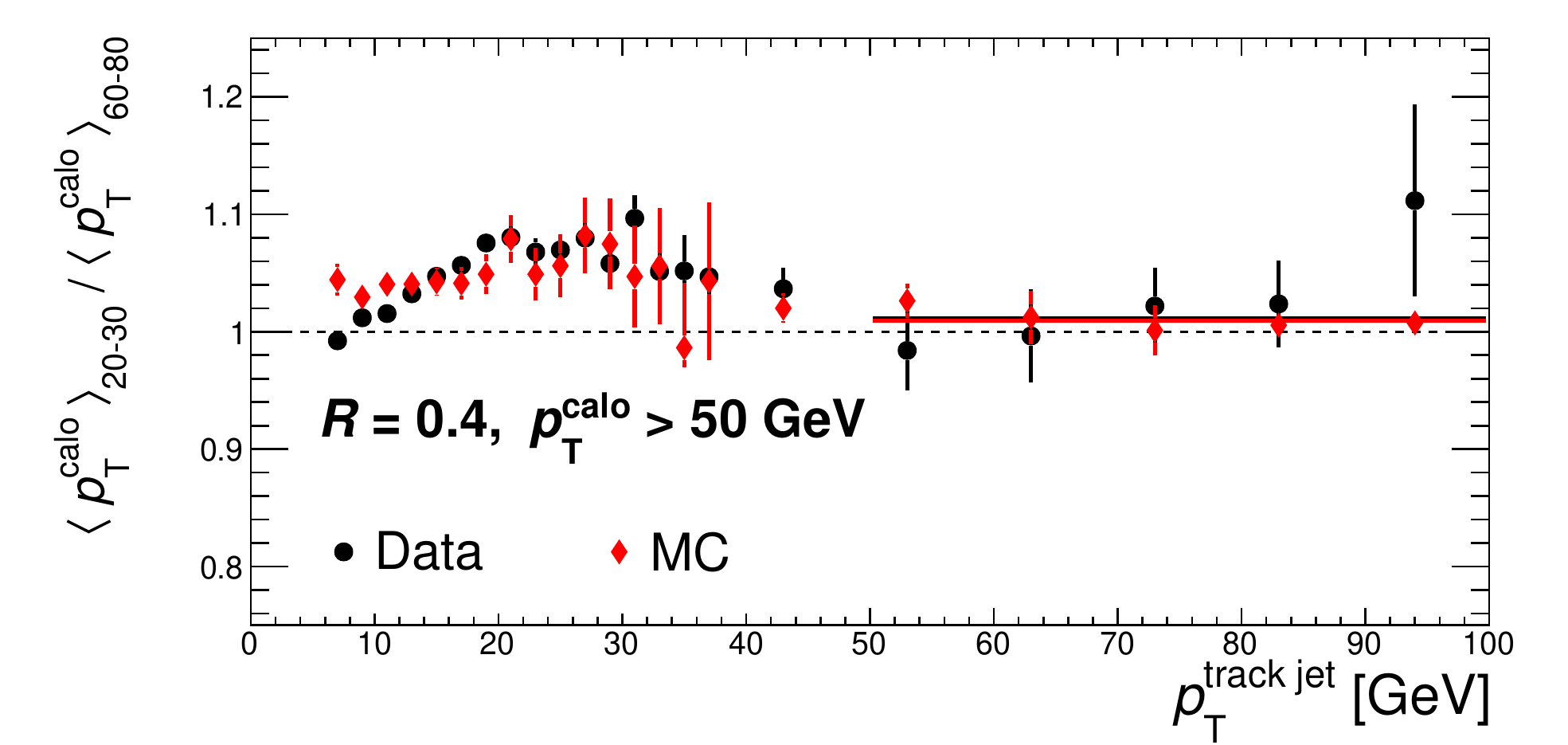}
\includegraphics[width=0.45\textwidth]{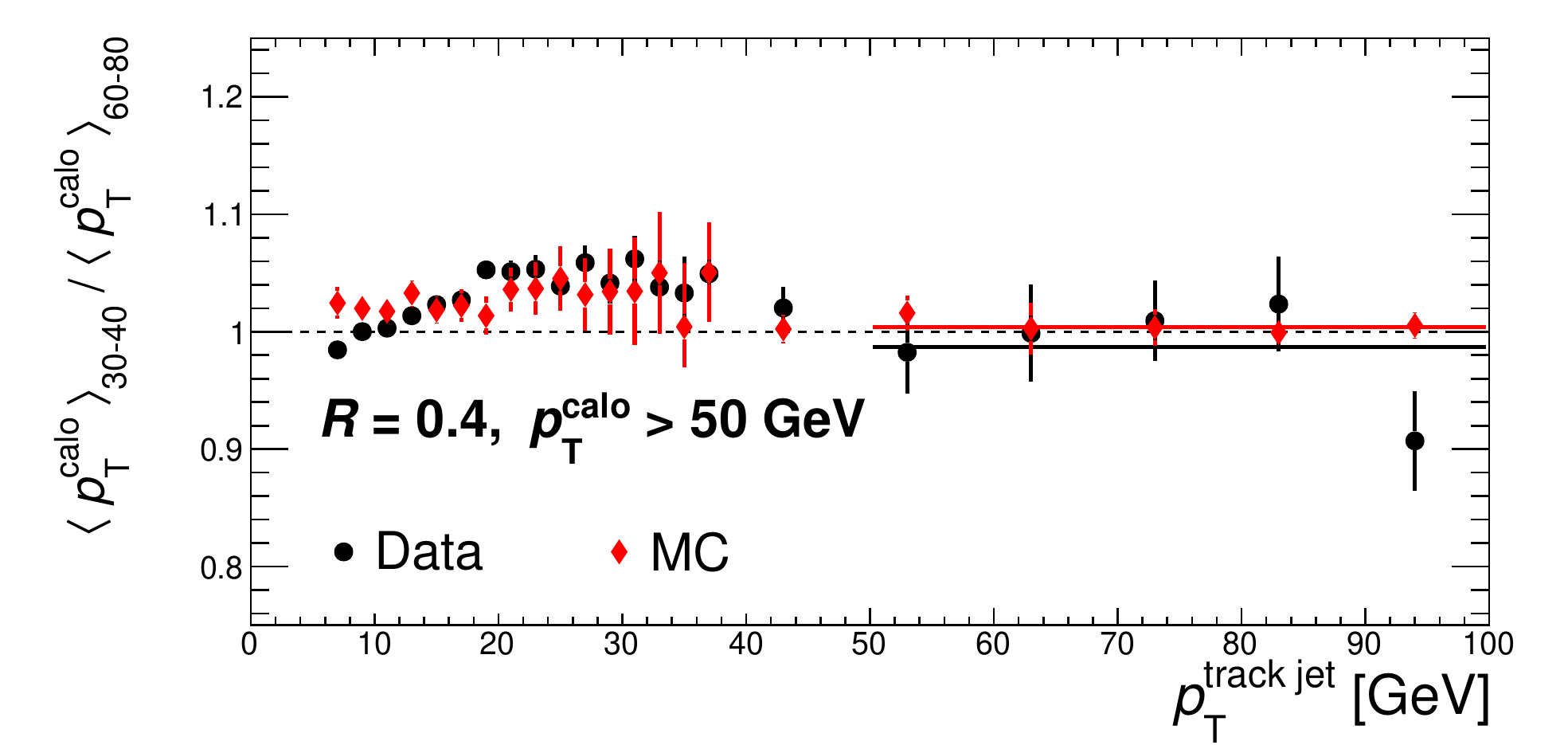}
\includegraphics[width=0.45\textwidth]{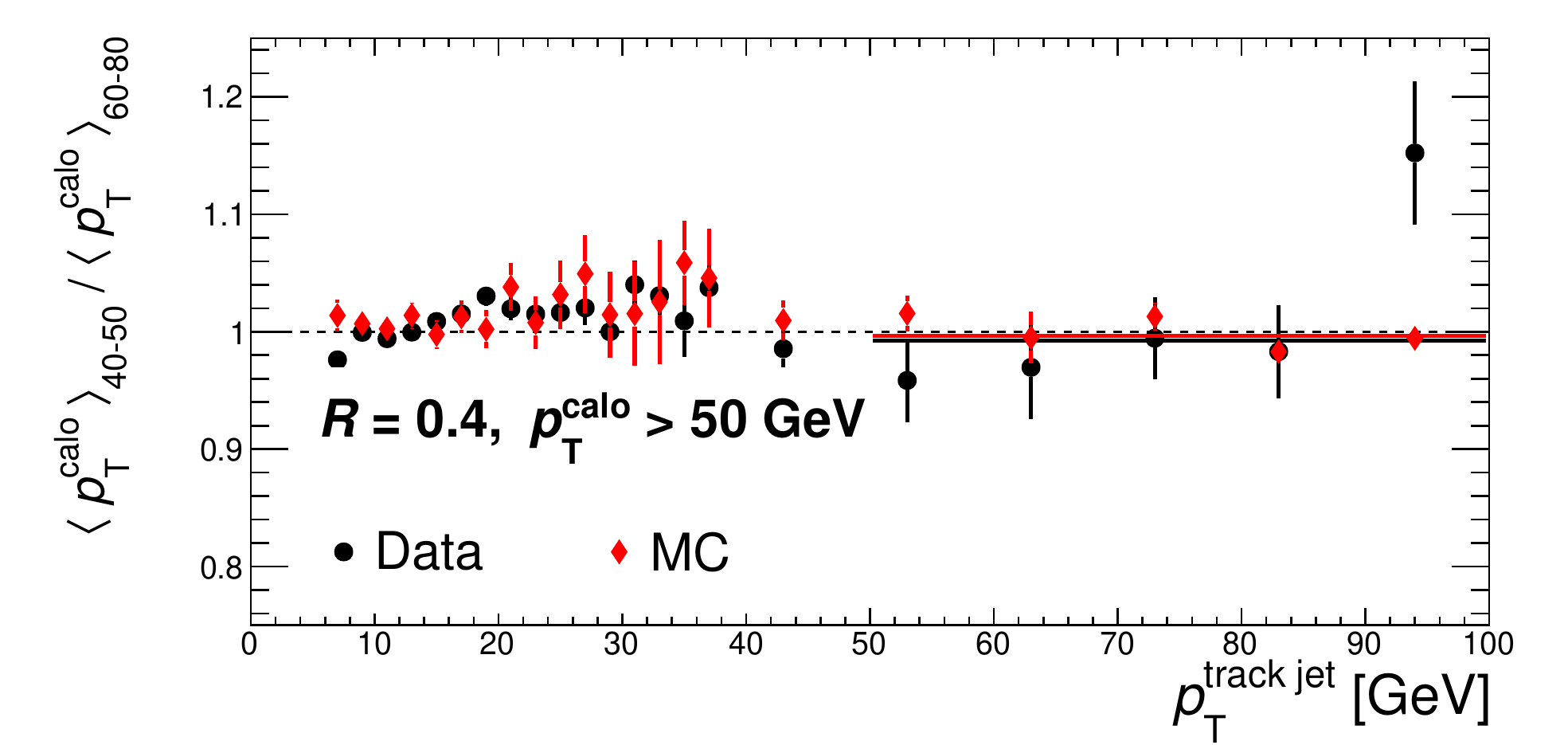}
\includegraphics[width=0.45\textwidth]{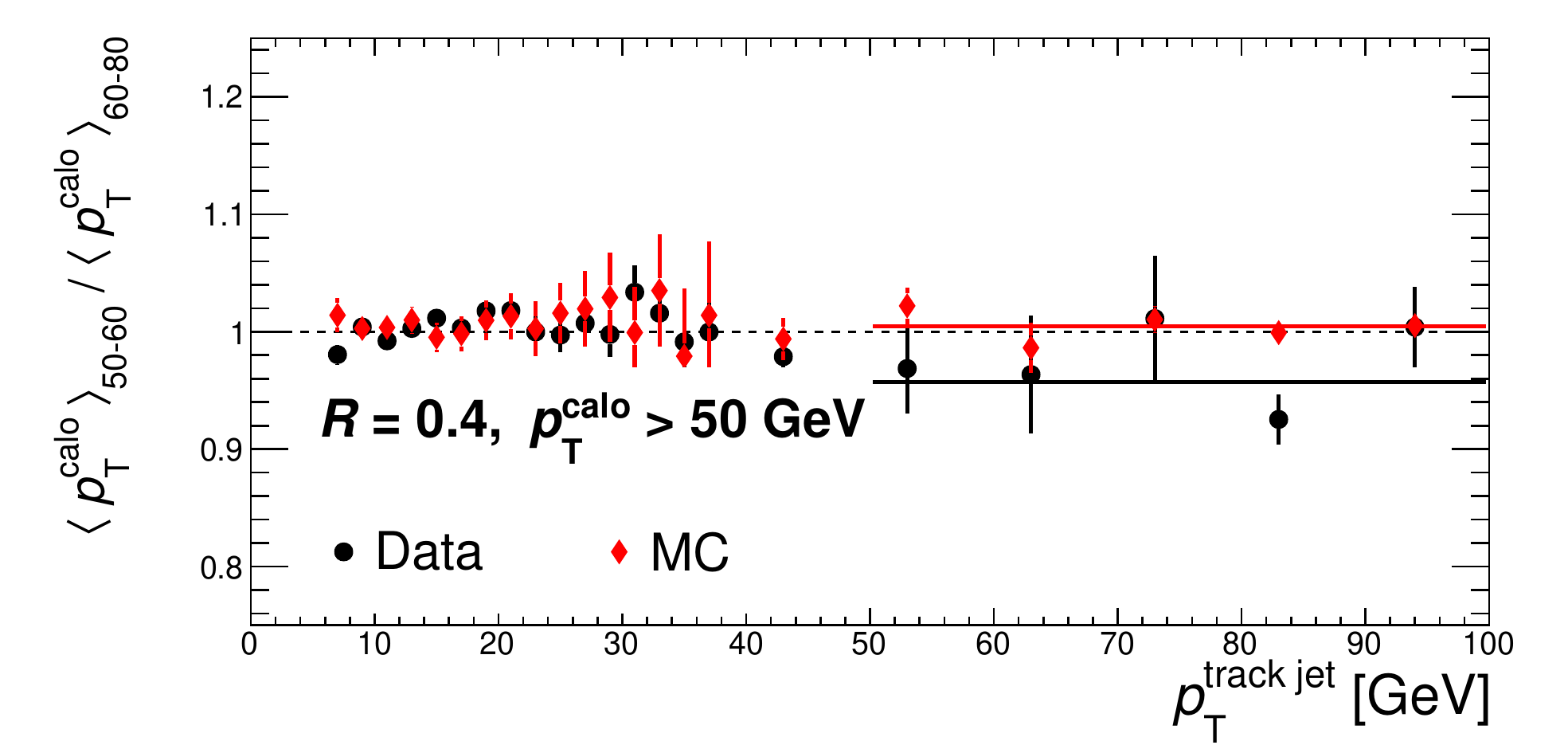}
\caption{The ratio $\langle p_{\mathrm{T}}^{\mathrm{calo}}
  \rangle_{\mathrm{cent}}/\langle p_{\mathrm{T}}^{\mathrm{calo}}
  \rangle_{60-80}$ as a function of $
  p_{\mathrm{T}}^{\mathrm{track}}$ for both data (black) and MC
  (red) for \RFour\ jets. Only jets with $p_{\mathrm{T}}^{\mathrm{calo}} > 50$~\GeV\
  are included in the average. A fit to a constant is shown for
  $p_{\mathrm{T}}^{\mathrm{track}} > 50$~\GeV.}
\label{fig:validation:insitu_CP}
\end{figure}
\begin{figure}[htb]
\centering
\includegraphics[width=0.49\textwidth]{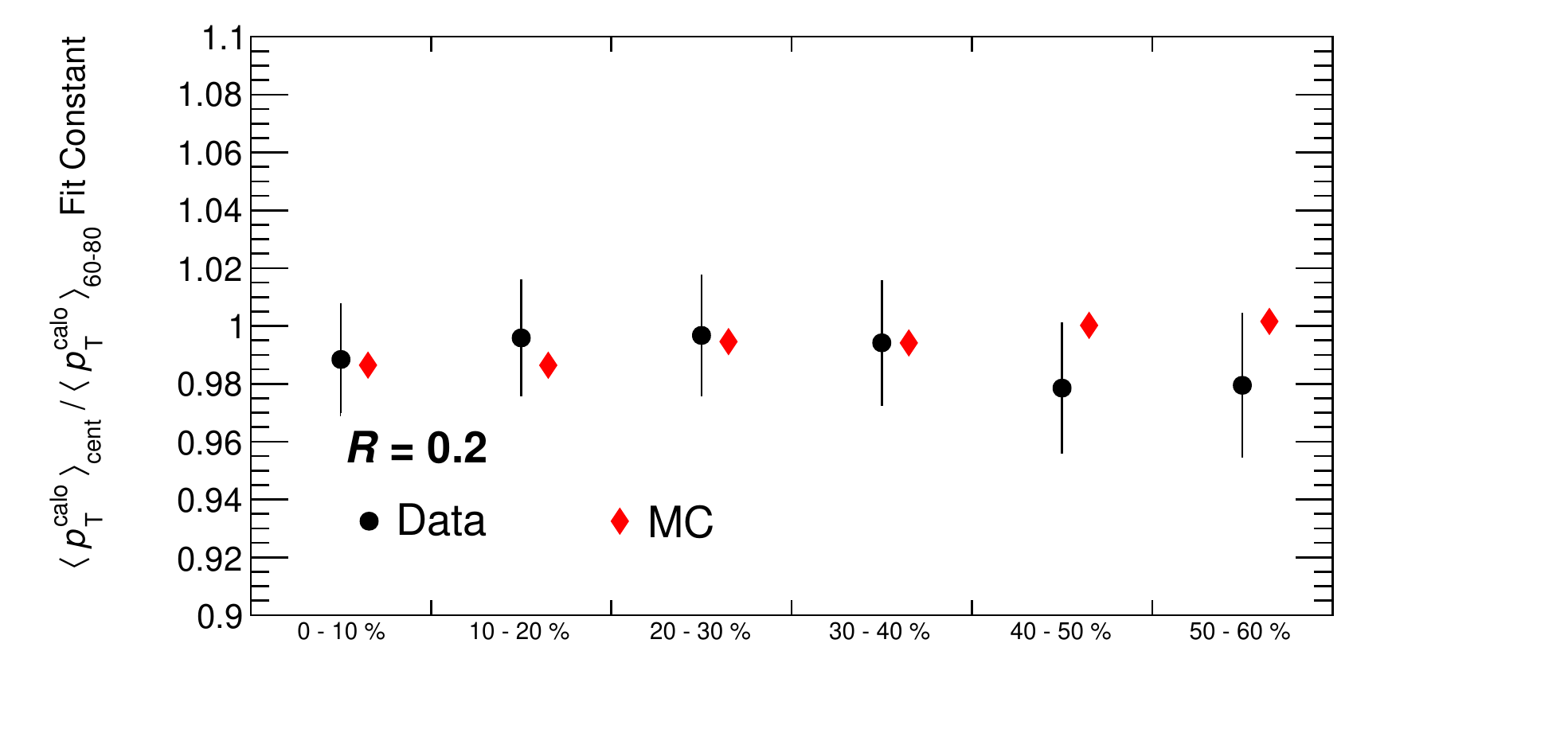}
\includegraphics[width=0.49\textwidth]{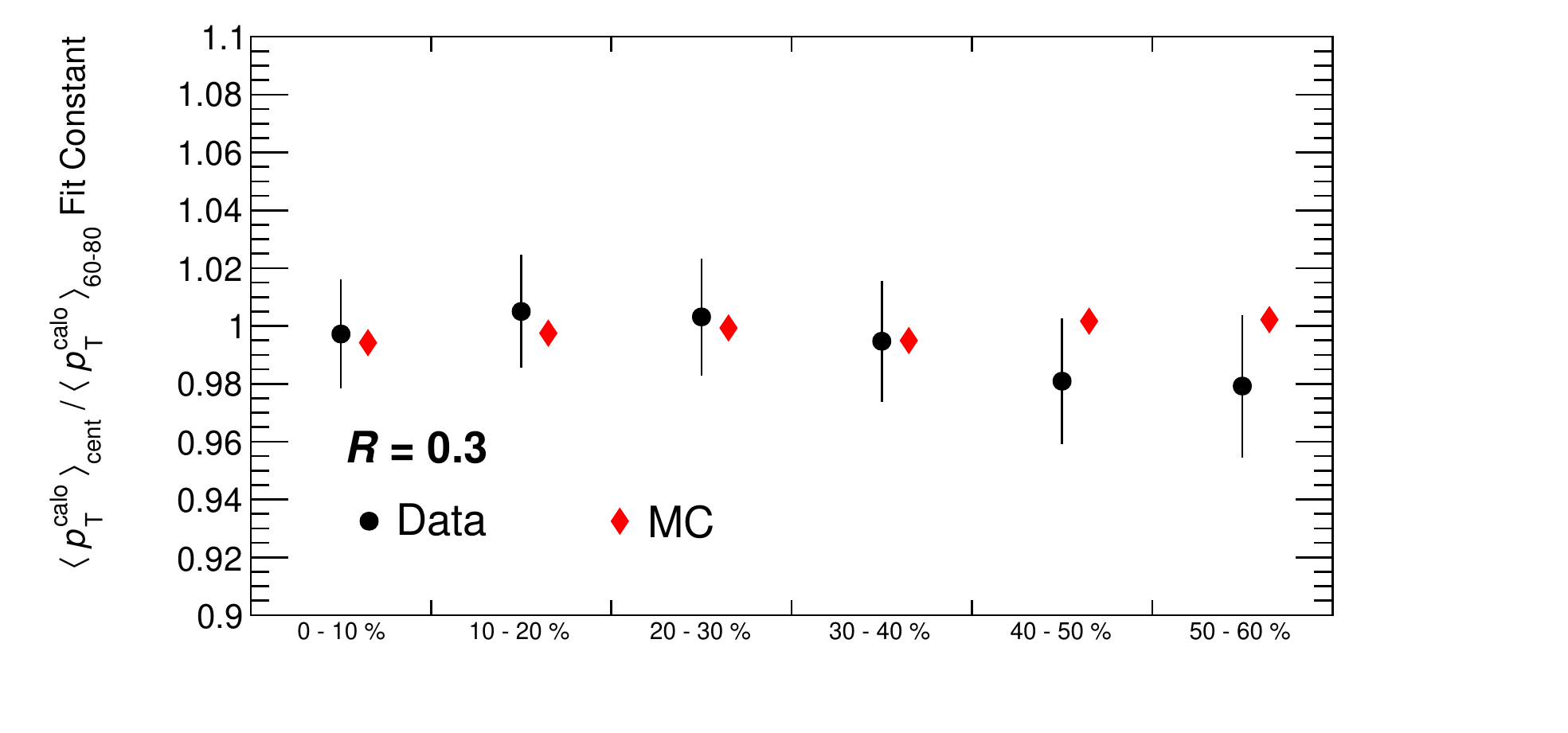}
\includegraphics[width=0.49\textwidth]{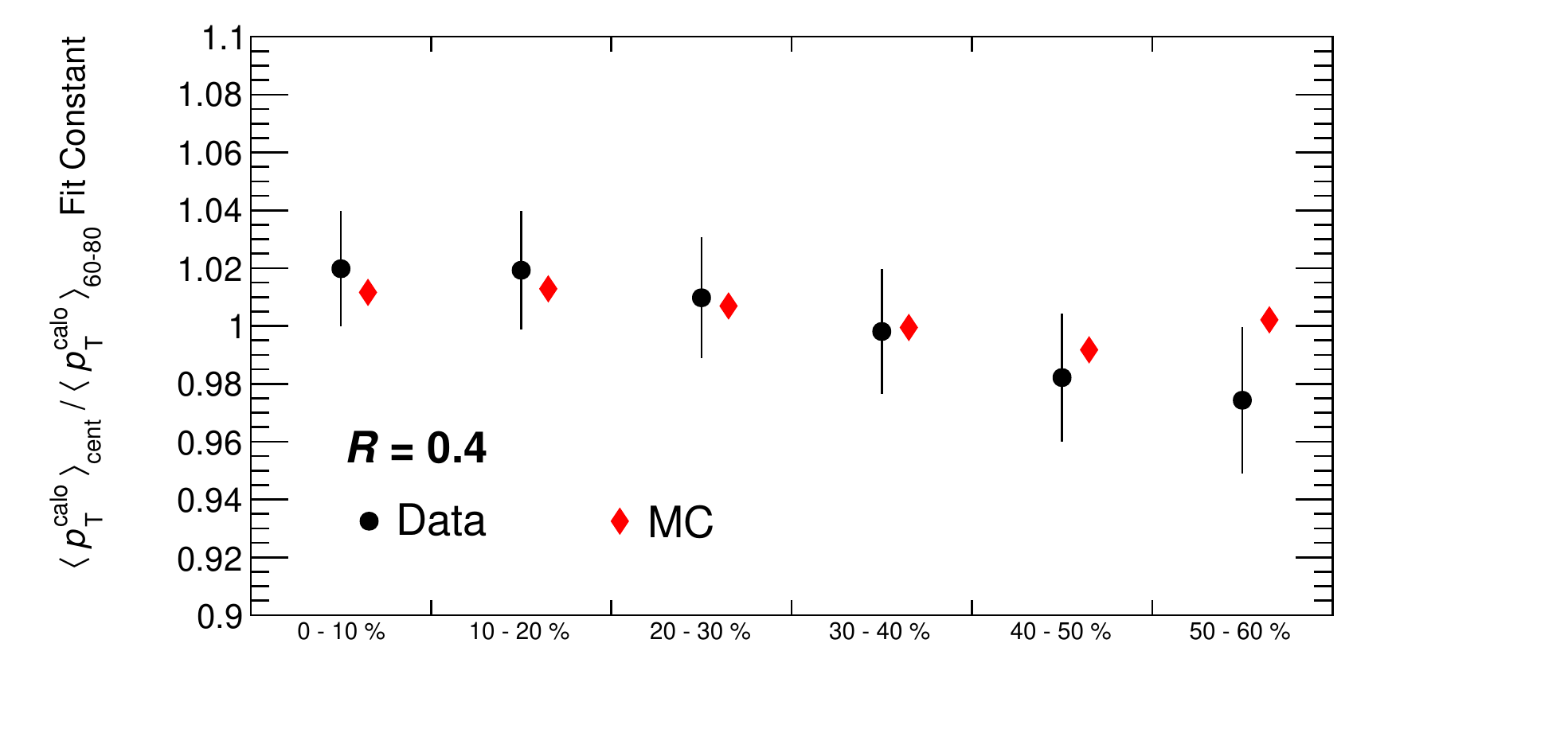}
\includegraphics[width=0.49\textwidth]{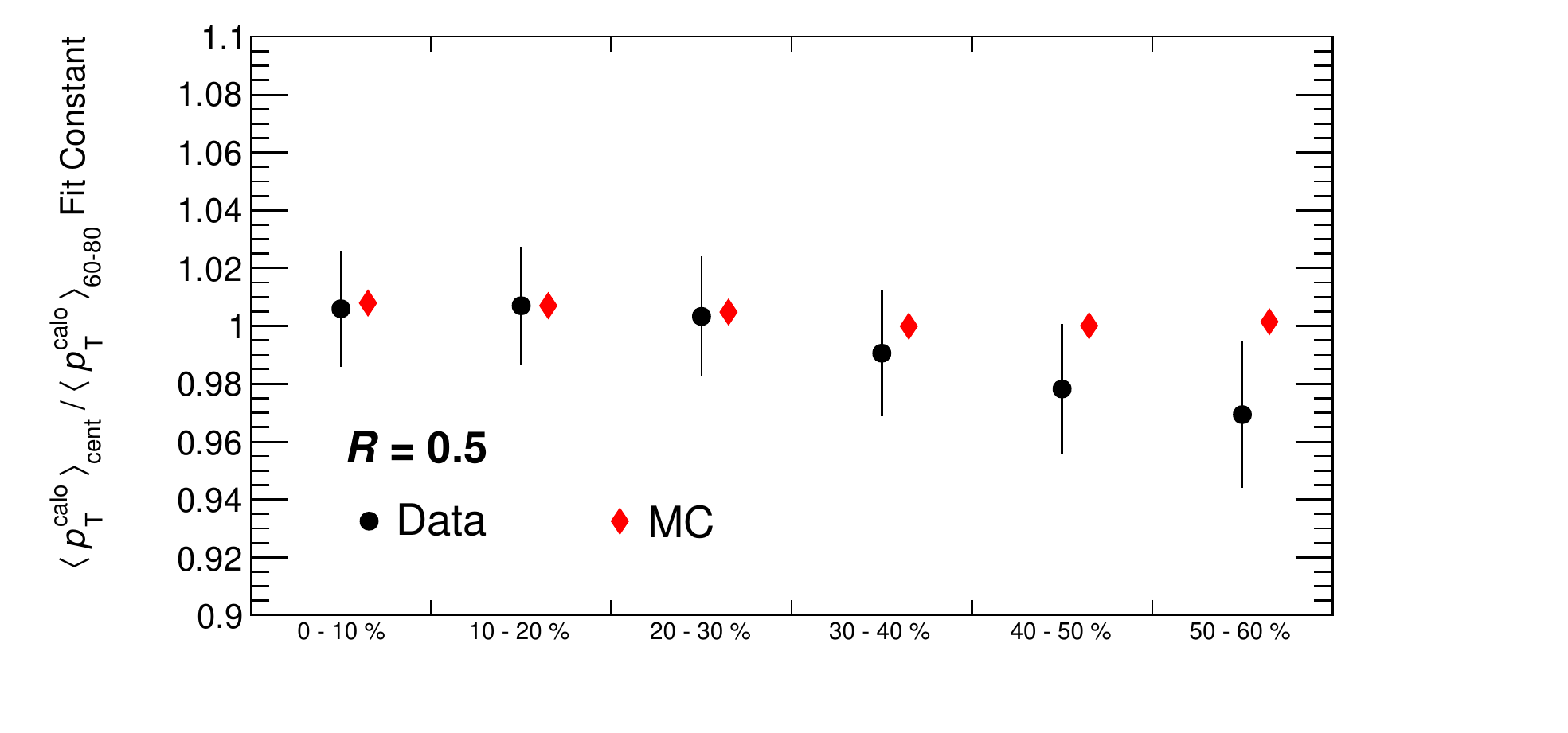}
\caption{Fit constants for $\langle p_{\mathrm{T}}^{\mathrm{calo}}
  \rangle_{\mathrm{cent}}/\langle p_{\mathrm{T}}^{\mathrm{calo}}
  \rangle_{60-80}$ for data (black) and MC (red) in different
  centrality bins are shown for different values of $R$.}
\label{fig:validation:insitu_fit}
\end{figure}
\begin{figure}[htb]
\centering
\includegraphics[width=0.49\textwidth]{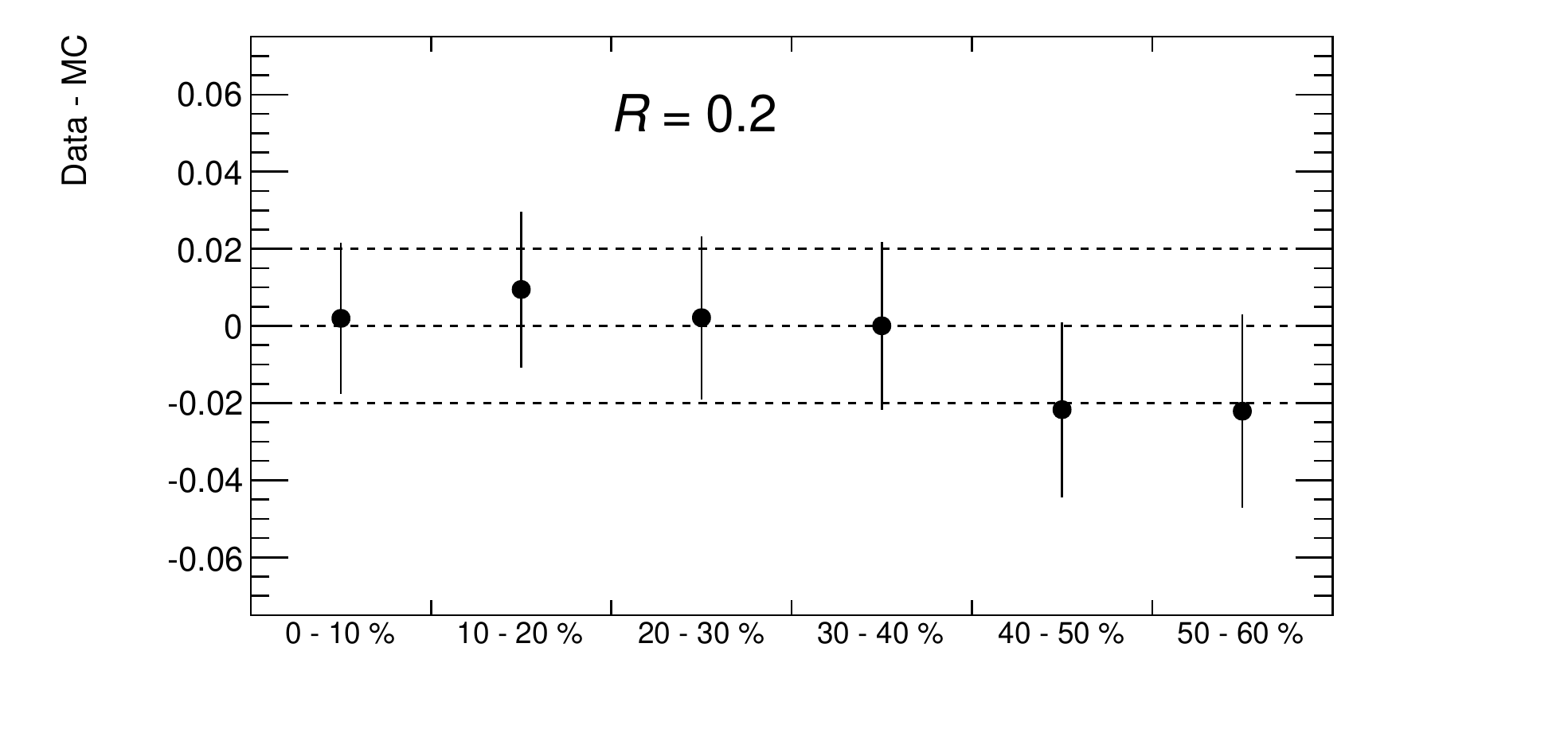}
\includegraphics[width=0.49\textwidth]{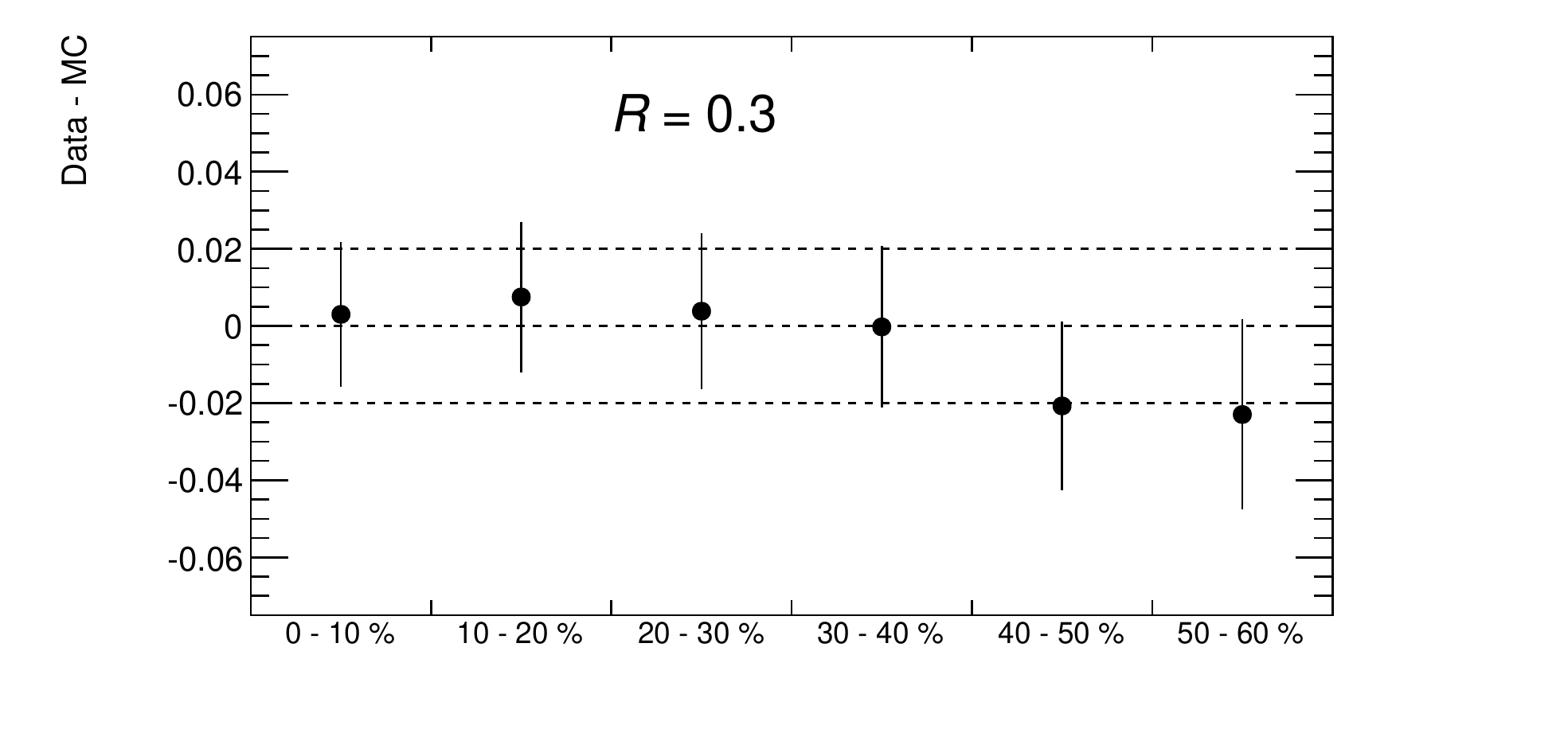}
\includegraphics[width=0.49\textwidth]{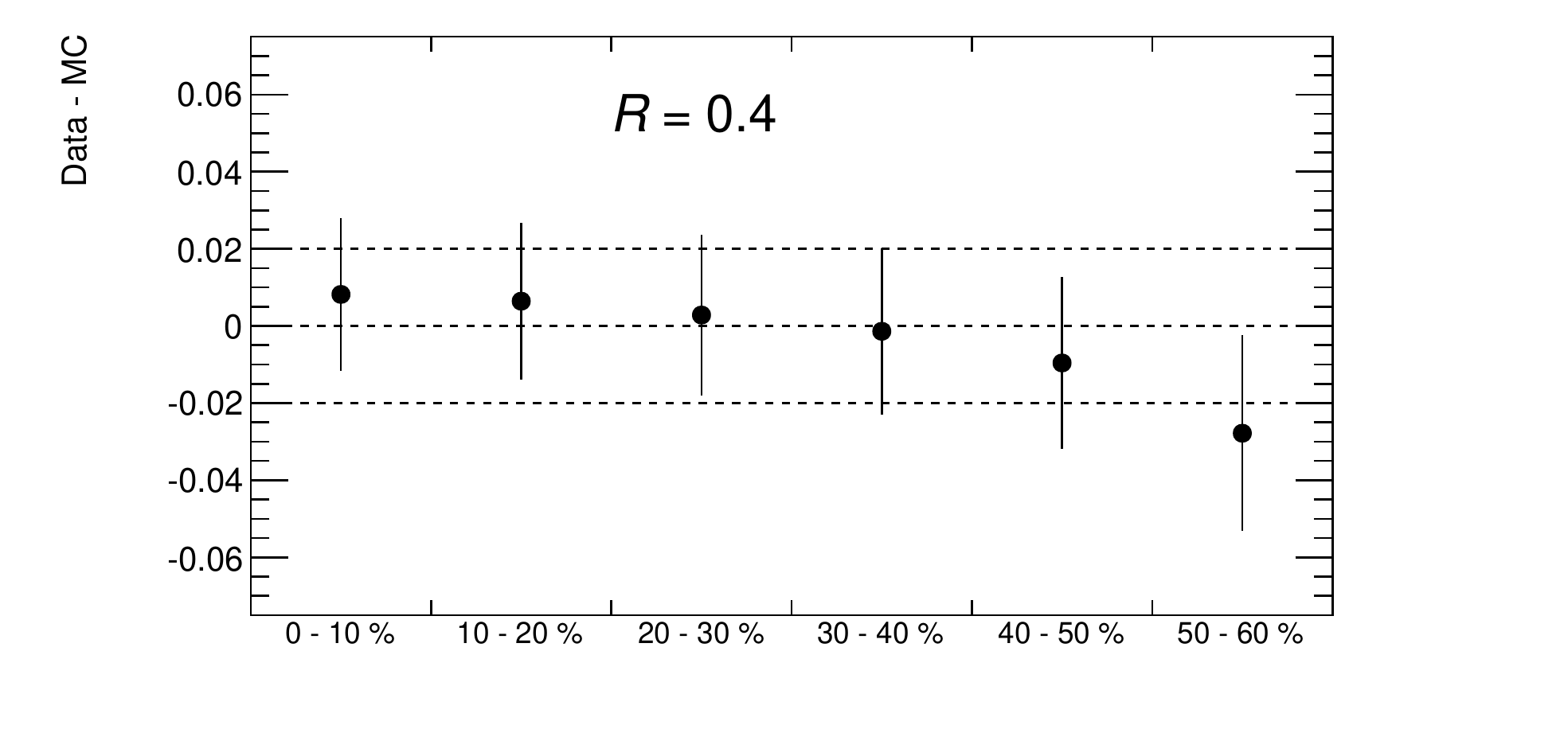}
\includegraphics[width=0.49\textwidth]{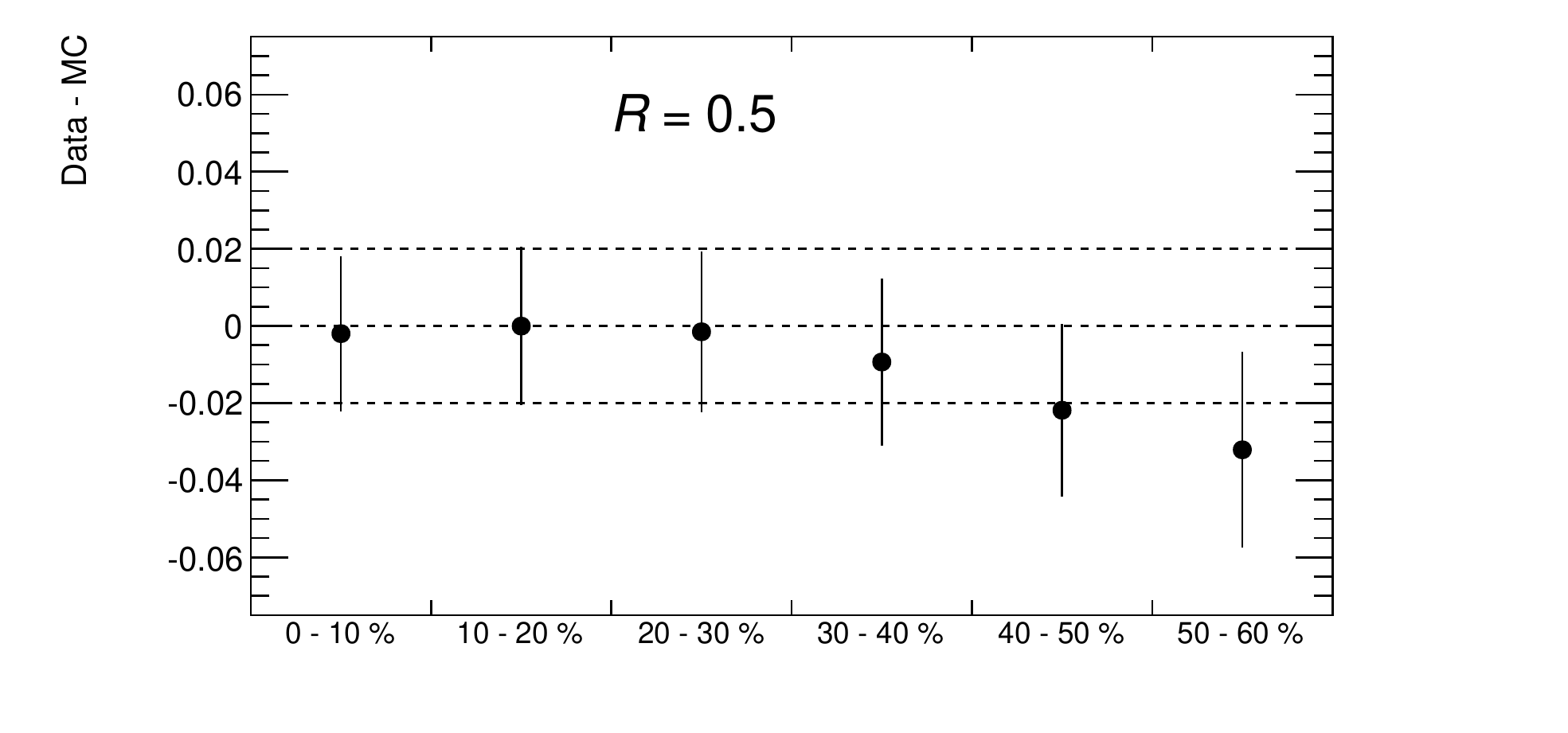}
\caption{The data-MC difference in constants extracted from $\langle p_{\mathrm{T}}^{\mathrm{calo}}
  \rangle_{\mathrm{cent}}/\langle p_{\mathrm{T}}^{\mathrm{calo}}
  \rangle_{60-80}$ fit for different $R$ values. Lines at $\pm2\%$ are
  shown for comparison.}
\label{fig:validation:insitu_err}
\end{figure}
The jet energy scale calibration is sensitive to the particle
composition and fragmentation of the jet. In particular it was found
in \pp\ events that the response was lower for broader jets~\cite{Aad:2011he}. One
concern is that quenching effects could result in broader jets
introducing a centrality dependence to the JES: a lower response in central collisions relative to
peripheral. However, such an effect is not consistent with the results of the in-situ
study as there is no evidence for the $\langle p_{\mathrm{T}}^{\mathrm{calo}}
  \rangle_{\mathrm{cent}}/\langle p_{\mathrm{T}}^{\mathrm{calo}}
  \rangle_{60-80}$ to be systematically below one.
\clearpage
\subsection{Jet Energy Resolution}
\label{sec:validation:JER}
\subsubsection{Fluctuations}
\label{sec:validation:JER:fluctuations}
To evaluate the accuracy of HIJING in describing the underlying event fluctuations seen in the data, 
the per-event standard deviation of the summed \ET\ for all $N\times M$ groups of 
towers has been evaluated. The numbers $M$ and $N$ were chosen so that
the area of the group of towers had approximately the same nominal
area as jets reconstructed with a given distance parameter $R$, and
values are given in Table~\ref{tbl:fluctuations}. A complete systematic study is presented
Ref.~\cite{ATLAS-CONF-2012-045}; examples of that study are shown here.
Figure~\ref{fig:jetetrms} shows the event-averaged standard deviation as a function of 
the \ETfcal\ in the event for $7\times7$ groups of towers, which have
approximately the same area as \RFour\ jets. The same quantity is
shown in both data and MC, with the MC \ETfcal\ scaled up by 12.6\% to 
match the energy scale of the measured distribution. This procedure is
equivalent to matching data to MC using fractions of the minimum-bias
cross-section. The re-scaled MC results in 
Fig.~\ref{fig:jetetrms} agree reasonably well with the results from the data over the entire\ETfcal\ 
range.
  \begin{figure}
\centerline{
\includegraphics[width=0.6\textwidth]{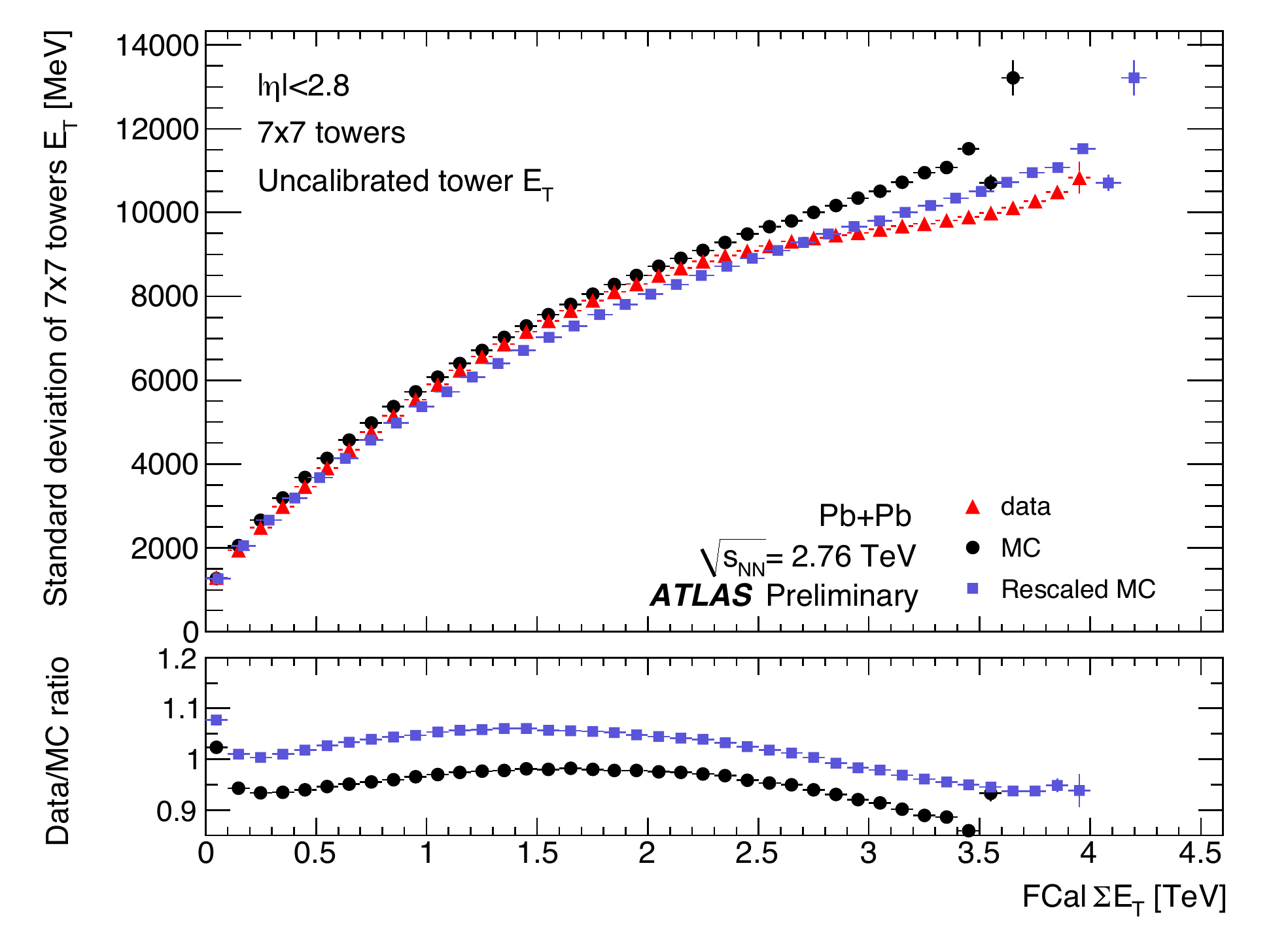}
}
\caption{Comparison of the per-event standard deviation of summed
  \ET\ for $7\times7$ groups of towers between data and the
  HIJING+GEANT MC simulated events as a function of FCal
  \sumet.}
\label{fig:jetetrms} 
\end{figure} 
Based on this data-driven test of the MC, it can be concluded that the HIJING+GEANT simulations of underlying event fluctuations can differ from the data by at 
most 10\%.

\begin{table}[h]
\centerline{
\begin{tabular}{|c||c|c|c|c|} \hline
$R$ & 0.2& 0.3&0.4&0.5 \\ \hline
$N \times M$  & $3 \times 4$ & $5 \times 5$ & $7 \times 7$  & $9 \times 9$ \\ \hline 
\end{tabular}
}
\caption{Correspondence between the size of a jet and the size of groups of towers.}
\label{tbl:fluctuations}
\end{table}

\subsubsection{Jet Energy Resolution Validation}
\label{sec:validation:JER:closure}

The knowledge of the size of the underlying event fluctuations can be used to validate the internal consistency of 
the jet energy resolution (JER) estimates. The jet energy resolution can be expressed in the following form

\begin{equation} 
\sigma(\et)/\et = a/\sqrt{\et} \oplus b/\et \oplus c \,,
\label{eqn:jer} 
\end{equation} 
referred to as the stochastic, noise and constant terms
respectively. The stochastic term represents fluctuations in the
calorimeter shower sampling while the constant term relates to effects
with $\Delta \ET\sim\ET$ such as dead regions. For the intrinsic energy resolution $\sigma(\et)$, the noise does not depend on the 
energy of incident particles and scales as $1/\et$ for the relative energy 
resolution $\sigma(\et)/\et$. The constant $b$ that quantifies the size of this term is the standard 
deviation of the noise energy and was evaluated in the fluctuation 
analysis described in
Sec.~\ref{sec:validation:JER:fluctuations}. The JER was fit with
the functional form of Eq.~\ref{eqn:jer} with $b$ taken from the
fluctuation analysis and $a$ and $c$ determined by the fit, and a sample of these
fits is shown in Fig.~\ref{fig:jetjerfit}. The $b$
values were adjusted slightly to account for the small differences in
area between the rectangular tower groups and the nominally circular jet size,
with area $\pi R^2$. The fit
parameters $a$ and $c$ are found to be independent of centrality,
confirming that the centrality dependence of the JER is related to the
underlying event fluctuations in a quantifiable way. The noise term extracted
from the fluctuation analysis is shown in the left panel of 
Fig.~\ref{fig:bterm} as a function of the jet area for different
centralities, along with a fit~\footnote{Although not directly relevant to jet performance,
  for all the centralities a fit value of $p_1 \approx 
0.59$ was obtained. This hints at some structure in the fluctuations
that should be investigated} $ b(A) = p_0A^{p_1}$. 
\begin{figure}
\centerline{
\includegraphics[width=0.425\textwidth]{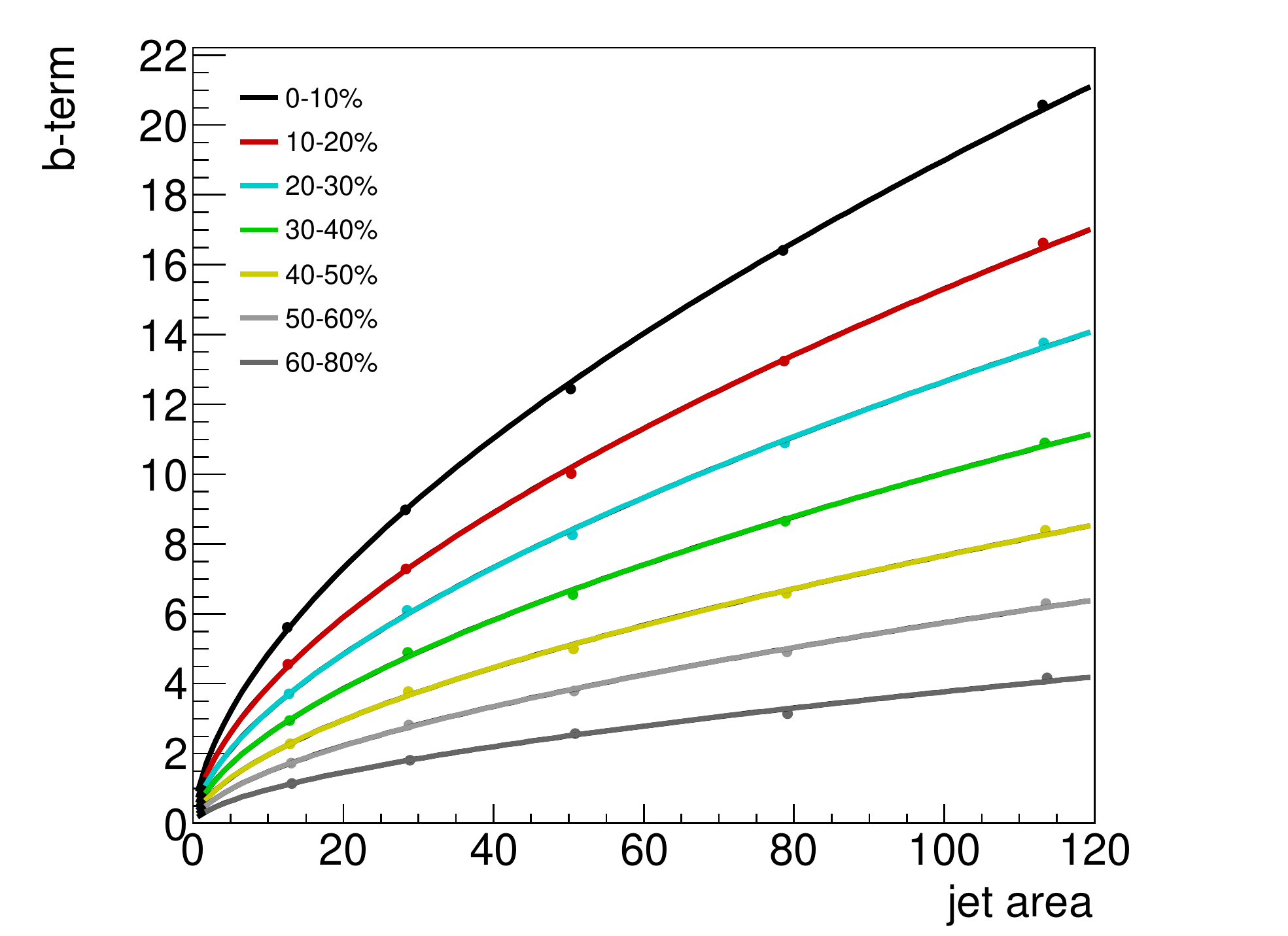}
\includegraphics[width=0.425\textwidth]{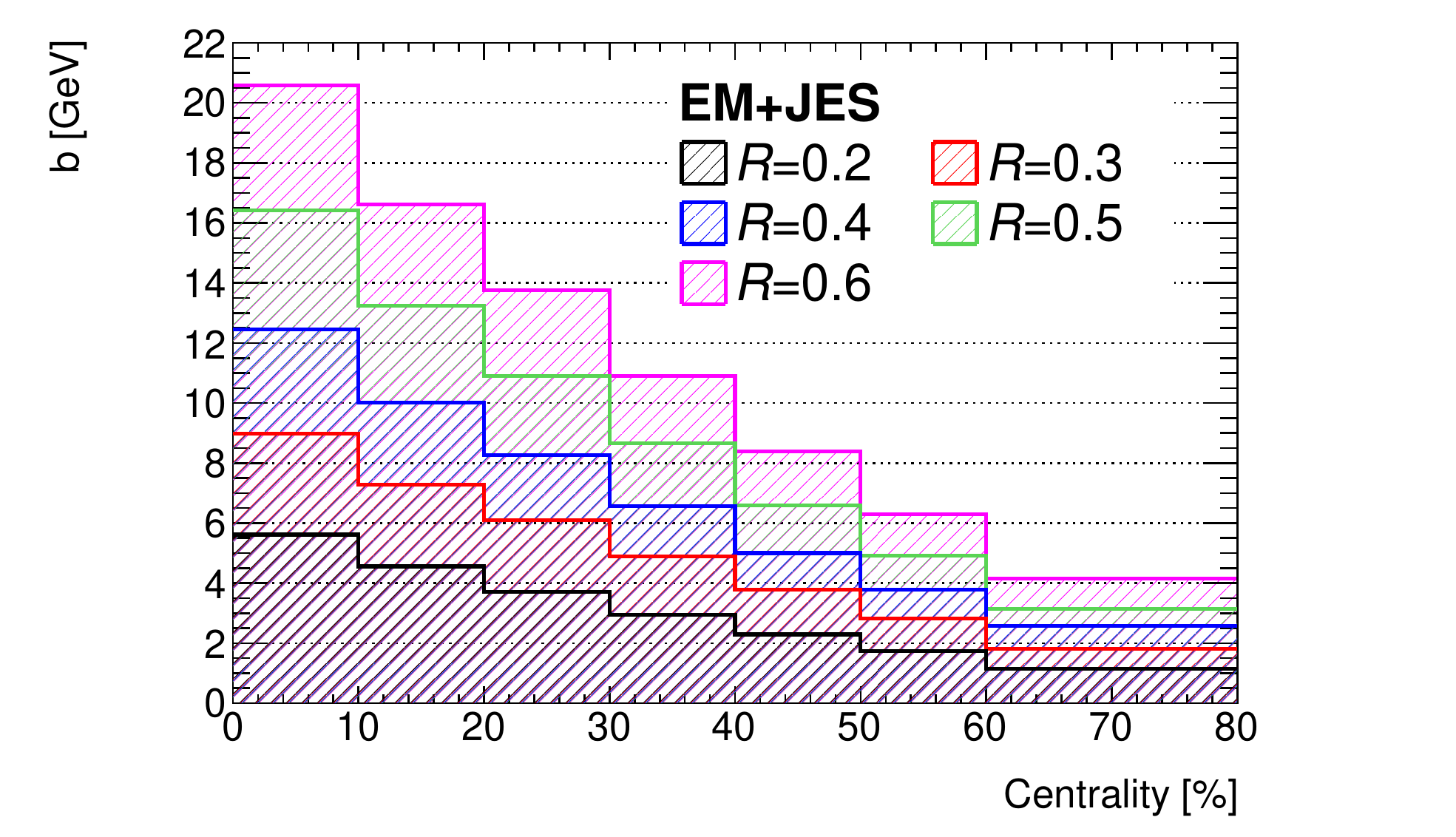}
}
\caption{The noise term obtained from the fluctuations analysis as a
  function of the jet area together with a power-law fit, $b(A) =
  p_0A^{p_1}$ is shown on the left. The $b$ values for fixed $R$ value
  as a function of centrality are shown on the right.} 
\label{fig:bterm} 
\end{figure} 
\begin{figure}[htb]
\centering
\includegraphics[width=0.375\textwidth]{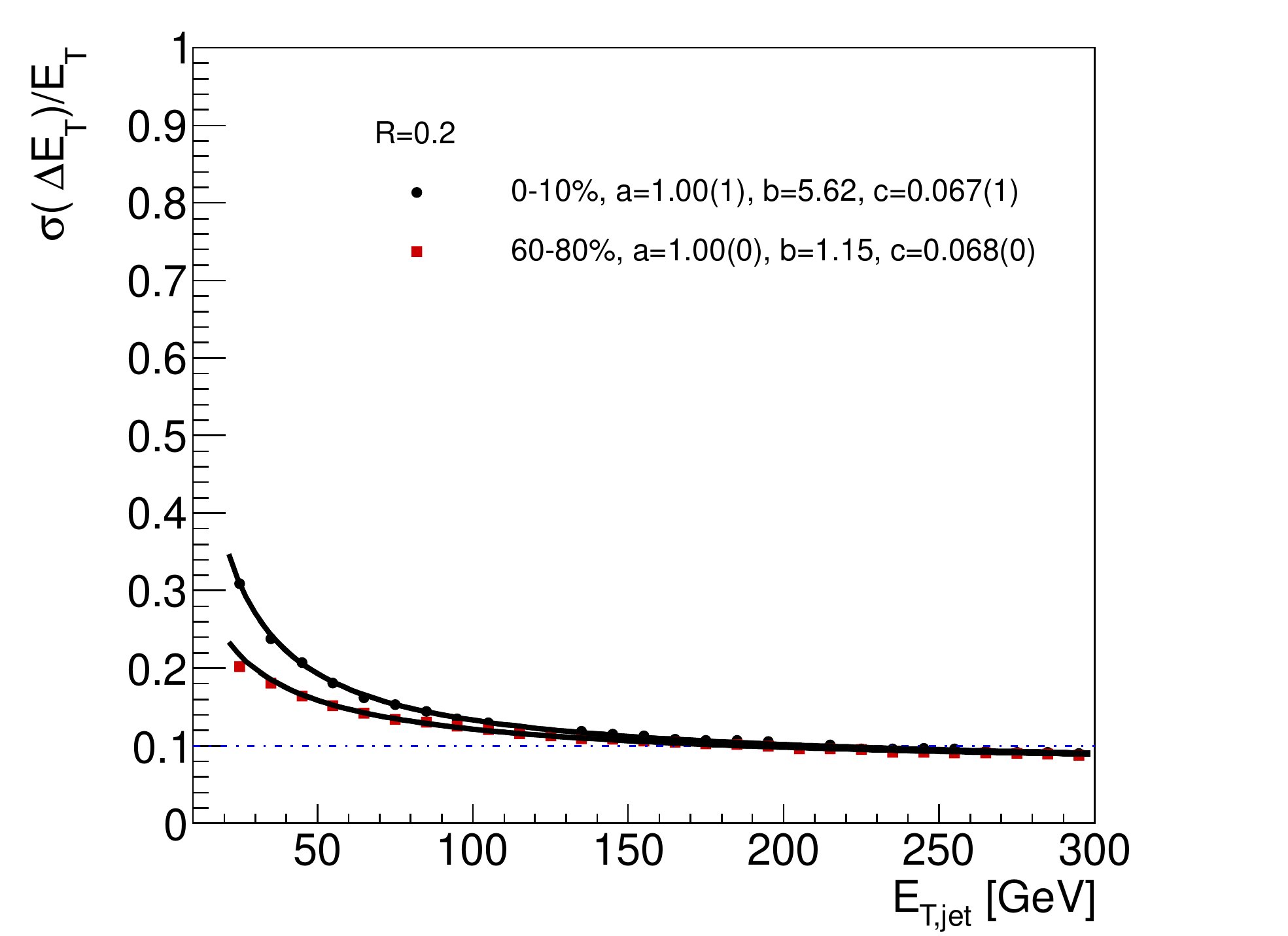}
\includegraphics[width=0.375\textwidth]{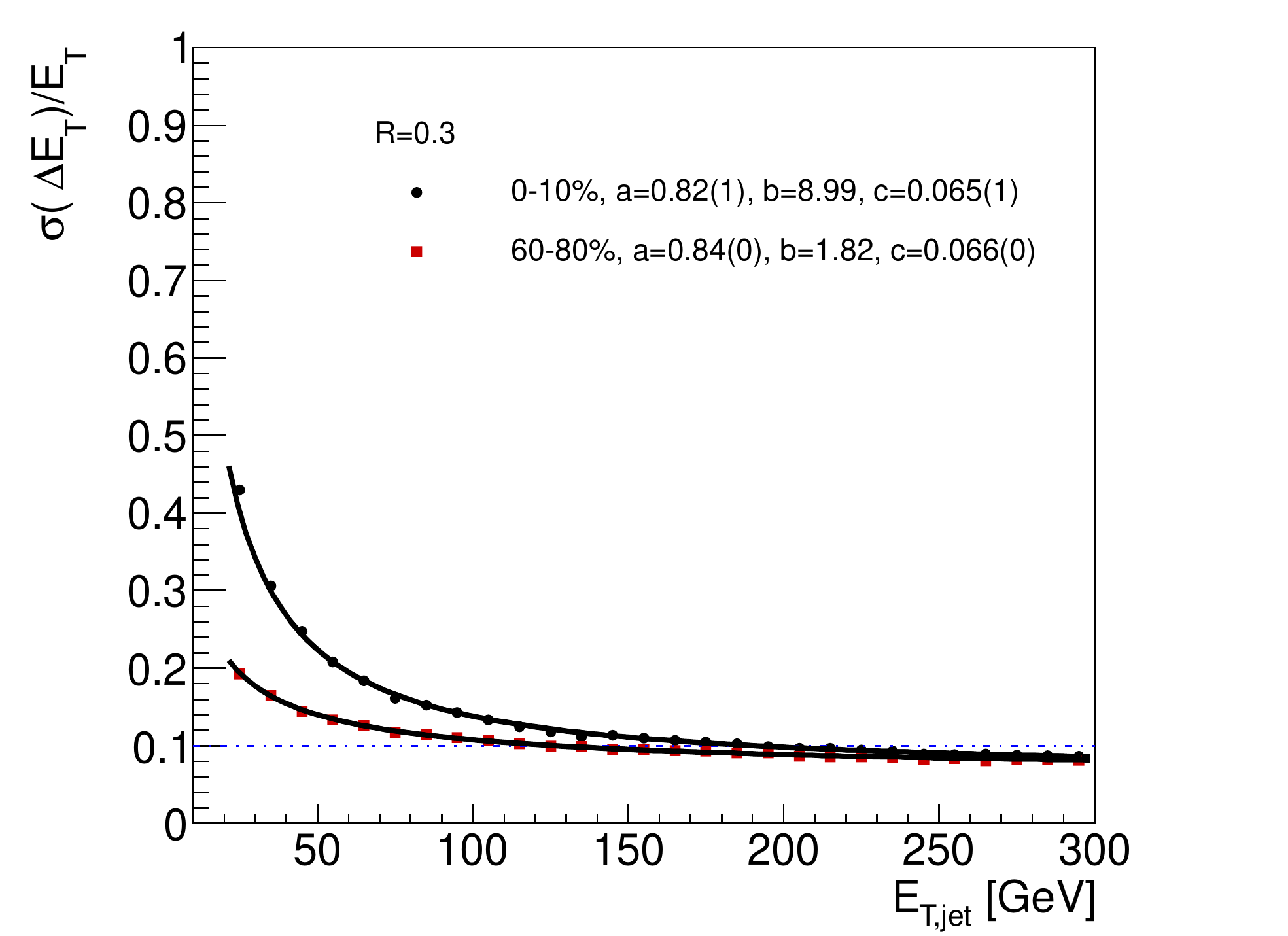}
\includegraphics[width=0.375\textwidth]{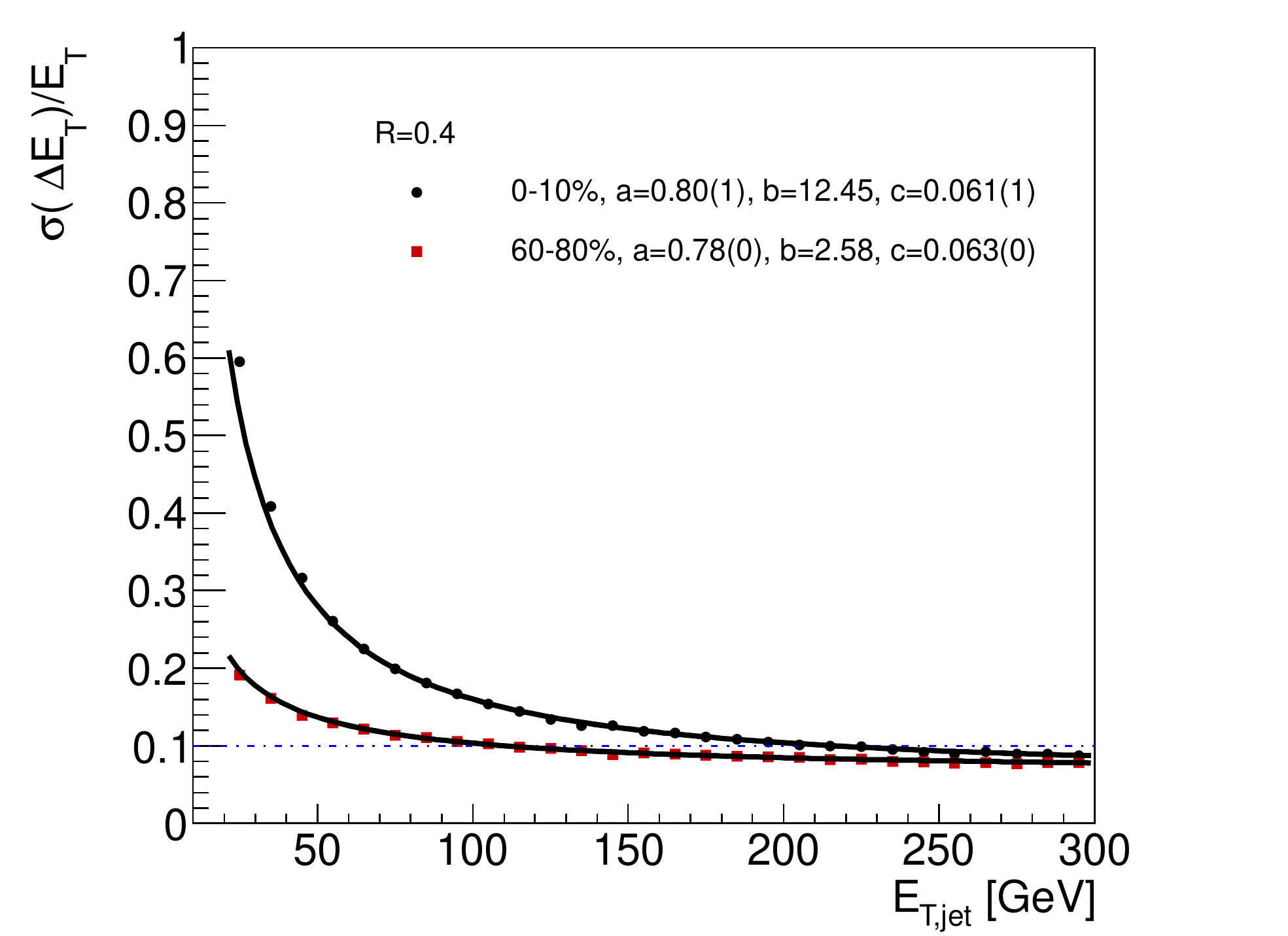}
\includegraphics[width=0.375\textwidth]{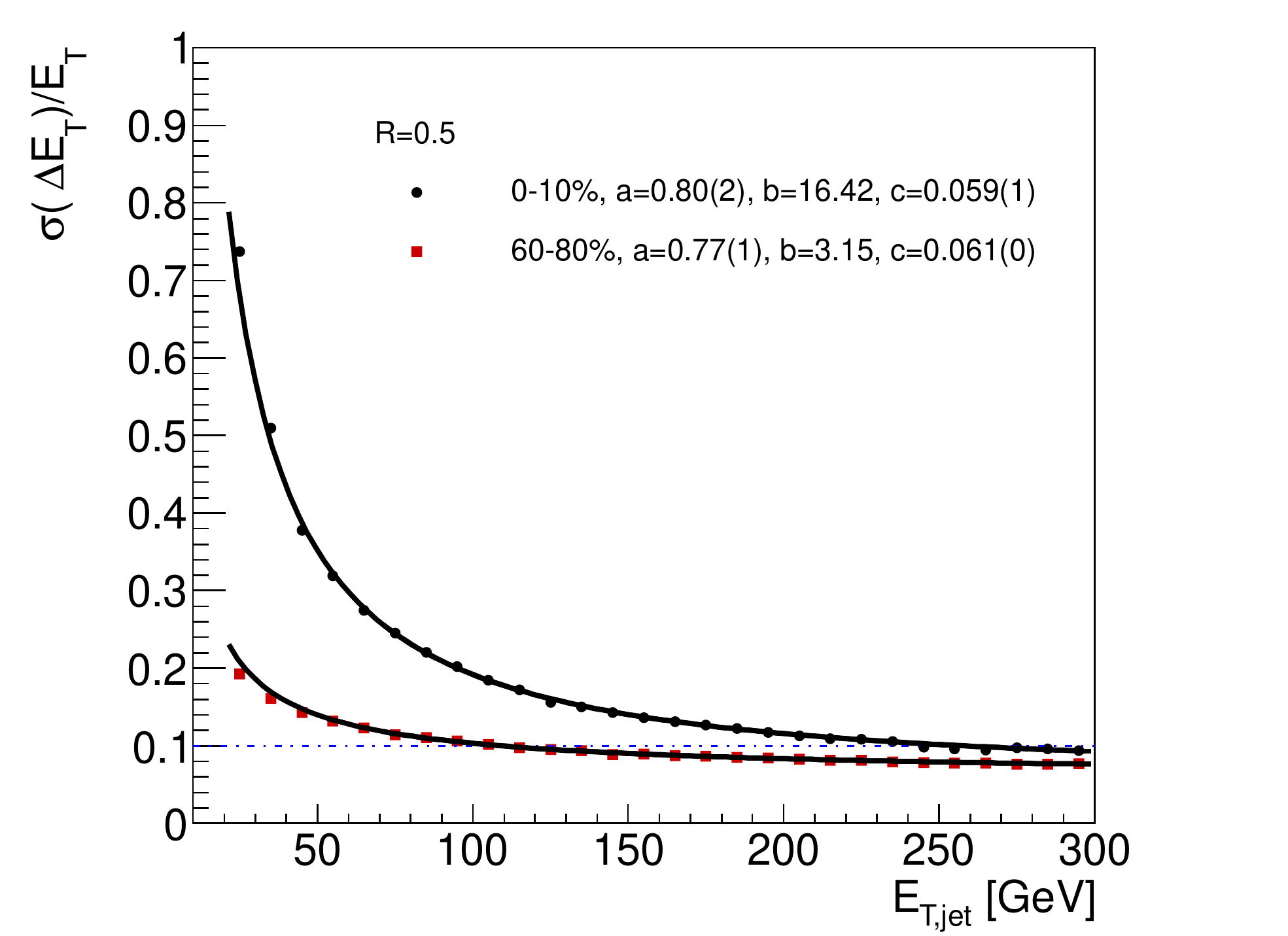}
\caption{The jet energy resolution in central and peripheral collisions for different jet sizes. 
Jet energy resolution is fit with the fixed noise term obtained from the fluctuation analysis.}
\label{fig:jetjerfit}
\end{figure}

\clearpage

\subsection{Data Overlay}
\label{sec:validation:overlay}

To supplement the data-driven checks discussed in the preceding
sections, samples using MC jets from PYTHIA overlaid onto \PbPb\ data were also studied. These samples were generated by
combining minimum bias \PbPb\ events with PYTHIA dijet samples using
the pileup overlay framework~\cite{Assamagan:1368189} and performing reconstruction on the combined
signal. The PYTHIA events are produced in different J samples as
discussed in Section~\ref{section:jet_rec:mc_sample}.
For each event the PYTHIA generation and subsequent GEANT simulation
is run with conditions matching data, including vertex position. The data events are taken from a dedicated \verb=MinBiasOverlay= stream in
the 2011 running which uses a L1 ZDC coincidence trigger and records
the data without zero suppression.

The validation of these samples is not yet complete, and high
statistics samples have not yet been produced, however aspects of
the ongoing validation studies provide additional support of the
data-driven checks presented in the preceding sections. The overlay
samples used embedded J3 and J4 PYTHIA jets. 
These results are only meaningful
over the jet \ET\ range for which the J3 and J4 samples are the
dominant contribution in the total cross section weighted sample
($80\lesssim \ET \lesssim 300$~\GeV). The JES as a function of \ET\
for \RFour\ jets is compared between 0-10\% and 60-80\% centrality
bins in Fig.~\ref{fig:validation:overlay_jes}. This comparison
indicates no dependence of the JES non-closure on centrality to the
level of statistical precision allowed by the sample ($\sim 1\%$). It is
consistent with the estimates of the relative centrality-dependent JES non-closure
derived in Section~\ref{performance:JES}, and summarized in
Fig.~\ref{fig:JES_summary}.
\begin{figure}[htb]
\centering
\includegraphics[width=0.8\textwidth]{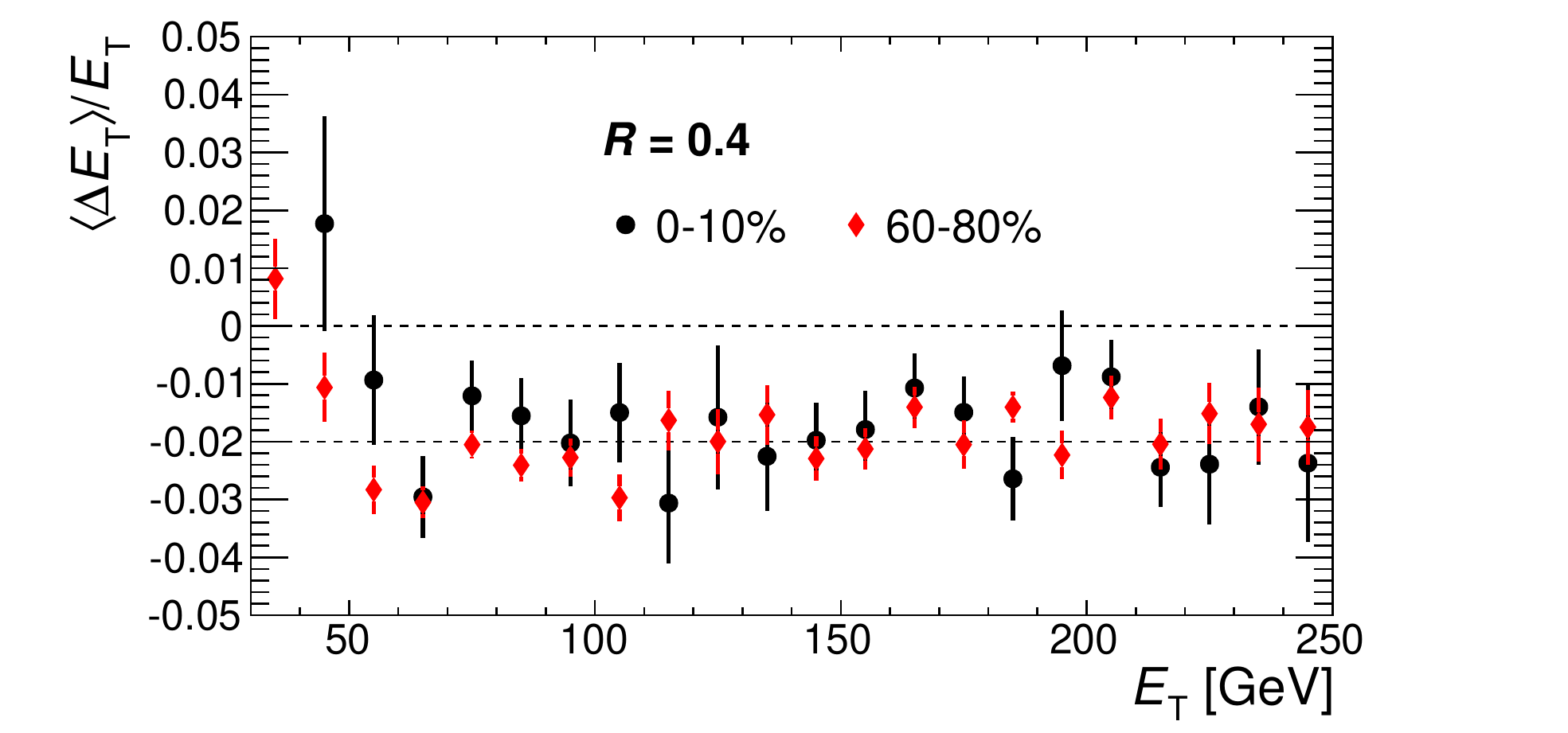}
\caption{JES in the overlay sample for \RFour\ jets in the 0-10\% (black)
  and 60-80\% (red) centrality bins. No systematic deviation is
  observed in the non-closure between the two centralities.}
\label{fig:validation:overlay_jes}
\end{figure}
Comparisons were made of different performance variables between these overlay samples
and the J3 and J4 embedded PYTHIA+HIJING samples in described in
Section~\ref{section:jet_rec:mc_sample}. A comparison of the JER between these two
samples in central and peripheral collisions is shown in
Fig.~\ref{fig:validation:overlay_jer}. Excellent agreement is found
between the overlay and HIJING samples, consistent with the data/MC
agreement found in the fluctuations analysis~\cite{ATLAS-CONF-2012-045}.
\begin{figure}[htbp]
\centering
\includegraphics[width=0.55\textwidth]{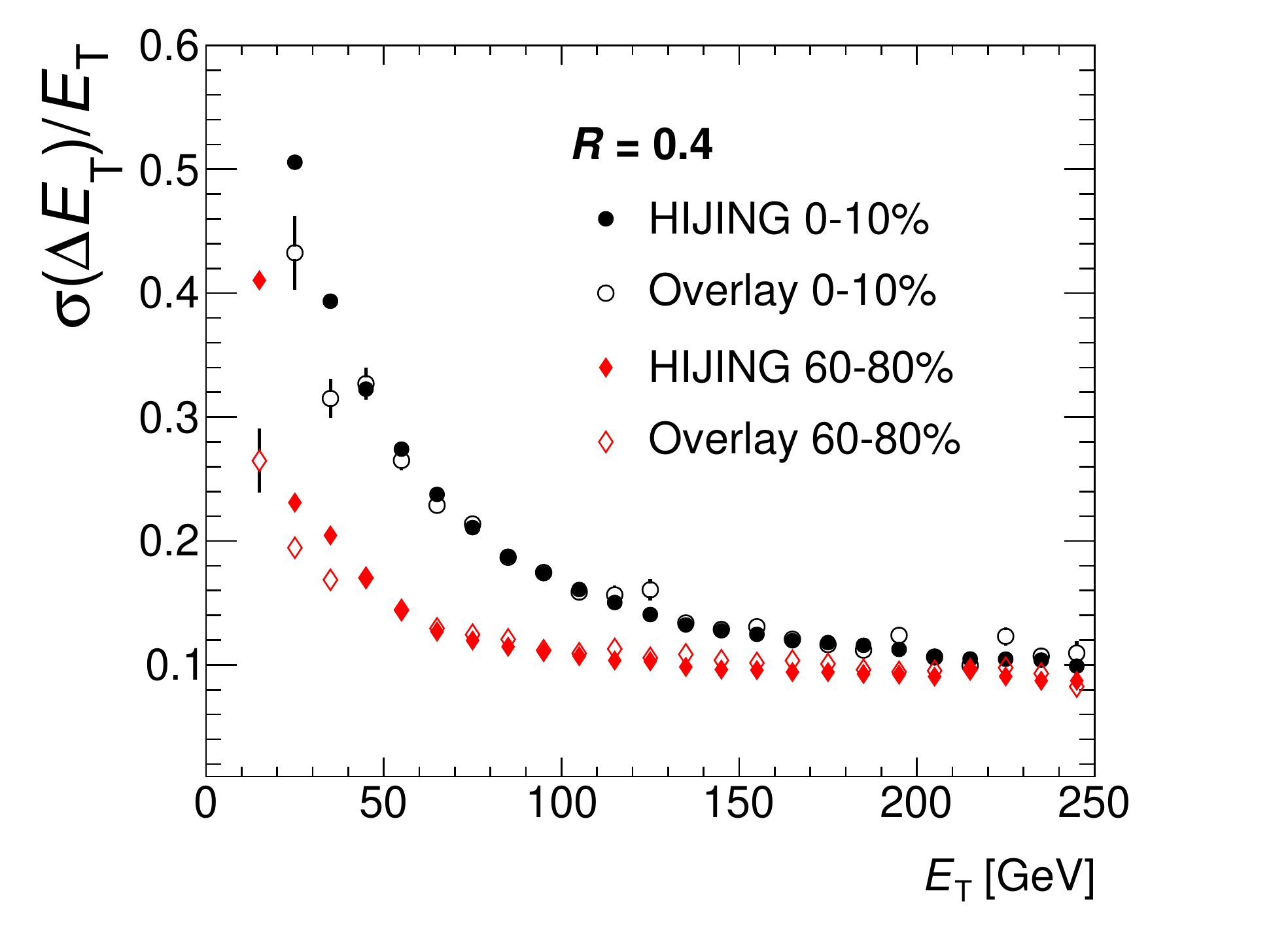}
\caption{Comparison of the JER in HIJING (solid) and the overlay
  sample (hollow) for \RFour\ jets in two centrality bins, 0-10\%
  (black) and 60-80\% (red).}
\label{fig:validation:overlay_jer}
\end{figure} A similar comparison of the jet reconstruction efficiency is shown in
Fig.~\ref{fig:validation:overlay_jre}. Above 50~\GeV\ the peripheral
efficiencies show good agreement, and there is a slightly greater
efficiency $(\sim5\%$) in the HIJING sample in central collisions than
in the overlay sample. As the J1 and J2 samples dominate at lower \pt,
the larger disagreement there is indicative of
significant differences between the data and MC.
\begin{figure}[htbp]
\centering
\includegraphics[width=0.55\textwidth]{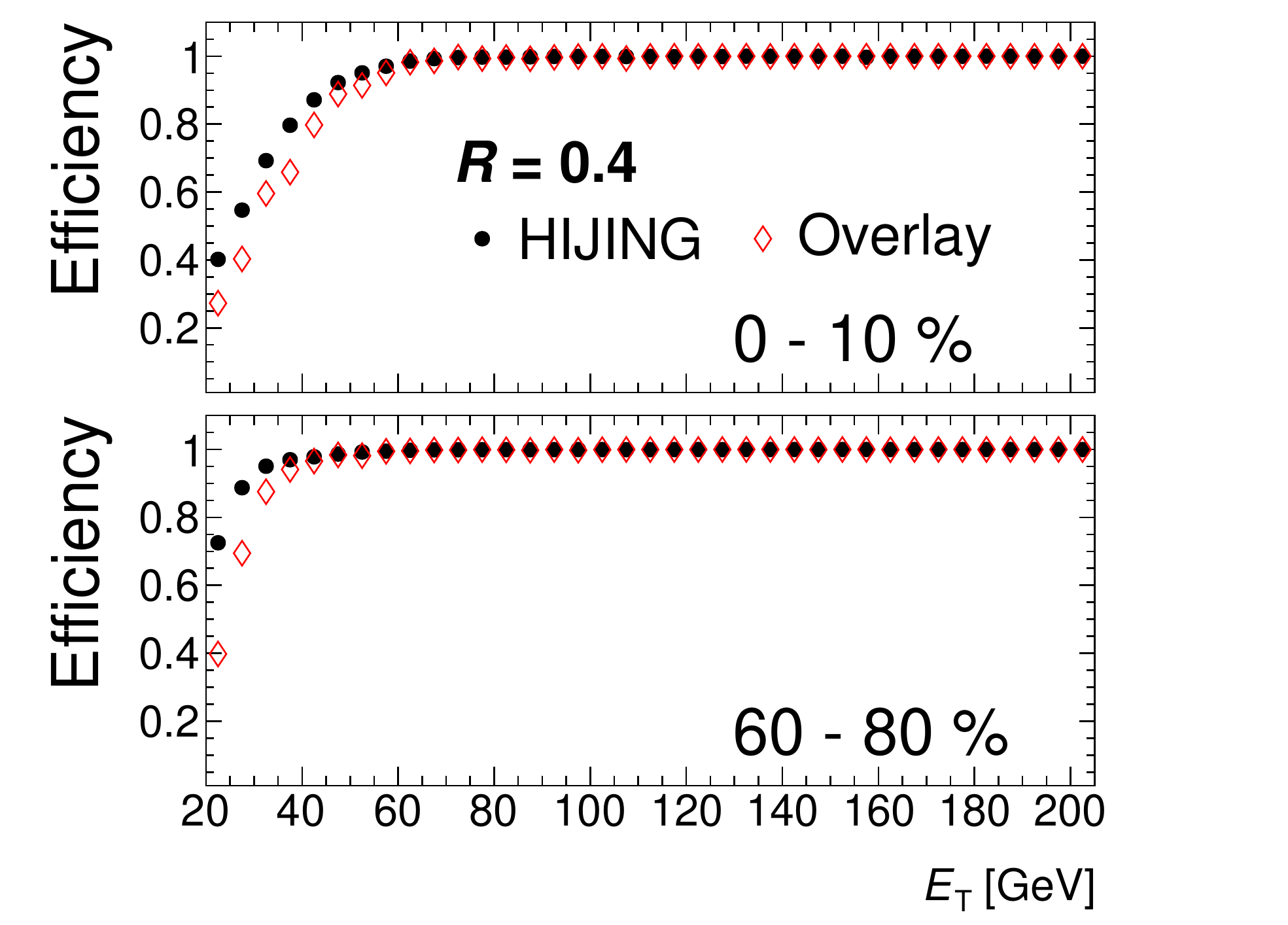}
\caption{Reconstruction efficiency in the overlay sample for \RFour\ jets in the 0-10\% (black)
  and 60-80\% (red) centrality bins. No systematic deviation is
  observed in the non-closure between the two centralities.}
\label{fig:validation:overlay_jre}
\end{figure}
\subsection{Jet Kinematics}
\label{sec:validation:jet_kinematics}
As an evaluation of the jet in terms of \pt\ and $y$ is sometimes preferred to
\et\ and $\eta$. For high-\et\ jets, it is expected that \pt\ and \et\
will be very similar as will $\eta$ and $y$. This was verified directly
in data. The two-dimensional correlations in $\et-\pt$ and
$\eta-y$ in different centrality bins were computed in fine bins and
are highly diagonal. The means of these
distributions, $\langle \pt \rangle$ vs \et\ and $\langle y \rangle$
vs $\eta$, are compared in central and peripheral collisions for
\RFour\ jets in
Fig.~\ref{fig:validation:et_pt}. No
significant deviation from the behavior $\pt\approx\et$ and $y\approx\eta$ is observed.
\begin{figure}
\centering
\includegraphics[width=0.49\textwidth]{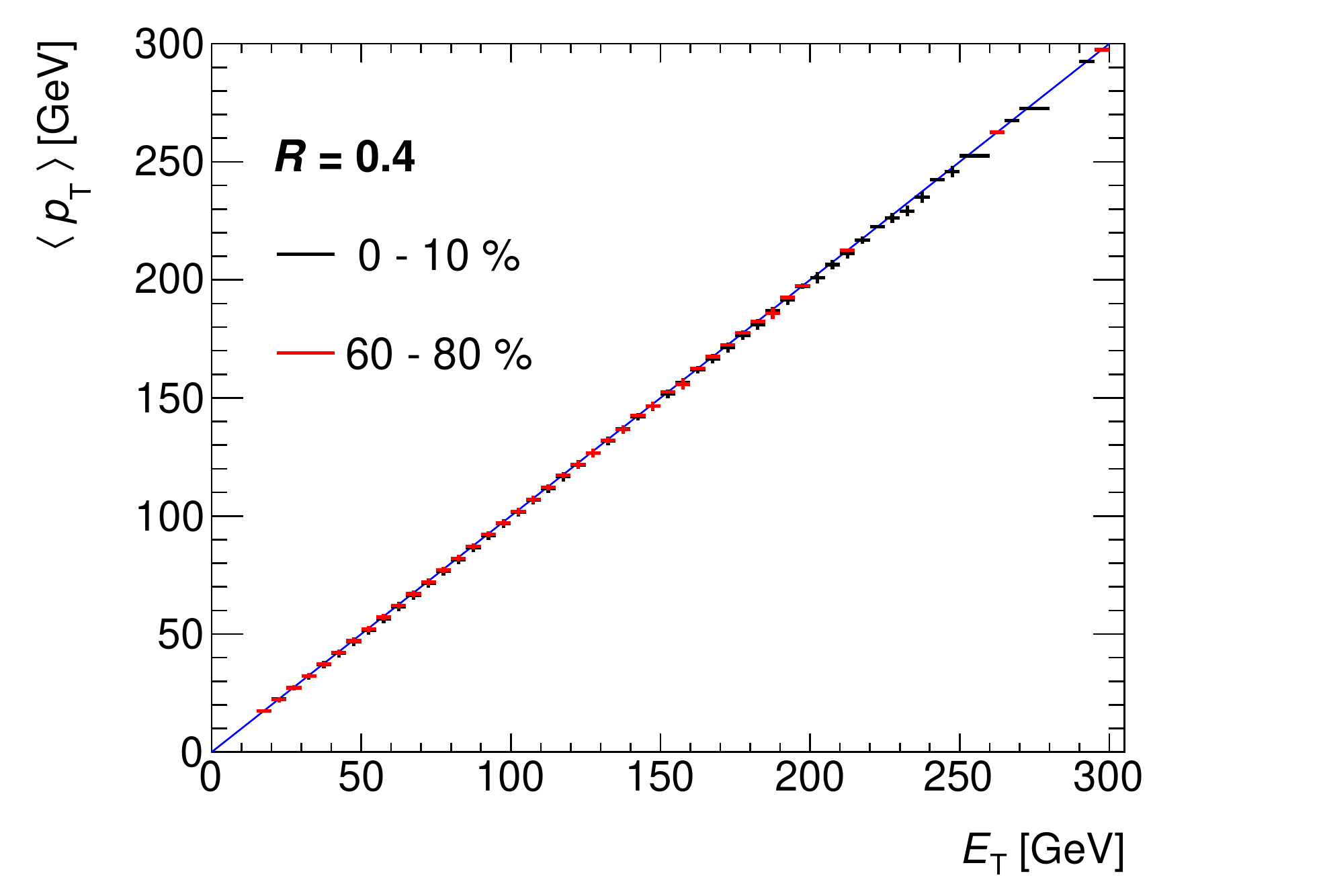}
\includegraphics[width=0.49\textwidth]{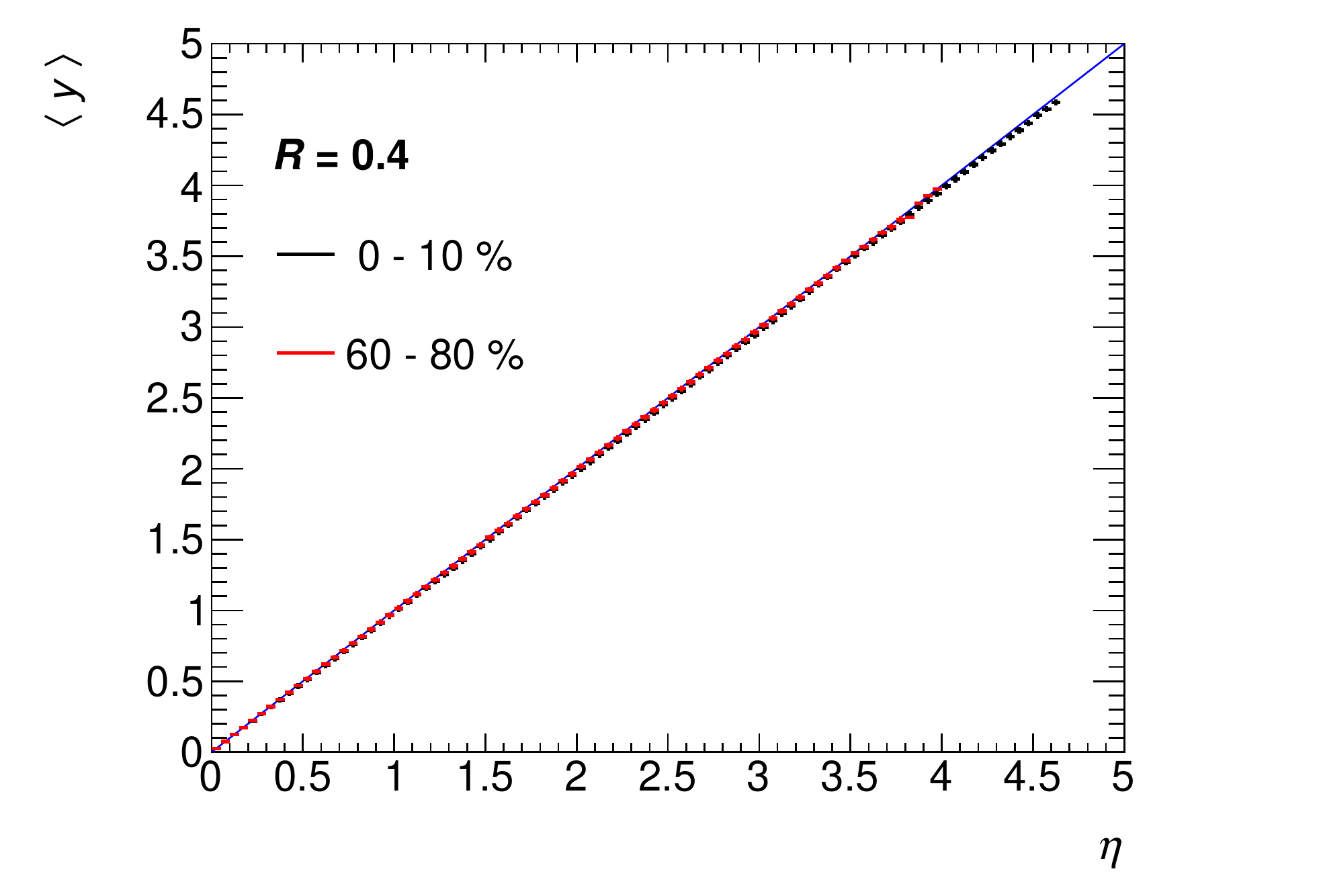}
\caption{Plots of $\langle \pt \rangle$ vs \et\ (left) and  $\langle y \rangle$
vs $\eta$ (right) comparing 0-10\% (black) and 60-80\% (red). The lines
$\pt=\et$ and $y=\eta$ are shown in blue for comparison.}
\label{fig:validation:et_pt}
\end{figure}
\section{Unfolding}
\label{section:analysis:unfolding}
\subsection{Observables}
The goal of this measurement is to study jet quenching by examining the centrality dependence of
the single inclusive jet spectrum. Collisions with different
centralities have different degrees of geometric overlap between the
colliding nuclei. This geometric effect translates into different per
collision effective nucleon luminosities. Thus comparisons between
different centrality bins should be scaled by \Ncoll, the effective
number of colliding nucleon pairs in a given centrality bin, to remove
the geometric enhancement. The ``central to
peripheral'' ratio, \Rcp, is defined as
\begin{equation}
\Rcp = \frac{
\frac{1}{\Ncollcent}
\frac{1}{\Nevt^{\mathrm{cent}}}
E\frac{d^3\Njet^{\mathrm{cent}}}{dp^3}
}{
\frac{1}{\Ncoll^{\mathrm{periph}}}
\frac{1}{\Nevt^{\mathrm{periph}}}
E\frac{d^3\Njet^{\mathrm{periph}}}{dp^3}
}\,.
\label{eq:rcpdef}
\end{equation}

The common phase space factors cancel, so the relevant
quantity to extract from the data is the per event yield in $p_{T}$ bin $i$,
\begin{equation}
Y_i\equiv\frac{1}{\Nevt}\frac{d\Njet^i}{d\pt^{i}}\,,
\end{equation}
in each centrality bin. The 60-80\% centrality bin is used as the
peripheral bin in all \Rcp\ and the notation
$\Rcollcent=\Ncollcent/\Ncollperiph$ is used to describe overall
geometric enhancement as determined by the Glauber Model.

Jets from the 2010 \PbPb\ data set were
reconstructed using the anti-\kt\ algorithm with $R=0.2-0.5$ using the
procedure described in Section~\ref{section:jet_rec:subtraction} over the rapidity
interval $|y| < 2.1$. All jets failing the fake rejection criteria
described in Sec.~\ref{corrections:fake} are excluded from the
analysis.  These spectra are shown in
Figs.~\ref{fig:analysis:unfolding:raw_spectra_23} and ~\ref{fig:analysis:unfolding:raw_spectra_45}.
\begin{figure}[htb]
\centering
\includegraphics[width =0.49\textwidth] {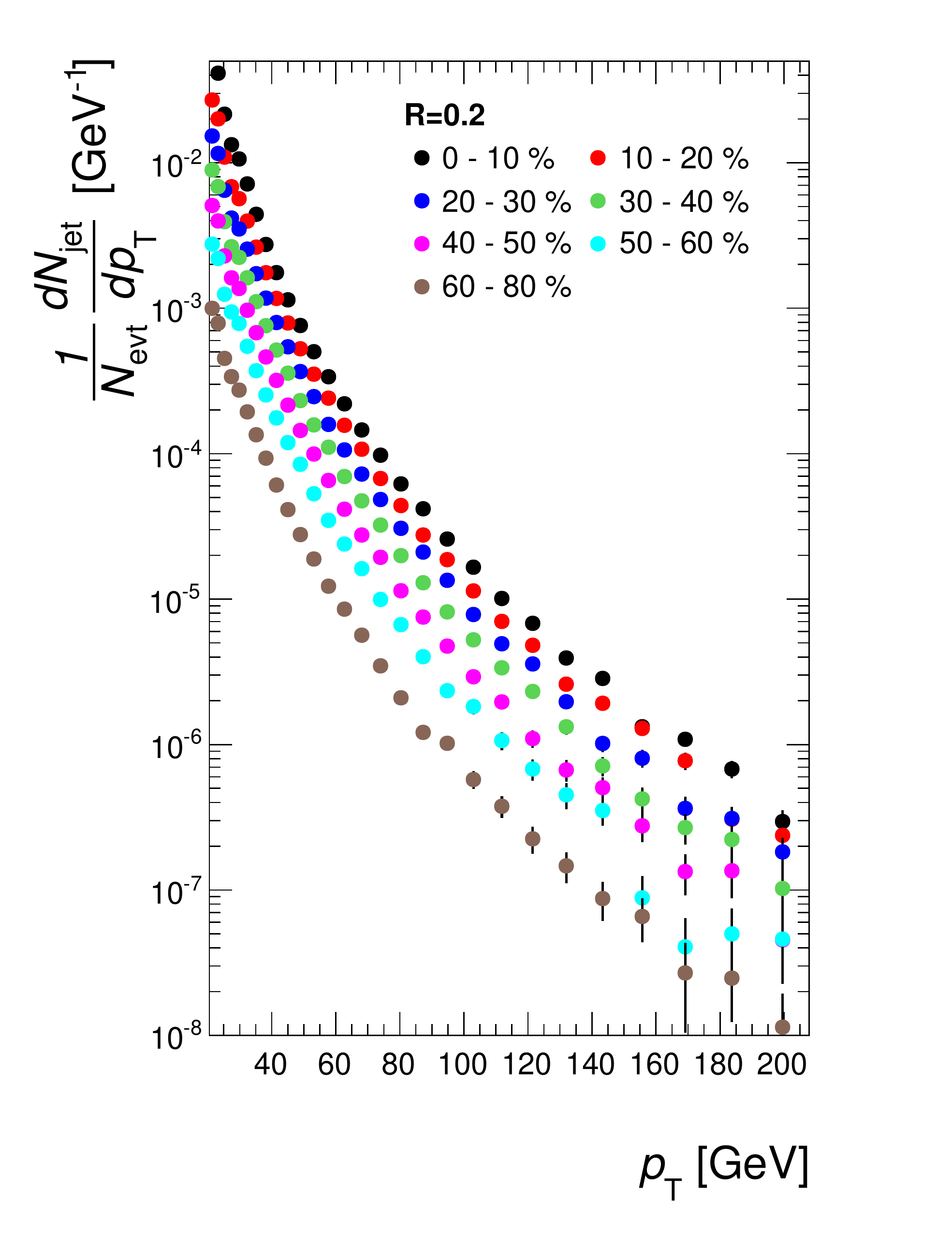}
\includegraphics[width =0.49\textwidth] {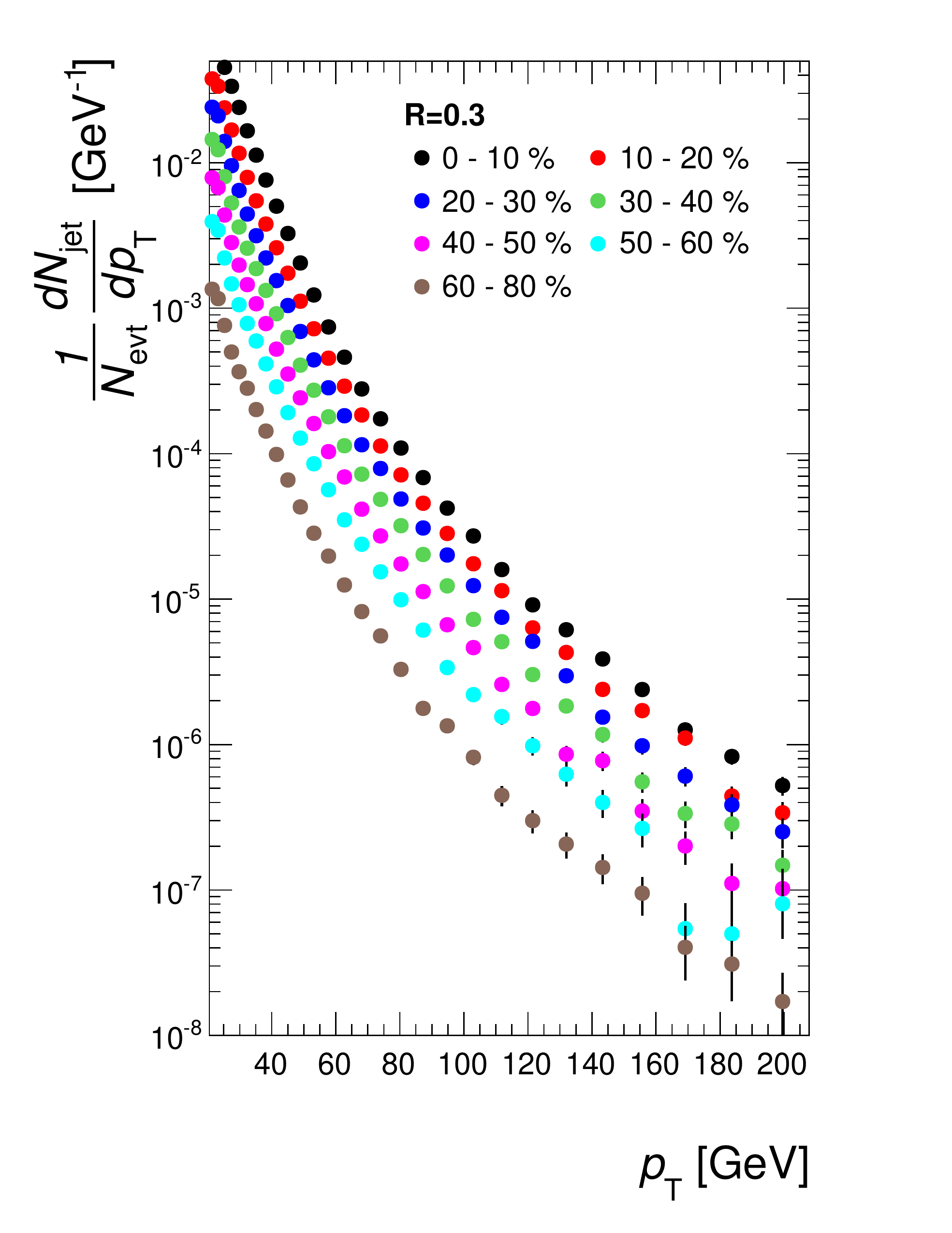}
\caption{Per event yields before correction for \RTwo\ (left) and
  \RThree\ (right) jets.}
\label{fig:analysis:unfolding:raw_spectra_23}
\end{figure}
\begin{figure}[htb]
\centering
\includegraphics[width =0.49\textwidth] {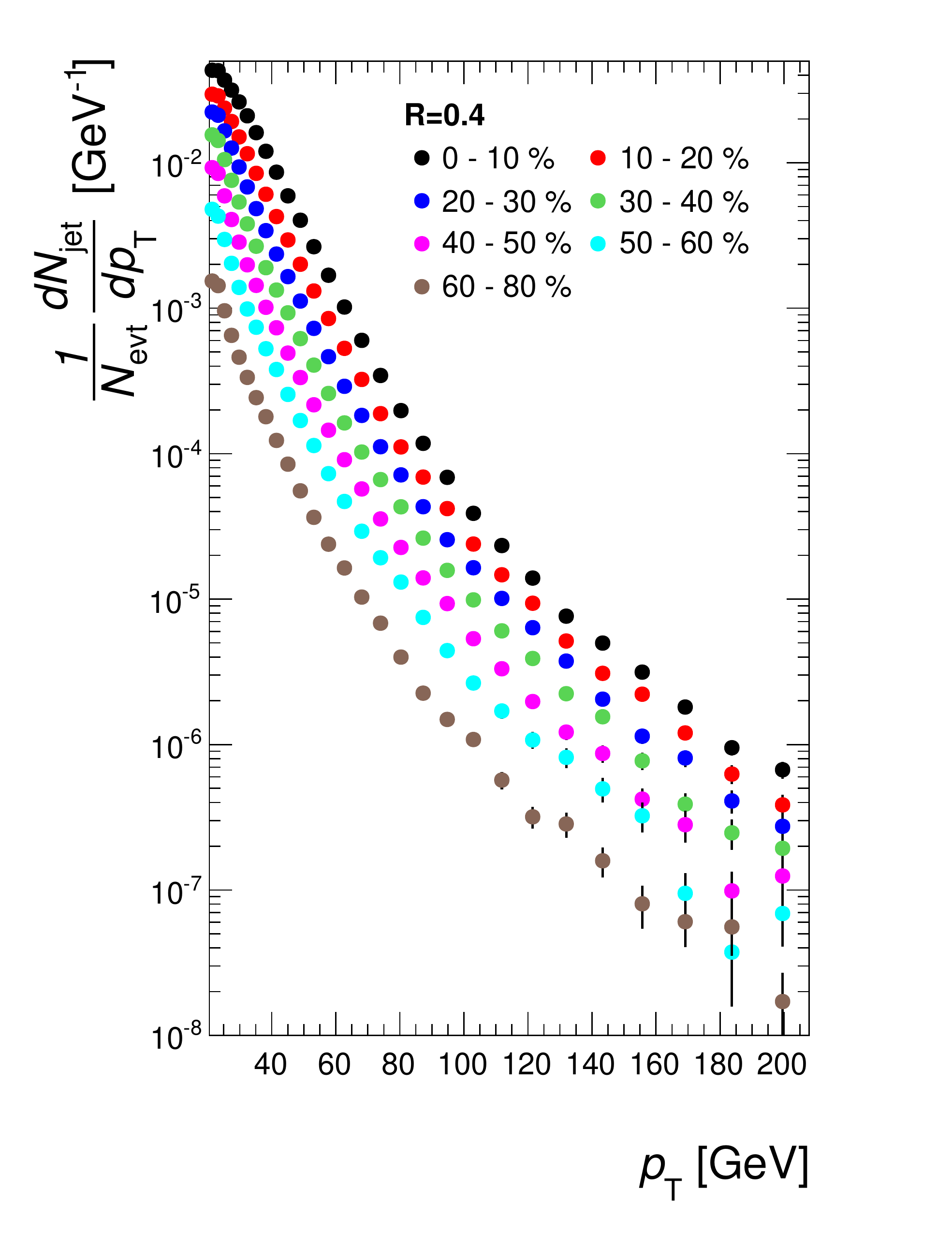}
\includegraphics[width =0.49\textwidth] {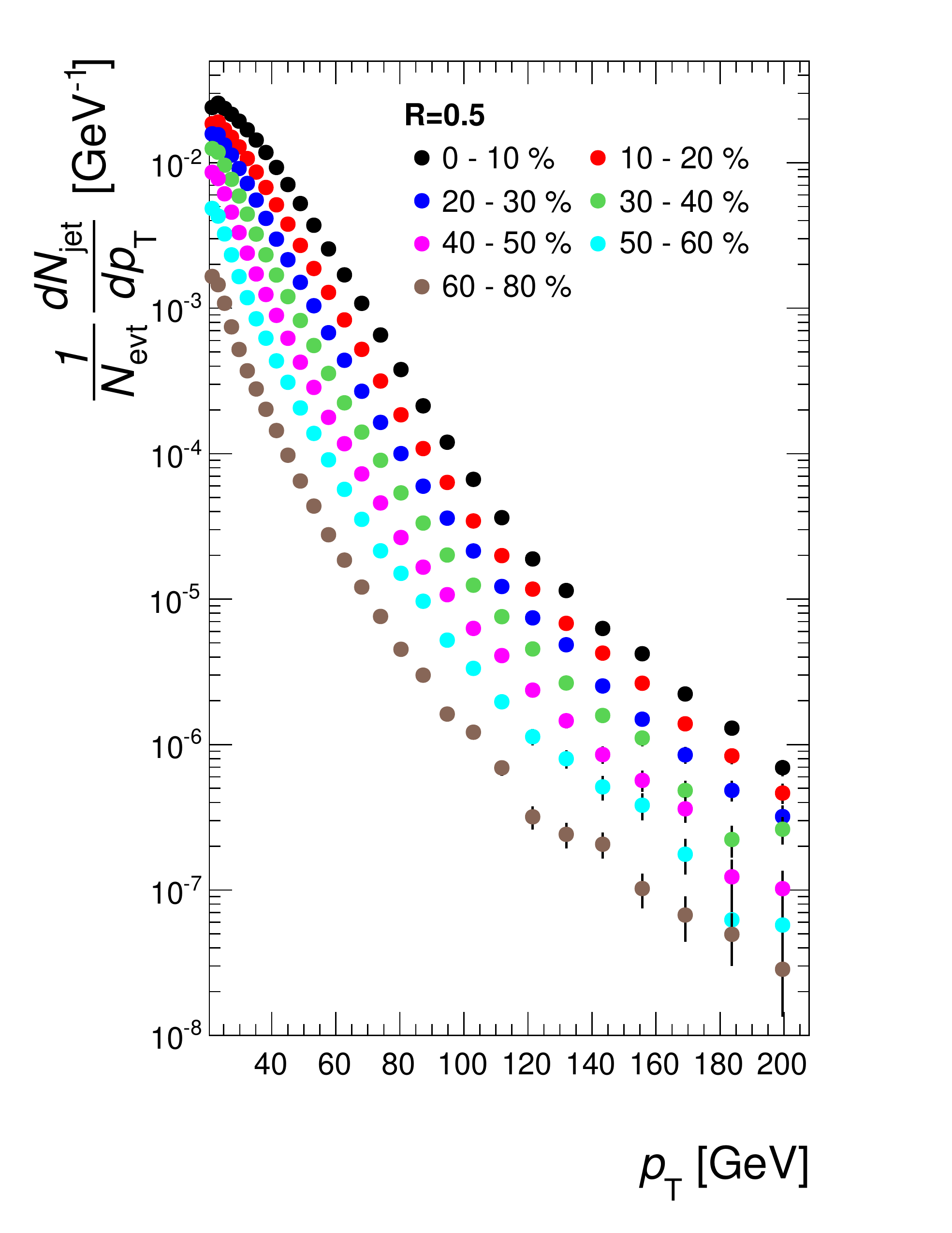}
\caption{Per event yields before correction for \RFour\ (left) and
  \RFive\ (right) jets.}
\label{fig:analysis:unfolding:raw_spectra_45}
\end{figure}

This measurement is a sampling of the per event jet multiplicity
distribution in a given \pt\ bin, which is given by the random
variable $n_i$ and is sampled $\Nevt$ times. The measured per event
yield is the mean of this variable,
\begin{equation}
Y_i=\frac{1}{d\pt^{i}}\frac{1}{\Nevt}\displaystyle \sum_{j=1}^{\Nevt} n^i_j\,.
\end{equation}
A single event can produce multiple jets; this introduces a
statistical correlation between bins and a full covariance matrix is
required to describe the statistical uncertainty of the
measurement described by
\begin{equation}
\mathrm{Cov}\left(Y_i,Y_j\right)\cong\frac{V_{ij}-Y_iY_j}{\Nevt}\,,
\end{equation}
where the quantity
\begin{equation}
V_{ij}\equiv
\frac{1}{\Nevt}\frac{d\Njet^{ij}}{d\pt^id\pt^j}=\frac{1}{d\pt^id\pt^j}\frac{1}{\Nevt}\displaystyle
\sum_{k=1}^{\Nevt} n^i_k n^k_k\,,
\end{equation}
must also be recorded on a per event basis in addition to $Y_i$. The measured covariances are shown in
Figs.~\ref{fig:analysis:unfolding:raw_cov_cent} and~\ref{fig:analysis:unfolding:raw_cov_R}, as the
variance-normalized correlation $\rho$
\begin{equation}
\rho_{ij}=\frac{\mathrm{Cov}(Y_i,Y_j)}{\sqrt{\mathrm{Var}(Y_i)}\sqrt{\mathrm{Var}(Y_j)}}\,.
\end{equation}
The matrix is strongly diagonal, indicating a weak correlation
between different bins. However, the unfolding procedure will correct
for bin migration and will introduce additional correlations between
bins, resulting in a non-trivial covariance matrix. 
\begin{figure}[htb]
\centering
\includegraphics[width =1\textwidth] {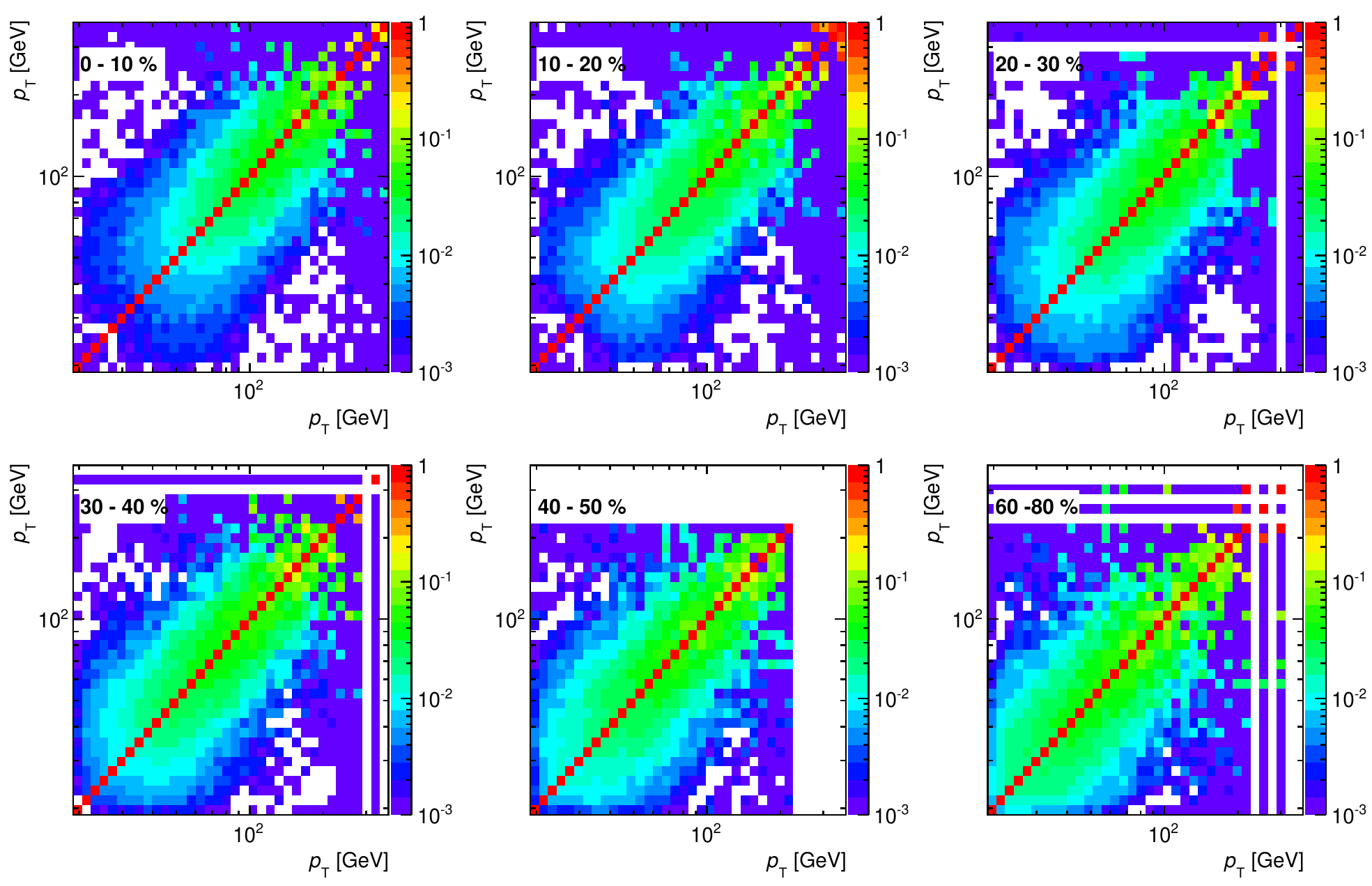}
\caption{Spectrum correlations for EM+JES \RFour\ jets.}
\label{fig:analysis:unfolding:raw_cov_cent}
\end{figure}
\begin{figure}[htb]
\centering
\includegraphics[width =1\textwidth] {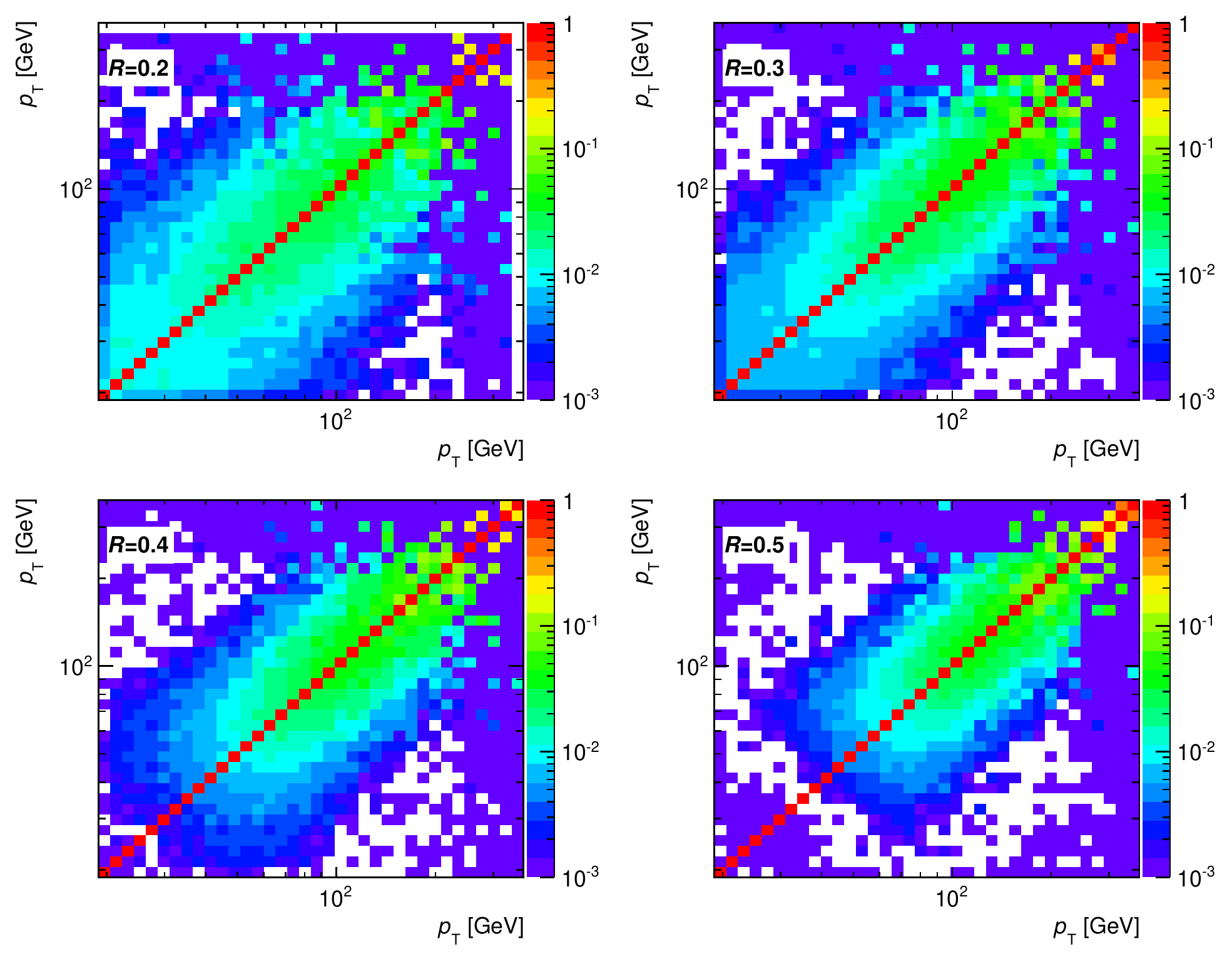}
\caption{Spectrum correlations for EM+JES $R=0.2,\,0.3,\,0.4$ and $0.5$
  jets for the 0-10\% centrality bin.}
\label{fig:analysis:unfolding:raw_cov_R}
\end{figure}

\subsection{Method}
\label{section:analysis:unfolding:unfolding}
The Monte Carlo is used to correct the raw spectrum for the finite jet
energy resolution introduced by underlying event fluctuations and
detector effects. For each truth jet, the nearest
matching reconstructed jet with $\Delta R < 0.2$ is found and the pair
$(\pt^{\mathrm{truth}},\pt^{\mathrm{reco}})$ is recorded to build the response matrix
$\mathbf{\hat{A}}$. Then the input truth, $T_i$, and output reconstructed, $R_i$, distributions are
related via a linear transformation
\begin{equation}
R_j=\displaystyle\sum_i \hat{A}_{ij} T_i\,,
\label{eqn:analysis:unfolding:response_matrix}
\end{equation}
where $T_i$ is the full truth~\pt\ spectrum, including truth jets that
had no matching reconstructed jet. If the Monte Carlo accurately
describes the underlying event and detector response, a similar bin
migration is assumed to apply to the measured spectrum, $b_i$ as well:
\begin{equation}
b_j=\displaystyle\sum_i \hat{A}_{ij} x_i\,.
\label{eqn:analysis:unfolding:response_matrix_data}
\end{equation}
$x_i$ is the desired spectrum corrected for bin migration. The
goal of an unfolding procedure is to invert this equation in a controlled
fashion and obtain the true spectrum. The full inverse of the response matrix may possess properties
introducing undesired features on the data, in particular the
inversion may amplify statistical fluctuations. Additionally, the
system of equations is over-determined and modifications to the linear
system, motivated by the specific application, must be applied to have a
well-determined solution. This study addresses these problems by using a singular value
decomposition (SVD) based approach to regularize the unfolding as
presented in Ref.~\cite{Hocker:1995kb}. The implementation of the SVD method in the \texttt{RooUnfold}
software package~\cite{roounfold} was used to perform the
unfolding~\footnote{\texttt{RooUnfold-1.1.1} was used with minor
  additions to allow for checks with continuous values of the
  regularization parameter.}.

The first modification to Eq.~\ref{eqn:analysis:unfolding:response_matrix_data} is
to recast the system in terms of a linear least squares problem, where
each equation in the system is inversely weighted by the uncertainty in the
measured spectrum $\Delta b_i$,
 \begin{equation}
\displaystyle \sum_i\left(\sum_j \frac{A_{ij}
  x_j-b_i}{\Delta b_i}\right)^2=\mathrm{min}\,.
\label{eqn:analysis:unfolding:lls_scaled}
\end{equation}
Thus equations with larger statistical uncertainty have less
significance in the solution. In this
framework the response matrix, $A'_{ij}=A_{ij}/\Delta b_i$, is expressed in terms of its SVD
\begin{equation}
\mathbf{A^{\prime}}=\mathbf{U}\mathbf{\Sigma} \mathbf{V}^T\,,
\label{eqn:analysis:unfolding:svd}
\end{equation}
where $\mathbf{\Sigma}$ is a diagonal matrix containing the singular
values,$s_i$, and $\mathbf{U}$ and $\mathbf{V}$ are matrices composed of left and right singular vectors which
each form orthonormal bases. For small singular values, $1/s_i$ can become large
and statistical fluctuations in the data become magnified resulting in distortions in the
unfolded distribution. To address this effect, the singular values are
regularized by imposing smoothness requirements on the unfolded
spectrum by expanding Eq.~\ref{eqn:analysis:unfolding:lls_scaled}
to include the simultaneous minimization of the curvature of the
spectrum,
\begin{equation}
(\mathbf{A^{\prime}}\mathbf{x}-\mathbf{\tilde{b}})^T(\mathbf{A^{\prime}}\mathbf{x}-\mathbf{\tilde{ b}})+\tau(\mathbf{Cx})^T(\mathbf{Cx})=\mathrm{min}\,,
\end{equation}
where $\tilde{b}_i=b_i/\Delta b_i$, $\mathbf{Cx}$ is the second derivative of $\mathbf{x}$ expressed in terms of
finite differences and $\tau$ is a regularization parameter. The
behavior for arbitrary $\tau$ is obtained by rewriting
Eq.~\ref{eqn:analysis:unfolding:response_matrix_data} as
\begin{equation}
\mathbf{A^{\prime}}\mathbf{C}^{-1}\mathbf{Cx}=\mathbf{b}\,,
\end{equation}
and decomposing $\mathbf{A^{\prime}C}$ as in Eq.~\ref{eqn:analysis:unfolding:svd}. 
The form of $\mathbf{C}$ used in~\cite{Hocker:1995kb} and~\cite{roounfold}
assumes $\mathbf{x}$ is filled using bins of fixed width. In this analysis,
variable bins are used so that $\ln \pt^{i} -\ln \pt^{i+1}$ is
constant, and thus $\mathbf{Cx}$ is a finite difference approximation of
$\frac{d^2x}{d(\ln\pt)^2}$. This measure of curvature is more
appropriate given the steeply falling nature of the
jet spectrum.

The behavior of the
system under statistical fluctuations can be understood by
analyzing the behavior of the data expressed in the basis supplied by
the SVD
\begin{equation}
\mathbf{d}=\mathbf{U}^T\mathbf{b}\,.
\label{eqn:analysis:unfolding:d_definition}
\end{equation}
The $d_i$ are normalized by the error and statistical fluctuations are
of order unity. The values of $i$ for which $|d_{i}|\gg 1$ are the statistically
significant equations in the linear system, and the number of such equations is the rank of
the system $k$. The remaining singular values correspond to
statistically irrelevant equations and are regularized via
\begin{equation}
\frac{1}{s_i}\xrightarrow{\tau}\frac{s_i}{s_i^2+\tau}\,,\,\,\,\,\,\,\,\,\,\tau=s_k^2\,.
\label{eqn:analysis:unfolding:regularized_sv}
\end{equation}

The final modification is to express the solution as a set of
weights relative to another distribution \xinibf, 
\begin{equation}
x_i=w_i/\xini_{i}\,,
\end{equation}
absorb the \xinibf\ into the response,
$\tilde{A}_{ij}=A'_{ij}/{\xini}_j$, solve for the weights
\begin{equation}
(\mathbf{\tilde{A} w}-\mathbf{\tilde{b})}^T(\mathbf{\tilde{A} w}-\mathbf{\tilde{b}})+\tau(\mathbf{Cw})^T(\mathbf{Cw})=\mathrm{min}\,,
\label{eqn:analysis:unfolding:lls_reg}
\end{equation}
and multiply back by \xinibf\ to obtain $\mathbf{x}$. 

For well-determined systems, these operations have no effect on the
solution. However when the system is overdetermined this is no longer true,
and \xinibf\ must be chosen carefully. In Ref.~\cite{Hocker:1995kb}
it is argued that the truth spectrum, $T$, used to generate $\mathbf{\hat{A}}$ is the
proper choice of \xinibf.

In this study it was found that the turn on of the efficiency, which
results from a combination of reconstruction efficiency and fake rejection,
introduced additional curvature and thus larger values of $w_i$ at
low \pt. This has the potential to propagate instabilities in the
unfolding at low \pt\, where the corrections are the least well known,
throughout the entire spectrum, including regions where no efficiency
corrections are required. Therefore the efficiency correction was
removed from the unfolding problem by constructing $\mathbf{\hat{A}}$ to not include inefficiency. Then the
unfolded spectrum $x$ is divided by $\varepsilon$ to correct for efficiency in the final
step after the unfolding is complete. To construct \xini\ the MC truth spectrum was re-weighted by a
smooth power-law to remove any statistical fluctuations in the MC then
multiplied by the efficiency,
\begin{equation}
\xini (\pt)= x_0\varepsilon(\pt)\pt^{-5}\,.
\label{eqn:xini_def}
\end{equation}

\subsection{Response Determination}
\label{section:analysis:unfolding:response_determination}

The SVD method relies on the curvature constraint to minimize the
effects of statistical fluctuations on unfolding. This means
that any non-smoothness in the response matrix or \xini\ due to
statistical fluctuations in the Monte Carlo sampling will distort the
unfolding. To address this issue, the response matrices obtained from
the Monte Carlo study were subjected to a series of smoothing
procedures before begin used in the unfolding.


For each J sample, a two-dimensional histogram, $M_{ij}$, was
populated from the
$(p_{\mathrm{T}}^{\mathrm{truth}},p_{\mathrm{T}}^{\mathrm{reco}})$ of
each truth-reconstructed matched jet pair. The $p_{\mathrm{T}}^{\mathrm{truth}}$ distribution, without requiring a
reconstructed jet match, is was also recorded in a
histogram $T_i$. The $M_{ij}$ and $T_i$ from the different J samples are then combined
using the cross section weighting discussed in
Section~\ref{section:jet_rec:mc_sample}. The response matrix was then constructed
from these two distributions
\begin{equation}
\hat{A}_{ij}=\frac{M_{ij}}{\displaystyle \sum_i M_{ij}}\,,
\label{eqn:analysis:unfolding:response_from_transfer}
\end{equation}
which satisfies Eq.~\ref{eqn:analysis:unfolding:response_matrix} if
there is no inefficiency. One-dimensional distributions, obtained by
projecting $M_{ij}$ in fixed $\pt^{\mathrm{truth}}$ slices, are typically Gaussian-like
with integrals equal to the total efficiency in that
bin. The tails of these distributions are often produced by cases where a single truth jet
is reconstructed as a pair of split jets and vice versa and can display
behavior that is qualitatively different than the rest of the
distribution. While this represents a small effect on the
bin migration, the changes in shape may cause problems with the
additional smoothness constraints in the SVD method. Thus
contributions to each slice less than $0.1$\% were removed and the
rest of the distribution rescaled to keep the integral the same. The
one dimensional slice of this distribution after smoothing for \RFour\ jets in the
0-10\% centrality bin are shown in
Fig.~\ref{fig:analysis:unfolding:response_slices}. Checks were performed to test the
sensitivity to this cutoff that showed no significant change in the
final results with this cut removed.
\begin{figure}[htb]
\centering
\includegraphics[width =1\textwidth] {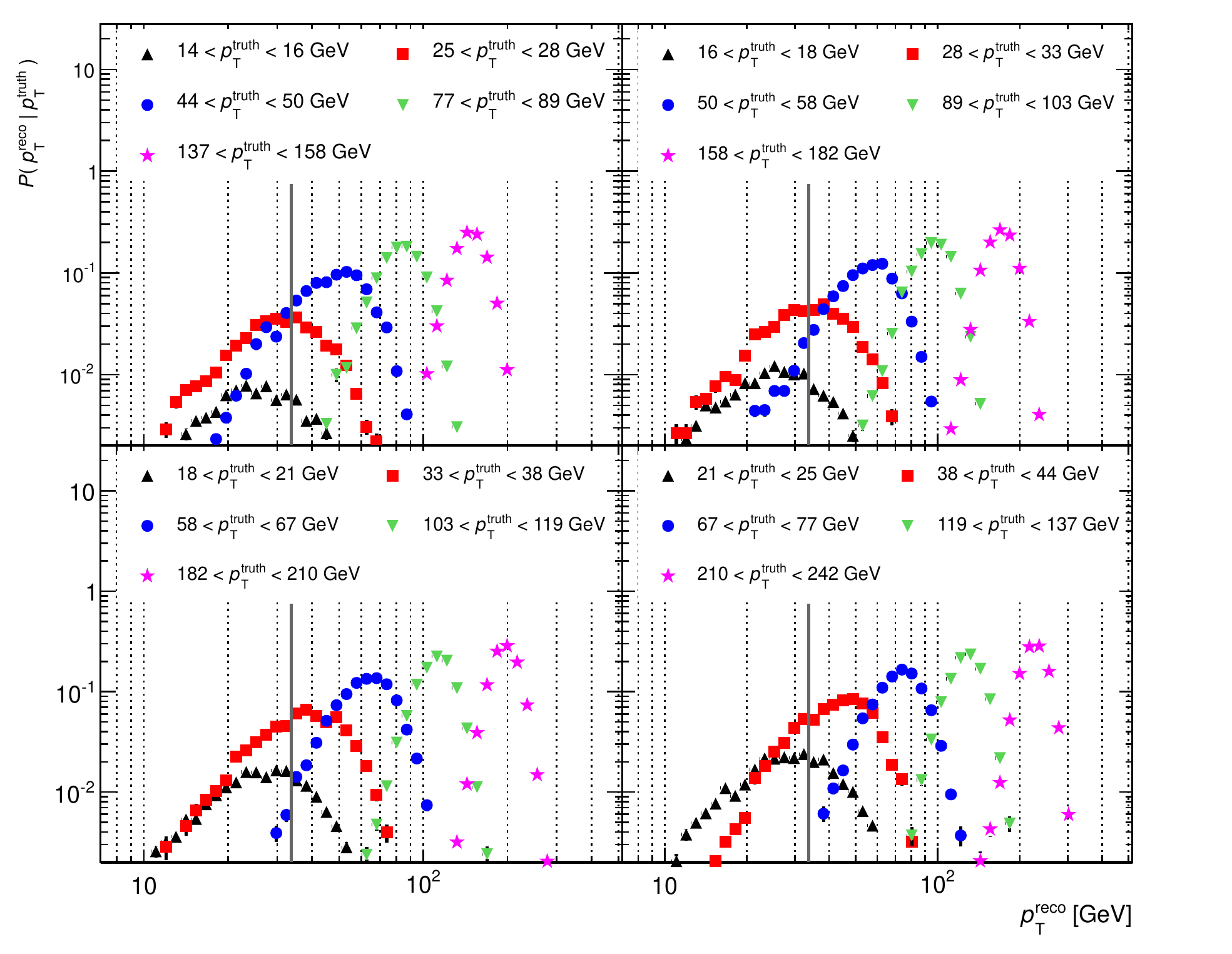}
\caption{1D projections at fixed $\pt^{\mathrm{truth}}$ of the response matrix
  after smoothing for EM+JES \RFour\ jets in the 0-10\% centrality
  bin. A vertical line at 34 \GeV\ is shown to indicate the minimum
  value of reconstructed \pt\ used in the unfolding.}
\label{fig:analysis:unfolding:response_slices}
\end{figure}

The binning used for the spectra and response matrices cover
the range $7 < \pt < 1000 \GeV$. The bin boundaries for the
$i^{\,\mathrm{th}}$ bin, ${\pt^{+}}_i$ and ${\pt^{-}}_i$, were chosen
such that $ {\pt^{+}}_i/{\pt^{-}}_i=\alpha $ is
independent of $i$ and fixed by the number of bins. The unfolding was
performed on an interval,
$[\pt^{\mathrm{low}},\pt^{\mathrm{high}} ]$, of this range chosen
based on requirements of the unfolding procedure. To allow
proper bin migration in both directions the range of \pt\ values allowed for
the unfolded spectrum must exceed the range of the reconstructed
values on both sides ($\pt^{\mathrm{low}\,\mathrm{truth}} <
\pt^{\mathrm{low}\,\mathrm{reco}}$ and $\pt^{\mathrm{high}\,\mathrm{truth}} >
\pt^{\mathrm{high}\,\mathrm{reco}}$). The unfolded result should then only be used
over the reconstructed \pt\ range, possibly excluding the most
extremal bins on either edge of this range due to bias. The linear
system is under-determined unless there are at least as many
reconstructed bins as truth (unfolded) bins on their respective
ranges. A bin selection satisfying these requirements is summarized in
Table~\ref{tbl:pt_binning}.

\begin{table}
\centerline{
\begin{tabular}{|c | c | c | c | c | c  | c |} \hline
& $N_{\mathrm{bins}}$ & $\alpha$ & Bin low & 
Bin high & $\pt^{\mathrm{low}}$ [\GeV] & $\pt^{\mathrm{high}}$ [\GeV] \\ \hline
Truth & 35 &1.15& 5 &  27 & 12 & 320 \\ \hline
Reco & 60 & 1.09&20 & 44 & 34 & 270 \\ \hline
\end{tabular}
}
\caption{Binning for truth and reconstructed spectra and response
  matrices showing the minimum and maximum \pt\ ranges and the
  corresponding histogram bins.}
\label{tbl:pt_binning}
\end{table}

\subsection{Statistical Uncertainties}
\label{section:analysis:unfolding:stat_errors}
The measured covariance is used to generate statistical uncertainties
in the unfolded spectrum by using a pseudo-experiment technique. A
pseudo-experiment consists constructing
a new input spectrum, which is then unfolded.  The procedure uses the measured covariance of the data
to generate a new spectrum consistent with correlated Poisson fluctuations of the measured spectrum.
1000 pseduo-experiments are performed, and the statistical covariance of the distribution over
all such pseudo-experiments is a measure of the effect of the statistical uncertainty of the
measured spectrum on the unfolded result.

The effect of statistical fluctuations in the MC used to obtain the
response matrix is computed in a similar fashion. In this case each
pseudo-experiment uses a different response matrix, which is a
possible statistical fluctuation of the input response matrix. These
fluctuations are generated by using the error on the matrix element to
generate a new value of the matrix element consistent with a Poisson
fluctuation of the original. The total statistical covariance is taken to be the sum of the 
covariance matrices obtained from these two procedures, which is
equivalent to summing the uncertainties in quadrature.

The unfolding method systematically associates features of
the unfolded spectrum with statistical fluctuations and damps their
contributions to the result through regularization. It is expected
that increases in the regularization parameter may artificially reduce the
sensitivity of the unfolding procedure to statistical fluctuations in
the input spectrum. This effect is shown in
Fig.~\ref{fig:analysis:unfolding:statistical_errors_kvals}, where the statistical
errors on the \Rcp\ have been evaluated using the pseudo-experiment technique for
values of $k$ from 4 to 10, with the statistical errors on the
measured spectrum before unfolding shown for reference. The maximal
value of $k$ was chosen to be 10, which is
approximately half the
number of bins in the reconstructed sample with finite counts. This
corresponds to the Nyquist frequency of the system: the highest $k$
value that can be included without introducing aliasing effects in the
signal decomposition and reconstruction. 
\begin{figure}[htb]
\centering
\includegraphics[width =1\textwidth] {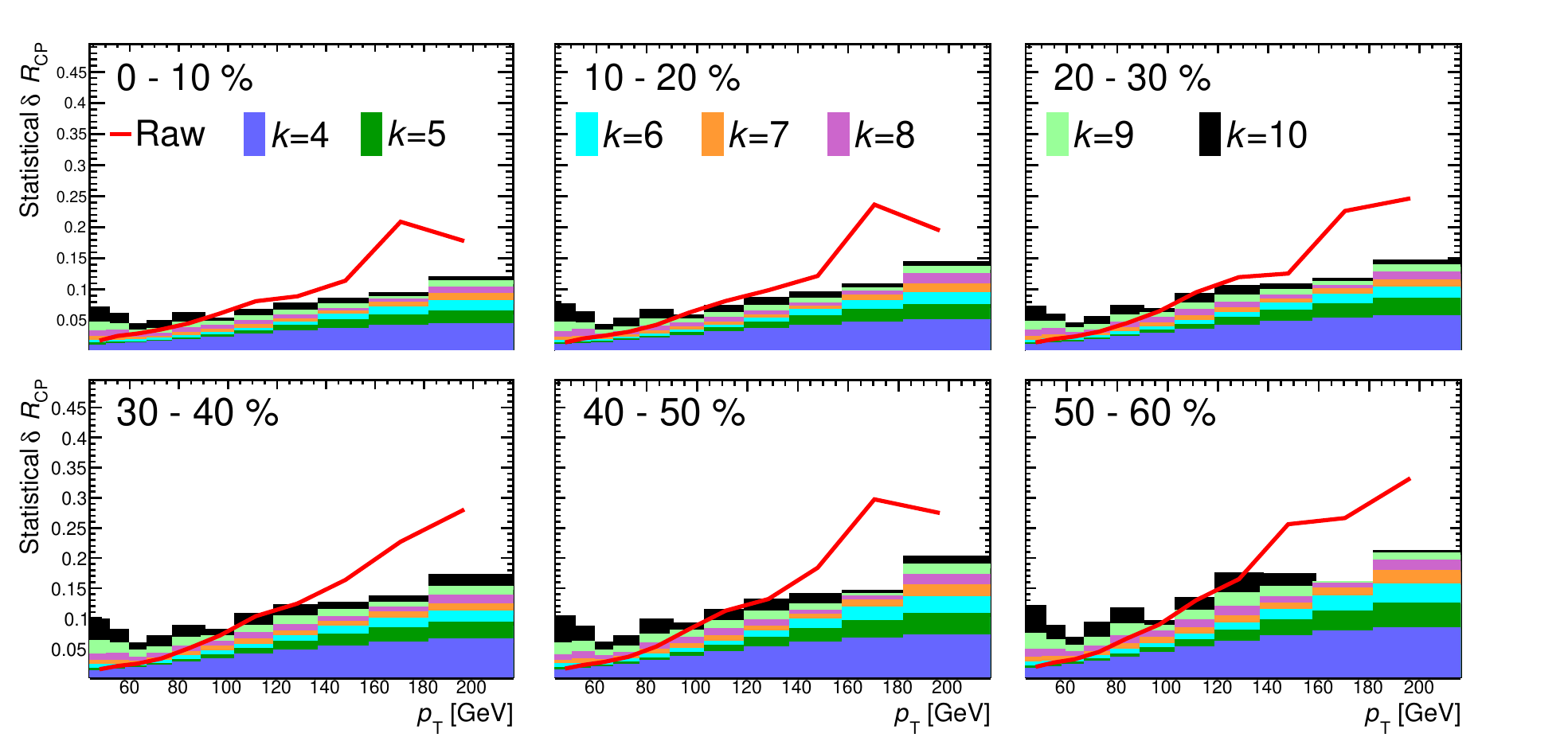}
\caption{Statistical errors on the unfolded \Rcp\ evaluated using the
  pseudo-experiment method for $4\leq k \leq 10$ in various centrality
bins for \RFour\ jets. Statistical errors on the raw spectrum before
unfolding are shown as a red curve.}
\label{fig:analysis:unfolding:statistical_errors_kvals}
\end{figure}

\subsection{$d$ Distributions and Refolding}
\label{section:analysis:unfolding:d_dist}
The choice of the regularization parameter is driven by the
distribution of $d$ values as defined in
Eq.~\ref{eqn:analysis:unfolding:d_definition}. The coefficient $d_i$ is the
projection of the data (error normalized) onto the direction
associated with the $i$th largest singular value. For values of
$|d_i|\gg 1$, the corresponding equation in the now diagonal linear
system is statistically well determined. Values of $|d_{i}|$ of order
unity are consistent with statistical fluctuations in the data that
select out a particular singular value. The contributions from these equations, which become
large when the matrix is inverted, must be regulated. From these
considerations, the value of $k$ is chosen where the $d$ distribution
transitions from statistically relevant equations to equations
consistent with statistical fluctuations. The $d$ distributions for
the \RFour\ jets are shown in Fig.~\ref{fig:analysis:unfolding:d_val_EM4}. $k=5$ was
chosen for all centrality bins and $R$ values.
\begin{figure}[htb]
\centering
\includegraphics[width =1\textwidth] {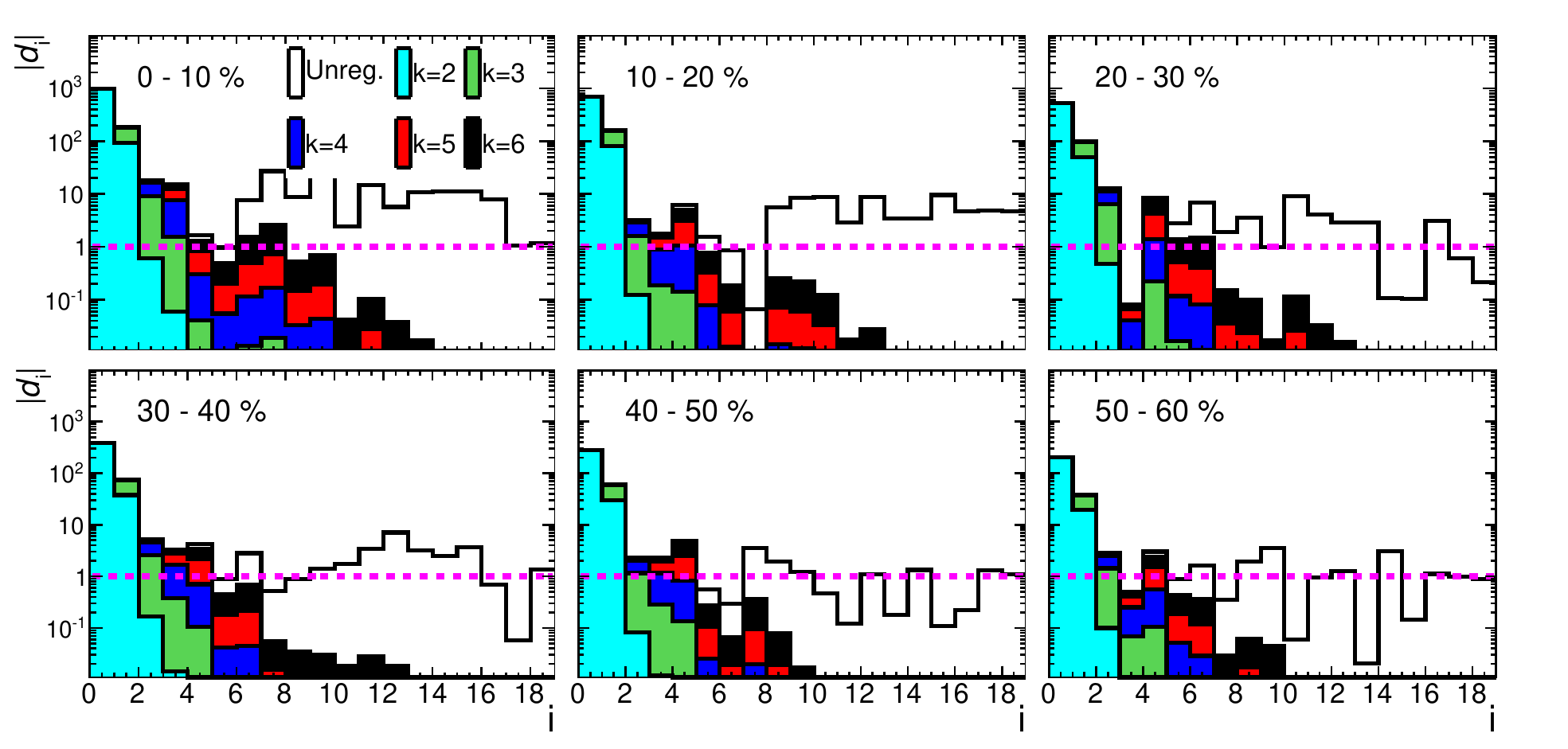}
\caption{$d$ distributions for EM+JES \RFour\ jets in various
  centrality bins, with regularized distributions overlaid with
  different values of $k$.}
\label{fig:analysis:unfolding:d_val_EM4}
\end{figure}

Choosing $k$ to be too small (or equivalently choosing $\tau$ too
large) will over-regularize the result by
removing relevant information from the response matrix. Thus the
choice of $k$ must be made in conjunction with a refolding test, where the unfolded spectrum is multiplied by the full,
unregulated response matrix and compared to the input
spectrum. To make this comparison more rigorous both in justifying
the choice of $k$ and providing a mechanism for analyzing the
systematics, Eq.~\ref{eqn:analysis:unfolding:regularized_sv} was extended to
consider continuous $\tau$ values. The refolded spectrum is compared to the original measured
spectrum, and the degree of closure of the operations is quantified by
a $\chi^2$,
\begin{equation}
\chi^2=(\mathbf{Ax}-\mathbf{b})^{T}\mathbf{B}^{-1}(\mathbf{Ax}-\mathbf{b})\,,
\end{equation}
where $\mathbf{B}$ is the measured covariance of the data. Since $\mathbf{x}$ is the
solution to the unfolding problem constructed from $\mathbf{b}$, the
fluctuations on $\mathbf{b}$ and the refolded spectrum $\mathbf{Ax}$ are not
statistically independent. This means that this $\chi^2$ is not a sum
of squares of independent variables with unit normal distribution, and
does not follow a $\chi^2$ probability distribution. Despite this
limitation, this variable, especially its behavior as a function of $\tau$, is still useful in describing the refolding
closure so long as it is not misinterpreted. 

The distribution of the refolded
$\chi^2$ as a function of $\tau$ is shown in
Fig.~\ref{fig:analysis:unfolding:chi2_vs_tau}. The growth in $\chi^2$ at large $\tau$
in the neighborhood of the smallest $k$ values shows that such
regularizations remove enough information to prevent closure in the
refolding. Small values of $\tau$ maintain a faithful refolding
procedure, with low $\chi^2$, at the expense of amplifying statistical
fluctuations in the unfolded spectrum.

Under-regularization will allow statistical fluctuations to become
amplified in the unfolding. Once an appropriate choice of \xini\ is
made, the curvature of the weights distribution can be used to
quantify the extent to which this effect occurs. There are two sources of curvature in
the weight distribution: statistical fluctuations and differences
between the unfolded spectrum and \xini. The optimal
regularization point occurs when the statistical fluctuations have
been suppressed, but differences between \xini\ and the unfolded
spectrum are not smoothed away. As the sources of curvature are
changing near this value of $\tau$ it is expected that the curvature
distribution will change shape in this region. The curvature as a
function of $\tau$ is shown overlaid with the $\chi^2$ distributions in
Fig.~\ref{fig:analysis:unfolding:chi2_vs_tau}. Regions of flatness in the $\chi^2$ overlap in $\tau$ with the qualitative changes in
the curvature distribution.

The procedure for extracting statistical uncertainties on the unfolded
spectrum is discussed in Section~\ref{section:analysis:unfolding:stat_errors}. This
addresses the question of how
sensitive the unfolded spectrum is to statistical fluctuations on the
measured spectrum for a given regularization procedure (i.e. choice of
$k$ or $\tau$). It is possible that were a different statistical
fluctuation of the data observed, a different regularization procedure
could have been chosen, which is more sensitive to statistical
fluctuations. This would occur if different fluctuations had different
$\chi^2$ and curvature distributions as functions of $\tau$. To check
the stability of these distributions, the pseudo-experiment method was
again used where these two distributions were generated for many
possible fluctuations of the data. The stability of the two
distributions under statistical fluctuations indicates that choice of
regularization would not change under statistical fluctuations, and
further justifies the procedure described in Section~\ref{section:analysis:unfolding:stat_errors} as a
full measure of the statistical uncertainty.

\begin{figure}[htb]
\centering
\includegraphics[width =1\textwidth] {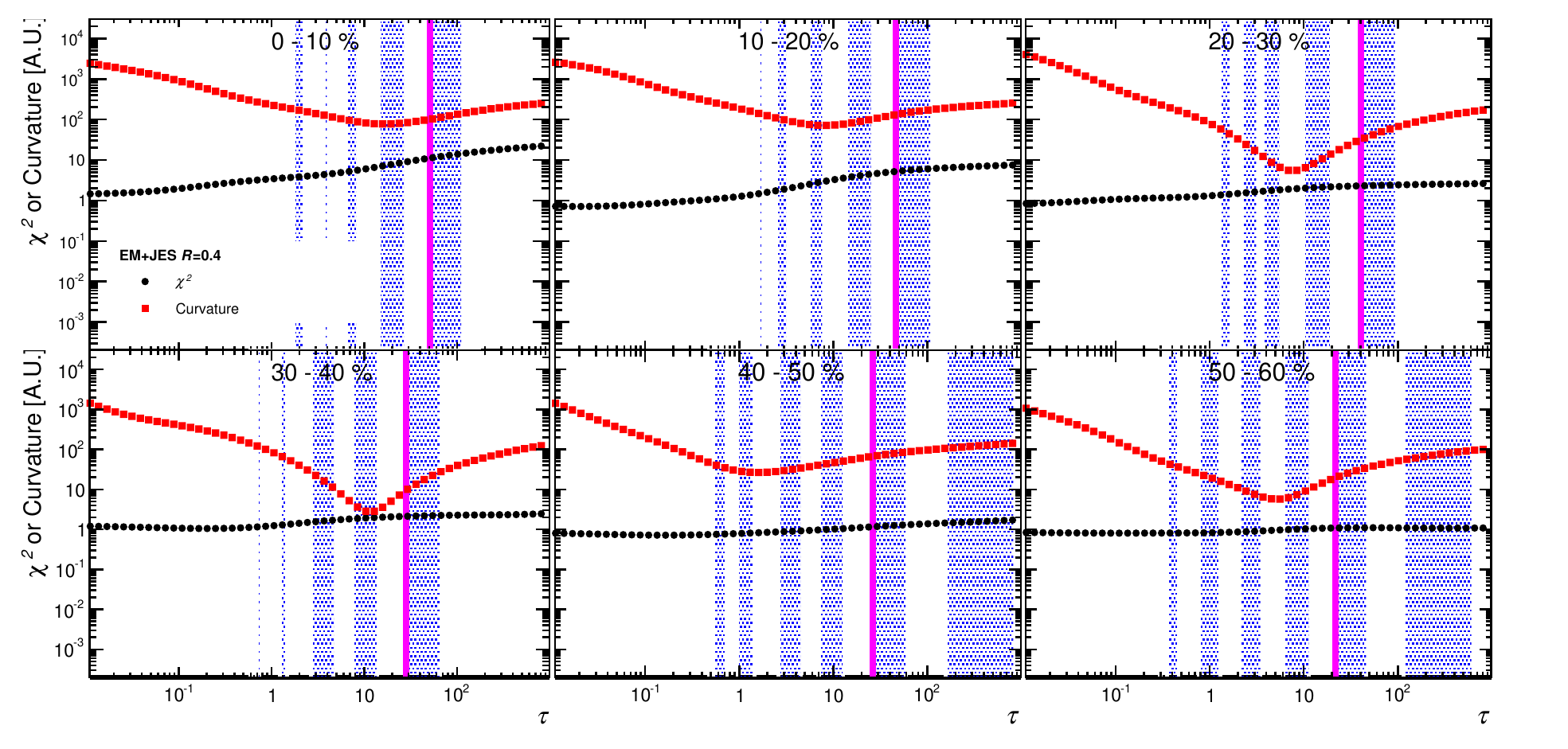}
\caption{$\chi^2$ and curvature distributions as a function of the regularization
  parameter $\tau$ for EM+JES \RFour\ jets in various
  centrality bins. In the discrete case, values of $\tau$ are fixed to
  be the singular values, $\tau_k=s^2_k$. The alternating colors indicate a range of $\tau$
  values between those allowed in the discrete case, with the $\tau$
  corresponding to a fixed $k$ value occurring at the boundaries of the
  colored regions.}
\label{fig:analysis:unfolding:chi2_vs_tau}
\end{figure}

\section{Systematic Uncertainties}
\label{section:analysis:systematics}
Systematic uncertainty on the final, unfolded result is estimated both
from ingredients from the MC into the unfolding and the unfolding
procedure itself. For some sources, the uncertainties in \Rcp\ are
point-wise correlated, e.g.~in \Rcp\ vs.~\pt\ changes in the JER
cause all points in the curve to move up/down together. Such
systematics are indicated by continuous lines drawn around the values
of the \Rcp. Which systematics show point-wise correlations depend
which variable the \Rcp\ is being plotted against and are summarized in
Table~\ref{tbl:sys_err_calc}. As the \Rcp\ is a ratio between two
different centrality bins, errors that are correlated in centrality are
taken from the variation of the \Rcp\ due to that
systematic. Errors on the \Rcp\ due to sources that are uncorrelated
in centrality are taken as independent errors on the
central and peripheral spectra and combined in
quadrature. Errors that show no point-wise correlation are shown
by shaded boxes and indicate the maximal variation of the points
independent of one another due to such systematics. The relative error
of each individual contribution to the systematic error on \Rcp\
vs \pt\ is shown in Figs.~\ref{fig:analysis:systematics:sys_err_23} and~\ref{fig:analysis:systematics:sys_err_45}.
\begin{figure}[tbph]
\centering
\includegraphics[width=1\textwidth] {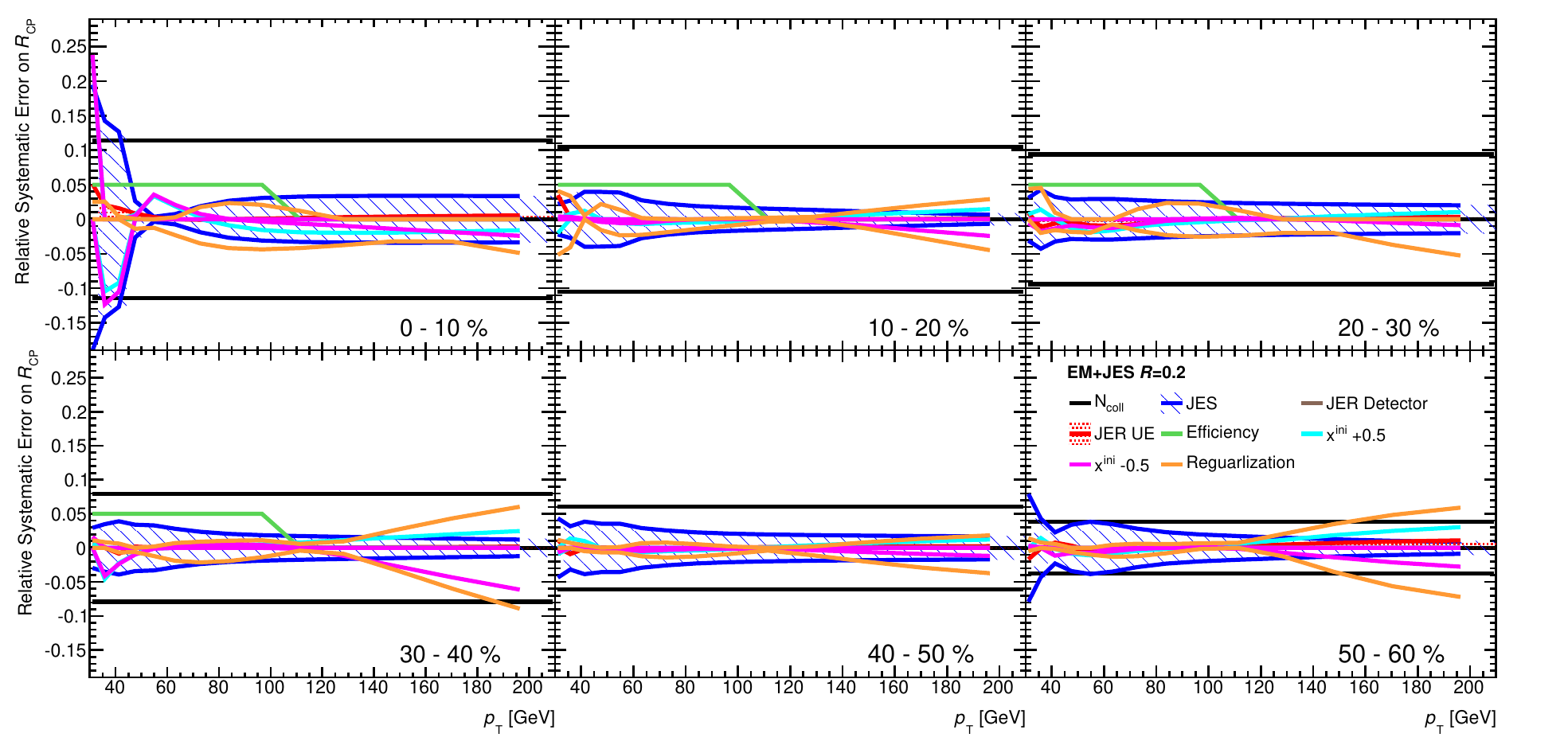}
\includegraphics[width=1\textwidth] {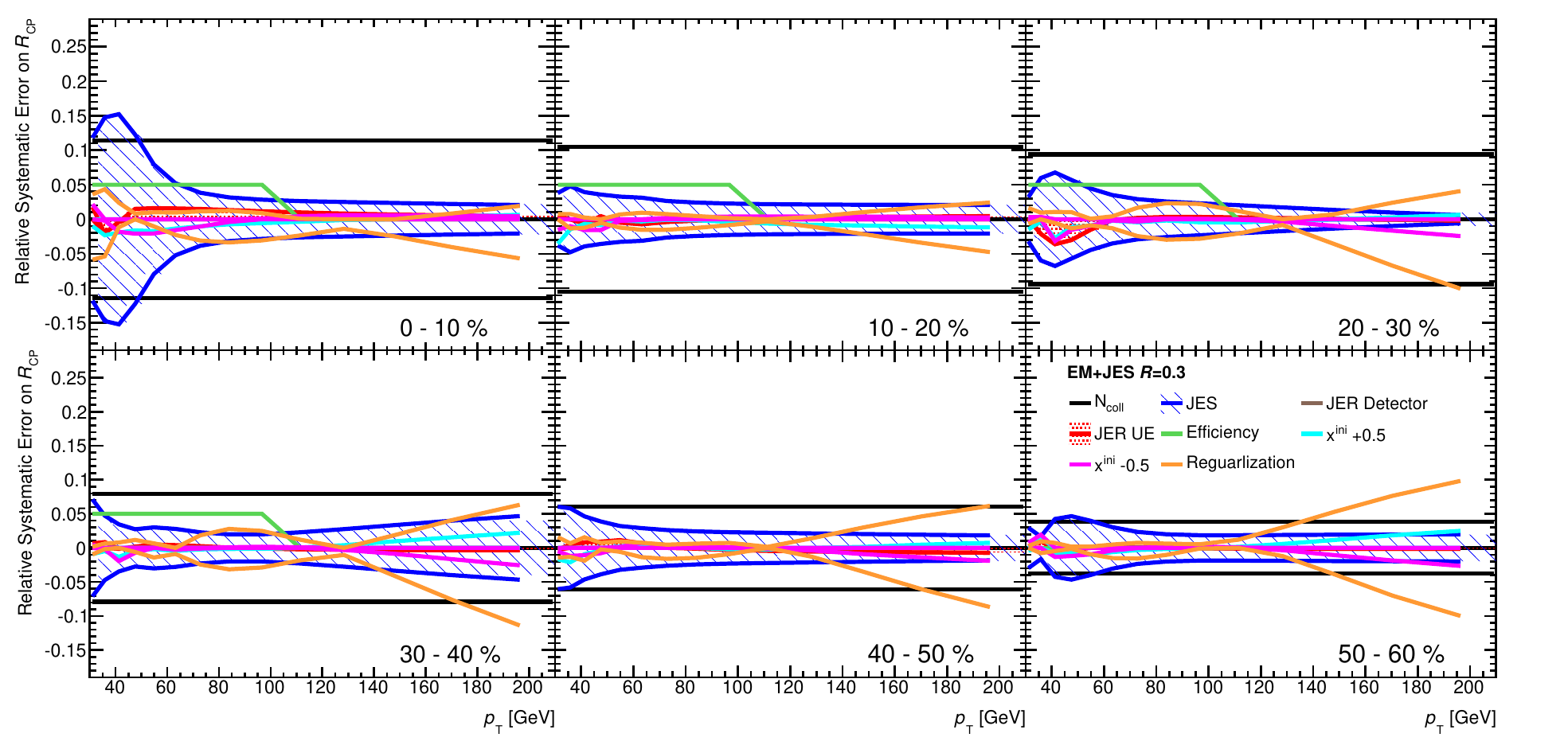}
\caption{Contributions of various systematic uncertainties on \Rcp\ as a function of jet \pt\ for all centrality
  bins expressed as a relative error for the \RTwo\ (top) and
  \RThree\ (bottom) jets.}
\label{fig:analysis:systematics:sys_err_23}
\end{figure}

\begin{figure}[tbph]
\centering
\includegraphics[width=1\textwidth] {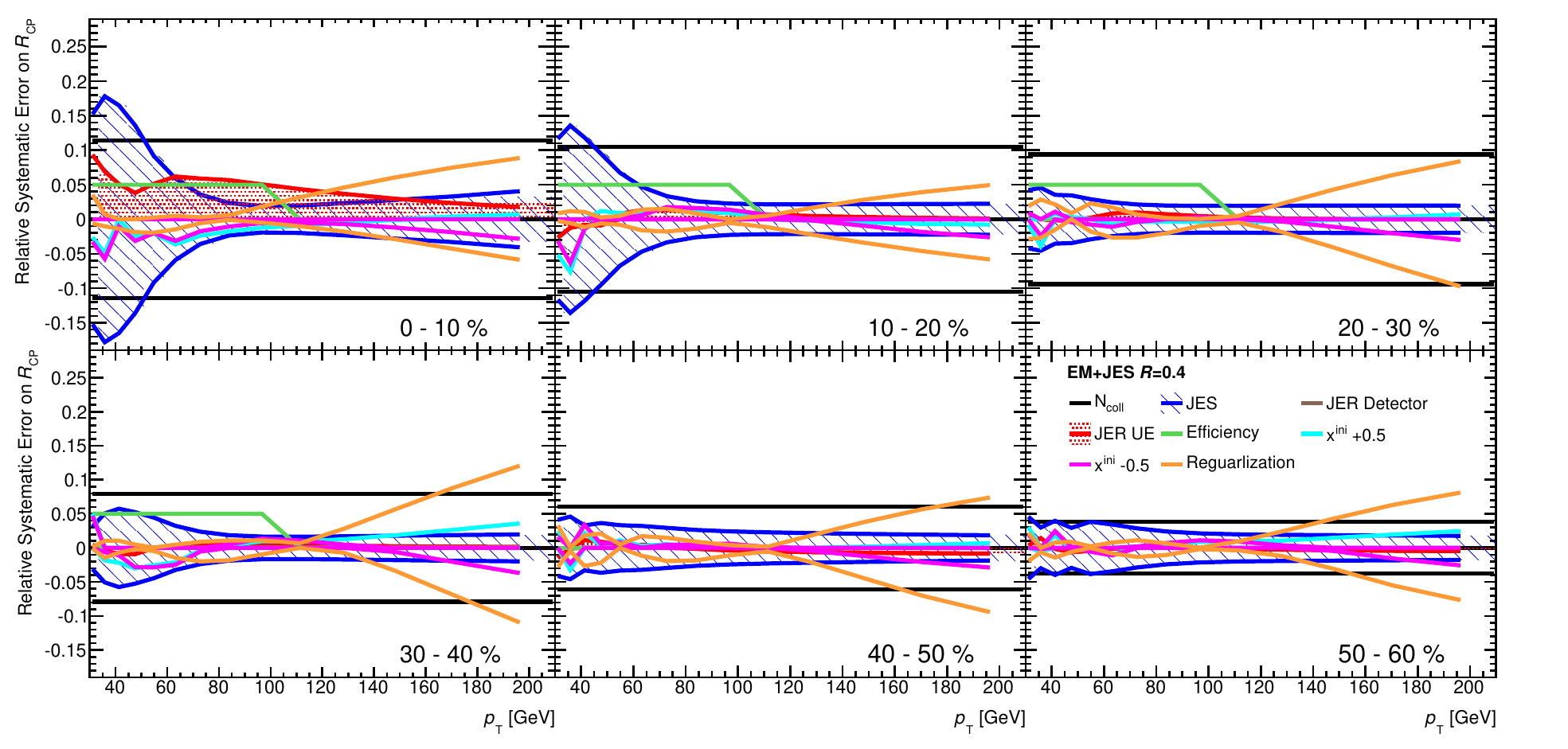}
\includegraphics[width=1\textwidth] {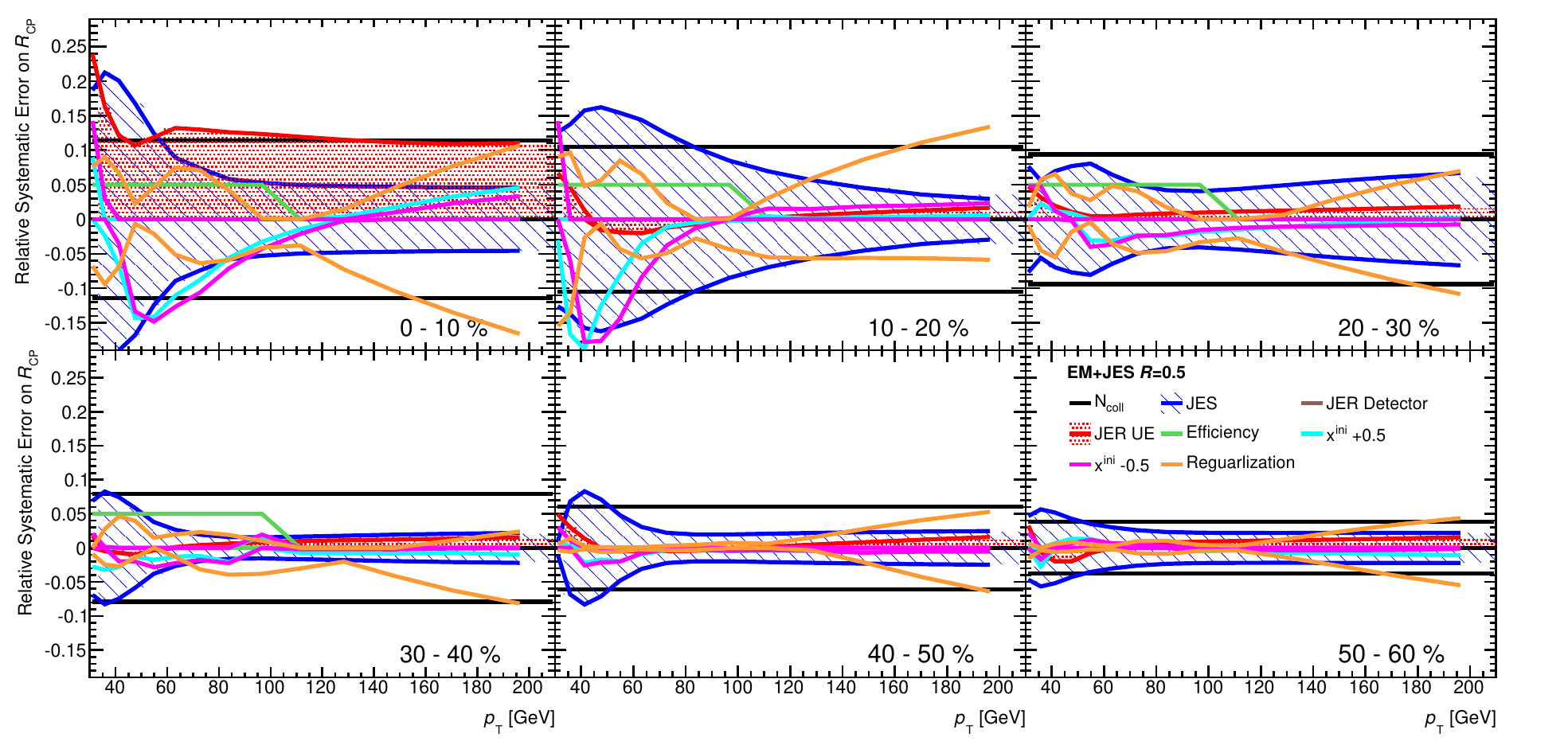}
\caption{Contributions of various systematic uncertainties on \Rcp\ as a function of jet \pt\ for all centrality
  bins expressed as a relative error for the \RFour\ (top) and \RFive\
(bottom).}
\label{fig:analysis:systematics:sys_err_45}
\end{figure}

\begin{table}
\centerline{
\begin{tabular}{|l || c | c | c |} \hline
Systematic & $\pt$ & Centrality & $R$ \\ \hline
Regularization & & & \\ \hline
\xini\ & $\times$ & $\times$ & \\ \hline
Efficiency & $\times$ & $\times$ & \\ \hline
JES & $\times$ & $\times$ & $\times$ \\ \hline
JER & $\times$ & & $\times$ \\ \hline
$\Ncoll$ & $\times$ & $\times$ & $\times$ \\ \hline
\end{tabular}
}
\caption{Systematics on \Rcp\ with point-wise correlations when
  plotted as a function of different variables are denoted by an
  $\times$. }
\label{tbl:sys_err_calc}
\end{table}

\subsection{Regularization}
The systematic uncertainties associated with the unfolding procedure
were estimated by considering the sensitivity of the procedure to the
regularization parameter. The unfolding was performed with a different
value of $k$ and the variation in the unfolded spectra were taken as
systematic errors. The errors on the central and peripheral spectra
were then combined to determine an error on the~\Rcp. Independent
systematics were generated for both increased and decreased
regularization by unfolding with $k=4$ and $k=6$.

An alternative estimate of the systematic uncertainty associated with
under-regularization was also tested and gave similar results as
regularizing with the $k=4$. This evaluation was motivated by linear least squares problems where the
systematics are commonly estimated by considering variation of the
$\chi^2$ by 1 about the minimum value
\begin{equation}
\Delta \chi^2 =\chi^2_{+}-\chi^2=\chi^2/NDF\,.
\end{equation}
The unfolding problem is not a strict minimization of the $\chi^2$ as $\chi^2$ monotonically
increases with $\tau$. Thus for the purposes of estimating the effect of
increasing $\tau$, the system is effectively at a minimum and the
above estimate was used to determine a $\tau$ value. The spectra were
unfolded with this $\tau$ and the variation in the spectra were used
as above to generate a systematic on the \Rcp.

\subsection{\xini\ and Efficiency}
One advantage of the SVD procedure is that the unfolded result is less
sensitive to the choice of \xini\ than the truth spectrum
used in a bin-by-bin unfolding procedure. To test this sensitivity the
power in Eq.~\ref{eqn:xini_def} was varied $\pm0.5$.

In peripheral collisions the fake jet rate is expected to be
vanishingly small above $40$~\GeV. The fraction of jets remaining after fake rejection as a function of
$\pt$ in peripheral data is consistent with the same quantity in all
centralities in the MC as shown in Fig.~\ref{fig:corrections:fake:data_ratio}. There is a small deviation at the highest
$\pt$ where the MC does not fully saturate. However, this disagreement
is present in all centrality bins and divides out in the \Rcp. No
systematic error was taken from this aspect of the efficiency.

The uncertainty on the jet reconstruction efficiency prior to any fake
rejection was estimated by comparing efficiencies in the HIJING sample
and the overlay. The uncertainty in the ratio of central to peripheral
efficiencies was taken conservatively to be $5\%$ for $\pt < 100$~\GeV\
for all centralities and $R$ values.

\subsection{Energy Scale}
Systematic uncertainties in the jet energy scale can arise both from
disagreement between the MC, where the numerical inversion constants
are derived, and data, as well as systematic shifts in the energy
scale between different centralities associated with the underlying
event subtraction. The former source only affects the absolute jet
energy scale uncertainty which divides out in the \Rcp. This is true
to the extent to which the changes in absolute scale do not distort the
unfolding differently in different
centralities.

Relative changes in the energy scale between central and peripheral
collisions are estimated by combining two studies. The difference in
closure in the MC JES analysis, as presented in
Sec.~\ref{section:jet_rec:performance} provides sensitivity to the effect of
the background subtraction on the JES. Additionally, the in-situ JES
validation comparing the track-calorimeter jet energy scales in
Sec.~\ref{sec:validation:JES:track_jet} provides further
constraints as it is sensitive to differences in jet fragmentation between
the two centralities.

Using these studies as input, systematic uncertainties were estimated
on the energy scale for each centrality relative to the peripheral and
are shown in Table~\ref{tbl:jes_uncertainty}. The MC samples were
processed adding an additional shift, $\pt\rightarrow \pt+\Delta \pt$, to the \pt\ of each
reconstructed jet when filling out the response matrices. A constant
shift was applied above 70~\GeV. Below the shift value was a linear
function of \pt, increasing to twice the high-\pt\ value at 40~\GeV,
\begin{equation}
\frac{\Delta \pt}{\pt}=\left\{\begin{array}{cc}f\left(1+\frac{70-\pt}{30}\right) & \pt < 70 \GeV \\f & \pt > 70 \GeV\end{array}\right.
\end{equation}
where $f$ is the value given in Table~\ref{tbl:jes_uncertainty}. The data was
unfolded using the new response matrices resulting in a modified
spectrum used to compute a new \Rcp. Variations from this \Rcp\ from
the original in each \pt\ bin were taken as symmetric errors due to
this systematic.
\begin{table}
\centerline{
\begin{tabular}{|r||r| r | r | r| r|r|} \hline
            $R$ &        0 - 10 \% &      10 - 20 \% &      20 - 30 \% &      30 - 40 \% &      40 - 50 \% &      50 - 60 \%\\ \hline\hline
             0.2 &            0.5  \% &            0.5  \% &            0.5  \% &            0.5  \% &            0.5  \% &            0.5  \%\\ \hline
             0.3 &            1.0  \% &            0.5  \% &            0.5  \% &            0.5  \% &            0.5  \% &            0.5  \% \\ \hline
             0.4 &            1.5  \% &            1.0  \% &            0.5  \% &            0.5  \% &            0.5  \% &            0.5  \%\\ \hline
             0.5 &            2.5  \% &            1.5  \% &
             1.0  \% &       0.5  \% &            0.5  \% &            0.5  \%\\    \hline
 \end{tabular}
 }
 \caption{Energy scale shifts relative to peripheral used to generate
   systematic uncertainties.}
 \label{tbl:jes_uncertainty}
 \end{table}

\begin{figure}[tbhp]
\includegraphics[width =1\textwidth] {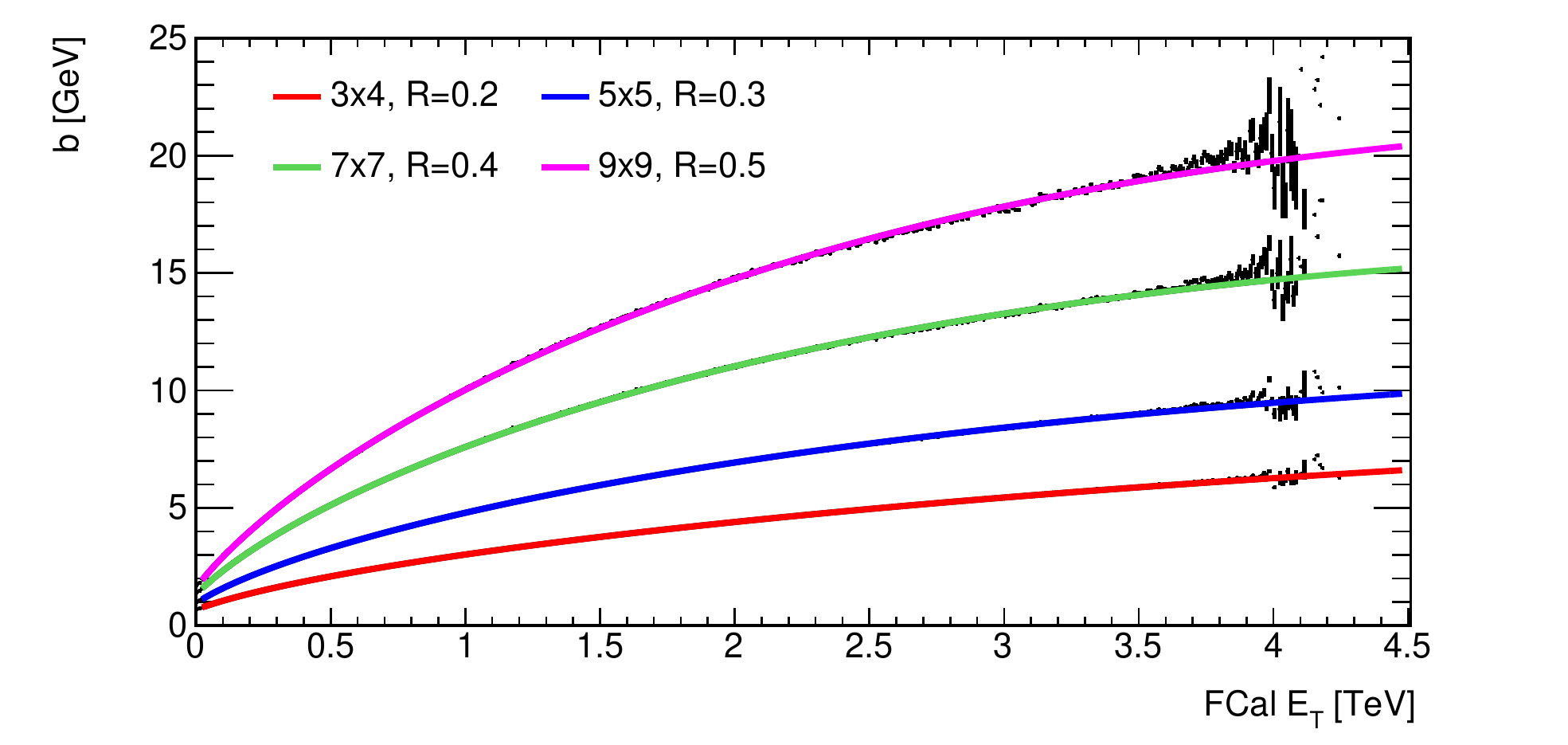}
\caption{$N\times M$ tower fluctuation distributions and
  fits. Values of $N$ and $M$ associated with jet $R$ values are
  indicated in the legend.}
\label{fig:analysis:systematics:flucutations_fit}
\end{figure}

\subsection{Energy Resolution}
To account for systematic uncertainties coming from disagreement
between the jet energy resolution in data and MC, the unfolding
procedure was repeated with a modified response matrix. This response
matrix was generated by repeating the MC study but with modifications
to the $\Delta \pt$ for each matched truth-reconstructed jet
pair. A ``detector'' systematic was constructed to account for
uncertainty between the data and MC without considering the effects of
the underlying event. This procedure follows the recommendation for
JER uncertainty in 2010 jet analyses in \pp. The
\verb=JetEnergyResolutionProvider=
tool~\cite{JERUncertaintyProvider} was used to retrieve the fractional
resolution, $\sigma^{\mathrm{DET}}_{\mathrm{JER}}$ as a function of jet $\pt$ and $\eta$~\cite{ATLAS-CONF-2010-054}. The jet $\pt$ 
was then smeared by
\begin{equation}
\pt \rightarrow \pt\times \mathcal{N}(1,\sigma^{\mathrm{DET}}_{\mathrm{JER}})\,,
\end{equation}
where $\mathcal{N}(0,1)$ is the unit normal distribution. 

A separate systematic uncertainty considering the effect of disagreement of underlying
event fluctuations in the data and MC was also determined.
As discussed in Section~\ref{sec:validation:JER:closure}, MC is
internally consistent between the $N\times N$ tower RMS \et\
distributions and the JER as a function of \et\ as expressed in Eq.~\ref{eqn:jer}. The $b$ term is sensitive to fluctuations in the underlying event,
while the $a$ and $c$ terms are independent of centrality. This is
shown in Fig.~\ref{fig:jetjerfit}, where fits of the JER with
this form only differ significantly in the value of the $b$
parameter. Additionally, the underlying event fluctuations, as
measured by the RMS of the \et\ distribution of $N\times N$
tower groups, is consistent with the $b$ value obtained from the
fit of the JER for jets with the same area. 

The MC fluctuations, which give $b$ as a function of
FCal \et\ are fit with the functional form
\begin{equation}
b(\et^{\mathrm{FCal}})=\left(\frac{\et^{\mathrm{FCal}}
    +\et^0}{\gamma}\right)^{\alpha-\beta \et^{\mathrm{FCal}}}\,,
\end{equation}
which is shown in Fig.~\ref{fig:analysis:systematics:flucutations_fit}, and the values
of the fit parameters $\et^0$, $\alpha$, $\beta$ and $\gamma$ for the different $R$ values are
presented in Table~\ref{tbl:analysis:systematics:fluctuations_fit_parameters} along
with maximal estimates of the MC/data deviation of these distributions
given by the quantity $f$.
\begin{table}
\centerline{
\begin{tabular}{|c||c|
r@{.}l|
r@{.}l|
r@{.}l|
r@{.}l|
r@{.}l|} \hline
$R$
&$N\times M$
&  \multicolumn{2}{|c|}{$f \,[\%]$} 
&  \multicolumn{2}{|c|}{$\et^0 \,\,[\GeV]$}
&   \multicolumn{2}{|c|}{$\alpha$}
& \multicolumn{2}{|c|}{$\beta$}
&   \multicolumn{2}{|c|}{$\gamma\,\, [\GeV]$}\\ \hline
0.2 &        $3\times4$ &      2&5 &               94&4 &     0&622 &   0&0083 &       181&0\\ \hline
0.3 &        $5\times5$ &      2&5 &               81&9 &     0&641 &    0&0133 &      89&3\\ \hline
0.4 &        $7\times7$ &      5&0 &               75&1 &      0&680 &    0&0168 &      50&5\\ \hline
0.5 &       $9\times9$ &       7&5 &             75&4 &     0&712 &    0&0176 &      38&8\\ \hline
\end{tabular}
}
\caption{Fit parameters for background fluctuations in $N\times M$ towers as a function of \ETfcal.}
\label{tbl:analysis:systematics:fluctuations_fit_parameters}
\end{table}

As in the detector systematic, a new response matrix was
constructed. The $\Delta\pt$ was rescaled by an amount
\begin{equation}
\Delta\pt\rightarrow\Delta\pt \frac{\sigma_{\mathrm{JER}}(b(1+g))}{\sigma_{\mathrm{JER}}(b)}\,.
\end{equation}
$b$ is determined from the \ETfcal\ and $g$ is chosen to account for
differences in the fluctuations between data and MC based on
the values of $f$  shown Table~\ref{tbl:analysis:systematics:fluctuations_fit_parameters}. In most central 10\% of
collisions the tower \et\ distributions in the MC are broader than the
data, so the effect of the fluctuations was symmetrically reduced by
using $g=-f$. This systematic was applied as a one-sided error to the
most central bin. Any other differences in the RMS results
from the data having larger downward fluctuations, but smaller upward
fluctuations. To account for this effect the $g=2f$ was chosen for
$\Delta\pt <0$ and $g=-f$ for $\Delta\pt>0$. This was applied as an
asymmetric error to all but the most central bin. In no region of
centrality does the data have larger upward fluctuations than the MC.

\subsection{\Ncoll\ Uncertainties}

The uncertainties in the \Ncoll\ values obtained from the centrality
analysis are shown in Table~\ref{tbl:centrality_bins}. The errors
reported for all centrality bins are partially correlated because
individual variations in the sensitive parameters in the Glauber
calculation cause all \Ncoll\ values to increase or decrease
together. To evaluate the systematic uncertainties in the ratios of
\Ncoll\ between two centrality bins,
$\Rcollcent=\Ncollcent/\Ncollperiph$, that appear in the
expression for \Rcp\ we have directly evaluated  how this ratio varies
according to variations of the sensitive parameters in the
Glauber Monte Carlo. We obtain from this study separate estimates for
the uncertainty in the \Ncoll\ ratios; the results are provided in
Table~\ref{tbl:centrality_bins}.

\subsection{Consistency Checks}
In addition to the systematic variations discussed previously, cross checks were performed by varying aspects of the
analysis. These checks serve to test self-consistency of
the procedure, not to quantify sensitivities to quantities expected to
be uncertain. One such example is the fake rejection scheme, where it
is natural to consider whether the default scheme imposes a
fragmentation bias on the jets included in the spectrum and the \Rcp. An
alternative fake rejection scheme was assessed where the track jet
matching requirement was
replaced by a single track match within $\Delta R < 0.2$ of the jet
axis, where the track is subject to the following selection criteria:
\begin{itemize}
\item $\pt>4$~\GeV
\item $d_0,\,\, z_0\sin\theta < 1.5$~mm
\item $\frac{d_0}{\sigma_{d_0}},\,\,\frac{z_0\sin\theta}{\sigma_{z_0\sin\theta}} < 5.0$
\item $N_{Pixel} \geq 2$
\item $N_{SCT} \geq 7$
\end{itemize}
These requirements are significantly looser than those required on the
tracks input into the track jet reconstruction.  Although these looser matching criteria may lead to
increased sensitivity to fake jets at low \pt, as long as the fake
rate is not dramatically enhanced, the lower \pt\ threshold (4~\GeV\ vs 7~\GeV\ for track jets), should address
any concerns regarding a fragmentation bias. An identical analysis was
performed using measured spectra and response matrices generated from
this alternate scheme. The ratio of the \Rcp\ values to those from the
default scheme is shown in Fig.~\ref{fig:analysis:systematics:alt_fakes}, with the
statistical errors taken from errors generated on the nominal \Rcp\
through the pseudo-experiment method. The \Rcp\ values show good
agreement to the level of a few percent. The residual difference, an
enhancement at low \pt\ that decreases with centrality, is consistent with the expected slight increase
in fakes.
\begin{figure}[tbph]
\centering
\includegraphics[width=0.9\textwidth]{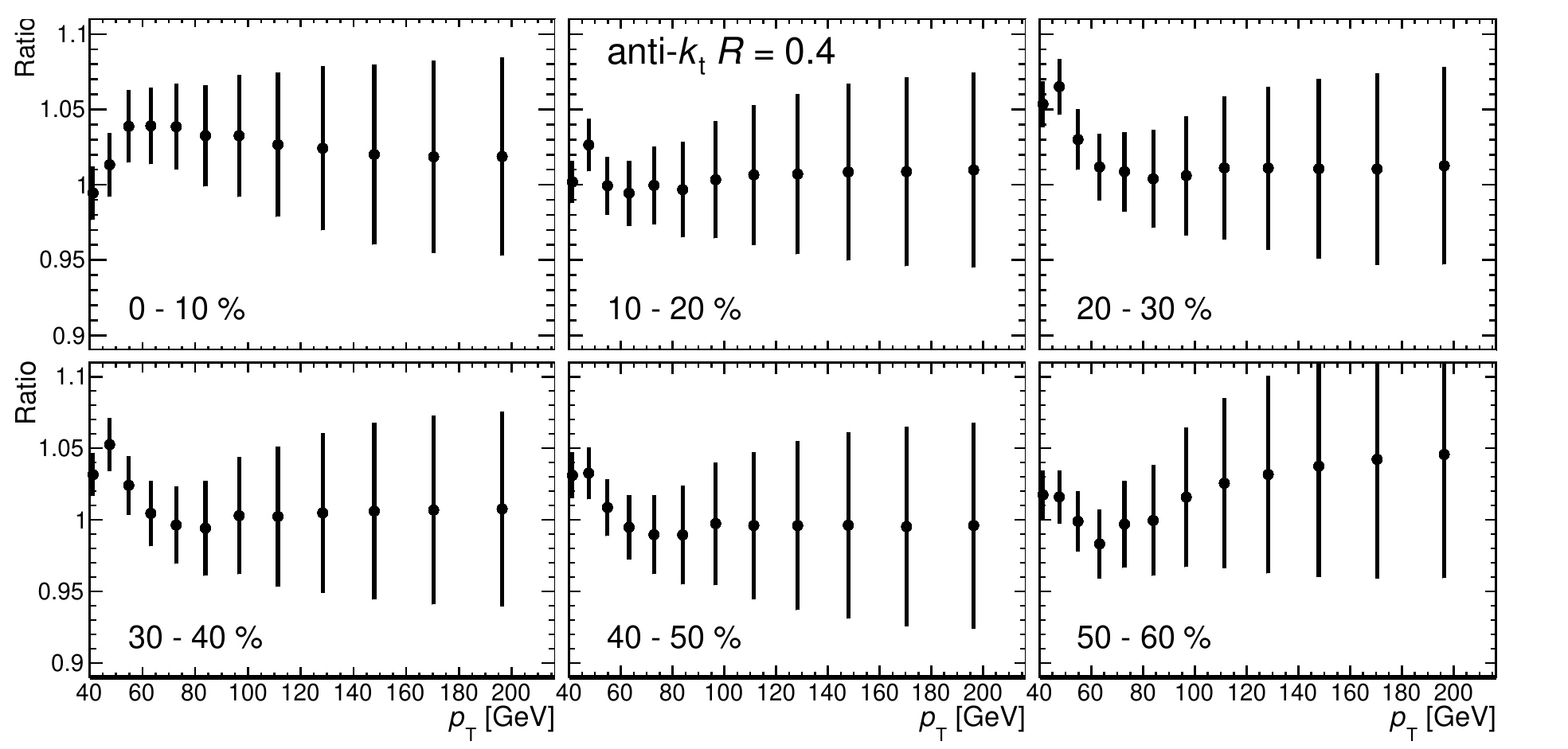}
\caption{Ratio of \Rcp\ values in the alternative fake rejection
  scheme (single track and cluster) to the nominal scheme (track jet
  and cluster) for \RFour\ jets in various centrality
  bins. Statistical errors are taken from the denominator only.}
\label{fig:analysis:systematics:alt_fakes}
\end{figure}

The GCW calibration scheme has a reduced sensitivity to jet width and
flavor content relative to the EM+JES~\cite{Aad:2011he}. A comparison between these two
calibration schemes was performed and the ratio of the \Rcp\ obtained
from the two schemes prior to unfolding is shown for \RFour\ jets in
Fig.~\ref{fig:analysis:systematics:gcw_jes_check}. The distribution shows good
agreement, indicating that the calorimeter response is not
significantly altered for quenched jets.
\begin{figure}[tbph]
\centering
\includegraphics[width=0.9\textwidth]{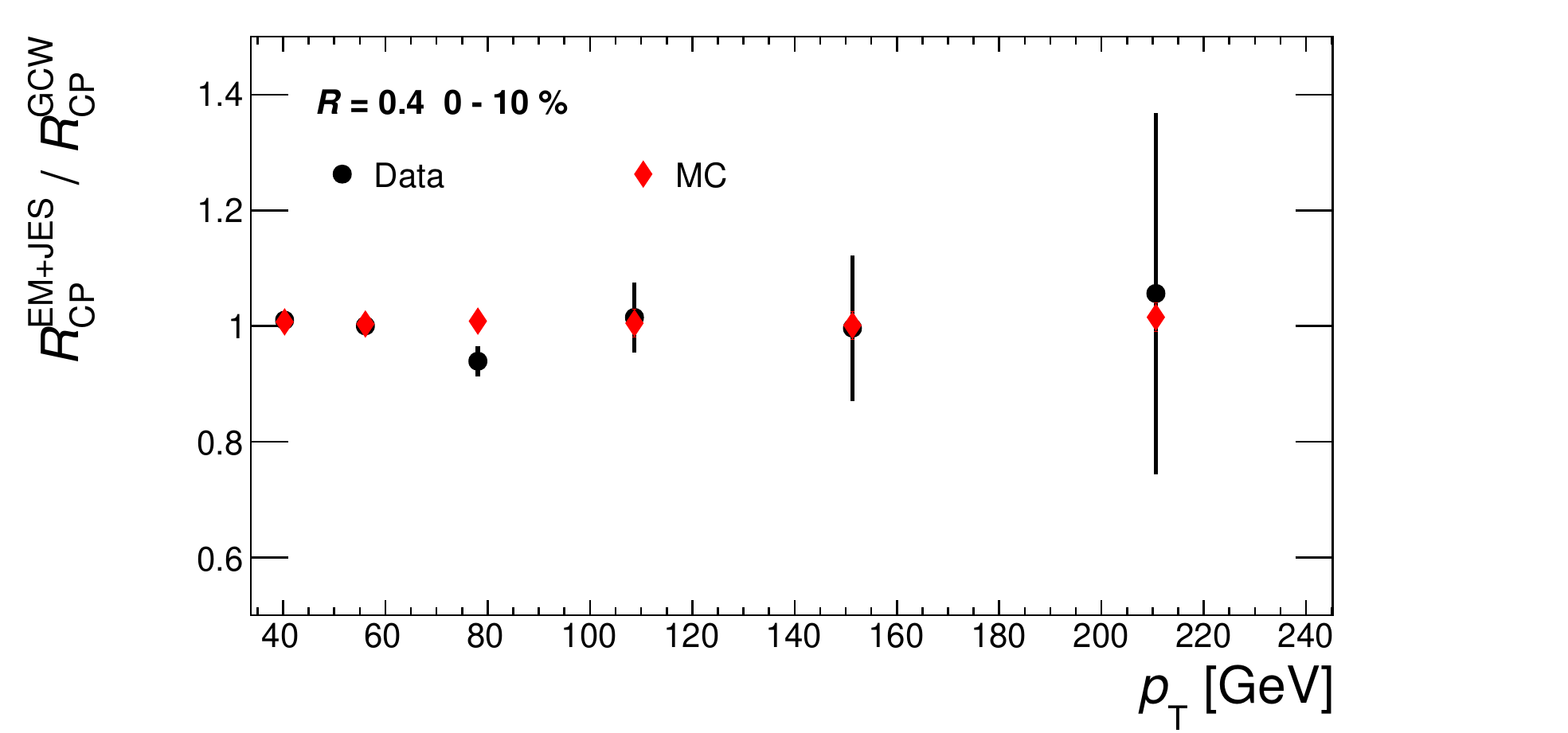}
\caption{The ratio of \Rcp\ for EM+JES and GCW \RFour\ jets before
  unfolding for data (black) and MC (red). Neither distribution shows
  a systematic difference between the two calibration schemes.}
\label{fig:analysis:systematics:gcw_jes_check}
\end{figure}

Consistency checks were performed to investigate any biases introduced by the unfolding procedure or the
choice of binning and \pt\ range. The reconstructed spectrum from the
MC sample was treated like the data and unfolded. The unfolded
spectrum was then compared directly to the MC truth, and the ratio is
taken to quantify the closure of the unfolding procedure. Since the final physics result is the
\Rcp, the bias on this quantity was also investigated. For each
centrality bin the unfolded spectrum was divided by the unfolded
spectrum in the 60-80\% centrality bin to form an \Rcp\ (with no
\Ncoll\ factor). The analogous quantity was constructed from the MC
truth and the ratio of the \Rcp's was taken. Finally, a refolding test
was preformed. When the unfolding was performed on the data, only this
last check was possible. Any relationship between the unfolding
non-closure in MC truth and MC refolding provides a context for
interpreting the refolding in data as a potential bias on the unfolded
data result.

The results of this test are shown in
Fig.~\ref{fig:analysis:systematics:mc_unfolding}, regularized with $k=5$, for the
\RFour\ jets. The MC truth non-closure shows some
structure at the 5\% level which is common to all centrality bins and
$R$ values (not shown). This structure is associated with the finite
regularization parameter, and the effects can be removed by increasing
$k$. Since the MC sample has much higher statistics than the data, the
optimal value of $k$ is likely to be much higher in the MC than in
the data, and $k$ can be increased without introducing strong effects
from statistical fluctuations in the reconstructed spectrum. To
facilitate the comparison with data, the value of $k$ was chosen to be near the
values appropriate for the data, and thus the MC unfolded spectrum is
slightly over-regularized. Nevertheless the comparison shows good
closure and insignificant systematic bias in the procedure. The \Rcp\
shows even less bias as some of the structure is common to all
centrality bins. The refolded reconstructed spectrum shows similar
structure, suggesting that this check will be sensitive to residual
bias in the data.
\begin{figure}[h]
\centering
\includegraphics[width =1\textwidth] {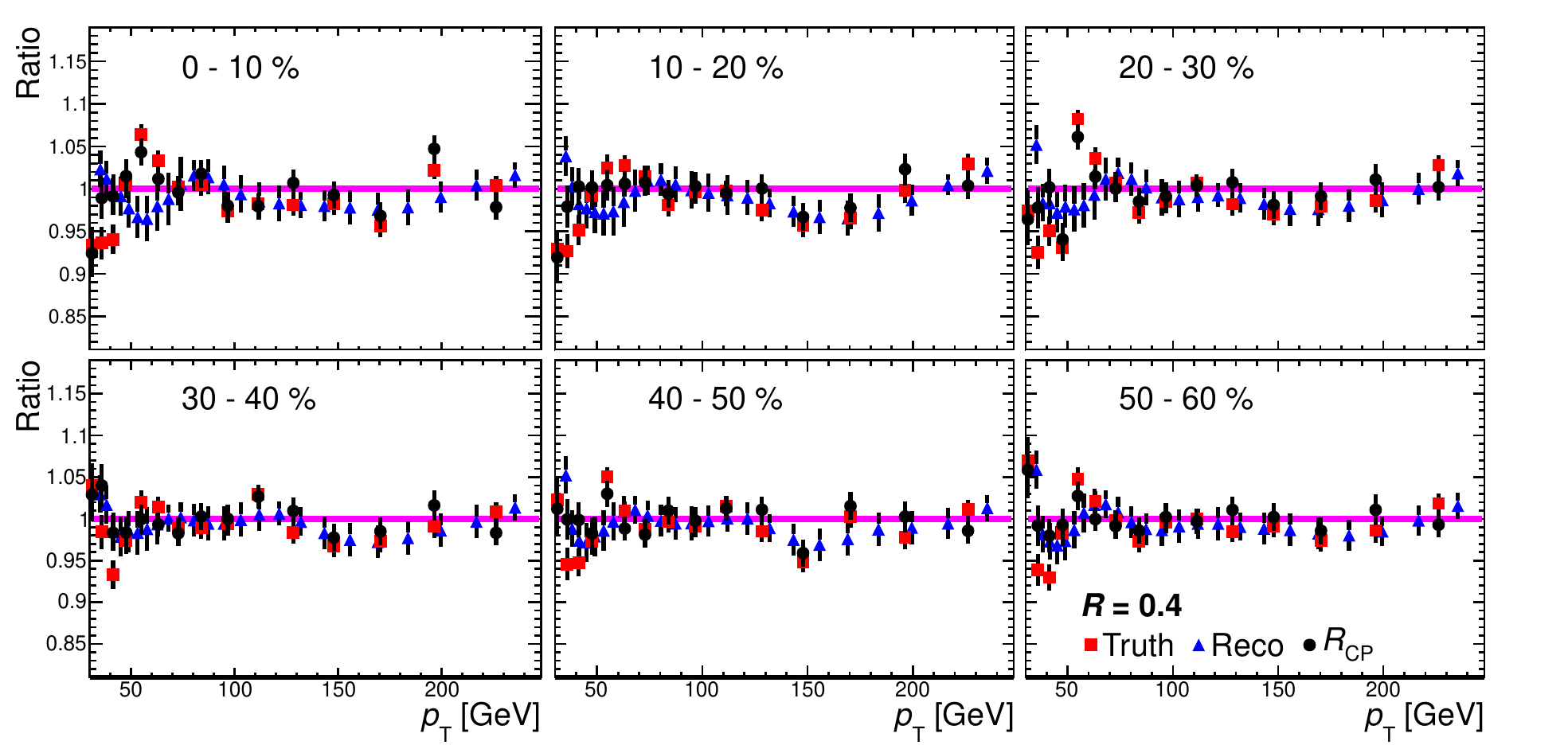}
\caption{Results of the unfolding self-consistency check performed on
  the MC for \RFour\ jets in different centrality bins. The ratio of
  unfolded to MC truth for both the spectrum and \Rcp\ is shown in red
and black respectively. The refolding on the reconstructed spectrum is
shown in blue to facilitate a comparison with the analogous quantity
in the data.}
\label{fig:analysis:systematics:mc_unfolding}
\end{figure}

\clearpage
\chapter{Results}
\label{section:results}
\section{Jet Spectra and \Rcp}
\label{section:results:rcp}

\subsection{Unfolded Spectra}
The unfolded spectra are shown in
Figs.~\ref{fig:results:rcp:unfolded_spectra_23} and~\ref{fig:results:rcp:unfolded_spectra_45} including the effect of the
efficiency correction. The ratios of these spectra to the raw
distribution as well as the refolding closure test are shown in
Figs.~\ref{fig:results:rcp:correction_ratios_23} and ~\ref{fig:results:rcp:correction_ratios_45}. The
unfolding procedure alters the measured statistical correlations
between bins resulting in large off-diagonal covariances. These
covariances were obtained using the method described in Section~\ref{section:analysis:unfolding:stat_errors} and are
shown in Figs.~\ref{fig:results:rcp:unfolded_cov_cent} and~\ref{fig:results:rcp:unfolded_cov_R}.
\begin{figure}[tbph]
\centering
\includegraphics[height=3.5 in] {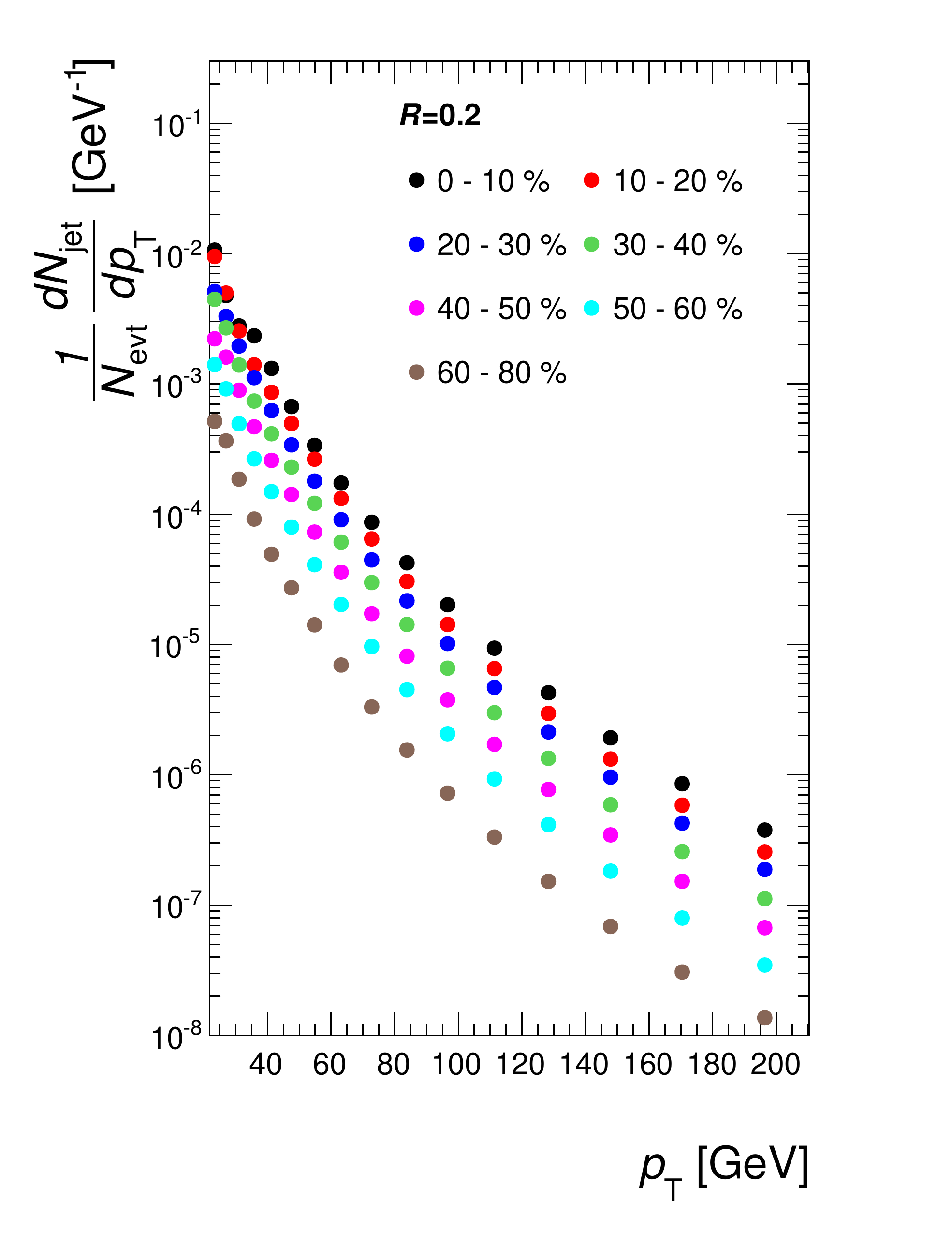}
\includegraphics[height=3.5 in] {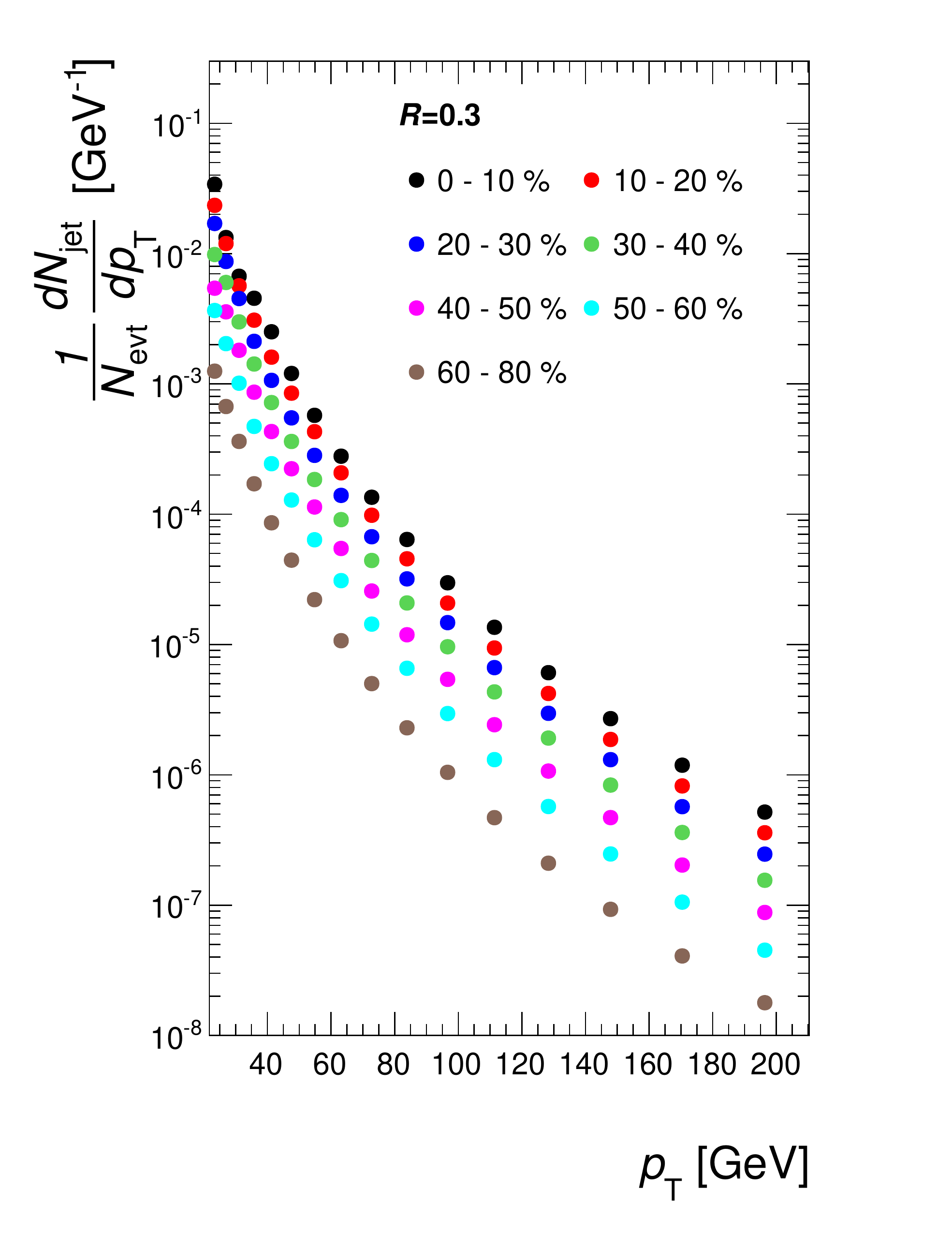}
\caption{Per event yields after unfolding and efficiency correction for  \RTwo\ (left) and
  \RThree\ (right) jets.}
\label{fig:results:rcp:unfolded_spectra_23}
\end{figure}
\begin{figure}[tbph]
\centering
\includegraphics[height=3.5 in] {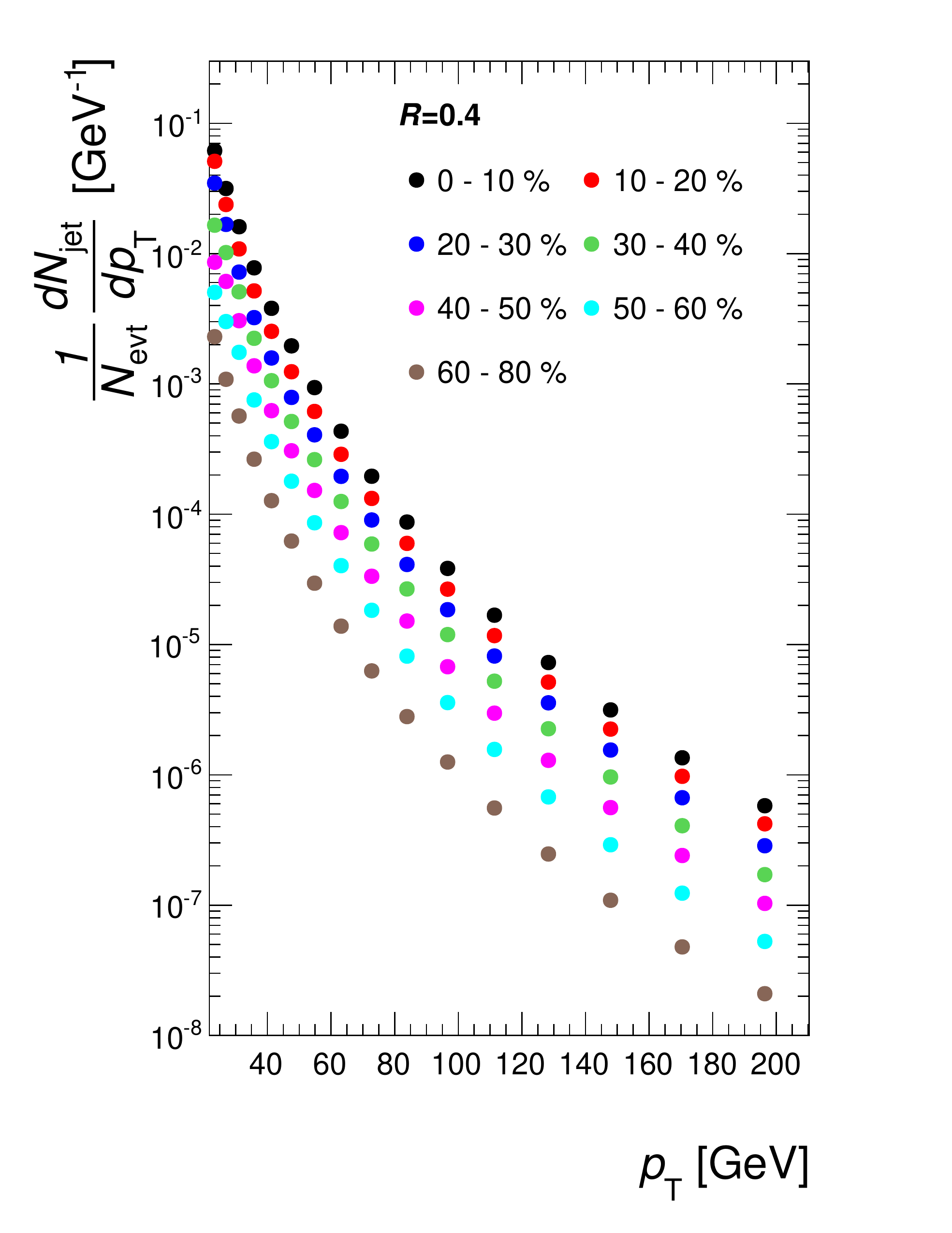}
\includegraphics[height=3.5 in] {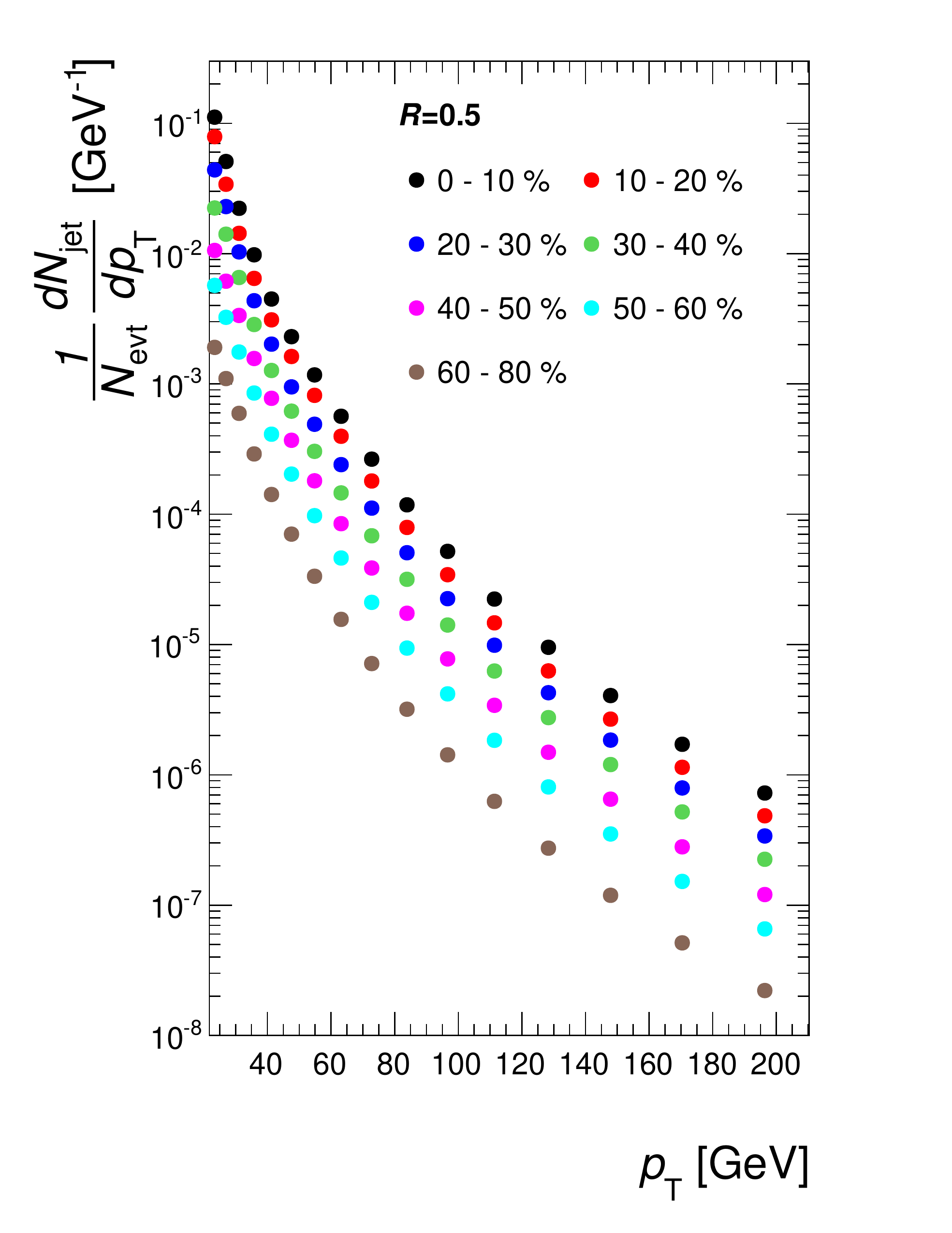}
\caption{Per event yields after unfolding and efficiency correction for \RFour\ (left) and
  \RFive\ (right) jets.}
\label{fig:results:rcp:unfolded_spectra_45}
\end{figure}
\begin{figure}[htb]
\centering
\includegraphics[width =1\textwidth] {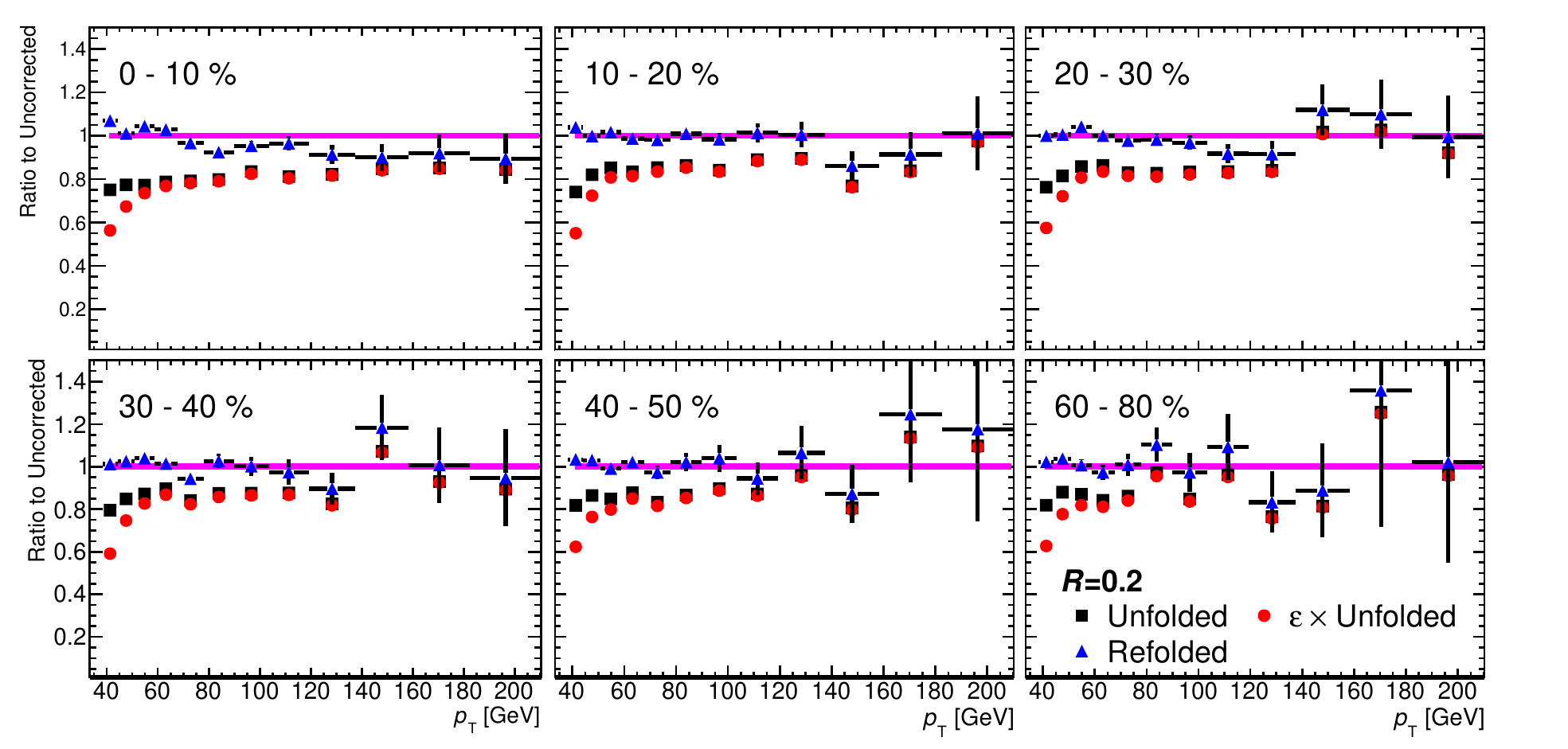}
\includegraphics[width =1\textwidth] {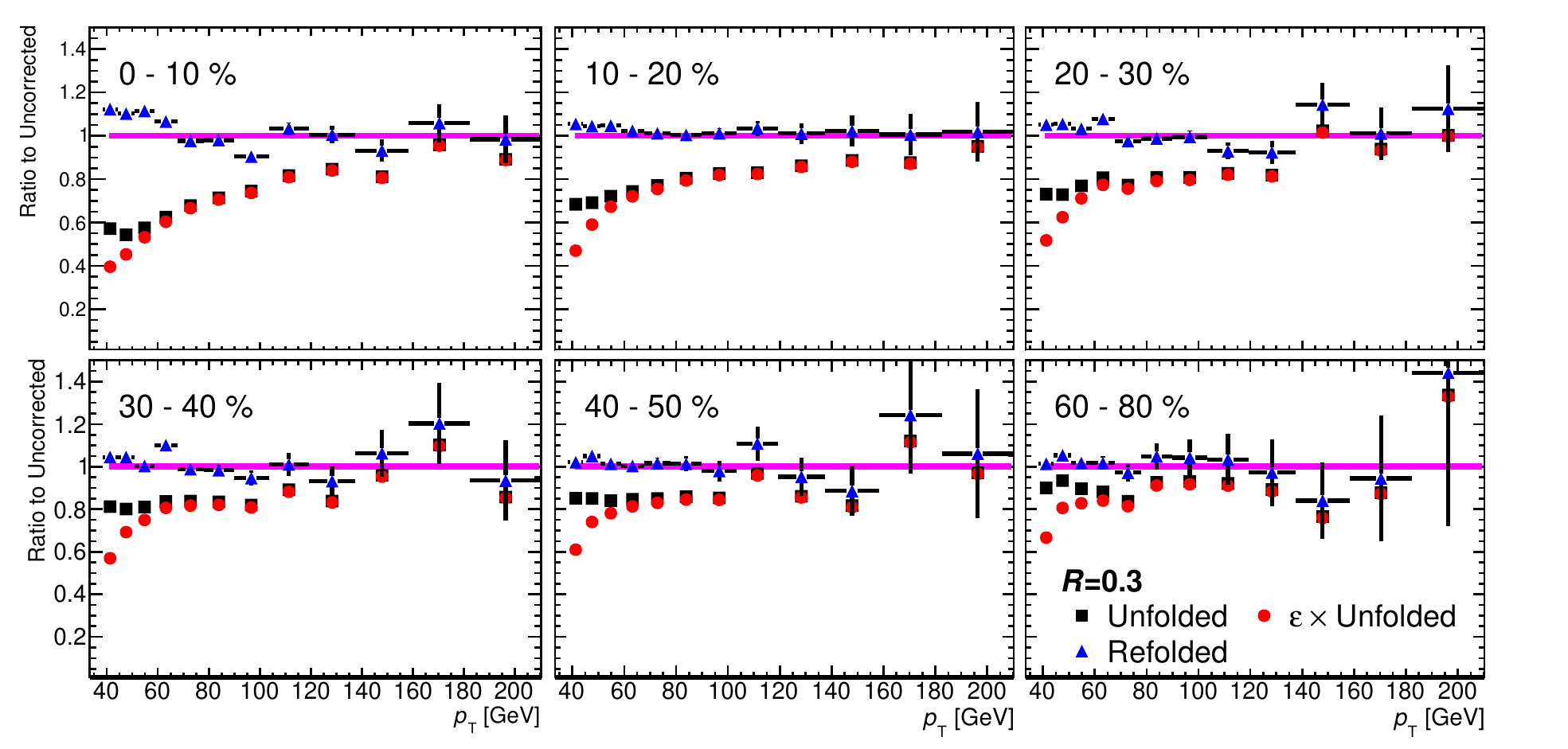}
\caption{Results of unfolding in centrality
  bins presented as various ratios to the raw spectrum for the \RTwo\ (top) and \RThree\
  (bottom) jets. The ratio of the unfolded spectrum
both before and after efficiency correction to the input spectrum are
shown in red and black respectively. The ratio of the refolded
spectrum to the input distribution is shown in blue with the error
bars indicating the relative error on the data. For all
centrality bins this ratio is near unity, indicating good closure in
the unfolding procedure.}
\label{fig:results:rcp:correction_ratios_23}
\end{figure}
\begin{figure}[htb]
\centering
\includegraphics[width =1\textwidth] {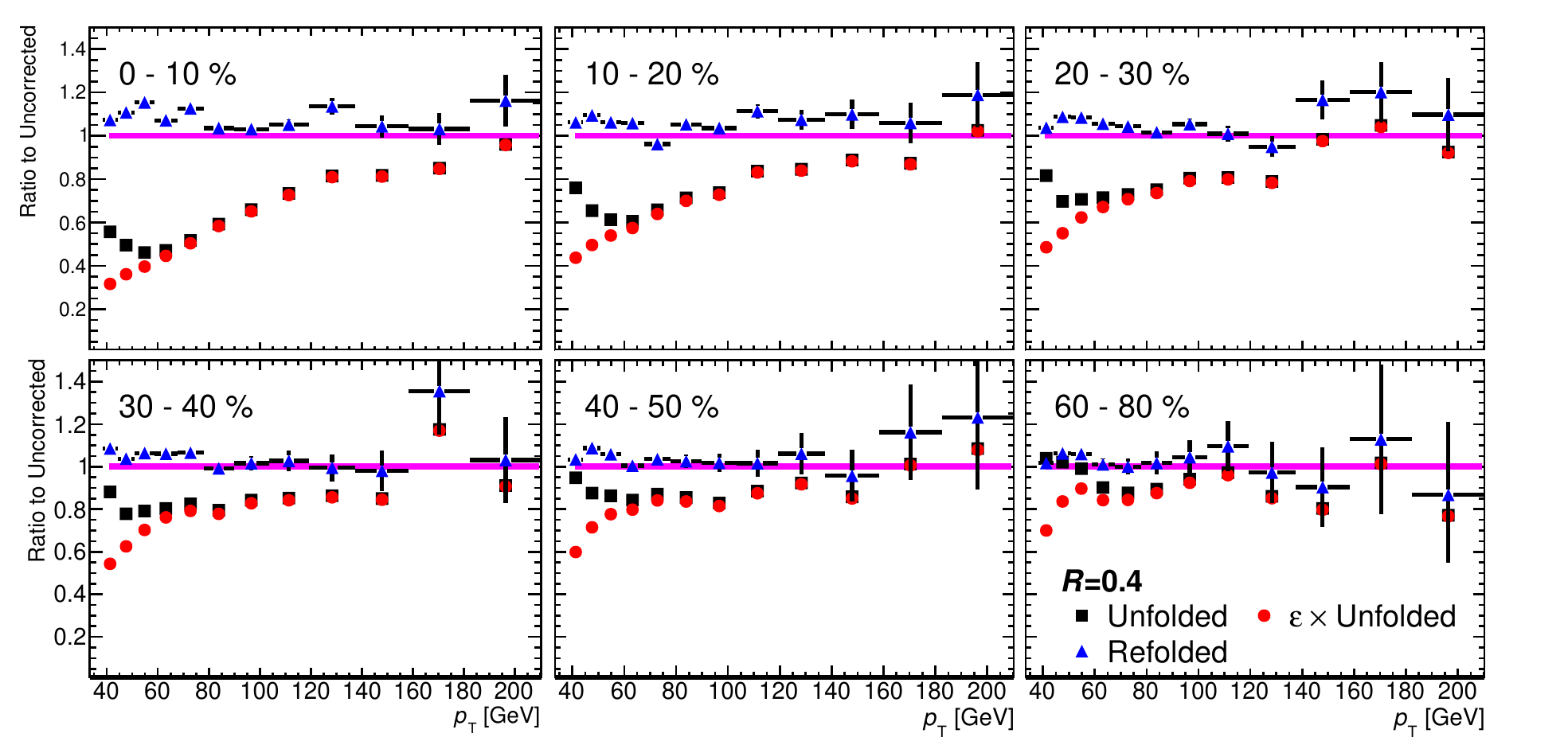}
\includegraphics[width =1\textwidth] {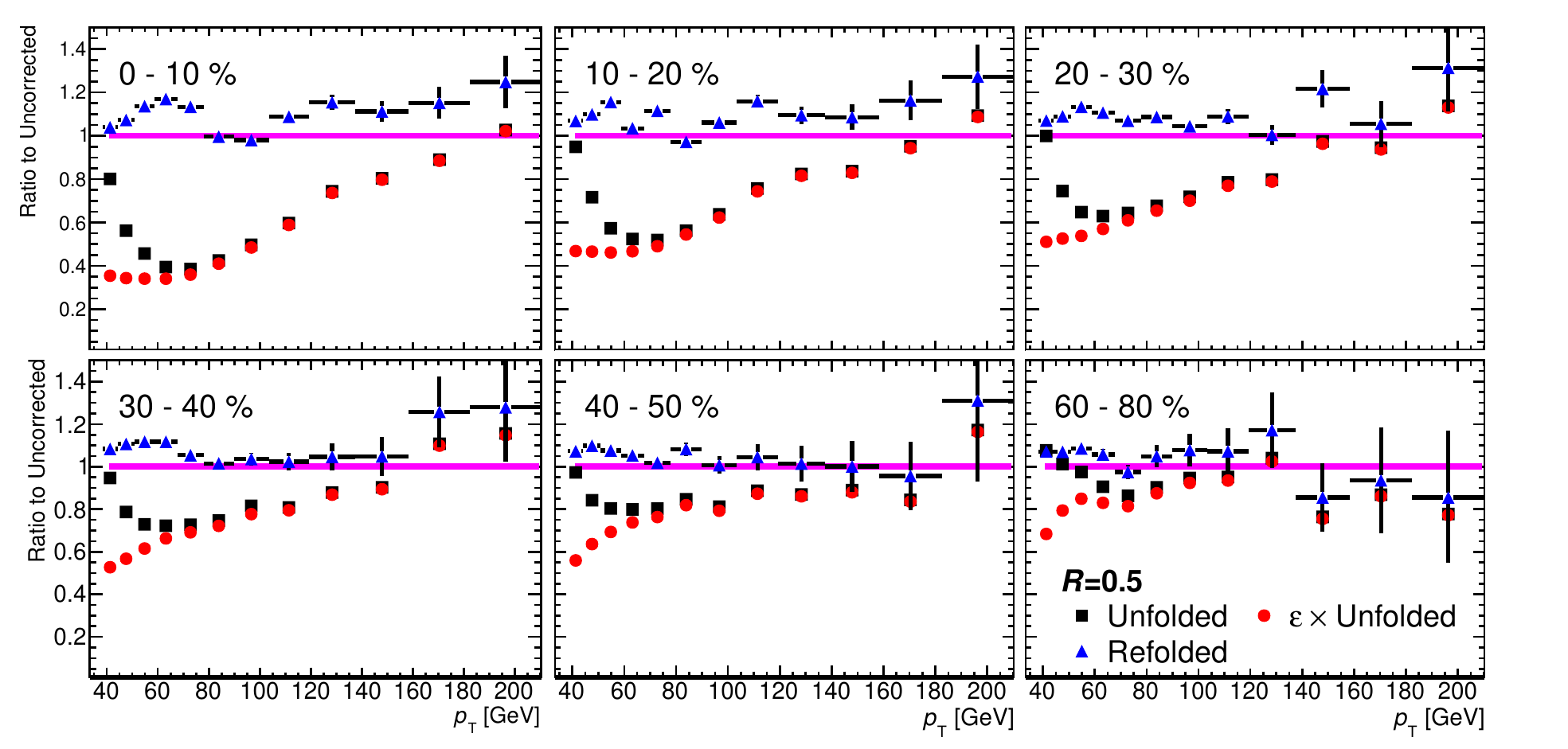}
\caption{Results of unfolding in centrality
  bins presented as various ratios to the raw spectrum for the \RFour\ (top) and \RFive\
  (bottom) jets. The ratio of the unfolded spectrum
both before and after efficiency correction to the input spectrum are
shown in red and black respectively. The ratio of the refolded
spectrum to the input distribution is shown in blue with the error
bars indicating the relative error on the data. For all
centrality bins this ratio is near unity, indicating good closure in
the unfolding procedure.}
\label{fig:results:rcp:correction_ratios_45}
\end{figure}

\begin{figure}[htb]
\centering
\includegraphics[width =0.45\textheight] {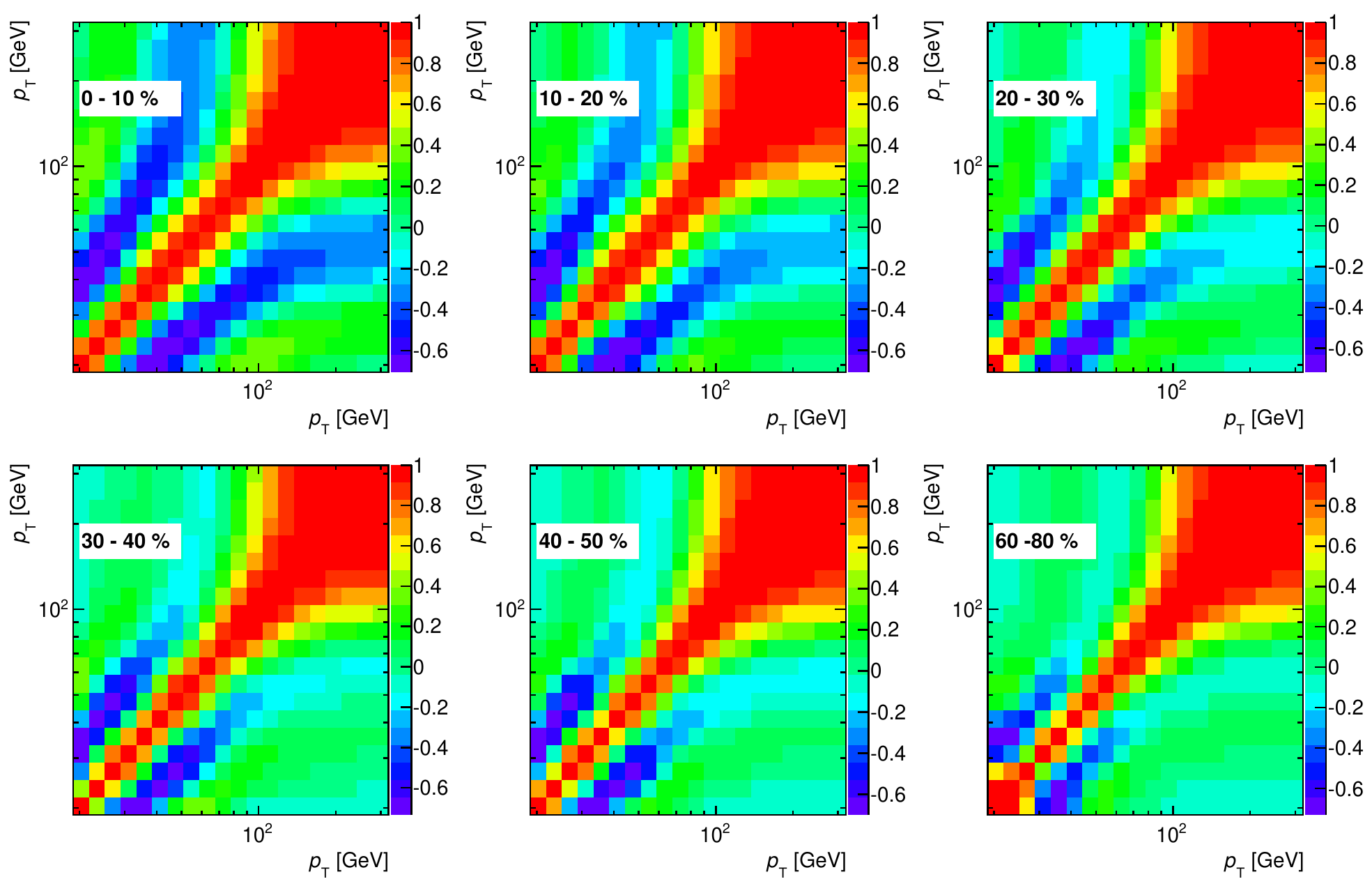}
\caption{Spectrum correlations after unfolding for \RFour\ jets.}
\label{fig:results:rcp:unfolded_cov_cent}
\end{figure}
\begin{figure}[htb]
\centering
\includegraphics[width =0.45\textheight] {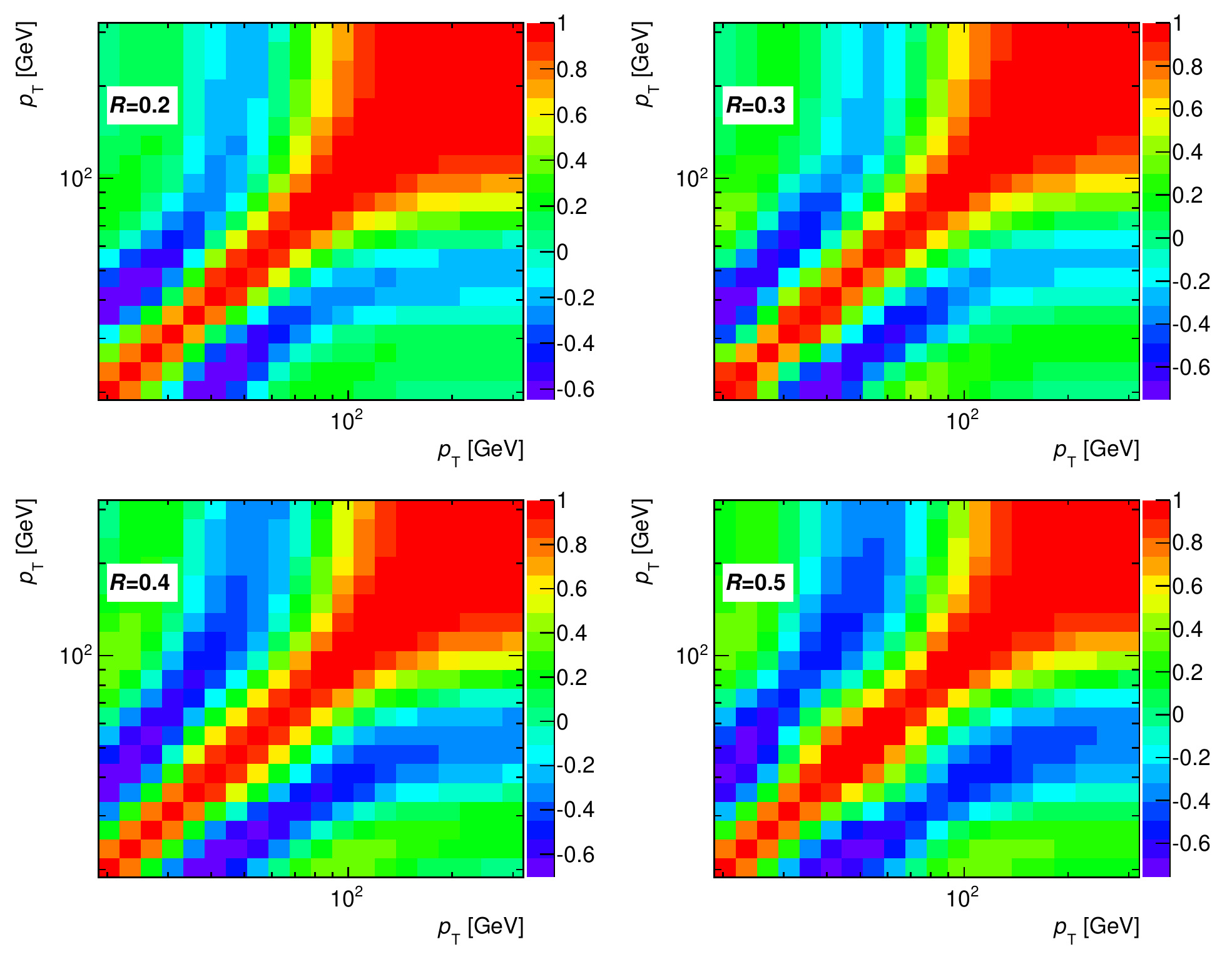}
\caption{Spectrum correlations after unfolding for $R=0.2,\,0.3,\,0.4$ and $0.5$
  jets for the 0-10\% centrality bin.}
\label{fig:results:rcp:unfolded_cov_R}
\end{figure}
\clearpage
\subsection{\Rcp}
The jet \Rcp\ as a function of \pt\ are shown in
Figs.~\ref{fig:results:rcp:rcp_vs_pt_R2}~-~\ref{fig:results:rcp:rcp_vs_pt_R5} with the
black bands and shaded boxes representing the correlated and
uncorrelated errors composed of the sources indicated in
Table~\ref{tbl:sys_err_calc}. A subset of the results for the \RTwo\ and \RFour\ jets
are presented in a slightly different fashion in
Figs.~\ref{fig:results:rcp:rcp_vs_pt_R2_paper}
and~\ref{fig:results:rcp:rcp_vs_pt_R4_paper}. In these figures the
\pt-dependences of selected centrality bins is compared in different
vertical panels. These results indicate that jets are suppressed in the
most central collisions by approximately a factor of two relative to
peripheral collisions. For all jet radii, the suppression shows at
most a weak dependence on jet \pt. The correlated systematic error
bands indicate that there may be a slight increase in \Rcp\ in the
range $40 \lesssim \pt \lesssim 100$~\GeV, although this is not
present in all centralities and radii. For some bins, the \Rcp\ shows
a mild decrease with increasing \pt\ at the highest \pt\ values, this
effect is also marginal within the systematic errors, and will be
discussed further in Chapter~\ref{section:conclusion}.

The \Rcp\ as a function of centrality,
as expressed by the average number of participants in each centrality
bin, is shown for different bins in jet \pt\ in
Figs.~\ref{fig:results:rcp:rcp_vs_npart_R2}~-~\ref{fig:results:rcp:rcp_vs_npart_R5}. Note
that the composition to the two types of systematic errors changes
when plotting \Rcp\ vs centrality as opposed to \Rcp\ vs \pt. In
general the \Npart\ dependence shows a weak dependence on jet
\pt. The \Npart\ dependence for the lowest \pt\ bins, shows a
linear drop in \Rcp\ in the most peripheral collisions before
flattening out. This dependence is slightly different than in the
highest \pt\ bins, where the dependence is more linear throughout the
full \Npart\ range.

Finally the $R$ dependence of the \Rcp\ is evaluated directly. This
comparison is shown for the 0-10\% centrality bin for different \pt\ bins
in Fig.~\ref{fig:results:rcp:rcp_vs_R_cent_1}. A comparison of this
dependence for different centrality bins but at fixed
\pt, $89 < \pt < 103$~\GeV, is shown in
Fig.~\ref{fig:results:rcp:rcp_vs_R_pt_15}. These results indicate an
at most weak reduction in the suppression with increasing jet
radius. There is essentially no reduction in suppression in the 50-60\%
bin, where there is a slight increase in the \Rcp\ with $R$ in the
most central bin.

An alternative evaluation of the dependence of the \Rcp\ on
jet radius is provided in Fig.~\ref{fig:results:rcpRratios} which shows the
ratio of \Rcp\ values between $R = 0.3, 0.4,$ and 0.5 jets and $R =
0.2$ jets, \RcpRatio, as a function of \pT\ for the 0-10\% centrality bin. When
evaluating the ratio, there is significant cancellation between the
assumed fully correlated systematic errors due to \Rcollcent, JES,
JER, and efficiency. The results in the figure indicate a significant
dependence of the \Rcp\ on jet radius for $\pt \lesssim 100$~\GeV\ in
the most central collisions. For lower \pT\ values, \RcpRatio\ for both
\RFour\ and \RFive\ differ from one 
beyond the statistical and systematic uncertainties. The maximal
difference between the \RTwo\ and \RFive\ jets occurs
$\pt\sim60$~\GeV, where \RcpRatio\ exceeds 1.3. The central
values for \RThree\ jets also differ from one, but that difference is
not significant when the 
systematic errors are accounted for.

\begin{figure}[htb]
\centering
\includegraphics[width =1\textwidth] {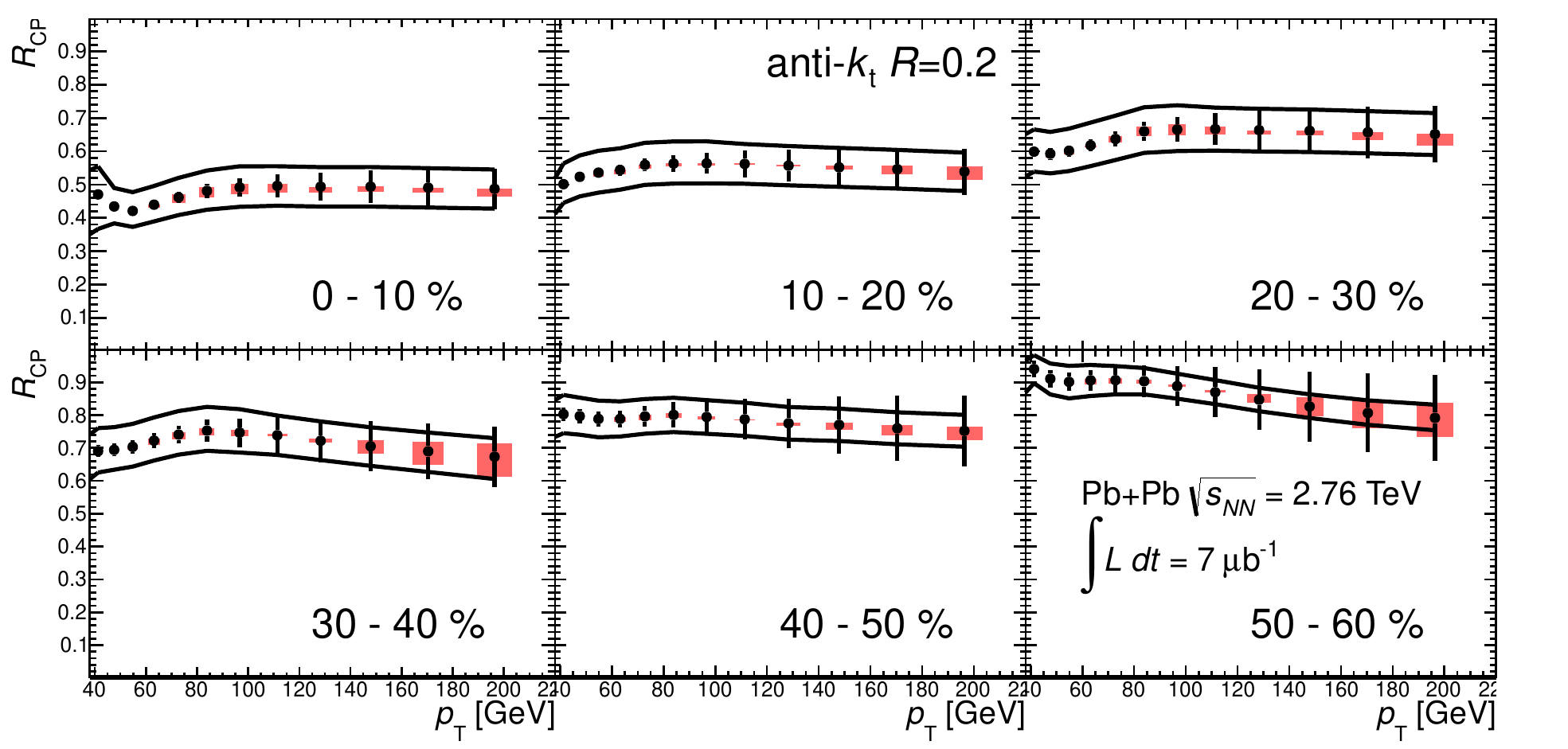}
\caption{Jet \Rcp\ vs \pt\ for \RTwo\ jets in different centrality bins.}
\label{fig:results:rcp:rcp_vs_pt_R2}
\end{figure}
\begin{figure}[htb]
\centering
\includegraphics[width =1\textwidth] {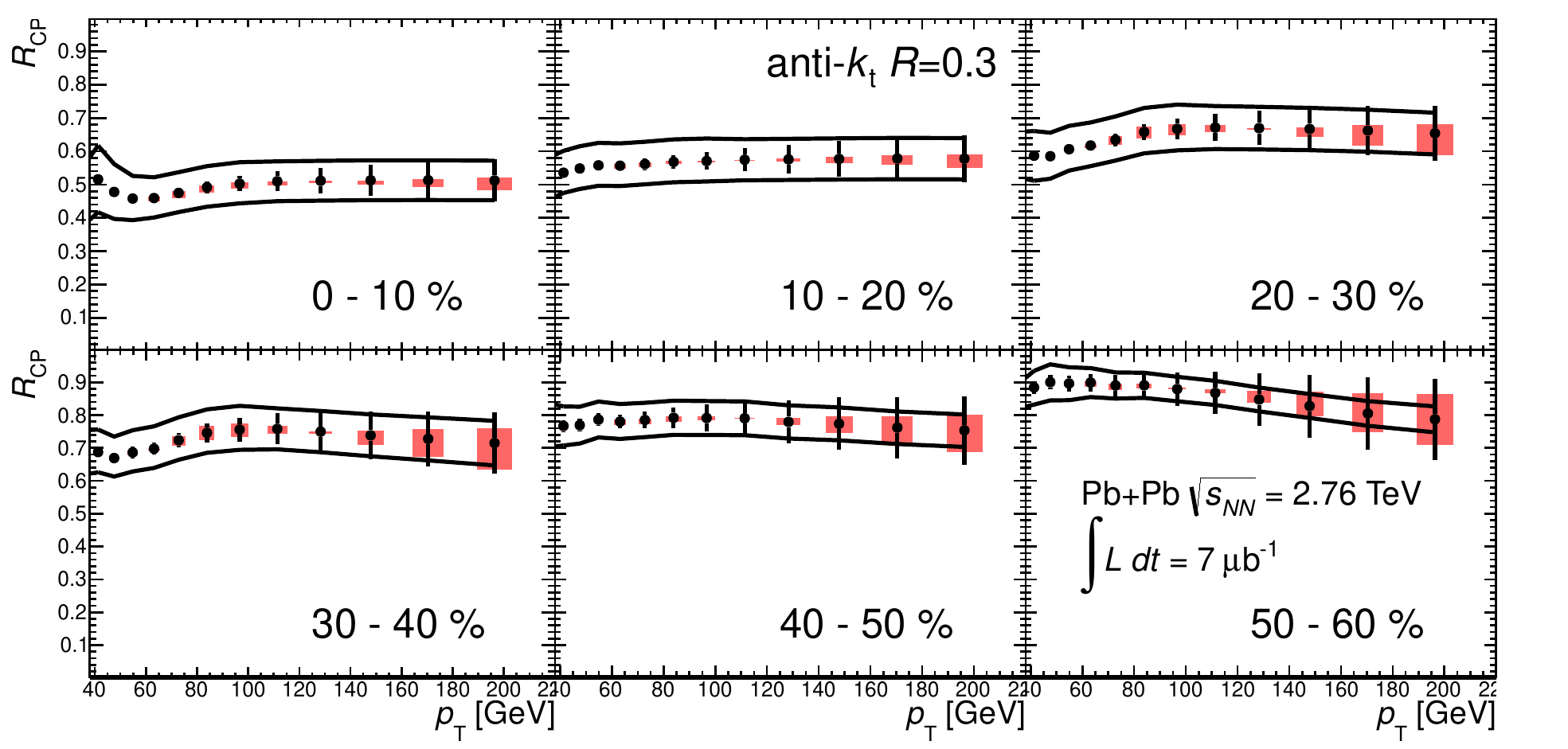}
\caption{Jet \Rcp\ vs \pt\ for \RThree\ jets in different centrality bins.}
\label{fig:results:rcp:rcp_vs_pt_R3}
\end{figure}
\begin{figure}[htb]
\centering
\includegraphics[width =1\textwidth] {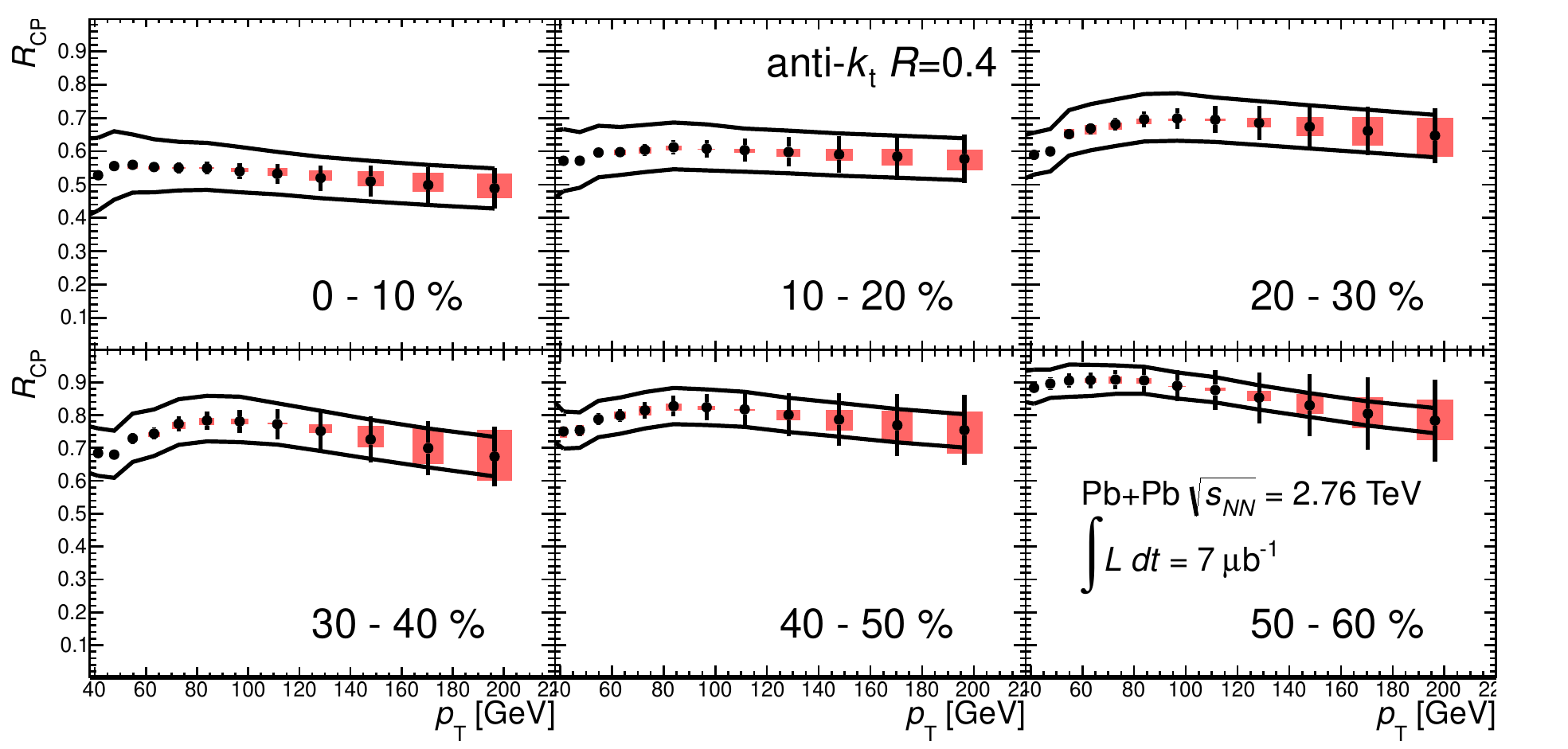}
\caption{Jet \Rcp\ vs \pt\ for \RFour\ jets in different centrality bins.}
\label{fig:results:rcp:rcp_vs_pt_R4}
\end{figure}
\begin{figure}[htb]
\centering
\includegraphics[width =1\textwidth] {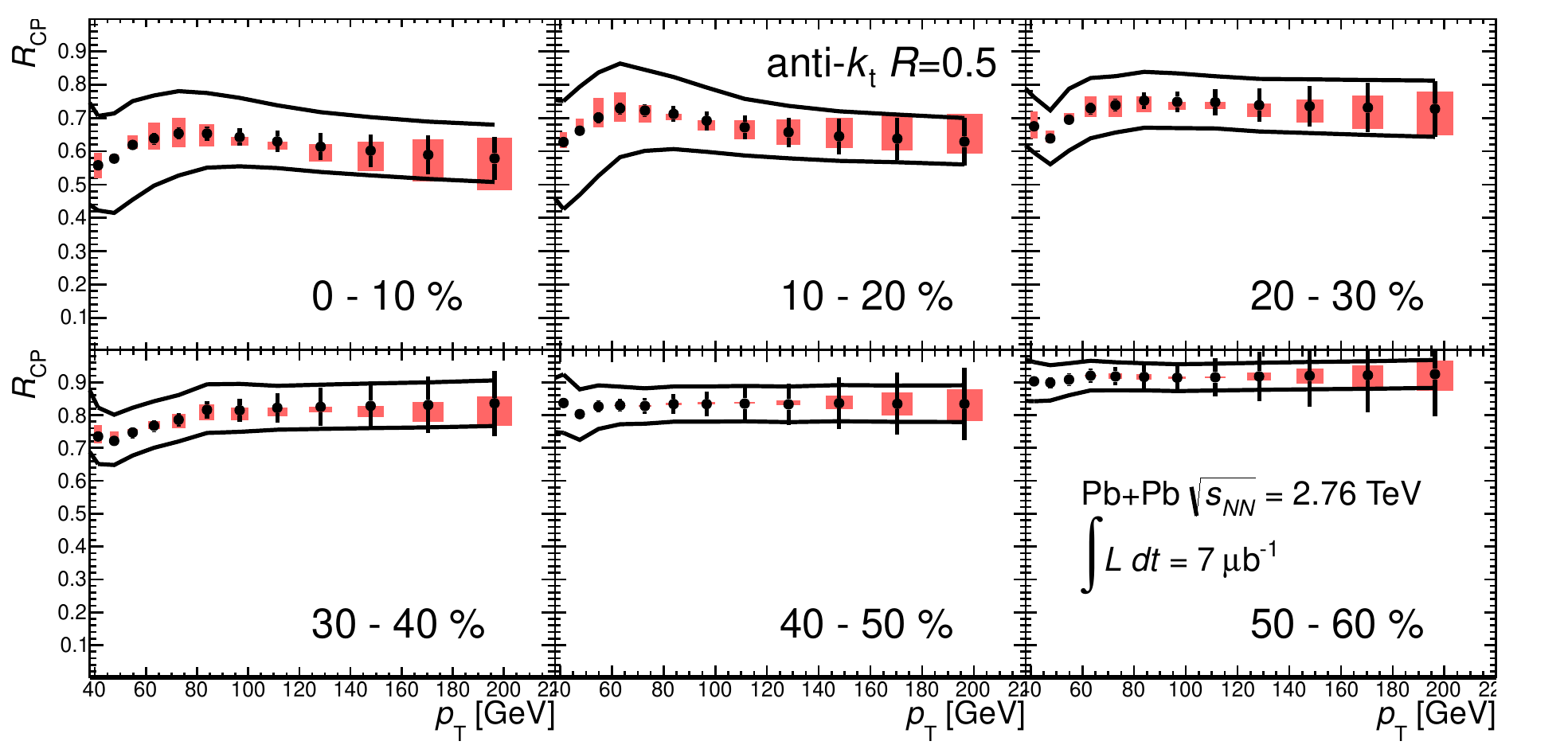}
\caption{Jet \Rcp\ vs \pt\ for \RFive\ jets in different centrality bins.}
\label{fig:results:rcp:rcp_vs_pt_R5}
\end{figure}

\begin{figure}[htb]
\includegraphics[width =0.9\textwidth] {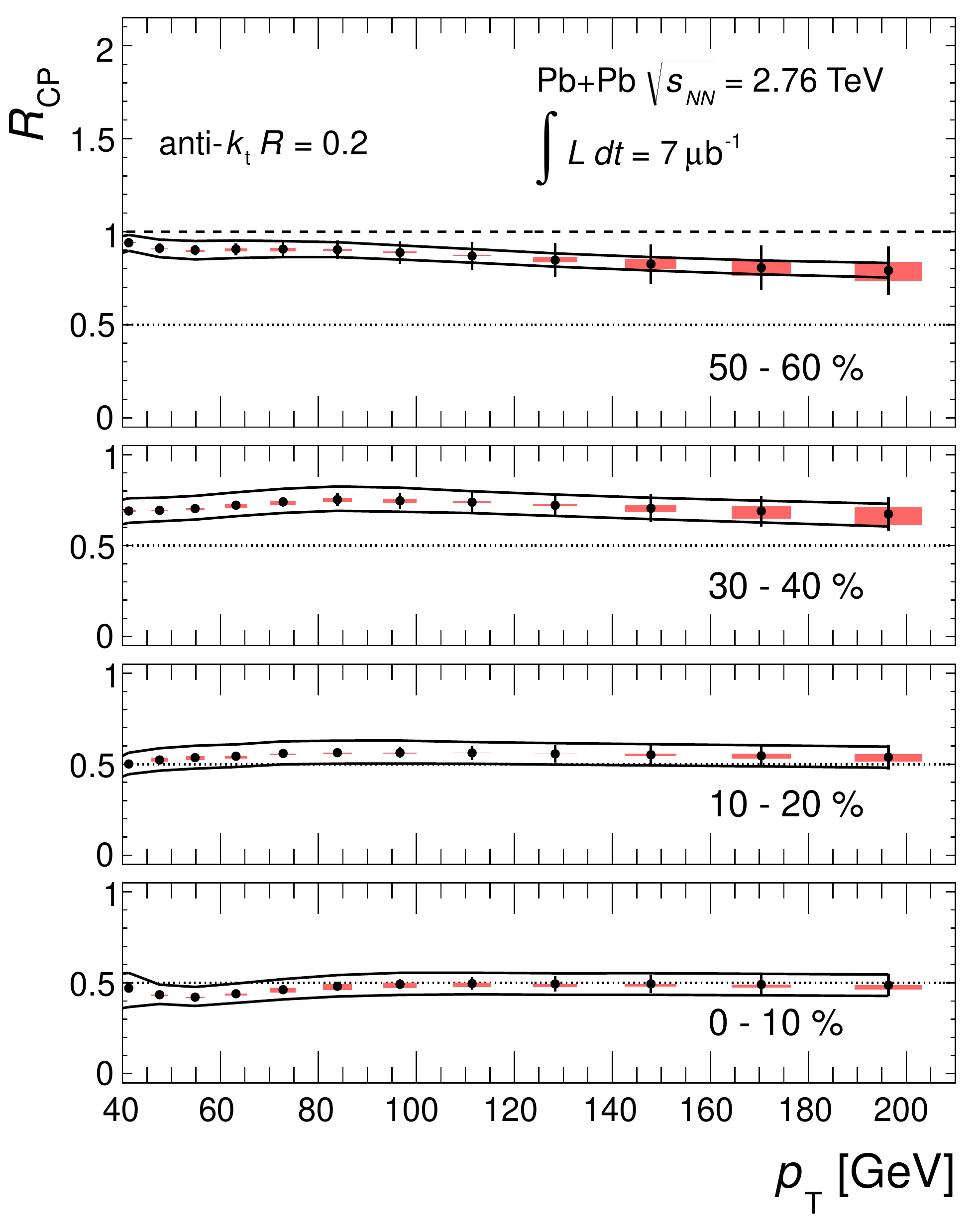}
\caption{Jet \Rcp\ vs \pt\ for \RTwo\ jets in different centrality bins.}
\label{fig:results:rcp:rcp_vs_pt_R2_paper}
\end{figure}
\begin{figure}[htb]
\includegraphics[width =0.9\textwidth] {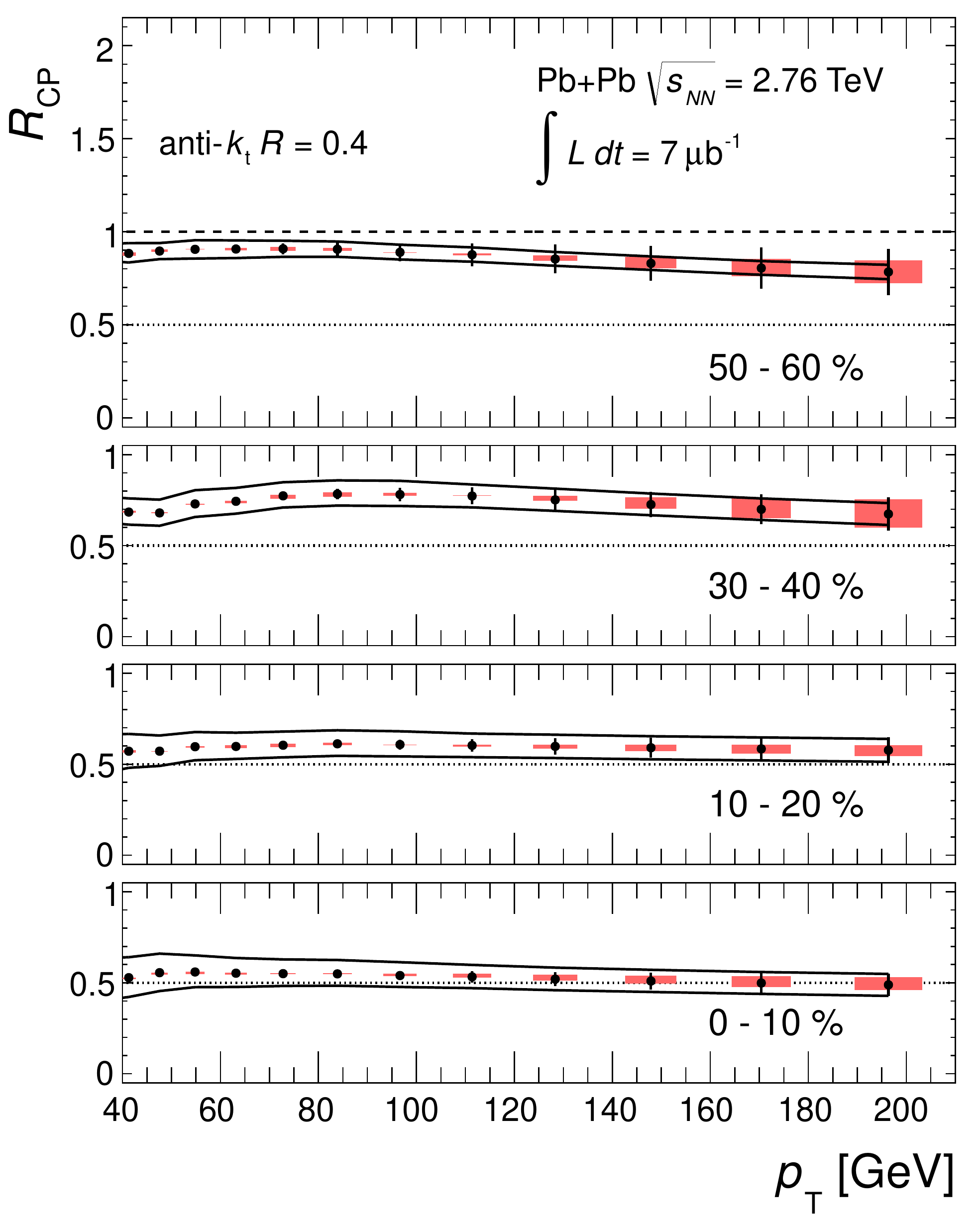}
\caption{Jet \Rcp\ vs \pt\ for \RFour\ jets in different centrality bins.}
\label{fig:results:rcp:rcp_vs_pt_R4_paper}
\end{figure}

\begin{figure}[htb]
\centering
\includegraphics[width =1\textwidth] {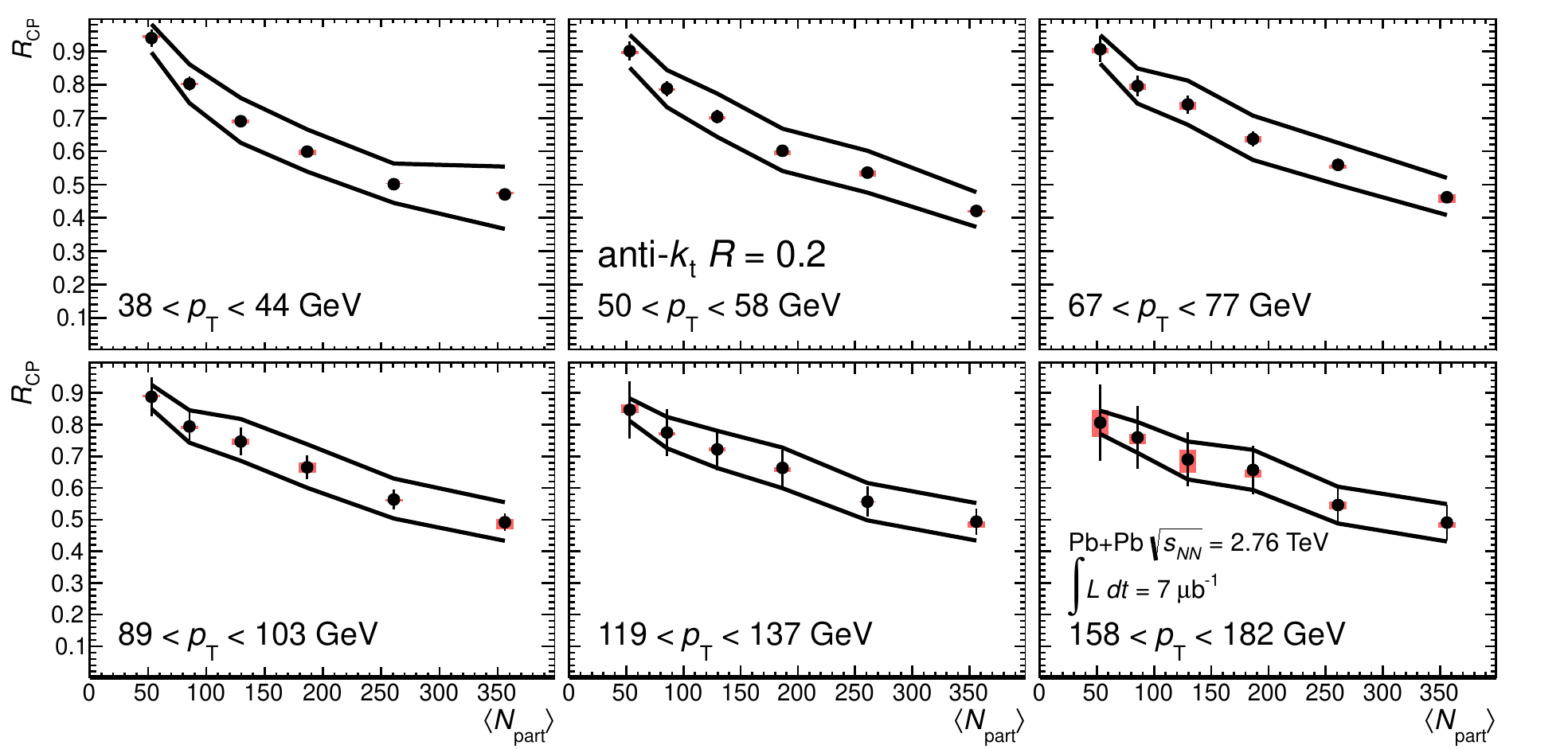}
\caption{Jet \Rcp\ vs \Npart\ for \RTwo\ jets in different \pt\ bins.}
\label{fig:results:rcp:rcp_vs_npart_R2}
\end{figure}
\begin{figure}[htb]
\centering
\includegraphics[width =1\textwidth] {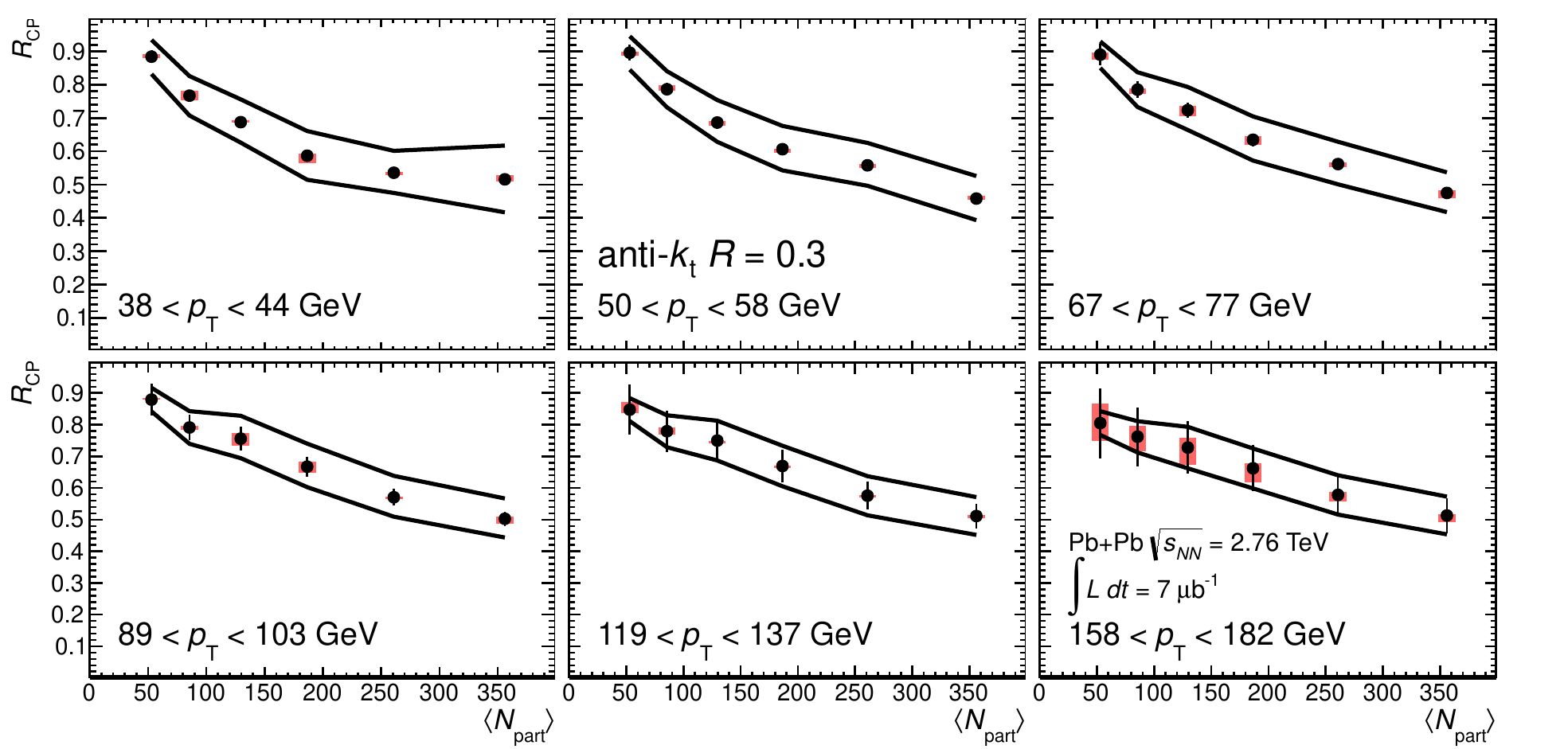}
\caption{Jet \Rcp\ vs \Npart\ for \RThree\ jets in different \pt\ bins.}
\label{fig:results:rcp:rcp_vs_npart_R3}
\end{figure}
\begin{figure}[htb]
\centering
\includegraphics[width =1\textwidth] {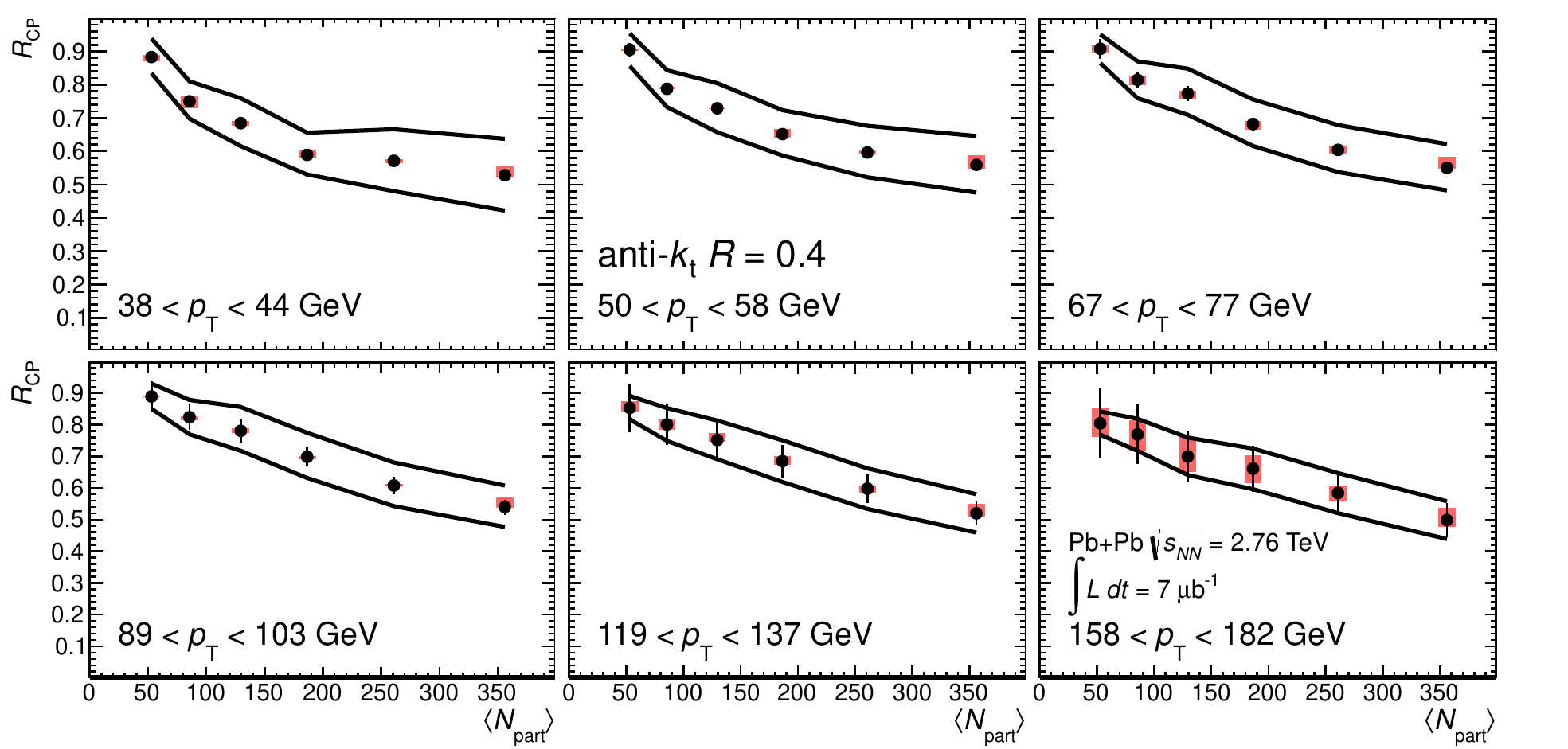}
\caption{Jet \Rcp\ vs \Npart\ for \RFour\ jets in different \pt\ bins.}
\label{fig:results:rcp:rcp_vs_npart_R4}
\end{figure}
\begin{figure}[htb]
\centering
\includegraphics[width =1\textwidth] {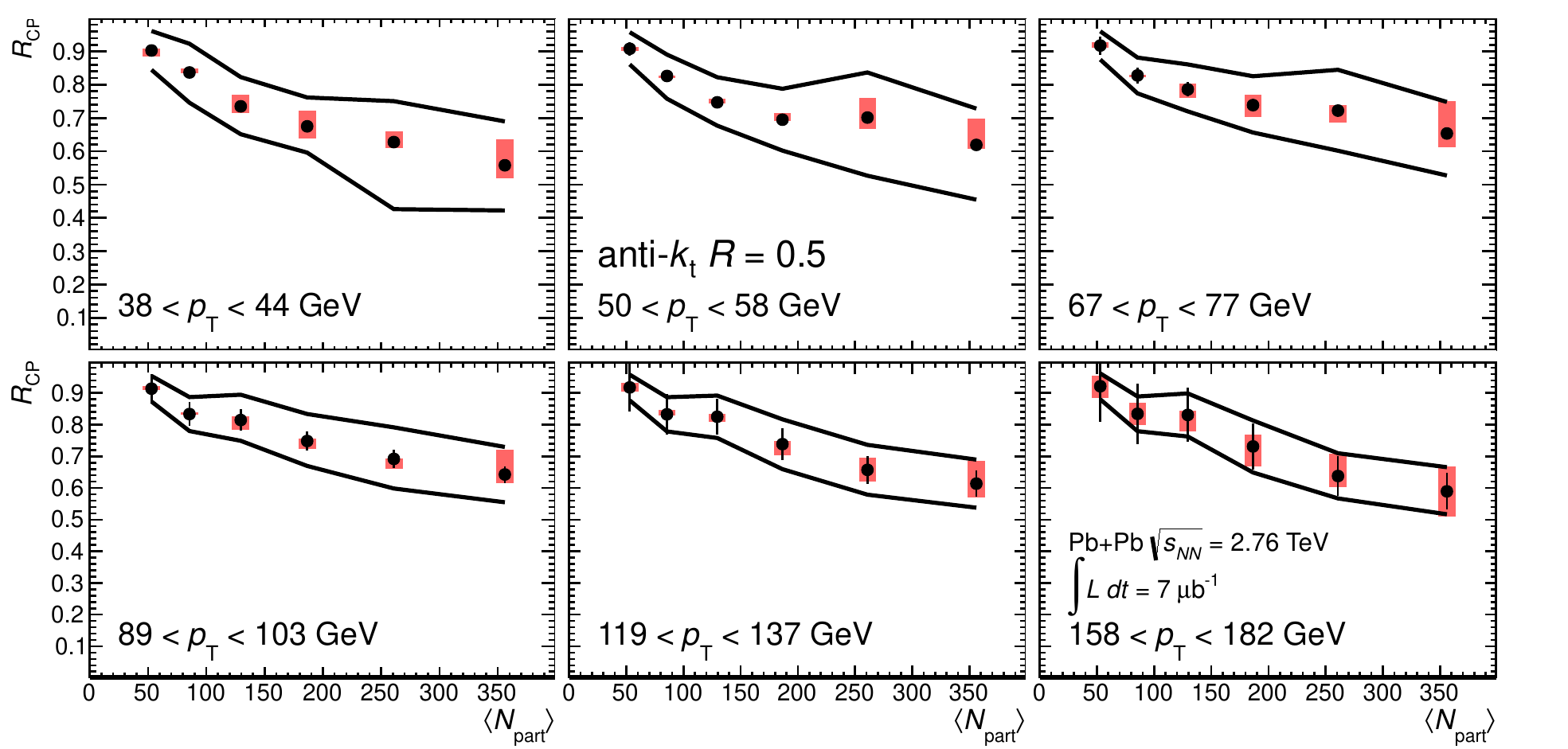}
\caption{Jet \Rcp\ vs \Npart\ for \RFive\ jets in different \pt\ bins.}
\label{fig:results:rcp:rcp_vs_npart_R5}
\end{figure}

\begin{figure}[htb]
\includegraphics[width =0.9\textwidth] {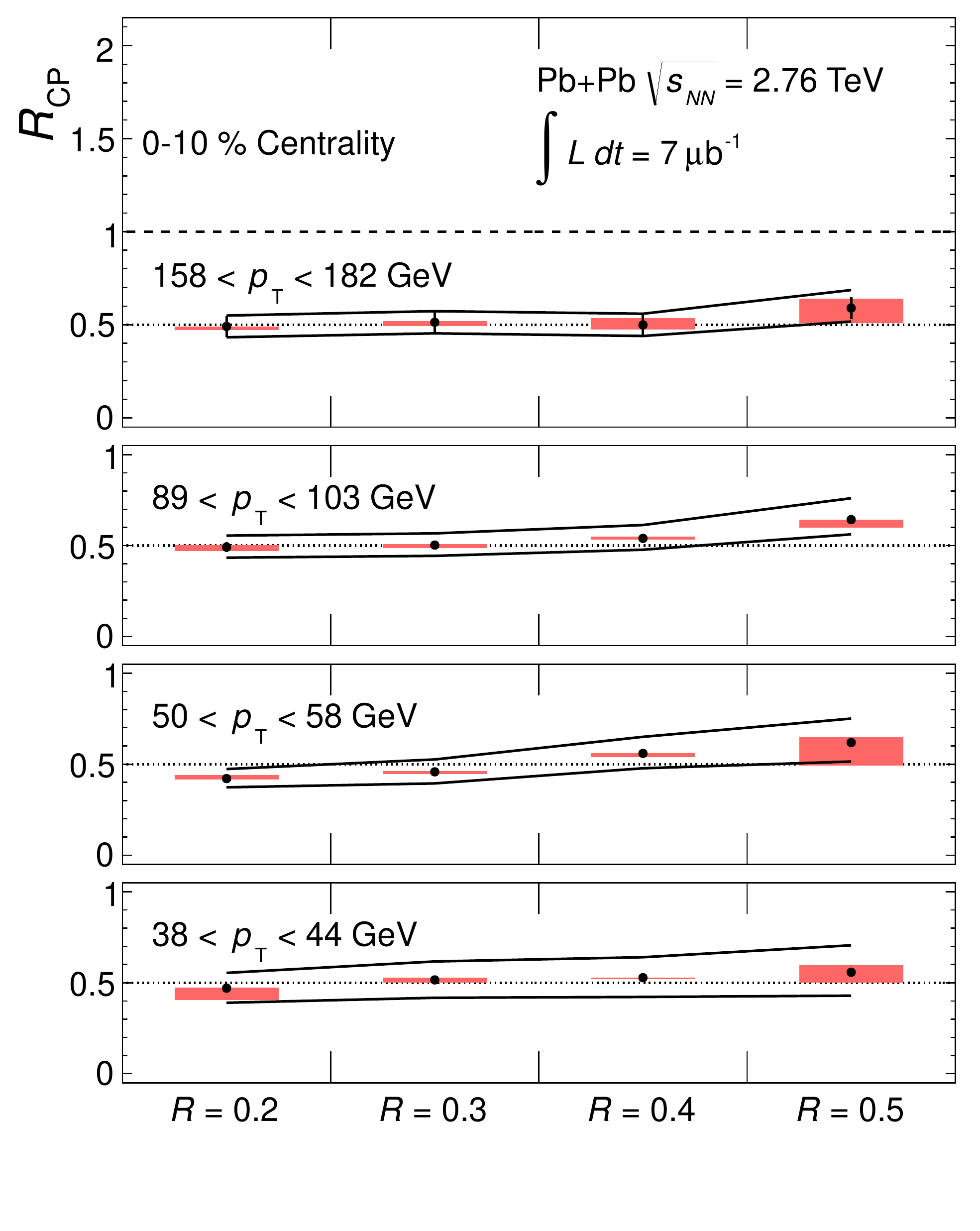}
\caption{Jet \Rcp\ vs $R$ for different \pt\ bins in the 0-10\%
  centrality bin.}
\label{fig:results:rcp:rcp_vs_R_cent_1}
\end{figure}

\begin{figure}[htb]
\includegraphics[width = 0.9\textwidth] {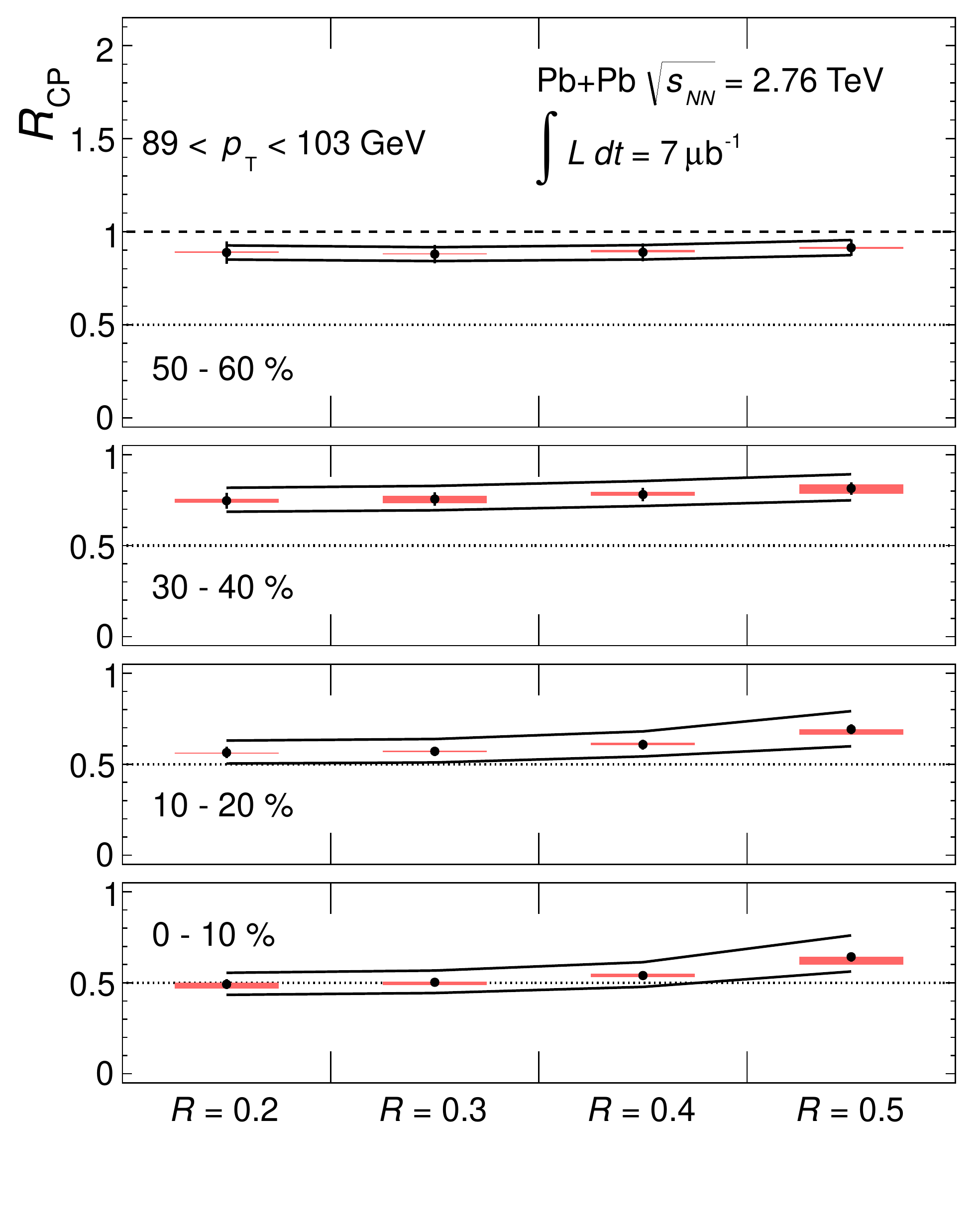}
\caption{Jet \Rcp\ vs $R$ for different centrality bins for jets with
  $89 < \pt < 103$~\GeV. }
\label{fig:results:rcp:rcp_vs_R_pt_15}
\end{figure}

\begin{figure}[htb]
\centering
\includegraphics[width =1\textwidth] {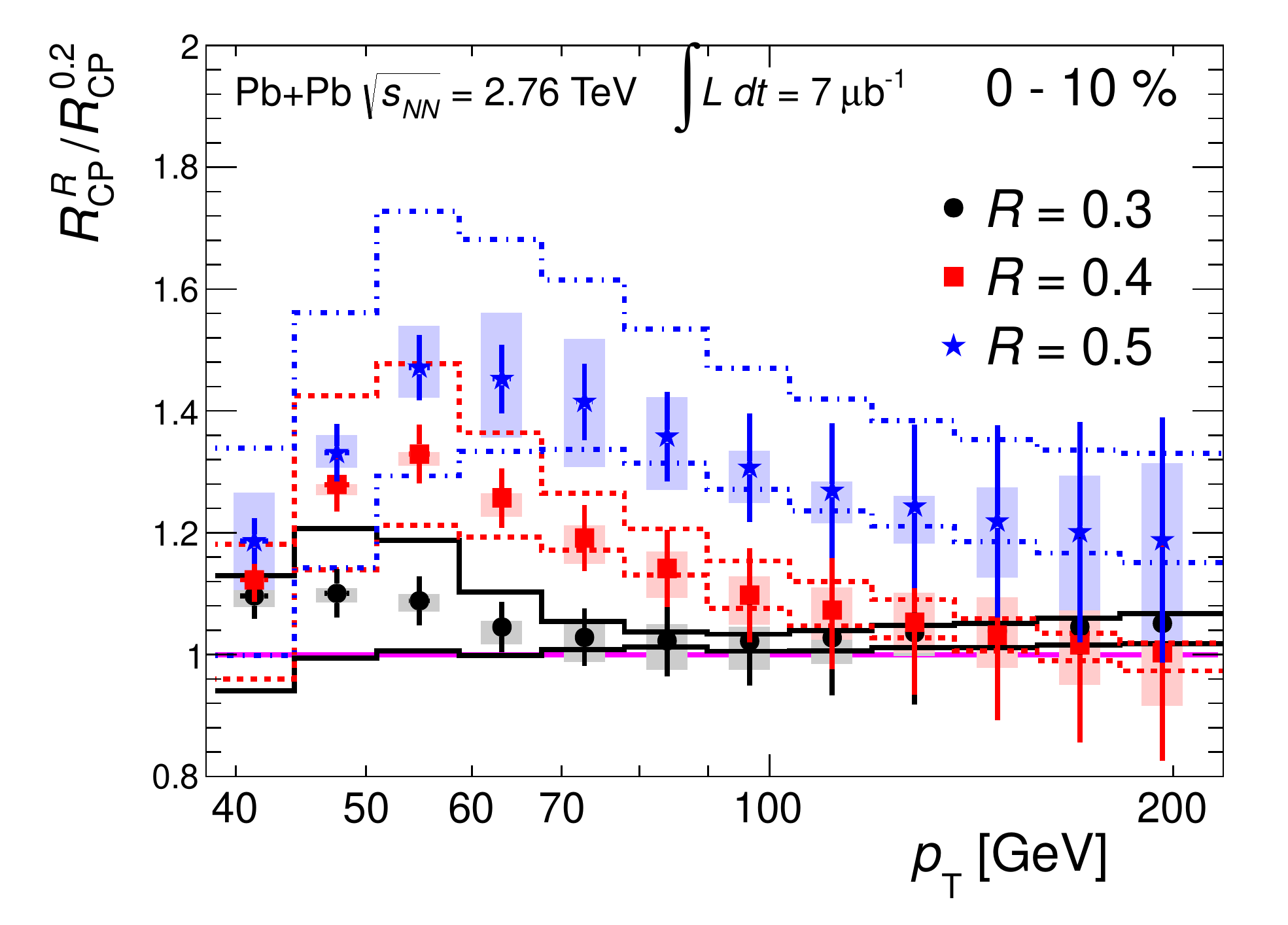}
\caption{
Ratios of \Rcp\ values between $R = 0.3, 0.4,$ and 0.5 jets and $R =
0.2$ jets as a function of \pT\ in the 0-10\% centrality bin. The error bars show statistical
uncertainties obtained by propagating the statistical uncertainties 
on the individual \Rcp\ values.  The shaded boxes indicate partially
correlated systematic errors. The solid lines indicate systematic
errors that are fully correlated between different \pT\ bins.
}
\label{fig:results:rcpRratios}
\end{figure}
\clearpage
\subsection{Energy Loss Estimate}
If a simple model of energy loss is assumed, the \pt-dependence of the
\Rcp\ can be isolated. Following the discussion in
Ref.~\cite{Adcox:2004mh}, an \Rcp\ that is independent of \pt\ can be
the consequence of energy loss that is proportional to the jet's
original \pt. The number
of jets measured at a final \pt\ after energy loss, \ptf, will be populated by jets that were
produced with an original momentum of \pti,
\begin{equation}
\ptf=(1-\Sloss)\pti
\end{equation}
If the unmodified (i.e. initial) jet spectrum is
described by a power law which is independent of centrality,
\begin{equation}
 \dfrac{dN}{d\pti} =\frac{A}{\pti^n}\,.
\end{equation}
The final measured spectrum is modified by the transformation from
\pti\ to \ptf:
\begin{equation}
\left.\dfrac{dN}{d\ptf}\right|_{\mathrm{cent}}=\dfrac{dN}{d\pti}\left.\dfrac{d\pti}{d\ptf}\right|_{\mathrm{cent}}\,.
\end{equation}
If the energy loss is defined relative to peripheral collisions then
the initial and measured spectra are equal,
\begin{equation}
\left.\dfrac{dN}{d\ptf}\right|_{\mathrm{periph}}=\dfrac{dN}{d\pti}=\frac{A}{\pti^n}\,.
\end{equation}
In central collisions the measured spectrum is related to the initial
by,
\begin{equation}
\left.\dfrac{dN}{d\ptf}\right|_{\mathrm{cent}}=\frac{A}{(1-\Sloss)^{-(n-1)}\ptf^n}=\Rcp \left.\dfrac{dN}{d\ptf}\right|_{\mathrm{periph}}\,.
\end{equation}
Therefore the \Rcp\ and lost energy fraction, \Sloss, are related by
\begin{equation}
\Rcp=(1-\Sloss)^{n-1}\,,\quad\quad \Sloss=1-\Rcp^{1/(n-1)}.
\end{equation}
The centrality and $R$ dependence of \Sloss\ have been extracted from
the measurements made here of \Rcp\ for $\pt > 70$~\GeV, where the
assumptions of power law spectrum and \pt-independent \Rcp\ are
appropriate. The power, $n$, was obtained from
fitting the \pt\ spectra in the 60-80\% centrality bin, and the
extracted fit constants and errors are shown in
Table~\ref{tbl:results:sloss}.

\begin{table}
\centering
\begin{tabular}{| c | c | c | c |} \hline
     $R$ &  $n\pm\delta_{n}$ &    $k\pm\delta_{k}$&  $S\pm\delta_{S}$ \\ \hline
         2 &          5.49 $\pm$ 0.07 & 0.90 $\pm$ 0.10 &       8.21e-04 $\pm$ 5.87e-05\\ \hline
         3 &          5.64 $\pm$ 0.07 & 0.88 $\pm$ 0.14 &       9.13e-04 $\pm$ 4.92e-04\\ \hline
         4 &          5.72 $\pm$ 0.06 & 0.88 $\pm$ 0.20 &       7.46e-04 $\pm$ 4.38e-04\\ \hline
         5 &          5.78 $\pm$ 0.06 & 0.74 $\pm$ 0.12 &       1.25e-03 $\pm$ 7.81e-04\\ \hline
\end{tabular}
\caption{Values of fit parameters and associated errors for different
  jet radii. The $n$ values were obtained from a power law fit of the
  jet spectrum in the 60-80\% centrality bin. $k$ and $S$ were
  obtained from fitting the \Npart\ dependence of \Sloss.}
\label{tbl:results:sloss}
\end{table}
In each centrality bin the \Rcp\ was
fit to a constant, $\overline{\Rcp}$, by minimizing the $\chi^2$ function,
\begin{equation}
\chi^2=\displaystyle \sum_i \sum_j ({\Rcp}_i-\overline{\Rcp})C_{ij}^{-1}({\Rcp}_j-\overline{\Rcp})\,.
\end{equation}
Here $C_{ij}$ is the full \pt\ covariance matrix of \Rcp, which
contains contributions from statistical correlations, uncorrelated
systematic errors and correlated systematic errors. The statistical
covariance was
obtained in Chapter~\ref{section:results}. For uncorrelated systematic
errors, the covariance matrix was taken as diagonal
$C_{ij}^{\mathrm{U}}=\sigma^{\mathrm{U}}_i \sigma^{\mathrm{U}}_j
\delta_{ij}$, and for fully correlated errors the covariance matrix
is $C_{ij}^{\mathrm{C}}=\sigma^{\mathrm{C}}_i
\sigma^{\mathrm{C}}_j$. The solution to the $\chi^2$ minimization, $\overline{\Rcp}$ is
expressed analytically in terms of the full covariance as
\begin{equation}
\overline{\Rcp}=\dfrac{\sum_{ij} C^{-1}_{ij}{\Rcp}_j}{\sum_{ij} C^{-1}_{ij}}\,.
\end{equation}
An uncertainty on this quantity was determined by varying $\overline{\Rcp}$
such that the $\chi^2$ per degree of freedom differed from its minimum
value by one. This analysis was performed in each centrality bin for all $R$
values with the resulting evaluation of \Sloss\ shown in Fig.~\ref{fig:results:sloss}. 
\begin{figure}[htbp]
\centering
\includegraphics[width=1\textwidth]{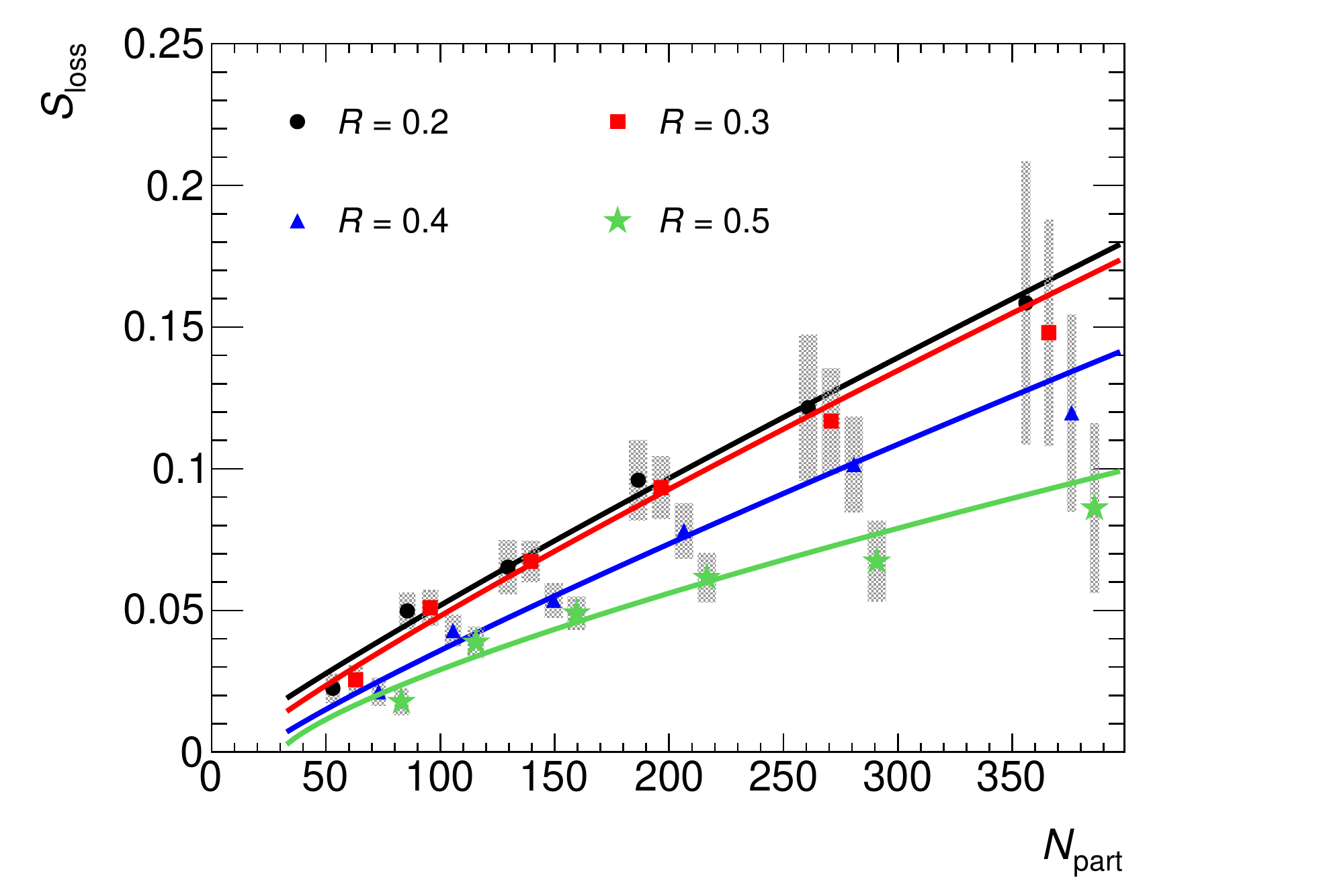}
\caption{\Sloss\ as a function of \Npart\ for different jet radii,
  with fits using the form, $\Sloss=S\Npart^k$ . The grey boxes indicate the
combined error due to uncertainties in both $\overline{\Rcp}$ and $n$
with the horizontal size determined by the uncertainty on \Npart; the points have been artificially
offset to facilitate comparison.}
\label{fig:results:sloss}
\end{figure}
The grey boxes indicate the
combined error due to uncertainties in both $\overline{\Rcp}$ and $n$
with the horizontal size determined by the uncertainty on \Npart\
given in Table~\ref{tbl:centrality_bins}; the points have been artificially
offset to facilitate comparison. The smaller radii show a nearly
linear growth in \Sloss\ with \Npart. However the rate of growth
diminishes with increasing $R$ value indicating that some of the lost
energy may be recovered with the larger jet definition. To provide a
quantitative estimate of the \Npart\ dependence, the \Sloss\ for each
$R$ value was fit with the functional form,
\begin{equation}
\Sloss=S\Npart^k\,,
\end{equation}
with the fit values of $S$ and $k$ and their associated fit errors
shown in Table~\ref{tbl:results:sloss} and the fit functions shown
with the points on Fig.~\ref{fig:results:sloss}.
\clearpage

\section{Dijet Asymmetry}
\label{section:results:asymmetry}
The observation of a centrality-dependent dijet asymmetry was the
first indication of jet quenching at the LHC and was reported by ATLAS in
Ref.~\cite{Aad:2010bu}. That publication was a snapshot of an ongoing, more
extensive asymmetry analysis which is one of the main results of this
thesis. The results presented here improve on the first published
result by using the increased statistics of the full collision data
set and benefiting from improvements in the jet
reconstruction performance. These results were presented at Quark
Matter 2011~\cite{Cole:2011zz,Angerami:2011is} and are presented more thoroughly in
Ref.~\cite{JetQM11}. It should be noted that the jet reconstruction procedure
has evolved between the completion of the asymmetry analysis and the
\Rcp\ analysis. In fact, these improvements were a significant
technical undertaking designed to improve the performance and enhance
the precision and \pt\ reach of the \Rcp\ measurement. The
reconstruction procedure and performance as documented in previous
chapters coincides with the \Rcp\ analysis. While main features of the
background subtraction (seed
finding, flow correction and iteration step) were also present when
the asymmetry analysis was performed, the details differ slightly
(flow iteration step, \ET\ cut on seeds). Those details and the
performance of that version of the reconstruction are documented
Ref.~\cite{JetQM11}, and generally will be omitted here. One important
distinction is the lack of jet energy scale calibration constants for $R\neq0.4$ jets
at the time of the asymmetry analysis. An analysis of the asymmetry
for \RTwo\ jets was performed using the \RFour\ calibration
constants. The result is a $\sim 5\%$ non-closure in the JES, which
drops to $\sim 10\%$ at the lowest \ET's. This effect, shown in
Fig.~\ref{fig:results:old_performance_R2} for the 0-10\% and 60-80\%
centrality bins, is independent of centrality and largely cancels in the asymmetry. The
\RFour\ jets do not suffer from this problem.
\begin{figure}[htb]
\centering
\includegraphics[width=0.49\textwidth]{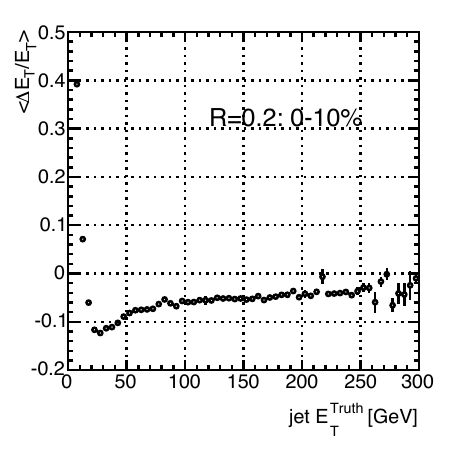}
\includegraphics[width=0.49\textwidth]{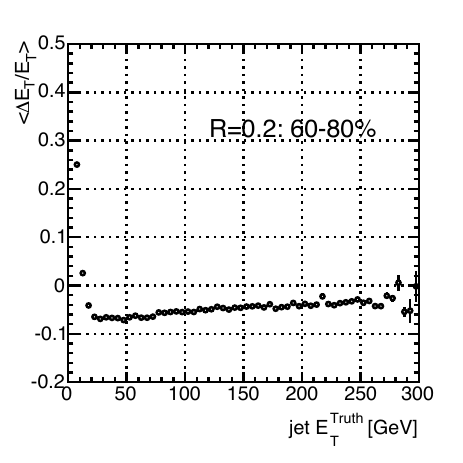}
\caption{The JES performance for \RTwo\ jets in the 0-10\% (left) and
  60-80\% (right) centrality bins with the older
version of jet reconstruction that was used in asymmetry analysis. As
no \RTwo\ calibration constants were available at the time, the
\RFour\ constants were used, resulting in a $5-10\%$ non-closure in
the energy scale.}
\label{fig:results:old_performance_R2}
\end{figure}

Following the procedures used in Ref.~\cite{Aad:2010bu} the
per-event dijet asymmetries are calculated according to 
\begin{equation}
\Aj =\frac{\ETOne -\ETTwo}{\ETOne+\ETTwo}
\label{eqn:asymmetry_definition}
\end{equation}
where \ETOne\ is the transverse energy of the leading
(highest \et) jet in the event and \ETTwo\ is the transverse energy of
the highest \et\ jet in the opposite hemisphere ($\Delta\phi > \pi/2$
with $\Delta \phi$ the azimuthal angle difference between the jets). 
Both jets are required to satisfy the $|\eta|<2.8$ requirement. 
Events for which the second highest
\et\ jet in the event fails the $\eta$ or $\Delta \phi$
requirements are not excluded, but rather the highest \et\ jet that
does satisfy these criteria is used. All presented results have the
requirement $\ETTwo > 25$~\GeV; events for which the second jet (passing the above
selections) fails this requirement are excluded from the analysis. 
\begin{figure}[htb]
\centering
\includegraphics[width=0.9\textwidth]{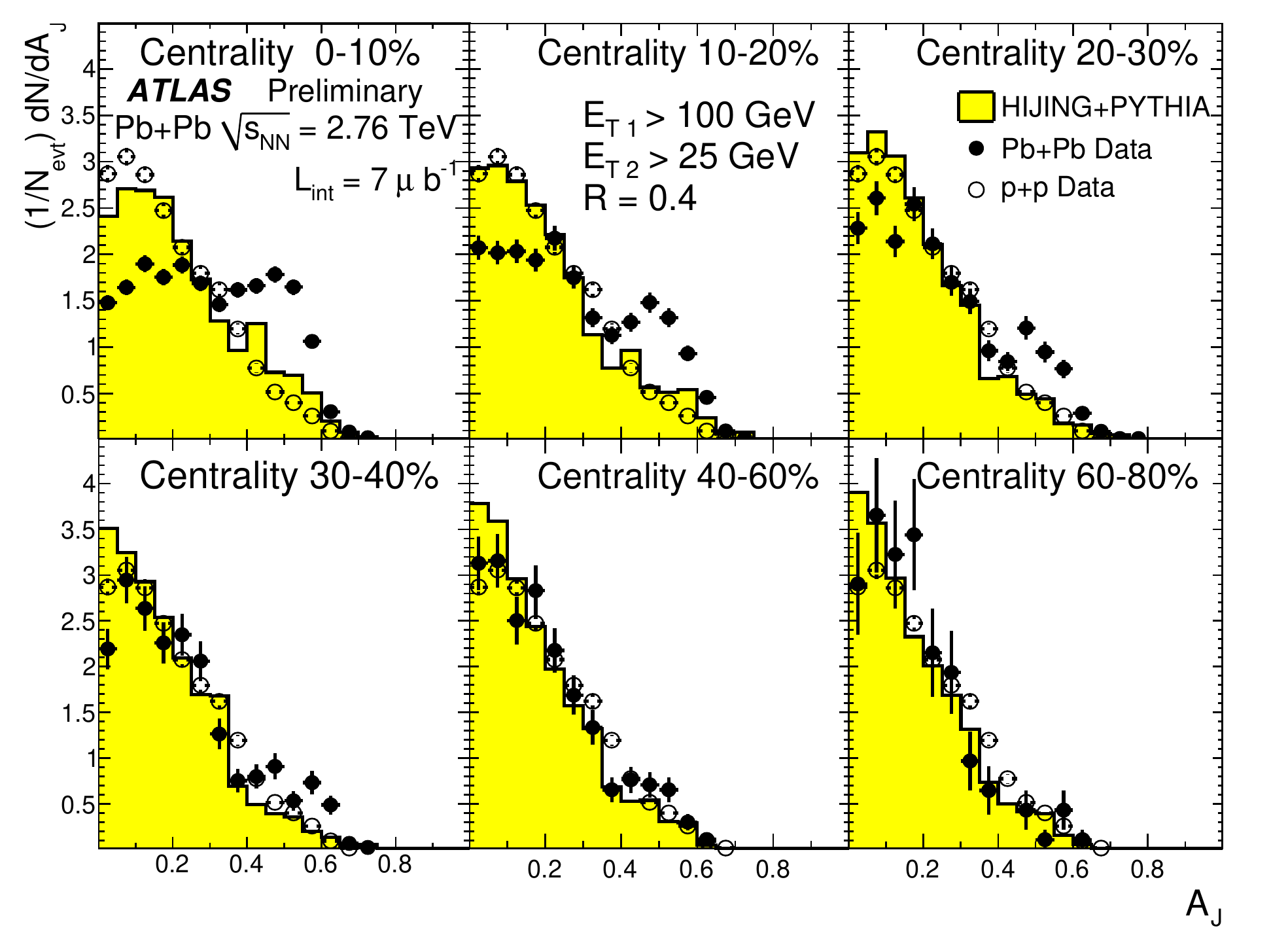}
\caption{Dijet asymmetry for~\RFour\ jets in six centrality bins in
  events with a leading jet with $\et > 100$~\GeV. A comparison to HIJING
  with embedded PYTHIA dijet events (yellow) and ATLAS
  $\sqrt{s}=7$~\TeV\ \pp\ data (open circles) is shown. }
\label{fig:antikt4HIItrFlow_asym_panel}
\end{figure}

Figure~\ref{fig:antikt4HIItrFlow_asym_panel} shows the asymmetry
distribution for \RFour\ jets with $\ETOne > 100$~\GeV\ for six bins
of collision centrality. Also shown for each centrality bin are
results obtained from \pp\ measurements at 7~\TeV~\cite{atlas:2010wv} and the
asymmetry distributions obtained from the HIJING+PYTHIA MC samples for the corresponding centrality bin using the above
described analysis procedures. The features observed in the original
dijet asymmetry publication are seen in
Fig.~\ref{fig:antikt4HIItrFlow_asym_panel}, namely good agreement
between \PbPb\ data, \pp\ data, and MC results in the more
peripheral (40-60\% and 60-80\%) centrality bins and an increasingly
strong modification of the asymmetry distributions for more central
collisions. The MC \Aj\ distributions show modest broadening
from peripheral to central collisions due to the increased underlying
event fluctuations. However, the modifications seen in the data are much
stronger than those seen in the MC for which the underlying
event fluctuations are shown to be consistent with those in \PbPb\ data
in Fig.~\ref{fig:jetetrms}.
\begin{figure}[htb]
\centering
\includegraphics[width=0.9\textwidth]{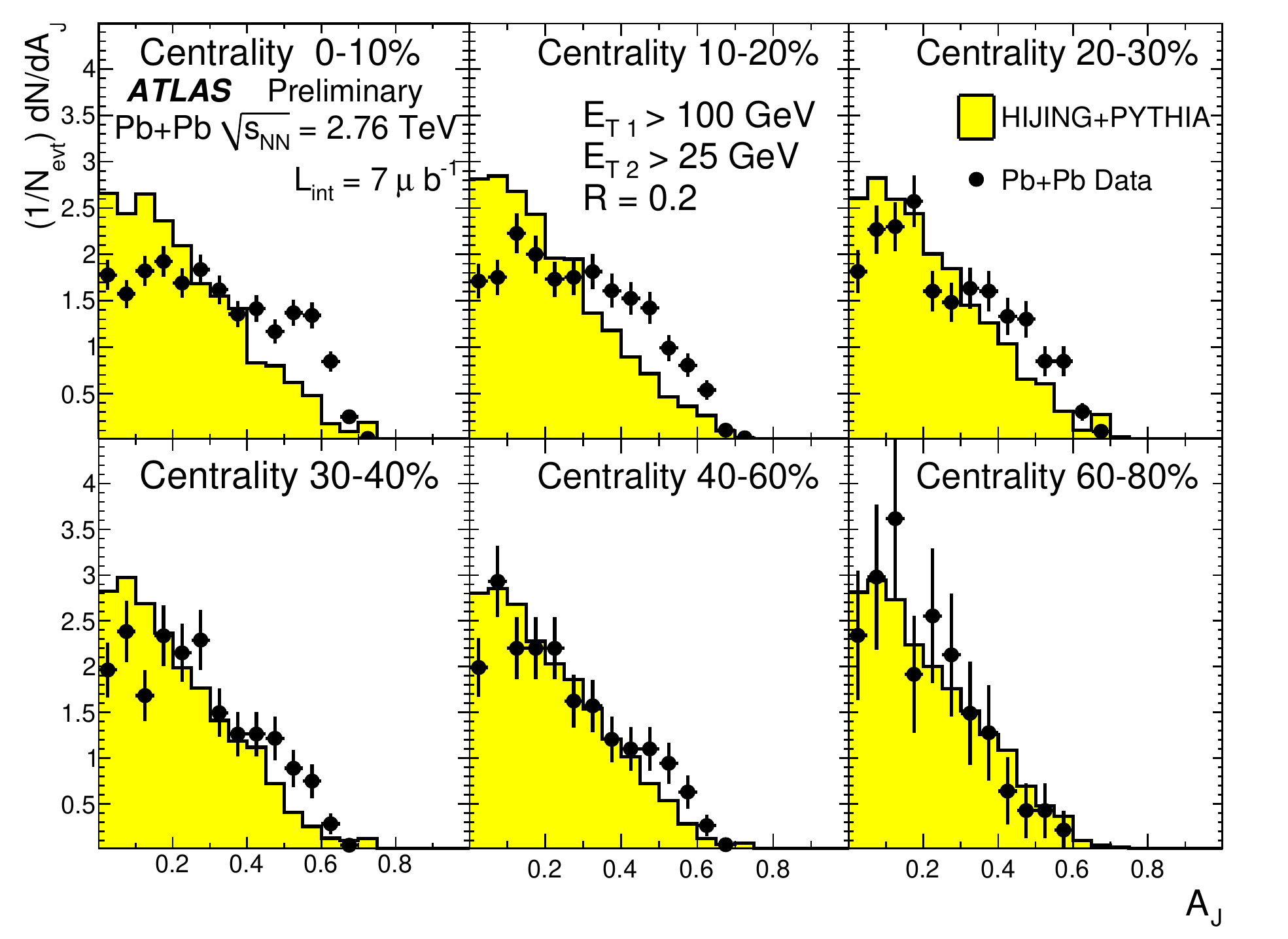}
\caption{Dijet asymmetry for~\RTwo\ jets in six centrality bins in
  events with a leading jet with $\et > 100\GeV$. A comparison to HIJING
  with embedded PYTHIA dijet events (yellow) is shown. }
\label{fig:antikt2HIAvg_100gev_asym_panel}
\end{figure}

Figure~\ref{fig:antikt2HIAvg_100gev_asym_panel} shows the measured
asymmetry distributions for \RTwo\ jets with $\ETOne >
  100$~\GeV\ compared to results from MC simulations. An
  analysis of \pp\ data with \RTwo\ is not yet available. The \RTwo\ asymmetry
  distributions show the same general features as seen in the
  \RFour\ results shown in
  Fig.~\ref{fig:antikt4HIItrFlow_asym_panel}. However, the 
  asymmetry distributions generally extend to slightly larger \Aj\ values for
  \RTwo\ jets than for \RFour\ jets. Another difference between the
  \RFour\ and \RTwo\ results can be seen in the 40-60\% centrality bin
  where a  modification of the asymmetry distribution
  is seen for \RTwo\ jets and not for \RFour\ jets.

One of the most important results presented in the original dijet
asymmetry letter was the observation that the dijet
\dphi\ distribution remains mostly unchanged in all centrality bins
while the asymmetry distribution is strongly modified in central
collisions. Figures~\ref{fig:antikt4HIItrFlow_dphi_panel} and
\ref{fig:antikt4HIItrFlow_dphi_panellogy} show updated
results for the \RFour\ \dphi\ distributions for dijets with $\ETOne >
100$~\GeV\ using the same dijet selection procedure described
above. Both linear and logarithmic vertical scales are shown. These figures confirm the original observation that the dijet
\dphi\ distributions are unmodified with the caveat
that a small combinatoric contribution is observed in the 0-10\% and
10-20\% centrality bins. Such a combinatoric contribution was also
seen in the original measurement. The dijet \dphi\ distributions for
\RTwo\ jets with $\ETTwo > 100$~\GeV\ are shown in
Figs.~\ref{fig:antikt2HIAvg_dphi_panel} and
\ref{fig:antikt2HIAvg_dphi_panellogy} with linear and logarithmic
vertical scales, respectively. The \RTwo\ \dphi\ distributions show no
modification and a much smaller combinatoric contribution.
 
\begin{figure}[htb]
\centering
\includegraphics[width=0.9\textwidth]{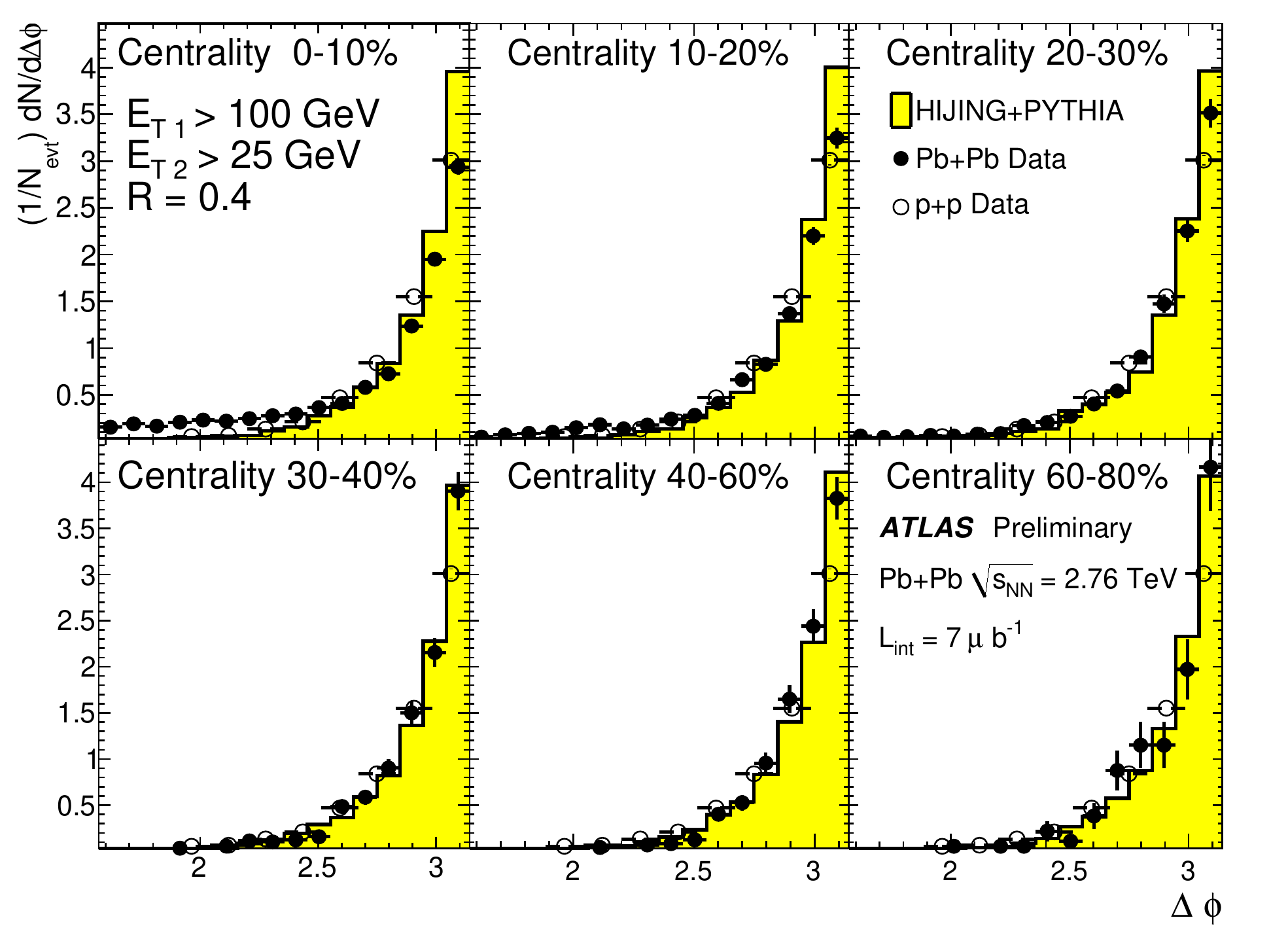}
\caption{Dijet \dphi\ distributions for~\RFour\ jets in $\sqrtsnn = 2.76$~\TeV\
  \PbPb\ collisions having a leading jet with $\et > 100$~\GeV\
  in six centrality bins. A comparison to HIJING
  with embedded PYTHIA dijet events (yellow) and ATLAS
  $\sqrt{s}=7\TeV$~\pp\ data (open circles) is shown. }
\label{fig:antikt4HIItrFlow_dphi_panel}
\end{figure}
\begin{figure}[htb]
\centering
\includegraphics[width=0.9\textwidth]{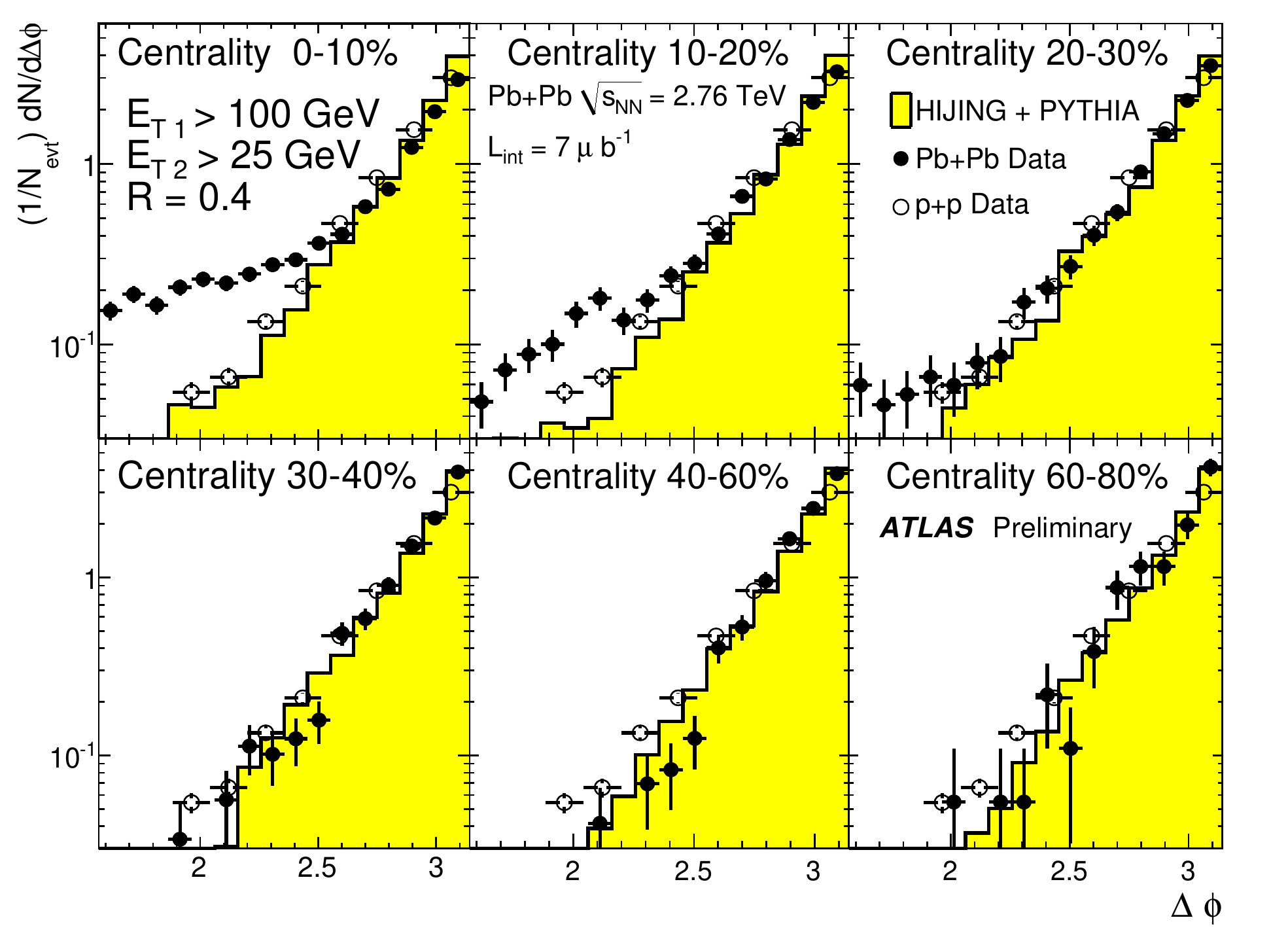}
\caption{Dijet \dphi\ distributions plotted with a logarithmic
  vertical scale for~\RFour\ jets in $\sqrtsnn = 2.76$~\TeV\
  \PbPb\ collisions having a leading jet with $\et > 100$~\GeV\
  in six centrality bins. A comparison to HIJING
  with embedded PYTHIA dijet events (yellow) and ATLAS
  $\sqrt{s}=7\TeV$~\pp\ data (open circles) is shown. }
\label{fig:antikt4HIItrFlow_dphi_panellogy}
\end{figure}

\begin{figure}[htb]
\centering
\includegraphics[width=0.9\textwidth]{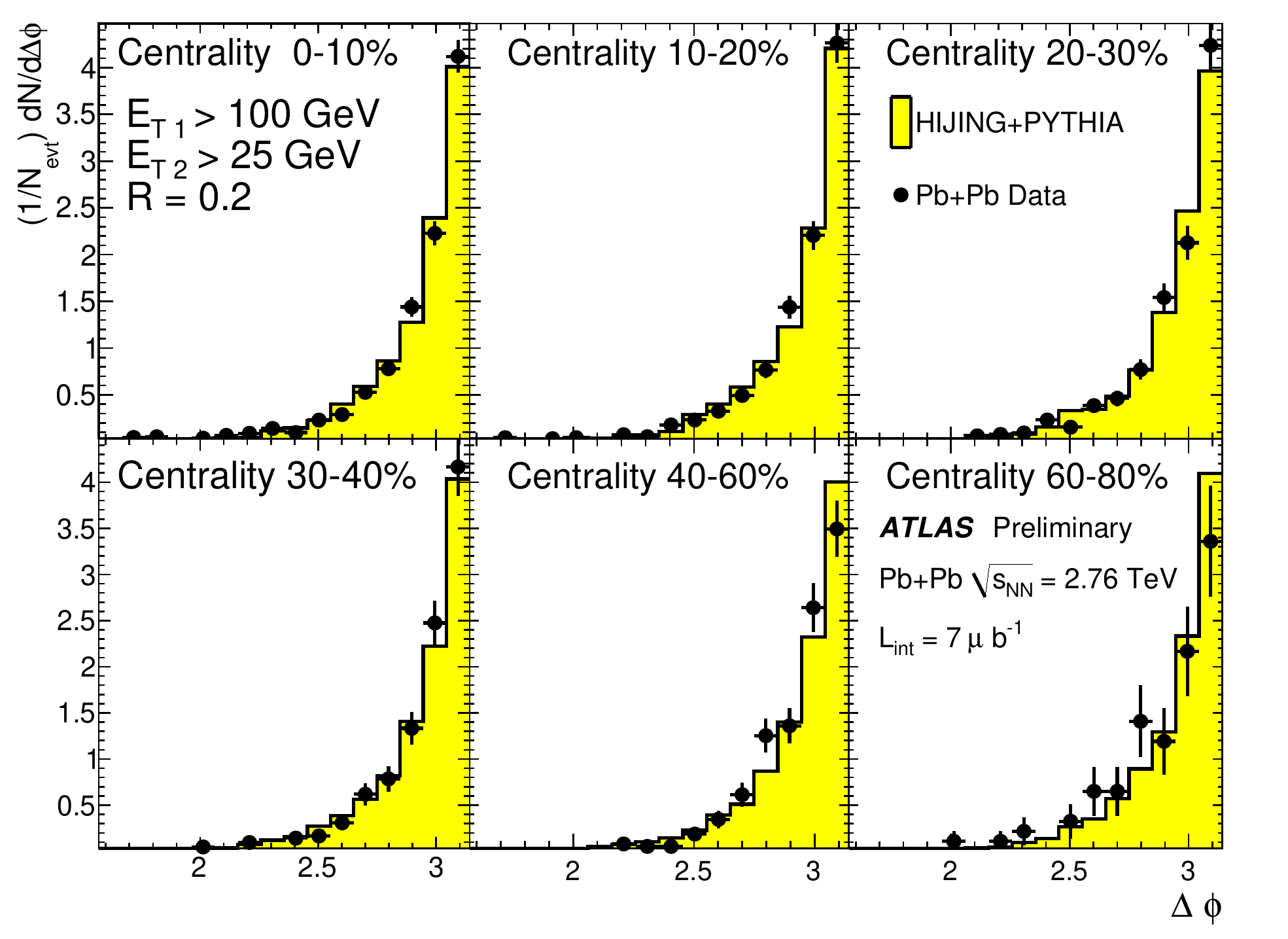}
\caption{Dijet \dphi\ distributions for~\RTwo\ jets in $\sqrtsnn = 2.76$~\TeV\
  \PbPb\ collisions having a leading jet with $\et > 100$~\GeV\ in six
  centrality bins. A comparison to HIJING
  with embedded PYTHIA dijet events (yellow) is shown. }
\label{fig:antikt2HIAvg_dphi_panel}
\end{figure}

\begin{figure}[htb]
\centering
\includegraphics[width=0.9\textwidth]{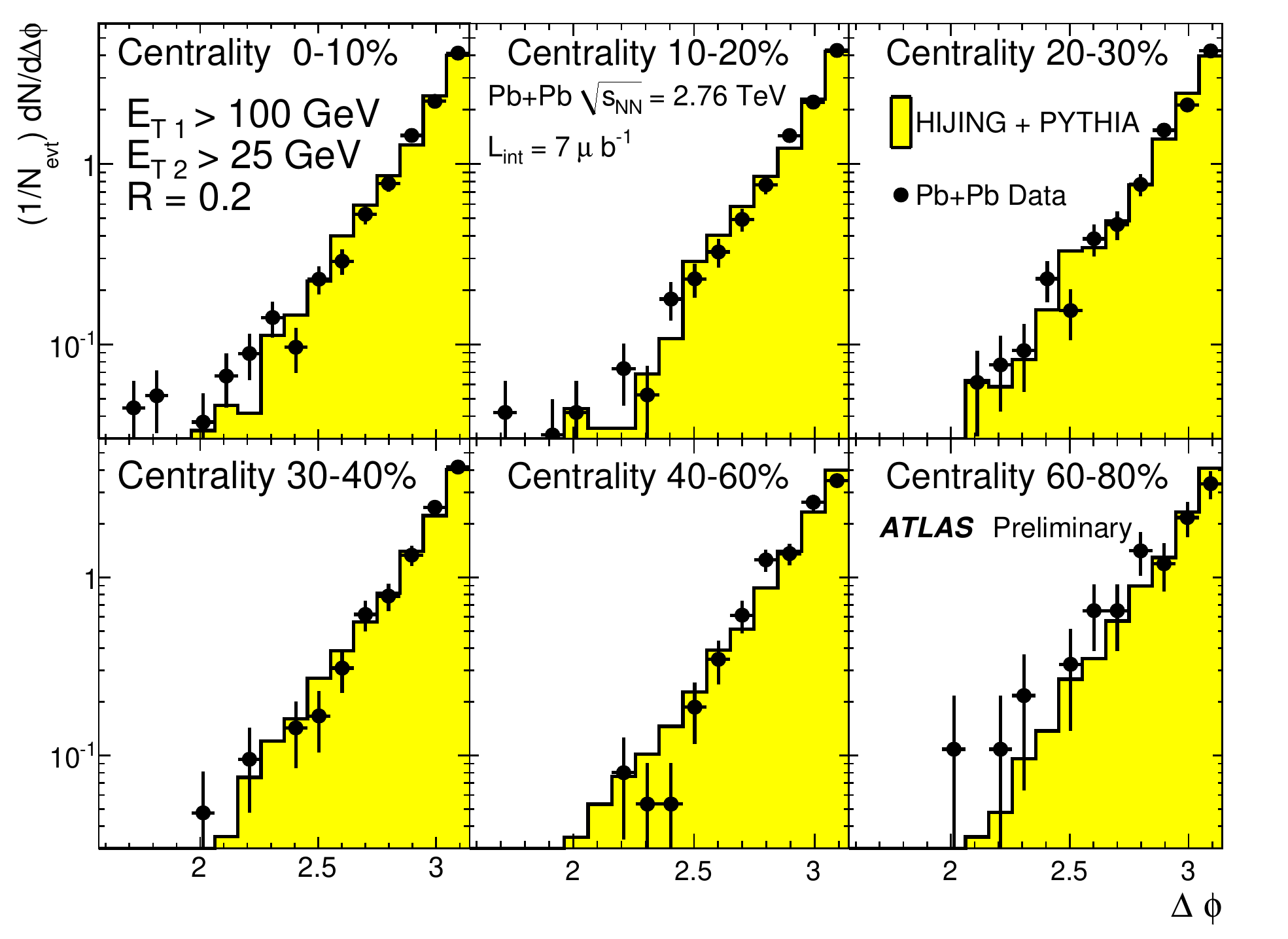}
\caption{Dijet \dphi\ distributions plotted with a logarithmic
  vertical scale for~\RTwo\ jets in $\sqrtsnn = 2.76$~\TeV\
  \PbPb\ collisions having a leading jet with $\et > 100$~\GeV\ in six
  centrality bins. A comparison to HIJING
  with embedded PYTHIA dijet events (yellow) is shown. }
\label{fig:antikt2HIAvg_dphi_panellogy}
\end{figure}

The evolution of the asymmetry with the \ETOne\ threshold is shown in
Figures~\ref{fig:RFourDijetAsymmEtonedep} and
\ref{fig:RTwoDijetAsymmEtonedep}, which contain distributions of dijet \Aj\ for
three different centrality bins, 0-10\% (top), 30-40\% (middle),
60-80\% (bottom), for three different ranges of energies (see figures)
for the leading jet. The peaking at larger values of \Aj\ becomes
more pronounced when \ETOne\ is restricted to a lower range and that
peaking becomes less pronounced as \ETOne\ increases. The peaking is
particularly prominent for \RTwo\ jets with $75 < \ETOne <
100$~\GeV\ where the peak at $\Aj \approx 0.55$ corresponds to $\ETTwo
\approx 25$~\GeV\ -- the minimum value for \ETTwo\ allowed in the
analysis. With increasing \ETOne\, the peak occurs at 
approximately the same \ETTwo\ value, but larger \Aj. The 
\Aj\ range accessible in the measurement is limited by the presence of low
\ET\ real or false soft jets in the event that are selected as
the second jet and by inefficiencies in the jet reconstruction at low
\ET. Because of the smaller jet size, the \RTwo\ jets have better
efficiency at low \ET\ and fewer fake jets. These
arguments would explain why the asymmetry distribution extends to larger values of
\Aj\ in the $100 < \ETOne < 125$~\GeV\ bin. To complete the survey of dijet asymmetry measurements, 
Fig.~\ref{fig:antikt2HIAvg_75gev_asym_panel} shows the full centrality
dependence of the \Aj\ distributions for \RTwo\ jets with $\ET > 75$~\GeV.
\begin{figure}[htb]
\centering
\includegraphics[width=0.9\textwidth]{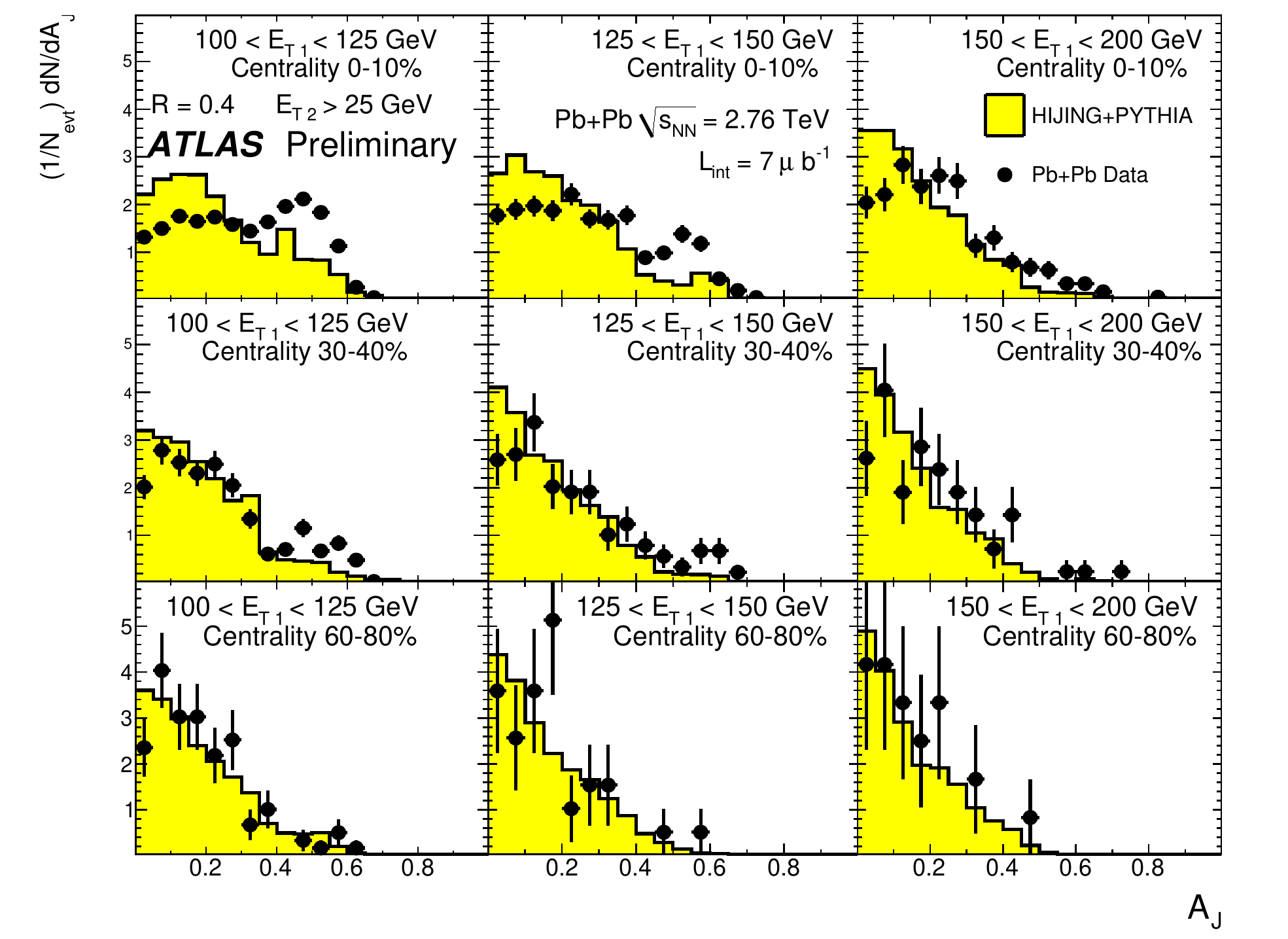}
\caption{Dijet asymmetry for~\RFour\ jets in three centrality bins:
0-10\% (top), 30-40\% (middle), 60-80\% (bottom)
 for $\sqrtsnn = 2.76$~\TeV\ \PbPb\ collisions with a leading jet with
 $100 < \ETOne < 125$~\GeV\ (left), 
$125 < \ETOne < 150$~\GeV\ (middle), $150 < \ETOne < 200$~\GeV\ (right).
For all of the plots, comparison to HIJING events with embedded PYTHIA dijets
for the same conditions on reconstructed jets as the data is shown
(yellow). }
\label{fig:RFourDijetAsymmEtonedep}
\end{figure}
\begin{figure}[htb]
\centering
\includegraphics[width=0.9\textwidth]{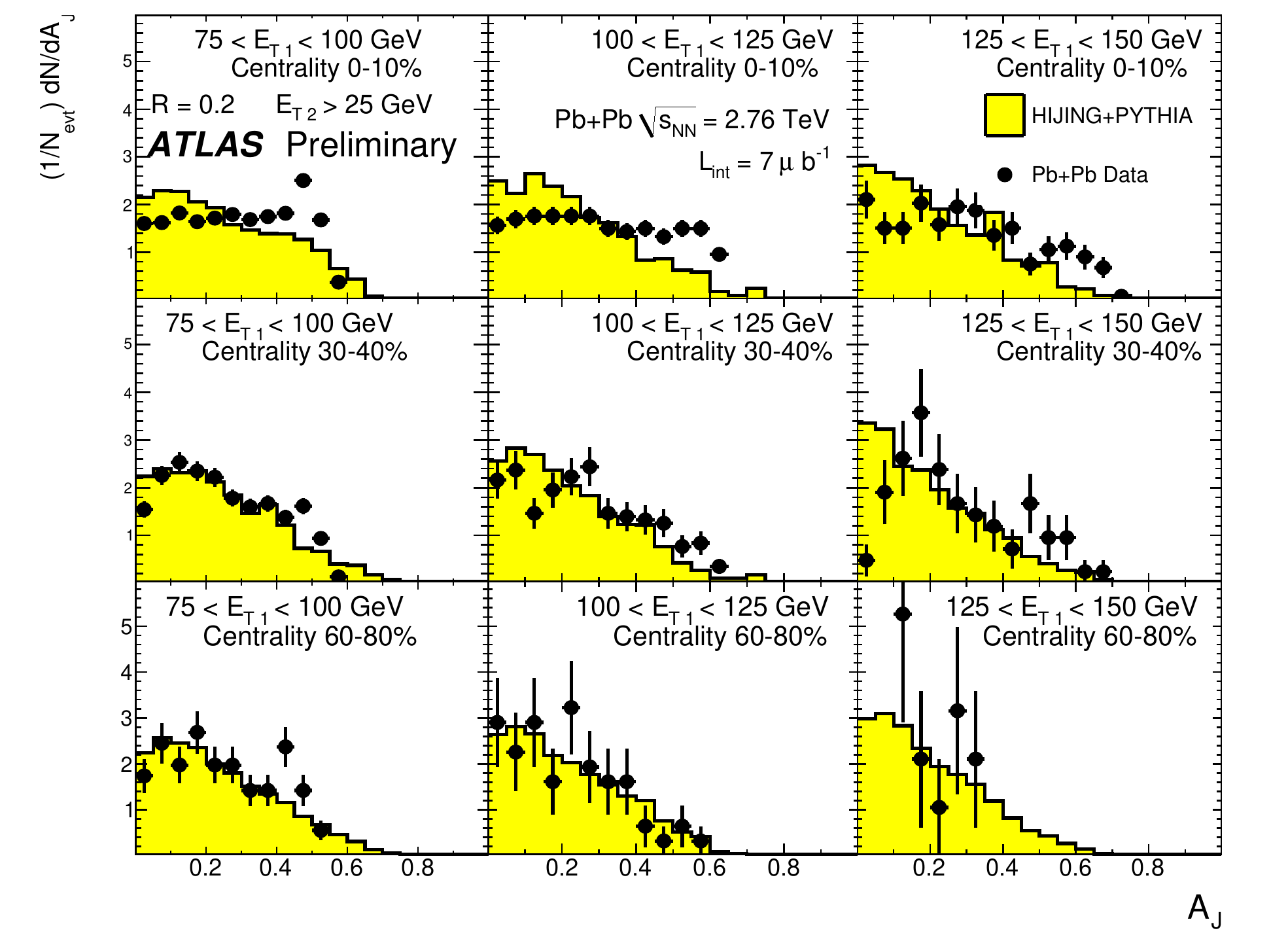}
\caption{Dijet asymmetry for~\RTwo\ jets in three centrality bins:
0-10\% (top), 30-40\% (middle), 60-80\% (bottom)
 for $\sqrtsnn = 2.76$~\TeV\ \PbPb\ collisions with a leading jet with
 $75 < \ETOne < 100$~\GeV\ (left), 
$100 < \ETOne < 125$~\GeV\ (middle), $125 < \ETOne < 150$~\GeV\ (right).
For all of the plots, comparison to HIJING events with embedded PYTHIA dijets
for the same conditions on reconstructed jets as the data is shown
(yellow). 
}
\label{fig:RTwoDijetAsymmEtonedep}
\end{figure}
\begin{figure}[htb]
\centering
\includegraphics[width=0.9\textwidth]{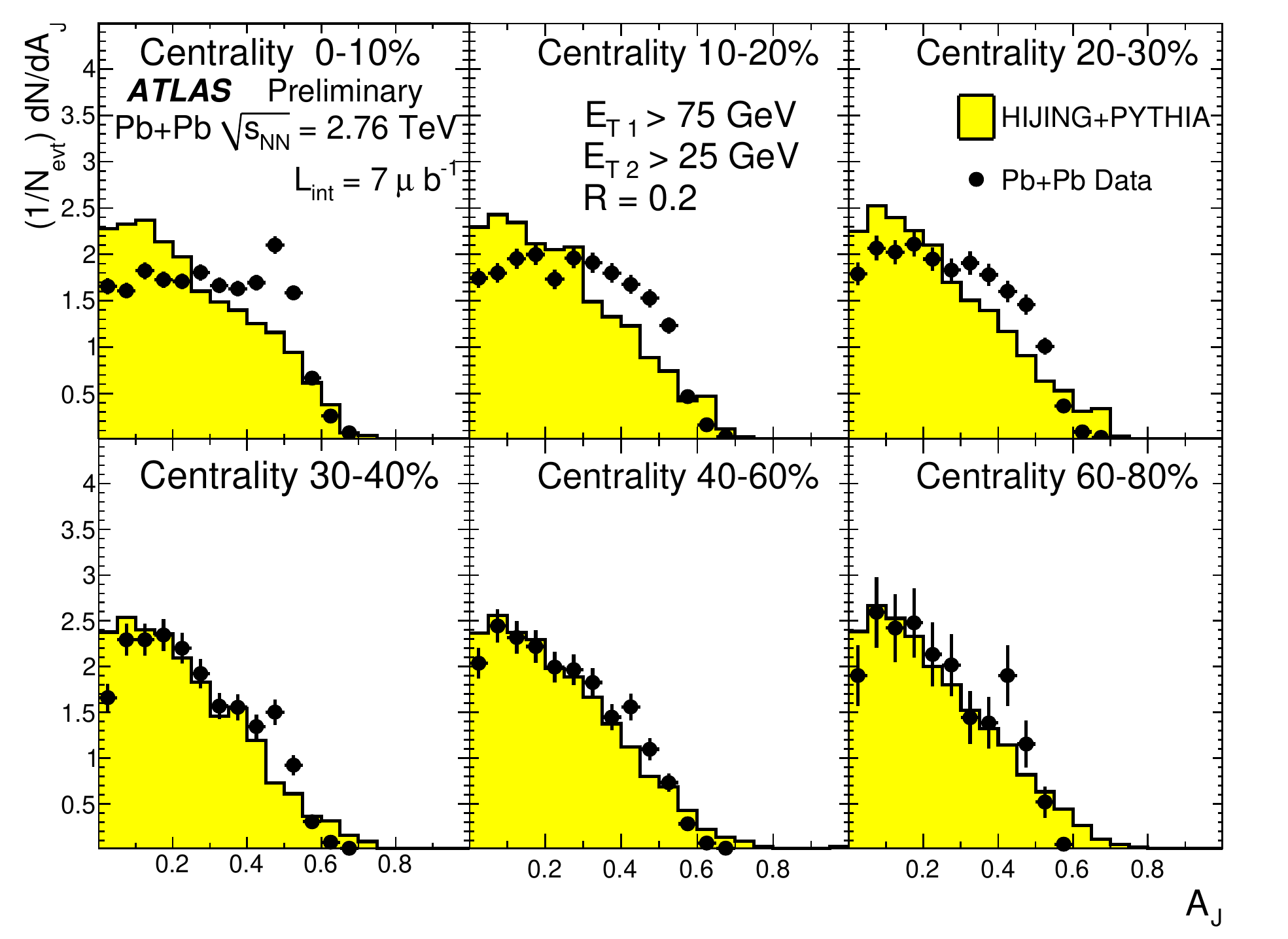}
\caption{Dijet asymmetry for~\RTwo\ jets in six centrality bins in
  events with a leading jet with $\et > 75\GeV$. A comparison to HIJING
  with embedded PYTHIA dijet events (yellow) is shown. }
\label{fig:antikt2HIAvg_75gev_asym_panel}
\end{figure}

\clearpage
\chapter{Conclusion}
\label{section:conclusion}
This thesis has presented results for the jet \Rcp\ and dijet
asymmetry. The asymmetry distributions indicate that there are a class of
events where there is a significant momentum imbalance between the two
leading jets in the event. The extent of this imbalance gradually increases with
centrality implying that the momentum balance between dijets is modified in the presence of the
medium. This is strongly suggestive of the interpretation that the asymmetry is
directly probing the differential energy loss of two partons
traversing different in-medium path lengths. Furthermore, these dijets
show an angular correlation that is consistent with vacuum dijet
production. When this result was first released, the lack of
modification of this angular correlation was considered surprising,
especially given the copious energy loss implied by the asymmetry.

The \Rcp\ results presented in this thesis show that jets are
suppressed in central collisions by approximately a factor of two
relative to peripheral collisions. This suppression is almost
completely independent of \pt. The approximate value of
$\Rcp\sim\frac{1}{2}$ and the slight increase with \pt\ are
qualitatively consistent with predictions using
the PYQUEN MC event generator~\cite{Abreu:2007kv}. As the angular size of the jets is
increased, by increasing the $R$ parameter, the \Rcp\ also
increases. This $R$-dependence of \Rcp\ is weak and is a marginal effect
above the measurement uncertainties.
For lower energy jets,
$50\lesssim \pt \lesssim 80$~\GeV, an increase in \Rcp\ with $R$ that is
greater than that for higher energy jets was predicted in
Ref.~\cite{He:2011pd}, shown in Fig.~\ref{fig:bkgr:vitev_raa}. While
the measurement uncertainties limit a strong conclusion, this trend is
present in the data as well.

The \pt-dependence can be compared
with that of the single particle \RAA\ measured at the LHC shown
in the summary plot in Fig.~\ref{fig:conclusion:cms_raa}.
\begin{figure}[hbt]
\centering
\includegraphics[width=0.6\textwidth]{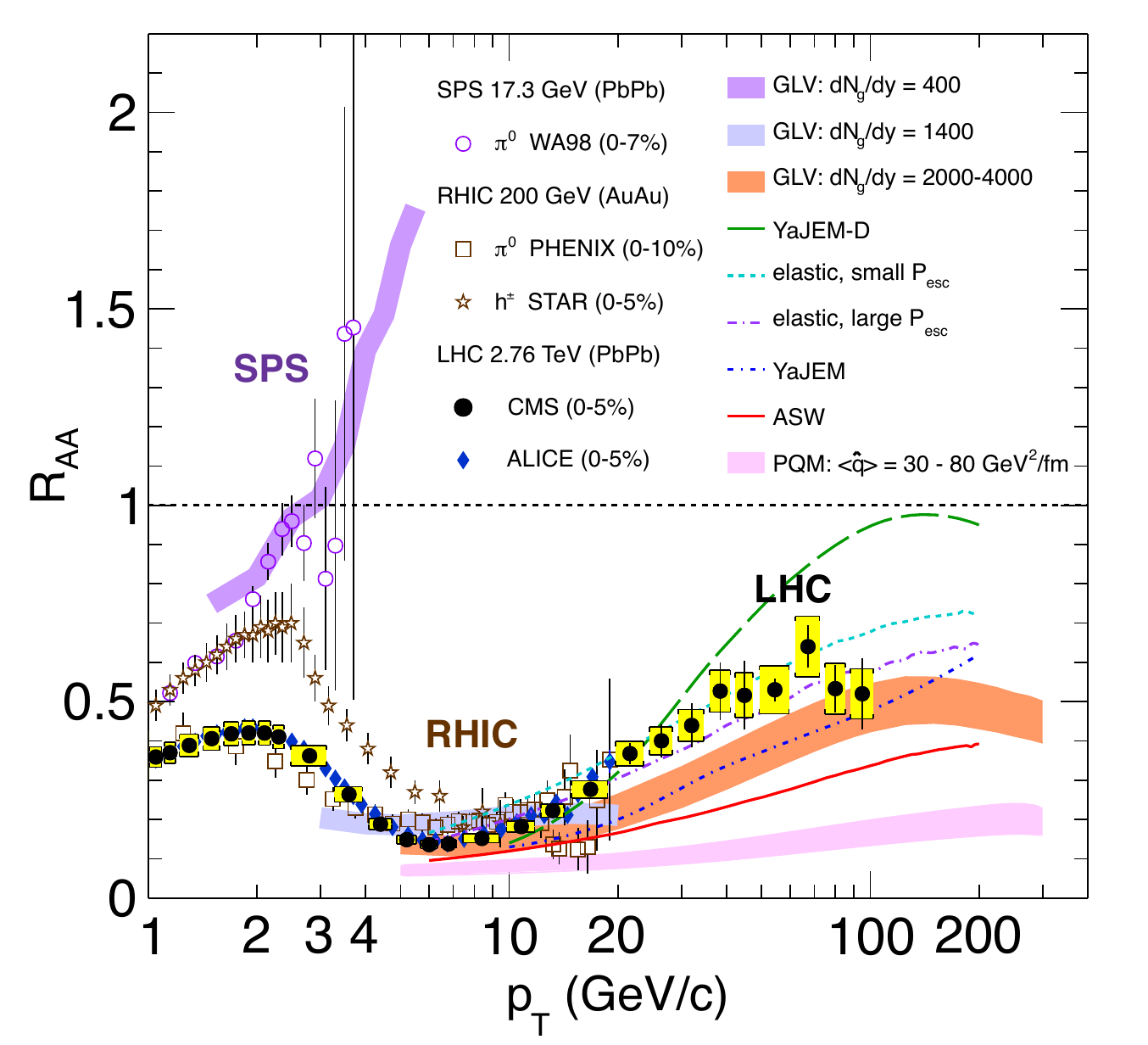}
\caption{Single hadron \RAA\ from various experiments combined with
  multiple theoretical calculations. The LHC results extending to high \pt\ show
  a clear reduction in the suppression that was discovered at RHIC. Figure adapted from Ref.~\protect\cite{CMS:2012aa}.}
\label{fig:conclusion:cms_raa}
\end{figure}
 The most recent RHIC results showed a constant suppression $\RAA\sim 0.1$ at
the highest available \pt. The LHC results have extended to higher
\pt\ where the \RAA\ is found to rise up to $\sim 0.5$, which is in
the same regime as the jet \Rcp\ shown here. While it has been argued
in this thesis that there is no requirement that the single particle
and jet suppression factors be related, the plateau in the
\pt-dependence exhibited by both observables suggests some common
feature of the high \pt\ energy loss regime.

Since the anti-\kt\ jets have
highly regular geometry, the $R$ parameter can be safely interpreted
as a cone radius. Changing the jet radius has the effect that medium induced radiation emitted at angles
between the old and new radii will be recovered by the jet and the energy
loss will be reduced. An additional effect is the change in the spectral
shape due to the change in definition, which is present even in the
case of no quenching. Therefore, evaluations of the energy loss as a
function of jet radius must be convolved with this second effect to be
comparable to the experimentally observed spectra or
\Rcp. This is normally the case in theoretical evaluations of the
\Rcp\ or \RAA\ where the per parton energy loss must be convolved with
the production spectrum. Equivalently, without the analogous deconvolution, a comparison
between the \Rcp\ at two radii (i.e the experimental observation in
this thesis) is indirectly sensitive to the energy
recovery.

A quantitative extraction of the energy loss was attempted in
Ref.~\cite{Adcox:2004mh} with single particle \RAA\ measurements from
PHENIX. The procedure used the unmodified inclusive single particle spectrum to remove the dependence
on the spectral shape under the ansatz of an energy loss scenario
where $\Delta \pt \propto \pt$. In the previous chapter, this procedure was applied to fully
reconstructed jets with $\pt > 70$~\GeV, with the extracted \Sloss\ values providing a
basis of comparison between the different $R$ values. The results of
this exercise, summarized in Fig.~\ref{fig:results:sloss}, indicate an
$R$ dependence of the energy loss that is significant beyond the
estimated uncertainty. At fixed
\Npart, the differences in \Sloss\ between the two $R$
values can be interpreted as the fractional recovered energy in the
annular region between the two radii. The \RTwo\ and \RThree\ values remain
relatively close, however increasing the radius to \RFour\ shows a
reduction in the energy loss fraction. The marginal decrease in
\Sloss\ is even greater when moving from \RFour\ to \RFive. This trend
indicates that the energy recovery grows faster than linearly with $R$.
The relative magnitudes of \Sloss\ evolve with \Npart, indicating that
the energy recovered by expanding the jet radius increases with
\Sloss.

To establish this behavior quantitatively, the \Npart\
dependence of \Sloss\ was fit with the form, $\Sloss=S\Npart^k$, with
the fit results given in Table~\ref{tbl:results:sloss}. This can be
compared with the single hadron analysis performed by PHENIX.
This analysis extracted larger values for \Sloss\ than those obtained here, which
is consistent with the general expectation that some of the energy
radiated by the parton may be recovered when using jets. Furthermore, the \Sloss\ was
found to scale approximately as $\Npart^{2/3}$, which is consistent
with the $L^2$ energy loss models discussed in
Section~\ref{section:bkgr:quenching}. The $k$ values calculated here
for full jets are much larger, with the $R=$0.2, 0.3 and 0.4 jets
having $k\sim0.9$. The \RFive\ shows a slower $\Npart^{3/4}$ growth.

The \Rcp\ is only sensitive to nuclear modification effects relative
to a peripheral reference spectrum in this case the 60-80\%
centrality bin. To quantify the full extent of these effects a \pp\
reference spectrum should be used. The LHC recorded \pp\ data at
2.76~\TeV\ during early 2011, and a full \RAA\ measurement using this
data is the obvious next step in a systematic study of jet quenching
at the LHC. A jet triggered sample of approximately
150~$\mathrm{nb}^{-1}$ was recorded by ATLAS, which provides ample
statistics for the jet spectrum measurement out to $\pt\simeq
250$~\GeV. As this reference spectrum will contain better statistics
than the 60-80\% \PbPb\ spectrum, the \RAA\ measurement can be
performed in bins of rapidity as well. Since the \RAA\ is sensitive to
the full extent of the quenching, this measurement will likely focus
on the maximal suppression observed in the 0-1\% or 0-5\% centrality bins.

To more firmly associate the jet suppression observation with the
quenching interpretation, suppression from the NPDF must be
constrained through future measurements. The jet \Rcp's tend to exhibit a decrease with increasing
\pt, particularly in the most central bins. While this effect is
marginally significant beyond systematic errors, a possible
explanation could be the sampling of the NPDFs in the EMC
region. This scenario must be investigated; any detailed
measurements of the quenching require the decomposition of the total
observed suppression into contributions from cold nuclear matter
effects and a quenching component. This can be checked with a measurement of the direct
photon \RAA, which is sensitive to hard scattering production rates
determined by the NPDFs, but not to quenching. 
Additionally the rates for \Zzero\ production can be used to constrain cold nuclear matter
effects, as the \Zll\ decay modes should also be immune to the
quenching. Both ATLAS~\cite{Aad:2010aa} and CMS~\cite{Chatrchyan:2011ua} have published \Zzero\ \Rcp\ values,
however these results are statistics limited and cannot provide
rigorous constraints on the NPDFs. This could be improved with results from the higher
statistics 2011 run. The upcoming \pPb\ run at the LHC should provide an opportunity to
perform full jet reconstruction in these collisions for the first
time. ATLAS should not only be able to provide precision measurements
for determination of the NPFs, but also measure the impact
parameter-dependence of the nuclear modification effects discussed in Section~\ref{section:bkgr:nuclear_modification}.

In the case of photon identification, a
significant background comes from \pizero's in jets. One of the primary
techniques at rejecting this background in analyzing \pp\ collisions is to require that the photon
be isolated; this is typically accomplished by placing a maximum threshold on the energy
deposited in a cone of fixed radius about the reconstructed photon
axis, excluding the energy of the photon. The subtraction procedures
developed for jet reconstruction in heavy ion collisions discussed in
this thesis have been adapted by ATLAS to be applied to the
isolation cones in the photon reconstruction.

As discussed previously, dijet observables provide complementary
information to the single inclusive measurements as they are sensitive
to the relative energy loss of the two jets. Furthermore, both gamma-jet and \Zzero-jet correlations can be used in
analogy with the asymmetry to determine single jet energy loss, and
are considered topics of future interest. The asymmetry results
presented here show that for systems with a 100~\GeV\ leading jet, a
significant fraction of events were found to have the sub-leading jet
correlated in $\Delta \phi\simeq \pi$ and have $\ETTwo\sim
25$~\GeV. This means that the quenching is strong enough to regularly
cause energy loss of $75\%$ of the jet's energy. Furthermore, the
relatively strong angular correlation indicates that this energy loss
must be incurred without significantly altering the direction of the
jet. Whether this feature is captured by various energy loss models is
not clear, as most models used the soft eikonal limit, which does not
fully describe the transverse deflection of
the leading parton.

The measured angular correlation can be used to test predictions of modified
acoplanarity distributions through elastic
scattering~\cite{Appel:1985dq,Blaizot:1986ma,Rammerstorfer:1990js}.  By rescaling the
$K_{\eta}$-dependence by the leading jet \pt, this quantity, defined
in Eq.~\ref{eqn:bkgr:acoplanarity}, can be related to the normalized dijet angular correlation,
\begin{equation}
\frac{1}{N}\frac{dN}{d\phi}\propto\frac{1}{\sigma_0(\pt)}\frac{d\sigma}{d\phi}=\pt\frac{dP}{dK_{\eta}}=\frac{dP}{d(K_{\eta}/\pt)}\,.
\end{equation}
This quantity shows almost no modification in heavy ion collisions,
and significantly less modification than the predicted acoplanarity
distributions, which generally show a reduction of 25-50\% in the
maximum of the correlation relative to \pp.

The measured angular correlation can be used to
constrain the magnitude of transverse momentum kicks, $\langle \kt^2 \rangle$, imparted to the
jet through random elastic scattering in the medium. For a dijet
system with back-to-back jets with momentum \pt\ before energy loss,
each jet can undergo random small angle multiple scattering resulting
in an RMS angular deflection of
\begin{equation}
\sigma_{\mathrm{jet}}\equiv\sqrt{\langle\Delta \phi_{\mathrm{jet}}^2\rangle} \simeq \frac{\sqrt{\langle \kt^2 \rangle}}{\pt}\,.
\end{equation}
The angular deflection of each jet is independent of the other so the
effect on the dijet distribution is an additional kick of
$\sigma_{\mathrm{dijet}}=\sqrt{2}\sigma_{\mathrm{jet}}$. An upper limit
on  $\sigma_{\mathrm{jet}}$ can be
placed by sampling the dijet $\Delta \phi$ distribution obtained from the MC including
additional Gaussian smearing with different values of $\sigma_{\mathrm{dijet}}$ and testing to see which values are
consistent with the
data. Figure~\ref{fig:conclusion:dijet_angle_smear} shows the data
along with the nominal MC distribution and several smeared variants of
the MC distribution for \RFour\ jets with $\ETOne > 100$~\GeV\ in the
0-10\% centrality bin. The distributions are scaled
  to have unit value in the highest $\Delta \phi_{\mathrm{dijet}}$ bin
  to facilitate comparisons of the shapes. The
  $\sigma_{\mathrm{jet}}=0.015$ distribution shown in green is clearly
  broader than the data, as is the $\sigma_{\mathrm{jet}}=0.01$
  distribution shown in blue except in one bin. If the former is taken
  as an estimate of the upper bound on $\sigma_{\mathrm{jet}}$ and
  100~\GeV\ is used as an estimate of the jet \ET, this corresponds to
  a limit of $\sqrt{\langle \kt^2 \rangle} \lesssim 1.5$~\GeV.
\begin{figure}[htb]
\centering
\includegraphics[width=0.75\textwidth]{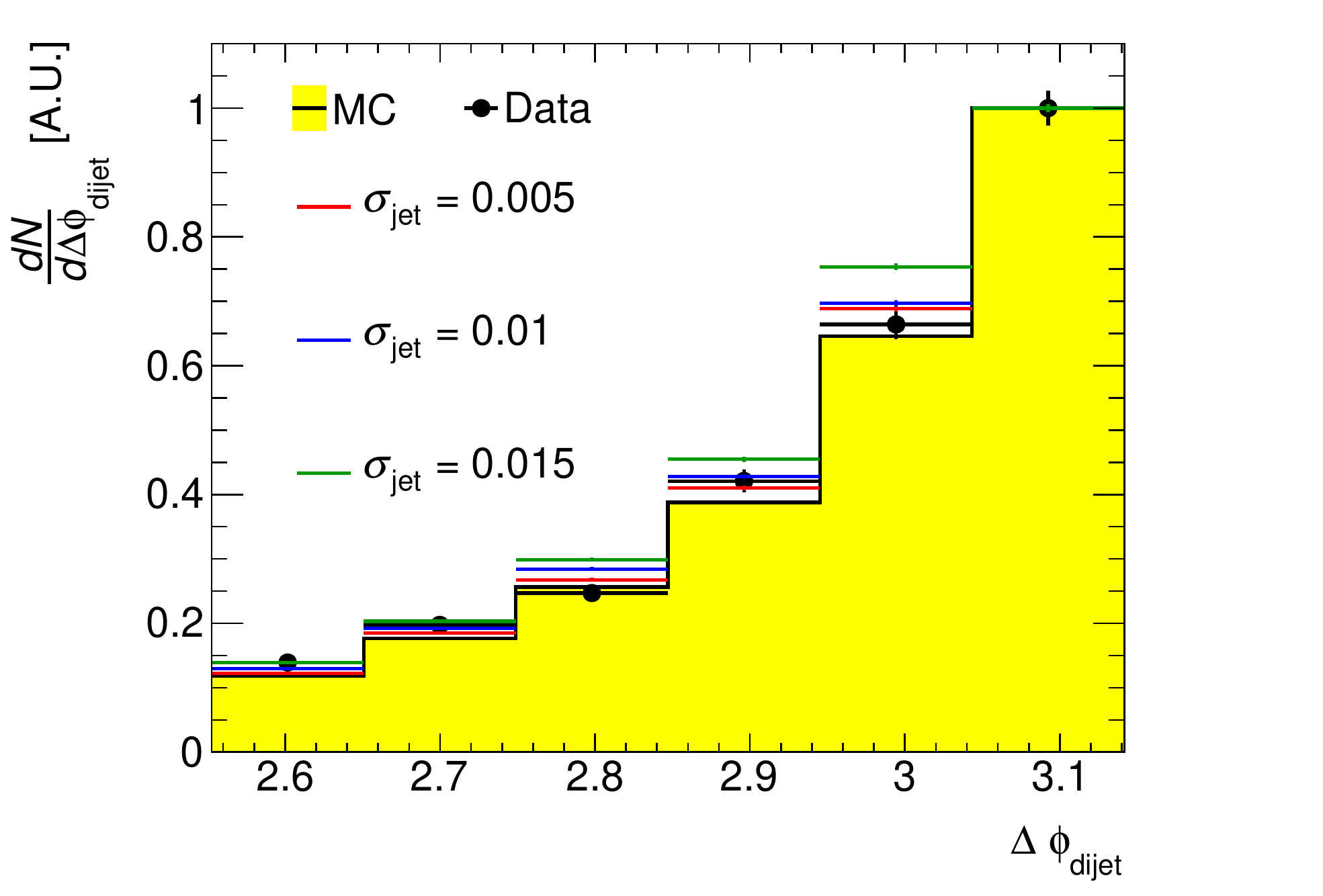}
\caption{The $\Delta \phi_{\mathrm{dijet}}$ distribution for data
  (black), MC (filled yellow) and MC with additional Gaussian smearing
  using three different $\sigma$ values. The distributions are scaled
  to have unit value in the highest $\Delta \phi_{\mathrm{dijet}}$ bin. The errors represent the
  statistical errors in the data or MC sampling.}
\label{fig:conclusion:dijet_angle_smear}
\end{figure}

The fact that the jet suffers essentially no angular deflection means
that there is a single well-defined jet axis. For
events with incomplete nuclear overlap the collision zone is oblong
and there are a variety of path lengths the jet could take through the
same medium. Furthermore, if the overlap is assumed to have
elliptical geometry, the orientation and eccentricity of the
overlap region are accessible on a per event basis through the flow
observables $\Psi_2$ and $v_2$ respectively. If coordinates are chosen
such that the impact parameter vector is aligned with the $y$ axis,
the jet's linear trajectory is given by $y=\cot\Delta\psi x + y^{\prime}-\cot\Delta\psi x^{\prime}$,
where $\Delta\psi$ is the angle between the jet axis and the event plane,
with the primed coordinates indicating the jet production location.
The geometry of this
scenario is shown in Fig.~\ref{fig:conclusion:geom}. The production point of the jet is
unknown on a per event basis, but the average of such positions could
be taken over many collisions with the same flow characteristics.
\begin{figure}[hbt]
\centering
\includegraphics[width=0.7\textwidth]{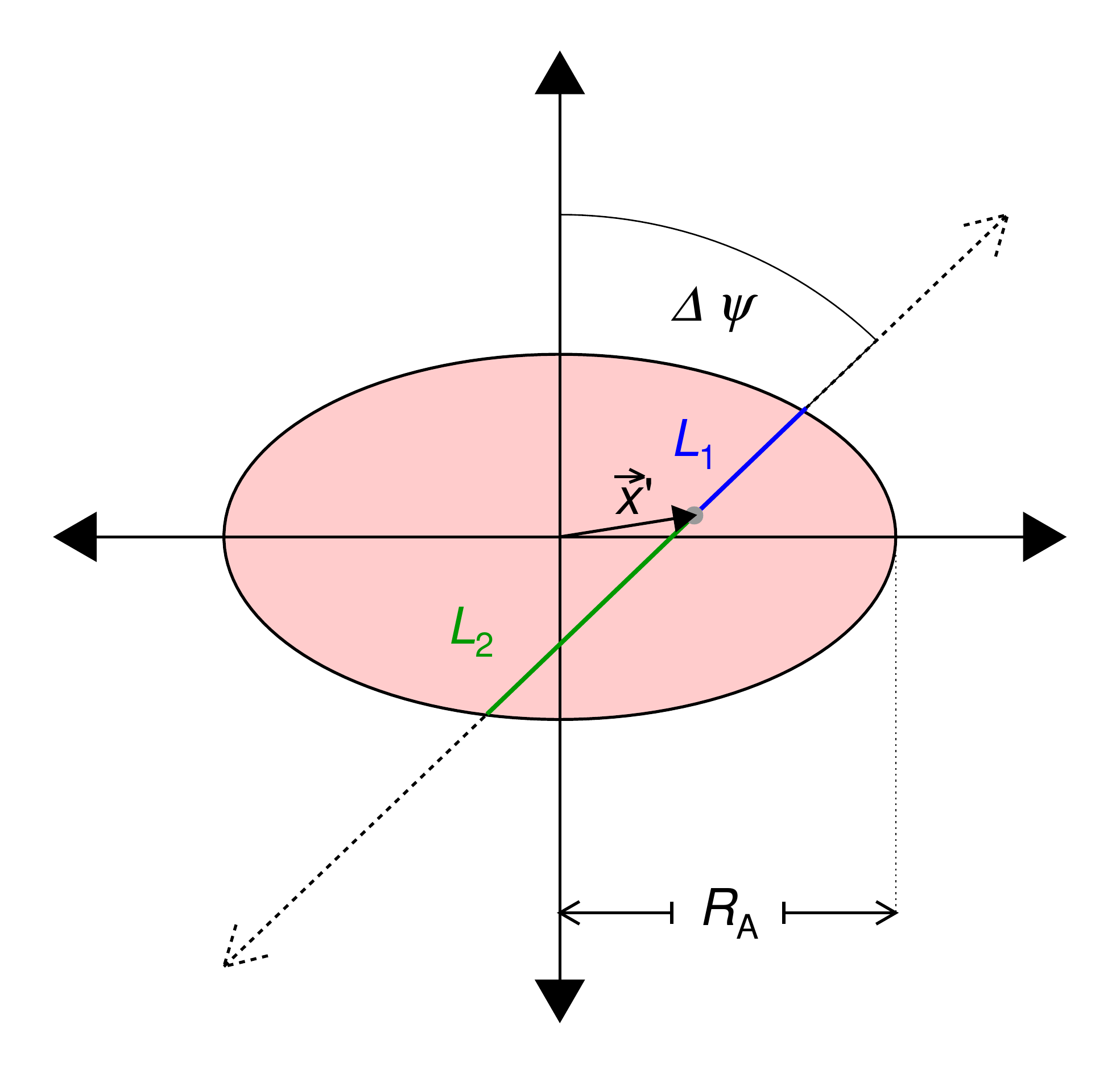}
\caption{Elliptical geometry of collision zone, with the impact
  parameter direction chosen to align with the $y$-axis. The jet
  production point is located by the vector
  $\mathbf{x}^{\prime}$. Jets travel path lengths $L_1$ and $L_2$
  through the medium which are denoted by blue and green lines
  respectively. The jet axis is oriented with angle $\Delta \psi$ with
  respect to $\Psi_2$.}
\label{fig:conclusion:geom}
\end{figure}
In this geometrical picture, two-jet differential and single inclusive
jet observables are sensitive to different moment-like quantities in
the energy loss. Consider the variable $\xi$, which is some quantity
sensitive to the energy loss. In a given event this should be a
function of the production position, initial parton kinematics and
collision geometry. The average of this variable over events of 
similar eccentricity and centrality should be related to the
differential energy loss, integrated over all different jet production
points inside an ellipse by the eccentricity,
\begin{equation}
\langle \xi \rangle_{\varepsilon} (\Delta \psi) = \int d^2x^{\prime}
\rho(\mathbf{x}^{\prime}) \xi(\mathbf{x}^{\prime},\varepsilon, \Delta \psi)\,,
\end{equation}
where $\rho$ is the transverse density of possible jet production
points. The dependence of $\xi$ on $\mathbf{x}^{\prime}$ comes from
the path length travelled in the medium,
i.e. $\xi(\mathbf{x}^{\prime},\epsilon,\Delta
\psi)=\xi(L(\mathbf{x}^{\prime},\varepsilon,\Delta \psi))$. The function $L$ can be found by determining
the coordinates of the jet-ellipse intersection and computing the path
length. The coordinates of where the jet leaves the medium are defined
by the solutions to the simultaneous set of equations for the jet axis
and the ellipse,
\begin{equation}
x^2+\frac{y^2}{1-\varepsilon^2}=R^2_{\mathrm{A}}\,,
\end{equation}
where $\varepsilon$ is the eccentricity and $R_{\mathrm{A}}$ is the nuclear
radius which forms the semi-major axis of the ellipse. The
eccentricity is assumed to be determined by centrality, but it may be
possible to use per event $v_2$ values as an additional handle to
constrain the range of eccentricity values in a class of events. Using
the jet reconstruction techniques described in Chapter~\ref{section:jet_rec}
the \ET-integrated flow harmonics and total \ET\ can be measured by
excluding energy deposition associated with jets on a per event basis. This should allow
better understanding of non-flow contributions to the harmonics as
well as the fraction of total \ET\ and particle production that comes
from hard processes.

When considering the single inclusive jet energy loss, $\xi$ is
sensitive to $L$ directly. For example, in the BDPMS-Z scheme the
energy loss for a  single jet is approximately $\Delta \ET \propto L^2$, using this
variable for $\xi$,
\begin{equation}
\langle \Delta \ET \rangle_{\mathrm{\varepsilon}} (\Delta \psi)\propto 
\int d^2x^{\prime}
\rho(\mathbf{x}^{\prime}) L^2(\mathbf{x}^{\prime},\varepsilon, \Delta
\psi)\equiv \overline{L^2} (\Delta \psi,\varepsilon)\,.
\end{equation}
This is contrasted with the case where the dijet differential energy is
the energy loss observable, $\Delta \ETOne -\Delta \ETTwo\propto \Delta
L^2$. This quantity has a different dependence on the collision
geometry since the path lengths of the two jets are correlated on a
per event basis. Thus
when averaging over similar events the quantity,
\begin{equation}
\langle \Delta \ETOne -\Delta \ETTwo \rangle_{\mathrm{\varepsilon}} (\Delta \psi)\propto 
\int d^2x^{\prime}
\rho(\mathbf{x}^{\prime}) \Delta L^2(\mathbf{x}^{\prime},\varepsilon, \Delta
\psi)\equiv \overline{\Delta L^2} (\Delta \psi,\varepsilon)\,,
\end{equation}
provides an independent handle on the geometrical dependence of the
energy loss. If put on a more rigorous theoretical footing,
calculations of these two variables in an energy loss framework would
provide a testable prediction of the path length dependence. The three-jet rate is high enough at the LHC that a similar approach
could be applied to events with three jets. These have additional per
event geometric correlations and would provide independent moments to
check the energy loss.

Relating a measured asymmetry to the partonic level energy loss can be
challenging due to the intrinsic width of the vacuum asymmetry
distribution. Furthermore, the distinction between multi-jet events
and splitting due to hard radiation is not always clear. Some of these
issues may be resolved by adding an additional step to the
experimental analysis where the final jets are used as input to
another round of clustering with a jet finding algorithm. For example,
\RTwo\ jets are robust against underlying event fluctuations, but are
prone to more splitting than larger jet definitions. A set of \RTwo\
jets could be used as input to a less geometrically restrictive
finding such as the \kt\ or Cambridge-Aachen algorithm with larger $R$
values, to build super-jet objects to be used in asymmetry-like analyses.

Jet structure and properties should also provide details on the
quenching mechanism. Measurements of jet shapes as well as the
transverse and longitudinal distributions of particles within jets are
sensitive to the angular pattern of the medium-induced radiation. A preliminary analysis of the jet fragmentation has been presented by
ATLAS~\cite{JetQM11}, and a more complete study using the 2011 data set
is presently underway. This analysis should provide constraints on
quenching mechanism as it relates to the medium modified
fragmentation functions discussed in
Section~\ref{section:bkgr:quenching}. Although experimentally
challenging, the analysis of jet fragmentation can be applied to
reconstructed photons. The fragmentation photons provide an even
cleaner probe of the angular radiation pattern; unlike the hadronic
fragments the photons do not continue to interact with the medium
after being emitted.

It is hoped that the measurements presented here will inspire theoretical calculations in
the context of the various paradigms discussed in
Section~\ref{section:bkgr:quenching} to consider the angular
dependence of the energy loss and its effect on \Rcp. Comparisons can
be made both in terms of radiation recovery between the different
radii as well as the total amount of large angle radiation ($R>5$) to
see whether this can give the expected total suppression of $\Rcp\sim
0.5$. Any plausible quenching scenario must simultaneously satisfy
both of these observations.

\clearpage



\addcontentsline{toc}{chapter}{Bibliography}
\singlespace
\bibliography{main}

\providecommand{\href}[2]{#2}\begingroup\raggedright\begin{thebibliography}{100}

\bibitem{Aad:2010bu}
{Atlas} Collaboration, {\em {Observation of a Centrality-Dependent Dijet
  Asymmetry in Lead-Lead Collisions at $\sqrtsnn=2.76$~\TeV\ with the ATLAS
  Detector at the LHC}\/},
  \href{http://dx.doi.org/10.1103/PhysRevLett.105.252303}{Phys.Rev.Lett. {\bf
  105} (2010)  252303},
\href{http://arxiv.org/abs/1011.6182}{{\tt arXiv:1011.6182}}.

\bibitem{Han:1965pf}
M.~Han and Y.~Nambu, {\em {Three Triplet Model with Double $SU(3)$
  Symmetry}\/},
\href{http://dx.doi.org/10.1103/PhysRev.139.B1006}{Phys.Rev. {\bf 139} (1965)
  B1006--B1010}.

\bibitem{GellMann:1962xb}
M.~Gell-Mann, {\em {Symmetries of baryons and mesons}\/},
\href{http://dx.doi.org/10.1103/PhysRev.125.1067}{Phys.Rev. {\bf 125} (1962)
  1067--1084}.

\bibitem{GellMann:1964nj}
M.~Gell-Mann, {\em {A Schematic Model of Baryons and Mesons}\/},
\href{http://dx.doi.org/10.1016/S0031-9163(64)92001-3}{Phys.Lett. {\bf 8}
  (1964)  214--215}.

\bibitem{Kokkedee:936332}
J.~J.~J. Kokkedee, {\em The quark model\/},
\newblock CERN-TH-757 (1967)  .

\bibitem{Zweig:352337}
G.~Zweig, {\em An $SU_3$ model for strong interaction symmetry and its
  breaking: Part I\/},
\newblock CERN-TH-401 (1964)  .

\bibitem{Zweig:570209}
G.~Zweig, {\em An $SU_3$ model for strong interaction symmetry and its
  breaking: Part II\/},
\newblock CERN-TH-412 (1964)  .

\bibitem{Barnes:1964pd}
V.~Barnes \textit{et al.}, {\em {Observation of a Hyperon with Strangeness
  $-3$}\/},
\href{http://dx.doi.org/10.1103/PhysRevLett.12.204}{Phys.Rev.Lett. {\bf 12}
  (1964)  204--206}.

\bibitem{Bjorken:1968dy}
J.~Bjorken, {\em {Asymptotic Sum Rules at Infinite Momentum}\/},
\href{http://dx.doi.org/10.1103/PhysRev.179.1547}{Phys.Rev. {\bf 179} (1969)
  1547--1553}.

\bibitem{Bloom:1969kc}
E.~D. Bloom \textit{et al.}, {\em {High-Energy Inelastic $ep$ Scattering at
  $6$-Degrees and $10$-Degrees}\/},
\href{http://dx.doi.org/10.1103/PhysRevLett.23.930}{Phys.Rev.Lett. {\bf 23}
  (1969)  930--934}.

\bibitem{Friedman:1972sy}
J.~I. Friedman and H.~W. Kendall, {\em {Deep inelastic electron scattering}\/},
\href{http://dx.doi.org/10.1146/annurev.ns.22.120172.001223}{Ann.Rev.Nucl.Part.Sci.
  {\bf 22} (1972)  203--254}.

\bibitem{Bjorken:1969ja}
J.~Bjorken and E.~A. Paschos, {\em {Inelastic Electron-Proton and
  $gamma$-Proton Scattering, and the Structure of the Nucleon}\/},
\href{http://dx.doi.org/10.1103/PhysRev.185.1975}{Phys.Rev. {\bf 185} (1969)
  1975--1982}.

\bibitem{Feynman:1969ej}
R.~P. Feynman, {\em {Very high-energy collisions of hadrons}\/},
\href{http://dx.doi.org/10.1103/PhysRevLett.23.1415}{Phys.Rev.Lett. {\bf 23}
  (1969)  1415--1417}.

\bibitem{Feynman:1969wa}
R.~P. Feynman, {\em {The behavior of hadron collisions at extreme energies}\/},
\newblock Invited paper at the Third Conference on High-Energy Collisions,
  Stony Brook, New York, 5-6 Sep 1969  .

\bibitem{GellMann:1954fq}
M.~Gell-Mann and F.~Low, {\em {Quantum electrodynamics at small distances}\/},
\href{http://dx.doi.org/10.1103/PhysRev.95.1300}{Phys.Rev. {\bf 95} (1954)
  1300--1312}.

\bibitem{Wilson:1969zs}
K.~G. Wilson, {\em {Non-Lagrangian models of current algebra}\/},
\href{http://dx.doi.org/10.1103/PhysRev.179.1499}{Phys.Rev. {\bf 179} (1969)
  1499--1512}.

\bibitem{Callan:1970yg}
C.~G. Callan, {\em {Broken scale invariance in scalar field theory}\/},
\href{http://dx.doi.org/10.1103/PhysRevD.2.1541}{Phys.Rev. {\bf D2} (1970)
  1541--1547}.

\bibitem{Symanzik:1970rt}
K.~Symanzik, {\em {Small distance behavior in field theory and power
  counting}\/},
\href{http://dx.doi.org/10.1007/BF01649434}{Commun.Math.Phys. {\bf 18} (1970)
  227--246}.

\bibitem{Callan:1973pu}
C.~G. Callan and D.~J. Gross, {\em {Bjorken scaling in quantum field
  theory}\/},
\href{http://dx.doi.org/10.1103/PhysRevD.8.4383}{Phys.Rev. {\bf D8} (1973)
  4383--4394}.

\bibitem{Christ:1972ms}
N.~H. Christ, B.~Hasslacher, and A.~H. Mueller, {\em {Light cone behavior of
  perturbation theory}\/},
\href{http://dx.doi.org/10.1103/PhysRevD.6.3543}{Phys.Rev. {\bf D6} (1972)
  3543}.

\bibitem{Yang:1954ek}
C.-N. Yang and R.~L. Mills, {\em {Conservation of Isotopic Spin and Isotopic
  Gauge Invariance}\/},
\href{http://dx.doi.org/10.1103/PhysRev.96.191}{Phys.Rev. {\bf 96} (1954)
  191--195}.

\bibitem{Politzer:1973fx}
H.~Politzer, {\em {Reliable Perturbative Results for Strong Interactions?}\/},
\href{http://dx.doi.org/10.1103/PhysRevLett.30.1346}{Phys.Rev.Lett. {\bf 30}
  (1973)  1346--1349}.

\bibitem{Gross:1973id}
D.~Gross and F.~Wilczek, {\em {Ultraviolet Behavior of Non-Abelian Gauge
  Theories}\/},
\href{http://dx.doi.org/10.1103/PhysRevLett.30.1343}{Phys.Rev.Lett. {\bf 30}
  (1973)  1343--1346}.

\bibitem{Gross:1973ju}
D.~Gross and F.~Wilczek, {\em {Asymptotically Free Gauge Theories. 1}\/},
\href{http://dx.doi.org/10.1103/PhysRevD.8.3633}{Phys.Rev. {\bf D8} (1973)
  3633--3652}.

\bibitem{Gross:1974cs}
D.~Gross and F.~Wilczek, {\em {Asymptotically Free Gauge Theories. 2}\/},
\href{http://dx.doi.org/10.1103/PhysRevD.9.980}{Phys.Rev. {\bf D9} (1974)
  980--993}.

\bibitem{Politzer:1974fr}
H.~Politzer, {\em {Asymptotic Freedom: An Approach to Strong Interactions}\/},
\href{http://dx.doi.org/10.1016/0370-1573(74)90014-3}{Phys.Rept. {\bf 14}
  (1974)  129--180}.

\bibitem{Georgi:1951sr}
H.~Georgi and H.~Politzer, {\em {Electroproduction scaling in an asymptotically
  free theory of strong interactions}\/},
\href{http://dx.doi.org/10.1103/PhysRevD.9.416}{Phys.Rev. {\bf D9} (1974)
  416--420}.

\bibitem{Drell:1970wh}
S.~Drell and T.~Yan, {\em {Massive Lepton Pair Production in Hadron-Hadron
  Collisions at High-Energies}\/},
\href{http://dx.doi.org/10.1103/PhysRevLett.25.316,
  10.1103/PhysRevLett.25.316}{Phys.Rev.Lett. {\bf 25} (1970)  316--320}.

\bibitem{Brock:1994er}
{CTEQ} Collaboration, {\em {Handbook of perturbative QCD; Version 1.1:
  September 1994}\/},
Rev. Mod. Phys. (1994)  .

\bibitem{Nakamura:2010zzi}
K.~Nakamura \textit{et al.}, {\em {Review of particle physics}\/},
\href{http://dx.doi.org/10.1088/0954-3899/37/7A/075021}{J.Phys.G {\bf G37}
  (2010)  075021}.

\bibitem{Callan:1969uq}
C.~G. Callan and D.~J. Gross, {\em {High-energy electroproduction and the
  constitution of the electric current}\/},
\href{http://dx.doi.org/10.1103/PhysRevLett.22.156}{Phys.Rev.Lett. {\bf 22}
  (1969)  156--159}.

\bibitem{Bodek:1979rx}
A.~Bodek \textit{et al.}, {\em {Experimental Studies of the Neutron and Proton
  Electromagnetic Structure Functions}\/},
\href{http://dx.doi.org/10.1103/PhysRevD.20.1471}{Phys.Rev. {\bf D20} (1979)
  1471--1552}.

\bibitem{Ezhela:2003pp}
V.~Ezhela, S.~Lugovsky, and O.~Zenin, {\em {Hadronic part of the muon g-2
  estimated on the $\sigma^{2003}_{\mathrm{tot}} (\ee\rightarrow hadrons)$
  evaluated data compilation}\/},
\href{http://arxiv.org/abs/hep-ph/0312114}{{\tt arXiv:hep-ph/0312114}}.

\bibitem{Hanson:1975fe}
G.~Hanson \textit{et al.}, {\em {Evidence for Jet Structure in Hadron
  Production by \ee\ Annihilation}\/},
\href{http://dx.doi.org/10.1103/PhysRevLett.35.1609}{Phys.Rev.Lett. {\bf 35}
  (1975)  1609--1612}.

\bibitem{Wu:1984ik}
S.~L. Wu, {\em {\ee\ Physics at PETRA: The First 5-Years}\/},
\href{http://dx.doi.org/10.1016/0370-1573(84)90033-4}{Phys.Rept. {\bf 107}
  (1984)  59--324}.

\bibitem{Bethke:2009jm}
S.~Bethke, {\em {The 2009 World Average of \alphas}\/},
  \href{http://dx.doi.org/10.1140/epjc/s10052-009-1173-1}{Eur.Phys.J. {\bf C64}
  (2009)  689--703},
\href{http://arxiv.org/abs/0908.1135}{{\tt arXiv:0908.1135}}.

\bibitem{Faddeev:1967fc}
L.~Faddeev and V.~Popov, {\em {Feynman Diagrams for the Yang-Mills Field}\/},
\href{http://dx.doi.org/10.1016/0370-2693(67)90067-6}{Phys.Lett. {\bf B25}
  (1967)  29--30}.

\bibitem{Sterman:1977wj}
G.~F. Sterman and S.~Weinberg, {\em {Jets from Quantum Chromodynamics}\/},
\href{http://dx.doi.org/10.1103/PhysRevLett.39.1436}{Phys.Rev.Lett. {\bf 39}
  (1977)  1436}.

\bibitem{Dokshitzer:1978hw}
Y.~L. Dokshitzer, D.~Diakonov, and S.~Troian, {\em {Hard Processes in Quantum
  Chromodynamics}\/},
\href{http://dx.doi.org/10.1016/0370-1573(80)90043-5}{Phys.Rept. {\bf 58}
  (1980)  269--395}.

\bibitem{Sterman:1978bi}
G.~F. Sterman, {\em {Mass Divergences in Annihilation Processes. 1. Origin and
  Nature of Divergences in Cut Vacuum Polarization Diagrams}\/},
\href{http://dx.doi.org/10.1103/PhysRevD.17.2773}{Phys.Rev. {\bf D17} (1978)
  2773}.

\bibitem{Wilson:1974sk}
K.~G. Wilson, {\em {Confinement of Quarks}\/},
\href{http://dx.doi.org/10.1103/PhysRevD.10.2445}{Phys.Rev. {\bf D10} (1974)
  2445--2459}.

\bibitem{Nielsen:1980rz}
H.~B. Nielsen and M.~Ninomiya, {\em {Absence of Neutrinos on a Lattice. 1.
  Proof by Homotopy Theory}\/},
\href{http://dx.doi.org/10.1016/0550-3213(81)90361-8}{Nucl.Phys. {\bf B185}
  (1981)  20}.

\bibitem{Nielsen:1981hk}
H.~B. Nielsen and M.~Ninomiya, {\em {No Go Theorem for Regularizing Chiral
  Fermions}\/},
\href{http://dx.doi.org/10.1016/0370-2693(81)91026-1}{Phys.Lett. {\bf B105}
  (1981)  219}.

\bibitem{Kogut:1974ag}
J.~B. Kogut and L.~Susskind, {\em {Hamiltonian Formulation of Wilson's Lattice
  Gauge Theories}\/},
\href{http://dx.doi.org/10.1103/PhysRevD.11.395}{Phys.Rev. {\bf D11} (1975)
  395}.

\bibitem{Symanzik:1983dc}
K.~Symanzik, {\em {Continuum Limit and Improved Action in Lattice Theories. 1.
  Principles and $\phi^4$ Theory}\/},
\href{http://dx.doi.org/10.1016/0550-3213(83)90468-6}{Nucl.Phys. {\bf B226}
  (1983)  187}.

\bibitem{Bazavov:2009bb}
A.~Bazavov \textit{et al.}, {\em {Nonperturbative QCD simulations with $2+1$
  flavors of improved staggered quarks}\/},
  \href{http://dx.doi.org/10.1103/RevModPhys.82.1349}{Rev.Mod.Phys. {\bf 82}
  (2010)  1349--1417},
\href{http://arxiv.org/abs/0903.3598}{{\tt arXiv:0903.3598}}.

\bibitem{Nambu:1970aaz}
Y.~Nambu. {Lectures at the Copenhagen symposium (1970)}.

\bibitem{Goto:1971ce}
T.~Goto, {\em {Relativistic quantum mechanics of one-dimensional mechanical
  continuum and subsidiary condition of dual resonance model}\/},
\href{http://dx.doi.org/10.1143/PTP.46.1560}{Prog.Theor.Phys. {\bf 46} (1971)
  1560--1569}.

\bibitem{Andersson:1983ia}
B.~Andersson \textit{et al.}, {\em {Parton Fragmentation and String
  Dynamics}\/},
\href{http://dx.doi.org/10.1016/0370-1573(83)90080-7}{Phys.Rept. {\bf 97}
  (1983)  31--145}.

\bibitem{Sjostrand:1993yb}
T.~Sjostrand, {\em {High-energy physics event generation with PYTHIA~5.7 and
  JETSET~7.4}\/},
\href{http://dx.doi.org/10.1016/0010-4655(94)90132-5}{Comput.Phys.Commun. {\bf
  82} (1994)  74--90}.

\bibitem{Collins:1985ue}
J.~C. Collins, D.~E. Soper, and G.~F. Sterman, {\em {Factorization for Short
  Distance Hadron-Hadron Scattering}\/},
\href{http://dx.doi.org/10.1016/0550-3213(85)90565-6}{Nucl.Phys. {\bf B261}
  (1985)  104}.

\bibitem{Collins:1988ig}
J.~C. Collins, D.~E. Soper, and G.~F. Sterman, {\em {Soft Gluons and
  Factorization}\/},
\href{http://dx.doi.org/10.1016/0550-3213(88)90130-7}{Nucl.Phys. {\bf B308}
  (1988)  833}.

\bibitem{Collins:1989gx}
J.~C. Collins, D.~E. Soper, and G.~F. Sterman, {\em {Factorization of Hard
  Processes in QCD}\/},  Adv.Ser.Direct.High Energy Phys. {\bf 5} (1988)
  1--91,
\href{http://arxiv.org/abs/hep-ph/0409313}{{\tt arXiv:hep-ph/0409313}}.

\bibitem{Brodsky:2002cx}
S.~J. Brodsky, D.~S. Hwang, and I.~Schmidt, {\em {Final-state interactions and
  single-spin asymmetries in semi-inclusive deep inelastic scattering}\/},
  \href{http://dx.doi.org/10.1016/S0370-2693(02)01320-5}{Phys.Lett. {\bf B530}
  (2002)  99--107},
\href{http://arxiv.org/abs/hep-ph/0201296}{{\tt arXiv:hep-ph/0201296}}.

\bibitem{Brodsky:2002ue}
S.~J. Brodsky \textit{et al.}, {\em {Structure functions are not parton
  probabilities}\/},
  \href{http://dx.doi.org/10.1103/PhysRevD.65.114025}{Phys.Rev. {\bf D65}
  (2002)  114025},
\href{http://arxiv.org/abs/hep-ph/0104291}{{\tt arXiv:hep-ph/0104291}}.

\bibitem{Collins:2002kn}
J.~C. Collins, {\em {Leading twist single transverse-spin asymmetries:
  Drell-Yan and deep inelastic scattering}\/},
  \href{http://dx.doi.org/10.1016/S0370-2693(02)01819-1}{Phys.Lett. {\bf B536}
  (2002)  43--48},
\href{http://arxiv.org/abs/hep-ph/0204004}{{\tt arXiv:hep-ph/0204004}}.

\bibitem{Boer:2003cm}
D.~Boer, P.~Mulders, and F.~Pijlman, {\em {Universality of $T$-odd effects in
  single spin and azimuthal asymmetries}\/},
  \href{http://dx.doi.org/10.1016/S0550-3213(03)00527-3}{Nucl.Phys. {\bf B667}
  (2003)  201--241},
\href{http://arxiv.org/abs/hep-ph/0303034}{{\tt arXiv:hep-ph/0303034}}.

\bibitem{Collins:1981uk}
J.~C. Collins and D.~E. Soper, {\em {Back-to-back jets in QCD}\/},
\href{http://dx.doi.org/10.1016/0550-3213(81)90339-4}{Nucl.Phys. {\bf B193}
  (1981)  381}.

\bibitem{Collins:1984kg}
J.~C. Collins, D.~E. Soper, and G.~F. Sterman, {\em {Transverse Momentum
  Distribution in Drell-Yan Pair and \Wboson\ and \Zboson\ Boson
  Production}\/},
\href{http://dx.doi.org/10.1016/0550-3213(85)90479-1}{Nucl.Phys. {\bf B250}
  (1985)  199}.

\bibitem{Collins:2007nk}
J.~Collins and J.-W. Qiu, {\em {\kt\ factorization is violated in production of
  high-transverse-momentum particles in hadron-hadron collisions}\/},
  \href{http://dx.doi.org/10.1103/PhysRevD.75.114014}{Phys.Rev. {\bf D75}
  (2007)  114014},
\href{http://arxiv.org/abs/0705.2141}{{\tt arXiv:0705.2141}}.

\bibitem{Rogers:2010dm}
T.~C. Rogers and P.~J. Mulders, {\em {No generalized transverse momentum
  dependent factorization in the hadroproduction of high transverse momentum
  hadrons}\/},  \href{http://dx.doi.org/10.1103/PhysRevD.81.094006}{Phys.Rev.
  {\bf D81} (2010)  094006},
\href{http://arxiv.org/abs/1001.2977}{{\tt arXiv:1001.2977}}.

\bibitem{Altarelli:1977zs}
G.~Altarelli and G.~Parisi, {\em {Asymptotic Freedom in Parton Language}\/},
\href{http://dx.doi.org/10.1016/0550-3213(77)90384-4}{Nucl.Phys. {\bf B126}
  (1977)  298}.

\bibitem{Gribov:1972ri}
V.~Gribov and L.~Lipatov, {\em {Deep inelastic $ep$ scattering in perturbation
  theory}\/},
Sov.J.Nucl.Phys. {\bf 15} (1972)  438--450.

\bibitem{Dokshitzer:1977sg}
Y.~L. Dokshitzer, {\em {Calculation of the Structure Functions for Deep
  Inelastic Scattering and \ee\ Annihilation by Perturbation Theory in Quantum
  Chromodynamics.}\/},
Sov.Phys.JETP {\bf 46} (1977)  641--653.

\bibitem{Mueller:1982cq}
A.~H. Mueller, {\em {Multiplicity and Hadron Distributions in QCD Jets:
  Nonleading Terms}\/},
\href{http://dx.doi.org/10.1016/0550-3213(83)90176-1}{Nucl.Phys. {\bf B213}
  (1983)  85}.

\bibitem{Dokshitzer:1991ej}
Y.~L. Dokshitzer, V.~A. Khoze, and S.~Troian, {\em {Inclusive particle spectra
  from QCD cascades}\/},
\href{http://dx.doi.org/10.1142/S0217751X92000818}{Int.J.Mod.Phys. {\bf A7}
  (1992)  1875--1906}.

\bibitem{Dokshitzer:238162}
Y.~L. Dokshitzer, V.~A. Khoze, A.~H. Müller, and S.~I. Troyan, {\em Basics of
  perturbative QCD}.
\newblock Basics of. Ed. Frontières, Gif-sur-Yvette, 1991.

\bibitem{Khoze:1996dn}
V.~A. Khoze and W.~Ochs, {\em {Perturbative QCD approach to multiparticle
  production}\/},
  \href{http://dx.doi.org/10.1142/S0217751X97001638}{Int.J.Mod.Phys. {\bf A12}
  (1997)  2949--3120},
\href{http://arxiv.org/abs/hep-ph/9701421}{{\tt arXiv:hep-ph/9701421}}.

\bibitem{Fong:1990nt}
C.~Fong and B.~Webber, {\em {One and two particle distributions at small $x$ in
  QCD jets}\/},
\href{http://dx.doi.org/10.1016/0550-3213(91)90302-E}{Nucl.Phys. {\bf B355}
  (1991)  54--81}.

\bibitem{Dokshitzer:1982xr}
Y.~L. Dokshitzer, V.~S. Fadin, and V.~A. Khoze, {\em {Double Logs of
  Perturbative QCD for Parton Jets and Soft Hadron Spectra}\/},
\href{http://dx.doi.org/10.1007/BF01614423}{Z.Phys. {\bf C15} (1982)  325}.

\bibitem{Florian:2007hc}
{de Florian, Daniel and Sassot, Rodolfo and Stratmann, Marco}, {\em {Global
  analysis of fragmentation functions for protons and charged hadrons}\/},
  \href{http://dx.doi.org/10.1103/PhysRevD.76.074033}{Phys.Rev. {\bf D76}
  (2007)  074033},
\href{http://arxiv.org/abs/0707.1506}{{\tt arXiv:0707.1506}}.

\bibitem{Florian:2007aj}
{de Florian, Daniel and Sassot, Rodolfo and Stratmann, Marco}, {\em {Global
  analysis of fragmentation functions for pions and kaons and their
  uncertainties}\/},
  \href{http://dx.doi.org/10.1103/PhysRevD.75.114010}{Phys.Rev. {\bf D75}
  (2007)  114010},
\href{http://arxiv.org/abs/hep-ph/0703242}{{\tt arXiv:hep-ph/0703242}}.

\bibitem{Hirai:2007cx}
M.~Hirai, S.~Kumano, T.-H. Nagai, and K.~Sudoh, {\em {Determination of
  fragmentation functions and their uncertainties}\/},
  \href{http://dx.doi.org/10.1103/PhysRevD.75.094009}{Phys.Rev. {\bf D75}
  (2007)  094009},
\href{http://arxiv.org/abs/hep-ph/0702250}{{\tt arXiv:hep-ph/0702250}}.

\bibitem{Webber:1983if}
B.~Webber, {\em {A QCD Model for Jet Fragmentation Including Soft Gluon
  Interference}\/},
\href{http://dx.doi.org/10.1016/0550-3213(84)90333-X}{Nucl.Phys. {\bf B238}
  (1984)  492}.

\bibitem{Marchesini:1991ch}
G.~Marchesini \textit{et al.}, {\em {HERWIG: A Monte Carlo event generator for
  simulating hadron emission reactions with interfering gluons. Version~5.1 -
  April 1991}\/},
\href{http://dx.doi.org/10.1016/0010-4655(92)90055-4}{Comput.Phys.Commun. {\bf
  67} (1992)  465--508}.

\bibitem{Catani:2001cc}
S.~Catani \textit{et al.}, {\em {QCD matrix elements + parton showers}\/},
  JHEP {\bf 0111} (2001)  063,
\href{http://arxiv.org/abs/hep-ph/0109231}{{\tt arXiv:hep-ph/0109231}}.

\bibitem{Alwall:2007fs}
J.~Alwall \textit{et al.}, {\em {Comparative study of various algorithms for
  the merging of parton showers and matrix elements in hadronic collisions}\/},
   \href{http://dx.doi.org/10.1140/epjc/s10052-007-0490-5}{Eur.Phys.J. {\bf
  C53} (2008)  473--500},
\href{http://arxiv.org/abs/0706.2569}{{\tt arXiv:0706.2569}}.

\bibitem{Gleisberg:2003xi}
T.~Gleisberg \textit{et al.}, {\em {SHERPA 1. alpha: A Proof of concept
  version}\/},  \href{http://dx.doi.org/10.1088/1126-6708/2004/02/056}{JHEP
  {\bf 0402} (2004)  056},
\href{http://arxiv.org/abs/hep-ph/0311263}{{\tt arXiv:hep-ph/0311263}}.

\bibitem{Mangano:2002ea}
M.~L. Mangano, M.~Moretti, F.~Piccinini, R.~Pittau, and A.~D. Polosa, {\em
  {ALPGEN, a generator for hard multiparton processes in hadronic
  collisions}\/},  JHEP {\bf 0307} (2003)  001,
\href{http://arxiv.org/abs/hep-ph/0206293}{{\tt arXiv:hep-ph/0206293}}.

\bibitem{Huth:1990mi}
J.~E. Huth \textit{et al.}, {\em {Toward a standardization of jet
  definitions}\/},
\newblock FERMILAB-CONF-90-249-E (1990)  .

\bibitem{Ellis:1990ek}
S.~D. Ellis, Z.~Kunszt, and D.~E. Soper, {\em {The One Jet Inclusive
  Cross-Section at Order $\alpha_{\mathrm{S}}^3$ Quarks and Gluons}\/},
  \href{http://dx.doi.org/10.1103/PhysRevLett.64.2121}{Phys.Rev.Lett. {\bf 64}
  (1990)  2121}.

\bibitem{Salam:2009jx}
G.~P. Salam, {\em {Towards Jetography}\/},
  \href{http://dx.doi.org/10.1140/epjc/s10052-010-1314-6}{Eur.Phys.J. {\bf C67}
  (2010)  637--686},
\href{http://arxiv.org/abs/0906.1833}{{\tt arXiv:0906.1833}}.

\bibitem{Cacciari:2008gp}
M.~Cacciari, G.~P. Salam, and G.~Soyez, {\em {The anti-\kt\ jet clustering
  algorithm}\/},  \href{http://dx.doi.org/10.1088/1126-6708/2008/04/063}{JHEP
  {\bf 0804} (2008)  063},
\href{http://arxiv.org/abs/0802.1189}{{\tt arXiv:0802.1189}}.

\bibitem{Bartel:1986ua}
{JADE} Collaboration, {\em {Experimental Studies on Multi-Jet Production in
  \ee\ Annihilation at PETRA Energies}\/},
\href{http://dx.doi.org/10.1007/BF01410449}{Z. Phys. {\bf C33} (1986)  23}.

\bibitem{Bethke:1988zc}
{JADE} Collaboration, {\em {Experimental Investigation of the Energy Dependence
  of the Strong Coupling Strength}\/},
\href{http://dx.doi.org/10.1016/0370-2693(88)91032-5}{Phys. Lett. {\bf B213}
  (1988)  235}.

\bibitem{Catani:1991hj}
S.~Catani \textit{et al.}, {\em {New clustering algorithm for multi-jet
  cross-sections in \ee\ annihilation}\/},
\href{http://dx.doi.org/10.1016/0370-2693(91)90196-W}{Phys.Lett. {\bf B269}
  (1991)  432--438}.

\bibitem{Catani:1993hr}
S.~Catani \textit{et al.}, {\em {Longitudinally invariant \kt\ clustering
  algorithms for hadron hadron collisions}\/},
\href{http://dx.doi.org/10.1016/0550-3213(93)90166-M}{Nucl.Phys. {\bf B406}
  (1993)  187--224}.

\bibitem{Ellis:1993tq}
S.~D. Ellis and D.~E. Soper, {\em {Successive combination jet algorithm for
  hadron collisions}\/},
  \href{http://dx.doi.org/10.1103/PhysRevD.48.3160}{Phys.Rev. {\bf D48} (1993)
  3160--3166},
\href{http://arxiv.org/abs/hep-ph/9305266}{{\tt arXiv:hep-ph/9305266}}.

\bibitem{Cacciari:2005hq}
M.~Cacciari and G.~P. Salam, {\em {Dispelling the $N^{3}$ myth for the \kt\
  jet-finder}\/},
  \href{http://dx.doi.org/10.1016/j.physletb.2006.08.037}{Phys.Lett. {\bf B641}
  (2006)  57--61},
\href{http://arxiv.org/abs/hep-ph/0512210}{{\tt arXiv:hep-ph/0512210}}.

\bibitem{Dokshitzer:1997in}
Y.~L. Dokshitzer \textit{et al.}, {\em {Better jet clustering algorithms}\/},
  JHEP {\bf 9708} (1997)  001,
\href{http://arxiv.org/abs/hep-ph/9707323}{{\tt arXiv:hep-ph/9707323}}.

\bibitem{Wobisch:1998wt}
M.~Wobisch and T.~Wengler, {\em {Hadronization corrections to jet
  cross-sections in deep inelastic scattering}\/},
\href{http://arxiv.org/abs/hep-ph/9907280}{{\tt arXiv:hep-ph/9907280}}.

\bibitem{ATL-PHYS-PUB-2009-012}
{ATLAS} Collaboration, {\em Jet Reconstruction Performance\/},
  ATL-PHYS-PUB-2009-012 (2009).
\newblock \url{http://cdsweb.cern.ch/record/1167330/}.

\bibitem{Kluge:2006xs}
T.~Kluge, K.~Rabbertz, and M.~Wobisch, {\em {FastNLO: Fast pQCD calculations
  for PDF fits}\/},
\href{http://arxiv.org/abs/hep-ph/0609285}{{\tt arXiv:hep-ph/0609285}}.

\bibitem{Fermi:1950jd}
E.~Fermi, {\em {High-energy nuclear events}\/},
Prog.Theor.Phys. {\bf 5} (1950)  570--583.

\bibitem{Landau:1953gs}
L.~Landau, {\em {On the multiparticle production in high-energy collisions}\/},
Izv.Akad.Nauk Ser.Fiz. {\bf 17} (1953)  51--64.

\bibitem{Hagedorn:1965st}
R.~Hagedorn, {\em {Statistical thermodynamics of strong interactions at
  high-energies}\/},
Nuovo Cim.Suppl. {\bf 3} (1965)  147--186.

\bibitem{Collins:1974ky}
J.~C. Collins and M.~Perry, {\em {Superdense Matter: Neutrons Or Asymptotically
  Free Quarks?}\/},
\href{http://dx.doi.org/10.1103/PhysRevLett.34.1353}{Phys.Rev.Lett. {\bf 34}
  (1975)  1353}.

\bibitem{Shuryak:1980tp}
E.~V. Shuryak, {\em {Quantum Chromodynamics and the Theory of Superdense
  Matter}\/},
\href{http://dx.doi.org/10.1016/0370-1573(80)90105-2}{Phys.Rept. {\bf 61}
  (1980)  71--158}.

\bibitem{Shuryak:1981fz}
E.~V. Shuryak, {\em {Two Scales and Phase Transitions in Quantum
  Chromodynamics}\/},
\href{http://dx.doi.org/10.1016/0370-2693(81)91158-8}{Phys.Lett. {\bf B107}
  (1981)  103}.

\bibitem{lebellac}
M.~LeBellac, {\em Thermal Field Theory}.
\newblock Cambridge University Press, Cambridge, 1996.

\bibitem{Blaizot:2001nr}
J.-P. Blaizot and E.~Iancu, {\em {The Quark gluon plasma: Collective dynamics
  and hard thermal loops}\/},
  \href{http://dx.doi.org/10.1016/S0370-1573(01)00061-8}{Phys.Rept. {\bf 359}
  (2002)  355--528},
\href{http://arxiv.org/abs/hep-ph/0101103}{{\tt arXiv:hep-ph/0101103}}.

\bibitem{Bazavov:2009zn}
A.~Bazavov \textit{et al.}, {\em {Equation of state and QCD transition at
  finite temperature}\/},
  \href{http://dx.doi.org/10.1103/PhysRevD.80.014504}{Phys.Rev. {\bf D80}
  (2009)  014504},
\href{http://arxiv.org/abs/0903.4379}{{\tt arXiv:0903.4379}}.

\bibitem{Braaten:1989mz}
E.~Braaten and R.~D. Pisarski, {\em {Soft Amplitudes in Hot Gauge Theories: A
  General Analysis}\/},
\href{http://dx.doi.org/10.1016/0550-3213(90)90508-B}{Nucl.Phys. {\bf B337}
  (1990)  569}.

\bibitem{Braaten:1991gm}
E.~Braaten and R.~D. Pisarski, {\em {Simple effective Lagrangian for hard
  thermal loops}\/},
\href{http://dx.doi.org/10.1103/PhysRevD.45.R1827}{Phys.Rev. {\bf D45} (1992)
  1827--1830}.

\bibitem{weldon}
H.~A. Weldon, {\em {Covariant Calculations at Finite Temperature: The
  Relativistic Plasma}\/},
\href{http://dx.doi.org/10.1103/PhysRevD.26.1394}{Phys. Rev. {\bf D26} (1982)
  1394}.

\bibitem{linde}
A.~D. Linde, {\em {Infrared Problem in Thermodynamics of the Yang-Mills
  Gas}\/},
\href{http://dx.doi.org/10.1016/0370-2693(80)90769-8}{Phys. Lett. {\bf B96}
  (1980)  289}.

\bibitem{gross:1981py}
D.~J. Gross, R.~D. Pisarski, and L.~G. Yaffe, {\em {QCD and Instantons at
  Finite Temperature}\/},
\href{http://dx.doi.org/10.1103/RevModPhys.53.43}{Rev. Mod. Phys. {\bf 53}
  (1981)  43}.

\bibitem{Lee:1974ma}
T.~Lee and G.~Wick, {\em {Vacuum Stability and Vacuum Excitation in a Spin 0
  Field Theory}\/},
\href{http://dx.doi.org/10.1103/PhysRevD.9.2291}{Phys.Rev. {\bf D9} (1974)
  2291}.

\bibitem{Lee:1974kn}
T.~D. Lee, {\em {Abnormal Nuclear States and Vacuum Excitations}\/},
\href{http://dx.doi.org/10.1103/RevModPhys.47.267}{Rev. Mod. Phys. {\bf 47}
  (1975)  267}.

\bibitem{Rajagopal:2000wf}
K.~Rajagopal and F.~Wilczek, {\em {The Condensed matter physics of QCD}\/},
\href{http://arxiv.org/abs/hep-ph/0011333}{{\tt arXiv:hep-ph/0011333}}.

\bibitem{Bjorken:1982qr}
J.~Bjorken, {\em {Highly Relativistic Nucleus-Nucleus Collisions: The Central
  Rapidity Region}\/},
\href{http://dx.doi.org/10.1103/PhysRevD.27.140}{Phys.Rev. {\bf D27} (1983)
  140--151}.

\bibitem{Adcox:2001ry}
{PHENIX} Collaboration, {\em {Measurement of the mid-rapidity transverse energy
  distribution from $\sqrtsnn=130$~\GeV\ \AuAu\ collisions at RHIC}\/},
  \href{http://dx.doi.org/10.1103/PhysRevLett.87.052301}{Phys.Rev.Lett. {\bf
  87} (2001)  052301},
\href{http://arxiv.org/abs/nucl-ex/0104015}{{\tt arXiv:nucl-ex/0104015}}.

\bibitem{Maldacena:1997re}
J.~M. Maldacena, {\em {The large $N$ limit of superconformal field theories and
  supergravity}\/},
  \href{http://dx.doi.org/10.1023/A:1026654312961}{Adv.Theor.Math.Phys. {\bf 2}
  (1998)  231--252},
\href{http://arxiv.org/abs/hep-th/9711200}{{\tt arXiv:hep-th/9711200}}.

\bibitem{Aharony:1999ti}
O.~Aharony \textit{et al.}, {\em {Large $N$ field theories, string theory and
  gravity}\/},
  \href{http://dx.doi.org/10.1016/S0370-1573(99)00083-6}{Phys.Rept. {\bf 323}
  (2000)  183--386},
\href{http://arxiv.org/abs/hep-th/9905111}{{\tt arXiv:hep-th/9905111}}.

\bibitem{Danielewicz:1984ww}
P.~Danielewicz and M.~Gyulassy, {\em {Dissipative Phenomena in Quark Gluon
  Plasmas}\/},
\href{http://dx.doi.org/10.1103/PhysRevD.31.53}{Phys.Rev. {\bf D31} (1985)
  53--62}.

\bibitem{Policastro:2001yc}
G.~Policastro, D.~Son, and A.~Starinets, {\em {The Shear viscosity of strongly
  coupled $N=4$ supersymmetric Yang-Mills plasma}\/},
  \href{http://dx.doi.org/10.1103/PhysRevLett.87.081601}{Phys.Rev.Lett. {\bf
  87} (2001)  081601},
\href{http://arxiv.org/abs/hep-th/0104066}{{\tt arXiv:hep-th/0104066}}.

\bibitem{Kovtun:2003wp}
P.~Kovtun, D.~T. Son, and A.~O. Starinets, {\em {Holography and hydrodynamics:
  Diffusion on stretched horizons}\/},  JHEP {\bf 0310} (2003)  064,
\href{http://arxiv.org/abs/hep-th/0309213}{{\tt arXiv:hep-th/0309213}}.

\bibitem{Glauber:1959wd}
{R.J. Glauber, in \textit{Lectures in Theoretical Physics}, ed. W.E. Brittin,
  L.G. Dunham, 1:315. New York: Interscience (1959)}.

\bibitem{Glauber:1970jm}
R.~Glauber and G.~Matthiae, {\em {High-energy scattering of protons by
  nuclei}\/},
Nucl.Phys. {\bf B21} (1970)  135--157.

\bibitem{Czyz:1969jg}
W.~Czyz and L.~Maximon, {\em {High-energy, small angle elastic scattering of
  strongly interacting composite particles}\/},
\href{http://dx.doi.org/10.1016/0003-4916(69)90321-2}{Annals Phys. {\bf 52}
  (1969)  59--121}.

\bibitem{Miller:2007ri}
M.~L. Miller \textit{et al.}, {\em {Glauber modeling in high energy nuclear
  collisions}\/},
  \href{http://dx.doi.org/10.1146/annurev.nucl.57.090506.123020}{Ann.Rev.Nucl.Part.Sci.
  {\bf 57} (2007)  205--243},
\href{http://arxiv.org/abs/nucl-ex/0701025}{{\tt arXiv:nucl-ex/0701025}}.

\bibitem{DeJager:1987qc}
H.~De~Vries, C.~W. De~Jager, and C.~De~Vries, {\em {Nuclear charge and
  magnetization density distribution parameters from elastic electron
  scattering}\/},
Atom. Data Nucl. Data Tabl. {\bf 36} (1987)  495--536.

\bibitem{Bialas:1976ed}
A.~Bialas, M.~Bleszynski, and W.~Czyz, {\em {Multiplicity Distributions in
  Nucleus-Nucleus Collisions at High-Energies}\/},
\href{http://dx.doi.org/10.1016/0550-3213(76)90329-1}{Nucl.Phys. {\bf B111}
  (1976)  461}.

\bibitem{Cronin:1974zm}
J.~Cronin \textit{et al.}, {\em {Production of Hadrons with Large Transverse
  Momentum at 200~\GeV, 300~\GeV, and 400~\GeV}\/},
\href{http://dx.doi.org/10.1103/PhysRevD.11.3105}{Phys.Rev. {\bf D11} (1975)
  3105}.

\bibitem{Antreasyan:1978cw}
D.~Antreasyan \textit{et al.}, {\em {Production of Hadrons at Large Transverse
  Momentum in 200~\GeV, 300~\GeV\ and 400~\GeV\ \pp\ and \pn\ Collisions}\/},
\href{http://dx.doi.org/10.1103/PhysRevD.19.764}{Phys.Rev. {\bf D19} (1979)
  764--778}.

\bibitem{Qiu:2004da}
J.~W. Qiu and I.~Vitev, {\em {Coherent QCD multiple scattering in
  proton-nucleus collisions}\/},
  \href{http://dx.doi.org/10.1016/j.physletb.2005.10.073}{Phys.Lett. {\bf B632}
  (2006)  507--511},
\href{http://arxiv.org/abs/hep-ph/0405068}{{\tt arXiv:hep-ph/0405068}}.

\bibitem{Qiu:2001hj}
J.~W. Qiu and G.~F. Sterman, {\em {QCD and rescattering in nuclear targets}\/},
   \href{http://dx.doi.org/10.1142/S0218301303001235}{Int.J.Mod.Phys. {\bf E12}
  (2003)  149},
\href{http://arxiv.org/abs/hep-ph/0111002}{{\tt arXiv:hep-ph/0111002}}.

\bibitem{Mueller:1985wy}
A.~H. Mueller and J.~W. Qiu, {\em {Gluon Recombination and Shadowing at Small
  Values of $x$}\/},
\href{http://dx.doi.org/10.1016/0550-3213(86)90164-1}{Nucl.Phys. {\bf B268}
  (1986)  427}.

\bibitem{Gribov:1981ac}
L.~Gribov, E.~Levin, and M.~Ryskin, {\em {Singlet Structure Function at Small
  $x$: Unitarization of Gluon Ladders}\/},
\href{http://dx.doi.org/10.1016/0550-3213(81)90007-9}{Nucl.Phys. {\bf B188}
  (1981)  555--576}.

\bibitem{Gribov:1984tu}
L.~Gribov, E.~Levin, and M.~Ryskin, {\em {Semihard Processes in QCD}\/},
\href{http://dx.doi.org/10.1016/0370-1573(83)90022-4}{Phys.Rept. {\bf 100}
  (1983)  1--150}.

\bibitem{Geesaman:1995yd}
D.~F. Geesaman, K.~Saito, and A.~W. Thomas, {\em {The nuclear EMC effect}\/},
\href{http://dx.doi.org/10.1146/annurev.ns.45.120195.002005}{Ann.Rev.Nucl.Part.Sci.
  {\bf 45} (1995)  337--390}.

\bibitem{Frankfurt:1988nt}
L.~Frankfurt and M.~Strikman, {\em {Hard Nuclear Processes and Microscopic
  Nuclear Structure}\/},
\href{http://dx.doi.org/10.1016/0370-1573(88)90179-2}{Phys.Rept. {\bf 160}
  (1988)  235--427}.

\bibitem{Eskola:2009uj}
K.~Eskola, H.~Paukkunen, and C.~Salgado, {\em {EPS09: A New Generation of NLO
  and LO Nuclear Parton Distribution Functions}\/},
  \href{http://dx.doi.org/10.1088/1126-6708/2009/04/065}{JHEP {\bf 0904} (2009)
   065},
\href{http://arxiv.org/abs/0902.4154}{{\tt arXiv:0902.4154}}.

\bibitem{Wang:1991hta}
X.~N. Wang and M.~Gyulassy, {\em {HIJING: A Monte Carlo model for multiple jet
  production in p p, p A and A A collisions}\/},
\href{http://dx.doi.org/10.1103/PhysRevD.44.3501}{Phys.Rev. {\bf D44} (1991)
  3501--3516}.

\bibitem{Capella:1992yb}
A.~Capella \textit{et al.}, {\em {Dual parton model}\/},
\href{http://dx.doi.org/10.1016/0370-1573(94)90064-7}{Phys.Rept. {\bf 236}
  (1994)  225--329}.

\bibitem{Ranft:1987xn}
J.~Ranft, {\em {Hadron production in Hadron-Nucleus and Nucleus-Nucleus
  Collisions in the Dual Monte Carlo Multichain Fragmentation Model}\/},
\href{http://dx.doi.org/10.1103/PhysRevD.37.1842}{Phys.Rev. {\bf D37} (1988)
  1842}.

\bibitem{Ranft:1986ut}
J.~Ranft, {\em {Transverse Energy Distributions in Nucleus-Nucleus Collisions
  in the Dual Monte Carlo Multichain Fragmentation Model}\/},
\href{http://dx.doi.org/10.1016/0370-2693(87)91401-8}{Phys.Lett. {\bf B188}
  (1987)  379}.

\bibitem{Andersson:1986gw}
B.~Andersson, G.~Gustafson, and B.~Nilsson-Almqvist, {\em {A Model for Low \pt\
  Hadronic Reactions, with Generalizations to Hadron-Nucleus and
  Nucleus-Nucleus Collisions}\/},
\href{http://dx.doi.org/10.1016/0550-3213(87)90257-4}{Nucl.Phys. {\bf B281}
  (1987)  289}.

\bibitem{NilssonAlmqvist:1986rx}
B.~Nilsson-Almqvist and E.~Stenlund, {\em {Interactions Between Hadrons and
  Nuclei: The Lund Monte Carlo, Fritiof Version 1.6}\/},
\href{http://dx.doi.org/10.1016/0010-4655(87)90056-7}{Comput.Phys.Commun. {\bf
  43} (1987)  387}.

\bibitem{Kajantie:1987pd}
K.~Kajantie, P.~Landshoff, and J.~Lindfors, {\em {Minijet Production in
  High-Energy Nucleus-Nucleus Collisions}\/},
\href{http://dx.doi.org/10.1103/PhysRevLett.59.2527}{Phys.Rev.Lett. {\bf 59}
  (1987)  2527}.

\bibitem{Eskola:1988yh}
K.~Eskola, K.~Kajantie, and J.~Lindfors, {\em {Quark and Gluon Production in
  High-Energy Nucleus-Nucleus Collisions}\/},
\href{http://dx.doi.org/10.1016/0550-3213(89)90586-5}{Nucl.Phys. {\bf B323}
  (1989)  37}.

\bibitem{Wang:1990qp}
X.~N. Wang, {\em {Role of multiple mini-jets in high-energy hadronic
  reactions}\/},
\href{http://dx.doi.org/10.1103/PhysRevD.43.104}{Phys.Rev. {\bf D43} (1991)
  104--112}.

\bibitem{Bjorken:1982tu}
J. D. Bjorken, fermilab-PUB-82/059-THY (1982).

\bibitem{Thoma:1990fm}
M.~H. Thoma and M.~Gyulassy, {\em {Quark damping and energy loss in the high
  temperature QCD}\/},
\href{http://dx.doi.org/10.1016/S0550-3213(05)80031-8}{Nucl.Phys. {\bf B351}
  (1991)  491--506}.

\bibitem{Braaten:1991we}
E.~Braaten and M.~H. Thoma, {\em {Energy loss of a heavy quark in the
  quark-gluon plasma}\/},
\href{http://dx.doi.org/10.1103/PhysRevD.44.2625}{Phys.Rev. {\bf D44} (1991)
  2625--2630}.

\bibitem{Thoma:1991ea}
M.~H. Thoma, {\em {Collisional energy loss of high-energy jets in the quark
  gluon plasma}\/},
\href{http://dx.doi.org/10.1016/0370-2693(91)90565-8}{Phys.Lett. {\bf B273}
  (1991)  128--132}.

\bibitem{Appel:1985dq}
D.~A. Appel, {\em {Jets as a probe of quark-gluon plasmas}\/},
\href{http://dx.doi.org/10.1103/PhysRevD.33.717}{Phys.Rev. {\bf D33} (1986)
  717}.

\bibitem{Blaizot:1986ma}
J.~Blaizot and L.~D. McLerran, {\em {Jets in expanding quark-gluon plasmas}\/},
\href{http://dx.doi.org/10.1103/PhysRevD.34.2739}{Phys.Rev. {\bf D34} (1986)
  2739}.

\bibitem{Rammerstorfer:1990js}
M.~Rammerstorfer and U.~W. Heinz, {\em {Jet acoplanarity as a quark-gluon
  plasma probe}\/},
\href{http://dx.doi.org/10.1103/PhysRevD.41.306}{Phys.Rev. {\bf D41} (1990)
  306--309}.

\bibitem{Armesto:2011ht}
N.~Armesto \textit{et al.}, {\em {Comparison of Jet Quenching Formalisms for a
  Quark-Gluon Plasma `Brick'}\/},
\href{http://arxiv.org/abs/1106.1106}{{\tt arXiv:1106.1106}}.

\bibitem{Landau:1953um}
L.~Landau and I.~Pomeranchuk, {\em {Limits of applicability of the theory of
  bremsstrahlung electrons and pair production at high-energies}\/},
Dokl.Akad.Nauk Ser.Fiz. {\bf 92} (1953)  535--536.

\bibitem{Migdal:1956tc}
A.~B. Migdal, {\em {Bremsstrahlung and pair production in condensed media at
  high-energies}\/},
\href{http://dx.doi.org/10.1103/PhysRev.103.1811}{Phys.Rev. {\bf 103} (1956)
  1811--1820}.

\bibitem{Gyulassy:1990ye}
M.~Gyulassy and M.~Plumer, {\em {Jet Quenching in Dense Matter}\/},
\href{http://dx.doi.org/10.1016/0370-2693(90)91409-5}{Phys.Lett. {\bf B243}
  (1990)  432--438}.

\bibitem{Gyulassy:1993hr}
M.~Gyulassy and X.~N. Wang, {\em {Multiple collisions and induced gluon
  bremsstrahlung in QCD}\/},
  \href{http://dx.doi.org/10.1016/0550-3213(94)90079-5}{Nucl.Phys. {\bf B420}
  (1994)  583--614},
\href{http://arxiv.org/abs/nucl-th/9306003}{{\tt arXiv:nucl-th/9306003}}.

\bibitem{Baier:1996kr}
R.~Baier, Y.~L. Dokshitzer, A.~H. Mueller, S.~Peigne, and D.~Schiff, {\em
  {Radiative energy loss of high-energy quarks and gluons in a finite volume
  quark-gluon plasma}\/},
  \href{http://dx.doi.org/10.1016/S0550-3213(96)00553-6}{Nucl.Phys. {\bf B483}
  (1997)  291--320},
\href{http://arxiv.org/abs/hep-ph/9607355}{{\tt arXiv:hep-ph/9607355}}.

\bibitem{Zakharov:1996fv}
B.~Zakharov, {\em {Fully quantum treatment of the Landau-Pomeranchuk-Migdal
  effect in QED and QCD}\/},  \href{http://dx.doi.org/10.1134/1.567126}{JETP
  Lett. {\bf 63} (1996)  952--957},
\href{http://arxiv.org/abs/hep-ph/9607440}{{\tt arXiv:hep-ph/9607440}}.

\bibitem{Baier:2002tc}
R.~Baier, {\em {Jet quenching}\/},
  \href{http://dx.doi.org/10.1016/S0375-9474(02)01429-X}{Nucl.Phys. {\bf A715}
  (2003)  209--218},
\href{http://arxiv.org/abs/hep-ph/0209038}{{\tt arXiv:hep-ph/0209038}}.

\bibitem{Salgado:2003gb}
C.~A. Salgado and U.~A. Wiedemann, {\em {Calculating quenching weights}\/},
  \href{http://dx.doi.org/10.1103/PhysRevD.68.014008}{Phys.Rev. {\bf D68}
  (2003)  014008},
\href{http://arxiv.org/abs/hep-ph/0302184}{{\tt arXiv:hep-ph/0302184}}.

\bibitem{Gyulassy:1999zd}
M.~Gyulassy, P.~Levai, and I.~Vitev, {\em {Jet quenching in thin quark gluon
  plasmas. 1. Formalism}\/},
  \href{http://dx.doi.org/10.1016/S0550-3213(99)00713-0}{Nucl.Phys. {\bf B571}
  (2000)  197--233},
\href{http://arxiv.org/abs/hep-ph/9907461}{{\tt arXiv:hep-ph/9907461}}.

\bibitem{Gyulassy:2000er}
M.~Gyulassy, P.~Levai, and I.~Vitev, {\em {Reaction operator approach to
  non-Abelian energy loss}\/},
  \href{http://dx.doi.org/10.1016/S0550-3213(00)00652-0}{Nucl.Phys. {\bf B594}
  (2001)  371--419},
\href{http://arxiv.org/abs/nucl-th/0006010}{{\tt arXiv:nucl-th/0006010}}.

\bibitem{Wiedemann:2000ez}
U.~A. Wiedemann, {\em {Transverse dynamics of hard partons in nuclear media and
  the QCD dipole}\/},
  \href{http://dx.doi.org/10.1016/S0550-3213(00)00286-8}{Nucl.Phys. {\bf B582}
  (2000)  409--450},
\href{http://arxiv.org/abs/hep-ph/0003021}{{\tt arXiv:hep-ph/0003021}}.

\bibitem{Armesto:2007dt}
N.~Armesto \textit{et al.}, {\em {Medium-evolved fragmentation functions}\/},
  \href{http://dx.doi.org/10.1088/1126-6708/2008/02/048}{JHEP {\bf 0802} (2008)
   048},
\href{http://arxiv.org/abs/0710.3073}{{\tt arXiv:0710.3073}}.

\bibitem{Polosa:2006hb}
A.~D. Polosa and C.~A. Salgado, {\em {Jet Shapes in Opaque Media}\/},
  \href{http://dx.doi.org/10.1103/PhysRevC.75.041901}{Phys.Rev. {\bf C75}
  (2007)  041901},
\href{http://arxiv.org/abs/hep-ph/0607295}{{\tt arXiv:hep-ph/0607295}}.

\bibitem{Zapp:2012nw}
K.~C. Zapp and U.~A. Wiedemann, {\em {Coherent Radiative Parton Energy Loss
  Beyond the BDMPS-Z Limit}\/},
\href{http://arxiv.org/abs/1202.1192}{{\tt arXiv:1202.1192}}.

\bibitem{Armesto:2009fj}
N.~Armesto, L.~Cunqueiro, and C.~A. Salgado, {\em {Q-PYTHIA: A Medium-modified
  implementation of final state radiation}\/},
  \href{http://dx.doi.org/10.1140/epjc/s10052-009-1133-9}{Eur.Phys.J. {\bf C63}
  (2009)  679--690},
\href{http://arxiv.org/abs/0907.1014}{{\tt arXiv:0907.1014}}.

\bibitem{Dainese:2004te}
A.~Dainese, C.~Loizides, and G.~Paic, {\em {Leading-particle suppression in
  high energy nucleus-nucleus collisions}\/},
  \href{http://dx.doi.org/10.1140/epjc/s2004-02077-x}{Eur.Phys.J. {\bf C38}
  (2005)  461--474},
\href{http://arxiv.org/abs/hep-ph/0406201}{{\tt arXiv:hep-ph/0406201}}.

\bibitem{Loizides:2006cs}
C.~Loizides, {\em {High transverse momentum suppression and surface effects in
  Cu+Cu and \AuAu\ collisions within the PQM model}\/},
  \href{http://dx.doi.org/10.1140/epjc/s10052-006-0059-8}{Eur.Phys.J. {\bf C49}
  (2007)  339--345},
\href{http://arxiv.org/abs/hep-ph/0608133}{{\tt arXiv:hep-ph/0608133}}.

\bibitem{Zapp:2008gi}
K.~Zapp \textit{et al.}, {\em {A Monte Carlo Model for 'Jet Quenching'}\/},
  \href{http://dx.doi.org/10.1140/epjc/s10052-009-0941-2}{Eur.Phys.J. {\bf C60}
  (2009)  617--632},
\href{http://arxiv.org/abs/0804.3568}{{\tt arXiv:0804.3568}}.

\bibitem{Lokhtin:2005px}
I.~Lokhtin and A.~Snigirev, {\em {A Model of jet quenching in ultrarelativistic
  heavy ion collisions and high-p(T) hadron spectra at RHIC}\/},
  \href{http://dx.doi.org/10.1140/epjc/s2005-02426-3}{Eur.Phys.J. {\bf C45}
  (2006)  211--217},
\href{http://arxiv.org/abs/hep-ph/0506189}{{\tt arXiv:hep-ph/0506189}}.

\bibitem{Arnold:2000dr}
P.~B. Arnold, G.~D. Moore, and L.~G. Yaffe, {\em {Transport coefficients in
  high temperature gauge theories. 1. Leading log results}\/},  JHEP {\bf 0011}
  (2000)  001,
\href{http://arxiv.org/abs/hep-ph/0010177}{{\tt arXiv:hep-ph/0010177}}.

\bibitem{Arnold:2001ba}
P.~B. Arnold, G.~D. Moore, and L.~G. Yaffe, {\em {Photon emission from
  ultrarelativistic plasmas}\/},  JHEP {\bf 0111} (2001)  057,
\href{http://arxiv.org/abs/hep-ph/0109064}{{\tt arXiv:hep-ph/0109064}}.

\bibitem{Arnold:2001ms}
P.~B. Arnold, G.~D. Moore, and L.~G. Yaffe, {\em {Photon emission from quark
  gluon plasma: Complete leading order results}\/},  JHEP {\bf 0112} (2001)
  009,
\href{http://arxiv.org/abs/hep-ph/0111107}{{\tt arXiv:hep-ph/0111107}}.

\bibitem{Arnold:2002ja}
P.~B. Arnold, G.~D. Moore, and L.~G. Yaffe, {\em {Photon and gluon emission in
  relativistic plasmas}\/},  JHEP {\bf 0206} (2002)  030,
\href{http://arxiv.org/abs/hep-ph/0204343}{{\tt arXiv:hep-ph/0204343}}.

\bibitem{Jeon:2003gi}
S.~Jeon and G.~D. Moore, {\em {Energy loss of leading partons in a thermal QCD
  medium}\/},  \href{http://dx.doi.org/10.1103/PhysRevC.71.034901}{Phys.Rev.
  {\bf C71} (2005)  034901},
\href{http://arxiv.org/abs/hep-ph/0309332}{{\tt arXiv:hep-ph/0309332}}.

\bibitem{Turbide:2005fk}
S.~Turbide, C.~Gale, S.~Jeon, and G.~D. Moore, {\em {Energy loss of leading
  hadrons and direct photon production in evolving quark-gluon plasma}\/},
  \href{http://dx.doi.org/10.1103/PhysRevC.72.014906}{Phys.Rev. {\bf C72}
  (2005)  014906},
\href{http://arxiv.org/abs/hep-ph/0502248}{{\tt arXiv:hep-ph/0502248}}.

\bibitem{MehtarTani:2010ma}
Y.~Mehtar-Tani, C.~A. Salgado, and K.~Tywoniuk, {\em {Anti-angular ordering of
  gluon radiation in QCD media}\/},
  \href{http://dx.doi.org/10.1103/PhysRevLett.106.122002}{Phys.Rev.Lett. {\bf
  106} (2011)  122002},
\href{http://arxiv.org/abs/1009.2965}{{\tt arXiv:1009.2965}}.

\bibitem{Wiedemann:2000tf}
U.~A. Wiedemann, {\em {Jet quenching versus jet enhancement: a quantitative
  study of the BDMPS-Z gluon radiation spectrum}\/},
  \href{http://dx.doi.org/10.1016/S0375-9474(01)00362-1}{Nucl.Phys. {\bf A690}
  (2001)  731--751},
\href{http://arxiv.org/abs/hep-ph/0008241}{{\tt arXiv:hep-ph/0008241}}.

\bibitem{Horowitz:2009eb}
W.~Horowitz and B.~Cole, {\em {Systematic theoretical uncertainties in jet
  quenching due to gluon kinematics}\/},
  \href{http://dx.doi.org/10.1103/PhysRevC.81.024909}{Phys.Rev. {\bf C81}
  (2010)  024909},
\href{http://arxiv.org/abs/0910.1823}{{\tt arXiv:0910.1823}}.

\bibitem{Arnold:2008vd}
P.~B. Arnold and W.~Xiao, {\em {High-energy jet quenching in weakly-coupled
  quark-gluon plasmas}\/},
  \href{http://dx.doi.org/10.1103/PhysRevD.78.125008}{Phys.Rev. {\bf D78}
  (2008)  125008},
\href{http://arxiv.org/abs/0810.1026}{{\tt arXiv:0810.1026}}.

\bibitem{CaronHuot:2008ni}
S.~Caron-Huot, {\em {O(g) plasma effects in jet quenching}\/},
  \href{http://dx.doi.org/10.1103/PhysRevD.79.065039}{Phys.Rev. {\bf D79}
  (2009)  065039},
\href{http://arxiv.org/abs/0811.1603}{{\tt arXiv:0811.1603}}.

\bibitem{Vitev:2008rz}
I.~Vitev, S.~Wicks, and B.~W. Zhang, {\em {A theory of jet shapes and cross
  sections: from hadrons to nuclei}\/},
  \href{http://dx.doi.org/10.1088/1126-6708/2008/11/093}{JHEP {\bf 0811} (2008)
   093},
\href{http://arxiv.org/abs/0810.2807}{{\tt arXiv:0810.2807}}.

\bibitem{Vitev:2009rd}
I.~Vitev and B.~W. Zhang, {\em {Jet tomography of high-energy nucleus-nucleus
  collisions at next-to-leading order}\/},
  \href{http://dx.doi.org/10.1103/PhysRevLett.104.132001}{Phys.Rev.Lett. {\bf
  104} (2010)  132001},
\href{http://arxiv.org/abs/0910.1090}{{\tt arXiv:0910.1090}}.

\bibitem{He:2011pd}
Y.~He, I.~Vitev, and B.~W. Zhang, {\em {Next-to-leading order analysis of
  inclusive jet and di-jet production in heavy ion reactions at the Large
  Hadron Collider}\/},
\href{http://arxiv.org/abs/1105.2566}{{\tt arXiv:1105.2566}}.

\bibitem{Ovanesyan:2011xy}
G.~Ovanesyan and I.~Vitev, {\em {An effective theory for jet propagation in
  dense QCD matter: jet broadening and medium-induced bremsstrahlung}\/},
  \href{http://dx.doi.org/10.1007/JHEP06(2011)080}{JHEP {\bf 1106} (2011)
  080},
\href{http://arxiv.org/abs/1103.1074}{{\tt arXiv:1103.1074}}.

\bibitem{Gubser:2006bz}
S.~S. Gubser, {\em {Drag force in AdS/CFT}\/},
  \href{http://dx.doi.org/10.1103/PhysRevD.74.126005}{Phys.Rev. {\bf D74}
  (2006)  126005},
\href{http://arxiv.org/abs/hep-th/0605182}{{\tt arXiv:hep-th/0605182}}.

\bibitem{Herzog:2006gh}
C.~Herzog, A.~Karch, P.~Kovtun, C.~Kozcaz, and L.~Yaffe, {\em {Energy loss of a
  heavy quark moving through $N=4$ supersymmetric Yang-Mills plasma}\/},
  \href{http://dx.doi.org/10.1088/1126-6708/2006/07/013}{JHEP {\bf 0607} (2006)
   013},
\href{http://arxiv.org/abs/hep-th/0605158}{{\tt arXiv:hep-th/0605158}}.

\bibitem{CasalderreySolana:2006rq}
J.~Casalderrey-Solana and D.~Teaney, {\em {Heavy quark diffusion in strongly
  coupled $N=4$ Yang-Mills}\/},
  \href{http://dx.doi.org/10.1103/PhysRevD.74.085012}{Phys.Rev. {\bf D74}
  (2006)  085012},
\href{http://arxiv.org/abs/hep-ph/0605199}{{\tt arXiv:hep-ph/0605199}}.

\bibitem{CasalderreySolana:2007qw}
J.~Casalderrey-Solana and D.~Teaney, {\em {Transverse Momentum Broadening of a
  Fast Quark in a $N=4$ Yang Mills Plasma}\/},
  \href{http://dx.doi.org/10.1088/1126-6708/2007/04/039}{JHEP {\bf 0704} (2007)
   039},
\href{http://arxiv.org/abs/hep-th/0701123}{{\tt arXiv:hep-th/0701123}}.

\bibitem{Liu:2006ug}
H.~Liu, K.~Rajagopal, and U.~A. Wiedemann, {\em {Calculating the jet quenching
  parameter from AdS/CFT}\/},
  \href{http://dx.doi.org/10.1103/PhysRevLett.97.182301}{Phys.Rev.Lett. {\bf
  97} (2006)  182301},
\href{http://arxiv.org/abs/hep-ph/0605178}{{\tt arXiv:hep-ph/0605178}}.

\bibitem{Liu:2006he}
H.~Liu, K.~Rajagopal, and U.~A. Wiedemann, {\em {Wilson loops in heavy ion
  collisions and their calculation in AdS/CFT}\/},
  \href{http://dx.doi.org/10.1088/1126-6708/2007/03/066}{JHEP {\bf 0703} (2007)
   066},
\href{http://arxiv.org/abs/hep-ph/0612168}{{\tt arXiv:hep-ph/0612168}}.

\bibitem{Armesto:2006zv}
N.~Armesto, J.~D. Edelstein, and J.~Mas, {\em {Jet quenching at finite `t Hooft
  coupling and chemical potential from AdS/CFT}\/},
  \href{http://dx.doi.org/10.1088/1126-6708/2006/09/039}{JHEP {\bf 0609} (2006)
   039},
\href{http://arxiv.org/abs/hep-ph/0606245}{{\tt arXiv:hep-ph/0606245}}.

\bibitem{Abelev:2008ae}
{{STAR}} Collaboration, {\em {Centrality dependence of charged hadron and
  strange hadron elliptic flow from $\sqrtsnn=200$~\GeV\ \AuAu\ collisions}\/},
   \href{http://dx.doi.org/10.1103/PhysRevC.77.054901}{Phys.Rev. {\bf C77}
  (2008)  054901},
\href{http://arxiv.org/abs/0801.3466}{{\tt arXiv:0801.3466}}.

\bibitem{Aamodt:2010pa}
{{ALICE}} Collaboration, {\em {Elliptic flow of charged particles in \PbPb\
  collisions at $2.76$~\TeV}\/},
  \href{http://dx.doi.org/10.1103/PhysRevLett.105.252302}{Phys.Rev.Lett. {\bf
  105} (2010)  252302},
\href{http://arxiv.org/abs/1011.3914}{{\tt arXiv:1011.3914}}.

\bibitem{Song:2011qa}
H.~Song, S.~A. Bass, and U.~Heinz, {\em {Elliptic flow in
  $200\,\mathrm{A}$~\GeV\ \AuAu\ collisions and $2.76\,\mathrm{A} $~\TeV\
  \PbPb\ collisions: insights from viscous hydrodynamics + hadron cascade
  hybrid model}\/},
  \href{http://dx.doi.org/10.1103/PhysRevC.83.054912}{Phys.Rev. {\bf C83}
  (2011)  054912},
\href{http://arxiv.org/abs/1103.2380}{{\tt arXiv:1103.2380}}.

\bibitem{Aad:2012bu}
{ATLAS} Collaboration, {\em {Measurement of the azimuthal anisotropy for
  charged particle production in $\sqrtsnn=2.76$~\TeV\ lead-lead collisions
  with the ATLAS detector}\/},  \href{http://arxiv.org/abs/1203.3087}{{\tt
  arXiv:1203.3087}}.
to appear in Phys.Rev.~\textbf{C}.

\bibitem{Adcox:2001jp}
{PHENIX} Collaboration, {\em {Suppression of hadrons with large transverse
  momentum in central \AuAu\ collisions at $\sqrtsnn = 130$~\GeV}\/},
  \href{http://dx.doi.org/10.1103/PhysRevLett.88.022301}{Phys.Rev.Lett. {\bf
  88} (2002)  022301},
\href{http://arxiv.org/abs/nucl-ex/0109003}{{\tt arXiv:nucl-ex/0109003}}.

\bibitem{Adler:2003qi}
{PHENIX} Collaboration, {\em {Suppressed \pizero\ production at large
  transverse momentum in central \AuAu\ collisions at $\sqrtsnn=200$~\GeV}\/},
  \href{http://dx.doi.org/10.1103/PhysRevLett.91.072301}{Phys.Rev.Lett. {\bf
  91} (2003)  072301},
\href{http://arxiv.org/abs/nucl-ex/0304022}{{\tt arXiv:nucl-ex/0304022}}.

\bibitem{Adler:2002xw}
{STAR} Collaboration, {\em {Centrality dependence of high \pt\ hadron
  suppression in \AuAu\ collisions at $\sqrtsnn = 130$~\GeV}\/},
  \href{http://dx.doi.org/10.1103/PhysRevLett.89.202301}{Phys.Rev.Lett. {\bf
  89} (2002)  202301},
\href{http://arxiv.org/abs/nucl-ex/0206011}{{\tt arXiv:nucl-ex/0206011}}.

\bibitem{Adams:2003kv}
{STAR} Collaboration, {\em {Transverse momentum and collision energy dependence
  of high \pt\ hadron suppression in \AuAu\ collisions at ultrarelativistic
  energies}\/},
  \href{http://dx.doi.org/10.1103/PhysRevLett.91.172302}{Phys.Rev.Lett. {\bf
  91} (2003)  172302},
\href{http://arxiv.org/abs/nucl-ex/0305015}{{\tt arXiv:nucl-ex/0305015}}.

\bibitem{Adler:2002tq}
{STAR} Collaboration, {\em {Disappearance of back-to-back high \pt\ hadron
  correlations in central \AuAu\ collisions at $\sqrtsnn = 200$~\GeV}\/},
  \href{http://dx.doi.org/10.1103/PhysRevLett.90.082302}{Phys.Rev.Lett. {\bf
  90} (2003)  082302},
\href{http://arxiv.org/abs/nucl-ex/0210033}{{\tt arXiv:nucl-ex/0210033}}.

\bibitem{Adams:2005ph}
{STAR} Collaboration, {\em {Distributions of charged hadrons associated with
  high transverse momentum particles in \pp\ and \AuAu\ collisions at
  $\sqrtsnn= 200$~\GeV}\/},
  \href{http://dx.doi.org/10.1103/PhysRevLett.95.152301}{Phys.Rev.Lett. {\bf
  95} (2005)  152301},
\href{http://arxiv.org/abs/nucl-ex/0501016}{{\tt arXiv:nucl-ex/0501016}}.

\bibitem{Adler:2003ii}
{PHENIX} Collaboration, {\em {Absence of suppression in particle production at
  large transverse momentum in $\sqrtsnn=200$~\GeV\ \dAu\ collisions}\/},
  \href{http://dx.doi.org/10.1103/PhysRevLett.91.072303}{Phys.Rev.Lett. {\bf
  91} (2003)  072303},
\href{http://arxiv.org/abs/nucl-ex/0306021}{{\tt arXiv:nucl-ex/0306021}}.

\bibitem{Adams:2003im}
{STAR} Collaboration, {\em {Evidence from \dAu\ measurements for final state
  suppression of high \pt\ hadrons in \AuAu\ collisions at RHIC}\/},
  \href{http://dx.doi.org/10.1103/PhysRevLett.91.072304}{Phys.Rev.Lett. {\bf
  91} (2003)  072304},
\href{http://arxiv.org/abs/nucl-ex/0306024}{{\tt arXiv:nucl-ex/0306024}}.

\bibitem{Adler:2005ig}
{PHENIX} Collaboration, {\em {Centrality dependence of direct photon production
  in $\sqrtsnn=200$~\GeV\ \AuAu\ collisions}\/},
  \href{http://dx.doi.org/10.1103/PhysRevLett.94.232301}{Phys.Rev.Lett. {\bf
  94} (2005)  232301},
\href{http://arxiv.org/abs/nucl-ex/0503003}{{\tt arXiv:nucl-ex/0503003}}.

\bibitem{Reygers:2008pq}
K.~Reygers, {\em {Characteristics of Parton Energy Loss Studied with High-\pt\
  Particle Spectra from PHENIX}\/},
  \href{http://dx.doi.org/10.1088/0954-3899/35/10/104045}{J.Phys.G {\bf G35}
  (2008)  104045},
\href{http://arxiv.org/abs/0804.4562}{{\tt arXiv:0804.4562}}.

\bibitem{Evans:2008zzb}
e.~Evans, Lyndon and e.~Bryant, Philip, {\em {LHC Machine}\/},
\href{http://dx.doi.org/10.1088/1748-0221/3/08/S08001}{JINST {\bf 3} (2008)
  S08001}.

\bibitem{Jackson:1998hs}
G.~P. Jackson, {\em {A dedicated hadronic B-factory: Accelerator
  considerations}\/},
\href{http://dx.doi.org/10.1016/S0168-9002(98)00336-2}{Nucl. Instrum. Meth.
  {\bf A408} (1998)  296--307}.

\bibitem{Bailey:691782}
R.~Bailey and P.~Collier, {\em Standard Filling Schemes for Various LHC
  Operation Modes\/},
\newblock LHC-PROJECT-NOTE-323  , \url{http://cdsweb.cern.ch/record/691782}.

\bibitem{Aad:2008zzm}
{{ATLAS}} Collaboration, {\em {The ATLAS Experiment at the CERN Large Hadron
  Collider}\/},
\href{http://dx.doi.org/10.1088/1748-0221/3/08/S08003}{JINST {\bf 3} (2008)
  S08003}.

\bibitem{Aad:2010zh}
{{ATLAS}} Collaboration, {\em {Drift Time Measurement in the ATLAS Liquid Argon
  Electromagnetic Calorimeter using Cosmic Muons}\/},
  \href{http://dx.doi.org/10.1140/epjc/s10052-010-1403-6}{Eur.Phys.J. {\bf C70}
  (2010)  755--785},
\href{http://arxiv.org/abs/1002.4189}{{\tt arXiv:1002.4189}}.

\bibitem{flowpaper}
{ATLAS} Collaboration, {\em {Measurement of the pseudorapidity and transverse
  momentum dependence of the elliptic flow of charged particles in lead-lead
  collisions at $\sqrtsnn = 2.76$~\TeV\ with the ATLAS detector}\/},
  Phys.Lett. {\bf B707} (2012)  330--348,
\href{http://arxiv.org/abs/1108.6018}{{\tt arXiv:1108.6018}}.

\bibitem{Agostinelli:2002hh}
S.~Agostinelli \textit{et al.}, {\em {GEANT4: A Simulation toolkit}\/},
\href{http://dx.doi.org/10.1016/S0168-9002(03)01368-8}{Nucl.Instrum.Meth. {\bf
  A506} (2003)  250--303}.

\bibitem{Allison:2006ve}
J.~Allison \textit{et al.}, {\em {GEANT4 developments and applications}\/},
\href{http://dx.doi.org/10.1109/TNS.2006.869826}{IEEE Trans.Nucl.Sci. {\bf 53}
  (2006)  270}.

\bibitem{Aad:2010ah}
{ATLAS} Collaboration, {\em {The ATLAS Simulation Infrastructure}\/},
  \href{http://dx.doi.org/10.1140/epjc/s10052-010-1429-9}{Eur.Phys.J. {\bf C70}
  (2010)  823--874},
\href{http://arxiv.org/abs/1005.4568}{{\tt arXiv:1005.4568}}.

\bibitem{Masera:2009zz}
M.~Masera \textit{et al.}, {\em {Anisotropic transverse flow introduction in
  Monte Carlo generators for heavy ion collisions}\/},  Phys.~Rev. {\bf C79}
  (2009)  064909.

\bibitem{Aad:2011he}
{ATLAS} Collaboration, {\em {Jet energy measurement with the ATLAS detector in
  proton-proton collisions at $\sqrts=7$~\TeV}\/},
  \href{http://arxiv.org/abs/1112.6426}{{\tt arXiv:1112.6426}}.
to appear in Eur.Phys.J.~\textbf{C}.

\bibitem{centralitynote}
{ATLAS} Collaboration, {\em Centrality Determination in the 2010 \PbPb\ physics
  data\/},  ATL-COM-PHYS-2011-427 (2011).

\bibitem{ATLAS-CONF-2012-045}
{ATLAS} Collaboration, {\em Transverse energy fluctuations in \PbPb\ collisions
  at $\sqrtsnn=2.76$~\TeV\ with the ATLAS detector at the LHC\/},
  ATLAS-CONF-2012-045 (2012).
\newblock \url{http://cdsweb.cern.ch/record/1440894}.

\bibitem{Assamagan:1368189}
{ATLAS} Collaboration, {\em Overlay for ATLAS Simulation\/},
  ATL-SOFT-INT-2011-001 (2011).

\bibitem{Hocker:1995kb}
A.~Hocker and V.~Kartvelishvili, {\em {SVD approach to data unfolding}\/},
  \href{http://dx.doi.org/10.1016/0168-9002(95)01478-0}{Nucl.~Instrum.~Meth.
  {\bf A372} (1996)  469--481}.

\bibitem{roounfold}
RooUnfold.
\newblock http://hepunx.rl.ac.uk/~adye/software/unfold/RooUnfold.html.

\bibitem{JERUncertaintyProvider}
JetEnergyResolutionProvider.
\newblock
  https://twiki.cern.ch/twiki/bin/view/Main/JetEnergyResolutionProvider.

\bibitem{ATLAS-CONF-2010-054}
{ATLAS} Collaboration, {\em Jet Energy Resolution and Selection Efficiency
  Relative to Track Jets from In-situ Techniques with the ATLAS Detector Using
  Proton-Proton Collisions at a Center of Mass Energy $\sqrts=7$~\TeV\/},
  ATLAS-CONF-2010-054 (2010).
\newblock \url{http://cdsweb.cern.ch/record/1281311}.

\bibitem{Adcox:2004mh}
{PHENIX} Collaboration, {\em {Formation of dense partonic matter in
  relativistic nucleus-nucleus collisions at RHIC: Experimental evaluation by
  the PHENIX collaboration}\/},
  \href{http://dx.doi.org/10.1016/j.nuclphysa.2005.03.086}{Nucl.Phys. {\bf
  A757} (2005)  184--283},
\href{http://arxiv.org/abs/nucl-ex/0410003}{{\tt arXiv:nucl-ex/0410003}}.

\bibitem{Cole:2011zz}
B.~A. Cole, {\em {Jet probes of $\sqrtsnn=2.76$~\TeV\ \PbPb\ collisions with
  the ATLAS detector}\/},
\href{http://dx.doi.org/10.1088/0954-3899/38/12/124021}{J.Phys.G {\bf G38}
  (2011)  124021}.

\bibitem{Angerami:2011is}
A.~Angerami, {\em {Measurement of Jets and Jet Suppression in
  $\sqrtsnn=2.76$~\TeV\ Lead-Lead Collisions with the ATLAS detector at the
  LHC}\/},  \href{http://dx.doi.org/10.1088/0954-3899/38/12/124085}{J.Phys.G
  {\bf G38} (2011)  124085},
\href{http://arxiv.org/abs/1108.5191}{{\tt arXiv:1108.5191}}.

\bibitem{JetQM11}
{ATLAS} Collaboration, {\em Centrality dependence of Jet Yields and Jet
  Fragmentation in Lead-Lead Collisions at $\sqrtsnn = 2.76$~\TeV\ with the
  ATLAS detector at the LHC\/},  ATLAS-CONF-2011-075 (2011).
\newblock \url{http://cdsweb.cern.ch/record/1353220}.

\bibitem{atlas:2010wv}
{ATLAS} Collaboration, {\em {Measurement of inclusive jet and dijet cross
  sections in proton-proton collisions at 7 \TeV\ centre-of-mass energy with
  the ATLAS detector}\/},
  \href{http://dx.doi.org/10.1140/epjc/s10052-010-1512-2}{Eur.Phys.J. {\bf C71}
  (2011)  1512}.

\bibitem{Abreu:2007kv}
e.~Armesto, N., e.~Borghini, N., e.~Jeon, S., e.~Wiedemann, U.A., S.~Abreu,
  \textit{et al.}, {\em {Heavy Ion Collisions at the LHC - Last Call for
  Predictions}\/},
  \href{http://dx.doi.org/10.1088/0954-3899/35/5/054001}{J.Phys.G {\bf G35}
  (2008)  054001},
\href{http://arxiv.org/abs/0711.0974}{{\tt arXiv:0711.0974}}.

\bibitem{CMS:2012aa}
{CMS} Collaboration, {\em {Study of high-\pT\ charged particle suppression in
  \PbPb\ compared to \pp\ collisions at $\sqrtsnn=2.76$~\TeV}\/},
\href{http://arxiv.org/abs/1202.2554}{{\tt arXiv:1202.2554}}.

\bibitem{Aad:2010aa}
{ATLAS} Collaboration, {\em {Measurement of the centrality dependence of \jpsi\
  yields and observation of \Zzero\ production in lead-lead collisions with the
  ATLAS detector at the LHC}\/},
  \href{http://dx.doi.org/10.1016/j.physletb.2011.02.006}{Phys.Lett. {\bf B697}
  (2011)  294--312},
\href{http://arxiv.org/abs/1012.5419}{{\tt arXiv:1012.5419}}.

\bibitem{Chatrchyan:2011ua}
{CMS} Collaboration, {\em {Study of \Zzero\ boson production in \PbPb\
  collisions at nucleon-nucleon centre of mass energy $= 2.76$~\TeV}\/},
  \href{http://dx.doi.org/10.1103/PhysRevLett.106.212301}{Phys.Rev.Lett. {\bf
  106} (2011)  212301},
\href{http://arxiv.org/abs/1102.5435}{{\tt arXiv:1102.5435}}.

\end{thebibliography}\endgroup
\bibliographystyle{atlasnote} 

\end{document}